%% file: Thesis.tex
\documentclass[12pt,twoside,openright]{report}
\include{misc/preamble}

\title{QCD resummation for high-p$_T$ jet shapes at hadron colliders}
\author{Kamel Khelifa-Kerfa}
\date{2012}
	
\usepackage[phd, fancyhdr]{misc/muthesis}
\begin{document}
 
\maketitle

\tableofcontents

\begin{center}
 \textbf{Word count:} 65,000 
\end{center}

\cleardoublepage
\phantomsection
\addcontentsline{toc}{chapter}{\listfigurename}
\listoffigures

\newpage

\cleardoublepage
\phantomsection
\addcontentsline{toc}{chapter}{\listtablename}
\listoftables
%
%

\include{misc/abs}
\declaration

\include{misc/copyright}
\include{misc/ack}
\include{misc/pub}
%
%
    \fancyhead[LO,RE]{\it\nouppercase\leftmark}
\include{ch1/chap1}
    \fancyhead[LO]{\it\nouppercase\rightmark}
    \fancyhead[RE]{\it\nouppercase\leftmark}
\include{ch2/chap2}
\include{ch3/chap3}
\include{ch4/chap4}
\include{ch5/chap5} 
\include{ch6/chap6} 
\include{ch7/chap7} 
\include{conc/conc}
%
\appendix
\input{ch2/ch2app}
\input{ch3/ch3app}

\input{ch5/ch5app}
\input{ch6/ch6app}
\input{ch7/ch7app}
%
%
\singlespacing
\cleardoublepage
\renewcommand{\bibname}{References}
\addcontentsline{toc}{chapter}{References}
\pagestyle{fancy}
\bibliographystyle{misc/JHEP}
\providecommand{\href}[2]{#2}\begingroup\raggedright\endgroup

\end{document}

%% file: misc/preamble.tex
\usepackage[dvips,linktocpage]{hyperref}
\hypersetup{pdfinfo={
		Title={PhDThesis-KK},
		Author={Kamel Khelifa-Kerfa},
		Subject={High energy physics},
		Keywords={QCD, jets},
		CreationDate={2012}
}}
\usepackage{amsmath,amssymb}
\usepackage[dvips]{graphicx}
\usepackage{epsfig}
\usepackage{enumerate}
\usepackage{setspace}
\usepackage{color}
\usepackage{fancyhdr}
\usepackage{rotating}
\usepackage{bbold}
\usepackage{algorithm, algorithmic}
\usepackage{breakurl}
\numberwithin{equation}{chapter}
\numberwithin{figure}{chapter}
\numberwithin{table}{chapter}

\newcommand\jhep[3]   {{\it J. High Energy Phys.\ }{\bf #1} (#2) #3}

\newcommand{\hepph}[1]{hep-ph/#1}

\newcommand{\A}{\alpha}
\newcommand{\B}{\beta}
\newcommand{\g}{\gamma}
\newcommand{\G}{\Gamma}
\newcommand{\om}{\omega}
\newcommand{\Om}{\Omega}
\newcommand{\ld}{\lambda}
\newcommand{\Ld}{\Lambda}
\newcommand{\s}{\sigma}
\newcommand{\Sg}{\Sigma}
\newcommand{\ro}{\rho}
\newcommand{\De}[0]{\Delta}
\newcommand{\de}{\delta}
\newcommand{\ep}{\epsilon}
\newcommand{\vep}{\varepsilon}

\newcommand{\herwig}{{\sc herwig} }
\newcommand{\pythia}{{\sc pythia} }
\newcommand{\ariadne}{{\sc ariadne} }
\newcommand{\sherpa}{{\sc sherpa} }
\newcommand{\event}{{\sc event}{\footnotesize 2} }
\newcommand{\fastjet}{{\sc FastJet} }
\newcommand{\spartyjet}{{\sc SpartyJet} }
\newcommand{\arclus}{{\sc arclus} }

\newcommand{\caesar}{{\sc caesar} }

\newcommand{\eerad}{{\sc eerad}{\footnotesize 3} }
\newcommand{\nlojet}{{\sc nlojet}{\footnotesize++} }
\newcommand{\mcfm}{{\sc mcfm} }

\newcommand{\MSbar}{\overline{\textrm{MS}}}
\newcommand{\MeV}{\text{ MeV}}
\newcommand{\GeV}{\text{ GeV}}
\newcommand{\TeV}{\text{ TeV}}

\newcommand{\NP}{\rm NP}

\newcommand{\as}{\alpha_s}
\newcommand{\asb}{\bar{\alpha}_s}
\newcommand{\aspi}{\frac{\as}{\pi}}
\newcommand{\astpi}{\frac{\as}{2\pi}}
\newcommand{\gs}{g_{s}}
\newcommand{\Lqcd}{\Lambda_{\text{QCD}}}

\newcommand{\akt}{\text{anti--$\rm k_{T} $}}
\newcommand{\ca}{\text{C/A} }
\newcommand{\kT}{\text{$\rm k_{T} $}}
\newcommand{\AKT}{anti-k$_\mathrm{T}$ }

\newcommand{\KT}{k$_\mathrm{T}$ }

\newcommand{\br}[1]{\overline{#1}}
\newcommand{\qqb}{q\,\bar{q}}
\newcommand{\qb}{\bar{q}}
\newcommand{\EE}{e^{+}e^{-}}
\newcommand{\hh}{hadron-hadron\,}

\newcommand{\CF}{\mathrm{C_F}}
\newcommand{\CFsq}{\mathrm{C_F^2}}
\newcommand{\TR}{\mathrm{T_R}}
\newcommand{\TF}{\mathrm{T_F}}

\newcommand{\CA}{\mathrm{C_A}}
\newcommand{\CAsq}{\mathrm{C_A^2} }
\newcommand{\Nc}{\mathrm{N_{c}}}
\newcommand{\Ncsq}{\mathrm{N_c^2}}
\newcommand{\nf}{n_f}

\newcommand{\WP}{W_\mathrm{P}}
\newcommand{\Hg}{H_g}
\newcommand{\Hq}{H_q}

\newcommand{\kt}[1]{k_{#1}}

\newcommand{\pt}[1]{p_{#1}}

\newcommand{\half}{\frac{1}{2}}

\newcommand{\sqs}{\sqrt{s}}

\newcommand{\sh}{\hat{\sigma}}

\newcommand{\ta}{\tau_{a}}
\newcommand{\bta}{\bar{\tau}_a}

\newcommand{\te}{\tau_{E_0}}

\def\rJ{\mathrm{J}}
\newcommand{\EJ}{E_{\mathrm{J}}}

\newcommand{\MJ}{M_{\mathrm{J}}}
\newcommand{\NGL}{\mathrm{NG}}
\def\Eo{E_0}
\def\tauo{\ta_{E_0}}
\def\Rs{R_s}
\def\uv{\hat{\mathbf{n}}}
\def\deta{\Delta\eta}
\def\tauo{\tau_{E_0}}
\def\Lng{L_\mathrm{ng}}
\def\inn{\mathrm{in}}
\def\outt{\mathrm{out}}

\DeclareMathOperator{\Tr}{Tr}

\DeclareMathOperator{\Li}{Li}

\renewcommand{\d}[0]{\textrm{\,d}}

\newcommand{\df}[2]{\frac{\textrm{d}#1}{\textrm{d}#2}}
\newcommand{\be}[0]{\begin{equation}}
\newcommand{\ee}[0]{\end{equation}}

\newcommand{\pa}{\partial}
\newcommand{\inner}[2]{#1\,.\,#2}
\newcommand{\cinner}[2]{\left(#1\,.\,#2\right)}

\newcommand{\eq}[1]{Eq.~\eqref{#1}}
\newcommand{\eqs}[2]{Eqs.~\eqref{#1} and \eqref{#2}}
\newcommand{\eqss}[3]{Eqs.~\eqref{#1}, \eqref{#2}, and \eqref{#3}}
\newcommand{\chap}[1]{Chapter~\ref{#1}}
\renewcommand{\sec}[1]{Sec.~\ref{#1}}
\newcommand{\ssec}[1]{Subsec.~\ref{#1}}
\newcommand{\fig}[1]{Fig.~\ref{#1}}

\newcommand{\app}[1]{Appendix~\ref{#1}}

\newcommand{\nn}{\nonumber}

\newcommand{\ra}{\rightarrow}
\newcommand{\Ra}{\Rightarrow}

\newcommand{\lra}{\leftrightarrow}
\newcommand{\LRa}{\Leftrightarrow}
\newcommand{\Or}{\mathcal{O}}

\newcommand{\mb}[1]{\mathbf{#1}}
\newcommand{\mbb}[1]{\mathbb{#1}}
\newcommand{\mc}[1]{\mathcal{#1}}
\newcommand{\mr}[1]{\mathrm{#1}}
\renewcommand{\Re}[0]{\mathfrak{Re}}
\renewcommand{\Im}[0]{\mathfrak{Im}}
\newcommand{\cSup}[2]{#1^{(#2)}}
\newcommand{\cSub}[2]{#1_{(#2)}}
\newcommand{\cbr}[1]{\left(#1\right)}
\newcommand{\sbr}[1]{\left[#1\right]}
\newcommand{\wt}[1]{\widetilde{#1}}
\newcommand{\fs}{\footnotesize}

\newcommand{\slashed}[2]{#1_#2\!\!\!\!\!\slash}
\newcommand{\Slashed}[3]{#1_{#2#3}\!\!\!\!\!\!\!\slash}
\newcommand{\SLashed}[4]{#1_{#2#3#4}\!\!\!\!\!\!\!\!\!\slash}
\def\Dslash{D\!\!\!\!\slash}

\def\pslash{p\!\!\!\slash}
\def\lslash{\ell\!\!\!\slash}
\def\qslash{q\!\!\!\slash}

\newcommand{\vect}[1]{\mathbf{#1}}
\newcommand{\abs}[1]{\left\lvert #1\right\rvert}
\newcommand{\bra}[1]{\left\langle #1\right\rvert}
\newcommand{\ket}[1]{\left\lvert #1\right\rangle}
\newcommand{\braket}[2]{\big<#1\big|#2\big>}

%% file: misc/abs.tex
\begin{abstract}
Exploiting the substructure of jets observed at the LHC to better understand and interpret the experimental data has recently been a very active area of research. In this thesis we study the substructure of high-$p_T$ QCD jets, which form a background to many new physics searches. In particular, we explore in detail the perturbative distributions of a certain class of observables known as non-global jet shapes. More specifically, we identify and present state-of-the-art calculations, both at fixed-order and to all-orders in the perturbative expansion, of a set of large logarithms known as non-global logarithms. Hitherto, these logarithms have been largely mis-treated, and in many cases ignored, in the literature despite being first pointed out more than a decade ago. Our work has triggered the interest of many groups, particularly Soft and Collinear Effective Theory (SCET) groups, and led to a flurry of papers on non-global logarithms and related issues.  

Although our primary aim is to provide analytical results for hadron-hadron scattering environments, it is theoretically instructive to consider the simpler case of $e^{+} e^{-}$ annihilation. We thus examine, in chapters \ref{ch:EEJetShapes1}, \ref{ch:EEJetShapes2} and \ref{ch:EEJetShapes3}, the the said jet shapes in the latter environment and compute the full next-to-leading logarithmic resummation of the large logarithms present in the distribution for various jet definitions. We exploit the gained experience to extend our calculations to the more complex hadronic environment in chapter \ref{ch:HHJetShapes1}. We provide state-of-the-art resummation of the jet mass observable in vector boson + jet and dijet QCD processes at the LHC up to next-to-leading logarithmic accuracy. The resultant distribution of the former (vector boson + jet) process agrees well, after accounting for hadronisation corrections, with standard Monte Carlo event generators and potential comparisons to data from the LHC will be made soon. 
\end{abstract}

%% file: misc/copyright.tex
\chapter*{Copyright Statement}
\begin{enumerate}
\item[i.]  The author of this thesis (including any appendices and/or
  schedules to this thesis) owns certain copyright or related rights
  in it (the ``Copyright'') and s/he has given The University of
  Manchester certain rights to use such Copyright, including for
  administrative purposes.
\item[ii.] Copies of this thesis, either in full or in extracts and
  whether in hard or electronic copy, may be made \textbf{only} in
  accordance with the Copyright, Designs and Patents Act 1988 (as
  amended) and regulations issued under it or, where appropriate, in
  accordance with licensing agreements which the University has from
  time to time. This page must form part of any such copies made.
\item[iii.] The ownership of certain Copyright, patents, designs,
  trade marks and other intellectual property (the ``Intellectual
  Property'') and any reproductions of copyright works in the thesis,
  for example graphs and tables (``Reproductions''), which may be
  described in this thesis, may not be owned by the author and may be
  owned by third parties. Such Intellectual Property and Reproductions
  cannot and must not be made available for use without the prior
  written permission of the owner(s) of the relevant Intellectual
  Property and/or Reproductions.
\item[iv.] Further information on the conditions under which
  disclosure, publication and commercialisation of this thesis, the
  Copyright and any Intellectual Property and/or Reproductions
  described in it may take place is available in the University IP
  Policy (see \url{http://documents.manchester.ac.uk/DocuInfo.aspx?DocID=487}), in any relevant Thesis restriction declarations deposited in the
   University Library, The University Library's regulations (see
   \url{http://www.manchester.ac.uk/library/aboutus/regulations}) and
   in The University's policy on presentation of Theses.
\end{enumerate}

%% file: misc/ack.tex
\chapter*{Acknowledgements}

First and foremost all praises are due to Almighty {\sc ALLAH}.

I would like to thank my supervisor, Mrinal Dasgupta, for his support and continuous effort throughout the course of my PhD. I am indebted to Yazid Delenda for intriguing discussions, particularly in chapter \ref{ch:EEJetShapes3}, reviewing and proofreading the thesis, and to Simone Marzani for conceptual and technical discussions as well as providing the numerical codes used in chapter \ref{ch:HHJetShapes1}. My thanks extend to Andrea Banfi for assistance in chapter \ref{ch:EEJetShapes2} and Mike Seymour for aid with the numerical program \event as well as helpful feedback on chapter \ref{ch:EEJetShapes2}. I would also like to thank Michael Spannowsky for collaborating on the work presented in chapter \ref{ch:HHJetShapes1}, and in particular for providing the results of the Monte Carlo event generators. The program of Ref. \cite{Dasgupta:2001sh} with the implementation of the \KT clustering in \cite{Delenda:2006nf}, which has been heavily used in this thesis, has not been made public yet and for that 
reason I would like to thank the authors, Mrinal Dasgupta, Gavin P. Salam and Andrea Banfi, for approving its usage in this thesis.

The encouragement of Apostolos Pilaftsis, my supervisor for the MPhys project, is highly appreciated, and so is the help of Jeff Forshaw and Fred Loebinger regarding administrative issues.

Moreover, I would like to express my gratitude to my brother Abdelkader for the enormous support he offered me throughout my educational career.

The work presented in this thesis has been sponsored by both the University of Manchester and the Algerian government. I would like to express my gratitude to the Faculty of Engineering and Physical Sciences, the ministry of higher education and scientific research and the Algerian consulate in London for their financial and administrative support during my stay in the United Kingdom.

%% file: misc/pub.tex
\chapter*{List of Publications}

\begin{itemize}

\item Y.~Delenda and K.~Khelifa-Kerfa, \emph{On the resummation of clustering logarithms for non-global observables}, \jhep{09}{2012}{109}  [\hepph{1207.4528}].

\item M.~Dasgupta, K.~Khelifa-Kerfa, S.~Marzani, M.~Spannowsky, \emph{On jet mass distributions in Z+jet and dijet processes at the LHC}, \jhep{10}{2012}{126} [\hepph{1207.1640}].

\item K.~Khelifa-Kerfa, \emph{Non-global logs and clustering impact on jet mass with a jet veto distribution}, \jhep{02}{2012}{072} [\hepph{1111.2016}]

\item A.~Banfi, M.~Dasgupta, K.~Khelifa-Kerfa and S.~Marzani, \emph{Non-global  logarithms and jet algorithms in high-pT jet shapes}, \jhep{08}{2010}{064} [\hepph{1004.3483}].
\end{itemize}

%% file: ch1/chap1.tex

\chapter{Introduction}
\label{ch:Intro}

With the two Large Hadron Collider (LHC) experiments, ATLAS and CMS, running and
producing a proliferation of valuable data, a new era of physics is being
unfolded. Already at a centre--of--mass energy $7\TeV$, the Standard Model (SM)
of particle physics has been ``re--discovered'' \cite{Herten:2011qx}. The
priority list of the LHC programme includes a plethora of long--standing
problems in High Energy Physics (HEP) and related physics domains. Perhaps,
at the top of the list is the search for yet the only unobserved SM particle;
the Higgs\footnote{Both ATLAS and CMS have recently reported the observation of
a Higgs--like particle at a mass of $125 \GeV$ \cite{Aad:2012gk,
Chatrchyan:2012gu}. Investigations on its properties, such as spin, are still
ongoing.}. The Beyond Standard Model (BSM) searches, expected to reach their
full potential as the machine hits its designed energy $14\TeV$, cover a variety
of fundamental ideas, such as SuperSymmetery (SUSY), Extra Dimensions (ED), Dark
Matter (DM), Dark Energy (DE) etc.

Whilst the prospects are high the challenges are alike. New particles are
generally expected to be heavy and likely to have too short of a lifetime to be ever directly detected before they decay. The corresponding decay products need be both stable and interact strongly enough to leave any signature at the detectors\footnote{Weakly interacting particles such as neutrinos escape direct detection and are treated as missing energy.}. The last two conditions are generic features of SM particles, which will thus form the bulk of the final state products. Therefore, for a proton--proton collision at the LHC the final state particles could have originated from either an ``ordinary'' SM process (\emph{background}) or a ``more interesting'' BSM process (\emph{signal}). A reliable discovery is only feasible once a detailed understanding of both processes is established.

\begin{figure}[!t]
  \begin{center}
  \includegraphics[width=0.7\textwidth]{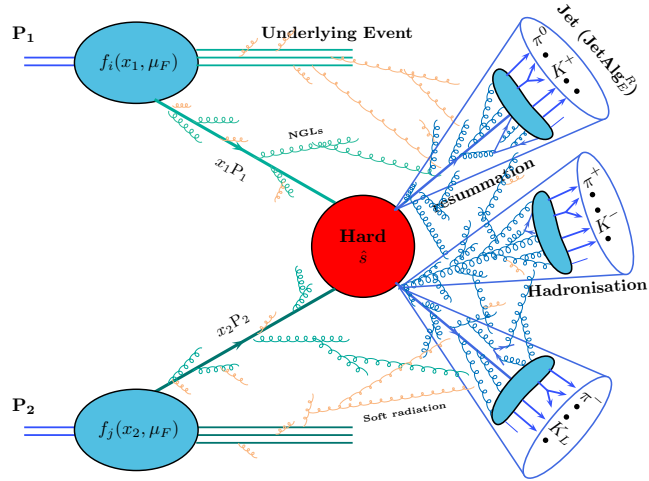}
  \caption{A typical LHC multijet event with associated various perturbative and non--perturbative effects. Former effects include: hard scattering subprocess, soft and collinear radiation (resummation), soft and wide--angle radiation, non-global logs (NGLs), effects of jet algorithms (JetAlg) with jet radius $R$ and recombination scheme $E$. Latter effects include: parton densities ($f(x,\mu_F)$), underlying event (UE) and hadronisation. This thesis concerns perturbative effects.}
 \label{fig:Intro:JetEvent}
 \end{center}  
\end{figure}
At energies as high as those probed by the LHC, many interesting physics signals will proceed via interactions involving low--$x$ partons\footnote{where $x$ is the fraction of longitudinal momentum of initial protons.}, which are predominantly gluons and sea quarks. As such they are characterised by both enormous phase space and high radiation activity (due to the large colour charge of gluons). Moreover, a significant fraction of those signals, which are challenging to measure, will involve in their decay chain production of quarks and gluons. The latter partons will emit radiation down to a characteristic scale, $\Lambda_{\NP}$,\footnote{For Quantum Chromo--Dynamics (QCD) this scale is $\Lambda_{\NP} \equiv \Lambda_{\rm QCD}^{\MSbar} \sim 200\MeV$ \cite{ellis2003qcd}.} before hadronising into sprays of nearly collimated hadrons, or \emph{jets}. \fig{fig:Intro:JetEvent} illustrates a typical multijet event at the LHC. In fact, nearly all final states expected in proton-proton collisions at $14 \TeV$ 
will involve production of jets, and for most channels they consist the dominant part of the detectable signal. On the other hand, jets in SM processes, such as QCD $2\to 2$ jets and Z/W + jets, will be produced with very high rates \cite{Ellis:2007ib} and may thus swamp any signal with similar signature. It is therefore necessary, particularly at the LHC more than any previous colliders, to improve our understanding and use of jets.

Although inclusive jet cross-sections fall off with transverse momentum $p_T$ \cite{Ellis:2007ib,Ellis:1988hv,Ellis:1989vm,Ellis:1990ek,Ellis:1992en}, high-$p_T$ jets have distinct features that compensate for such smallness and thus offer an alternative promising channel for discovery. For instance, there is a great possibility for the production of heavy BSM particles that could subsequently decay to ``boosted'' lighter SM particles, such as W, Z, top, Higgs (H) etc. The latter particles, having transverse momenta that far exceed their rest masses, would then decay to even lighter particles that are collimated and thus more likely to be clustered by jet algorithms into a single ``fat'' jet. This is due to the fact that the angular distance between decay products, $b$ and $c$, of a particle $a$ is proportional to the inverse of the transverse momentum of $a$. Precisely
\be
 \De R_{bc}^2 = \frac{m_a^2}{p_{Tb}\, p_{Tc}} = \frac{m_a^2}{z(1-z) p_{Ta}^2},
\ee
where $z$ is the momentum fraction of, say, particle $b$. Thus for $p_{Ta}^2 \gg m_a^2$, and given that the decay is not too asymmetric, i.e., $z$ neither too close to $0$ nor $1$, $\De R_{bc}$ will be small and $bc$ will end up in a single jet. Moreover, owing to the fact that non--perturbative corrections, particularly hadronisation, fall off with $p_T$ \cite{Dokshitzer:1995zt, Dokshitzer:1995qm}, high-$p_T$ jets are less sensitive to such corrections, making them ideal for clean perturbative investigations.

\begin{figure}[!t]
 \begin{center}
 \psfig{file=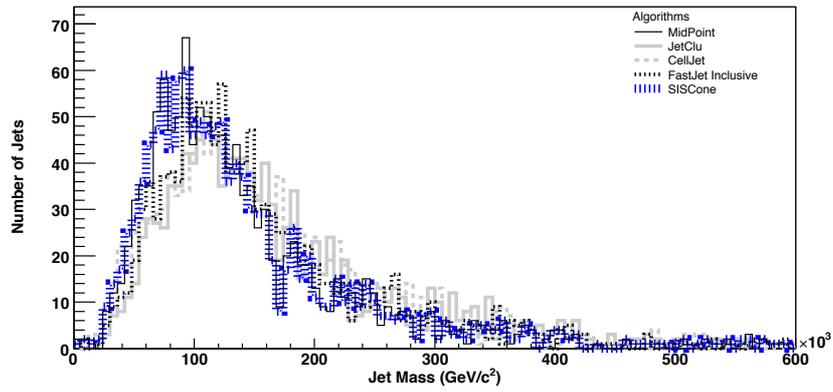, width=0.72 \textwidth}
 \caption{Jet mass distribution for an inclusive QCD jet sample generated for the LHC with $p_{T,\min}$ for the hard scattering of $2\TeV$. Jets are defined using different jet algorithms with jet radius of $0.7$. The peak is around $125 \GeV$ (in natural units, $c=\hbar=1$) (figure from \cite{Ellis:2007ib}).}
 \label{fig:Intro:QCDJetMassDist}
 \end{center}
\end{figure}

The above mentioned high-$p_T$ (boosted) signal jets have \emph{shape} and \emph{substructure} that are distinct from those of plain QCD jets initiated by light quarks and gluons. Exploring such a rich substructure may prove very useful in situations such as that depicted in \fig{fig:Intro:QCDJetMassDist}, where QCD events may have jet mass distributions that peak around $125 \GeV$ (the current experimentally measured Higgs' mass) and thus form a strong background to Higgs searches based on the use of jet masses. Such very high-$p_T$ jet events are not uncommon at the LHC given the high momentum probed and hence pose a great challenge. Therefore new and more powerful jet substructure techniques may offer an indispensable tool in boosted signal searches (see e.g \cite{Butterworth:2008iy, Agashe:2006hk, Lillie:2007yh, Agashe:2007ki, Agashe:2007zd, Brooijmans:2008se, Brooijmans:1077731, Butterworth:2002tt}). Note that the peak region in \fig{fig:Intro:QCDJetMassDist} is well within the perturbative domain of 
QCD and \emph{analytical ``resummation''} calculations are very efficient, as we shall be illustrating in this thesis.
The aforementioned substructure techniques may be divided into two broad classes: grooming techniques and jet shapes.

Jets, the footprint of the underlying QCD partons, are not intrinsically
well--defined, and they thus ought to be ``defined'' before they can be used. A
modern jet definition consists of a jet algorithm and a recombination scheme
\cite{Flaugher:1990rv, PhysRevLett.39.1436, springerlink:10.1007/BF01410449,
Bethke1988235}. We discuss these in more details in Chapter \ref{ch:Jets}.
Grooming techniques, reviewed for instance in \cite{Abdesselam:2010pt, Altheimer:2012mn}, are based on jet algorithms and primarily utilised to
identify \emph{subjets}, within a fat jet, as well as to mitigate isolated
diffuse soft radiation coming from the underlying event (UE) and pile--up (PU)\footnote{Underlying event refers to multiple soft/semi-hard parton--parton interactions while pile--up refers to multiple soft/semi-hard proton--proton interactions in the same bunch crossing.}. Filtering \cite{Butterworth:2008iy}, pruning \cite{Ellis:2009su, Ellis:2009me} and trimming \cite{Krohn:2009th} are amongst the widely used jet grooming algorithms. Applied to final state ``ungroomed'' jets, these algorithms produce groomed jets with greatly enhanced features, such as high mass resolution. In general, such tools and techniques have been shown to significantly reduce contaminations from the aforementioned sources and, in many cases, recovering otherwise buried search channels \cite{Butterworth:2008iy, Altheimer:2012mn}.
We shall briefly revisit this topic in Chapter \ref{ch:Jets}. 
%
%

Jet shapes are event shape--like observables\footnote{See e.g., \cite{ellis2003qcd, Banfi:2004yd, Banfi:2010xy} and
\href{http://home.fnal.gov/~zanderi/Caesar/caesar.html}{CAESAR homepage} (\url{http://home.fnal.gov/~zanderi/Caesar/caesar.html}) for a list of some of the most extensively studied event shapes.} that measure the energy flow within individual jets in an event. A comprehensive list of recently introduced jet shapes may be found in e.g., \cite{Abdesselam:2010pt, Altheimer:2012mn}. The field of jet shapes, and more generally jet substructure, has seen an intensive activity over the very recent years from both theory and experimental groups, with a wealth of impressive measurements coming from the LHC, e.g., \cite{Altheimer:2012mn, ATLAS:2012am, Aad:2012jf, Hinzmann:2012zz}. Substructure techniques have been used in various applications such as tagging jets from decay of boosted particles, particularly in the Higgs and SUSY searches \cite{Butterworth:2008iy,ATL-PHYS-PUB-2009-088, Plehn:2010st}. In this thesis we rather focus on establishing a better understanding of QCD jets through jet shapes, which would allow 
for an efficient background subtraction. Specifically we 
study two observables: angularities \cite{Berger:2002ig, Berger:2004xf} and jet mass, in typical QCD processes whereby final states are clustered using a variety of infrared and collinear (IRC) safe jet algorithms\footnote{More on IRC safe jet algorithms in Chapter \ref{ch:Jets}}.

In fact, much emphasis will be given to jet mass calculations. This is due to a number of reasons. First, the jet mass is the simplest, yet phenomenologically the most useful, shape observable to employ in the search for decay of heavy particles \cite{Ellis:2007ib, Altheimer:2012mn}. The first clue to a jet's identity is its mass, which is directly related to the mass of the particle initiating it. Second, it has played a central role in the analytical \emph{resummation} programme of numerous observables, including the most intensively studied ones such as thrust~\cite{Catani:1992ua}. Further, there has been a flurry of work on jet masses in the recent years, both within traditional QCD and some of its modern effective approximations such as soft and collinear effective theory (SCET \cite{Bauer:2000yr, Bauer:2001yt, Beneke:2002ph}) \cite{Ellis:2009wj,Ellis:2010rwa, Becher:2008cf,Cheung:2009sg, Jouttenus:2009ns,Chien:2010kc,Kelley:2011tj, Kelley:2011aa, Jouttenus:2011wh, Li:2011hy, Li:2012bw}. Once a firm 
understanding of jet mass distributions, resummation and other related issues is established, it will provide a natural setting for similar work on other jet shapes and jet substructure in general. However, such a task is delicate, especially in complex environments like those typical at hadron colliders. As a matter of fact, recent papers that addressed the latter task, such as Refs. \cite{Li:2011hy, Li:2012bw}, have omitted important contributions, as shall be explained in Chapter \ref{ch:HHJetShapes1}. In this thesis, we set-out to accomplish the said task by carrying out careful calculations of the many complex aspects involved. We focus on perturbative calculations of jet shape \emph{distributions} in QCD.

There are two complementary theoretical tools to study shape distributions:
analytical calculations and numerical Monte Carlo (MC) simulations. The former
are inclusive over the final state and in most cases only deal with a single
observable at a time. The calculations are performed in the perturbative region
of QCD and have so far only treated a limited set of observables (see for
instance \cite{Banfi:2004yd, Banfi:2001bz} and references therein). Numerical
simulations, on the other side, are completely exclusive over the final state
and thus any number of observables can be measured (simultaneously). In addition
to the perturbative region, MC \emph{event generators} (cf. fixed--order
Monte Carlos discussed below) probe the non--perturbative region\footnote{The perturbative and non--perturbative regions of QCD correspond to kinematical regions where the strong coupling constant is small, $\as \ll 1$, and large, $\as\sim 1$, respectively.} of QCD through phenomenological models that parametrise the effects  of processes such as hadronisation, UE and PU, in terms of a number tunable parameters. The state--of--the--art is that analytical calculations, both at fixed-- and all--orders (resummation), often have higher precision than event generators\footnote{This stems from the fact that the formal precision of event generators is leading logarithm (LL) accuracy. No higher accuracy is guaranteed beyond that due to higher order and non--perturbative contaminations.}. As such, they can play a crucial role in advancing the development of these generators as well as providing an indispensable tool for the interpretation of new physics signals, should they show up at the LHC. Moreover, analytical 
calculations are cleaner, offer a deeper insight into the dynamics of perturbative QCD and can systematically be improved. For these, and other reasons, the central focus in this thesis is on \emph{analytical} calculations of jet shape distributions. Comparisons of the analytical findings to the output of various MCs 
are, whenever possible, provided though.

Perturbative analytical calculations of a generic jet shape $v$ of high--$\pt{T}$ QCD jets may be divided into two regimes: small $v$ and large $v$. In the large $v$ region, $v \lesssim 1$, fixed--order calculations are sufficient to capture the full features of the measured jet shape distribution. Fixed--order predictions for an arbitrary IRC safe observable may be obtained with the aid of fixed--order Monte Carlos (FOMCs), such as \cite{Catani:1996jh, Nason:2004rx, Nagy:2001fj}. The small $v$ region, $v \ll 1$, on the other hand, receives large enhancements (usually logarithmic) in ratios of the energy scales present in the process. In an environment such that of the LHC (\fig{fig:Intro:JetEvent}), the presence of many scales is common for the majority of events. Such scales involve, for instance: jet $\pt{T}$, the measured value of the jet shape, jet radius $R$ and other parameters of the jet algorithm, non--perturbative scale $\Lqcd$, and any other kinematical and selection cuts. The said large 
logarithmic terms render fixed--order calculations unreliable and, more seriously, threaten the validity of the perturbative expansion. To restore the latter expansion, a summation of these logarithmic enhancements to all--orders is inevitable. Such a task is dubbed \emph{resummation}, which may be regarded as a consequence of more profound characteristic features of QCD \cite{Sterman:1995fz}. Namely coherence \cite{Gribov19831, Bassetto1983201} (and factorisation \cite{Collins:1985ue, Collins:1989gx, Collins:1988ig}). The small $v$ regime is of prime interest, in this thesis, and will thus form the bulk of the discussion presented herein.

In perturbative QCD (pQCD), resummation of large logarithms that occur in the
distribution of sufficiently inclusive shape observables, or \emph{global}
observables, has become a typical exercise in modern phenomenology \cite{Banfi:2004nk, Dasgupta:2003iq}. For many such observables, the corresponding resummation has even been automated \cite{Banfi:2004yd, Banfi:2001bz}. Resummation is most effective when the logarithms \emph{exponentiate}. This exponentiation, of jet shapes, follows primarily from the corresponding (exponentiation) property of QCD matrix--elements in the soft and collinear regions \cite{Catani:1992ua, Dokshitzer1979234, Parisi1979427}. If we define the \emph{jet shape fraction} (or integrated shape cross-section) $\Sg (v)$ as
%
\be
 \Sg(v) = \int^{v}_{0} \d v'\,\frac{1}{\cSup{\s}{0}} \df{\s}{v'},
\label{eq:def:shape_fraction}
\ee
where $\cSup{\s}{0}$ is the Born cross-section, then $\Sg(v)$ exponentiates\footnote{Not all shape observables exponentiate though. Examples of observables that have been proven inconsistent with exponentiation include the JADE jet--resolution thresholds \cite{springerlink:10.1007/BF01410449}. For such type of observables even an  LL resummation does not exist \cite{Banfi:2004yd, Brown1990657}.} precisely means that it assumes the following form at small $v$
\be
\Sg(v) = C(\as) \exp\sbr{L\,g_{1}(\as L) + g_{2}(\as L) + \as g_{3}(\as L) + \cdots} + D(\as, v),
\label{eq:Intro:exponentiation}
\ee 
with $L = \ln(1/v)$, $C(\as)$ sums the loop constants and the remainder function $D(\as,v)$ vanishes in the limit $v \ra 0$. The function $g_{1}(\as L)$ resums leading logarithms (LL), $\as^{n}\,L^{n+1}$, $g_{2}(\as L)$ resums next--to--leading logarithms (NLL), $\as^{n}\,L^{n}$, $g_{3}(\as L)$ resums next--to--next--to--leading logarithms (NNLL), $\as^{n}\,L^{n-1}$, and so on. Moreover, one must establish, in addition to the matrix elements, that the phase space defining the shape fraction for a particular shape variable factorises (and hence exponentiate), possibly after a suitable integral transformation \cite{ellis2003qcd}. We elaborate on this in more detail in Chapter \ref{ch:Jets} (\sec{ssec:Jets:Resummation}). For the aforementioned global observables, $g_{1}$ is simply given by the exponentiation of the single soft gluon emission result (known as ``Sudakov'' form factor). Further, the resummation of subleading logarithms. i.e., calculation of $g_{2}, g_{3}, \cdots$ can in principle be achieved 
through, for example, inclusion of the running coupling at higher loops, proper treatment of multiple emissions etc \cite{Banfi:2004yd}. The current state--of--the--art is N$^3$LL resummation for thrust and other event/jet shapes \cite{Becher:2008cf,Chien:2010kc}.

Nearly every measurement\footnote{Not all of them require resummation though. Often fixed--order calculations suffice.} at the LHC will involve \emph{non--global} observables \cite{Dasgupta:2001sh, Dasgupta:2002dc}. Such observables are sensitive only to specific regions of phase space. e.g., the invariant jet mass is only sensitive to radiation into the jet region. Restricting the jet mass observable in the latter region to be less than a specific value, say $\rm M_{J}$, induces large \emph{non--global} (single) logarithms (NGLs) in the ratio $\pt{T}^{2} R^{2}/\rm M_{J}^{2}$, with $R$ being the jet size determined by the jet algorithm used. The resummation of these logarithms, which contribute to $g_{2}$ and higher functions ($g_{3}, g_{4}, \rm etc$) and start at $\Or(\as^{2})$ in the perturbative expansion, have not been possible analytically due to the non--factorisation of the \emph{non--abelian} part of the multiple gluon emission matrix--element. 
Nonetheless, two alternative methods have been developed to resum NGLs. Namely: a Monte Carlo program \cite{Dasgupta:2001sh} and a non--linear evolution equation \cite{Banfi:2002hw}. Their formal accuracy is leading NGLs, which is NLL for the jet mass, and are only valid in the large-$\Nc$ limit\footnote{$\Nc$ is the number of colour degrees of freedom. For $\rm SU(3)$, the symmetry group of QCD, $\Nc = 3$.}. Neglected subleading colour corrections contribute at the $10\%$ level. Phenomenological impact of NGLs have been shown to lead to a reduction in the peak of the corresponding Sudakov form factor that can be significant for some observables \cite{Dasgupta:2001sh, Dasgupta:2002dc, Banfi:2003jj}. Therefore, any reliable resummation of non--global observables aiming at NLL accuracy, or beyond, should necessarily account for NGLs.

Jet clustering algorithms, other than cone--type \cite{Flaugher:1990rv} and
\AKT \cite{Cacciari:2008gp} algorithms, when applied to final state partons lead to a more complicated phase space due to the non--trivial role of clustering amongst soft partons. Generally the resultant phase space is not factorisable (into single particle phase space for each parton). Performing the integral in \eq{eq:def:shape_fraction} with modified phase space due to a jet algorithm yields two effects. First, new large single logarithms in the abelian sector. These are termed \emph{clustering} logarithms (CLs) and were first pointed out in \cite{Banfi:2005gj} for the interjet energy flow distribution. Second, a reduction in the impact of NGLs, as was shown in \cite{Appleby:2002ke} for the same distribution. Since the phase space becomes increasingly complex as the number of final state partons increases, analytical  resummation of CLs seems to be highly intricate. The MC program of Ref. \cite{Dasgupta:2001sh} is at the moment the only available method to resum both CLs and NGLs in the presence of a jet 
algorithm. 
Nevertheless, it has been shown, through explicit fixed--order calculations up to $\Or(\as^{4})$ in the perturbative expansion for the interjet energy flow distribution in $\EE$ annihilation, that CLs exponentiate \cite{Delenda:2006nf}. The analytical results agreed well with the output of the said MC program, indicating that the all--orders result is dominated by the first few orders. We carry out analogous calculations for the jet mass in Chapter \ref{ch:EEJetShapes3}.

As previously mentioned, resummation calculations are accurate for the dominant part of the shape distribution, specifically in the small $v$ (peak) region. For phenomenological analyses, including comparison to measurements, other contributions and corrections ought to be taken into account. These include (a) fixed--order calculations which can, as stated above, reliably reproduce the tail of the jet shape distribution (large $v$ region). Hence, a \emph{matching} of the two, resummation and fixed--order, calculations is necessary for an accurate prediction over the full kinematical range of the jet shape $v$; (b) non--perturbative (NP) corrections, such as hadronisation, UE and PU, which affect the small $v$ region of the shape distribution. The latter are non--calculable in pQCD and at present can only be quantified through models, best implemented in Monte Carlo event generators (see \cite{Buckley:2011ms} for a review). Meanwhile, theoretical progress on estimating the impact of non--perturbative 
corrections using perturbative methods such as renormalon--inspired techniques and related approaches \cite{Dokshitzer:1995zt, Dokshitzer:1995qm, Gardi:2001di, Ball:1995ni} have yielded promising results \cite{Dasgupta:2003iq, Beneke:1998ui, Dasgupta:2007wa, Banfi:2001aq, Dasgupta:2007hr}. Typically, these corrections, particularly hadronisation, may be included as a shift, $v \ra v + \langle\de v\rangle_{\rm NP}$ where $\langle\de v\rangle$ is the mean of the change in $v$, in the jet shape distribution\footnote{Strictly speaking, the shift is only valid to the right of the distribution peak, as shall be discussed in Chapter \ref{ch:Jets}.} \cite{Dasgupta:2002dc, Dokshitzer:1997ew}.

This thesis is divided into three main parts: physics background, results from $\EE$ annihilation and results from \hh collisions. The first part includes Chapters \ref{ch:QCDReview} and \ref{ch:Jets}. Chapter \ref{ch:QCDReview} surveys the theoretical basis of QCD and serves as a background to the rest of the thesis. Important ingredients which are essential to later discussions such as factorisation and matrix--elements are provided. Most noticeably, the \emph{eikonal} approximation is explained and the corresponding Feynman rules explicitly given. Since our pivotal object in this thesis are jets, Chapter \ref{ch:Jets} is devoted to theoretical exploration of jets, jet algorithms and jet shapes. A generic formulation of the latter is presented and related topics such as IRC safety addressed. At this point, the main conceptual basis is laid down and the computational machinery is set up.

The actual calculations are then carried out in the other two parts. Whilst our aim is to produce analytical results that can be compared to present hadron collider measurements, it is theoretically instructive to start with the simpler and cleaner $\EE$ annihilation environment. Indeed, the latter calculations form a natural setting for hadron collision calculations, as we shall see. Consequently, the second part, which includes Chapters \ref{ch:EEJetShapes1}, \ref{ch:EEJetShapes2} and \ref{ch:EEJetShapes3}, concerns jet shapes, specifically angularities, in chapter \ref{ch:EEJetShapes1}, and the invariant jet mass, in the remaining two chapters \ref{ch:EEJetShapes2} and \ref{ch:EEJetShapes3}, in $\EE$ annihilation. Notice that Chapter \ref{ch:EEJetShapes1} is similar in content to \cite{Banfi:2010pa} but treats the more general angularities shape observable. Chapter \ref{ch:EEJetShapes2} largely expands \cite{KhelifaKerfa:2011zu} and Chapter \ref{ch:EEJetShapes3} is simply drawn from \cite{Delenda:2012mm}.
The final part, which includes Chapter \ref{ch:HHJetShapes1}, extends the $\EE$ annihilation findings to \hh collisions, particularly addressing Z+jet and dijet processes at the LHC. Its content is essentially the same as \cite{Dasgupta:2012hg}. Once a next-to-leading order matching is performed, the latter will represent the state-of-the-art calculations for non--global observables at hadron colliders. 

We consider final state clustering by various jet algorithms in $\EE$ annihilation, and only one jet algorithm in \hh scattering. For each jet algorithm, we perform the following:
\begin{itemize}
 \item Compute the full leading--order shape distribution accounting for soft wide--angle radiation with full jet-radius and colour dependence in both $\EE$ and \hh scattering.

 \item Compute the full next--to--leading order (two-gluon emission) shape distribution up to NNLL in the \emph{expansion} (equivalent to $\as^2\,L^2$), with full jet-radius and colour dependence. The two-gluon emission calculations are generally performed within the eikonal framework. The latter is sufficient to achieve the said NNLL accuracy in both $\EE$ and \hh scattering.

 \item In $\EE$ annihilation, we additionally compute the full N$^3$LL terms in the expansion (equivalent to $\as^2\,L$), for which a more accurate emission amplitude is computed in \app{app:QCDReview}, and hence the only missing piece to obtain the full two-gluon distribution is the two--loop constant $C_2$. Our calculations represent the state-of-the-art, with full jet-radius dependence of both NGLs and CLs coefficients.
 
 \item Compute the full NLL resummation including numerical estimates of NGLs and CLs (only in the $\EE$ case) to all--orders with full jet-radius dependence in the large-$\Nc$ limit.
 
\item In the hadron-hadron scattering case we perform matching to (leading) fixed--order, estimate non--perturbative hadronisation corrections and compare to MC event generators.
\end{itemize}
Finally, in Chapter \ref{ch:conc} we draw our conclusions and highlight areas where more work is needed in light of upcoming LHC measurements.

In the next chapter, we begin with a review of QCD phenomenology.

%% file: ch2/chap2.tex

\chapter{QCD phenomenology}
\label{ch:QCDReview}

The contemporary Standard Model for elementary particles and forces encompasses three, out of four, fundamental forces of nature: Electromagnetic, Weak and strong forces. These interactions are elegantly described by the mathematically consistent gauge field theory (GFT)\footnote{Quantum field theory with the requirement of gauge invariance.}. Within this framework, Quantum Electrodynamics (QED) describes electromagnetic interactions, Electro--Weak (EW) describes the unified electromagnetic and weak interactions and Quantum Chromodynamics (QCD) describes the strong interactions. In this thesis we are concerned with QCD. The latter theory is formulated,  at the Lagrangian level, in terms of quarks and gluons. Quarks are described in GFT by the Dirac spinor fields and can exist in three different states of colour (usually referred to as red, green and blue) and six flavours; up (u), down (d), strange (s), charm (c), bottom (b) and top (t). Their electric charges are, in units of the electron charge ($e$): $+2/
3$ (u,c,t) and $-1/3$ (d,s,b). One striking difference between QED and QCD is that quarks have never been seen in isolation, i.e., as asymptotic states, as have electrons and photons. This QCD phenomenon is known as \emph{confinement}. What is instead observed at detectors are hadrons, both baryons and mesons. The former are composites of three spin--$\frac{1}{2}$ quarks, $q q' q''$, while the latter are composites of spin--$\frac{1}{2}$ quark--antiquark, $q \br{q}'$, pairs\footnote{The idea that hadrons are not themselves elementary but bound states of other more elementary states dates back to the early work of Fermi and Yang \cite{Fermi19491739}.}. 

Gluons, in analogy to photons, are vector fields which arise as a result of demanding the quark Lagrangian be invariant under local, or \emph{gauge}, rotations in the colour space. The corresponding mathematical group of such rotations for QCD is the special unitary group $\rm SU(3)$ (see below). Another sharp contrast between QED and QCD is the fact that gluons, which mediate the strong interaction, are colourful, i.e., they are charged. Consequently they can interact among themselves (in addition to their interactions with quarks).

In this chapter, we begin by briefly reviewing the main features of the general symmetry group $\rm SU(\Nc)$, of which $\rm SU(3)$ is a special case, on the way recalling some of the important relations and identities which will be useful in later sections (and chapters). Most of the material presented in this section is based on the references listed at the end of this introductory section. After that we write the general form of the QCD Lagrangian and discuss the various terms in it. To make contact with experiments, it is necessary that one is able to extract predictions from the latter abstract Lagrangian. At small coupling, $\gs$, this is achieved through perturbation theory, equipped with some essential features of QCD such as factorisation, to which we turn in \sec{sec:QCD:Running}. Calculations in perturbative QCD are carried out (in the traditional method) via the use of Feynman diagrams and Feynman rules. An important result of perturbation theory is the running of the coupling $\gs$. Unlike QED, 
the QCD strong coupling decreases as the energy scale at which it is measured increases. Therefore at small and medium energies, the coupling approaches unity and perturbation theory breaks down. Furthermore, when considering soft gluons dynamics, one may utilise a practically simpler effective theory known as \emph{eikonal} theory \cite{PhysRev.186.1656}. We elaborate on this, including the eikonal version of Feynman rules, in \app{sec:app:QCD:EikonalApprox}. Finally, in \app{sec:app:QCD:ee_example}, we present detailed perturbative calculations of $\EE$ annihilation into hadrons which will form the backbone of Chapters \ref{ch:EEJetShapes1}, \ref{ch:EEJetShapes2} and \ref{ch:EEJetShapes3}.
 
Further detailed discussions and in--depth treatment of QCD, and generally of GFT, may be found in standard textbooks, e.g., \cite{ellis2003qcd, RevModPhys.67.157, sterman1993introduction, zeidler2008quantum, peskin1995introduction, weinberg2000quantum, zinn2002quantum}. Reviews such as those in \cite{Sterman:2004pd, Sterman:2008kj, Soper:2000kt} are also useful. Other references specific to certain sections will be given therein.

\section{$\rm SU(\Nc)$ colour}
\label{sec:QCD:Colour}

In group theory (see e.g., Refs. \cite{georgi1999lie, jones1998groups, cornwell1997group}), $\rm SU(\Nc)$ is the group of special unitary $\Nc \times \Nc$ complex matrices $\rm U$ such that:
\be
\det \rm U = e^{\Tr \ln \rm U} = 1 \;\;\;\; \rm{and} \;\;\;\;\; U^{\dagger} U = \mbb 1.
\label{eq:Color:SUNc}
\ee
The above relations leave $\Ncsq -1$ independent real parameters to characterise $\rm U$. The matrices $\rm U$ may be parametrised by the exponential form
\be
\rm U = e^{\imath\, \om^{a}\, t^{a}}, \;\;\; a = 1,\cdots, \Ncsq-1,
\label{eq:Color:UExpForm}
\ee
where $\om^{a}$ are colour rotation angles, $t^{a}$ are the \emph{generators} of the the group and summation over repeated indices is (henceforth always) implied. From \eqs{eq:Color:SUNc}{eq:Color:UExpForm}, it follows that the generators are both hermitian and traceless,
\be
t^{a} = \left(t^\dagger \right)^{a},\;\;\; \Tr\left(t^{a}\right) = 0.
\label{eq:Color:Generatorsproperties}
\ee 
The set of generators $\{t^a\}$ forms a basis of a vector space over the field of complex numbers $\mbb C$ (or simply a Lie algebra \cite{georgi1999lie}):
\begin{subequations}
\begin{eqnarray}
 [t^a,t^b] = t^a .\,t^b - t^b .\,t^a &=& \imath f_{abc} t^c, \label{eq:Color:LieProduct}\\
 \left[t^a,[t^b,t^c]\right] + \left[t^c,[t^a,t^b]\right] + \left[t^b,[t^c,t^a]\right] &=& 0.
\label{eq:Color:LieAlgebra} 
\end{eqnarray}
\end{subequations}
The real expansion coefficients $f_{abc}$ are known as the \emph{structure} constants of the group and the second equality \eqref{eq:Color:LieAlgebra} is known as the \emph{Jacobi} identity. The structure constants satisfy analogous relations to \eqs{eq:Color:LieProduct}{eq:Color:LieAlgebra}. The generators $t^a$ can be \emph{represented} by either $D_{F}(t^a) = \Nc \times \Nc$ matrices, in the \emph{fundamental} representation, or by the structure constant, $[D_{A}(t^a)]_{bc} \equiv \left[T^a\right]_{bc} = -\imath f_{abc} = (\Ncsq-1) \times (\Ncsq-1)$ matrices, in the \emph{adjoint} representation\footnote{Note that for the sake of brevity we shall write $D_F(t^a) = t^a$ and $D_{A}(t^a) = T^a$ throughout. A widely used representation of $\rm SU(3)$ is that provided by Gell--Mann matrices. Their explicit expressions can be found in, e.g., \cite{ellis2003qcd}.}. For a given representation, $R$, the normalisation of the corresponding matrices, $D_R(t^a)$, is such
\be
 \Tr\left[D_{R}(t^a)\, D_{R}(t^b)\right] = T_{R}\,\delta^{ab}.
\label{eq:Color:GeneratorRepNormalisatn}
\ee
The $\rm SU(\Nc)$ (and hence $\rm SU(3)$) fundamental and adjoint representations have $T_{F} = 1/2$ and $T_{A} = \Nc$, respectively . For a given representation, $R$, the casimir operator (or ``colour charge''), $T_R^2$, which commutes with all other generators in the representation $R$ is given by \cite{georgi1999lie}
\be
(T_R^2)_{ij} = \sum_{a=1}^{\Ncsq-1} \sum_{k=1}^{d(R)} \left[D_{R}(t^a)\right]_{ik}  \left[D_{R}(t^a)\right]_{kj} = C_{R}\, \delta_{ij},
\label{eq:Color:CasimirOperator}
\ee
where $d(R)$ is the dimension of the representation $R$, namely $d(F) = \Nc$ and $d(A) = \Ncsq-1$. \eqs{eq:Color:GeneratorRepNormalisatn}{eq:Color:CasimirOperator} yield
\be
(T_R^2)_{ii} = T_R\,\left(\Ncsq-1\right) = C_R\, d(R).
\label{eq:Color:CasimirFormula}
\ee
We can then read off the $\rm SU(\Nc)$ casimirs (for the fundamental and adjoint representations),
\be
C_F = \frac{\Ncsq-1}{2\Nc},\;\;\; C_A = \Nc.
\label{eq:Color:SUnCasimirs}
\ee
In \app{app:ColorAlgebra} we provide some important colour identities as well as some useful relations that will prove indispensable when performing calculations in QCD.

\subsection{Large $\Nc$ limit}
\label{ssec:QCD:large-Nc}

\begin{figure}[h]
 \centering
 \includegraphics[width=0.75\textwidth]{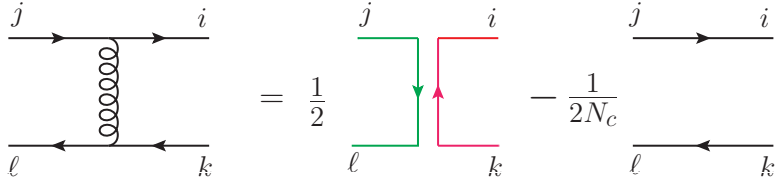}
 \caption{Graphical representation of the Fierz identity for a one--gluon exchange between a quark--antiquark pair ($t$--channel exchange).}
\label{fig:FI:FierzID_t}
\end{figure}

An alternative way to view the Fierz identity \eqref{eq:FI:FierzID} is as a one--gluon exchange between two quarks, or quark and antiquark, expressed in terms of plain quark lines (\fig{fig:FI:FierzID_t}),
\be
 \left(t^a\right)^i_j \left(t^a\right)^\ell_k = \half\,\de^i_k \,\de^\ell_j - \frac{1}{2 \Nc}\,\de^i_j\,\de^\ell_k.
\label{eq:FI:FierzID_t}
\ee 
If one considers the limit $\Nc \ra +\infty$, with $\left(\Nc\,\gs^2\right)$ held fixed, then \eq{eq:FI:FierzID_t} (and \fig{fig:FI:FierzID_t}) suggests that the colour structure of a gluon is equal to that of a quark and an antiquark. This approximation\footnote{It is an approximation because for QCD the number of colours is $\Nc =3$, which is hardly large enough.}, first introduced by 't~Hooft \cite{Hooft1974461} in the context of  strong interactions, turned out to be a powerful one in simplifying perturbative calculations by limiting the number of graphs contributing to a specific process. To leading order in $\Nc$ only topologically \emph{planar} graphs need to be computed. \emph{Non--planar} graphs' contributions to physical cross sections are suppressed by $1/\Ncsq$. 

\begin{figure}[t]
 \centering
 \includegraphics[width=0.6\textwidth]{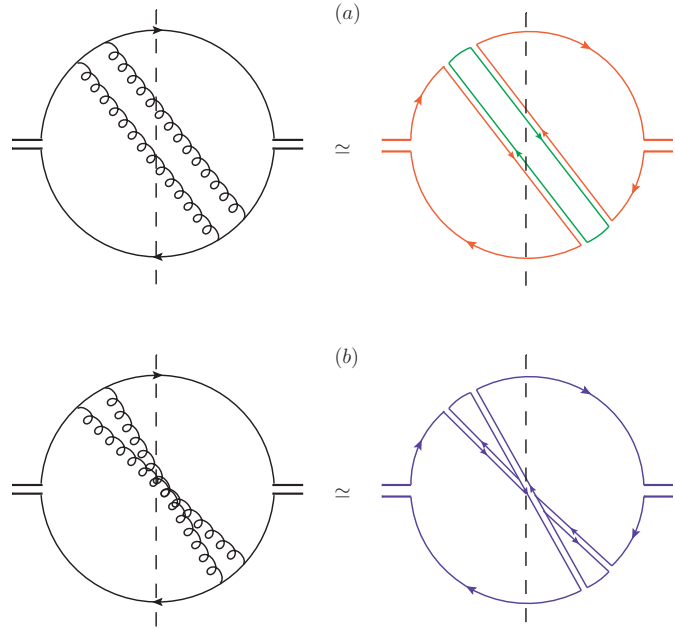}
 \caption{Graph $(a)$ is an example of a planar graph which contributes at leading $\Nc$. Graph $(b)$ is an example of non--planar graph which contributes at subleading $\Nc$. The graphs represent \emph{Feynman diagrams} for the amplitude squared of $\EE$ anihilation into $q\qb gg$ (to be discussed in \sec{sec:QCD:Lagrangian} and \app{app:QCDReview}).}
\label{fig:LNc:PlanarGraph}
\end{figure}

An example of planar and non--planar graphs is shown in \fig{fig:LNc:PlanarGraph}, which occurs in the $\EE$ annihilation into $\qqb$ accompanied with the emission of two soft gluons. Utilising the colour machinery developed in the previous section (\sec{sec:QCD:Colour}) and \app{app:ColorAlgebra}, in particular \fig{fig:FI:ColourAlgebra} and \eq{eq:FI:M-examples-a}, one obtains, to leading $\Nc$, an overall factor of $(\Nc\,\gs^2)^2$ and $(\gs^2)^2$ for graphs $(a)$ and $(b)$ respectively. In other words, graph $(b)$'s contribution to the amplitude squared is suppressed by $1/\Ncsq$ factor relative to graph $(a)$'s contribution. 

It is worthwhile noting that this approximation is the basis of the ``colour dipole evolution'' formalism that stands at the heart of some numerical Monte Carlo programs such as \ariadne \cite{Lonnblad:1992tz} and the program developed in \cite{Dasgupta:2001sh}. In the said formalism, the probability of a (soft) gluon emission off an ensemble of harder partons (quarks, antiquarks and gluons) is simply proportional to the sum of the independent colour dipoles formed by the ensemble (see eikonal method in \app{sec:app:QCD:EikonalApprox}).

\section{Lagrangian and gauge invariance}
\label{sec:QCD:Lagrangian}

The full QCD Lagrangian density may be written as (see e.g., Refs. \cite{ellis2003qcd, Sterman:2005vn}):
\be
\mc L_{\rm QCD} = \mc L_{\rm Dirac} + \mc L_{\rm Yang-Mills} + \mc L_{\rm gauge-fixing} + \mc L_{\rm ghost},
\label{eq:LQCD:QCDFullLagrangian}
\ee
where the various terms are defined below:
\begin{itemize}
 \item The quark content of QCD is described by the Dirac Lagrangian density
\be
\mc L_{\rm Dirac} = \sum_{f=1}^{\nf}\, \br{\mb{q}}_{f} \left(\imath\,\Dslash - m_{f}\right) \mb{q}_{f},
\label{eq:LQCD:DiracLagrangian}
\ee
where $\mb q$ labels a quark field of flavour $f$ and mass $m_{f}$. In QCD there are $\nf = 6$ flavours of quark fields and their conjugates, $\br{\mb{q}}_f = \mb q_f^\dagger\, \g^0$. The quark field forms a three--vector in the colour space; $\mb q = \left(q_{1}, q_{2}, q_{3}\right)$ with $q_{a}$ being Dirac spinor fields\footnote{The three values of the colour index $a$ are usually referred to as red, blue and green, as mentioned in the introductory part.}. The symbol $\Dslash$ denotes the product $\g^\mu\,D_{\mu}$ where $\g^\mu$ are the traceless Dirac matrices satisfying the Clifford algebra $\{\g^\mu, \g^\nu\} = 2\,g^{\mu\nu}$, with the latter metric given by $g^{\mu\nu} = \rm{diag}\left(+1,-1,-1,-1\right)$\footnote{Especially useful representation of the Dirac  $\g$ matrices is the Weyl (or chiral) representation. Explicit expressions of the latter may be found in e.g., \cite{peskin1995introduction}.}, and $D_{\mu}$ is the covariant derivative given by
\be
D_{\mu} = \partial_\mu - \imath\,\gs\,A_\mu.
\label{eq:LQCD:CovariantDerivative}
\ee
The $\rm SU(3)$ gluon fields are given by $A_\mu = \sum_{a=1}^8 A_{\mu}^a\,t^a$ when $D_\mu$ is acting on the quark field (colour triplet) and $A_\mu = \sum_{a=1}^8 A_{\mu}^a\,T^a$ when acting on the gluon field (colour octet), with $t^a$ and $T^a \left(\equiv \left(T^a\right)_{bc}\right)$ being, respectively, the generators of $\rm SU(3)$ in the fundamental and adjoint representations discussed previously. The strong coupling parameter, $\gs$, measures the strength of the strong interaction between quarks and gluons or gluons amongst themselves. 

\item The Yang--Mills Lagrangian density describes the free propagation of the gluon field and reads
\be
\mc L_{\rm Yang-Mills} = -\frac{1}{2} \Tr\left[F_{\mu\nu}^2(A)\right]^2,
\label{eq:LQCD:YMLagrangian}
\ee
where the non--abelian, i.e., non--commutative, $\rm SU(3)$ field strength tensor,  $F_{\mu\nu} = F_{\mu\nu}^a \,t^a$, is defined by the commutator
\be
 \left[D_\mu, D_\nu \right] = -\imath\,\gs F_{\mu\nu}.
\label{eq:LQCD:FieldTensor}
\ee
Explicitly written in terms of the gluon fields, $F_{\mu\nu} = \partial_\mu A_\nu - \partial_\nu A_\mu -\imath\,\gs \left[A_\mu, A_\nu\right]$. The sum of the Dirac and Yang--Mills Lagrangians is the \emph{classical} QCD Lagrangian. The latter is invariant under the local gauge transformations (represented by the $\rm SU(3)$ matrix $\rm U$ given in \eq{eq:Color:UExpForm}):
\begin{eqnarray}
\nn \mb{q}(x) &\ra& \mb{q}'(x) = \mathrm{U} (x)\, \mb{q}(x),\\
 A_\mu(x)  &\ra& A_\mu'(x) = \mathrm{U}(x) \left[A_\mu(x) + \frac{\imath}{\gs}\, \partial_\mu \right] \mathrm{U}^\dagger (x).
 \label{eq:LQCD:GaugeTransformations}
\end{eqnarray}
The field strength tensor \eqref{eq:LQCD:FieldTensor} thus transforms under the latter finite transformations \eqref{eq:LQCD:GaugeTransformations} as
\be
 F_{\mu\nu} \ra F'_{\mu\nu}  = \mathrm{U}\,F_{\mu\nu}\,\rm U^\dagger.
\label{eq:LQCD:FieldTensorGaugeTrans}
\ee
Considering infinitesimal gauge transformations, for which the $\rm U$ matrix may be written as an expansion around the unit matrix; $\rm U = \mbb 1 + \imath\,\om^a t^a$, then it follows from \eq{eq:LQCD:FieldTensorGaugeTrans} that $F_{\mu\nu}$ is not gauge invariant,
\be
 F_{\mu\nu}^a \ra F_{\mu\nu}^{a\,'} = F_{\mu\nu}^a - f_{abc}\,F_{\mu\nu}^b\,\om^c + \Or(\om^2).
\label{eq:LQCD:FieldTensorInfGaugeTrans}
\ee
This result is in contrast to the QED case where the photon field strength tensor is gauge invariant. The immediate consequence of this is that gluons are coloured and cannot thus be seen as asymptotic states. Notice also that since a mass term $A_\mu\,A_\nu$ breaks the gauge invariance then it cannot be included in the Lagrangian density \eqref{eq:LQCD:YMLagrangian}. It follows that gluons are massless. 

\item  To eliminate the gauge freedom in the Yang--Mills Lagrangian one must fix a gauge to work in. Without the gauge fixing term the gluon propagator is not well defined (see e.g., Refs. \cite{sterman1993introduction, bjorken1964relativistic}). One possible class of gauge conditions is that of \emph{covariant} gauges, defined by
\be
 \mc L_{\rm gauge-fixing} = -\frac{1}{2\,\xi} \left(\pa_\mu\,A^\mu\right)^2,
\label{eq:LQCD:GaugeFixing}
\ee
where $\xi$ is the gauge parameter. Any value of $\xi$ can be used and any gauge--invariant calculation is independent of the choice of latter. Amongst the commonly used choices are $\xi =1,0$ for Feynman and Landau gauges respectively.

\item To remove the unphysical polarisation states of the gluon vector field, one introduces an unphysical Faddeev--Popov \emph{ghost} term in the QCD Lagrangian \cite{Faddeev196729}. This is best illustrated using the path integral formalism of quantum field theory (QFT) \cite{Abers:1973qs}. For the above mentioned covariant gauges the ghost Lagrangian reads
\be
 \mc L_{\rm ghost} = \pa_\mu\, \eta^{a \dagger} D^\mu_{ac}\,\eta^c,
\label{eq:LQCD:GhostLagrangian}
\ee 
with $\eta$ being a complex scalar field with anti--commuting properties of a spinor field\footnote{A sign that is an unphysical field. Such a term is not even needed in certain gauges such as axial gauges.}.
\end{itemize}

\subsection{Perturbation theory and Feynman rules}
\label{ssec:QCD:FeynmanRules&PT}

The QCD Lagrangian\footnote{We use the terminologies Lagrangian and Lagrangian density interchangeably. The former is the three--volume integral of the latter.} \eqref{eq:LQCD:QCDFullLagrangian} may be written as the sum of \emph{free}, $\mc L_0$, and \emph{interaction}, $\mc L_{\rm I}$, Lagrangians:
\be
 \mc L = \mc L_{0} + \mc L_{\rm I}.
\label{eq:LQCD:Free+InterLagrangians}
\ee
$\mc L_0$ contains terms that are bilinear in fields whilst $\mc L_{\rm I}$ include all remaining terms of the full Lagrangian. Perturbation theory is constructed on the basis that the interaction part, $\mc L_{\rm I}$, can be treated as small modifications, or \emph{perturbations}, to the free theory, described by $\mc L_0$. We can conceptually sketch a scattering process $i \ra f$ as starting with an initial state $\ket{i}$ of freely propagating particle waves at time $t_0$ and ending with a final state $\ket{f}$ of freely propagating particle waves at time $t$. The interaction of the free particles with the force carrying particles take place at a very short instant within the interval $\left(t-t_0\right)$. The corresponding time evolution operator of the state $\ket{i}$ is given by\footnote{We work in the \emph{interaction} picture of QFT throughout. For more details on the latter consult, for instance, \cite{peskin1995introduction}.}
\be
 \mc T\left(t,t_0\right) = \mathrm{T}\left\{ \exp\left[ -\imath \int_{t_0}^t \d^4 x\, \mc L_{\rm I}(x)\right]\right\},
\label{eq:LQCD:TimeEvolutionOperator}
\ee
where $\rm T \{\cdots\}$ stands for the \emph{time ordered} product of fields in $\mc L_{\rm I}$. In the limit where the initial and final states take place at a long time in the past, $t_0 \ra -\infty$, and a long time in the future, $t \ra +\infty$, respectively, the evolution operator \eqref{eq:LQCD:TimeEvolutionOperator} is simply the $S$--matrix,
\be
 S \equiv \lim_{t_0/t \ra \mp\infty} \mc T(t,t_0).
\label{eq:LQCD:Smatrix}
\ee
The $S$--matrix encodes a plethora of information about the scattering process. The entries of the $S$--matrix, or the matrix--elements $S_{fi}$, are related to the $i \ra f$ transition amplitude (or invariant matrix elements) $\mc M_{fi}$ through,
\be
 S_{fi} = \bra{f} S\ket{i} = \de_{fi} -\imath T_{fi},
\label{eq:LQCD:SmatrixEntries}
\ee
with the $T$--matrix elements given by
\be
 \imath T_{fi} = \imath \left(2\pi\right)^4 \de^{(4)}\left(P_i - P_f \right)\, \mc M_{fi},
\label{eq:LQCD:TransitionAmplitude}
\ee
where $P_{i(f)}$ is the total four--momentum of initial (final) state particles. The delta term $\de_{fi}$ in \eqref{eq:LQCD:SmatrixEntries} is the zeroth order in the expansion of \eqref{eq:LQCD:Smatrix} and describes the state where no interaction happens. The transition amplitude $\imath T_{fi}$, or $\imath \mc M_{fi}$, consists of the timed--ordered expansion of $S$, starting at first order and adding an extra interaction term $\mc L_{\rm I}(x)$ at each higher order.

\begin{figure}[!t]
 \centering
 \includegraphics[width=0.8\textwidth]{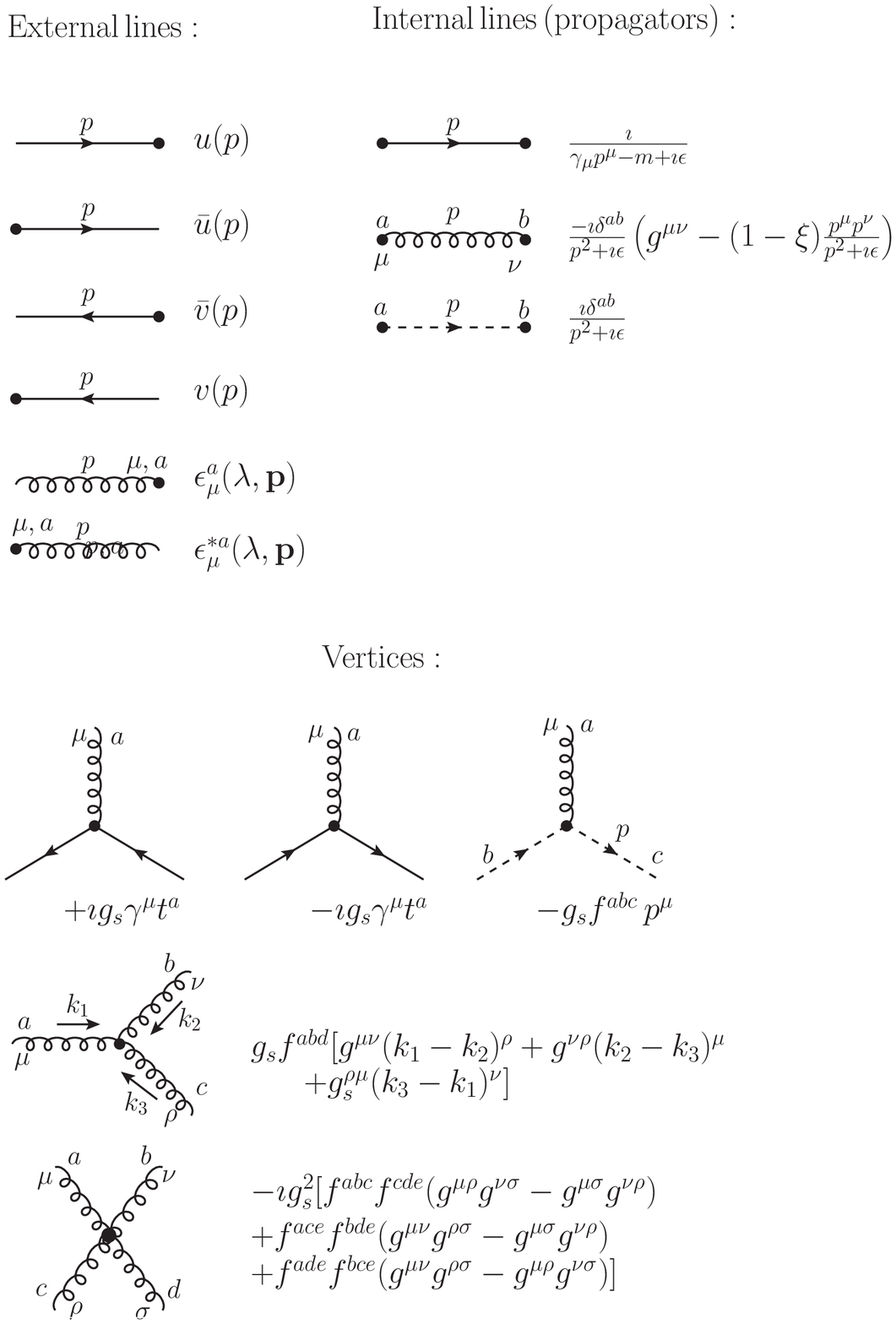}
 \caption{Feynman rules for QCD in a covariant gauge: gluons (curly lines), fermions (solid lines) and ghosts (dashed lines).}
 \label{fig:LQCD:FeynmanRulesQCD}
\end{figure}

To compute $\mc M_{fi}$ one employs a set of \emph{Feynman rules}, along with a graphical visualisation of the process via \emph{Feynman graphs}, derived from the Lagrangian \eqref{eq:LQCD:QCDFullLagrangian}. These rules are represented by \emph{external} lines for incoming (initial) or outgoing (final) particles, \emph{internal} lines for propagators (that are neither initial nor final) and \emph{vertices} for the interaction terms. The set of Feynman rules for QCD are depicted in \fig{fig:LQCD:FeynmanRulesQCD}.

\subsection{Optical theorem and cross-sections}
\label{ssec:QCD:OpticalTheorem}

An important feature of the $S$--matrix that is useful in checking the consistency of Feynman amplitude calculations is \emph{unitarity}. This follows from the fact that the probability of producing a final state $\ket{f}$ given an initial state $\ket{i}$ is simply $\left|S_{fi} \right|^2$. Summing over all possible final states, the total probability must be equal to unity. Schematically
\be
 \sum_{f} \left|S_{fi}\right|^2 = 1\;\; \Ra \;\; S^\dagger S = 1. 
\label{eq:LQCD:Unitarity}
\ee
Substituting for $\imath T$ from \eq{eq:LQCD:SmatrixEntries}, the unitarity relation \eqref{eq:LQCD:Unitarity} implies
\be
 -\imath \left(T_{i' i} - T^{\star}_{i' i}\right) = \sum_{f} T_{f i'}^{\star} T_{f i}.
\label{eq:LQCD:TransAmplUnitarity}
\ee 
In the special case of elastic, $\ket{i'}=\ket{i}$, two--particle scattering, \eq{eq:LQCD:TransAmplUnitarity} leads to the optical theorem:
\be
  \Im\left(\mc M_{ii}\right) = \ld^{1/2}(s, m_a^2, m_b^2)\,\s_{\rm tot}\left(i \ra X \right),
\label{eq:LQCD:OpticalTheorem}
\ee
which relates the forward scattering amplitude to the total cross section for production of all possible final states $X$. The normalisation factor $\ld(s,m_a^2,m_b^2)$ given in terms of the centre--of--mass energy squared $s$ and the masses of the scattering particles $a$ and $b$ has the form $\ld (x,y,z) = x^2 + y^2 + z^2 -2\,x y -2\,x z - 2\,y z$. The total cross-section $\s_{\rm tot}$ is given by:
\be
 \s_{\rm tot}(i \ra X) = \sum_{f}\, \int\, \mathrm{F}\,\abs{\br{\mc M_{fi}}}^2\,\d\Pi_{f}, 
\label{eq:LQCD:2To2TotalCrossSection}
\ee
with the sum being over all possible final states $\ket{f}$, $\rm F$ is the flux of initial particles, $\abs{\br{\mc M_{fi}}}^2$ is the invariant matrix elements squared summed (averaged) over final (initial) state spin, polarisation and colour and $\d\Pi_{f}$ is the corresponding final state phase space. Let $\ket{f}$ be composed of $n$ particles with momenta $p_\ell^\mu = \left(E_\ell, \vect{p}_\ell\right)$, then the $n$--particle Lorentz invariant phase space is defined by
\begin{eqnarray}
 \d \Pi_n &=& \left(2\pi\right)^4 \de^{(4)}\left(P_f - p_a - p_b\right) \prod_{\ell=1}^{n} \frac{\d^3 \vect{p}_{\ell}}{\left(2\pi\right)^3 2 E_\ell}\, \frac{1}{\rm m\, !},
\label{eq:LQCD:GeneralPhaseSpaceFactor}
\\
 \rm F  &=& \left(4 E_a E_b \left|\vect{v}_a - \vect{v}_b\right|\right)^{-1} = \frac{1} {2 \ld^{1/2}(s,m_a^2,m_b^2)},
\label{eq:LQCD:FluxOfInitialParticles} 
\end{eqnarray}
where $\vect{v}_{a}$ and $\vect{v}_b$ are the three--velocities of the two particles ($a$ and $b$) in the initial state and $\rm m$ is the number of identical particles in the final state.

\section{Running of $\as$}
\label{sec:QCD:Running}

\begin{figure}[!t]
 \centering
 \includegraphics[width=0.6\textwidth]{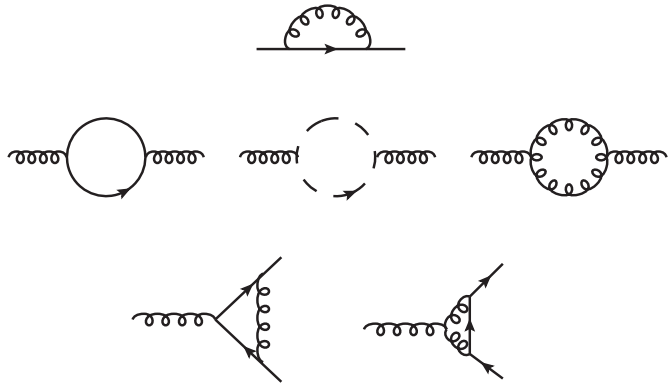}
 \caption{Feynman diagrams contributing to QCD $\B$ function. (top) fermion self--energy diagrams, (middle) gluon self--energy diagrams and (bottom) fermion--gluon vertex diagrams.}
 \label{fig:RunAlpha:QCDBeta}
\end{figure}

QCD is a \emph{renormalisable}\footnote{Renormalisation may be summarised, very briefly, as a procedure for treating ultra--violet (UV) divergences of Feynman amplitudes. For an intensive discussion of renormalisation the reader is referred to e.g., Ref. \cite{collins1986renormalization}.} field theory. This implies that the coupling $\as \equiv \gs^2/4\pi$ (defined with analogy to the fine structure constant of QED $\A \equiv e^2/4\pi$) must be defined at the renormalisation scale, which we denote $\mu$. The renormalisation flow of the coupling $\as(\mu)$ is determined by the renormalisation group equation (RGE), which can be cast in the form \cite{ellis2003qcd, zinn2002quantum}
\be
 \mu^2\frac{\d\as(\mu)}{\d\mu^2} = \B(\as(\mu)).
\label{eq:RunAlpha:RenormGroupEqu}
\ee
The $\B$ function is computed in perturbation theory as a power series in $\as(\mu)$, from Feynman diagrams involving fermion self--energy, gluon self--energy and fermion--gluon vertex, as shown in \fig{fig:RunAlpha:QCDBeta}.
We thus write
\be
 \B(\as) = -\as^2 \cbr{\B_0 + \B_1\,\as + \cdots},
\label{eq:RunAlpha:BetaFunctionQCD}
\ee
where $\B_0 = \left(11\CA -2\nf\right)/12\pi$, $\B_1 = \left(17\CA^2-5\CA\nf-3\CF\nf\right)/24\pi^2$ and the minus sign on the right--hand side of \eqref{eq:RunAlpha:BetaFunctionQCD} distinguishes QCD from QED, for which one has (see e.g., \cite{Kataev:2012rf})
\be
 \B^{\rm QED}(\A) = +\A^2 \cbr{\B_0^{\rm QED} + \B_1^{\rm QED} \A + \cdots},
\label{eq:RunAlpha:BetaFunctionQED}
\ee
with $\B_0^{\rm QED} = \nf/3\pi$ and $\B_1^{\rm QED} = 1/4\pi^2$. The solution to \eq{eq:RunAlpha:RenormGroupEqu} determines the value of the coupling at a scale $\mu_2$ in terms of its value at another scale $\mu_1$\footnote{Usually the magnitude of $\as$ in QCD is given at the mass of the $Z$ boson: $\as(\mu = m_Z) \sim 0.118$ \cite{Bethke:2009jm}. The magnitude at any other scale is then deduced from the solution of \eq{eq:RunAlpha:RenormGroupEqu}.}. At lowest order, one obtains for QCD an expression of the form
\be
 \as(\mu_2) = \frac{\as(\mu_1)}{1 + \B_0\,\as(\mu_1)\ln\left(\mu_2^2/\mu_1^2\right)} = \frac{1}{\B_0\,\ln\left(\mu_2^2/\Lqcd^2\right)},
\label{eq:RunAlpha:CouplingAtOneLoop}
\ee
where the last equality follows from rewriting \eq{eq:RunAlpha:RenormGroupEqu} as,
\be
 \ln\left(\frac{\mu_2^2}{\mu_1^2}\right) = \int_{\as(\mu_1)}^{\as(\mu_2)} \frac{\d\as}{\B(\as)},
\label{eq:RunAlpha:RenormGroupEquV2}
\ee
and setting $\mu_1 \ra \Lqcd$. Provided the first coefficient $\B_0$ is positive, which is only true if the number of flavours $\nf$ is less than about $16.5$, then the coupling weakens (diverges) logarithmically at larger (smaller) renormalisation masses $\mu$. Indeed, $\as$ approaches zero in the limit $\mu \ra \infty$ and blows up in the limit $\mu \ra \Lqcd$, as can be seen from \eq{eq:RunAlpha:CouplingAtOneLoop}. The former limit is known as \emph{asymptotic} freedom \cite{PhysRevD.8.3633, Politzer:1974fr} and justifies the validity of perturbation theory in this regime. The latter limit signals the failure of the lowest--order approximation \eqref{eq:RunAlpha:CouplingAtOneLoop} where the theory becomes strongly coupled and thus essentially \emph{non--perturbative}. To the contrary, the QED coupling $\A$ is small at low energies, $\A(\mu = 0) \sim 1/137$, and grows logarithmically at high energies, $\A(\mu = m_Z) \sim 1/128$.

\subsection{Confinement and factorisation}
\label{ssec:QCD:Factorisation}

Probed at sufficiently high energies, or momentum transfer $\gg \Ld_{\rm QCD}$, hadrons may reliably be considered as ``bags'' of weakly interacting quarks and gluons. Consequently, collisions between hadrons can effectively be regarded as scattering amongst their ``free'' constituent partons (quarks and gluons). The \emph{hard} scattering process, to which mainly two partons, one from each of the colliding hadrons, participate with the rest of the ``spectator'' partons interfering at lower energies and thus forming hadron remnants, is fully calculable in perturbation theory. Such a hard scattering occurs, by the uncertainty principle, at very short distances.

As the energy falls to $\sim \Lqcd$ or below, the quarks and gluons inside a hadron become strongly coupled to the extent that it is impossible to knock an individual quark out of the hadron. This phenomenon is know as \emph{confinement}. Starting with a meson (quark--antiquark bound state), if one attempts to pull apart its constituents then instead of the meson breaking into two quarks it breaks into two mesons. The new quark--antiquark pair is created from quantum mechanical fluctuations once the energy in the ``flux tube''\footnote{This concept arises in the investigation of QCD confinement from the point of view of a non--abelian analogue to the Meissner effect in superconductors \cite{Sterman:2005vn}.}, formed between the original quark--antiquark pair, becomes higher than the rest mass of the pair (recall that the masses of the light quarks, $\Or(5 \MeV)$, are far below the QCD non--perturbative scale $\Lqcd \sim 200 \MeV$). Therefore at low energies, or equivalently long distances, individual quarks 
cannot be observed as they immediately confine, or \emph{hadronise}, into hadronic bound states.

A typical event at the LHC proceeds as follows: First, two protons are made to collide head--on at a high centre--of--mass energy. The actual hard scattering occurs between the constituent partons, one from each proton. As the quarks travel towards the outer parts of the detector they lose energy by emitting gluons, which in turn radiate more gluons and/or split into quark--antiquark pairs. At each step the energy falls and the number of quarks and gluons grows up leading to the formation of a ``shower'' of partons. Eventually, at low energies, quarks and gluons undergo the inevitable process of hadronisation (confinement) into mesons and baryons, which are observed at detectors as collimated energy depositions, or equivalently ``jets''. The latter will be the pivotal subject of the next chapter (Chapter \ref{ch:Jets}).

Two processes that take place at two disparately different scales, long and short distances, are quantum mechanically incoherent. i.e., their interference vanishes, up to inverse powers of the hard scale of the system, as shown in \eq{eq:RunAlpha:Factorisation}. This allows for a separate treatment of each process. Such a separation is known as \emph{factorisation}. The general form of a  factorised physical cross section for a process initiated by two hadrons with momenta $P_1$ and $P_2$ is \cite{Collins:1985ue, Collins:1989gx, Sterman:2008kj, Bodwin:1984hc}
\be
\s\cbr{P_1,P_2} = \sum_{i,j} \int \d x_1 \d x_2\, f_i(x_1,\mu_F) f_j(x_2, \mu_F)\, \sh_{ij}\cbr{p_1,p_2,\as(\mu_F),Q/\mu_F} + \Or(1/Q^p), 
\label{eq:RunAlpha:Factorisation}
\ee
where $p_i = x_i P_i$ is the momentum of parton $i$ participating in the hard scattering, as schematically depicted in \fig{fig:Intro:JetEvent}. $\sh$ is a short distance, hard scattering partonic cross section, which is \emph{infrared safe}\footnote{Infrared (and collinear) safety, which refers to the insensitivity to soft emissions and collinear splittings, is crucial to the finiteness of perturbative calculations. This topic will be addressed in the next chapter (\chap{ch:Jets}).} and calculable in perturbation theory, and $f$ is a long distance function that is not calculable in perturbation theory. It is however ``universal'' and can be measured experimentally once and for all. In essence, the function $f$ describes the distribution (and dynamics) of partons in the colliding hadrons. As such $f$ is termed the \emph{parton distribution function} (PDF)\footnote{Considering final state hadrons, $f$ may be taken to be the \emph{fragmentation} function (usually denoted $D$) \cite{ellis2003qcd}.}, and the 
subscripts $i$ and $j$ in \eq{eq:RunAlpha:Factorisation} refer to the specific parton flavour (including gluons). The symbols $Q$ and $\mu_F$ denote, respectively, the partonic hard and factorisation scales. The latter marks the boundary between short distance and long distance, or equivalently between perturbative and non--perturbative, dynamics. Thus a parton emitted at a scale less than $\mu_F$ will be absorbed into the parton distributions, while a parton emitted at a scale larger than $\mu_F$ will be considered as part of  $\sh$. The more orders included in $\sh$ the less dependence of $\sh$ on the scale $\mu_F$. The last term in \eq{eq:RunAlpha:Factorisation}, with $p$ being a real number, is there to indicate that factorisation holds up to corrections that behave as inverse powers of $Q$. It is worth noting that the factorisation formula \eqref{eq:RunAlpha:Factorisation} has been proven for sufficiently inclusive observables in a variety of processes, e.g.,  Drell--Yan process (hadron + hadron $\to \
mu^+ \mu^-$ + X) and scattering of colourless hadrons (hadron + hadron $\ra$ hadron + X), where X denotes ``anything else'' in the final state \cite{Collins:1985ue, Collins:1989gx, Collins:1988ig, Collins:1987pm}.

Since physical quantities, such as $\s$, cannot depend on the (arbitrary) factorisation scale $\mu_F$ then
\be
\frac{\pa\,\ln\s}{\pa\ln\mu_F} = 0.
\label{eq:RunAlpha:muIndependence}
\ee 
Although parton distributions are not calculable from first principles of QCD, as stated above, \eqref{eq:RunAlpha:muIndependence} allows for perturbative calculations of their dependence on the scale $\mu_F$. Substituting \eq{eq:RunAlpha:Factorisation} into \eq{eq:RunAlpha:muIndependence}, we schematically have
\be
 \frac{\pa\ln f}{\pa\ln\mu_F} \propto P, \qquad f=q_i,g.
\label{eq:RunAlpha:EvolutionEqu}
\ee
where the function $P$ can only depend on the shared variables between $\sh$ and $f$; the coupling and convolution variables. The solution of \eq{eq:RunAlpha:EvolutionEqu} describes the ``evolution'' of the PDFs between two scales, say, $\mu_F$ and $\mu'_F$. Once measured experimentally at a scale $\mu_F$, they can be perturbatively predicted at arbitrary high scales, $\mu'_F \gg \Lqcd$, (provided the coupling remains small over the evolution range). The function $P$ actually refers to the Altarelli-Parisi splitting functions (or evolution kernels) and \eq{eq:RunAlpha:EvolutionEqu} is the Dokshitzer--Gribov--Lipatov--Altarelli--Parisi (DGLAP) evolution equation \cite{Gribov:1972ri, Lipatov:1974qm, Altarelli:1977zs, Dokshitzer:1977sg}. It is one of the fundamental equations in pQCD (analogous to the $\B$ function equation \eqref{eq:RunAlpha:RenormGroupEqu}). Written in the standard form, including explicit dependence on the partons' (longitudinal) momentum fractions, the latter evolution equation reads \cite{
ellis2003qcd}.
\be
\frac{\pa f_i(x, \mu_F)}{\pa\ln \mu_F} = \frac{\as\cbr{\mu_F}}{2\pi}\,\sum_j \int_x^1 \frac{\d z}{z}\,f_j(z, \mu_F)\, P_{ij}\left(\frac{x}{z}, \as(\mu_F)\right).  
\label{eq:RunAlpha:DGLAPEquation}
\ee
The splitting functions $P_{ij}$ essentially correspond to the rate of production of parton $i$ from parton $j$ as the process evolves in $\ln\mu_F$. They have the following perturbative expansion in the running coupling:
\be
 P_{ij}\cbr{z,\as} = \cSup{P_{ij}}{0}(z) + \frac{\as}{2\pi} \cSup{P_{ij}}{1}(z) + \cdots,
\label{eq:RunAlpha:SplittingFunExpansion}
\ee
where the leading splitting functions are
\begin{eqnarray}
\nn \cSup{P_{qq}}{0}(z) &=& \CF\sbr{\frac{1+z^2}{(1-z)_+} + \frac{3}{2}\,\de(1-z)},
\\
\nn \cSup{P_{qg}}{0}(z) &=& \TF \sbr{z^2 + \left(1-z\right)^2},
\\
\nn \cSup{P_{gq}}{0}(z) &=& \CF\sbr{\frac{1 + \left(1-z\right)^2}{z}},
\\
\nn \cSup{P_{gg}}{0}(z) &=& 2\,\CA\sbr{\frac{z}{(1-z)_+} + \frac{1-z}{z} + z(1-z)} + \sbr{\frac{11\,\CA - 4\,\nf\TR}{6}} \de\left(1-z\right).\\
\label{eq:RunAlpha:SplittingKernelsLO}
\end{eqnarray}
Additionally $\cSup{P_{qq'}}{0} = \cSup{P_{q\qb}}{0} = 0$. The ``plus'' distribution used in the above equations is defined by
\be
 \int_0^1 \d x\; \frac{h(x)}{(1-x)_+} = \int_0^1 \d x\; \frac{h(x) - h(1)}{(1-x)},
\label{eq:RunAlpha:PlusDist}
\ee
where $h(x)$ is an arbitrary smooth function. The state--of--the--art of the splitting kernels is up to $\Or(\as^3)$ \cite{Moch:2004pa, Vogt:2004mw}. 

Lastly, notice that the DGLAP equation \eqref{eq:RunAlpha:DGLAPEquation} can be solved numerically either in $x$-space or in its conjugate Mellin space:
\be
 f(N,\mu) = \int_0^1 \d x\, x^{N-1} f(x,\mu).
\ee
In the latter space it may be possible to solve the equation analytically. Inverting back to the $x$--space is not, except in very special cases, however analytically possible. Throughout this thesis we shall adopt the standard choice $\mu_F = Q$ and thus do not worry about the evolution of parton densities. There is a plethora of numerical programs that straightforwardly handle such an evolution, see e.g., \cite{Salam:2008qg} (and references therein). Lastly note that a conventional way of estimating PDF uncertainties is to choose the renormalisation scale $\mu_R$, which has so far been implicitly taken to be equal to $\mu_F$, to be $\mu_R^2 = (x_\mu \mu_F)^2$ and vary $x_\mu$ in the range $1/2 < x_\mu < 2$ (often just takes the values $x_\mu = 1/2,1,2$) \cite{Salam:2010zt}. 

In the next chapter, we address more specific issues that are of direct concern in the current thesis. Namely the phenomenology of ``jets''.

%% file: ch3/chap3.tex

\chapter{Jet phenomenology}
\label{ch:Jets}
 
\begin{figure}[t]
 \centering
 \includegraphics[width=0.5\textwidth]{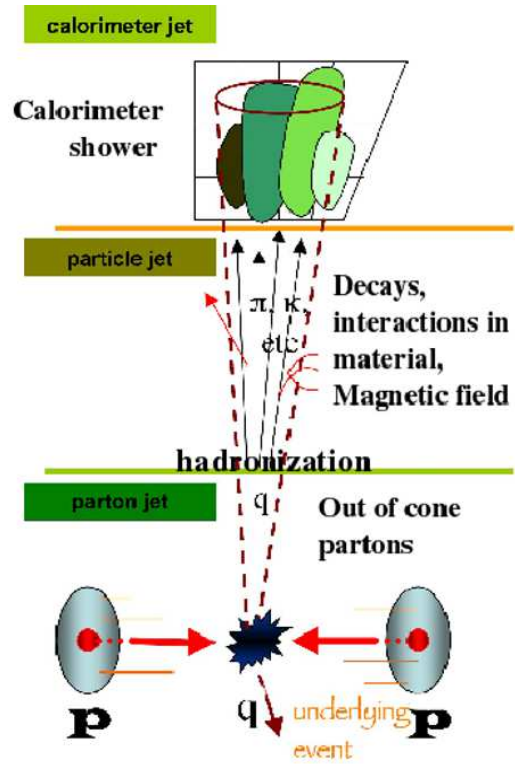}
 \caption{A schematic view of jet production at hadron colliders. The figure,
modified from \cite{Ellis:2007ib}, shows a proton--proton collision at the LHC.}
\label{fig:Jets:JetFormation}
\end{figure}

Ideally, theoretical predictions and/or interpretation of experimental data of
final states at colliders are to be performed in terms of short distance
colour--charged partons. However, what is observed instead at these colliders are long distance colour--neutral hadrons (along with other particles such as leptons), which are composite states of the latter partons.
More precisely, the final state hadrons, and particles, are observed in
collimated sprays, rather than large--angle separated individual particles,
forming ``jets'' that to some extent resemble the final state of the underlying
partonic process. \fig{fig:Jets:JetFormation} is a schematic illustration of jet
formation at hadron colliders\footnote{For $\EE$ machines, only the schematic
view of the final state applies. The initial--state is simply point--like
electrons and positrons.}. The whole process, from initial colliding beam
particles to final state detected particle jets, may be thought of as,
approximately, three  disentangled stages according to the energy scales
(distance) probed: low (long), high (short) and then low (long). Such a
separation is a product of factorisation discussed in the previous chapter
(\ssec{ssec:QCD:Factorisation}).   

We summarise the three stages, bottom--up in \fig{fig:Jets:JetFormation}, as
follows\footnote{We have already described hadronic collisions in
\ssec{ssec:QCD:Factorisation} in a broad sense. Here we provide more details and
set the scene for the subsequent sections.}. As the circulating bunches of
protons in the LHC beam--pipe collide head--on, a single ``event'' takes place
between a proton--proton pair. A single parton from each proton participates in
the short--distance hard scattering process. The distribution of partons within
each proton is determined through PDFs which depend, among other variables, on
the factorisation scale $\mu_F$ (usually set equal to the hard process scale)
and the parton's momentum fraction $x$ \cite{Soper:1996sn}. The insofar
described stage is considered non--perturbative due to the fact that PDFs are 
not calculable in pQCD but rather determined from global fits to data collected
from various experiments (primarily Deep Inelastic Scattering (DIS), see e.g.,
\cite{Martin:2009iq, Thorne:2009ky}). On their way to the hard interaction
point, the incident partons will radiate. Such a radiation is dubbed (for
obvious reasons) initial state radiation (ISR), and will contribute either to
the evolution of the PDFs (DGLAP equation \eqref{eq:RunAlpha:DGLAPEquation}), if
it is collinear radiation, or to the final state, if it is wide--angle radiation\footnote{It is still not entirely clear whether the ISR collinear radiation completely factorises out from the final state for non-inclusive observables in hadron-hadron collisions. See for instance Refs.
\cite{Catani:2011st, Collins:2007jp, Collins:2007nk, Forshaw:2012bi} for
detailed discussion on QCD coherence and factorisation breaking.}. The
spectators, the other constituent partons of the incident protons, also interact
with each other and emit radiation, but at a much softer scale relative to the
hard scattering scale, contributing to the non--perturbative underlying event
(UE).

At the interaction point the short--distance perturbative stage takes place. It
is characterised by large momentum transfers and may result in changing the
flavour of the scattering partons and/or producing more partons or other BSM
particles. The corresponding scattering cross section is calculable in pQCD. While leaving the interaction point the final state coloured partons emit radiation in the form of low energy and/or collinear (to the emitting parton) gluons and quark--antiquark pairs. This is the final state radiation (FSR), which forms the showering phase and leads to large logarithmic contributions to the final cross section that may be ``resummed'' either analytically or numerically via parton shower Monte Carlos (PSMCs), to a certain level of accuracy (see for instance \cite{Banfi:2010xy, Buckley:2011ms}). This issue -resummation- stands at the heart of this thesis and will be examined in detail in \sec{sec:Jets:JetShapes}. It should be emphasised, as stated previously, that the above picture, that ISR and FSR are disentangled from each other, is strictly speaking a probabilistic (approximate) picture, which is implemented in PSMCs. Physically, there will always be ``quantum interferences'' between the two radiations leading 
to contributions of the type investigated in \cite{Dokshitzer:2005ek} (``fifth'' form factor).

As the process evolves away from the hard interaction point towards lower
energies (longer distances) the coloured partons resulting from the showering as
well as those coming from softer interactions, both from UE and PU\footnote{See definitions in introductory chapter \ref{ch:Intro}.}, undergo a non--perturbative hadronisation phase yielding colour singlet hadrons.
The latter (massive) hadrons are produced either at ground or excited state. The
excited state hadrons, or ``resonances'', decay to lighter ones, at ground
state, with high (low) longitudinal (transverse) momentum, with respect to the
direction of the outgoing resonance. At high centre--of--mass energies, $\sqs \gg$ EW scale $\sim 90 \GeV$, massive particles, both elementary and bound--state
resonances in SM and BSM, decay into lighter particles that are highly boosted
in the lab frame. The resultant final jets have a different topology to the
unboosted case. This field has seen substantial developments in the very recent
years (see e.g., the {\sc boost\fs11} review \cite{Abdesselam:2010pt}).

At leading--order, parton jets are infinitely narrow and only start to build non--vanishing width at next--to--leading order. The showering and hadronisation phases contribute towards the increase of the latter width, thus smearing out the energy distribution within the jet (a ``splash--out'' effect \cite{Ellis:2007ib}). The UE and PU, on the other hand, give rise to a ``splash--in'' effect whereby extra (soft) energy is added to the jets. The final hadron (jets) eventually pass through the various layers of the detector where they undergo some interactions resulting in -yet another level of complexity- ``calorimeter'' jets, shown at the top of \fig{fig:Jets:JetFormation}, which embodies the detector response. The latter (detector-level jets) is beyond the scope of the present thesis and we refer the interested reader to Ref. \cite{Ellis:2007ib} for more information. Hitherto, we have presented no formal definition of a jet that applies at all levels; parton, shower, hadron and detector. 
We do so in \sec{sec:Jets:JetAlgortihms}.
Once a jet is properly defined, we then exploit its usage in understanding QCD through exploring its ``internal'' and ``external'' properties in \sec{sec:Jets:JetShapes}. We succinctly summarise the main points in \sec{sec:Jets:Summary}.

It is worth noting that the field of jets (and related topics) has seen a
substantial growth over the last few years, mainly due to the fact that the
final state at the LHC is predominantly jets. Different research groups tackled
the issue from different angles and with different aims (understanding QCD,
background--signal discrimination, $\cdots$). A sample of literature reviews on the topic may be found in Refs. \cite{Ellis:2007ib, Abdesselam:2010pt, Altheimer:2012mn, Banfi:2004yd, Dasgupta:2007wa, Seymour:1995gq, Campbell:2006wx, Salam:2009jx, Zhang:2012rz} (and references therein).

\section{Jet clustering algorithms}
\label{sec:Jets:JetAlgortihms}

The measured kinematics (four--vectors) of ``jets'' at the detector level (from energy depositions into calorimeter cells and tracks of charged particles) provide a measure of the kinematics of the underlying short--distance theory partons and/or decaying massive (boosted) particles\footnote{After effects of FSR, ISR, hadronisation, UE and PU are properly estimated and corrected for at a sufficiently reliable precision level.} \cite{Ellis:2007ib, ATLAS-TDR-015, Froidevaux:2006rg, Aad:2009wy, Bayatian:2006zz}. Jets are not, however, intrinsically well--defined objects, just as partons are not well--defined due to their infinite branching probabilities in pQCD, and thus a prescription for defining them is required. A wide variety of ``jet definitions'' have been proposed in the literature (for a review see, for instance, Refs. \cite{Buttar:2008jx, Salam:2009jx}). A typical jet definition consists of a \emph{jet algorithm} and a \emph{recombination scheme}. A jet algorithm is essentially a set of mathematical 
rules that details the procedure of merging (``daughter'') objects into ``protojets'' (parent objects) based on a measure of their ``closeness''. The objects, described by their $4$--vectors, refer to partons in perturbative final state, final state hadrons in the simulated Monte Carlo events or calorimeter output at detectors. The recombination scheme specifies how the kinematics of the parent protojets are constructed from those of the daughters. 

Different jet definitions implement different jet algorithms and/or recombination schemes with different strengths and weaknesses. Some important general properties that should be fulfilled by a jet definition for it to be practically useful were set out by the ``Snowmass accord'' \cite{Huth:1990mi}. They include, along others, the crucial requirement of infrared and collinear (IRC) safety of the jet algorithm. The presently available jet algorithms may be divided into two main categories, depending on the manner they operate: cone algorithms and sequential recombination algorithms. In what follows below, we first provide a brief description of the first type and take it as an example to address the issue of IRC safety. We then consider the second type of jet algorithms. Notice that our calculations in this thesis are confined to the latter type of algorithms and they are thus discussed in more detail (than cone algorithms).

\subsection{Cone algorithms}
\label{sssec:Jets:ConeAlgs}

The first appearance of a jet algorithm was in the pioneering work of Sterman
and Weinberg \cite{Sterman:1977wj}, wherein they calculated the (partial) cross
section for $\EE$ hadronic events, in which all but a fraction $\ep \ll 1$ of the
total $\EE$ energy is emitted within some pair of oppositely directed \emph{cones} of half--angle $\de \ll 1$. This was the first basic definition of a cone algorithm in terms of two parameters; energy and angle.

A modern cone algorithm operates in two distinct steps. In the first step it searches for \emph{stable} cones\footnote{A stable cone is a cone whose axis coincides with the axis of the sum of all three--momenta of particles it contains.}. In the second step it runs a \emph{split--merge} procedure to resolve any overlapping stable cones. The outcome of this procedure is that each object, in the initial list, is assigned to a single jet in the final list. Insofar the only IRC safe cone algorithm available is the seedless infrared safe cone (SISCone) algorithm developed in Ref. \cite{Salam:2007xv}. A detailed description of various cone algorithms may be found in the review \cite{Salam:2009jx}.

\subsection{IRC safety of jet algorithms}
\label{ssec:Jets:IRCSafety}

As mentioned in the outset, a generic jet algorithm clusters objects that are
close to each other (in phase space). This feature guarantees that divergent
perturbative contributions from virtual and soft and collinear real emissions
cancel out. This implies that the number of hard jets found by the algorithm
must remain intact when an object undergoes a collinear splitting or a soft
emission is added to the ensemble of the initial objects. It is only through
this real--virtual cancellation that the algorithm can serve to define finite
quantities, such as cross section. Furthermore, being insensitive to collinear
splittings means that the jet algorithm will not significantly be affected by non--perturbative processes such as fragmentation and decay of hadrons (collinear splittings). Also, being insensitive to soft radiation helps in minimising exposure to contaminations from UE and PU \cite{Salam:2009jx}.    

\begin{figure}[t]
 \centering
 \includegraphics[width=7.4cm]{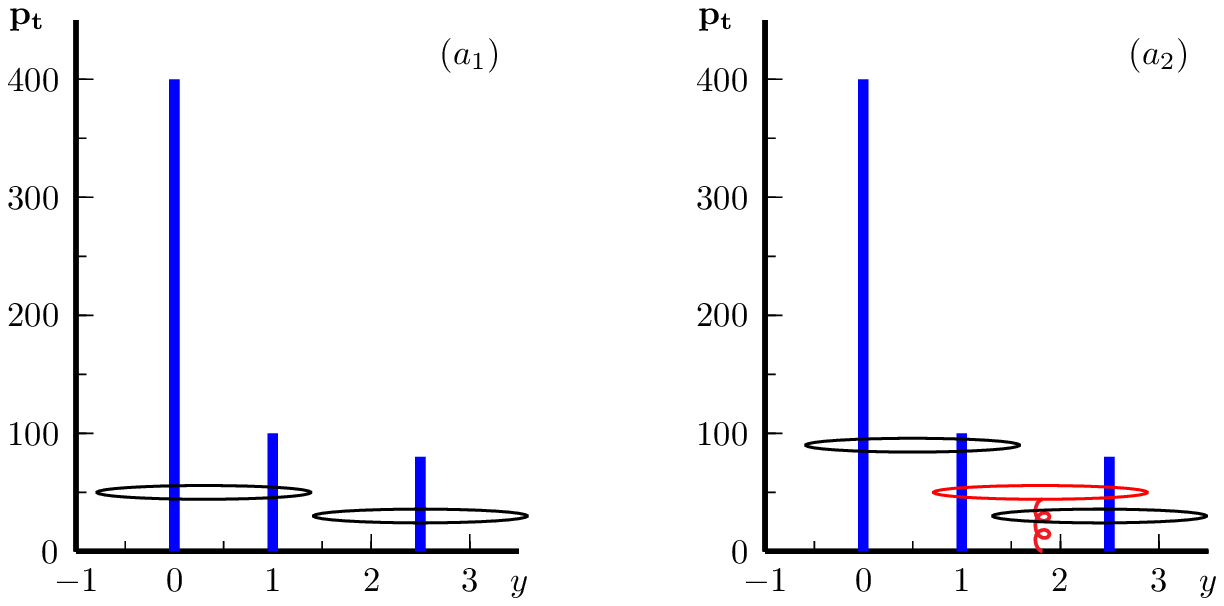}
 \includegraphics[width=7.4cm]{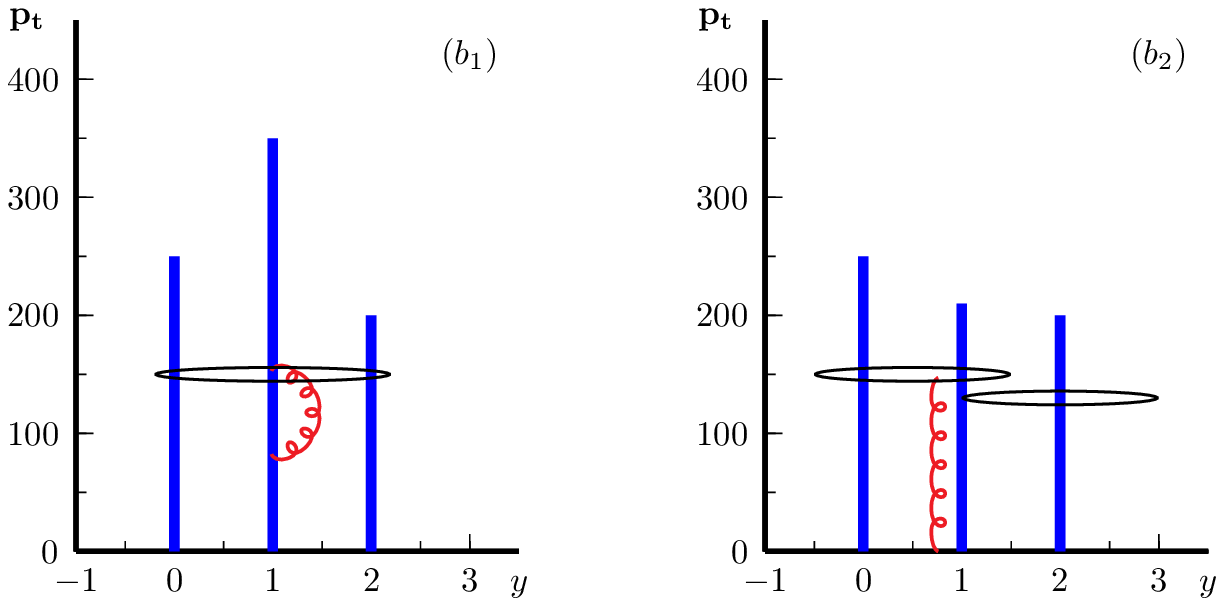}
 \caption{Figures $(a_1), (a_2)$ illustrate the output of a an IR unsafe cone
algorithm and figures $(b_1), (b_2)$ illustrate the output of a collinear unsafe
cone algorithm. The plane $(\mb{p_ t}, y)$ is the 2D projection of the 3D phase
space $(\mb p_t, y, \phi)$.}
\label{fig:Jets:IRCSafetyOfJetAlgs}
\end{figure}

\fig{fig:Jets:IRCSafetyOfJetAlgs} schematically illustrates IRC unsafe jet
algorithms of cone type, where stable cones are found by iterating from \emph{all} objects (seeds) and then running a split--merge procedure, as described above. The unsafety is manifested in the change of the number of final jets found by the algorithm. In \fig{fig:Jets:IRCSafetyOfJetAlgs} $(a_1)$, the leftmost and middle objects are within a distance $R$ while the rightmost is more than a distance $R$ away from the middle object. The said cone algorithm, with radius $R$, finds two stable cones with no overlapping and hence two final jets. Adding a soft object to the ensemble, as in \fig{fig:Jets:IRCSafetyOfJetAlgs} $(a_2)$, leads to three overlapping stable cones. The split--merge procedure may then easily lead to the identification of three final jets. This is IR unsafety.

Now consider a cone algorithm that iterates from the hardest object first. In \fig{fig:Jets:IRCSafetyOfJetAlgs} $(b_1)$, the middle object is the hardest and hence sets the first seed. Given that the other two objects are within a distance $R$, the algorithm finds a single stable cone and thus one final jet. If the middle object undergoes a collinear splitting, as in \fig{fig:Jets:IRCSafetyOfJetAlgs} $(b_2)$, then the hardest object becomes the leftmost and hence sets the first seed. Iteration leads to a jet containing the leftmost and middle objects only. The remaining rightmost object provides a new seed and goes on to form a separate second jet. This is collinear unsafety. 
It is crucial to note that perturbative calculations based on IRC unsafe jet
algorithms yield infinite answers at some order, say $\as^n$, in the
perturbation expansion. In our example, \fig{fig:Jets:IRCSafetyOfJetAlgs} $(b_1)$ and $(b_2)$, the algorithm fails real-virtual cancellation at the first order, $\as$. Thus, at and beyond this particular order ($n$) the perturbative series is not meaningful since all terms are as large as the $(n-1)^{\rm th}$ term and cannot be ignored. Therefore we shall exclusively work with IRC safe algorithms when performing any calculation throughout this thesis. Moreover, all the jet algorithms that we shall be discussing and many others are implemented in the \fastjet package \cite{Cacciari:2011ma}. 
 
Next we discuss the second type of jet algorithms, which are both much simpler and IRC safe.

\subsection{Sequential recombination algorithms}
\label{sssec:Jets:SeqRecAlgs}

The JADE algorithm \cite{Bartel:1986ua} was the first practically
useful\footnote{As mentioned in \cite{Salam:2009jx} earlier original, and
somewhat complicated, work on sequential recombination algorithms and related
properties appeared in e.g., Refs. \cite{Dorfan:1980gc, Sjostrand:1982am}.}
version of a sequential recombination (SR) algorithm applied on $\EE$ hadronic
multijet final states. SR algorithms proceed onto building final jets by
recursively recombining pairs of objects in a $2 \ra 1$ sequence. At the
perturbative level they work in a ``reverse'' fashion to parton branching ($1
\ra 2$ sequence). This -clustering--branching- analogy is fostered further by the
fact that the clustering procedure, of an SR algorithm, takes on an approximate
probabilistic picture, just as the branching process does. These and other
discernible characteristics have made SR algorithms a powerful tool not only to
find jets but to carry out detailed analyses on the internal structure of jets.
The latter  ``jet substructure'' analyses, based on smart utilisations of SR algorithms along with some other methods, have grown sufficiently large and deserve a separate section (\ssec{ssec:Jets:jetGrooming}).

Modern SR algorithms come in a rich spectrum. They share a common strategy
though. In Algorithm~\ref{Alg:SRAlgInclusive} we present the details of an
\emph{inclusive} variant of a typical SR algorithm that depends on a radius $R$
and a continuous parameter $\rm p$. Such algorithm is referred to in
\cite{Cacciari:2011ma} as ``generalised $k_t$'' algorithm. The \emph{exclusive}
variant of Algorithm~\ref{Alg:SRAlgInclusive} is similar except that:
\begin{enumerate}[1.]
 \item After step $5$ (and before step $6$): If $d_{\min} < d_{\rm
cut}$, where $d_{\rm cut}$ is a resolution cut--off parameter, then continue to step $6$. Otherwise, declare all remaining objects to be final jets and terminate the iteration.
 \item Step $9$: if $d_{\min}$ is a $d_{iB}$ then object $i$ is considered to be
part of the ``beam jet'' and is thus discarded. Return to step $1$.
\end{enumerate} 
\begin{algorithm}[!t]
\caption{Inclusive sequential recombination algorithm (generalised $k_{\rm T}$
\cite{Cacciari:2011ma})} \label{Alg:SRAlgInclusive}
\begin{algorithmic}[1]
\REQUIRE list of objects $\{p_i^\mu\}$, jet-radius $R$, real parameter $\rm p$.
\ENSURE $p_i^\mu$ is in terms of ($\pt{ti}, y_i, \phi_i$), all measured with respect to the beam direction.
\medskip
\STATE Put the set of current objects equals to the set of all objects in the
event.
\REPEAT 
 \STATE For each object compute its distance to the beam $B$
        \begin{equation}
            d_{iB} = \pt{ti}^{2\rm p}.
        \end{equation}
 \STATE For each pair of objects $i, j$ compute the pairwise distance metric
        \begin{equation}
           d_{ij} = \min\cbr{\pt{ti}^{2\rm p}, \pt{tj}^{2\rm p}} \frac{\De
R^2_{ij}}{R^2},
        \end{equation}
       where $\De R_{ij}^2 = \cbr{y_i-y_j}^2 + \cbr{\phi_i-\phi_j}^2$.
 \STATE Find the minimum $d_{\min}$ of all $d_{ij}$ and $d_{iB}$.
 \IF{$d_{\min}$ is a $d_{ij}$}
    \STATE merge objects $i$ and $j$ into a single protojet using a
recombination scheme and add (remove) the protojet (objects $i$ and $j$) to
(from) the current list of objects.
 \ELSE\IF{$d_{\min}$ is a $d_{iB}$}
       \STATE declare object $i$ to be a final jet and remove it from the list of
current objects. 
   \ENDIF
 \ENDIF
\UNTIL{No more objects left.}
\RETURN list of final jets.
\end{algorithmic}
\end{algorithm}
Note that the radius parameter $R$ is usually taken to be of order $1$ in the exclusive variant. 

The inclusive SR algorithms that will be considered in this thesis correspond to the following special choices of the parameter $\rm p$:
\begin{itemize}
 \item $\rm p = -1$: \AKT algorithm \cite{Cacciari:2008gp}.
 \item $\rm p = \;\;\;0$: Cambridge/Aachen (C/A) algorithm
\cite{Dokshitzer:1997in, Wobisch:1998wt}.
 \item $\rm p = \;\;\;1$: \KT algorithm \cite{Catani:1993hr}.
\end{itemize}
It should be noted that unlike the \KT (and C/A), which proceeds by clustering
soft objects first (or via energy--independent clustering in C/A), the \AKT proceeds by first clustering soft objects to the harder ones in the neighbourhood. As such the \AKT algorithm is similar in spirit to a cone algorithm with the exception that it maintains IRC safety due to its hierarchical nature. While offering a valuable replacement for the cone algorithm\footnote{Note that calculations with cone algorithms are far simpler than with typical SR algorithms. The \AKT retains the simplicity of the cone algorithms while dispelling their IRC culprit.}, the substructure defined by the \AKT algorithm cannot be mapped onto that of QCD jets resulting from parton branching. As a consequence of this, care should be exercised when carrying out substructure studies using the \AKT algorithm.

Each of the above listed algorithms has some strengths and weaknesses, which are
highlighted in \cite{Salam:2007xv}, and using a variety of them for physics
analyses is highly desirable as concluded in \cite{Ellis:2007ib}. Lastly it is worth mentioning that a flurry of variants of SR algorithms\footnote{Some with extra parameters that need, in most cases, to be tuned to the specific physics analysis being carried out.} have been developed over the past few years. These include, for instance, flavoured $k_t$ algorithm \cite{Banfi:2006hf}, variable--$R$ algorithm \cite{Krohn:2009zg} and \arclus ($3 \ra 2$ clustering sequence) algorithm \cite{Lonnblad:1992qd}.

\subsection{Jet grooming techniques}
\label{ssec:Jets:jetGrooming}

Jet grooming algorithms, which may be regarded as variants of SR algorithms,
were largely motivated by UE and PU investigations. Soft radiation coming from
the latter non--perturbative processes and which end up in jets may be
``groomed'' (mitigated) away through modifying the substructure of the
respective jets. Each grooming algorithm operates on a single jet and produces a
``groomed'' jet. The latter contains less junk--radiation and is thus of
improved resolution, leading to more accurate extractions of its properties. Popular jet grooming algorithms include: filtering \cite{Butterworth:2008iy, Rubin:2010fc}, trimming \cite{Krohn:2009th} and pruning \cite{Ellis:2009me}. Although all three methods aid in better identifying the origin of a particular jet, trimming was originally proposed for use on jets from QCD light quarks while filtering and pruning have typically been used on jets from decay of (boosted) heavy particles \cite{Abdesselam:2010pt}. 

As mentioned earlier, jet grooming is one of the main jet substructure techniques devised to boost new physics searches in a jetty environment such as that experienced at LHC. Other substructure tools and methods, aside from jet shapes which we address below, include: Y--splitter \cite{Brooijmans:1077731}, various taggers \cite{Plehn:2010st, Kaplan:2008ie, Giurgiu:2009wv, Kribs:2009yh,  Kribs:2010hp, Krohn:2011zp}, shower deconstruction \cite{Soper:2011cr}, three--body kinematic variables \cite{Thaler:2008ju} and many others (see Refs. \cite{Abdesselam:2010pt, Altheimer:2012mn}\footnote{See also the {\sc boost\fs12} page at ``\url{http://ific.uv.es/~boost2012/}''.}). Most of these substructure tools have been incorporated into the \fastjet package\footnote{\spartyjet \cite{Delsart:2012jm} is another software/package built around \fastjet and provides a variety of other jet--oriented tools including measurements of some event and jet observables.}. This package, written in a ``jet--oriented'' manner, has 
served as a framework for the development of a whole branch of study that is primarily concerned with jets: ``jetography''. The reader is 
referred to Ref. \cite{Salam:2009jx} for more details.

\section{Jet shapes}
\label{sec:Jets:JetShapes}

Event shape observables (or simply event shapes), such as thrust, characterise
the geometrical properties of the energy flow in QCD final state events. For example, a pencil--like event produces a value of the (thrust) event shape that is different from that produced by a sphere--like event. Event shapes are constructed by taking a weighted sum of the four--momenta, transverse momenta, energies etc of final state particles\footnote{Particles here refer, as previously discussed in
\sec{sec:Jets:JetAlgortihms}, to: calorimeter towers, clusters or charged tracks
at the detector level, particles at hadronic level, many partons at the shower
level and few partons at perturbative level.}. While imparting information on the \emph{global} shape of an event, hence dubbed \emph{global (event) shapes}, event shapes convey no information regarding the jet content of the event, such as energy distribution within individual jets.

Event shape--like observables that are sensitive to the properties of single jets in an event are known as \emph{jet shapes}\footnote{The term ``jet shapes'' originally devised for the observable $\psi(r;R)$ which measures the fraction of energy confined in an annulus of width $dr = R-r$ \cite{Seymour:1997kj}.}. They can serve to quantify a variety of topologies in the final state, from pure QCD--like to EW bosons' two--pronged decays to top quark (and other proposed new particles) three--pronged decays\footnote{More information on two-- and three--pronged decays may be found in \cite{Salam:2009jx}.}. Each of the latter mentioned topologies yields a different value for the jet shape. Jet shapes are by construction of \emph{non--global} nature, as they only measure specific parts, rather than the whole, of the the event. Such a non--global property, besides making jet shapes more delicate, opens up the door for a better scrutiny of QCD dynamics. 

A trend of jet observables, under the broader class of jet substructure, have
been devised in the recent years. These include, for instance: N--subjettiness
\cite{Thaler:2010tr}, jet width and jet eccentricity \cite{Chekanov:2010vc}, jet pull \cite{Gallicchio:2010sw} and many others, e.g., \cite{Hook:2011cq, Almeida:2008yp, Almeida:2008tp, Chekanov:2010gv, Almeida:2010pa, Jankowiak:2011qa, Gallicchio:2011xq} (see Refs. \cite{Abdesselam:2010pt, Altheimer:2012mn} for a review). To facilitate later discussions, we concentrate in this introductory review on the jet shape named \emph{angularities} \cite{Ellis:2010rwa, Almeida:2008yp} (originally introduced in \cite{Berger:2002ig, Berger:2004xf} for two-jets in $\EE$ annihilation), which embodies a spectrum of other jet shapes through a continuous parameter $a$ and has the functional form:
\begin{eqnarray}
 \ta &=& \frac{1}{2 \EJ} \sum _{i\in\,\rm J} \kt{ti}\, e^{-\eta_i(1-a)},
\label{eq:Jets:Angularities}
\end{eqnarray}
where $\kt{t} = \abs{\vect{k}_t}$ is the transverse momentum relative to the jet direction and $\eta = -\ln\tan(\theta/2)$ is the (pseudo-)rapidity measured from the jet axis. $\EJ \sim Q/2$ is the energy of the jet $\rm J$ (which will be replaced by the transverse momentum of the jet in hadron--hadron scattering) and $Q$ is the hard scale. The parameter $a$ runs over the range $-\infty < a < 2$ due to IRC safety (see below). For $a=0$ the angularities jet shape \eqref{eq:Jets:Angularities} reduces to the jet mass fraction, defined by
\be
 \rho = \frac{\MJ^2}{Q^2} = \frac{1}{Q^2} \cbr{\sum_{i\in\,\rm J} k_i}^2,
\label{eq:Jets:JetMassDefn}
\ee
where $k_i$ is the four--momentum of particle $i$. The case $a=1$ corresponds to jet broadening ($B$) \cite{Rakow:1981qn}. It has been shown in \cite{Dokshitzer:1998kz} that recoil of the primary quarks (in $\EE$ two-jet events)\footnote{Recall that the quark transverse momentum induced via recoil is (assuming the quark to be in the right hemisphere): $\vect{p}_t = -\sum_{i\in \mc H_R} \vect{k}_{ti}/Q$, where $\mc H_R$ is the right hemisphere ($\eta_i>0$).} against soft gluons contributes to the $B$-distribution at single logarithmic (SL) accuracy. As such for $a = 1$ in $\ta$ \eqref{eq:Jets:Angularities} recoil effects cannot be neglected should we aim to compute the $\ta$ distribution at the said accuracy. In fact, for this value of $a$ ($a=1$) $\ta$ is independent of $\eta$ and contributions of hard collinear radiation to the angularities jet shape are equal to contributions from soft wide-angle radiation (both contribute at SL accuracy). For $a>1$ the jet shape $\ta$, which becomes $\propto e^{+\eta_i}$, 
receives \emph{larger} contributions from hard collinear radiation than from soft wide-angle radiation. Since a full treatment of recoil effects is beyond the scope of this thesis\footnote{We refer the reader to eg., \cite{Banfi:2004yd, Dokshitzer:1998kz} for details on the treatment of recoil effects in event/jet shape studies.}, we restrict our analyses to the case $a<1$ throughout.

A somewhat related general form of jet shapes was introduced in \cite{Banfi:2004yd}:
\be
  V\cbr{\{\wt{p}\}, k} = d_\ell \cbr{\frac{\kt{t}^{(\ell)}}{Q}}^{a_\ell} e^{-b_\ell \eta^{(\ell)}} g_\ell\cbr{\phi^{(\ell)}},
 \label{eq:Jets:GenJetShape}
\ee
where $\{\wt{p}\}$ represents the Born momenta after recoil against emission
$k^\mu$, with transverse momentum $\kt{t}^{(\ell)}$, (pseudo-)rapidity $\eta^{(\ell)}$ and azimuthal angle $\phi^{(\ell)}$ measured with respect to the emitting parton leg $\ell$. IRC safety implies that $a_\ell > 0$ and $a_\ell + b_\ell >0$ \cite{Banfi:2004yd, Berger:2002ig, Berger:2003iw}\footnote{Clearly if $a_\ell < 0$ then $V \propto Q/k_t^{(\ell)}$ and hence the jet shape is IR unsafe. Moreover, the contribution of any particle to the jet shape behaves as $ k_t^{(\ell)} e^{-b_\ell\eta^{(\ell)}} = \om \sin\theta^{(\ell)} e^{-b_\ell \eta^{(\ell)}} \sim \om e^{-(a_\ell + b_\ell) \eta^{(\ell)}}$ in the collinear limit. Collinear safety ($V$ should stay finite as $\eta^{(\ell)} \ra +\infty$) then implies that $a_\ell + b_\ell >0$.}. The angularities \eqref{eq:Jets:Angularities} then corresponds to the specific choice: $a_\ell = d_\ell = g_\ell(\phi) = 1, Q = 2\EJ$ and $b_\ell = 1-a$ with emissions restricted to be within the jet reach. The said IRC conditions on $V$ then translate into $a < 2$ for $\ta$ \
eqref{eq:Jets:Angularities}.

\subsubsection{IRC and recursive IRC safety}
\label{sssec:Jets:rIRCsafety}

In a similar fashion to IRC safety of a jet algorithm, discussed in
\ssec{ssec:Jets:IRCSafety}, IRC safety of an observable means that
\cite{Banfi:2004yd} given an ensemble of partons then the addition of a
relatively much softer and/or collinear splitting of one (or more) of those
partons should not significantly change the value of the observable. Precisely,
the change should not be more than a positive power of the softness/collinearity
of the splitting(s), normalised to the hard scale. Recursive IRC (rIRC) safety conditions slightly, but crucially, extend those of ``plain'' IRC and may be formulated as in Eqs.~(3.3), (3.4) and (3.5) of Ref. \cite{Banfi:2004yd}\footnote{Instead of rewriting the full expressions here we rather use them to carry out an explicit example below.}. rIRC safety not only ensures the validity of perturbative expansion of the corresponding jet shape distribution but also ensures the ``resummability'', or equivalently exponentiation, of the latter perturbative series (see below, \ssec{ssec:Jets:Resummation}).

Let us verify, for instance, that the angularities $\ta$ jet shape
\eqref{eq:Jets:Angularities} is an rIRC safe observable. If we introduce the variable
\be
\wt{\tau}_a\cbr{\{\wt{p}\}, k} = \frac{k_t}{Q} e^{-\eta(1-a)},
\label{eq:Jets:AngularsGenForm_1}
\ee
and use Eq.~(3.3) of \cite{Banfi:2004yd}, which states that $\wt{\tau}_a(\{\wt{p}\}, k_i(\ld_i)) = \ld_i$ where $k_i(\ld_i)$ are momentum functions that depend on the parameters $\ld_i$, then the angularities $\ta$ in \eqref{eq:Jets:Angularities} reads:
\be
\ta \equiv \bta\cbr{\{\wt{p}\}, k_1(\ep\ld_1), \cdots, k_m(\ep\ld_m)} = \sum_{i\in\rm J}^m \wt{\tau}_a\cbr{\{\wt{p}\}, k_i(\ep\ld_i)} = \ep \sum_{i\in\rm J}^m \ld_i,
\label{eq:Jets:AngularsGenForm}
\ee
where $\ep \ll 1$ is a small parameter. Applying the conditions of rIRC safety given in Eqs. (3.4), (3.5a) and (3.5b) in \cite{Banfi:2004yd} we find: (a) for the first condition given in (3.4) we have
\be
\lim_{\ep\ra 0} \frac{1}{\ep}\,\bta\cbr{\{\wt{p}\}, k_1(\ep\ld_1),\cdots,
k_m(\ep\ld_m)} = \sum_{i\in\rm J}^m \ld_i,
\label{eq:Jets:rIRCLimitsOnAngulars_1}
\ee
(b) for the second condition given in (3.5a) we have
\be
\lim_{\ld_{m+1}\ra 0} \lim_{\ep\ra 0} \frac{1}{\ep}\, \bta\cbr{\{\wt{p}\},
k_1(\ep\ld_1),\cdots, k_{m+1}(\ep\ld_{m+1})} = \lim_{\ld_{m+1}\ra 0}
\sum_{i\in\rm J}^{m+1} \ld_i = \sum_{i\in\rm J}^m \ld_i.
\label{eq:Jets:rIRCLimitsOnAngulars}
\ee
and finally (c) for the third condition given in (3.5b) we have 
\begin{multline}
\lim_{\mu\ra 0}\lim_{\ep\ra 0} \frac{1}{\ep}\, \bta\cbr{\{\wt{p}\},
k_1(\ep\ld_1),\cdots, \{k_{a_\ell},k_{b_\ell}\}(\ep\ld_\ell,\mu),\cdots, k_m(\ep\ld_m)}
\\ = \lim_{\mu\ra 0} \Bigg[\sum_{i\in\rJ, i\neq\ell}\ld_i + \mu\ld_\ell + (1-\mu)\ld_\ell\Bigg] = \sum_{i\in\rJ}\ld_i,
\label{eq:Jets:rIRCLimitsOnAngulars2}
\end{multline}
where $\mu^2=\cbr{k_{a_\ell} + k_{b_\ell}}^2/k_{t\ell}^2$ and $\lim_{\mu\ra 0} k_{a_\ell} + k_{b_\ell} = k_\ell$. \eqss{eq:Jets:rIRCLimitsOnAngulars_1}{eq:Jets:rIRCLimitsOnAngulars}{eq:Jets:rIRCLimitsOnAngulars2} proves that $\ta$ is an rIRC safe observable. Consequently the $\ta$ \emph{perturbative distribution} can be resummed into the form \eqref{eq:Intro:exponentiation}. 

In the next section we first define the (integrated) jet shape distribution and  present an explicit calculation at one-gluon level. Then we elaborate on some of the important new contributions that crop up at two-gluon level and are related to the non-global nature of $\ta$.

\subsection{Jet shape distributions}
\label{sssec:Jets:JetShapeDistribn}

\begin{figure}[!t]
 \centering
 \includegraphics[width=0.45\textwidth]{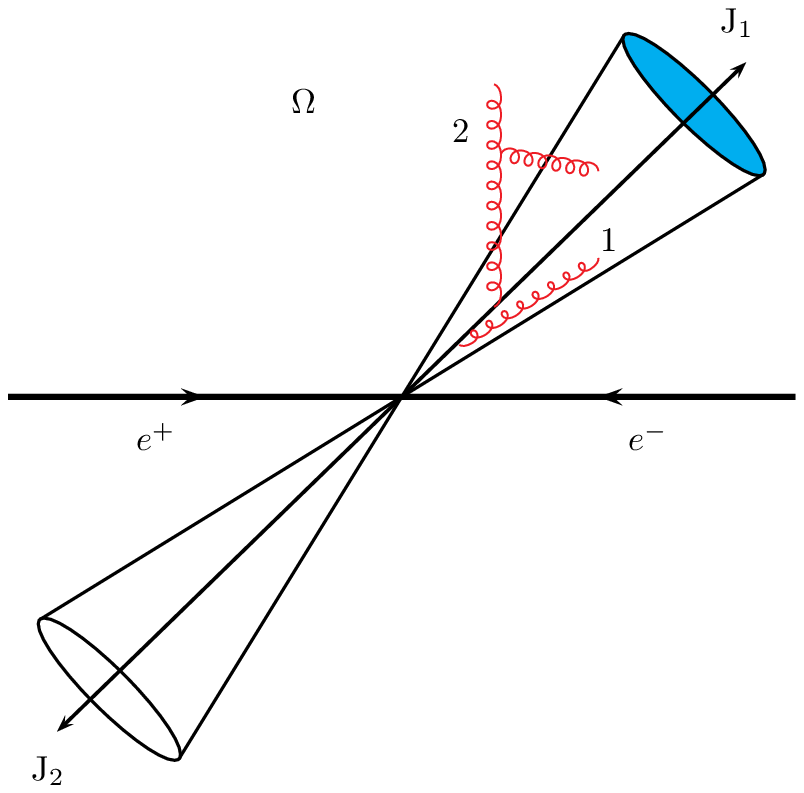}
 \hspace{0.5cm}
 \includegraphics[width=0.45\textwidth]{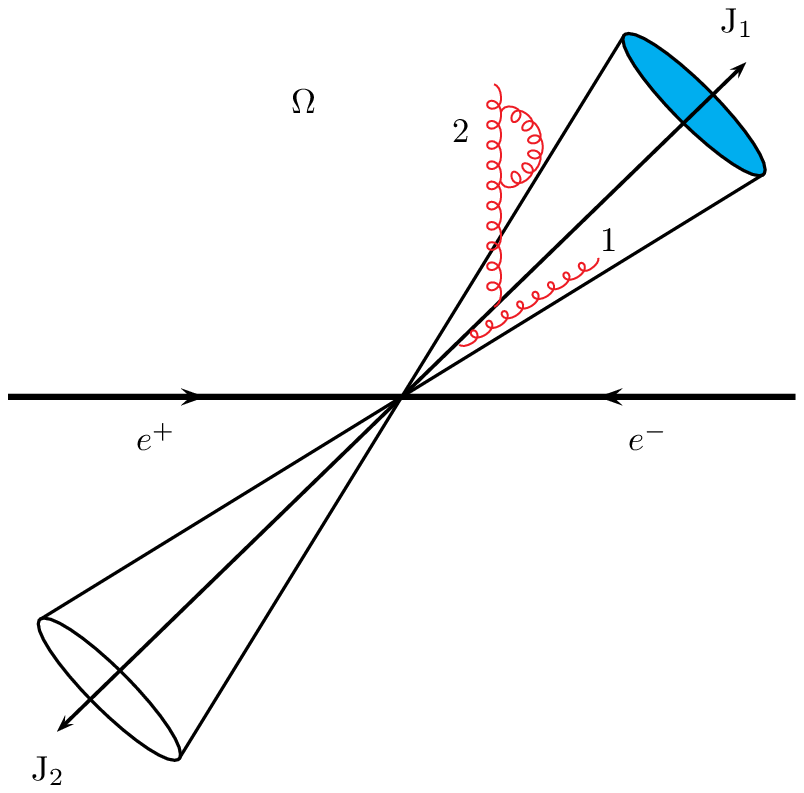}
 \caption{Sources of global (configuration 1) and non--global (configuration 2) logarithms in dijet events at fixed--order. For configuration 2, we show a real secondary emission (left) and its virtual correction (right).}
 \label{fig:Jets:EEdijets}
\end{figure}

We define the \emph{integrated jet shape distribution}, or simply the
\emph{shape fraction}\footnote{Also called jet shape cross section.}, $f(\ta)$, as the fraction of events for which the observable is smaller than a given value $\ta$ \cite{ellis2003qcd}:
\be
 f\cbr{\ta} = \frac{1}{\cSup{\s}{0}} \int^1_0 \d\s \,\Theta\sbr{\ta - \bta\cbr{\{\wt{p}\}, k_1,
\cdots, k_n}} \Xi^{\rm JA}(k_1,\cdots,k_n), 
\label{eq:Jets:ShapeFractionDefn}
\ee
where $\cSup{\s}{0}$ the is Born cross section and $\d\s$ is the total differential cross section for producing $n$ partons in the final state. The function $\Xi^{\rm JA}$ represents the resultant phase space constraint due to applying the jet algorithm ($\rm JA$) to cluster final state particles in the event. To study some of the features of the shape fraction $f(\ta)$, it is instructive to consider a specific example. For simplicity we consider two-jet production in $\EE$ annihilation process, schematically depicted in \fig{fig:Jets:EEdijets}, where the angularities shape of only one jet is measured ($\rm J_1$). Final state particles are clustered using the \AKT jet algorithm.

There are two sources of large logarithmic contributions to the shape fraction
in the region of small $\ta$ ($\ta \ll 1$): ``primary'' (or independent)
emissions, such as gluon $1$ in \fig{fig:Jets:EEdijets}, which are radiated off the hard parton (initiating the jet) into the jet vicinity. For such emissions, the corresponding Born $+$ first order correction to the shape fraction \eqref{eq:Jets:ShapeFractionDefn} reads
\be
 f(\ta^\ell) = 1 + \int\d\Phi_1 \abs{\mc M(k_1)}^2\,\sbr{\Theta\cbr{\ta^\ell -
\kt{t1} e^{-\eta_1(1-a)}/Q} - 1}\,\Xi^{\akt}\cbr{k_1},
\label{eq:Jets:AngularsGenAlpha1}
\ee
where $\ta^\ell$ represents the angularities shape of the $\ell^{\rm th}$ jet in the final state and $\d\Phi_1$ is the one--gluon phase space \eqref{eq:app:Eik:GluonPhaseSpace}. The \AKT function $\Xi^{\akt}\!$, which ensures that gluon $1$ is clustered with $\rJ_\ell$, is given by:
\be
\Xi^{\akt}\cbr{k_1} \equiv \Theta_{k_1\in\rm J_\ell} = \Theta\cbr{R^2 - \De R_{\ell i}^2},
\label{eq:Jets:AKTXiFun}
\ee
where $\De R_{ij}^2$ is defined in Algorithm~\ref{Alg:SRAlgInclusive} with $y\equiv \eta$. The $-1$ term in \eqref{eq:Jets:AngularsGenAlpha1} stands for the virtual contribution, whose matrix--element is identical (modulo a minus sign) to that of the real contribution \eqref{eq:app:Eik:EikMESquared} in the eikonal limit, as discussed in \app{sec:app:QCD:ee_example} (\ssec{sssec:app:EE:VirtualGluonCorrections}). Specialising to the centre-of-mass frame\footnote{In which $\Theta_{k_1\in\rJ_\ell} = \Theta\cbr{\eta_1 \pm \ln\tan(2/R)}$, where the pseudorapidity $\eta$ is defined in \eq{eq:app:Eik:Azimuth2Polar} and $+(-)$ is for the right (left) jet.}, it is a straightforward exercise to show that at LL accuracy and assuming small-$R$ jets (full details in the next chapter \ref{ch:EEJetShapes1}), \eq{eq:Jets:AngularsGenAlpha1} yields
\be
 f(\ta^\ell) \propto 1 - \CF\aspi\,\ln^2\cbr{\frac{1}{\ta^\ell}}.
 \label{eq:Jets:AngularsGenAlpha1Final}
\ee
Therefore primary emissions give rise to double (soft and collinear) logarithms (DL) contributions to the shape fraction. At the $n^{\rm th}$ order of the perturbative \emph{expansion} of $f(\ta^\ell)$, the contribution is of the form $\as^n L^{2n}, L = \ln(1/\ta^\ell)$. i.e., up to two logarithms per coupling $\as$. These logarithms thus exponentiate and may be resummed to all--orders in a rather simple fashion, as shall be elaborated on in
\ssec{ssec:Jets:Resummation}.

\subsection{Non--global logarithms}
\label{ssec:Jets:NGLs}

The second source of large logarithms is due to ``secondary'' (or correlated)
emissions originating from the complementary region of phase space, $\Om$
(out--of--jets region), as illustrated by configuration $2$ in
\fig{fig:Jets:EEdijets}. While real emissions contribute to the value of the jet
shape, $\ta^\ell$, their corresponding virtual corrections do not. This creates
a mismatch between real--virtual divergent contributions leaving behind
non--vanishing large logarithms in the ratio of the energy scales of
$\mathrm{J}_\ell$ and $\Om$. Such logarithms, dubbed \emph{non--global}
logarithms (NGLs) and appearing at $\Or(\as^n , n\geq 2)$, were first pointed
out in \cite{Dasgupta:2001sh}. For our specific angularities example, we have
the leading NGLs contribution to the shape fraction at $\Or(\as^2)$, in the soft
limit and including the virtual corrections, given by
\be
 f^{\NGL}(\ta^\ell) = -\int\d\Phi_1\d\Phi_2\,\abs{\mc
M(k_1,k_2)}^2\,\Theta\cbr{\kt{t2} e^{-\eta_2(1-a)}- Q\,\ta^\ell} \Xi^{\akt}\cbr{k_1, k_2},
\label{eq:Jets:AngularsAlpha2NG}
\ee
where $\Xi^\akt\cbr{k_1,k_2} = \Theta_{k_1\notin\rm J_\ell} \Theta_{k_2\in\rm
J_\ell}$ and the soft part of the matrix--element squared for the emission of
two gluons is given by the term $2\,S$ in
\eq{eq:app:EE:SoftNonAbelianAmpAlphas2}. The minus sign in \eqref{eq:Jets:AngularsAlpha2NG} is due to the fact real emissions, which contribute to the shape fraction only in the region $\bta(\{\wt{p}\}, k_2) < \ta^\ell$, completely cancel against virtual corrections, thus leaving only virtual corrections to contribute in the complementary region ($\bta(\{\wt{p}\}, k_2) > \ta^\ell$). The detailed evaluation of the phase space integral in \eq{eq:Jets:AngularsAlpha2NG} will be postponed to Chapter \ref{ch:EEJetShapes1}. For the time being, we simply state the final result. In the case where only one jet, say $\rJ_\ell$, is measured and the other is left unmeasured, we have for small-$R$ jets and assuming $\kt{t1}^2 \gg \kt{t2}^2$ (which suffices to extract the leading non--global logarithms) the leading NGLs contribution reads:
\be
 f^{\NGL}(\ta^\ell) = - S_2\,\sbr{\astpi\,\ln\cbr{\frac{1}{\ta^\ell}}}^2,
\label{eq:Jets:AngularsAlpha2NGFinal}
\ee
where we have denoted the NGLs coefficient at $\as^2$ by $S_2$, which is given by $S_2 = \CF\CA\,\pi^2/3$, i.e., independent of the jet radius (in the small-$R$ limit). \eq{eq:Jets:AngularsAlpha2NGFinal} shows that NGLs (which originate from soft emissions at large angles) are single logarithms (SL), contributing at next--to--leading log (NLL) level to the angularities shape variable. Due to the minus sign in \eqref{eq:Jets:AngularsAlpha2NGFinal}, NGLs impact on the shape fraction is thus to reduce the primary emission contribution. 

\begin{figure}[!t]
 \begin{center}
 \includegraphics[width=0.45\textwidth]{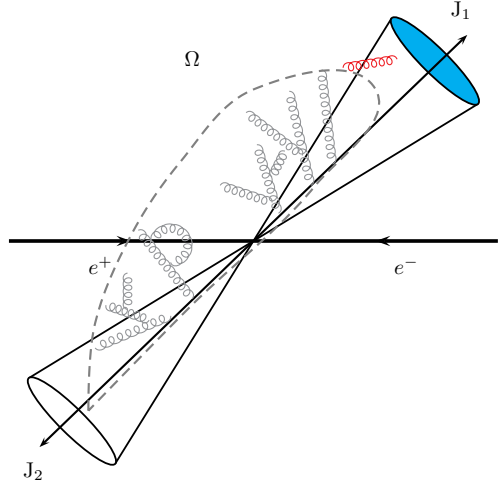}
 \caption{An into-jet $\rm J_1$ coherent radiation from an ensemble of soft large-angle gluons (that are outside $\rm J_1$). Such configurations contribute to NGLs at higher-orders.}
\label{fig:Jets:NGLsHigherOrders}
\end{center}
\end{figure}

At the $n^{\rm th}$ order in the perturbative expansion of $f(\ta^\ell)$ the non-global contribution $f^{\NGL}(\ta^\ell)$ corresponds to configurations where an ensemble of $(n-1)$ soft, wide-angle particles\footnote{To extract the leading NGLs it suffices to consider these particles to be energy-ordered; $\om_1 \gg \om_2 \gg \cdots \gg \om_{n-1}$.} outside the jet coherently emit a single softest gluon into the jet, as depicted in \fig{fig:Jets:NGLsHigherOrders}. The ensemble may assume any complicated geometry. As such it has not been possible \cite{Dasgupta:2001sh, Dasgupta:2002bw} to \emph{analytically} resum these logs to all--orders, as has the case been with primary (global) logs. However, employing the large-$\Nc$ limit one can either write an evolution equation \cite{Banfi:2002hw} or resort to a dipole evolution code \cite{Dasgupta:2001sh} to \emph{numerically} compute the resummation. The solution to the said all--orders evolution equation adheres a \emph{factorised} form \cite{Banfi:2002hw, 
Dokshitzer:2003uw}:
\be
 f(\ta^\ell) = \Sg_P(\ta^\ell)\, S(t)
\label{eq:Jets:FactorisatnOfNGLs}
\ee
where the first term $\Sg_P$ is the Sudakov form factor given by bremsstrahlung emission from primary hard partons and is discussed below. The second factor arises from successive soft secondary branchings (as described above). The parameter $t$ is known as the evolution parameter and is given by
\be
 t = \int_{e^{-L}}^{1} \frac{\d x}{x} \frac{\as(x Q)}{2\pi},
\label{eq:Jets:EvolutionParamDef}
\ee
where $x= 2\,k_t/Q$ and the logarithm $L \propto \ln(1/\ta^\ell)$ (with the exact formula of $L$ depending on the configuration under study). A widely used parametrisation of $S(t)$ is given in \cite{Dasgupta:2001sh}
\be
S(t) = \exp\sbr{ -\frac{\CAsq}{2}\,S_2\,\cbr{\frac{1 + (at)^2}{1 + (bt)^c} } t^2 + \Or(1/\Ncsq) },
\label{eq:Jets:StFittingForm}
\ee
where $S_2$ is the $\mc O(\as^2)$ coefficient and the parameters $a, b$ and $c$ are determined from fitting \eqref{eq:Jets:StFittingForm} to the output of the MC program of Ref. \cite{Dasgupta:2001sh}. Actually, for use in phenomenology one replaces the large-$\Nc$ colour factor $\CA/2$ by the corresponding full colour factor such that when expanding \eqref{eq:Jets:StFittingForm} one recovers the full $\mc O(\as^2)$ result. Note that the neglected $1/\Ncsq$ non-planar part contribute at the $10\%$ level.

In--depth discussion of this and other related features of NGLs is presented in the subsequent chapters.  It is in fact one of the central aims of this thesis to compute, as accurately as possible, NGLs at fixed--order and quantify their all--orders impact on the total resummed jet shape distribution.

\subsection{Clustering logarithms}
\label{ssec:Jets:CLs}

So far we have only dealt with jet shape distributions where final state jets
are defined with the \AKT algorithm. As emphasised earlier (see
\sec{sec:Jets:JetAlgortihms}) the \AKT algorithm works, particularly in the soft
limit, like a rigid cone. Soft partons are clustered independently to their
nearest hard neighbours. This simple picture is spoiled by other SR algorithms
such as k$_{\rm T}$, C/A as well as the SISCone algorithm, where soft partons are
iteratively clustered together, based on their closeness, and may well form
separate jets from hard partons' jets. The impact of this on the final form of
the jet shape distribution is two fold:
\begin{itemize}
 \item Significant reduction in the NGLs contribution due to tighter phase space
constraints. This was first shown in \cite{Appleby:2002ke} for
gaps--between--jets distribution.
 \item Breakdown of the naive single--gluon exponentiation (at NLL) associated with primary emissions due to the appearance of a tower of large single logarithms. This was first pointed out in \cite{Delenda:2006nf, Banfi:2005gj} for the same distribution (gaps--between--jets). We refer to these logarithms as
\emph{clustering logarithms} (CLs). They are a type of non--global logarithm as they only show--up for non--global observables. As a result of this some authors,
e.g., of Ref. \cite{Kelley:2012kj}, refer to them as \emph{abelian} NGLs.
\end{itemize}
Postponing the discussion of both effects at fixed--order and all--orders to later
chapters, we note that the leading CLs contribution to angularities shape fraction at $\Or(\as^2)$ is, for the case of one measured jet $\rJ_\ell$, of the form
\be
 f^{\rm JA}(\ta^\ell) = C_2^{P, \rm JA}\,\CFsq\cbr{\astpi}^2\, L^2,
\label{eq:Jets:AngularsAlpha2CLs}
\ee
where $L = \ln(1/\ta^\ell)$ and $C_2^{P, \rm JA}$ is the outcome of the phase space integral. For small jet radii, $R \ll 1$, $C_2^{P, \rm JA}$ only depends on the geometry of the final state (and neither $R$ nor the jet shape). As $R$ increases $C_2^{P, \rm JA}$ dependence on $R$ is restored as shall be explicitly shown in Chapter \ref{ch:EEJetShapes2}. For the \KT and \ca algorithms: $C_2^{P, \kT} = C_2^{P, \ca} \simeq \pi^2/27 = S_2/\Ncsq$ as $R \ra 0$. This relation between $C_2^{P, \kT(\ca)}$ and $S_2$ further fosters the non-global nature of CLs.

Just like NGLs, the SL clustering logs persist at every order in the perturbative expansion of the non--global jet shape distribution and must be resummed for a given phenomenological calculation to be meaningful. Although the all--orders result can currently only be obtained via the MC program of \cite{Dasgupta:2001sh}, the situation is slightly more satisfactory than with NGLs. It has been possible, for both gaps--between--jets and jet mass as discussed in Chapter \ref{ch:EEJetShapes3}, to analytically compute the first few orders  in the corresponding $g_2$ (see \eq{eq:Intro:exponentiation}) CLs--tower and evidently show, by explicit comparison to the output of the said MC program, that they capture the full behaviour of the all--orders result. Full details are presented in Chapter \ref{ch:EEJetShapes3}.

Below we outline the necessary steps for a typical resummation calculation of primary global logarithms (GLs).

\subsection{Resummation}
\label{ssec:Jets:Resummation}
             
Fixed--order perturbative calculations usually involve a small number of
additional partons to the Born configuration and can produce accurate results at
the large--$\ta$ tail region of the jet shape distribution
($\ta \lesssim 1$). Such regions of phase space describe large departures from
the Born--event energy flow pattern due to the extra partons being energetic and
at large angles. Their contributions are free from large logarithms, suppressed
by powers of the coupling $\as$ and comply with the convergence of the perturbative series. Fixed--order calculations are usually carried out with the help of Fixed--order Monte Carlos (FOMCs) such as \event \cite{Catani:1996vz}, \eerad \cite{GehrmannDeRidder:2007jk, GehrmannDeRidder:2008ug}, for $\EE$ annihilation, and \mcfm \cite{Campbell:2002tg}, \nlojet \cite{Nagy:2001fj}, for hadron-hadron scattering, to mention just a few. These programs are fully exclusive over the final state and can produce predictions for arbitrary IRC observables. The prevalent accuracy across the majority of the available FOMCs is next--to--leading order (NLO), with the exception of \eerad which reaches next--to--next--to--leading order (NNLO) for observables in $\EE$ annihilation events. 
\begin{figure}[t]
 \begin{center}
 \includegraphics[width=0.60\textwidth]{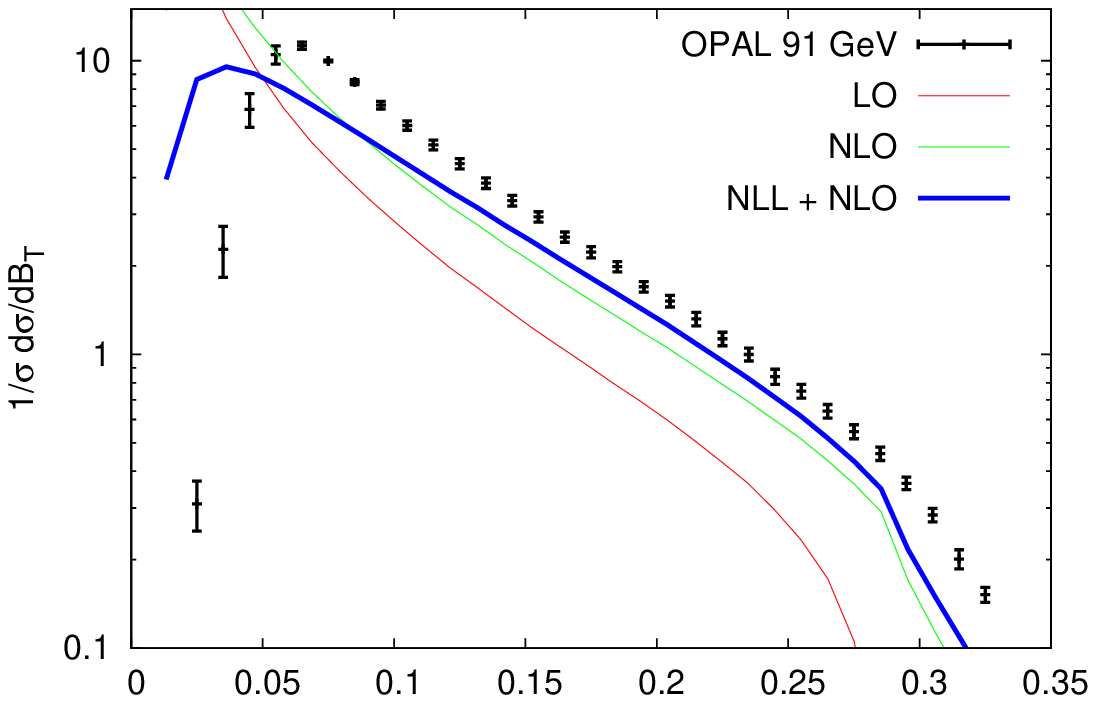}
 \includegraphics[width=0.61\textwidth]{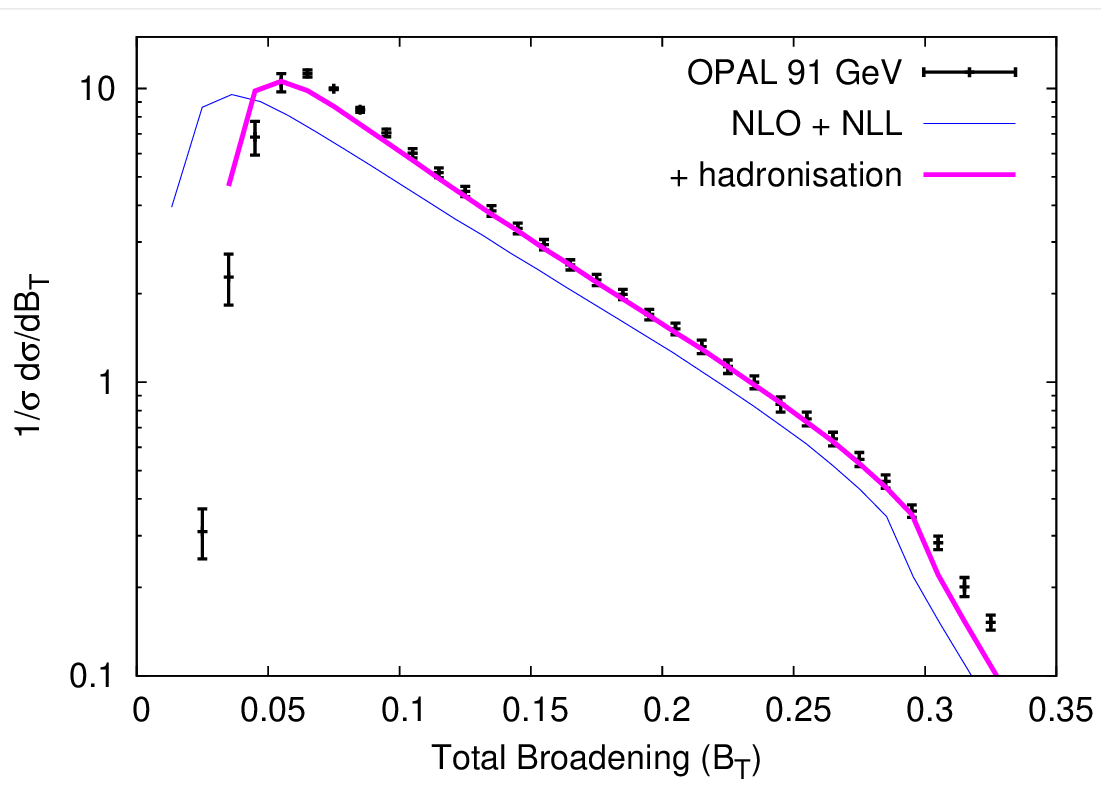}
 \caption{Fixed--order vs (matched) resummed calculations and comparison to
experimental data for the total broadening shape variable (left) without
hadronisation effects and (right) with hadronisation effects (figure from \url{http://www.lpthe.jussieu.fr/~salam/talks/}).}
\label{fig:Jets:FOvsResum}
\end{center}
\end{figure}

In the small $\ta$ region, $\ta \ll 1$, in which departures from the Born
energy--flow configuration are small due to the extra partons being soft and/or
collinear to the initial hard Born partons, and which constitutes the bulk of
final state configurations, fixed--order perturbative calculations are unreliable, as illustrated in \fig{fig:Jets:FOvsResum} (left). This
is due to the fact that each power of the coupling $\as$ is accompanied by up to
two powers of large logarithms associated with soft and collinear divergences,
as exemplified in the previous subsections. The all--orders perturbative series
of the shape fraction $f(\ta)$ \eqref{eq:Jets:ShapeFractionDefn} may be cast in
the form
\be
f(\ta) = 1 + \sum_{n=1}^\infty \sbr{\cSup{A_n}{1}\,\as^n\,L^{2n} +
\cSup{A_n}{2}\,\as^n\,L^{2n-1} + \cSup{A_n}{3}\,\as^n\,L^{2(n-1)} + \cdots },
\qquad \ta \ll 1,
\label{eq:Jets:ResumAngularsGen}
\ee
where the coefficients $\cSup{A_n}{m}$ are determined from
phase space integrals and $L$ is defined as before. In the above equation $\cSup{A_n}{1}$ is the LL coefficient, $\cSup{A_n}{2}$ the NLL coefficient and so on (in the \emph{expansion}). For many jet shape observables, such as the angularities $\ta$, the dominant LL terms \emph{exponentiate}. Similarly, the NLL terms factorise and can be resummed to all--orders \cite{Banfi:2002hw, Catani:1993hr} as shown above. However achieving this NLL accuracy is a delicate question and requires detailed understanding of the observable's analytical properties as well as the phase space integrals. We have already illustrated how correlated emissions and jet algorithms can non-trivially give rise to NLL contributions. The exponential form of $f(\ta)$ is given by
\be
f(\ta) \propto \exp\sbr{L\,g_1\cbr{\as\,L} + g_2\cbr{\as\,L} +
\as\,g_3\cbr{\as\,L} + \cdots},
\label{eq:Jets:ResumAngularsGenExp}
\ee
where the $g_i$ functions were defined in Chapter \ref{ch:Intro}. N$^m$LL resummation precisely means the analytical computation of the functions $g_1, g_2,\cdots, g_m$. While $g_1$ can, as stated above, be fully determined, only a subset of the logs building--up $g_2$ can be analytically calculated for non--global observables. NGLs and CLs, which contribute to $g_2$, are presently only obtainable (to all--orders) through numerical estimates.

\subsubsection{Factorisation and exponentiation in Mellin space}
\label{sssec:Jets:MellinSpcaseResum}

Resummation was originally performed (to NLL accuracy in $\EE$ annihilation) by factorising the jet shape fraction in Mellin space and then exponentiating the logarithmically enhanced corrections to all--orders using eikonal techniques \cite{Catani:1992ua}. Importantly, in Mellin space the total shape fraction can be obtained in the eikonal (soft) limit by exponentiating the single--parton emission shape fractions. This is a direct consequence of the factorisation of the $n$--particle (eikonal) matrix element and phase space.

The jet shape fraction \eqref{eq:Jets:ShapeFractionDefn} may be rewritten explicitly to all--orders as
\be
 f(\ta) = \sum_{n=2}^\infty \frac{1}{\cSup{\s}{0}} \int \d \cSup{\s}{n}(k_1,\cdots,k_n) \Theta\sbr{\ta - \bta\cbr{\{\wt{p}\},k_1,\cdots,k_n}}\,\Xi^{\rm JA}(k_1,\cdots,k_n),
\label{eq:Jets:ShapeFracAllOrders}
\ee
where $\d\cSup{\s}{n}$ is the exclusive cross section for the production of $n$ partons in the final state. In the eikonal limit\footnote{Recall that the eikonal, or soft, limit corresponds to $k_i \ll p_{q}, p_{\qb}$, or equivalently $(k_i p_{q}), (k_i p_{\qb}) \ll (p_{q} p_{\qb})$. The latter is equivalent to $\om_i \ll E_{q} \simeq E_{\qb} \simeq Q/2$ where $E_q, E_{\qb}$ and $\om_i$ are the energies in the centre-of-mass of the dipole $(q\qb)$ \cite{Dokshitzer:2008ia}.}, it reads for \emph{primary} emissions, off a primary hard quark--antiquark pair, (see \app{sec:app:QCD:EikonalApprox} and Refs. \cite{Bassetto1983201, Dokshitzer:2008ia})
\be
 \d\cSup{\s}{n} = \d\Phi_{m+n}\braket{m+n}{m+n} = \d\Phi_m \braket{m}{m}\,\frac{1}{n!} \prod_{i=1}^n \frac{\d\kt{ti}^2}{\kt{ti}^2}\,\d\eta_i\,\frac{\d\phi_i}{2\pi}\,\frac{\as(\kt{ti}^2)}{2\pi}\,\CF\,\om_{q\qb}(k_i),
\label{eq:Jets:NthOrderEikXsec}
\ee
where $\d\Phi_m\braket{m}{m}$ is the differential Born cross section, which factors out of the rest of the integrand and when integrated will cancel the denominator in \eq{eq:Jets:ShapeFracAllOrders}, and the factor $1/n!$ ensures that no phase space region is double counted when the partons' transverse momenta assume no particular ordering. Notice that the argument of the running $\as$ is the transverse momentum squared $k_t^2$ \cite{Dokshitzer:1995qm, Gribov:1972ri, Amati:1980ch}.
The classical antenna function $\om_{ij}(k)$ is defined in \eq{eq:app:Eik:AntennaFun}. Here we replace the energy $\om$ by the transverse momentum $k_t$\footnote{Note that different definitions of the antenna function will appear in Chapters \ref{ch:EEJetShapes2} and \ref{ch:HHJetShapes1}. Nonetheless, the corresponding calculations are, once a particular definition is chosen, consistent.};
\be
 \om_{ij}(k) = \frac{k_t^2\,\cinner{p_i}{p_j}}{\cinner{p_i}{k} \cinner{p_j}{k}}.
\label{eq:Jets:AntennaFun}
\ee
To account for hard collinear emissions, we simply replace the antenna function $\om_{q\qb}$ with the full $q \to q g$ splitting kernel
\be
 \d\eta\frac{\d\phi}{2\pi}\,\om_{q\qb}(k) \to \d z\, P_{gq}(z),
\label{eq:Jets:AntenaFun2SplitiingKernels}
\ee
where $P_{gq}(z)$ is defined in \eqs{eq:RunAlpha:SplittingFunExpansion}{eq:RunAlpha:SplittingKernelsLO} and for a gluon emitted sufficiently collinear\footnote{Such that $\theta_{\ell k} \ll \theta_{\ell\ell'}$, with $\ell,\ell' = q,\qb$.} to leg $\ell\,(\ell = q,\qb)$: $k \simeq z p_\ell + k_\perp = \kt{t}(\cosh\eta,\cos\phi,\sin\phi,\sinh\eta)$. Then, in the $q\qb$ dipole centre-of-mass system, with $p_\ell$ in the $+ z$ direction, we have $z =  \kt{t} e^{\eta}/Q$. The jet algorithm function is, for $n$ partons contributing to the shape observable of the $\ell^{\rm th}$ jet,
\be
 \Xi^{\rm JA}(k_1,\cdots,k_n) = \prod_{i=1}^n\, \Theta_{k_i \in \rm J_\ell} = \prod_{i=1}^n\, \Theta\cbr{z_i - \frac{2\kt{ti}}{Q R}},
\label{eq:Jets:JAFunAllOrders}
\ee
where the last equality follows, in the small-$R$ limit, from the fact that, for the above kinematic set-up, 
$\Theta_{k\in \rm J_\ell} = \Theta\cbr{\eta -\ln\tan(R/2)} \simeq \Theta(\eta - \ln(2/R))$. 
It only remains to write the jet shape step function in a factorised form. This is achieved through \emph{Mellin transformation} \cite{Lindelof}
\be
 \int_0^\infty \d\ta^\ell\, e^{-\ta^\ell\, N_\ell} \Theta\cbr{\ta^\ell -  \sum_{i \in \rm J_\ell} \frac{(k_{ti}/Q)^{2-a}}{z_i^{1-a}} } = \prod_{i=1} \exp\sbr{- \frac{N_\ell (k_{ti}/Q)^{2-a}}{z_i^{1-a}}},
\label{eq:Jets:ThetaMellinTrans}
\ee
where $N_\ell$ is the Mellin conjugate of $\ta^\ell$. Therefore, taking the Mellin transform of the shape fraction, $f(N_\ell)$, and accounting for virtual corrections, we write the resummed jet shape distribution in Mellin space as
\be
 \ln f(N_\ell) = \CF \int\frac{\d\kt{t}^2}{\kt{t}^2} \frac{\as(\kt{t}^2)}{2\pi} \int_0^1 \d z\, P_{gq}(z)\, \Theta\cbr{z - \frac{2\kt{t}}{Q R}} \sbr{e^{-\frac{N_\ell (k_{t}/Q)^{2-a}}{z^{1-a}} } - 1}.
\label{eq:Jets:ShapeFracAllOrdersMellinSpace}
\ee
To evaluate the above exponent to NLL accuracy we note that \cite{Catani:1992ua} both the coupling and splitting kernel must be computed to two--loops order in the $\overline{\rm MS}$ scheme, or equivalently to one--loop order in the Catani-Marchesini-Webber (CMW) scheme\footnote{It is sometimes referred to as ``Monte Carlo'' \cite{Catani:1990rr} or ``physical'' \cite{Dokshitzer:1995qm} scheme.} \cite{Catani:1990rr}, where
\be
 \as^{\rm CMW} = \as^{\MSbar}\cbr{1 + K \frac{\as^{\MSbar}}{2\pi}}.
\label{eq:Jets:CMW-MSbarRelations}
\ee
The factor $K$ is given in the $\MSbar$ scheme by\footnote{Note that the $K$ factor is related to the \emph{cusp anomalous dimension} $\G_{\rm cusp}$ \cite{Korchemsky:1987wg, Korchemskaya:1992je, Korchemsky:1993uz}, which has a perturbative expansion given by: $\G_{\rm cusp}(\as) = \CF\,(\as/\pi) + \frac{1}{2}\CF\,K\,(\as/\pi)^2+\cdots$.} \cite{Kodaira:1981nh, Davies:1984sp, Catani:1988vd}
\be
K = \CA \left(\frac{67}{18}- \frac{\pi^2}{6} \right) - \frac{5}{9} \nf.
\label{eq:Jets:KFactor}
\ee
We substitute for $\as$ in \eqref{eq:Jets:ShapeFracAllOrdersMellinSpace} the relation \eqref{eq:Jets:CMW-MSbarRelations} and evaluate the integral for $\as^{\MSbar}$ at two-loops.
Moreover, at the said accuracy one can use the approximation \cite{Catani:1990rr, Catani:1989ne}
\be
 \exp\sbr{-\frac{N_\ell (k_t/Q)^{2-a}}{z^{1-a}} } - 1 \simeq -\Theta\cbr{\frac{(k_t/Q)^{2-a}}{z^{1-a}} - \br{N}_\ell^{-1}},
\label{eq:Jets:LLAprox}
\ee
where $\br{N}_\ell = N_\ell\, e^{-\g_E}$ and $\g_E \simeq 0.577\cdots$ is the Euler gamma constant. To invert the resultant expression back into the $\ta^\ell$ space we use the inverse Mellin transformation \cite{Catani:1992ua}
\be
f(\ta^\ell) = \frac{1}{2\pi\imath} \int_C \frac{\d N_\ell}{N_\ell}\, e^{\ta^\ell\, N_\ell}\, f(N_\ell) = \frac{\exp\sbr{\mc R(\as, L) - \g_E \mc R'(\as, L)}}{\G(1 + \mc R'(\as, L))},
\label{eq:Jets:ShapeFracAllOrdersInvMellinSpace}
\ee
where the contour of integration $C$ runs parallel to the imaginary axis, to the right of all singularities of the integrand and $\G$ is the Euler Gamma function. The last equality in \eq{eq:Jets:ShapeFracAllOrdersInvMellinSpace} is valid at NLL accuracy \cite{Catani:1992ua} with $\mc R$ being analogous to $\ln f(N_\ell)$ with $N_\ell \to 1/\ta^\ell$, $\mc R' = \pa_L \mc R$ and $L$ is as before equals to $\ln(1/\ta^\ell)$. We implement these techniques to arrive at the form reported in Chapter \ref{ch:EEJetShapes1} (\eq{eq:EEJS1:RadiatorTau_B}).

\subsubsection{Matching}
\label{sssec:Jets:Matching}

To obtain predictions for the jet shape distribution that are accurate at both
ends, $\ta \ll 1$ and $\ta \lesssim 1$, one combines resummed and fixed--order
results with the aid of a ``matching'' procedure. Various matching prescriptions
have been proposed in the literature (see e.g., \cite{Dasgupta:2002dc, Catani:1993hr}). An $\rm N^m LL + N^{m'}LO$ matching procedure should fulfil the following requirements \cite{Banfi:2010xy}:   
\begin{itemize}
 \item The matched result should be correct up to $\rm N^m LL$ terms in the
exponent and the expanded matched result should be correct up to and including
$\Or\cbr{\as^n\,L^{2(n-m)}}$.
 \item The expanded matched result should reproduce the fixed--order result up
to and including $\rm N^{m'}LO$ terms.
 \item The jet shape distribution should tend to one at the maximum allowed
value of the jet shape, $\ta^{\max}$, while the corresponding differential
distribution should vanish:
 \be
   f(\ta^{\max}) = 1,\qquad \frac{\d f(\ta)}{\d\ta}\Big|_{\ta = \ta^{\max}} = 0.
  \label{eq:Jets:MatchingRequirements}
 \ee 
\end{itemize}
The matched resummed result will then be of the form stated in the introductory
chapter, \eq{eq:Intro:exponentiation}. The current state--of--the--art is $\rm
N^3LL + N^2LO$, e.g., thrust and heavy jet mass distributions \cite{Becher:2008cf, Chien:2010kc}. 

Owing to the considerable importance of resummed (and matched) results, special efforts have been devoted over the last decade to alleviate some of the limitations that are inherent in resummed calculations, for instance the need for considerable understanding of the details of the observable being resummed, manually repeating cumbersome calculations for each new observable etc. Notably the automation of the resummation programme in a semi--analytical approach with state--of--the--art resummation accuracy, clean separation between $g_i$ functions without spurious contamination from uncontrolled higher--orders and more exclusiveness over the final state. {\sc NumResum} \cite{Banfi:2001bz}, {\sc DISresum} \cite{Dasgupta:2002dc} and \caesar \cite{Banfi:2004yd} are examples of numerical programs developed specifically to achieve the aforementioned (automated resummation) goal. Despite these remarkable achievements, the above--listed programs can \emph{only} treat global observables, which comprise a subset of 
the 
observables measured in real experiments. Including the other subset, non-global observables, into the automation programme requires more effort. We hope that this thesis will be a contribution to such effort.

\subsection{Non-perturbative effects}
\label{ssec:Jets:NonPTEffects}

Analytical studies, based on the renormalon--inspired model \cite{Dokshitzer:1995zt} (and related approaches \cite{Dokshitzer:1995qm}), revealed that event shape distributions receive power corrections of the order of $1/Q$ (more precisely $1/\tau Q$ for an event shape observable $\tau$), e.g., \cite{Monni:2011gb, Webber:1994cp, Beneke:1995pq, Akhoury:1995sp, Korchemsky:1994is, Korchemsky:1997sy, Korchemsky:1999kt, Gardi:2001ny}, in the small $\ta \ll 1$ region. The coefficient of this correction is typically a product of two factors: the first is perturbatively calculable and observable dependent, and the other is fundamentally non--perturbative but universal across a range of processes. In
effect, the correction leads to a shift in the exponent of the resummed
distribution (in other words it induces a change in the logarithm of the
observable, $L \ra \wt{L}$), as indicated in \fig{fig:Jets:FOvsResum} (right).
Hadronisation, as well as UE, corrections to jet observables have also been
studied in \cite{Dasgupta:2007wa} using a variant of the aforementioned model.

The jet mass distribution, and more generally jet shape distributions,
will have to be corrected, in analogy with event shapes, by a non--perturbative
\emph{shape function} in the endpoints $\MJ^2 \ra 0\,(\ta \ra 0)$. A general relation between the perturbative distribution $f(\ta)$ and the full distribution, $f_{\rm full}$, including non-perturbative effects, $f_{\rm NP}$, may be written as
\be
 f_{\rm full}(\ta) = \int \d x f_{\rm NP}(\ta, x, \as(Q), Q) f \cbr{\ta - x/Q}.
\label{eq:Jets:PT-NPTConv}
\ee
In the two-jet region and for observables that are linear in the momenta of multiple soft particles, \ref{eq:Jets:PT-NPTConv} can be cast in a simpler form \cite{Korchemsky:1994is, Korchemsky:1997sy, Korchemsky:1999kt}
\be
f_{\rm full}(\ta) \simeq \int \d x f_{\rm NP}(x)\, f \cbr{\ta - x/Q}.
\label{eq:Jets:PT-NPTConvSimple}
\ee
In other words the non-perturbative shape function is independent of $\ta$ and $\as$. A physically interesting approximation to the latter shape function is a $\de$-function \cite{Dokshitzer:1997ew, Korchemsky:1994is}, which simply \emph{shifts} the perturbative distribution
\be
 f_{\rm full}(\ta) = f(\ta - \langle \de\tau_{a,\rm had}\rangle),
\label{eq:Jets:ShiftDiestrib}
\ee
where $\langle \de\tau_{a,\rm had}\rangle$ is the mean value of the hadronisation contribution to $\ta$\footnote{See for instance Ref. \cite{Dasgupta:2002dc} for examples on how to calculate such means for event shapes.} and $f$ is the resummed shape fraction \eqref{eq:Jets:ShapeFracAllOrdersInvMellinSpace}. Note that the shift is strictly valid to the right of the Sudakov peak. It should be emphasised that for more accurate estimations of hadronisation care should be taken regarding the effect of hadron masses, which are usually proportional to powers of $\ln Q$ (or precisely $\ln \tau Q$), for event shapes, and multiply the same perturbatively calculable
factor as the $1/Q$ corrections \cite{Salam:2001bd}. In Chapter \ref{ch:HHJetShapes1}, when estimating hadronisation corrections to the total resummed distribution we confine ourselves to the shift approximation.

Lastly it is worth mentioning that numerical Monte Carlo event generators (MCEGs), which include all the necessary ingredients for a theory--data comparison and whose predictions can be applied to any arbitrary observable, will continue to play a central role in jet (and generally QCD) phenomenology.  Popular present--day general purpose MCEGs include: \herwig \cite{Marchesini:1987cf, Bahr:2008pv}, \pythia \cite{Sjostrand:1982am, Sjostrand:2006za, Sjostrand:2007gs}, \sherpa \cite{Gleisberg:2003xi, Gleisberg:2008ta} and \ariadne \cite{Lonnblad:1992tz}. A thorough review of the physics basis of MCEGs, their main features and uses may be found in, e.g., \cite{Buckley:2011ms}. In Chapter \ref{ch:HHJetShapes1} we compare our analytical results to the first three MCEGs (mentioned above).

\section{Summary}
\label{sec:Jets:Summary}

In this chapter, and the previous one, we have outlined some of the concepts,
tools and methods applicable to QCD processes in general and to jet related
studies in specific. Asymptotic freedom and factorisation theorems of QCD
equipped with Feynman graphs, spin and colour combinatorics allow for the
perturbation calculations of a wealth of QCD processes in $\EE$ annihilation,
DIS and hadron--hadron collisions. Final states in the said processes are not simple QCD partons, but rather their footprints; jets. Jet algorithms not only provide a framework for defining jets but constitute, with the aid of jet shape observables, an indispensable tool for detailed understanding of jet substructure, which in turn allows for an optimal use of jets in new physics searches. In the  subsequent chapters we attempt to shed some light on this very last issue from the perspective of exploring the structure of QCD jets. We begin with a study of jet shapes in the theoretically favoured $\EE$ annihilation processes and then use the experience gained there to address the theoretically more challenging, and presently more interesting, hadron-hadron scattering processes.

%% file: ch4/chap4.tex

\chapter{Resummation of angularities in $e^+ e^-$ annihilation}
\label{ch:EEJetShapes1}

\section{Introduction}
\label{sec:EEJS1:intro}

It has long been known that probing the shape and structure of high-$p_T$ jets is potentially of great value in searches for new particles at collider experiments \cite{Seymour:1993mx}. With the advent of the LHC and much activity in improving and  developing  jet algorithms \cite{Cacciari:2008gp, Salam:2007xv, Dokshitzer:1997in, Wobisch:1998wt, Catani:1993hr, Ellis:1993tq}, studies of this nature have received considerable impetus. In particular, much recent attention has been focused on using jet studies for the identification of boosted massive particles which decay to hadrons forming a collimated jet,  see for instance Refs. \cite{Butterworth:2008iy, Butterworth:2002tt, Ellis:2009su, Ellis:2009me, Kaplan:2008ie, Kribs:2009yh, Thaler:2008ju, Chekanov:2010vc, Almeida:2008yp, Almeida:2008tp, Skiba:2007fw, Holdom:2007nw,  Krohn:2009wm, Plehn:2009rk, Butterworth:2009qa, Baur:2008uv, FileviezPerez:2008ib, Bai:2008sk, Brooijmans:2009xa}.

In the same context a method has been suggested to study the shapes of one or more jets produced in multi-jet events at fixed jet multiplicity \cite{Ellis:2009wj, Ellis:2010rwa}. The precise details of the observable suggested in those references involve defining a jet shape/energy flow correlation similar to that introduced in Ref. \cite{Berger:2003iw}. Specifically the proposal was to measure the shapes of one or more jets in an event leaving other jets unmeasured and introducing a cut on hadronic activity outside high--$p_T$ (hard) jets, to hold the hard-jet multiplicity fixed. This is in contrast to for instance hadronic event shapes \cite{Banfi:2010xy, Banfi:2001bz, Banfi:2004nk, Dasgupta:2003iq} which by construction are sensitive to the shape of the overall event rather than an isolated jet.

In the present chapter we wish to use this observable as a case study to make several points that we believe will be useful both within and outside the specific context. The main points that we wish to address concern the resummation of soft gluon effects that become important in describing the observable distribution for small values of the shape variable $\ta$ and the energy cut $E_0$.

In particular in this chapter we shall study the structure of non-global logarithms (NGLs) \cite{Dasgupta:2001sh, Dasgupta:2002bw} as well as compute them  for jets defined using various jet algorithms. We remind the reader that observables that are sensitive to radiation in a limited phase-space region such as the interior of a particular jet are non-global in the sense that they receive logarithmic contributions from correlated soft emission, which are highly non-trivial to treat to all-orders. Existing resummations of NGLs have been confined to a few special cases \cite{Dasgupta:2001sh, Dasgupta:2002dc, Banfi:2002hw, Rubin:2010fc, Dasgupta:2002bw} and to the large--$\Nc$ limit. Given that the observable we study in the current chapter is non-global, it is worth examining in detail the precise structure of NGLs, which by definition start at $\mathcal{O}(\alpha_s^2)$ in the perturbative expansion and are of the same size as the logarithms resummed in  Refs. \cite{Ellis:2009wj, Ellis:2010rwa}. Thus they need 
to be taken into account in order to achieve next--to--leading log accuracy.

We find non-global logarithms arise both in the ratio of the shape variable $\ta$ and the energy cut-off $E_0$ as well as in $E_0/Q$ where $Q$ is the hard scale of the process, which is naturally of the order of the hard jet $p_T$\footnote{The potential presence of such logarithms was also mentioned in \cite{Ellis:2009wj, Ellis:2010rwa}.}.  We argue that in the limit of narrow well-separated jets a simple picture emerges for NGLs. The simplicity in the non-global structure is to do with the fact of QCD coherence \cite{QCD_coherence}. Narrow well-separated jets do not affect each  other's evolution even in the non-global component which arises individually as an edge effect from the boundary of each jet, precisely as the NGLs in the case of a hemisphere mass in $\EE$ annihilation arise from the edge separating the observed and unobserved hemispheres \cite{Dasgupta:2001sh}.  Hence the resummation of non-global logarithms arising at each jet boundary can simply be  taken from the existing result for a 
hemisphere\footnote{This statement should be qualified as it is correct only for the case of the \AKT jet algorithm \cite{Cacciari:2008gp},  which is the one we recommend for study of such observables.} up to corrections that vanish as powers of the jet radius. This simple structure of non-global effects in turn provides us with an ansatz that can be used for any jet event of arbitrary jet multiplicity.

We also assess here the numerical contribution of the non-global logarithms and find that while limiting the value of $E_0$ is of some use in diminishing
their size the effect is still of order twenty percent as far as the peak height of shape distributions is concerned. 
In fact we find that changing the value of $E_0$ is not particularly useful as a means of reducing the non-global contribution. Specifically following the original proposal in Ref.~\cite{Berger:2002ig} it was suggested in Ref. \cite{Ellis:2010rwa} that one may take the value of $E_0/Q$ to be of the same order as the jet shape variable $\ta$, which we agree eliminates the non-global contribution from the measured jet. However in this case the contribution from the unmeasured jet becomes as significant as the contribution we are attempting to eliminate and hence the overall effect of this choice turns out to actually increase the overall non-global component. With the resummed results of the current chapter however one does not have to be too concerned about the precise choice of $E_0$ as the non-global terms should be accounted for, at least within the large-$\Nc$ approximation and up to corrections vanishing as powers of the jet radius\footnote{These effects would amount to perhaps a ten percent change in 
the non-global term which we do not expect to be of significant phenomenological consequence.}.

Another point we wish to make is concerning the role of the different jet algorithms. The computation of NGLs can actually be carried out in any jet algorithm, analytically for the first non--trivial term in the perturbative expansion and by means of the numerical Monte Carlo code developed in \cite{Dasgupta:2001sh, Appleby:2002ke, Rubin:2010fc} in the leading--$\Nc$ limit. Indeed it was found \cite{Appleby:2002ke}, in gaps--between--jets studies, that the use of certain sequential recombination algorithms (such as the \KT or Cambridge--Aachen C/A) can significantly reduce the non-global logarithms due to the soft gluon clustering inherent in such algorithms. It was however later demonstrated \cite{Delenda:2006nf, Banfi:2005gj} that one pays a price for this reduction in the non-global component in the form of a more complicated result for the independent emission terms. While independent emission is commonly associated with the exponentiation of the single-gluon result, this association is spoiled by 
the application of jet algorithms other than the \AKT algorithm. As we shall show in this chapter the results of soft gluon clustering in the \KT and C/A algorithms modify the independent emission (global term) which deviates from the naive exponentiation of a single gluon at a relevant single logarithmic (SL) accuracy.  Moreover the effect of the clustering near the boundary of a collinear jet no longer produces logarithms suppressed in the jet radius $R$ as was the case for small central rapidity gaps discussed in Refs. \cite{Delenda:2006nf, Banfi:2005gj} but rather pure single-logarithms. We refer to these logarithms as \emph{clustering} logs (CLs). These effects are absent for the \AKT algorithm as already pointed out in Ref. \cite{Cacciari:2008gp}, since that algorithm clusters soft gluons independently to the hard parton and hence produces circular jets in the soft limit, i.e., it can be regarded in this limit as a rigid cone. Hence for the present moment and postponing a resummation of the clustering 
logarithms to Chapter \ref{ch:EEJetShapes3}\footnote{We actually carry out the resummation for the jet mass. Results for the angularities $\ta$ should be analogous.} we confine our studies to the \AKT algorithm. We do however provide an explicit fixed-order computation of the single-logarithmic corrections in the independent emission piece, that arise in other algorithms as we believe this point deserves some stress.

The chapter is organised as follows. In \sec{sec:EEJS1:HighPtJetShapes} we define our observable, choosing the angularities shape variable in dijet events as an example of a jet-shape observable, while imposing a cut $E_0$ in the inter-jet energy flow.
In \sec{sec:EEJS1:SoftLimitCalcs} we perform the leading and next-to leading order calculation of such observable in the soft limit, which elucidates the structure of the logarithms arising from independent soft gluon emissions as well as non-global logarithms from correlated emissions. We use these results to construct an argument which culminates with the resummation of these logarithms in \sec{sec:EEJS1:Resummation}. We also present a study which assesses the numerical significance of the non-global logarithms as a function of the parameters $\ta$ and $E_0$. In \sec{sec:EEJS1:OtherJetAlgs} we highlight the fact that for algorithms other than the \AKT exponentiation of the single gluon result is not sufficient to capture the next-to--leading logarithms \emph{even in the independent emission piece, let alone the non-global terms}. Finally we draw our conclusions in \sec{sec:EEJS1:Conc}.

\section{High-$p_T$ jet-shapes/energy-flow correlation}
\label{sec:EEJS1:HighPtJetShapes}

We wish to examine a situation where one studies the shapes of one or more high-$p_T$ jets in jet events with definite multiplicity. From the results we shall obtain below for such events one can draw conclusions also about the jet shape distribution  for instance for the process $pp \to j+X$, where one can demand the production of a jet $j$ setting a value for a particular jet-shape, while summing over everything else denoted by $X$.

For the points we address in this chapter we can for illustrative purposes and without loss of generality consider high--$p_T$ dijet events. In order to restrict the jet multiplicity we can place a cut $E_0$ whereby we veto the interjet activity such that the sum of transverse energies of emissions outside the two high--$p_T$ jets is less than this value. This was also the definition adopted in Refs. \cite{Ellis:2009wj, Ellis:2010rwa} where the parameter $\Ld$ indicated a cut on additional jet activity along the above lines. 

Moreover, in this chapter we are interested in physics at the boundary of the triggered hard jets and specifically in the non--global logarithms that arise at these boundaries. Hence we can for our discussion ignore the effects of initial state radiation which can simply be accommodated once the structure of the results is understood. Since it is this structure that we focus on, it proves advantageous to consider as an analogy the production of dijets in $\EE$ annihilation. Hence all our points can be made in full generality by considering two hard jets in $\EE$ processes where one measures the shape of one of the jets leaving the other jet unmeasured as prescribed in Refs. \cite{Ellis:2009wj, Ellis:2010rwa}. Our results should also then be directly comparable to those obtained by other authors using soft-collinear effective theory \cite{Ellis:2009wj, Ellis:2010rwa}.

\subsection{Observable definition}
\label{ssec:EEJS1:ObservableDefn}

We shall pick the angularities, $\ta$, introduced in \cite{Berger:2002ig, Berger:2004xf} and analysed in \cite{Ellis:2009wj, Ellis:2010rwa} with a mis--treatment of NGLs, as a specific simple example of a jet-shape variable. We defined $\ta$ in the previous chapter (precisely \eq{eq:Jets:Angularities}) in terms of the hadronic variables: transverse momentum ($k_t$), pseudorapidity ($\eta$) and azimuth ($\phi$). An equivalent definition in terms of $\EE$ variables\footnote{See \eq{eq:app:Eik:Azimuth2Polar} for the relations between the two sets of variables.} is given by \cite{Berger:2004xf}
\be
 \ta = \bta (\{\wt{p}\},k_1,\cdots,k_n) = \sum^n_{i\in\rJ} \frac{\om_i}{Q} \sin^a\theta_i \cbr{1-\cos\theta_i}^{1-a}; \;\; -\infty < a < 1,
\label{eq:EEJS1:AngulDef}
\ee
where $\theta_i$ is the polar angle of gluon $i$ with respect to the jet axis, $\om_i$ is its energy and $E_\rJ \sim Q/2$ is the jet energy with $Q$ being the hard scale. 

We study the integrated and normalised shape distribution (equivalently the shape cross--section or shape fraction) given by the general expression
\be
\Sg(\ta, E_0) = \int_{0}^{\ta} \d\ta^1 \int_{0}^{E_0} \d E_3 \; \frac{1}{\cSup{\s}{0}} \frac{\d^4 \sigma}{\d\ta^1 \d E_3 \d\vect{p}_1\d\vect{p}_2},
\label{eq:EEJS1:GenShapeFrac}
\ee
where $\cSup{\s}{0}$ is the Born cross-section and $\ta^1$ denotes the angularity of the jet with momentum $\vect{p}_1$. The above equation indicates that we are restricting the shape of the jet with three momentum $\vect{p}_1$ to be less than $\ta$ leaving the shape of the other jet with momentum $\vect{p}_2$ unmeasured. Likewise, the interjet energy flow $E_3$ is restricted to be less than $E_0$ hence our observable definition above is in precise accordance with the definition in Refs. \cite{Ellis:2009wj, Ellis:2010rwa}. We shall in future leave the dependence on jet momenta $\vect{p}_1, \vect{p}_2$ unspecified and to be understood. 

We carry out a calculation for the above observable which includes a resummation of large logarithms in $R^{2-a}/\ta$ to next-to--leading logarithmic (equivalently single logarithmic (SL))  accuracy  in the exponent. We include a description of non-global single logarithms in the leading-$\Nc$ limit. Additionally we resum the logarithmic dependence on $Q/E_0$ to single logarithmic accuracy again accounting for the non-global contributions. Our main aim is to study the effect of the non-global logarithms neglected for instance in previous calculations of jet shapes \cite{Ellis:2009wj, Ellis:2010rwa, Almeida:2008yp} on the shape fraction \eq{eq:EEJS1:GenShapeFrac}. While resumming logarithms in $\ta$ and $E_0$ we shall neglect those logarithms that are suppressed by powers of the jet radius $R$  which shall enable us to treat non-global logarithms straightforwardly\footnote{More specifically we shall neglect corrections varying as $R^2/\Delta_{ij}$ where $\Delta_{ij}=1-\cos\theta_{ij}$ is  a measure of the 
angular separation between the hard jets. This parameter emerges naturally in fixed--order computation of NGLs for energy flow outside jets \cite{Banfi:2003jj} and it was also treated as negligible in \cite{Ellis:2009wj, Ellis:2010rwa}.}. Hence our calculation addresses the range of study where $E_0/Q \gg \ta$ and is valid in the limit of relatively small jet radius $R$. We shall not however resum terms varying purely as $\alpha_s \ln R$ which for the values of $R$ we consider can safely be ignored from a phenomenological viewpoint. We thus aim to resum large logarithms in $\ta$ and $E_0/Q$ in what one may call the approximation of narrow well separated jets. 
According to our estimates this approximation and our consequent resummation should enable relatively accurate phenomenological studies of jet shapes.  

The perturbative expansion of the shape fraction $\Sg$ \eqref{eq:EEJS1:GenShapeFrac} in terms of coupling $\as$ may be cast in the form
\be
\Sigma = \cSup{\Sg}{0} + \cSup{\Sg}{1}  + \cSup{\Sg}{2} + \cdots,
\label{eq:EEJS1:PTExpandShapeFrac}
\ee
where $\cSup{\Sg}{0}$ refers to the Born contribution and is equal to $1$. In the next section we shall carrying out a calculation of the logarithmic structure that emerges at the one and two gluon levels, in the limit of soft gluon emission. 
These calculations will help us to identify the full logarithmic structure and point the way towards a resummed treatment. We start below with a leading order calculation in the soft limit.

\section{Soft limit calculations}
\label{sec:EEJS1:SoftLimitCalcs}

We start by considering the effect of a single soft emission $k$ by a hard $q\qb$ pair, produced in $\EE$ annihilation. At this level all infrared and collinear (IRC) safe jet algorithms will yield the same result. We can write the parton momenta as 
\begin{align}
\nn p_1 &= \frac{Q}{2}(1,0,0,1), \\
\nn p_2 &= \frac{Q}{2}(1,0,0,-1), \\
k  & = \omega\left(1,\sin \theta \cos \phi, \sin \theta \sin \phi, \cos \theta \right),
\label{eq:EEJS1:PartonMomenta}
\end{align}
where $p_1$ and $p_2$ are the hard partons and we have neglected recoil against the soft gluon emission $k$, which is irrelevant at the logarithmic accuracy we seek. Let $p_1$ corresponds to the measured jet. Then if the gluon is combined with the parton $p_1$ one restricts the angularity of the resulting jet to be below $\ta$ while if combined with $p_2$ the angularity is unrestricted. Likewise one can consider the parton $p_2$ to be in the measured jet direction, which will give an identical result. We thus have
\begin{multline}
\cSup{\Sg}{1}(\ta,E_0) =  \frac{1}{\cSup{\s}{0}} \int_0^{\ta^{\max}} \d\ta^1\, \frac{\d\cSup{\s}{1}}{\d\ta^1} \Theta\cbr{\ta - \ta^1} \Theta_{k\in\rJ_1} +
\\+ 
\int_0^{Q/2} \d E_3\, \frac{\d\cSup{\s}{1}}{\d E_3} \Theta\cbr{E_0 - E_3} \Theta_{k\notin\rJ_1,\rJ_2} 
\label{eq:EEJS1:ShapeFracAlpha1_A}
\end{multline}
Note that there is no constraint on the gluon energy when it is combined with the jet $\rJ_2$. Moreover, one notices that \eqref{eq:EEJS1:ShapeFracAlpha1_A} is actually a sum of two distributions. In the limit $E_0\ra Q/2$, \eqref{eq:EEJS1:ShapeFracAlpha1_A} corresponds purely to the angularity distribution. While in the limit $\ta\ra\ta^{\max}$ it reduces to the interjet energy flow (or gaps-between-jets) distribution.

In all the commonly used IRC safe jet algorithms the soft gluon $k$ will form a jet with a hard parton if it is within a specified distance $R$ of the hard parton. The distance is measured for hadron collider processes in the $(\eta,\phi)$ plane as $\Delta \eta^2 +\Delta \phi^2$ where $\Delta \eta$ is the separation in rapidity and $\Delta \phi$ is the separation in $\phi$ between the hard parton and the gluon $k$. In the limit of small angles, relevant for small-$R$ values $R \ll 1$,  which we shall consider here, the distance measure reduces to $\theta_{pk}^2$ where $\theta_{pk}$ is the angle between the gluon $k$ and hard parton $p$.  Thus $k$ and $p_1$ form a jet if $\theta_{p_1k}^2<R^2$. Otherwise the gluon is outside the jet formed by $p_1$ which at this order remains massless. If the gluon does not also combine with hard parton $p_2$ to form a jet, one restricts its energy to be less than $E_0$ as required by the definition of the observable. Differences between the various algorithms shall emerge in 
the following section where we examine the emission of two soft gluons.

Thus we can write for the contribution of the real soft gluon $k$ with $\theta_{p_1k}^2 \equiv \theta_1 < R^2$ to the shape fraction
\be
\cSup{\Sg_r}{1} = \frac{\CF\as}{\pi} \int\frac{\d\om}{\om} \frac{\d\theta_1^2}{\theta_1^2}\, \Theta \left(\ta - \ta^1\right).
\label{eq:EEJS1:RealAngShapeFracAlpha1}
\ee
where the superscript $r$ denotes the real emission piece. In this same soft region virtual corrections are exactly minus the real contributions, but unconstrained; therefore we can cancel the real emission result above entirely against the virtual piece and we are left with
\be
\cSup{\Sg_{\rm in}}{1} = -\frac{\CF\as}{\pi} \int_0^{Q/2}\frac{\d\om}{\om} \int_0^{R^2}\frac{\d\theta_1^2}{\theta_1^2} \Theta \left (\frac{\om\,\theta_1^{2-a}}{2^{1-a}\,Q} -\ta\right),
\label{eq:EEJS1:ShapeFracAlpha1Ang_A}
\ee
where we used the small-angle approximation to simplify the expression \eq{eq:EEJS1:AngulDef} for a single gluon since $\theta^2 < R^2 \ll 1$.  The suffix ``$\mathrm{in}$'' denotes the contribution to $\cSup{\Sg}{1}$ from the region where the gluon is in the measured jet. Performing the integral over angles with the specified constraint results in
\begin{multline}
\cSup{\Sg_{\rm in}}{1} = - \frac{\CF\as}{\pi} \int_{\frac{2\ta Q}{R^{2-a}} }^{Q/2} \frac{d\om}{\om} \ln\sbr{R^2 \cbr{\frac{\om}{2^{1-a}\ta Q}} ^{\frac{2}{2-a}}} \\
= - \frac{\CF\as}{\pi}\,\frac{1}{2-a}\,\ln^2\cbr{\frac{(R/2)^{2-a}}{\ta}} \Theta\cbr{\frac{(R/2)^{2-a}}{\ta}-1}.
\label{eq:EEJS1:ShapeFracAlpha1Ang_B}
\end{multline}
Notice that using $\eta, \kt{t}$ variables one would find that the argument of the logarithm is $\tan^{2-a} (R/2)$, to which $(R/2)^{2-a}$ is the small-$R$ approximation. 

Next we consider the region where the soft emission flies outside either hard jets, with the corresponding contribution $\cSup{\Sg_{\rm out}}{1}$. Since here we are no longer confined to the small angle approximation we use $k_t$ and $\eta$ with respect to the jet axis as integration variables where $\eta$ is the gluon rapidity. In these variables one can represent the contribution of the gluon $k$ after real-virtual cancellation as 
\be
\cSup{\Sg_{\rm out}}{1} = -\frac{2\,\CF\as}{\pi} \int^{Q/2} \frac{\d\kt{t}}{\kt{t}} \int_{-\ln 2/R}^{\ln 2/R}d\eta \,\Theta \left(\kt{t} \cosh\eta - E_0\right).
\label{eq:EEJS1:ShapeFracAlpha1E0_A}
\ee
where the limits on the rapidity integral reflect the out of jet region.
Performing the integrals one obtains, to the required single-logarithmic accuracy,
\be
\cSup{\Sg_{\rm out}}{1} = -\frac{2\,\CF\as}{\pi} \ln\frac{Q}{E_0} \left(2 \ln \frac{2}{R}\right).
\label{eq:EEJS1:ShapeFracAlpha1E0_B}
\ee

The full soft result at leading order is $\cSup{\Sg}{1} = \cSup{\Sg_{\rm in}}{1} + \cSup{\Sg_{\rm out}}{1}$. We see that the angularity distribution receives double logarithmic corrections which in the present case are in the ratio $R^{2-a}/\ta$.  Taking account of hard collinear emissions one would obtain also single logarithms in $R^{2-a}/\ta$, which we shall account for in our final results. 

We shall now move to considering two-gluon emission and the structure of the non-global logarithms that arise at this level.

\subsection{Two-gluon calculation and non-global logarithms}
\label{ssec:EEJS1:TwoGluonsAndNGLs}

Going beyond a single soft emission to the two gluon emission case the precise details of the jet algorithm start to become important. In what follows below we shall consider only the \AKT algorithm since in the soft limit the algorithm functions essentially as a perfect cone algorithm \cite{Cacciari:2008gp}. In particular this implies that soft gluons are recombined with the hard partons independently of one another (one can neglect soft gluon clustering effects)
which considerably eases the path to a resummed prediction. The logarithmic structure for other jet algorithms is also interesting and we shall discuss it in a later section.

Here we carry out an explicit two-gluon calculation to obtain the structure of NGLs for the observable at hand. Referring to the non-global contribution to $\Sg(\ta,E_0)$ as $S$, we compute below $S_2$ the first non-trivial term of $S$ (see the previous chapter, \eq{eq:Jets:AngularsAlpha2NGFinal}). Our results shall indicate a way forward towards a resummed result incorporating these effects. As in  Refs. \cite{Dasgupta:2001sh, Dasgupta:2002bw} we shall consider the emission of gluons $k_1$ and $k_2$ such that $\om_1 \gg \om_2$, i.e., strong energy ordering\footnote{Working in this regime simplifies the calculations and is sufficient to extract the full leading NGLs.}. In this limit the squared matrix element can be split into an independent  emission term $\propto \CFsq$ and a correlated emission term $\propto \CF\CA$. The former is incorporated in the standard resummed results based on exponentiation of a single gluon, which we discuss later.

Let us concentrate on the $\CF\CA$ term missed by the single gluon exponentiation, and which generates the NGLs we wish to study and resum. Consider the following kinematics:
\begin{align}
\nn p_1 &=\frac{Q}{2} \left(1,0,0,1\right), \\
\nn p_2 &= \frac{Q}{2} \left(1,0,0,-1 \right), \\ 
\nn k_1 &= \omega_1 \left(1,\sin\theta_1,0,\cos\theta_1\right),\\
k_2 &= \omega_2 \left(1,\sin\theta_2 \cos\phi, \sin \theta_2 \sin \phi, \cos \theta_2 \right),
\label{eq:EEJS1:PartonMomentaAlpha2}
\end{align}
with $\omega_1 \gg \omega_2$ as stated above.

\begin{figure}
\begin{center}
\epsfig{file=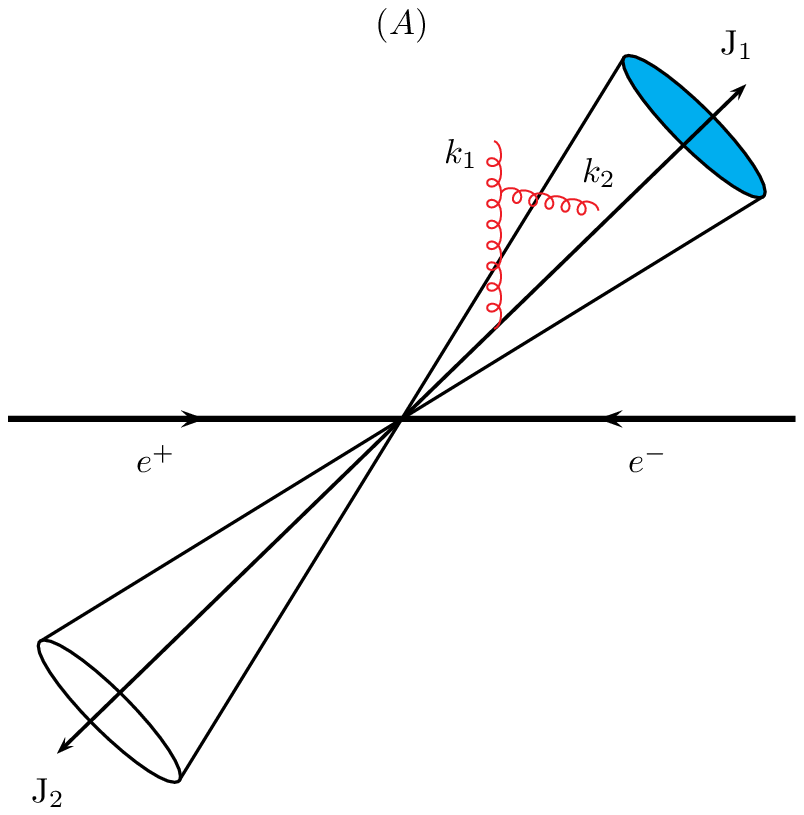,    width = 0.32 \textwidth}
\epsfig{file=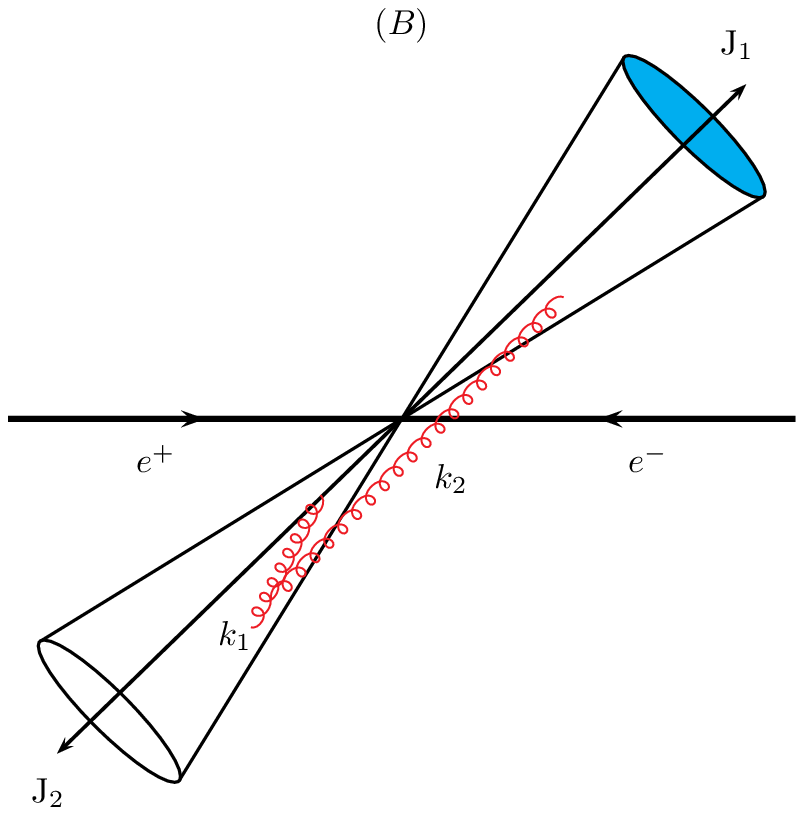, width = 0.32 \textwidth}
\epsfig{file=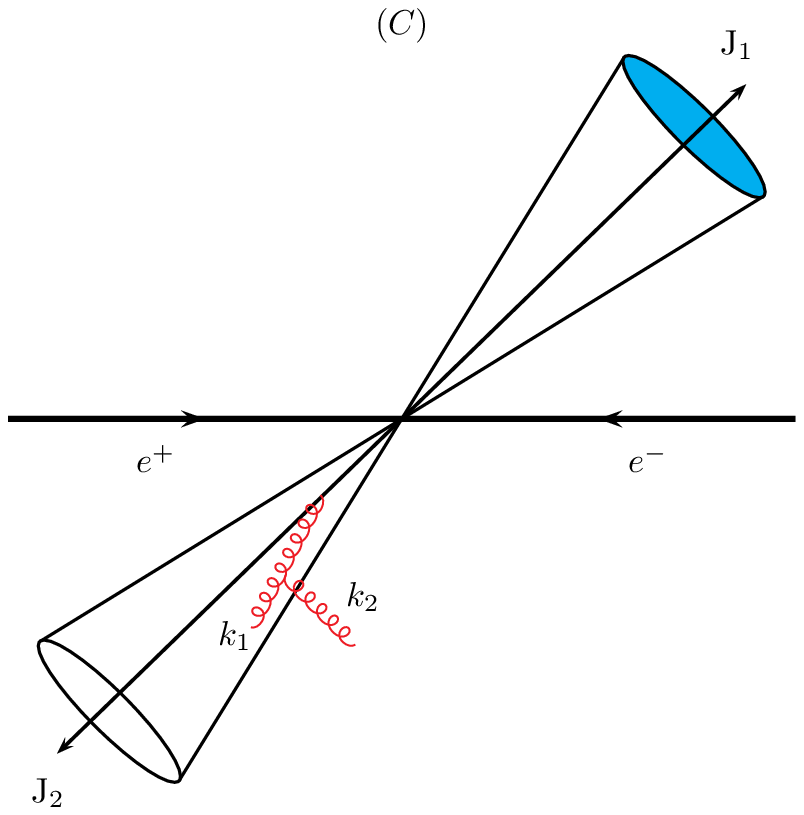,   width = 0.32 \textwidth}
\caption{Diagrams representing the correlated emissions which give rise to the lowest-order non-global logarithms. $(A)$: the harder gluon $k_1$ lies outside both jets and the softest one $k_2$ is recombined with the measured jet and contributes to the angularity distribution. $(B)$: the harder gluon is inside the unmeasured jet and emits a softer gluon into the measured jet. $(C)$: the harder gluon is inside the unmeasured jet and emits a softer gluon outside both jets, which contributes to the $E_0$-distribution. }
\label{fig:EEJS1:NGLs}
\end{center}
\end{figure}

Consider the situation where the harder gluon $k_1$ is not recombined with either jet but the softest emission $k_2$ is recombined with $\rJ_1$ (initiated by hard parton $p_1$). This situation corresponds to the diagram $(A)$ of \fig{fig:EEJS1:NGLs}.
In the small-angle limit, which applies for the case $R \ll 1$, the condition for $k_2$ to be recombined with $p_1$ is simply $\theta_2^2< R^2$ or equivalently $1-R^2/2 < \cos\theta_2 < 1$ while one has $-1+R^2/2 < \cos\theta_1 <1-R^2/2$, which ensures that $k_1$ is outside both jets. We integrate the squared matrix element for ordered soft emission \cite{Dokshitzer1992675} over the azimuth of gluon $k_2$ to get the angular function (in the eikonal approximation) \cite{Dasgupta:2001sh}
\be
\Om(\theta_1,\theta_2) = \frac{2}{\cbr{\cos\theta_2-\cos\theta_1} \cbr{1-\cos\theta_1}\cbr{1+\cos \theta_2} }.
\label{eq:EEJS1:OmegaFun}
\ee
Then defining the energy fractions $x_i = 2 \omega_i/Q$, the required integral for NGLs reads
\begin{multline}
\Sg^{(1), \NGL} = -4\,\CF\CA \cbr{\astpi}^2 \int_0^1 \frac{\d x_1}{x_1} \int_0^1 \frac{\d x_2}{x_2}\; \Theta \left(\frac{2E_0}{Q}-x_1\right) \Theta\cbr{x_1 - x_2}
\\ \int_{1-R^2/2}^{1} \d\cos\theta_2  \int_{-1+R^2/2}^{1-R^2/2} \d\cos\theta_1 \, \Om(\theta_1,\theta_2) \\ \Theta\sbr{x_2\sin^a\theta_2 (1-\cos\theta_2)^{1-a}-2\ta},
\label{eq:EEJS1:S2_A}
\end{multline}
where we note the constraints on $k_1$ and $k_2$ imposed by the observable definition. Note that as in Ref. \cite{Dasgupta:2001sh} the constraint on $k_2$ emerges after including the term where $k_2$ is a virtual gluon such that the divergence of real emission is cancelled and the piece we retain above is the virtual leftover. Integrating over $x_1, x_2$ we then obtain
\begin{multline}
\Sg^{(1), \NGL} = - 2\,\CF\CA \cbr{\astpi}^2\int_{1-R^2/2}^{1} \d\cos\theta_2 \int_{-1+R^2/2}^{1-R^2/2} \d\cos\theta_1  \, \Om(\theta_1, \theta_2)\,
\\ \ln^2\sbr{\frac{\ta\, Q}{E_0\sin^a\theta_2(1-\cos\theta_2)^{1-a}}} \Theta\left(1- \frac{\ta\, Q}{E_0\sin^a\theta_2(1-\cos\theta_2)^{1-a}}\right).
\label{eq:EEJS1:S2_B}
\end{multline}
The step function in the above equation yields, in the small-angle limit,
\be
 \cos\theta_2 \simeq 1-\frac{\theta_2^2}{2} < 1- 2^{\frac{a}{a-2}} \cbr{\frac{Q\ta}{E_0}}^{\frac{2}{2-a}},
\label{eq:EEJS1:S2_C}
\ee
which leads to the constraint $Q\,\ta/2E_0 < (R/2)^{2-a}$. Moreover, we have from the above mentioned recombination condition, $2(1-\cos\theta_2) < R^2$, that
\be
 \sin^a\theta_2 \cbr{1-\cos\theta_2}^{2-a} \lesssim \frac{R^{2-a}}{2^{1-a}}.
\label{eq:EEJS1:S2_D}
\ee
We can then write, to leading--log accuracy,
\be
 \ln^2\sbr{\frac{Q\,\ta}{E_0 \sin^a\theta_2(1-\cos\theta_2)^{1-a}}} \simeq \ln^2\sbr{\frac{Q\,\ta}{2 E_0\,(R/2)^{2-a}}} + \Or\cbr{\ln\sbr{\frac{Q\,\ta}{2 E_0\,(R/2)^{2-a}}}},
\ee
and neglect the remaining subleading logs of $R^{2-a}$. Thus we have in the small-$R$ limit
\be
 \Sg^{(1), \NGL} = - S_2 \cbr{\astpi}^2\,\ln^2\sbr{\frac{Q\,\ta}{2 E_0\,(R/2)^{2-a}}} \Theta\cbr{\cbr{R/2}^{2-a} - \frac{Q\,\ta}{2 E_0}},
\label{eq:EEJS1:S2_Final} 
\ee
where $S_2$ is the (pure) angular integral in \eq{eq:EEJS1:S2_B},
\begin{eqnarray}
\nn S_2 &=& 2\,\CF\CA\,\int_{1-R^2/2}^{1} \d\cos\theta_2 \int_{-1+R^2/2}^{1-R^2/2} \d\cos\theta_1  \, \Om(\theta_1, \theta_2),\\
 &=& \CF\CA \sbr{\frac{\pi^2}{3} - \mc O(R^4)}.
\label{eq:EEJS1:S2_Coeff}
\end{eqnarray}
We recall that as stated before we neglect logarithms suppressed by powers of $R$ and also that our resummation will be valid when $\ta/(R/2)^{2-a} \ll E_0/Q$ and hence can ignore the corrections to $\pi^2/3$. We note that one also can receive a contribution to the non-global logs from the case where $k_1$ is part of the unmeasured jet, \fig{fig:EEJS1:NGLs}($B$), but this configuration produces a coefficient that varies as $R^4$ and hence can be ignored, consistently with our approximation. 

Lastly carrying out the integration with the harder gluon $k_1$ inside the measured jet and the softest one $k_2$ outside does not give us large logarithms in the region we are interested in, hence \eq{eq:EEJS1:S2_Final} is our final result for the first non-global piece affecting the $\ta$ distribution.

Next we consider the case that the harder gluon $k_1$ is in the unobserved jet and emits $k_2$ outside both jets, as depicted in \fig{fig:EEJS1:NGLs}($C$). In this case repeating the calculation in the same way produces to our accuracy 
\be
 \Sg^{(1), \NGL} = -S_2 \cbr{\astpi}^2\;\ln^2\cbr{\frac{Q}{2 E_0}},
\label{eq:EEJS1:S2_E0}
\ee
where $S_2$ is given above (\eq{eq:EEJS1:S2_Coeff}).

The above results are noteworthy in many respects. Note that the result \eqref{eq:EEJS1:S2_Final} corresponds, in the small-$R$ limit, to the result already obtained for the hemisphere jet-mass in Ref. \cite{Dasgupta:2001sh} provided one replaces $1/\ta$ in that result by $\frac{2 E_0}{Q\ta} (R/2)^{2-a}$. This is because the non-global evolution takes place from energies of order $Q \ta/(R/2)^{2-a}$ up to those of order $E_0$, whereas for the hemisphere mass the relevant energy for the harder gluon was of order $Q$. More interestingly the coefficient of $S_2$ for small-$R$, $\pi^2/3$, is the same as was obtained there. The origin of this is the fact that the collinear singularity between $k_1$ and $k_2$ dominates the angular integral. As has been noted before \cite{Dasgupta:2002bw} as one separates the gluons in rapidity  the contribution to the non-global term, which represents correlated gluon emission, falls exponentially as gluons widely separated in rapidity are emitted essentially independently. Thus 
in the 
small-$R$ limit, and up to corrections suppressed by $R^4$ the results for the $\ta$ distribution arise from the edge of the measured jet independently of the evolution of the unobserved jet. Likewise there are NGLs given by \eq{eq:EEJS1:S2_E0} which affect purely the inter-jet energy flow $E_0$ distribution. These arise, in the small-$R$ limit, purely from the edge of the unmeasured jets and are independent of the evolution of the measured jet which is well separated in rapidity (similar results were obtained in the work of Refs. \cite{Banfi:2003jj, Rubin:2010fc}).

Thus a simple picture arises for NGLs where each jet evolves independently and the effects arise from the edges of the jet. For the measured jet, the non-global logarithms involve the ratio of the shape variable $(R/2)^{2-a}/\ta$ to the energy flow variable $E_0/Q$, while for the unmeasured jets, they involve the ratio $Q/E_0$. The coefficients of NGLs will be identical within our accuracy (small-$R$ limit) to those computed for the hemisphere mass (where the effect is again an edge effect coming from the hemisphere boundary) and hence the resummation of the non-global effects from each jet can simply be taken from the resummation carried out in Ref. \cite{Dasgupta:2001sh} simply modifying the evolution variable. This will be done in the next section. 

To conclude we wish to draw attention to the fact that we have determined, with a fixed order calculation,  the precise non-global structure which was not included in Refs.~\cite{Ellis:2009wj, Ellis:2010rwa} and knowledge of the nature of these logarithms should pave the way for more accurate phenomenological studies. We remind the reader that our study above is valid only for the case of the \AKT algorithm. Other jet algorithms will give different non-global pieces as discussed in Refs.~\cite{Delenda:2006nf, Banfi:2005gj, Appleby:2002ke}. In fact even the resummation of independent emission terms will be different in other algorithms, a fact that is not widely appreciated and that we shall stress in a later section. 

In the following section we turn to resummed results and provide a simple ansatz which will be valid for arbitrarily complex processes involving jet production.

\section{Resummation}
\label{sec:EEJS1:Resummation}

Having observed the key feature of the non-global logarithms (independent contributions from each jet) that allow us to write a resummed result we shall now focus on the resummation in more detail. The main point to note is that NGLs provide a factor that corrects straightforward single-gluon exponentiation (as discussed in \ssec{ssec:Jets:NGLs}):
\be
\Sg\cbr{\frac{R^{2-a}}{\ta},\frac{Q}{E_0}} = \Sg^{\rm ind} \cbr{\frac{R^{2-a}}{\ta},\frac{Q}{E_0}}  S^{\rm ng} \cbr{\frac{E_0 R^{2-a}}{Q\ta},\frac{Q}{E_0}}.
\label{eq:EEJS1:FullResumForm}
\ee
Thus we shall first provide the result for the single-gluon exponentiation taking account of hard-collinear emission and the running coupling, which contains leading and next-to--leading logarithms in $R^{2-a}/\ta$ as well as leading logarithms in $Q/E_0$.

\subsection{Independent emission contribution}
\label{ssec:EEJS1:IndepResum}

The resummation of independent emission contributions based on a squared matrix element that has a factorized structure for multi-gluon emission is by now a standard procedure and we shall avoid listing these details (see for instance \cite{Banfi:2004yd} for a detailed study of these techniques). We shall provide here only details of the final result for independent emission valid for the \AKT algorithm only. We stress once again that even the independent emission piece will differ at next-to--leading logarithmic accuracy from that reported below if using another jet algorithm.

The result for the independent emission contribution can be written in the usual form (described in \ssec{ssec:Jets:Resummation}, \eq{eq:Jets:ShapeFracAllOrdersInvMellinSpace})
\be
\Sg^{\rm ind} \cbr{\frac{R^{2-a}}{\ta},\frac{Q}{E_0}} = \frac{\exp 
\left[-\mc R_{\ta} -\gamma_E \mc R'_{\ta} \right]}{\Gamma\left(1+\mc R'_{\ta}\right)} \exp \left[-\mc R_{E_0}\right].
\label{eq:EEJS1:IndepResumForm}
\ee
Here $\mc R_{\ta}$ and $\mc R_{E_0}$ are functions of $R^{2-a}/\ta$ and $Q/E_0$ respectively, representing the exponentiation of the one gluon result. They describe the resummation of large (global) logarithms to next-to--leading logarithmic accuracy in $R^{2-a}/\ta$ and leading logarithmic accuracy in $Q/E_0$. Note that the factorised form \eqref{eq:EEJS1:IndepResumForm} can be proven straightforwardly by following the steps outlined in \ssec{ssec:Jets:Resummation} for the specific distribution
\begin{multline}
 \Sg(\ta,E_0) = \sum_{n,m} \int \d\ta\d E_0 \frac{1}{\cSup{\s}{0}} \frac{\d^2\s}{\d\ta \d E_0}\, \Theta\cbr{\ta - \sum_{i\in\rJ_1}^n \frac{\om_i}{Q} \sin^a\theta_i \cbr{1-\cos\theta_i}^{1-a}}\\ \Theta\cbr{E_0 - \sum_{i\notin \rJ_1,\rJ_2}^m \om_i}.
\label{eq:EEJS1:CorrDistribnAtAllOrders}
\end{multline}

With inclusion of running coupling effects and the effects of hard collinear emission the function $\mc R_{\ta}$ can be written as (see \eqs{eq:Jets:ShapeFracAllOrdersMellinSpace}{eq:Jets:LLAprox})
\be
\mc R_{\ta} = \frac{\CF}{\pi} \int \frac{\d k_t^2}{k_t^2} \as(k_t^2) \mc F(k_t^2),
\label{eq:EEJS1:RadiatorTau_A}
\ee
where we defined
\be
 \mc F (k_t^2) = \int_0^1 \d z\, P_{gq}(z)\, \Theta\cbr{z - \frac{2\kt{t}}{Q R}} \Theta\sbr{\frac{(k_{t}/Q)^{2-a}}{z^{1-a}} - \ta}.
\label{eq:EEJS1:FkFun}
\ee
Carrying out the integration for the soft and hard phase space regions one obtains
\begin{multline}
\mc F (k_t^2) = \ln \left(\frac{Q\, R\,e^{-\frac{3}{2}}}{2 k_t}\right) \Theta \left(\frac{QR}{2}-k_t\right) \Theta\sbr{\cbr{\frac{k_t}{Q}}^{2-a} - \ta} 
\\ 
+ \frac{1}{1-a} \ln\left(\frac{k_t (R/2)^{1-a}}{Q\,\ta} \right) \Theta\sbr{\ta - \cbr{\frac{k_t}{Q}}^{2-a}} \Theta\cbr{\frac{k_t^2}{Q^2} - \frac{\ta^2}{(R/2)^{2(1-a)}}},
\label{eq:EEJS1:RadiatorTau_B}
\end{multline}
where the factor $e^{-3/2}$ in the argument of the logarithm in the first term above takes account of the hard collinear region $2\omega/Q \to 1$, for a primary hard quark\footnote{In order to obtain this one replaces as usual $\d x/x \to \d x \sbr{1+(1-x)^2}/2x$ where $x=2 \omega/Q$ in the integral over gluon energy, which is essentially introducing the full splitting function instead of its soft singular term.}. For a primary hard gluon one makes the changes: $\CF\to\CA$ and $e^{-3/2}\to e^{-2\pi\B_0/\CA}$, where $\B_0$ is given below.

Carrying out the integral over $k_t$ one obtains the familiar \cite{Catani:1992ua} results
\begin{subequations}
\begin{eqnarray}
\mc R_{\ta} &=& -L f_1(\ld)- f_2(\ld),
\\
\mc R'_{\ta} &=& - \frac{\pa}{\pa L } \cbr{L f_1(\ld)} .
\end{eqnarray}
\label{eq:EEJS1:RadiatorTau_Final}
\end{subequations}
The functions $f_1$ and $f_2$ are listed below:
\begin{subequations}
\be
f_1(\ld) = - \frac{\CF}{2\pi\B_0 (1-a)\ld} \sbr{\left(1- 2\ld \right) 
\ln(1-2\ld) - \left(2 - a - 2\ld \right) \ln\cbr{\frac{2-a-2\ld}{2-a}} },
\ee
and 
\begin{eqnarray}
\nn f_2(\ld) &=& - \frac{\CF K}{4\pi^2 \B_0^2 (1-a)} \sbr{(2-a) 
\ln\cbr{\frac{2-a-2\ld}{2-a}} - \ln(1 - 2\ld) } \\
\nn & & - f_{\rm coll}\ln\cbr{\frac{2-a-2\ld}{2-a}} - \frac{\CF\B_1}{2\pi\B_0^3 (1-a)} \Bigg[ \ln \left (1-2\lambda \right ) + \frac{1}{2} \ln^2(1-2\ld) \\
& & -\frac{2-a}{2} \ln^2\cbr{\frac{2-a-2\ld}{2-a}} -\cbr{2 - a} \ln\cbr{\frac{2-a-2\ld}{2-a}} \Bigg],
\end{eqnarray}
\label{eq:EEJS1:RadiatorFuncs-f1f2_quark}
\end{subequations}
where
\be
 f_{\rm coll}^q = \frac{3\,\CF}{4\pi\B_0}, \qquad f_{\rm coll}^g = 1,
\label{eq:EEJS1:fCollq&g}
\ee
$\lambda = \beta_0 \alpha_s L, \; L = \ln\sbr{(R/2)^{2-a}/\ta}$ and $\as =\as\cbr{Q R /2}$ is the $\rm\overline{MS}$ strong coupling. In the above results the $\beta$ function coefficients $\beta_0$ and $\beta_1$ are given by
\be
\B_0 = \frac{11 \CA - 2 \nf}{12 \pi}, \; \B_1 = \frac{17 \CAsq - 5 \CA\nf -3 \CF \nf}{24 \pi^2},
\label{eq:EEJS1:BetaFunCoeffs}
\ee
and the $K$ factor is defined in \eq{eq:Jets:KFactor}.

Likewise for the function $\mathcal{R}_{E_0}$ we have 
\be
\mc R_{E_0} = -\frac{2\,\CF}{\pi\B_0} \ln\frac{2}{R}\; \ln(1- 2\ld'), 
\label{eq:EEJS1:RadiatorE0}
\ee
where here $\ld' =  \B_0 \as L', \; L' = \ln\cbr{Q/2 E_0} $ and $\as = \as \cbr{Q/2}$. Note that the function $\mc R_{\ta}$ contains both a leading logarithmic term $Lf_1(\ld)$ and a next-to--leading or single logarithmic term $f_2 (\ld)$. The leading logarithms in $\mc R_{E_0}$ are single logarithms and next-to--leading logarithms in this piece are beyond our control. The term $\G\cbr{1+\mc R'_{\ta}}$ in \eq{eq:EEJS1:FullResumForm} arises as a result of multiple emissions contributing to a given value of the angularity and is purely single-logarithmic. The corresponding function for the $E_0$ resummation would be beyond our accuracy and hence is not included. We note the results presented here for $\mc R_{\ta}$ are identical to the ones for the $\EE$ hemisphere jet-mass, $\ro$, \cite{Catani:1992ua}, with the replacement $\rho \to \ta/(R/2)^{2-a}$ and $\as(Q) \to \as(Q R/2)$.

\subsection{Non-global component}
\label{ssec:EEJS1:NGLs}

The non-global terms arise independently from the boundary of individual 
jets in the approximation of narrow well-separated jets. The result for an individual jet is the same as that for energy flow into a semi-infinite rapidity interval which was computed in the large-$\Nc$ limit in Ref. \cite{Dasgupta:2001sh}.

In our two-jet example the contribution of NGLs can thus be written as 
\be
S^{\rm ng}\cbr{\frac{E_0\,R^{2-a} }{Q\,\ta}, \frac{Q}{E_0}} = S\left(t_{\rm{meas}}\right) S\left(t_{\rm{unmeas}}\right),
\label{eq:EEJS1:Sng}
\ee
where the function $S(t)$ was computed in Ref. \cite{Dasgupta:2001sh} and the factorisation holds in the limit of narrow well-separated jets (as discussed in \ssec{ssec:EEJS1:TwoGluonsAndNGLs}). From that reference one notes that for primary hard quarks
\be
S(t) = \exp \sbr{-\CF\CA \frac{\pi^2}{3} \left(\frac{1+(at)^2}{1+(bt)^c} \right)t^2},
\label{eq:EEJS1:Sng_hemi}
\ee
where $a=0.85\,\CA, \, b=0.86\,\CA, \, c=1.33$. For primary hard gluons one replaces $\CF$ in the equation above by $\CA$ (see \ssec{ssec:Jets:NGLs}). The single logarithmic evolution variables for the measured and unmeasured jet contributions read
\begin{eqnarray}
t_{\rm{meas}} &=&\frac{1}{2\pi} \int_{\frac{\ta Q}{2E_0 (R/2)^{2-a}}}^{1} \frac{\d x}{x} \as(x E_0),
\\
t_{\rm{unmeas}} &=& \frac{1}{2\pi} \int_{\frac{2E_0}{Q}}^1 \frac{\d x}{x} \as\left(xQ/2\right),
\label{eq:EEJS1:EvolutnVariables}
\end{eqnarray} 
which represent the evolution of the softest gluon with a running coupling that depends on the gluon energy. For the measured jet the softest gluon evolves between scales of order $Q \ta/(R/2)^{2-a}$ and $E_0$ while for the unmeasured jet the evolution is from $E_0$ up to the jet energy $Q/2$.

Carrying out the integrals (one-loop running coupling is sufficient here) gives
\begin{eqnarray}
t_{\rm meas} &=&-\frac{1}{4\pi\B_0} \ln\sbr{1-\B_0 \as\left(E_0\right) \ln \frac{2 E_0 (R/2)^{2-a}}{Q \ta}}, 
\\
t_{\rm unmeas} &=& -\frac{1}{4\pi\B_0} \ln\sbr{1-\B_0 \as\left(Q/2\right)\ln \frac{Q}{2E_0} }.
\label{eq:EEJS1:EvolutnVariablesOneLoop}
\end{eqnarray}
In the following sub-section we shall illustrate the effects of NGLs on the shape-variable distributions for different values of $E_0$.

\subsection{Numerical studies}
\label{ssec:EEJS1:NumericStudies}

\begin{figure}[!t]
\centering
\epsfig{file=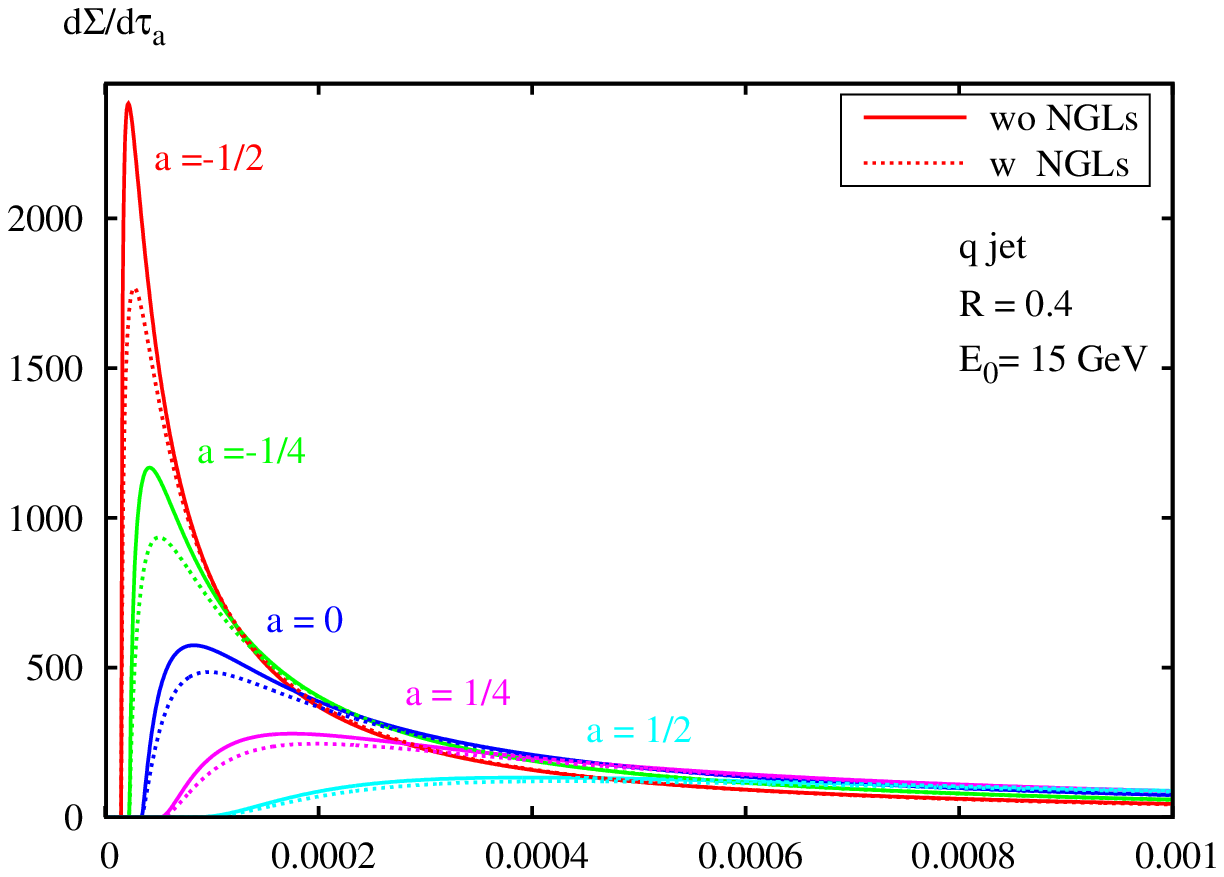, width = 0.48 \textwidth}
\epsfig{file=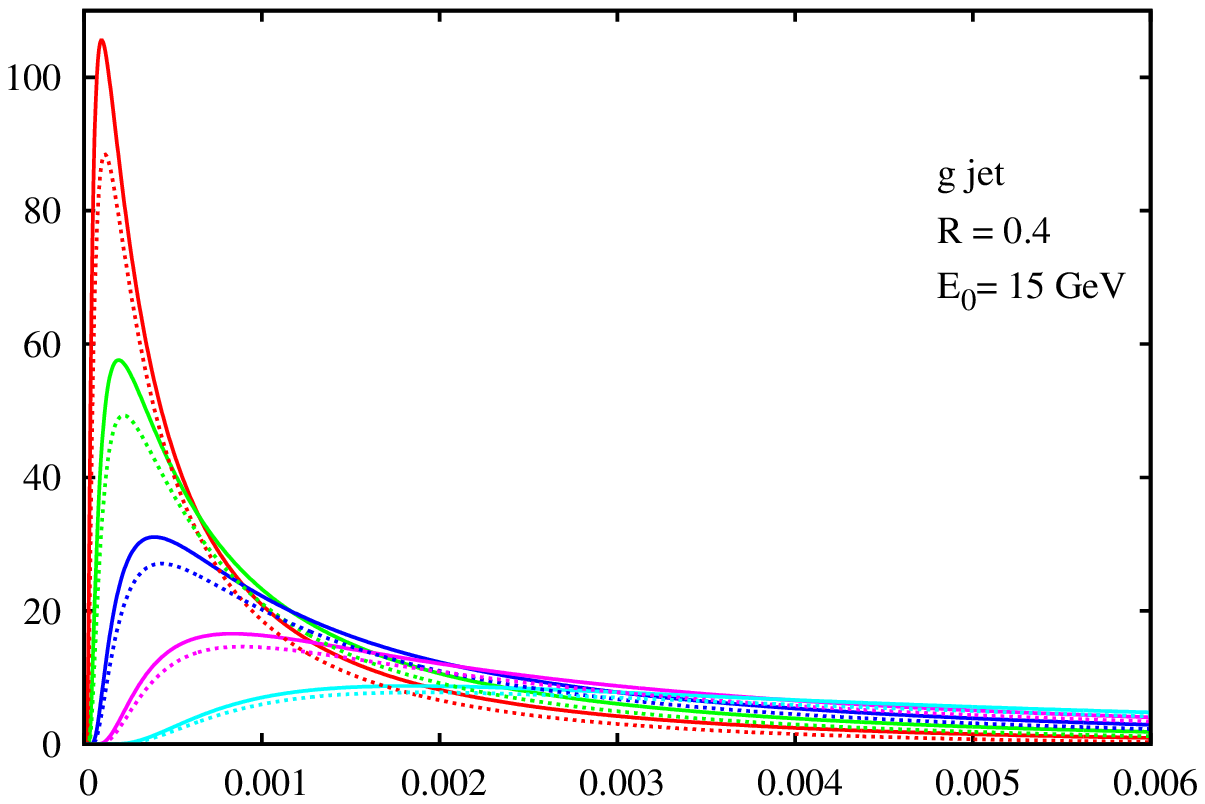, width = 0.48 \textwidth}
\epsfig{file=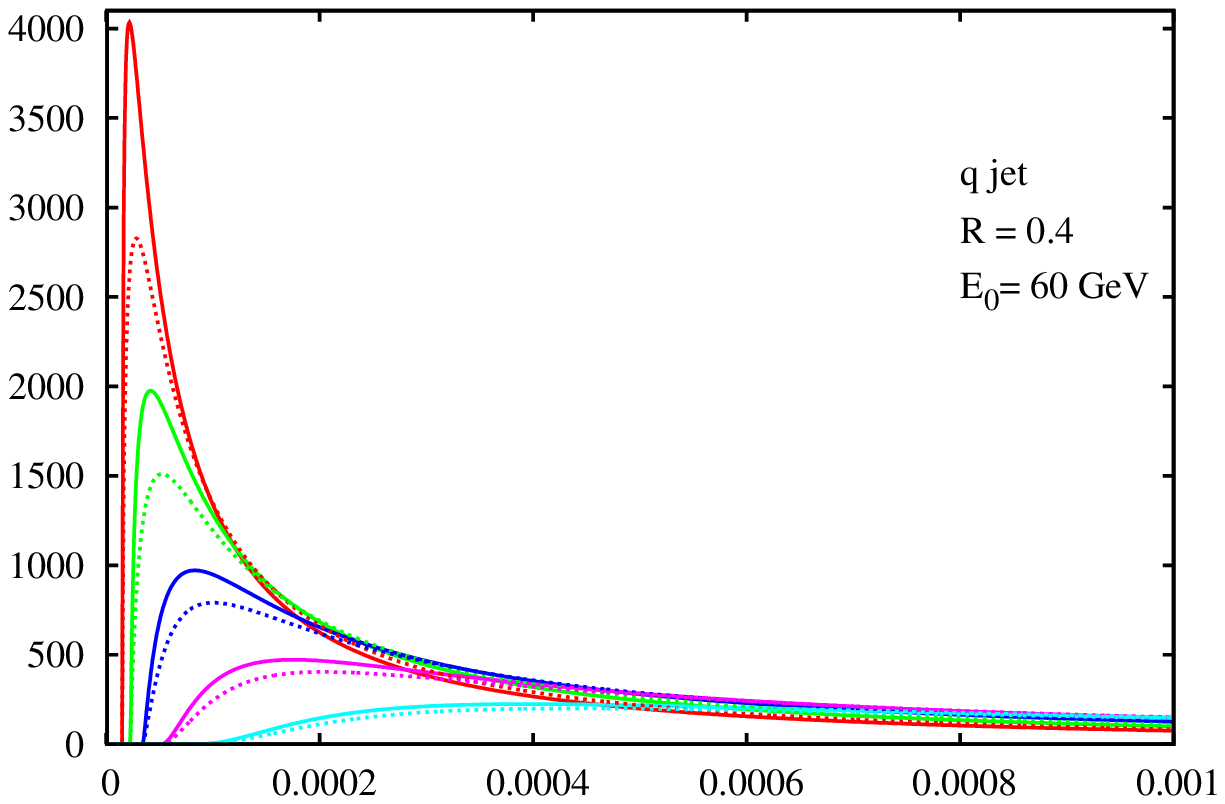, width = 0.48 \textwidth}
\epsfig{file=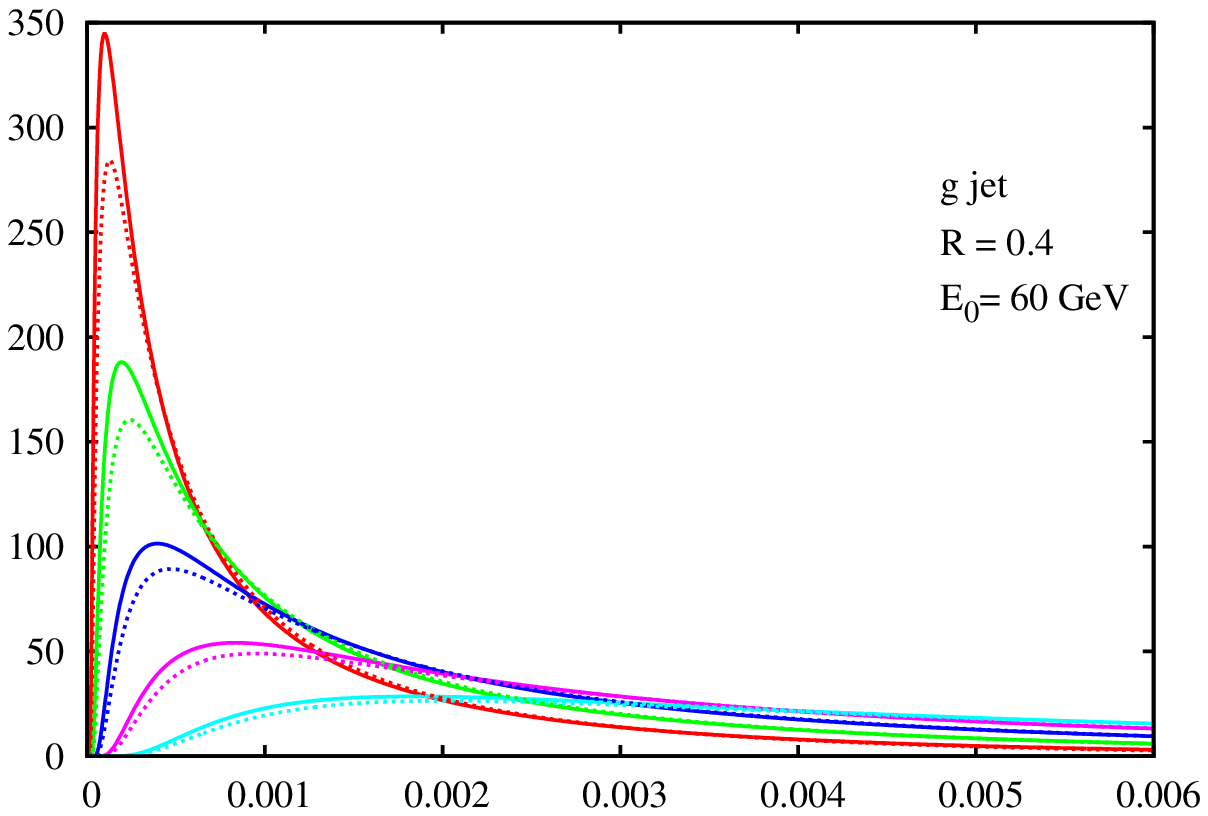, width = 0.48 \textwidth}
\epsfig{file=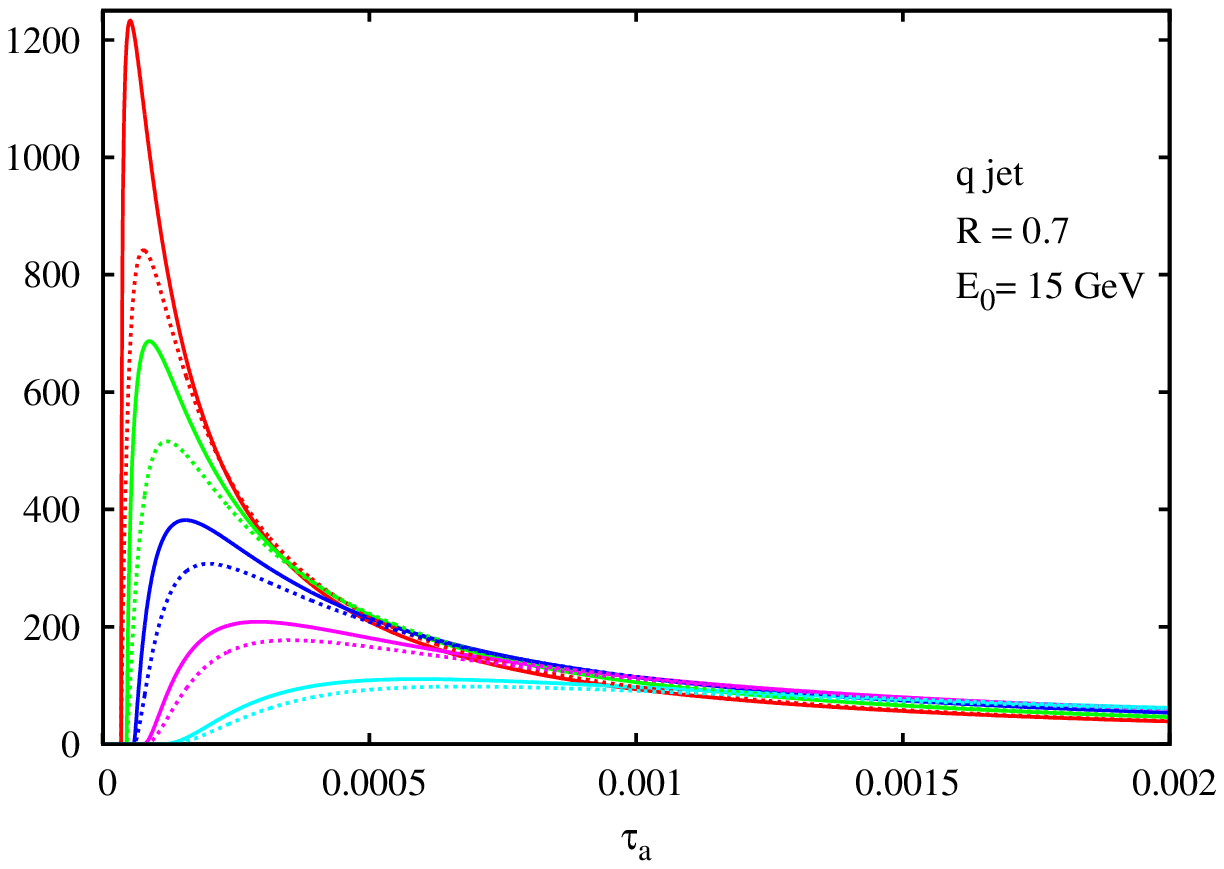, width = 0.48 \textwidth}
\epsfig{file=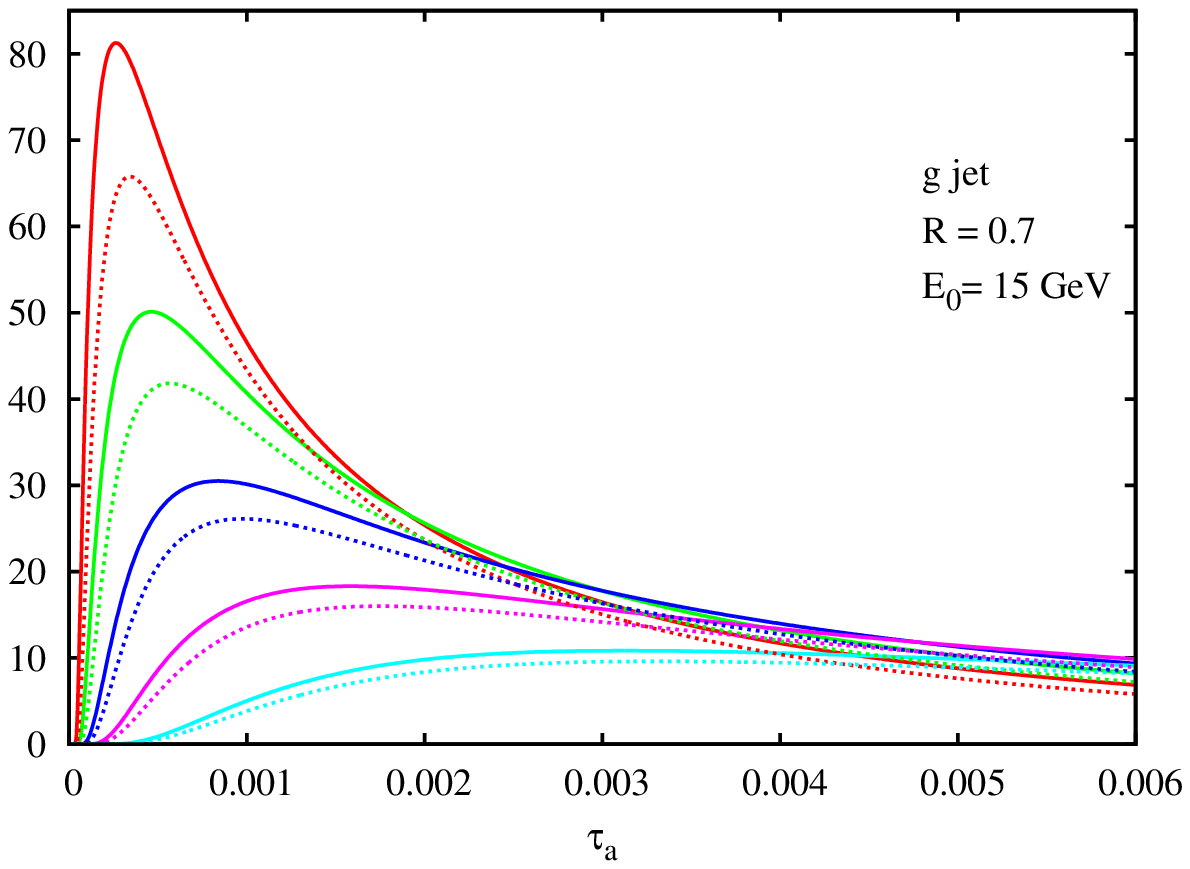, width = 0.48 \textwidth}
\epsfig{file=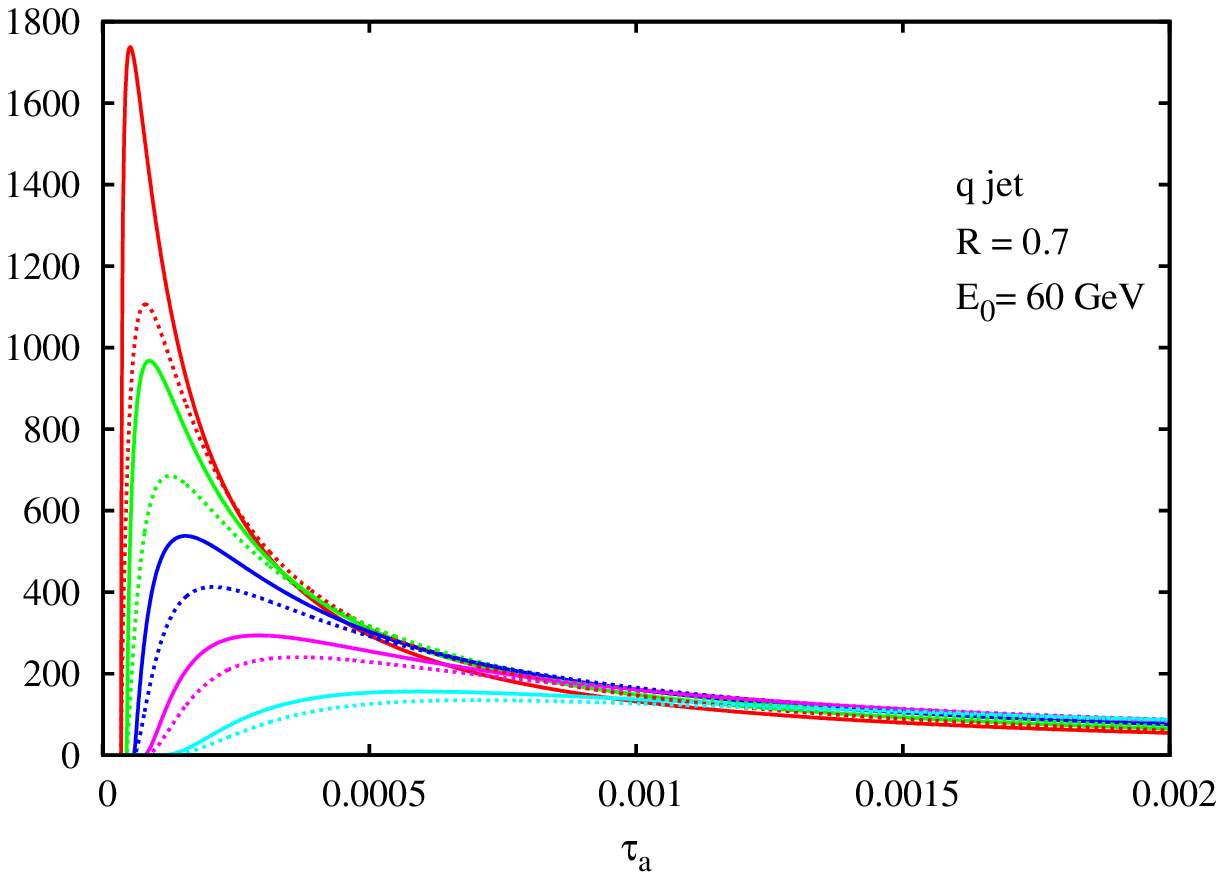, width = 0.48 \textwidth}
\epsfig{file=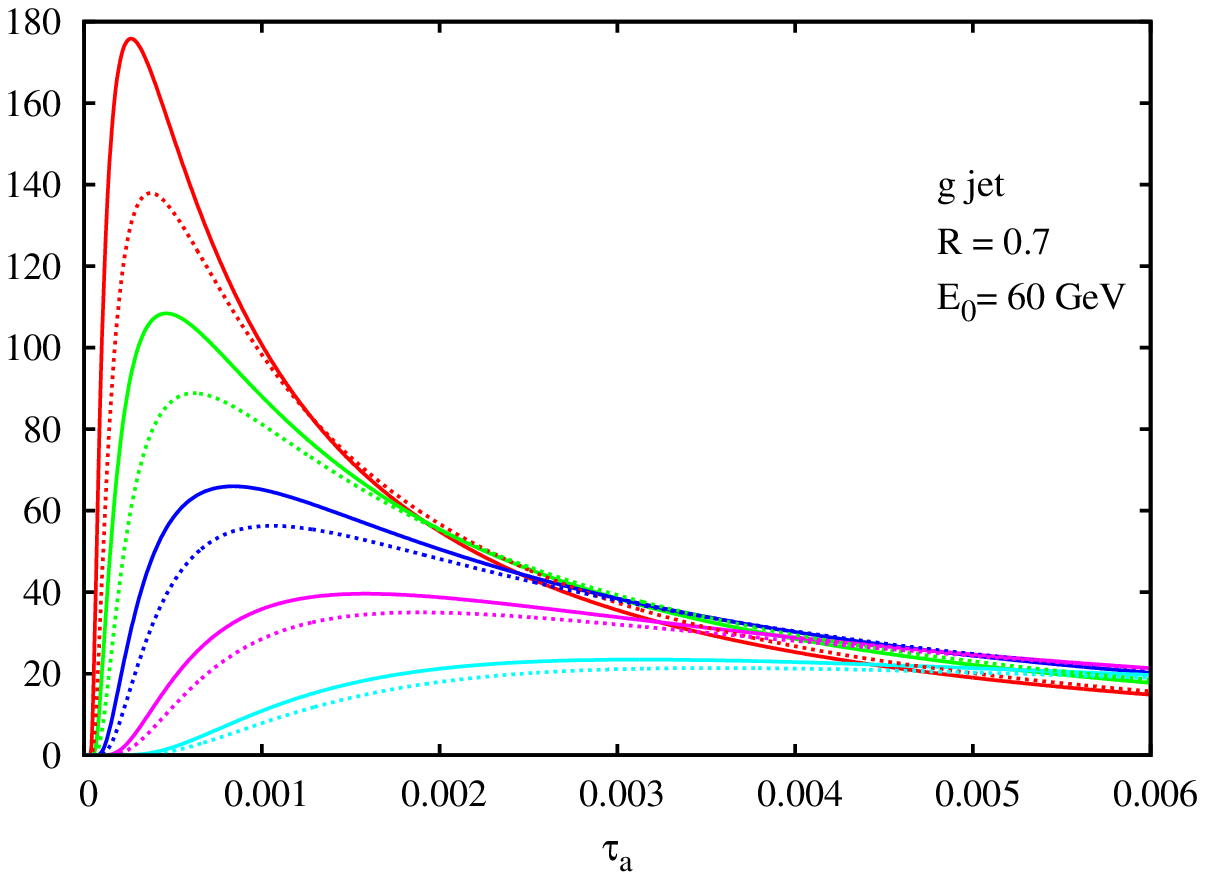, width = 0.48 \textwidth}
\caption{The angularity distribution $\d\Sg/\d\ta$ for five values of the parameter $a$ given at $Q=500\GeV$ (about $250 \,\mathrm{GeV}$ or so at a hadron collider) and different values of the jet radius and energy cut. The left column is for quark jets, while the right is for gluon gets. As shown in the legend, solid lines correspond to neglecting NGLs while dotted lines  take them into account.}
\label{fig:EEJS1:EffectOfSng}
\end{figure}
Let us examine the impact of NGLs on the differential angularity distribution, divided by the inclusive rate;  at our level of accuracy we have:
\be
\frac{1}{\cSup{\s}{0}} \frac{\d\s}{\d\ta} = \frac{\d\Sg}{\d\ta},
\ee
with $\Sg$ given by \eq{eq:EEJS1:FullResumForm}. From \fig{fig:EEJS1:EffectOfSng} one can see that for each value of $a$, NGLs do not change significantly the position of the peak of the distribution. However, their inclusion leads to a reduction in the peak height that increases with decreasing values of $a$. For example, for quark jets and with $1/2 > a > -1/2$ the reduction ranges between $10\%(a=\half)$ and $21\%(a=-\half)$ at $E_0=15 \GeV$ and between $12\%(a=\half)$ and $25\%(a=-\half)$ at $E_0 = 60 \GeV$, both with jet radius $R = 0.4$. Increasing the jet radius to $R = 0.7$ leads to further reduction in the \emph{peak} of about $5\%$ for each value of $a$ at both energies. For gluon jets, the distribution is generally broader, the peak height much smaller, the peak position ($\ta^{\rm peak}$) higher and the peak reduction due to NGLs is less pronounced than for quark jets.

Moreover increasing $E_0$ further one will observe that the effect of NGLs on the peak height can be as significant as about $30 \%$ and higher for negative values of $a$. Aside from the effect of the 
cut $E_0$ and jet radius $R$, we observe that as one lowers the value of the angularity parameter $a$, from $1$ towards negative values, NGLs become more significant and thus the reduction in the Sudakov peak gets increasingly more severe.

\begin{figure}[!t]
\centering
\epsfig{file=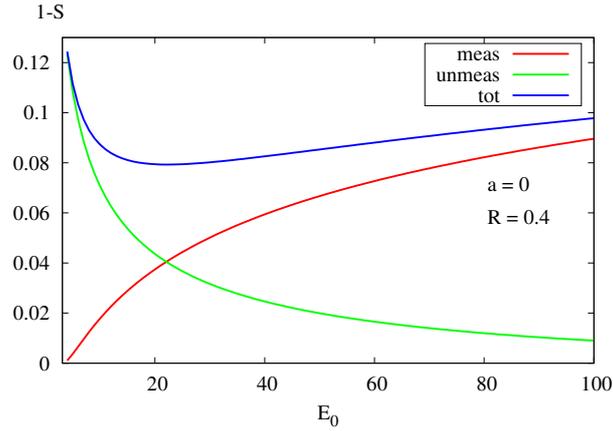, width = 0.6 \textwidth}
\caption{Non global contribution $1-S(t)$ from the measured jet (meas), the unmeasured jet (unmeas) and overall (tot) as a function of $E_0$ for $a=0, \tau_0 = 5\times 10^{-4}$ and $R=0.4$.} 
\label{fig:EEJS1:NGSplit}
\end{figure}

On the other hand it has been suggested \cite{Ellis:2010rwa} that one may eliminate NGLs by choosing $E_0/Q$ of order $\ta$. In our approximation, small-$R$, this prescription amounts to the choice $2E_0 (R/2)^{2-a} = Q\,\ta$. While this rids us of NGLs from the measured jet boundary, the contribution from the unmeasured jet boundary becomes increasingly important. This is reflected in \fig{fig:EEJS1:NGSplit} which plots separately the factors $1-S(t_{\rm meas})$, $1-S(t_{\mr{unmeas}})$ and $1-S^{\rm ng}$, where $S^{\rm ng}$ is defined in \eq{eq:EEJS1:Sng} as the product of the $S$ factors . The plots are presented as functions of $E_0$ for the illustrative values of $a=0, \tau_0 = 5 \times 10^{-4}$. Other parameters are the same as for the previous plots. As one can readily observe increasing the value of $E_0$ leads to a growth of the non-global contribution from the measured jet while the contribution from the unmeasured jet is somewhat diminished. Lowering $E_0$ leads to the opposite effect and the 
unmeasured jet contributions become increasingly significant. It is noteworthy that changing the value of $E_0$ in the range indicated has no significant effect on the size of the non-global effect overall. Also worth noting however is that the choice $E_0 = \ta Q/(2 (R/2)^{2-a})$ (the lowest value of $E_0$ shown in \fig{fig:EEJS1:NGSplit} for $a=0$) which eliminates the contribution from the measured jet (i.e., the red curve goes to zero) is not very helpful as the overall contribution stemming from the unmeasured jet entirely is more significant than for the higher values of $E_0$ discussed before. From this one realises that progressively decreasing the value of $E_0$ is not a way to eliminate the non-global contribution, for the observable at hand.

In the following section we shall show that for algorithms other than the \AKT even the independent emission resummed result is not equivalent at next-to--leading logarithmic level to the exponentiation of the single-gluon result.

\section{Other jet algorithms}
\label{sec:EEJS1:OtherJetAlgs}

Let us now consider the situation in other jet algorithms where the clustering or recombination of soft gluons amongst themselves may be an important effect. 
One such algorithm is the inclusive \KT algorithm discussed for the case of central gaps-between-jets in Refs. \cite{ Delenda:2006nf, Banfi:2005gj}. For such algorithms, starting from the two-gluon level, we need to revisit the independent emission calculations  and correct the naive exponentiation of a single gluon.  Note that in Refs. \cite{ Delenda:2006nf, Banfi:2005gj} the single logarithms obtained as a result of clustering were proportional to powers of the jet radius which would make them beyond our control here. However, as we shall see, in the collinear region we are concerned with here, this power suppression does not emerge, making these logarithms relevant to our study. To illustrate the role of soft gluon clustering and recombination we focus on the single inclusive angularity distribution and we ignore the cut corresponding to $E_0$. Placing this cut does not affect the conclusions we draw here.

\subsection{Revisiting the \AKT algorithm at two-gluons}
\label{ssec:EEJS1:RevisitAKT2Loop}

\begin{figure}
\centering
\epsfig{file=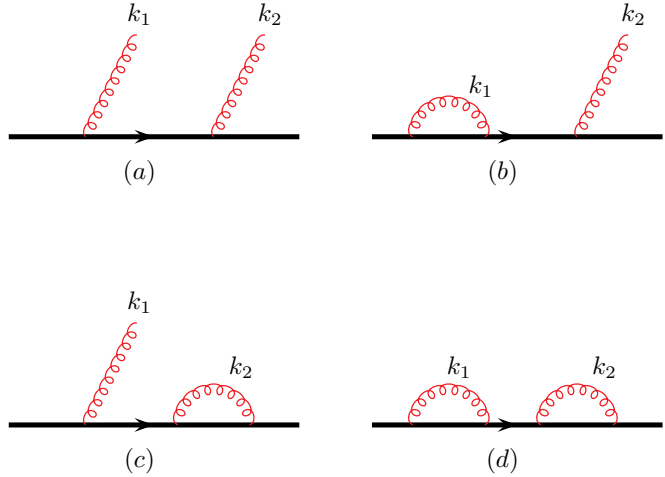, width = 0.6\textwidth}
\caption{Diagrams contributing to independent two-gluon emission from a hard parton line.}
\label{fig:EEJS1:Emission}
\end{figure}

To set the scene let us first carry out the independent emission calculation corresponding to two-gluon emission in the \AKT algorithm which in the soft limit works like a perfect cone. At the two-gluon level we have four terms corresponding to the independent emission of soft gluons in the energy ordered regime $x_1 \gg x_2$. These contributions are depicted in \fig{fig:EEJS1:Emission}.
The contribution to the squared matrix element for ordered two-gluon emission is the same for each of the diagrams in \fig{fig:EEJS1:Emission}, up to a sign. The double real (labelled (a)) and double virtual contributions (labelled (d)) can be expressed as 
\be
W(k_1,k_2) = 4 \CFsq \gs^4 \frac{\cinner{p_1}{p_2}^2}{\cinner{p_1}{k_1} \cinner{p_1}{k_2} \cinner{p_2}{k_1} \cinner{p_2}{k_2}},
\label{eq:EEJS1:Two_gluonEmissionAmp}
\ee
which in terms of the energy fractions $x_1$ and $x_2$ introduced in \ssec{ssec:EEJS1:TwoGluonsAndNGLs} becomes simply
\be
W(k_1,k_2) = 256\,\gs^4\,\frac{\CFsq}{Q^4} \frac{1}{x_1^2 x_2^2} \frac{1}{\left(1-\cos^2\theta_1 \right)\left(1-\cos^2 \theta_2 \right)}.
\label{eq:EEJS1:Two_gluonEmissionAmp_x1x2}
\ee
Since the calculation that follows below is intended for highly collimated jets, $R \ll 1$, we shall take the small angle limit of the above result, $\theta_1, \theta_2 \ll 1$. A similar result holds for the one-real one-virtual terms (b) and (c) in \fig{fig:EEJS1:Emission} with a relative minus sign. We are now in a position to compute the angularity distribution at the two gluon level for the independent emission $\CFsq$ term. 

We start by noting that the integration region for all graphs can be divided according to whether the real gluons $k_1$ and $k_2$ are inside or outside the measured jet. We have four distinct regions: $k_1,k_2$ both outside the measured jet, $k_1,k_2$ both inside the measured jet or either of the gluons inside and the other outside the jet. The condition for a given gluon to end up inside or outside the measured jet depends on the jet algorithm we choose to employ. In the \AKT algorithm the condition is particularly simple when considering only soft emissions; such an emission $k$ is inside the jet if it is within an angle $R$ of the hard parton initiating the jet, else it is outside.

Given this fact let us consider how the various diagrams (a)--(d) in \fig{fig:EEJS1:Emission} combine in the different regions mentioned above. 
Since we are computing the angularity distribution $\d\Sg/\d\ta$ for a fixed angularity $\ta$, the pure virtual diagram (d) makes no contribution and hence we shall omit all reference to it in what follows. In the region where both emissions are in the jet we shall treat the sum of graphs (a)--(c). Where the harder emission $k_1$ is in the jet and $k_2$ is out, graphs (a) and (c) cancel since the real $k_2$ does not contribute to the angularity exactly like the virtual $k_2$. This leaves diagram (b) which gives zero since the in-jet gluon $k_1$ is virtual and hence does not generate a value for the angularity. Hence the region with $k_1$ in and $k_2$ out gives no contribution. 

Now we consider $k_2$ in and $k_1$ out. The contributions with $k_2$ real (a) and (b) cancel as the graphs contribute in the same way to the angularity. The diagram with $k_2$ virtual (c) cannot contribute to the angularity as the real emission $k_1$ lies outside the jet. Hence we only need to treat the region with both gluons in and we shall show that this calculation correctly reproduces the result based on exponentiation of the single gluon result. The summed contribution of graphs (a) to (c) can be represented as 
\be
\frac{\d\Sg_{2}}{\d\ta} \sim \int \d\Phi\, W \left[\de\left(\ta -\frac{x_1\theta_1^{2-a}}{2^{2-a}}- \frac{x_2\theta_2^{2-a}}{2^{2-a}}\right) - \de\left(\ta -\frac{x_1\theta_1^{2-a}}{2^{2-a}}\right)- \de\left(\ta -\frac{x_2\theta_2^{2-a}}{2^{2-a}}\right)\right],
\label{eq:EEJS1:ShapeDistAKT_A}
\ee
where we wrote the contribution to the angularity from an emission with energy fraction $x$ and angle $\theta$ with respect to the hard parton as $x \sin^a\theta \left(1-\cos\theta \right)^{1-a}/2 \approx  x (\theta/2)^{2-a}$. 

To compute the leading double-logarithmic contribution and show that it corresponds to the exponentiation of the order $\alpha_s$ double-logarithmic term one can write $\de\left(\ta-x_1(\theta_1/2)^{2-a}- x_2(\theta_2/2)^{2-a} \right)$ as $\pa\sbr{\Theta \left(\ta-x_1 (\theta_1/2)^{2-a}- x_2(\theta_2/2)^{2-a} \right)}/\pa\ta$ and make the leading-logarithmic approximation 
\be
 \Theta \left(\ta-x_1(\theta_1/2)^{2-a}-x_2(\theta_2/2)^{2-a} \right) \to \Theta \left(\ta-x_1(\theta_1/2)^{2-a} \right) \Theta\left(\ta-x_2(\theta_2/2)^{2-a} \right),
\label{eq:EEJS1:ShapeDistAKT_B}
\ee
which allows us to make the replacement 
\begin{multline}
\de\left(\ta-x_1(\theta_1/2)^{2-a}- x_2(\theta_2/2)^{2-a}\right) \to \de\left(\ta-x_1(\theta_1/2)^{2-a}\right) \Theta\left(\ta-x_2(\theta_2/2)^{2-a} \right) \\ + 1 \lra 2\,.
\label{eq:EEJS1:ShapeDistAKT_C}
\end{multline}
Doing so and using the explicit forms of $W$ and the phase space $d \Phi$ in the small angle limit we get
\begin{multline}
\frac{d\Sg_{2}}{d\ta} = - 4\,\CFsq \cbr{\astpi}^2\, \int \frac{\d\theta_1^2}{\theta_1^2} \frac{\d\theta_2^2}{\theta_2^2} \frac{\d\phi}{2\pi} \frac{\d x_1}{x_1} \frac{\d x_2}{x_2} \Bigg[\de\left(\ta-x_1(\theta_1/2)^{2-a}\right)
\\
\Theta\left(x_2(\theta_2/2)^{2-a} -\ta \right) + 1 \leftrightarrow 2 \Bigg]
\Theta \left(R^2-\theta_1^2 \right) \Theta\left(R^2-\theta_2^2\right) \\ \Theta \left(x_1-x_2\right) \Theta(1-x_1).
\label{eq:EEJS1:ShapeDistAKT_D}
\end{multline}
Carrying out the integrals we straightforwardly obtain
\be
\frac{\d\Sg_{2}}{\d\ta} = -2 \left(\frac{\CF\as}{\pi(2-a)} \right)^2\, \frac{1}{\ta} \ln^3 \left(\frac{(R/2)^{2-a}}{\ta} \right),
\label{eq:EEJS1:ShapeDistAKT_Final}
\ee
which is precisely the result obtained by expanding the exponentiated double-logarithmic one-gluon result to order $\as^2$ and differentiating with respect to $\ta$.
Thus the standard double-logarithmic result for the angularity distribution arises entirely from the region with both gluons in the jet. Contributions from soft emission arising from the other regions cancel in the sense that they produce no relevant logarithms. 
\vfill

\subsection{The \KT algorithm and clustering logarithms}
\label{ssec:EEJS1:CLsKTandCA}

We shall now argue that for algorithms other than the \AKT relevant single-logarithmic contributions shall appear from the regions which cancelled in the argument above, although of course the leading double-logarithms are still precisely the same as for the \AKT case. An analysis of such miscancelling contributions is therefore necessary for a resummation aiming at next-to--leading logarithmic accuracy in the angularity. The logarithms we compute below correct the one-gluon exponentiated result for the angularity distribution at the single-logarithmic level starting from order $\as^2$. 

Let us consider the situation in, for instance, the \KT algorithm (Algorithm~\ref{Alg:SRAlgInclusive} with $\rm p=1$). When both $k_1$ and $k_2$ are within an angle $R$ of the hard parton both soft gluons get combined into the hard jet and this region produces precisely the same result as the \AKT algorithm, corresponding to exponentiation of the one-gluon result. Moreover, when both $k_1$ and $k_2$ are beyond an angle $R$ with respect to the hard parton there is no contribution from either to the angularity. However, when $k_1$ is beyond an angle $R$ and $k_2$ is inside an angle $R$ the situation changes from the \AKT case. This is because in the \KT algorithm when the two soft partons are separated by less than $R$ in angle they can be clustered together. The resulting soft jet has four-momentum $k_1+k_2$, when we use the four-momentum recombination scheme (E--scheme), and lies essentially along the harder gluon $k_1$. Thus when $k_1$ is beyond an angle $R$ it can pull $k_2$ out of the hard jet since the 
soft 
jet $k_1+k_2$ which replaces $k_2$ lies outside an angle $R$ of the hard parton. Hence such a configuration gives no contribution to the angularity distribution. 

In precisely the same angular region the virtual $k_1$, real $k_2$ diagram (b) (obviously unaffected by clustering) does however give a contribution whereas in the \AKT case it had cancelled the double real contribution (a). The graph with $k_1$ real and $k_2$ virtual gives no contribution as before. Thus a new uncancelled contribution arises for the \KT (and indeed the Cambridge--Aachen (C/A)) algorithm from the region where the two real gluons $k_1$ and $k_2$ are clustered, which can be given by computing the $k_1$ virtual $k_2$ real graph in the same angular region.

We now carry out this calculation explicitly. We consider the angles $\theta_1^2$, $\theta_2^2$ and $\theta_{12}^2$ as the angles between $k_1$ and the hard parton, $k_2$ and the hard parton and $k_1$ and $k_2$ respectively. Applying the \KT algorithm (Algorithm~\ref{Alg:SRAlgInclusive}, $\rm p=1$) in inclusive mode means constructing the distances $\omega_1^2 \theta_1^2$, $\omega_2^2 \theta_2^2$ and $\omega_2^2 \theta_{12}^2$ along with the distances (from the ``beam'') $\omega_1^2 R^2, \omega_2^2 R^2$, where, for the $e^{+}e^{-}$ case we consider here, the energy $\omega$ plays the role of the $k_t$ with respect to the beam in a hadron collider event. Now since $\theta_1^2 > R^2$, $\theta_2^2 <R^2$ the only quantities that can be a candidate for the smallest distance are $\omega_2^2 \theta_2^2$ and $\omega_2^2 \theta_{12}^2$. Thus the gluons are clustered and $k_2$ is pulled out of the jet if $\theta_{12}<\theta_2<R$. Otherwise $k_2$ is in the jet and cancels against virtual corrections. We write this as
\begin{equation}
 \Theta^{\rm clus}(k_1,k_2) = \Theta\left(\theta_1^2-R^2\right) \Theta \left(\theta_2^2-\theta_{12}^2\right)\Theta \left(R^2-\theta_{2}^2\right).
\label{eq:EEJS1:KT-CA-ClusFunAlpha2}
\end{equation}
Note that repeating the above discussion for the C/A algorithm yields an identical expression to \eq{eq:EEJS1:KT-CA-ClusFunAlpha2}. We can then write the contribution of graph (b) of \fig{fig:EEJS1:Emission} in the clustering region
\begin{multline} 
\frac{\d\Sg^{\rm clus}}{\d\ta} = -4\,\CFsq \cbr{\astpi}^2 \int \frac{\d\theta_1^2}{\theta_1^2} \frac{\d\theta_2^2}{\theta_2^2} \frac{\d\phi}{2\pi} \frac{\d x_1}{x_1} \frac{\d x_2}{x_2}\, \de\left[\ta -x_2(\theta_2/2)^{2-a} \right]\\ \Theta(x_1-x_2) \Theta^{\rm clus}(\theta_1,\theta_2).
\label{eq:EEJS1:ShapeDistKT_A}
\end{multline}
Using the fact that in the small-angle approximation relevant to our study 
\be
\theta_{12}^2 = \theta_1^2 +\theta_2^2-2 \theta_1 \theta_2 \cos \phi\,, 
\label{eq:EEJS1:SmallAngleApprox}
\ee
integrating over $x_1$ and $x_2$ and using $t= (\theta_2/2)^{2-a}/\ta$ 
one obtains
\begin{multline} 
\frac{\d\Sg^{\rm clus}} {\d\ta} = -4\,\CFsq \cbr{\astpi}^2 \frac{2}{(2-a)\ta} \int \frac{\d\theta_1^2}{\theta_1^2} \frac{\d t}{t} \frac{\d \phi}{2 \pi} \ln t \\
 \Theta \left(t-1\right) \Theta\left(\theta_1^2-R^2\right) \Theta\left(4 (\ta t)^{\frac{2}{(2-a)}} \cos^2\phi-\theta_1^2\right) \Theta \left(R^{2-a}/\ta - t\right).
\label{eq:EEJS1:ShapeDistKT_B}
\end{multline}
Carrying out the integral over $\theta_1^2$ results in
\begin{multline} 
\frac{\d\Sg^{\rm clus}} {\d\ta} = -4\,\CFsq \cbr{\astpi}^2 \frac{2}{(2-a)\ta} \int \frac{\d t}{t} \frac{\d\phi}{2 \pi} \ln\left(\frac{4 (\ta t)^{\frac{2}{2-a}} \cos^2 \phi}{R^2}\right) \ln t 
\\ \Theta \left(t-1\right) \Theta\left(4 (\ta t)^{\frac{2}{2-a}} \cos^2\phi-R^2\right) \Theta\left(R^{2-a}/\ta - t\right).
\label{eq:EEJS1:ShapeDistKT_C}
\end{multline}
Now we need to carry out the $t$ integral for which we note $t > \max\sbr{1, \sbr{R/(2\cos\phi)}^{2-a}/\ta }$. In the region of large logarithms 
which we resum one has however that $R^{2-a} \gg \ta$ and hence $\sbr{R/(2\cos\phi)}^{2-a}/\ta > 1$. This condition is reversed only when $\ta \sim R^{2-a}$; a region not enhanced by large logarithms and hence beyond our accuracy.

It is then straightforward to carry out the $t$ integral and doing so and 
extracting the leading singular behaviour in $\ta$ produces the result
\begin{multline} 
\frac{\d\Sg^{\rm clus}} {\d\ta} = -4\,\CFsq \cbr{\astpi}^2 \frac{1}{\ta}\ln \frac{1}{\ta} \int \frac{\d\phi}{\pi} \ln^2 \left(2 \cos \phi \right)\Theta \left (\cos \phi -\frac{1}{2} \right) \\ 
= -0.728\,\CFsq \cbr{\astpi}^2 \frac{1}{\ta} \ln\frac{1}{\ta}.
\label{eq:EEJS1:ShapeDistKT_Final}
\end{multline} 
This behaviour in the distribution translates into a next-to--leading logarithmic $\as^2 \ln^2(1/\ta)$ behaviour in the integrated cross-section, which is relevant for resummations aiming at this accuracy. We refer to the latter logarithms as clustering logs (CLs). As we mentioned before the above finding of single logarithmic corrections generated by clustering has also been reported before for the case of gaps-between-jets studies \cite{Banfi:2005gj}. Note however that the logarithms found there had coefficients that depended on powers of the jet radius starting at the $R^3$ level. In the present case however the presence of collinear singularities near the boundary of a jet of small radius $R$ ensures that there is no power suppression in $R$ and hence the logarithms generated are formally comparable to those we aim to control here and indeed those resummed in Refs. \cite{Ellis:2009wj, Ellis:2010rwa}. Note also that at this logarithmic level, the CLs coefficient does not depend on the parameter $a$. The 
dependence on $a$, just like the shape variable itself, $\ta$, is confined to the argument of the logarithm. This is in-line with our finding that the non-global coefficient, $S_2$, is independent of $a$. Thus non-global and clustering coefficients are only sensitive to the geometry of the final state (and not to the shape variable or parameters that enter in its definition). 

Likewise the clustering will also generate leading logarithms in the $E_0$ variable, which are again unsuppressed by any powers of $R$ and hence ought to be controlled. Lastly we point out that the logarithms generated by independent emission and subsequent \KT clustering were actually resummed in Ref. \cite{Delenda:2006nf} and that possibility also exists here. The explicit calculations will be presented to Chapter \ref{ch:EEJetShapes3}. Moreover, we mention that calculations of NGLs and CLs contributions at two-gluon level in the C/A algorithm produce results that are identical to the above \KT algorithm results reported here.

\section{Conclusion}
\label{sec:EEJS1:Conc}

We would like to conclude by emphasising the main points of our study. 
Given the current interest in the study of jet shapes and substructure for the purposes of discovering new physics at the LHC, it is worth examining the theoretical state of the art when it comes to looking at individual jet profiles in a multi-jet event. A step in this direction was taken for instance in Refs.~\cite{Ellis:2009wj, Ellis:2010rwa}. In the present chapter we have noted 
\begin{itemize}
\item Observables where one picks out for study one or more jets in multi-jet events are in principle non-global. The non-global logarithms will arise at next-to--leading or single-logarithmic accuracy in the jet-shape distributions.
If one studies jet events with a fixed multiplicity by imposing a cut $E_0$ on hadronic activity outside the high-$p_T$ jets, there are non-global logarithms 
involving the ratio of the shape variable $\ta$ and the energy flow $E_0$, as was first anticipated in \cite{Berger:2002ig}. Moreover, there are also non-global logarithms in $E_0/Q$ where $Q$ is the hard scale of the process. These logarithms are leading as far as the distribution in $E_0$ for a fixed $\ta$ is concerned. 

\item In the limit of narrow jets $R \to 0$, one may naively expect the non-global contributions to the jet-shape distributions to vanish with $R$ due to the apparently limited phase-space available for soft emission inside the jet. Here  we have pointed out that the non-global logarithms do not vanish in the small cone approximation  as mentioned for instance in Ref. \cite{deFlorian:2007fv}. One finds instead an effect (at small-$R$) that is independent of $R$ and arises from the edge of the jet. However, in the limit of narrow well-separated jets $R^2 \ll (1-\cos \theta_{ij})$, where $\theta_{ij}$ is the inter-jet separation, one can simplify the non-global contribution. In this limit, owing to QCD coherence and the nature of correlated multiple soft emissions, one can regard the non-global logarithms to arise independently from the boundary of each jet up to corrections that vanish as $R^2/\left(1-\cos \theta_{ij}\right)$. For a measured jet one picks up logarithms in $2E_0 (R/2)^{2-a}/(Q \ta)$ while for 
each unmeasured jet one has logarithms in $2E_0/Q$. The resummation of these logarithms yields a factor $S^j$ for each jet $j$, which is the factor computed, in the large-$\Nc$ limit, for the hemisphere jet-mass in $e^{+}e^{-}$ annihilation in Ref. \cite{Dasgupta:2001sh}, again up to corrections vanishing as $R^2/\left(1-\cos \theta_{ij} \right)$. 

\item The overall size of non-global logarithms depends on the precise values one chooses for $E_0/Q$, $R$, $a$ and $\ta$. However, broadly speaking, we find the contribution not to vary significantly with $E_0$ and to yield corrections of order $10-25 \%$ in the peak region of the $\ta$ distribution. Choosing $E_0/Q$ of order $\ta/(R/2)^{2-a}$ eliminates the non-global contributions from the measured jet but steeply enhances the contributions from the unmeasured jet and it is not an optimal choice for reducing the overall non-global contribution to this observable.

\item We emphasise that the above observations are valid for the \AKT algorithm in which our ansatz for resummation of jet shapes in an arbitrarily complex event is to correct the one-gluon exponentiation with a product of independent non-global factors from each jet. We further have emphasised that switching to algorithms other than the \AKT gives relevant next-to--leading logarithms in the shape distribution as well as leading logarithms in the $E_0$ distribution, even within the independent emission approximation. Thus predictions for observables such as the one discussed in this chapter, in those algorithms are prone to more uncertainty than our current study in the \AKT algorithm at least until such logarithms are also resummed.
\end{itemize}

We would like to stress that the general observations in this chapter are of applicability in a variety of other contexts. For instance, the issue of threshold resummation addresses limited energy flow outside hard jets, along the lines of the $E_0$ distribution here. The consequent non-global logarithms and the issue of the jet algorithm have not been addressed to any extent in the existing literature.  The same issues crop up in the case of resummation in the central  jet veto scale for the important study of Higgs production in association with two jets.

We hope that an awareness of the nature and size of the non-global contributions, the simplification that occurs in the small-$R$ limit and our comments about the situation in other jet algorithms will help to generate more accurate phenomenological studies for these important observables at the LHC. In particular  we shall address in more detail, in Chapter \ref{ch:HHJetShapes1}, the role of soft gluon effects and especially non-global logarithms on QCD predictions relevant to new physics searches at the LHC.

In the next chapter we present ``precision'' calculations of both non-global and clustering logarithms in various jet algorithms. The jet mass distribution in $\EE$ annihilation will be taken as a generic example of non-global jet shape observables. 

%% file: ch5/chap5.tex

\chapter{Non-global structure of jet shapes beyond leading log accuracy}
\label{ch:EEJetShapes2}

\section{Introduction}
\label{sec:EEJS2:Intro}

Within the same context of $\EE$ multijet events considered in the previous chapter, we study in this chapter the \emph{jet-thrust} $\te$, introduced in \cite{Kelley:2011tj}, where one measures the mass of the two highest--energy jets with a veto $E_0$ on energy flow into the interjet region $\Om$. We shall compute the individual colour ($\CFsq, \CF\CA$ and $\CF\TF\nf$) contributions to the total differential $\te$ distribution (at $\Or(\as^2)$) as well as the effect of the C/A (and indeed the k$_{\rm T}$) algorithm clustering at N$^3$LL (that is up to $\as^2\,L$) in the \emph{expansion}. The work presented in the current chapter may be regarded as an extension to that of the the previous chapter. 

Such an extension includes: (a) computing the full $R$ dependence of the leading and next-to-leading NGLs coefficients in the ant-k$_{\rm T}$, C/A and \KT algorithms, (b) computing the full $R$ dependence of the leading and next-to-leading CLs coefficients in the C/A and \KT algorithms, (c) providing a next-to-leading log (in the \emph{exponent}) resummation, including numerical estimates of all--orders NGLs and CLs in the large-$\Nc$ limit, for both \AKT and \KT algorithms with full $R$ dependence using the Monte Carlo program developed in \cite{Dasgupta:2001sh} with the clustering option added in \cite{Delenda:2006nf} and (d) checking our findings, and hence implicitly those of the previous chapter, for $a=0$, (which are valid in the small $R$ limit) against the NLO program \event \cite{Catani:1996vz}. The C/A and \KT jet algorithms are actually identical for both NGLs and CLs at $\Or(\as^2)$ as mentioned previously (\ssec{ssec:EEJS1:CLsKTandCA}), and hence we shall only explicitly discuss the C/A 
algorithm. However, owing to the fact that the said MC 
program \cite{Dasgupta:2001sh, Delenda:2006nf} only implements the \KT algorithm, all--orders treatment of NGLs and CLs shall therefore be addressed in the \KT algorithm.

The organisation of this chapter is as follows. In \sec{sec.fixed_order_1} we compute the full logarithmic part of the LO jet-thrust distribution. We then consider, in \sec{sec.fixed_order_2}, the fixed--order NLO  distribution in the eikonal limit and compute the NGLs coefficients up to $\as^2\,L$, in both \AKT and C/A jet algorithms. In the same section we derive an expression for the CLs leading and next-to-leading terms as well. Note that our calculations in this section include full jet--radius dependence. \sec{sec.resummation_in_QCD} is devoted to NLL resummation of our jet shape including determination of the coefficients of the logs in the exponent which will be used to compute the full N$^3$LL fixed--order differential distribution to be compared with the output of \event. We additionally investigate in the same section the impact of NGLs on the Sudakov resummation and discuss various factors that play a role in enhancing or reducing the latter impact, including clustering effects. Numerical 
distributions of 
the jet-thrust obtained using the program \event are compared to analytical results and the findings discussed in \sec{sec.numerical_results}. In light of this discussion, we draw our main conclusions in \sec{sec.conclusion}.

\section{Jet-thrust distribution at $\Or(\as)$}
\label{sec.fixed_order_1}

After briefly reviewing the definition of the jet-thrust, or the jet mass with a jet veto, observable presented in \cite{Kelley:2011tj,Kelley:2010qs}, the $\EE$ version of the general formula for sequential recombination jet algorithms (Algorithm~\ref{Alg:SRAlgInclusive}) is presented. We then move on to compute the LO integrated  distribution of this shape variable. At this order, all jet algorithms are identical. Note that partons (quarks and gluons) are assumed on--mass shell throughout.

\subsection{Observable definition}
\label{subsec.observable_def}

Consider $\EE$ annihilation into multijet events. First, one clusters events into jets of size (radius) $R$ with a jet algorithm. After clustering, one labels the momenta of the two hardest jets $p_{R}$ and $p_{L}$ and the energy of the third hardest jet $E_{3}$. The jet-thrust is then given by the sum of the two leading jets' masses after events with $E_{3} > \Eo$ are vetoed \cite{Kelley:2011tj},
\begin{equation}\label{eq.tau_omega}
 \te = \frac{m^{2}_{R} + m_{L}^{2}}{Q^{2}} = \frac{\rho_{R} + \rho_{L}}{4},
\end{equation}
where $\rho_{R}$ and $\rho_{L}$ are the jet mass fractions, studied in the previous chapter ($\ro = \tau_0$), for the two leading jets and $Q$ is the hard scale.

A general form of sequential recombination algorithms at hadron colliders is
presented in Algorithm~\ref{Alg:SRAlgInclusive}. The adopted version for $\EE$ machines corresponds simply to the replacements: $\pt{ti}\to E_i$ and $\De R_{ij}^2 = 2( 1-\cos\theta_{ij})$, where $E_i$ is the energy of parton, or pseudojet, $i$ and $\theta_{ij}$ the polar angle between two partons $i$ and $j$. Recall that two partons, $i$ and $j$, are merged together if
\begin{equation}\label{clust_cond}
 \De R_{ij}^2 < R^{2}.
\end{equation}
With regard to notation, we follow Ref. \cite{Kelley:2011tj} and work with the jet radius $\Rs$\footnote{The motivation for using $\Rs$ instead of $R$ was to have results that can directly and straightforwardly be compared to those found in \cite{Kelley:2011tj}, to which our paper \cite{KhelifaKerfa:2011zu} was a clarification regarding NGLs.}, which is given in terms of the above $R$ by:
\begin{equation}\label{R_Rs_rel}
\Rs = R^{2}/4.
\end{equation}

It is worth noting that the jet-thrust, defined in \ref{eq.tau_omega} is just the thrust in the threshold (dijet) limit, hence the name. To verify this we begin with the general formula of the thrust $T$ (or rather $1-T \equiv \tau$),
\begin{equation}\label{thrust_def}
 \tau = 1 - \max_{\uv} \frac{\sum_{i} |\mathbf{p}_{i}.\uv |}{\sum_{i}
|\mathbf{p}_{i}|},
\end{equation}
where the sum is over all final state three--momenta $\mathbf{p}_i$ and the maximum is over directions (unit vectors) $\uv$. In the threshold limit, enforced
by applying a jet veto $E_0$, $\EE$ annihilates into two back--to--back
jets and the \emph{thrust axis}, the maximum $\uv$, coincides with jet
directions. At LO, an emission of a single gluon, $k$, that is close to, and hence clustered with, say $p_{R}$, produces the following contribution to the thrust
\begin{equation}\label{eq.T^2}
 \tau \simeq \frac{E_{R} \omega}{Q}(1-\cos\theta_{k p_{R}}) + \frac{E_{L}
\omega}{Q}(1-\cos\theta_{k p_{L}}) + \frac{\omega^{2}}{Q^{2}}(1-\cos\theta_{k
p_{R}})(1-\cos\theta_{k p_{L}}),
\end{equation}
where $E_{R(L)}$ is the energy of the hard leg $p_{R(L)}$, $\omega$ the gluon
energy and we have discarded $\Or(\tau^{2})$ terms. Recalling that the first two
terms in the RHS of \eq{eq.T^2} are just the mass fractions $\rho_{R}$
and $\rho_{L}$, respectively, at LO and neglecting the third term (quadratic in
$\omega$) one concludes that 
\begin{equation}\label{tau_tauo}
\tau \simeq \te. 
\end{equation}
This relation can straightforwardly be shown to hold to all--orders in the two-jet limit.

\subsection{LO distribution}
\label{sec.Fixed-PT-antikt}

In the previous chapter  we computed the LO distribution of the jet mass fraction, $\rho = \tau_0$, in the small $R$ (hence small $\Rs$) limit using the matrix--element squared in the eikonal approximation (see \app{sec:app:QCD:EikonalApprox}). In this section, we use the full QCD matrix--element to restore the complete $\Rs$ dependence of the singular part of the $\te$ distribution. The general expression for the integrated and normalised $\te$ distribution, or equivalently the $\te$ shape fraction, is given by 
\begin{equation}\label{S1_tauo_dist}
\Sigma(\te,\Eo) = \int_{0}^{\te} \d\te' \int_{0}^{\Eo} \d
E_{3}\;\frac{1}{\cSup{\s}{0}} \frac{\d^{2} \s}{\d\te' \d E_{3}},
\end{equation}
where $\cSup{\s}{0}$ is the Born cross-section. The perturbative expansion of the shape fraction $\Sigma$ in terms of QCD coupling $\as$ may be cast in the form \eqref{eq:EEJS1:PTExpandShapeFrac}. The derivation of the first order correction, $\Sigma^{(1)}$, to the Born approximation is presented in \app{sec:app:EEJS2:LOShapeFrac}. The final result reads
\begin{multline}\label{R1_full-b}
\Sigma^{(1)}(\te,\Eo) = \frac{\CF \as}{2\pi} \left[-2\,\ln^{2}\te +\left(-3
+ 4\,\ln\frac{\Rs}{1-\Rs}\right)\,\ln\te \right] \Theta\left(2\,\te^{\max}(\Rs) -\te\right) \\+ \frac{\CF \as}{2\pi} \Bigg[ - 1 + \frac{\pi^{2}}{3} -
4\,\ln\frac{\Rs}{1-\Rs} \,\ln\frac{2\Eo}{Q} + f_{\Eo}(\Rs)\Bigg],
\end{multline}
where\footnote{Notice that for a single gluon emission, if the gluon is clustered with, say, $R$--jet then $m_L^2 = 0$ and $\te < \te^{\max}$ (and not $\te < 2\te^{\max}$). }
\be
 \te^{\max}(\Rs) = \frac{2}{\Rs} \cbr{1-\sqrt{1-\Rs}}-1,
\label{eq:EEJS2:tau_E0_max}
\ee
and the function $f_{\Eo}(\Rs)$ is given by 
\begin{multline}\label{f_omeg}
f_{\Eo}(\Rs) = -2\,\ln\Rs\,\ln\frac{\Rs}{1-\Rs} + 2\,\Li_{2}(\Rs) -
2\,\Li_{2}(1-\Rs) + \\ \frac{8\Eo}{Q}\cbr{1 + \frac{3}{4}\ln\frac{\Rs}{1-\Rs}} +
\Or\left(\frac{\Eo^{2}}{Q^{2}}\right).
\end{multline}
It is worthwhile to note that in the limit $\Rs \rightarrow 1/2$ the $\te$ distribution \eqref{R1_full-b} reduces, as expected from \eqref{tau_tauo}, to the well-known thrust distribution \cite{ellis2003qcd, Catani:1992ua} with upper limit $\tau < 1/3$. For $\Rs < 1/2$ the jet-thrust distribution includes, in addition to thrust distribution and other single logarithms due to finite $\Rs$ jet size, the interjet energy flow distribution \cite{Oderda:1998en} too,
\begin{equation}
\Sigma^{(1)}_{\mathrm{E\,flow}}(\Eo) = \frac{\CF \as}{2\pi} \left[-
4\,\ln\frac{\Rs}{1-\Rs} \ln\left(\frac{2\Eo}{Q}\right) +
\Or\left(\frac{\Eo}{Q}\right) \right],
\end{equation}
Here the interjet region, or rapidity gap referred to in literature as $\deta$,
is defined by the edges of the jets. Specifically, it is related to the
jet--radius $\Rs$ by
\begin{equation}\label{eta-Rs}
\deta = -\ln\left(\frac{\Rs}{1-\Rs}\right).
\end{equation}

The important features of the $\te$ distribution that are of concern to the
present chapter are actually contained in the second order correction term
$\Sigma^{(2)}$, which we address in the next section.

\section{Jet-thrust distribution at $\Or(\as^2)$}
\label{sec.fixed_order_2}

The $\Or(\as^2)$ matrix--element squared for $\EE$ annihilation, which includes the processes $\EE \to q(p_{a}) + \qb(p_{b}) + g_{1}(k_{1})+ g_{2}(k_{2})$ and $\EE \to q(p_{a}) + \qb(p_{b}) + q_{1}(k_{1})+ \qb_{2}(k_{2})$, has been derived in the eikonal approximation\footnote{The recoil effects are negligible in this regime and are hence ignored throughout. This limit also allows one to consider the produced jets to be back--to--back, hence the parametrisation         \eqref{4-momenta}.}, $Q \gg \cinner{Q}{k_i}$ for $i=1,2$, in \app{ssec:app:EE:AlphasSquareCorrections}. Let us first define the final state partons' momenta, in the two-jet limit, as
\begin{subequations}
\begin{eqnarray}
\nn p_{a} &=& \frac{Q}{2}(1,0,0,1),\\
    p_{b} &=& \frac{Q}{2}(1,0,0,-1),
\end{eqnarray}
and
\be
    k_{i} =  \kt{ti} (\cosh\eta_{i},\cos\phi_{i}, \sin\phi_{i},\sinh\eta_{i}),
\ee
\label{4-momenta}
\end{subequations}
where the hadronic variables, $(k_{ti}, \eta_{i}, \phi_{i})$, are measured with respect to the incoming beam direction. The final formula of the above mentioned matrix--element, \eq{eq:app:EE:MESAlphas2}, involves both soft and hard emissions (derived in the limit stated above). The corresponding differential cross-section can be cast in the form:
\begin{eqnarray}\label{W_2}
\nn \d\cSup{\s}{2} &=& \cSup{\s}{0}\,\d\Phi_2(k_1,k_2)\, S_{ab}(k_1,k_2), \\
 S_{ab}(k_{1}, k_{2}) &=& 4\,\CFsq W_{P} + \CF\CA \left(W_{S}+ H_{g}\right) + \CF\TF\nf H_{q},
\end{eqnarray}
where $W_{P}$ stands for primary (Abelian) term and $W_{S}$ for secondary soft (non--Abelian) leading term. If we define the antenna function $w_{ij}(k)$ by \eq{eq:Jets:AntennaFun} then the latter two amplitudes are given by
\begin{eqnarray}\label{W_P}
 W_{P} &=& w_{ab}(k_{1}) w_{ab}(k_{2}) = 4,
\end{eqnarray}
and $W_{S} = 2\,S$ with
\begin{eqnarray}\label{W_S}
\nonumber S &=& w_{ab}(k_{1}) \left[w_{a1}(k_{2})+w_{b1}(k_{2}) - w_{ab}(k_{2})\right],
\\
&=&  4 \left[\frac{\cosh(\eta_{1}- \eta_{2})}{\cosh(\eta_{1}- \eta_{2}) - \cos(\phi_{1} - \phi_{2})} - 1\right]\label{Ws_eta}.
\end{eqnarray}
$H_{g}$ and $H_{q}$ in \eq{W_2} are responsible for parton configurations with energies of the same order ($\kt{t1} \sim \kt{t2}$). The ``hard'' parts of the two-gluon $H_{g}$ and the two-quark $H_{q}$ contributions read
\begin{eqnarray}\label{Hg_Hq}
\nn H_{g} &=& 2\,\mc I^{2} - S\,\mc J - 4\frac{w_{ab}(k)}{(k_{1}k_{2})},\\
H_{q} &=& - 4\,\mc I^{2} + 4\frac{w_{ab}(k)}{(k_{1} k_{2})},
\end{eqnarray}
where $k = k_{1} + k_{2}$,
\begin{eqnarray}\label{J-R_ampl}
\nonumber \frac{\mc I}{\kt{t1}^2\kt{t2}^2} &=& \frac{(p_{a} k_{1}) (k_{2} p_{b}) - (p_{a} k_{2})(k_{1} p_{b})}{(k_{1} k_{2})(p_{a} k)(k p_{b})},
\\
\nn &=& \frac{2\sinh(\eta_{1}-\eta_{2})}{[\cosh(\eta_{1}-\eta_{2}) - \cos(\phi_{1}-\phi_{2})] [\kt{t1}^{2}+\kt{t2}^{2} + 2\kt{t1}\kt{t2} \cosh(\eta_{1}-\eta_{2})]},
\end{eqnarray}
and
\begin{eqnarray}
\frac{\mc J}{\kt{t1}^2 \kt{t2}^2} &=& \frac{(p_{a} k_{1})(k_{2} p_{b}) + (p_{a} k_{2})(k_{1} p_{b})}{(p_{a} k)(k p_{b})} = \frac{2\,\kt{t1}\kt{t2}\,\cosh(\eta_{1}-\eta_{2})}{[\kt{t1}^{2}+\kt{t2}^{2} + 2\kt{t1}\kt{t2} \cosh(\eta_{1}-\eta_{2})]}.
\end{eqnarray}
Notice that the virtual correction matrix-element in the eikonal approximation is simply minus the real emission one (we explicitly showed this in \app{sssec:app:EE:VirtualGluonCorrections} for a single gluon emission). The two--parton phase space measure, $\d\Phi_2$, in \eq{W_2} is given by
\begin{eqnarray}\label{PS_2}
\d\Phi_{2}(k_{1}, k_{2}) &=& \prod_{i=1}^{2} \frac{\d k_{ti}}{\kt{ti}} \d\eta_{i} \frac{\d\phi_{i}}{2\pi} \left(\frac{\as}{2\pi}\right)^{2}. 
\end{eqnarray}

It is worth noting that the primary emission, $W_{P}$, contribution to the $\te$ distribution is only fully accounted for by the single--gluon exponentiation in the \AKT algorithm case. If the final state is clustered with a jet algorithm other than the latter, $W_{P}$ integration over the modified phase space, due to clustering, leads to (see below) new logarithmic terms that escape the naive single--gluon exponentiation.
On the other hand, the secondary amplitude $W_{S}$ contribution is completely
missing from the latter Sudakov exponentiation in both algorithms.

First we outline the full $\as^{2}$ structure of the $\te$ distribution up
to $\as^2\,L$ level in the \AKT including the computation of the leading and next-to-leading NGLs coefficients. After that, we investigate the effects of final state partons' clustering on both primary and secondary emissions. The C/A algorithm is taken as a case study to illustrate the main points. Calculations where the final state is clustered with other jet algorithms should proceed in an analogous way to the C/A case. 

Note that whenever we deal with fixed-order $\te$ distribution then we use LL and N$^m$LL to refer to leading logs ($\as^n\,L^{2n}$) and $\underbrace{\rm next-\cdots-next}_{m=1,\cdots}$-to-leading log ($\as^n\,L^{2n-m}$) in the expansion. However, when dealing with resummed $\te$ distribution we use LL and N$^m$LL to refer to leading logs ($\as^n\,L^{n+1}$) and $\underbrace{\rm next-\cdots-next}_{m=1,\cdots}$-to-leading log ($\as^n\,L^{n+1-m}$) in the exponent.
We stress that our calculations are strictly valid in the two-jet limit, a region where $\te$ is small and large logarithms crop up.

\subsection{The \AKT algorithm}
\label{subsec.FO_anti-kt}

Considering all possible angular distances between $(k_{1},k_{2})$ and $(p_{a},p_{b})$ we compute below the corresponding contributions of primary and secondary emissions to the $\te$ distribution.

\subsubsection{Independent emission}
\label{subsec.anti-kt_CF2}

The LL contribution to the $\te$ distribution comes from diagrams corresponding to two-jet final states. That is diagrams where both real gluons, $k_{1}$ and $k_{2}$, are clustered with the hard partons $p_{a}$ and $p_{b}$. Diagrams where one of the two gluons is in the interjet region, and hence not clustered with either hard parton, contribute at N$^2$LL level. Other gluonic configurations lead to contributions that are N$^3$LL and beyond. The $\CFsq$ part of the $\Or(\as^{2})$ jet-thrust distribution may be found by expanding the exponential of the LO result \eqref{R1_full-b}. The full expression including the running coupling at two--loop in the $\overline{\mr{MS}}$ will be presented in \sec{sec.resummation_in_QCD}. For the sake of comparison to the clustering case, we only report here the the LL term, which reads
\begin{equation}\label{FO_2_CF2}
\Sigma^{(2)}_{P}(\te,\Eo) = 2\,\CFsq
\left(\frac{\as}{2\pi}\right)^{2}\,\ln^{4}(\te).
\end{equation} 

Next we consider the derivation of the secondary emissions contribution to the jet-thrust distribution up to N$^3$LL including the full jet--radius dependence.

\subsubsection{Correlated emission and NGLs}
\label{subsec.anti-kt-NGL}

In the \AKT algorithm the non--global logarithmic contribution to the
$\te$ distribution in the small $\Rs$ limit is simply twice that of the single jet mass with a jet veto, $\ro=\tau_0$, distribution studied in the previous chapter. This is in line with the near--edge nature of non--global enhancements. In two-jet events, the well-separated\footnote{such that the jet--radius is much smaller than the jets' separation; $\Rs \ll (1- \cos\theta_{ij})$, where $\theta_{ij}$ is the angle between jets $i$ and $j$.} jets receive the latter enhancements independently of each other.
\begin{figure}
\centering
\epsfig{file=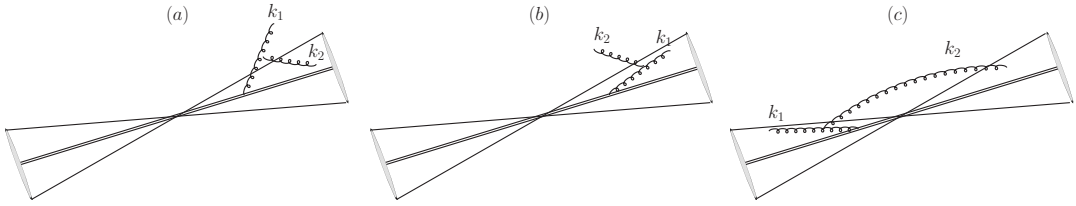, width = 0.95\textwidth}
\caption{Schematic representation of gluonic arrangement giving rise to NGLs. Here we assume that $\kt{t1} > \kt{t2}$. ($a$): harder gluon $k_1$ in interjet region emits softer gluon $k_2$ into $p_R$--jet, thus giving rise to leading NGLs contribution to the $p_R$--jet mass distribution. ($b$): $k_1$ inside $p_R$--jet emits $k_2$ into interjet region, thus giving rise to leading NGLs contribution to $E_0$ distribution. ($c$): $k_1$ inside $p_L$--jet emits $k_2$ into $p_R$--jet, thus leading to subleading NGLs contribution to $p_R$--jet mass. We have only shown the NGLs contributions to the $p_{R}$--jet, identical contributions apply to the $p_{L}$--jet.}
\label{fig.NGLs_akt}
\end{figure}
Possible final state gluonic arrangements relevant to NGLs at second order are
depicted in \fig{fig.NGLs_akt}. In this chapter we go beyond the small jet radius approximation assumed in the previous chapter when computing NGLs coefficients.

The all--orders resummed NGLs distribution may be written, including N$^3$LL non--global logs, in the form \cite{Dasgupta:2001sh}
\begin{equation}\label{S_t_gen}
\mc S(t) = 1+ \mc S_{22}\,\asb^2\,L^2 + \mc S_{21}\,\asb^2\,L + \cdots = 1+ \sum_{n=2}^{\infty} \sum_{m=1}^{n} \mc S_{nm}\,\asb^n\,L^{m},
\end{equation}
where $\asb\equiv \as/2\pi$ and $L = \ln(1/\te)$. We define the usual evolution parameter $t$ in terms of the coupling $\asb$ by
\begin{eqnarray}\label{t_param_akt}
 \nonumber t &=& \int_{\kt{t}^{\min}}^{\kt{t}^{\max}} \frac{\d\kt{t}}{\kt{t}} \;\asb(\kt{t}),\\
 &=& \asb\; \,\ln\left(\frac{\kt{t}^{\max}}{\kt{t}^{\min}}\right) = \asb\,\Lng,
\end{eqnarray}
where $\Lng$ is the non-global logarithm, the exact form of the upper and lower limits, $\kt{t}^{\max}$ and $\kt{t}^{\min}$, depend on the gluonic configuration and the second line in \eqref{t_param_akt} assumes a fixed coupling. We begin with the calculation of the leading non-global logs with the coefficient $\mc S_{22}$ for the configuration $(a)$ in \fig{fig.NGLs_akt}. While real gluon $k_2$ contributes to the mass of the $p_R$--jet, which then reads (using momentum parametrisation \eqref{4-momenta}) $\te = \ro_R = \kt{t2}\,e^{-\eta_2}/Q$, its corresponding virtual correction does not. Similar effect occurs at the $p_L$--jet. This leads to real-virtual mis-cancellation in the jet-thrust fraction \eqref{S1_tauo_dist} and thus giving rise to non-global logs.

Using the soft part of secondary emission eikonal amplitude $S_{ab}$, i.e., the term $W_S = 2\,S$ where $S$ is given in \eq{W_S}, the leading non--global contribution to the jet-thrust fraction, $ \cSup{\Sg_{\rm NG}}{2}$, from configuration $(a)$ in \fig{fig.NGLs_akt}, after adding virtual corrections, is
\begin{equation}\label{S2_akt_gen}
 \cSup{\mc S_{22}}{a}\,t_a^2 = -2\,\CF\CA\;\int \cSup{\d\Phi_2}{a}\, W_S\, \Theta\left( \ln\frac{\kt{t2}}{Q\te} - \eta_{2}\right) \Theta\sbr{\Eo - \kt{t1}\cosh\eta_{1}},
\end{equation}
where the factor $2$ accounts for the $p_L$--jet and the phase space measure, $\d\Phi_2^{(a)}$, is 
\begin{equation}
 \int\d\cSup{\Phi_2}{a} = \cbr{\astpi}^2 \int^{Q/2} \kt{t1} \d\kt{t1} \int^{Q/2} \kt{t2} \d\kt{t2} \int_{-\frac{\deta}{2}}^{\frac{\deta}{2}} \d\eta_1 \int_{\frac{\deta}{2}}^{+\infty} \d\eta_2 \int_0^{2\pi}\frac{\d\phi_1}{2\pi} \frac{\d\phi_2}{2\pi},
\label{S2_PS_a}
\end{equation}
where the interjet (gap) region, $\deta$ is given in \eq{eta-Rs} and due to boost invariance of rapidity variables it has been centred at $\eta=0$.
To account for the fact that the two gluons $k_1$ and $k_2$ may be of similar hardness (relevant at N$^3$LL), we insert the following unitary relation
\begin{equation}
 1 = \Theta\cbr{\kt{t1} - \kt{t2}} + \Theta\cbr{\kt{t2} - \kt{t1}},
\label{UnitarityRelation} 
\end{equation}
into the phase space measure \eqref{S2_PS_a} and multiply by a factor of $1/2$ to avoid double counting of overlapping regions of phase space. Therefore $\d\Phi^{(a)}_2$ reads
\begin{multline}
 \int\d\Phi_2^{(a)} = \cbr{\astpi}^2 \half\Bigg\{\int^{\frac{Q}{2}} \kt{t1}\d\kt{t1} \int^{\kt{t1}} \kt{t2}\d\kt{t2} \int_{-\frac{\deta}{2}}^{\frac{\deta}{2}} \d\eta_1 \int_{\frac{\deta}{2}}^{+\infty} \d\eta_2 \int_0^{2\pi}\frac{\d\phi_2}{2\pi}
\\  + 1 \lra 2\Bigg\},
\label{PS_NNLL}
\end{multline}
where we have used our freedom to set $\phi_1 = 0$. Due to the $1\lra 2$ symmetry of the soft amplitude \eqref{W_S}, the above phase space reduces to considering only one particular ordering (we choose $\kt{t1} > \kt{t2}$). Since there are no phase space constraints on the azimuthal angles then we average the soft amplitude $W_S$, \eq{W_S}, over $\phi_{2}$ to find
\begin{equation}
 \langle W_S \rangle = \int_0^{2\pi} \frac{\d\phi_2}{2\pi}\,W_S = -\frac{8}{\kt{t1}^2 \kt{t2}^2} \sbr{1 + \coth\cbr{\eta_1 - \eta_2}}.
\label{eq:WS_averaged}
\end{equation}
Performing the straightforward momentum integrals one obtains
\begin{multline}
 \mc S_{22}^{(a)}\,t_a^2 = 8\,\CF\CA\,\cbr{\astpi}^2\, \int_{-\frac{\deta}{2}}^{\frac{\deta}{2}} \d\eta_1 \int_{\frac{\deta}{2}}^{+\infty} \d\eta_2\, \ln^2\cbr{\frac{\Eo\,e^{-\eta_2}}{Q\,\te \cosh\eta_1}} \\ \sbr{1+\coth\cbr{\eta_1-\eta_2}} \Theta\cbr{\cosh\eta_1 - \frac{2\Eo}{Q}} \Theta\cbr{\ln\frac{\Eo}{Q\,\te \cosh\eta_1} - \eta_2}.
\label{S2_akt_integrated}
\end{multline}
Owing to the fact that it is non-trivial to analytically carry out the rapidity integrals in the form given above, we extract the leading non-global logarithms, compute their coefficient $\mc S_{22}$ analytically and provide a numerical estimate of the coefficient of the subleading non-global log (which contributes to $\mc S_{21}$). To this end we write
\begin{equation}
 \ln^2\sbr{\frac{\Eo\, e^{-\eta_2}}{Q\,\te \cosh\eta_1}} = \ln^2\sbr{\frac{2\Eo\,e^{-\deta}}{Q\, \te}} + 2\,\ln\sbr{\frac{2\Eo\,e^{-\deta}}{Q\,\te}}\,\cbr{\deta - \eta_2 -\ln2\cosh\eta_1} + \cdots,
\label{LeadingLog_expansion}
\end{equation}
where $\cdots$ denote terms that are beyond N$^3$LL. Thus the evolution parameter $t$ has the following formula for configuration $a$ at $\Or(\as^2)$:
\begin{equation}
 t_a^2 = \cbr{\astpi}^2\, \ln^2\cbr{\frac{2\Eo\,e^{-\deta}}{Q\, \te}} = \cbr{\astpi}^2\, L_{\rm ng}^2\,\Theta\sbr{\frac{2 E_0}{Q\,\te} -e^{\deta}},
\label{t_param_a}
\end{equation}
where the last step-function comes from $ E_0/(Q\te e^{\deta/2}) > \cosh\eta_1 > \cosh(\deta/2) \simeq e^{\deta/2}/2$ for $\deta > 0$. The expression of the leading NGLs coefficient in terms of $\ln(E_0/Q\te)$ is tedious and will not be reported here. However, in the limit $\te \rightarrow 0$ (region of large logs), it reduces to
\begin{multline}\label{S2_akt_a1}
\mc S_{22}^{(a)} (\deta) = -2\,\CF\CA \Bigg[\frac{\pi^{2}}{6} +
2\deta^{2} - 2\deta\,\ln\left(e^{2\deta}-1\right) - \Li_{2}\left(e^{-2\deta}\right) -\\ - \Li_{2}\left(1-e^{2\deta}\right)
\Bigg],
\end{multline}
where $\Li_2$ is the dilogarithm function \cite{olver2010nist}. An identical expression was found for the NGLs coefficient in the interjet energy flow distribution in \cite{Dasgupta:2002bw}\footnote{Our jet--radius, $\Rs$, is given in terms the parameter $c$, used in \cite{Dasgupta:2002bw}, by the relation: $1 - c = 2\Rs$.}. The fact that $\mc S_{22}^{(a)}$ is the same for $\te$ and interjet energy flow distributions means that the NGLs coefficient only depends on the geometry of the phase space and not on the observable itself. This is of course only true in the limit where the jet shape variable goes to zero. The difference between the jet shape variables amounts only to a difference in the logarithm's argument. We discuss the $R$-dependence of $\mc S_{22}^{(a)}$ below.

The remaining subleading coefficient is given, in the $\te\to 0$ limit, by
\begin{equation}
 \mc S_{21}^{(a)} = 16\,\CF\CA \int_{-\frac{\deta}{2}}^{\frac{\deta}{2}} \d\eta_1 \int_{\frac{\deta}{2}}^{+\infty} \d\eta_2\,\sbr{1+\coth\cbr{\eta_1-\eta_2}}  \cbr{\deta - \eta_2 -\ln2\cosh\eta_1}.
\label{S21a_akt}
\end{equation}
It is plotted in \fig{fig.NGLs_DL_SL} as a function of the jet-radius $\Rs$. Note that unlike the other N$^3$LL coefficients, computed below, which are maximum in the asymptotic region $\Rs\to 0$, $\mc S_{21}^{(a)}$ vanishes in the latter region.

Now consider configuration $(b)$ in \fig{fig.NGLs_akt}. Adding up the
corresponding virtual corrections, one obtains the following phase space
constraint
\begin{equation}\label{PS_const_b}
\Theta\left(\eta_{1} -\ln\left(\frac{\kt{t1}}{Q\te} \right) \right) \Theta\left(\kt{t2} - \frac{\Eo}{\cosh(\eta_{2})}\right).
\end{equation}
The phase space measure $\d\Phi^{(b)}_{2}$ is analogous, in the regime $\kt{t1} > \kt{t2}$, to the first line of $\d\Phi^{(a)}_{2}$ in \eqref{PS_NNLL}, with $\eta_1 \lra \eta_2$ swapped and the two $\Theta$--functions in \eqref{S2_akt_gen} replaced by those in \eq{PS_const_b}. The transverse momenta integrals yield
\begin{multline}
 \int_{\frac{\Eo}{\cosh\eta_2}}^{Q\te e^{\eta_1}} \frac{\d\kt{t1}}{\kt{t1}} \int_{\Eo}^{\kt{t1}} \frac{\d\kt{t2}}{\kt{t2}} = \half\,\ln^2\sbr{\frac{\Eo\, e^{-\eta_1}}{Q\te \cosh\eta_2}} \Theta\cbr{\eta_1 - \ln\frac{\Eo}{Q\te \cosh\eta_2}} \\ \times \Theta\cbr{1-2\te e^{\eta_1}}.
\label{S2_akt_integrated_b}
\end{multline}
The limits on $\eta_{1}$ are then $\ln(1/2\te) > \eta_{1} > \max[\deta/2,\ln(\Eo/Q\te\cosh\eta_2)]$. Since \eqs{S2_akt_integrated_b}{S2_akt_integrated} are symmetric under $1 \lra 2$ exchange but with opposite step--functions (since $k_1$ is still harder than $k_2$) then if we impose the constraint given in \eq{t_param_a}, i.e., $2\Eo/Q \gg \te e^{\deta}$, the lower limit becomes $\eta_{1} > \ln(\Eo/Q\te\cosh\eta_2)$. The contribution from configuration $(b)$ thus reads
\begin{multline}\label{S2_akt_b}
\cSup{\mc S_{22}}{b} t^{2}_b = 8\,\CF\CA \cbr{\astpi}^2 \int_{\ln\frac{\Eo}{Q\te \cosh\eta_2}}^{\ln(1/2\te)} \d\eta_{1} \int_{-\frac{\deta}{2}}^{\frac{\deta}{2}}\d\eta_{2}\, \ln^2\cbr{\frac{\Eo\, e^{-\eta_1}}{Q\te\cosh\eta_2}} \\ \left[1 + \coth(\eta_{1}-\eta_{2})\right] \Theta\cbr{\cosh\eta_2 - \frac{2\Eo}{Q}} \Theta\cbr{\ln\frac{\Eo}{Q\te\cos\eta_2} - \frac{\deta}{2}},
\end{multline}
where we have averaged the eikonal amplitude $W_S$ over $\phi_{2}$, after setting $\phi_1 = 0$. We can now proceed along similar lines to calculations in configuration $(a)$. The $t_b$ parameter is identical to $t_a$ given in \eq{t_param_a}. However, we see from the above equation that in the limit $\te\to 0$ both upper and lower limits on $\eta_1$ approach $\infty$, hence the coefficients $\cSup{\mc S_{22}}{b}$ and $\cSup{\mc S_{21}}{b}$ vanish in the latter limit.

The last contribution to NGLs at $\Or(\as^{2})$ comes from configuration $(c)$ in \fig{fig.NGLs_akt}. Upon the addition of the virtual correction, one is
left with the constraint
\begin{equation}\label{PS_const_c}
\Theta\left(Q\te - \kt{t1} e^{\eta_{1}} \right) \Theta\left(\kt{t2} e^{-\eta_{2}} - Q\te\right).
\end{equation}
Integrating out the transverse momenta one find for the evolution parameter
\begin{equation}
 t_c^2 = \cbr{\astpi}^2 \int_{Q\te e^{\eta_2}}^{Q\te e^{-\eta_1}} \frac{\d\kt{t1}}{\kt{t1}}  \int_{Q\te e^{\eta_2}}^{\kt{t1}} \frac{\d\kt{t2}}{\kt{t2}} = \half \cbr{\astpi}^2 \cbr{\eta_1 +\eta_2}^2.
\end{equation}
i.e., independent of $\te$ and thus beyond N$^3$LL for the $\te$ distribution.

We conclude that in the regime $2\Eo/Q \gg \te\, e^{\deta}$, which is equivalent to (\eq{eta-Rs}) $\Eo/Q \gg \te (1-\Rs)/\Rs$, the only non--vanishing
contribution to the NGLs comes from the phase space configuration $(a)$. Other
configurations either vanish  in the limit $\te \rightarrow 0$ (configuration $(b)$) or are subleading (configuration $(c)$). Hence 
\begin{equation}\label{S2_akt_tot}
 \mc S_{22} = \cSup{\mc S_{22}}{a},\;\;\; t = t_{a}.
\end{equation}
In \fig{fig.NG_coff_CA} (red curve) we plot $\mc S_{22}$ (for the anti-k$_{\rm T}$) as a function of the jet--radius $\Rs$. At the asymptotic limit $\deta \rightarrow +\infty$ (or equivalently $\Rs \rightarrow 0$) $\mc S_{22}$ saturates at $ -\CF\CA\; 2\pi^{2}/3$. This value (or rather half of it) is used as an approximation to $S_2 \equiv \mc S_{22}$ in the previous chapter. From \eq{S2_akt_a1}, we can see that the correction to such an approximation is less than $10\%$ for jet--radii smaller than $\Rs \sim 0.3$, which is equivalent to $R \sim 1$.  Furthermore, \eq{S2_akt_a1} confirms the observation made there (Chapter \ref{ch:EEJetShapes1}) that NGLs do not get eliminated when the jet--radius approaches zero. One may naively expects that when the jet size shrinks down to $0$ ($\Rs\to 0$), $\te$ becomes inclusive and hence $\mc S_{22}$ vanishes. To the contrary, $\mc S_{22}$ reaches its maximum in this limit. This is related to the nature of NGLs which originate mainly from the \emph{boundary} of the jet. 
Therefore, as long as the jet has a 
boundary (which is true even in the case $\Rs\to 0$) NGLs will be present. Further, in the limit $\Rs\to 0$ the rapidity separation between $k_1$ and $k_2$ (in \fig{fig.NGLs_akt}($a$)) can reach zero. This means that the amplitude is most singular and hence the phase space integration yields the largest value of $\mc S_{22}$.

Few important points are in order:
\begin{itemize}
  \item If we choose to order the energy scales in the $\Theta$--functions of \eq{t_param_a} the opposite way, i.e., $2\Eo/Q \ll \te\, e^{\deta}$ then configuration $(b)$ becomes leading, in NGLs, while the contribution from configuration $(a)$ vanishes. That is $t_{b}^{2}$ reads
   \begin{equation}\label{t_param_akt_b_no-ordering}
     t_{b}^{2} = \left(\frac{\as}{2\pi}\right)^{2}\ \ln^{2}\sbr{\frac{2\Eo\, e^{-\deta}}{Q\te}}\,\Theta\left(e^{\deta} - \frac{2\Eo}{Q\,\te}\right).
   \end{equation}
  and $\cSup{\mc S_{22}}{b}$ takes the form \eqref{S2_akt_a1} while $\cSup{\mc S_{22}}{a}$ vanishes. We do not consider this regime here though. 

  \item If, on the other hand, we do not restrict ourselves to any particular ordering of the scales, as is done in Refs. \cite{Kelley:2011tj} and \cite{Hornig:2011tg}, then both configurations $(a)$ and $(b)$ would contribute to the leading NGLs. Adding up $t_{a}^{2}$, in \eqref{t_param_a}, and $t_{b}^{2}$, in \eqref{t_param_akt_b_no-ordering}, the $\Theta$--functions sum up to unity and one recovers the result reported in the above mentioned references.
 
 \item Setting the veto scale $\Eo \sim \te Q$ in $t_{a}$, \eq{t_param_a}, and $t_{b}$, \eq{t_param_akt_b_no-ordering}, would diminish NGLs coming from both configurations $(a)$ and $(b)$ and the jet-thrust becomes essentially a global observable. This is unlike the observation made in the study of the angularities distribution ($\ta$) in the previous chapter (Chapter \ref{ch:EEJetShapes1}), where the above choice of $\Eo$ diminishes the NGLs near the measured jet but introduces other equally significant NGLs near the unmeasured jet.
\end{itemize}
Below we turn to the computation of the N$^3$LL contribution to the jet-thrust shape fraction. We only treat configuration $(a)$ in \fig{fig.NGLs_akt} and do not attempt to address the other subleading configurations.

\subsubsection{N$^3$LL non-global structure}
\label{sss.NNLLNGLs}

In analogy to the leading N$^2$LL NGLs contribution \eqref{S2_akt_gen}, the subleading N$^3$LL NGLs contribution to the jet-thrust distribution is given by
\begin{equation}
 \mc S_{21}\, \asb\,t = -2\int\d\Phi_2 \sbr{\CF\CA\, H_g + \CF\TF\nf\,H_q} \Theta\cbr{\ln\frac{\kt{t2}}{Q\,\te} - \eta_2} \Theta\sbr{E_0 - \kt{t1}\cosh\eta_1},
\label{S21_gen}
\end{equation}
where the hard emission amplitudes $H_g$ and $H_q$ are defined in \eq{Hg_Hq}. We follow the steps highlighted in calculating $\cSup{\mc S_{22}}{a} t^2_a$. However, we should bear in mind that the hard amplitudes $H_g$ and $H_q$ are not symmetric under the exchange of $\kt{t1} \lra \kt{t2}$, and hence we use the full expression \eqref{PS_NNLL}. Performing the energy integral one obtains an identical logarithm to that of the leading soft contribution, and hence similar evolution parameter to \eq{t_param_a}. In what follows below we report the phase space coefficient $\mc S_{21}$ of $\ln(1/\te)$:
\begin{equation}
 \mc S_{21} = \half\sbr{\int_{-\frac{\deta}{2}}^{\frac{\deta}{2}} \d\eta_1 \int_{\frac{\deta}{2}}^{+\infty} \d\eta_2 \int_0^{2\pi}\frac{\d\phi_2}{2\pi}  + 1 \lra 2}\, \mc A(\eta_1,\eta_2,\phi_2),
\label{S21_gen_b}
\end{equation}
where $\mc A$ is the amplitude obtained after integrating out the transverse momenta. Starting with the two-gluon hard amplitude, $H_g$, we have (leaving out a $-4\,\CF\CA$ pre-factor): 
\begin{enumerate}[i.]
 \item The $\mc I^2$ term
 \begin{equation}
  \mc A = W_{\mc I} = \sbr{\frac{1 + (\eta_2-\eta_1)\coth(\eta_2-\eta_1)}{\cbr{\cosh(\eta_1-\eta_2) - \cos\phi_2}^2 } }.
 \label{Amp_Isq_term}
 \end{equation}
and 
\begin{multline}\label{S21_J2}
  \mc S_{21}^{(\rm i)}(\deta) = \frac{1}{3}\Bigg[\frac{\pi^{2}}{6} - \Li_{2}\left(e^{-2\deta}\right) - \deta^{2} + \frac{\coth(\deta)}{2} - \frac{\deta}{2\sinh^{2}(\deta)} + \\+ \deta\,\ln\left[2\sinh(\deta)\right] \Bigg].
\end{multline}

 \item The $S \mc J$ term
 \begin{equation}
  \mc A = W_{S\mc J} = 2\sbr{\frac{(\eta_2-\eta_1)\cos\phi_2 \coth(\eta_2-\eta_1)}{\cosh(\eta_1-\eta_2) - \cos\phi_2} }.
 \label{Amp_SJ_term}
 \end{equation}
and
\begin{multline}\label{S21_SR}
  \mc S_{21}^{(\rm ii)}(\deta) = 2 \Bigg[\frac{\pi^{2}}{6} - \Li_{2}\left(e^{-2\deta}\right) - 2\deta^{2} - \frac{4\deta^{3}}{3} + \deta\,\ln\left(e^{2\deta} -1\right) \\- \deta\left[\frac{1}{2}\Li_{2}\left(e^{-2\deta}\right) + \Li_{2}\left(e^{2\deta}\right) \right] + \frac{\Li_{3}\left(e^{2\deta}\right)}{2} - \frac{\zeta_{3}}{2}\Bigg].
\end{multline}

 \item The $w_{ab}(k)$ term
 \begin{equation}
  \mc A = W_{w_{ab}(k)} = \half \sbr{\frac{(\eta_2-\eta_1)\coth(\eta_2-\eta_1)}{\cosh(\eta_1-\eta_2) - \cos\phi_2 } }.
 \label{Amp_w_term}
 \end{equation}
and
\begin{equation}\label{S21_k}
  \mc S_{21}^{(\rm iii)}(\deta) = \frac{1}{2}\left[\frac{\pi^{2}}{6} - \Li_{2}\left(e^{-2\deta}\right) - \deta^{2} + \deta\,\ln\left(2\sinh(\deta\right)\right].
\end{equation}  
\end{enumerate}
Adding up the above three contributions according to \eq{Hg_Hq}, i.e., 
\begin{equation}
\mc S_{21}^{\CA} = -4\,\CF\CA\cbr{2\mc S_{21}^{(\rm i)} - \mc S_{21}^{(\rm ii)} - 4 \mc S_{21}^{(\rm iii)} }, 
\end{equation}
one obtains for the N$^3$LL NGLs coefficient, in the $\CF\CA$ channel,
\begin{multline}\label{S21_Hg}
  \mc S_{21}^{\CA}(\deta) = -2\,\CF\CA\Bigg\{-\frac{11\pi^{2}}{9} + 2\zeta_{3} +\frac{16\deta^{3}}{3} + \frac{70\deta^{2}}{3} - 2\deta \Bigg[\frac{1}{6\sinh^{2}(\deta)}
  \\
   + 8\,\ln\left(e^{2\deta}-1\right) - \frac{11}{3}\,\ln\left(2\sinh(\deta)\right) -\Li_{2}\left(e^{-2\deta}\right) -
  \\
  -2\,\Li_{2}\left(e^{2\deta}\right) \Bigg] + \frac{\coth(\deta)}{3} + \frac{22}{3}\,\Li_{2}\left(e^{-2\deta}\right) - 2 \Li_{3}\left(e^{2\deta}\right)\Bigg\}.
\end{multline} 
The asymptotic values of $\mc S^{\CA}_{21}$ are given by (recall from \eq{eta-Rs} that $\deta \rightarrow +\infty$ is equivalent to $\Rs \rightarrow 0$ and $\deta \rightarrow 0$ is equivalent to $\Rs \rightarrow 1/2$),
\begin{eqnarray}
\nn \lim_{\deta \rightarrow +\infty} \mc S_{21}^{\CA} &=& -4\,\CF\CA\sbr{-\frac{11\pi^{2}}{18} + \zeta_{3} + \frac{1}{6}},\label{S_21_Rs-0_limit}
\\
\lim_{\deta \rightarrow 0} \mc S^{\CA}_{21} &=& 0.\label{S_21_Rs-half_limit}
\end{eqnarray}
where $\zeta_3 \simeq 1.202$.

Since $H_{q}$ is defined in terms of $\mc I$ and $w_{ab}(k)$, we find that it also contributes to the $\tauo$ non--global distribution. The evolution parameter is analogous to \eq{t_param_a} and the corresponding coefficient is simply given by
\begin{multline}\label{S21_Hq}
 \mc S_{21}^{\nf}(\deta) = -2\,\CF\TF\nf\Bigg\{\frac{4\pi^{2}}{9} - \frac{8}{3} \,\Li_{2}\left(e^{-2\deta}\right) + \frac{2\coth(\deta)}{3} + \frac{16\deta^{2}}{3} 
\\
 + \frac{4\deta}{3} \Bigg[\frac{1}{2\sinh^{2}(\deta)} + 6\,\ln\left(1-e^{-2\deta}\right) + 4\,\ln\left(\frac{1}{2\sinh(\deta)}\right) \Bigg]\Bigg\}.
\end{multline}
At the aforementioned $\deta$ limits, it reduces to
\begin{eqnarray}
\nn   \lim_{\deta \rightarrow +\infty} \mc S_{21}^{\nf} &=& -2\,\CF\TF\nf\sbr{\frac{4\pi^{2}}{9} - \frac{2}{3}},\label{S_21_Rs-0_limit_nf}
\\
 \lim_{\deta \rightarrow 0} \mc S_{21}^{\nf} &=& 0.\label{S_21_Rs-half_limit_nf}
\end{eqnarray}
We plot $\mc S_{21}^{\CA}$ and $\mc S_{21}^{\nf}$ coefficients as functions of the jet--radius $\Rs$ in \fig{fig.NGLs_DL_SL}. Notice that identical formulae to Eqs.\eqref{S21_Hg}, \eqref{S_21_Rs-0_limit}, \eqref{S21_Hq} and \eqref{S_21_Rs-half_limit_nf} have been found in \cite{Kelley:2011aa, Kelley:2011ng}\footnote{Note that in \cite{Kelley:2011aa, Kelley:2011ng} one has $\asb = \as/4\pi$.} using SCET.

In the next subsection we recompute both primary and correlated emission contributions to the jet-thrust distribution in the C/A algorithm up to N$^3$LL accuracy.

\subsection{The C/A algorithm}
\label{sec.FO_CA}

The definition of the C/A algorithm is given in Algorithm~\ref{Alg:SRAlgInclusive}
with $\rm p = 0$ and the corresponding $\EE$ version is discussed in \ssec{subsec.observable_def}. Unlike the \AKT algorithm, which successively merges soft gluons with the nearest hard parton, the  C/A algorithm proceeds by successively clustering soft gluons amongst themselves. Consequently, a soft parton may in many occasions be dragged into (away from) a jet region and hence contributing (not contributing) to the invariant mass of the latter. The jet mass, and hence $\te$, distribution is then modified. It is these modifications, due to soft--gluons self--clustering, that we shall address below.

Any clustering--induced contribution to the $\te$ distribution will only arise
from phase space configurations where the two soft gluons, $k_{1}$ and $k_{2}$,
are initially (that is, before applying the clustering) in different regions of
phase space. Configurations where both gluons are within the same jet region,
gluon $k_{1}$ is in one of the two jet regions and gluon $k_{2}$ is in the other, or both gluons are within the interjet region are not altered by clustering and calculations of the corresponding contributions will yield identical results to the \AKT algorithm. We can therefore write the $\te$ distribution in the C/A algorithm, at $\Or(\as^{2})$, as
\begin{equation}\label{Sig2_CA_akt}
\Sigma^{(2)}_{\ca}(\te,\Eo) = \Sigma^{(2)}_{\akt}(\te,\Eo) + 
\cSup{\Sg_{\rm clus}}{2}(\te,\Eo).
\end{equation}
It is the last term in \eq{Sig2_CA_akt} that we compute in the present
section.

\subsubsection{Independent emission}
\label{subsec.CA-LL}

Since we are aiming at N$^3$LL accuracy we will be taking into account hard emission contributions through employing the full Altarelli--Parisi splitting kernel for the emission of a gluon, with momentum fraction $x$, off a quark; $P_{gq}(x) = \sbr{1+(1-x)^2}/x$, and integrating over the full phase space including regions with $x_1 \sim x_2$. We follow the steps outlined in the previous subsection, \ssec{subsec.anti-kt-NGL} and \eq{UnitarityRelation}, to achieve this. Like $W_S$ \eqref{W_S}, the independent emission amplitude \eqref{W_P} is symmetric under the exchange of the two gluons and we simply consider one ordering, say $x_1 > x_2$, and multiply the final answer by two, which cancels against the factor $1/2$ that was introduced to avoid double counting. In fact, in the two-jet limit we can impose the stronger ordering $x_1 \gg x_2$ and focus on phase space regions that lead to large logarithms\footnote{Comparisons to the NLO program {\sc event}{\scriptsize 2} in \sec{sec.numerical_results} indicates 
that 
corrections to this 
approximation, if any, are negligible.}.

Let the two gluons' momenta, using the $\EE$ variables ($E,\theta,\phi$), be written as
\begin{eqnarray}\label{4-momenta_theta}
\nn k_1 &=& \om_1 \cbr{1,\sin\theta_1\cos\phi_1, \sin\theta_1\sin\phi_1, \cos\theta_1}, \\
k_2 &=& \om_2 \cbr{1,\sin\theta_2\cos\phi_2, \sin\theta_2\sin\phi_2, \cos\theta_2},
\end{eqnarray}
and keep the same parametrisation \eqref{4-momenta} for the hard $q\qb$ pair. Then assuming that $x_1 \gg x_2$, where $x_i = 2\om_i/Q$, and focusing on one jet, say $p_R$--jet (we multiply by two to account for the second jet) we have the following gluon configurations, which are absent in the \AKT case:
\begin{enumerate}[(A)]
\item The harder gluon $k_1$ is in the interjet region and the softer one $k_2$ is inside the jet. Applying the jet algorithm, $k_1$ pulls $k_2$ out of the jet if $d_{12} < d_{2j}$, where $j$ stands for the jet and $d_{ij}$ is the distance measure discussed in \ssec{subsec.observable_def}, to form a third jet, which is then vetoed to have energy less than $E_0$ (depicted in \fig{fig.CLs_CF2}). In the latter case gluon $k_2$ does not contribute to the $\te$ observable. Configurations whereby gluon $k_1$ is virtual and thus cannot pull $k_2$ out the of the jet do however contribute to the value of $\te$. Hence a real--virtual mis-cancellation occurs leading to the appearance of large logs. We translate this configuration into the following step functions on the phase space integral of the shape fraction,
 \begin{figure}[t]
 \centering
 \includegraphics[width=15cm]{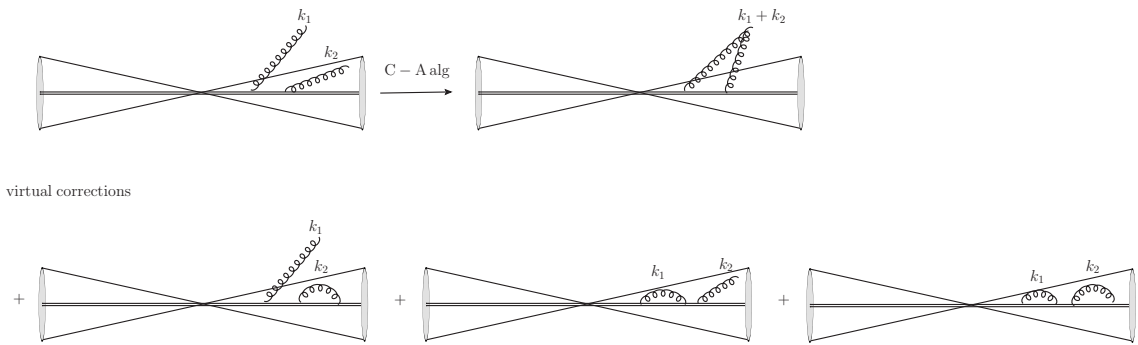}
 \caption{A schematic representation of a three--jet final state after applying
 the C/A algorithm on real emission along with virtual correction diagrams. The
 two gluons are clustered in the E--Scheme (see \sec{sec.fixed_order_1}).
 Identical diagrams hold for the $p_{L}$--jet.}
 \label{fig.CLs_CF2}
 \end{figure}
\begin{equation}\label{clust_cond_A}
\sbr{\Theta\cbr{\frac{2\Eo}{Q} - x_1 - x_2} -\Theta\cbr{\frac{2\Eo}{Q} - x_1} + \Theta\left(\frac{x_2 \theta_{2}^{2}}{4} - \te\right)} \Theta^{\rm clus}(k_1, k_2).
\end{equation}
where, in the small--angle limit, the last clustering function is given by \eq{eq:EEJS1:KT-CA-ClusFunAlpha2}

\item The harder gluon $k_1$ is in the interjet region and the softer one $k_2$ is inside the jet. Applying the jet algorithm, gluon $k_1$ \emph{does not} pull $k_2$ out if $d_{12} > d_{2j}$. This configuration is then identical to the \AKT case but with a more restricted phase space. If we write $\Theta(d_{12}-d_{2j}) = 1-\Theta(d_{2j} -d_{12})$ then the resultant phase space constraint after adding virtual corrections explicitly reads 
\begin{equation}\label{clust_cond_B}
\Theta\left(\frac{x_2 \theta_{2}^{2}}{4} - \te\right) \Theta\cbr{x_1 - \frac{2\Eo}{Q}} \left[1 - \Theta^{\rm clus}(k_1, k_2)\right].
\end{equation}
The first term in the square bracket is actually part of the \AKT contribution, $\Sg^{\akt}$ in \eq{Sig2_CA_akt}. Thus we only consider the second term (involving the clustering function $\Theta^{\rm clus}$).

\item The harder gluon $k_1$ is inside the jet and the softer one $k_2$ is in the interjet region. Applying the jet algorithm gluon $k_1$ pulls $k_2$ inside the jet if $d_{12} < d_{1j}$. Upon adding virtual corrections, depicted in \fig{fig.CLs_2}, one obtains the following phase space constraint
\begin{equation}\label{clust_cond_C}
\sbr{-\Theta\left(\te -\frac{x_1 \theta_{1}^{2}}{4} \right)
\Theta\left(\frac{x_2 \theta_{2}^{2}}{4}  - \te\right) +
\Theta\cbr{x_2 - \frac{2\Eo}{Q}}} \Theta_C^{\rm clus}(k_1, k_2),
\end{equation}
where we have assumed small angles limit and employed the LL accurate
approximation
\begin{equation}
\Theta\left(\te - \frac{x_1 \theta_1^2}{4} - \frac{x_2 \theta_2^2}{4}\right) \simeq \Theta\left(\te - \frac{x_1 \theta_1^2}{4}\right) \Theta\left(\te - \frac{x_2 \theta_2^2}{4}\right).
\end{equation}
The clustering function $\Theta^{\rm clus}_C$ is identical to
\eq{eq:EEJS1:KT-CA-ClusFunAlpha2} with $1 \lra 2$ swapped. We shall see below that this configuration yields subleading (beyond N$^3$LL) contribution and thus we should not worry about the above LL approximation.
\begin{figure}[t]
\centering
\includegraphics[width=15cm]{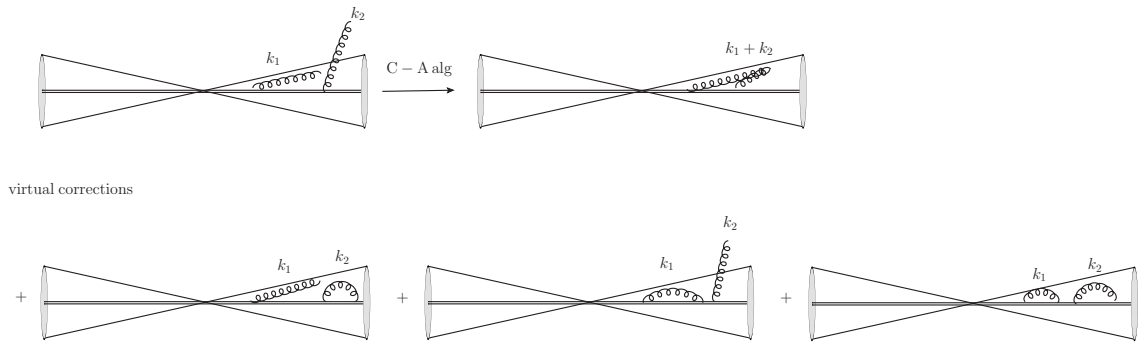}
\caption{A schematic representation of a two-jet final state after applying the C/A algorithm on real emission along with virtual correction diagrams. The two gluons are clustered in the E--Scheme (see \sec{sec.fixed_order_1}).
Identical diagrams hold for the left ($p_{L}$--) jet.}
\label{fig.CLs_2}
\end{figure}

\item The harder gluon $k_1$ is inside the jet and the softer one $k_2$ is in the interjet region. Applying the jet algorithm gluon $k_1$ \emph{does not} pull $k_2$ into the jet if $d_{12} > d_{1j}$. Adding virtual corrections yields the phase space constraint
\begin{equation}\label{clust_cond_D}
 \Theta\cbr{x_1 \theta_1^2 - 4\te} \Theta\cbr{x_2 - \frac{2\Eo}{Q}} \sbr{1 - \Theta_C^{\rm clus}(k_1, k_2)}.
\end{equation}
In analogy to \eq{clust_cond_B} the first term in the square bracket is part of the \AKT contribution and we only consider the second term.
\end{enumerate}
Adding up the contributions, to the clustering piece $\cSup{\Sg_{\rm clus}}{2}$, from cases (A) and (B) and recalling that we are considering the energy ordering: $x_1 \gg x_2$ and thus $x_1 + x_2 \sim x_1$ then we are left with
\begin{equation}\label{clust_cond_A+B}
 \Theta\cbr{x_2\theta_2^2 - 4\te} \Theta\cbr{\frac{2\Eo}{Q} - x_1} \Theta^{\rm clus}(k_1,k_2).
\end{equation}
The new contribution from the above configuration, (A)+(B), to the $\CFsq$ piece of the $\te$ distribution is then given by
\begin{multline}\label{eq:C2_CA_A}
 \cSup{\Sg_{\rm clus}}{2} = 8\,\CFsq\cbr{\astpi}^2\int_0^1 \d x_1\frac{ 1 + (1-x_1)^2}{x_1} \int_0^{x_1} \d x_2\frac{1+(1-x_2)^2}{x_2} \int_{-\frac{\pi}{3}}^{\frac{\pi}{3}}\frac{\d\phi_{2}}{2\pi}
\\ 
 \int_{2\sqrt{\Rs}}^{2\sqrt{\Rs}\cos\phi_{2}} \frac{\d\theta_{1}}{\theta_{1}} \int^{2\sqrt{\Rs}}_{\theta_1/2\cos\phi_{2}} \frac{\d\theta_{2}}{\theta_{2}}\, \Theta\cbr{x_2 \theta_2^2 - 4\te} \Theta\cbr{\frac{2\Eo}{Q} - x_1},
\end{multline}
Performing the energy integrals we have
\begin{multline}\label{eq:C2_CA_B}
 \cSup{\Sg_{\rm clus}}{2}  \propto 16\,\CFsq\cbr{\astpi}^2 \left\{\ln^2\cbr{\frac{\Eo\,\theta_2^2}{2 Q\,\te}} - \sbr{\frac{4\Eo}{Q} + \Or\cbr{\frac{\Eo^2}{Q^2}}} \ln\cbr{\frac{\Eo\,\theta_2^2}{2 Q\,\te}}  + \cdots \right\} \\ \Theta\cbr{\frac{2\Eo}{Q} - \frac{4\te}{\theta_2^2}},
\end{multline}
where $\cdots$ represent subleading terms. Up to N$^3$LL accuracy we write the above expression as
\begin{equation}\label{eq:C2_CA_C}
  \cSup{\Sg_{\rm clus}}{2}  \propto 16\,\CFsq\cbr{\astpi}^2 \left\{\ln^2\cbr{\frac{2 E_0\Rs}{Q\,\te}} + \ln\cbr{\frac{2 E_0 \Rs}{Q\,\te}}  \sbr{2\,\ln\frac{\theta_2^2}{4\Rs} -\frac{4 \Eo}{Q}} + \cdots \right\}.
\end{equation}
If we define the primary evolution parameter, $t_p$, at $\Or(\as^2)$ as
\begin{equation}\label{t_params_p}
 t_p^2 = \cbr{\astpi}^2\, \ln^2\cbr{\frac{2\Eo\Rs}{Q\,\te}}\,\Theta\cbr{\frac{2\Eo}{Q} - \frac{4\te}{\Rs}},
\end{equation}
which we note is identical to the non-global evolution parameter \eqref{t_param_a}, then the leading CLs coefficient is given by
\begin{equation}\label{C2_CA}
 C^P_{22} = 16\,\CFsq \int_{-\frac{\pi}{3}}^{\frac{\pi}{3}}\frac{\d\phi_{2}}{2\pi}
\int_{2\sqrt{\Rs}}^{2\theta_{2}\cos\phi_{2}} \int_{2\sqrt{\Rs}}^{2\sqrt{\Rs}\cos\phi_{2}} \frac{\d\theta_{1}}{\theta_{1}} \int^{2\sqrt{\Rs}}_{\theta_1/2\cos\phi_{2}}  \frac{\d\theta_{2}}{\theta_{2}} = 0.73\,\CFsq.
\end{equation}
This result is identical to\footnote{It is actually twice, since here we are measuring both jets.} that found in the previous chapter for the angularities (without a jet veto) distribution. The argument of the two logarithms are different though. The reason for this is that the clustering requirement only affects the distribution to which the softest gluon contributes, which in both cases is the jet mass distribution. Thus the clustering coefficient is, like the non-global coefficient, only dependent on the geometry of the final state (and not on the jet shape). The N$^3$LL subleading coefficient, of $\asb\, t_p$, is 
\begin{equation}\label{C21_CA}
 C^P_{21} = \wt{C}^P_{21} -  \frac{4\Eo}{Q}\,C^P_{22},
\end{equation}
where
\begin{equation}
\wt{C}^P_{21} = 32\,\CFsq \int_{-\frac{\pi}{3}}^{\frac{\pi}{3}}\frac{\d\phi_{2}}{2\pi}
\int_{2\sqrt{\Rs}}^{2\theta_{2}\cos\phi_{2}} \int_{2\sqrt{\Rs}}^{2\sqrt{\Rs}\cos\phi_{2}} \frac{\d\theta_{1}}{\theta_{1}} \int^{2\sqrt{\Rs}}_{\theta_1/2\cos\phi_{2}}  \frac{\d\theta_{2}}{\theta_{2}}\,\ln\frac{\theta_2^2}{4\Rs}
 = -0.59\,\CFsq.
\end{equation}
We can actually compute the above clustering coefficients beyond the small-angle limit. The full details for the leading coefficient $C^P_2 \equiv C^P_{22}$ are in \app{Sec:F2}. The final result reads
\begin{equation}\label{C2_CA_FullR}
 C^P_{22} = 4\,\CFsq \sbr{0.183 + 0.024\,\Rs^2 + 0.00183\,\Rs^4 + 0.000125\,\Rs^6 +
\Or(\Rs^{8})}.
\end{equation}
The $\wt{C}^P_{21}$ coefficient is evaluated numerically and the final results, together with those of $C_{22}^P$, for various $\Rs$ values, are reported in Table \ref{tab:CLs_Coeffs_FullR}. Note that we replace $\Rs$ in the argument of the clustering logarithm in \eq{C2_CA} by $\Rs/(1-\Rs)$ for large values of the jet radius when comparing to \event in \sec{sec.numerical_results}.
\begin{table}[t]
 \centering
\begin{tabular}{|c|c|c|c|c|c|}
\hline
 $\Rs$                 & $0.0025$ & $0.04$  & $0.12$  & $0.30$ & $0.50$\\
\hline
 $C^P_{22}/\CFsq$      & $0.73$   & $0.74$  & $0.76$  & $0.89$ & $1.30$ \\
\hline
 $\wt{C}^P_{21}/\CFsq$ & $-0.59$  & $-0.63$ & $-0.72$ & $-0.95$& $-1.45$ \\
\hline
\end{tabular}
\caption{Full $\Rs$ numerical results for the leading and next-to-leading clustering logs' coefficients.}
\label{tab:CLs_Coeffs_FullR}
\end{table}

We now address the other two cases. Adding up the phase space constraints (C) + (D) we have
\begin{equation}\label{clust_cond_C+D}
 \Theta\cbr{4\te - x_1\theta_1^2} \Theta_C^{\rm clus} \sbr{\Theta\cbr{x_2 - \frac{2\Eo}{Q}}-\cbr{x_2\theta_2^2 - 4\te} }.
\end{equation}
The first step function in the square bracket yields results that are exactly identical to case (A)+(B) discussed above. The only difference, however, is that the ordering in the evolution parameter is now reversed. That is
\begin{equation}\label{t_params_p_C+D}
 t_p^{'2} = \cbr{\astpi}^2\, \ln^2\cbr{\frac{2\Eo\Rs}{Q\,\te}}\,\Theta\cbr{\frac{4\te}{\Rs} - \frac{2\Eo}{Q}}.
\end{equation}
In this chapter we work in the regime of \eq{t_params_p} and thus \eq{t_params_p_C+D} is subleading and will be neglected. Likewise, the second step function in \eqref{clust_cond_C+D} yields upon energy integration logarithms purely in the ratio of $\theta_1$ and $\theta_2$, i.e., beyond N$^3$LL level and hence will be neglected too. 

Next we compute the $\CF\CA$ and $\CF\TF\nf$ pieces of $\cSup{\Sg_{\rm clus}}{2}$.

\subsubsection{Correlated emission}
\label{subsec.CA-NGL}

\begin{figure}[t]
\centering
\epsfig{file=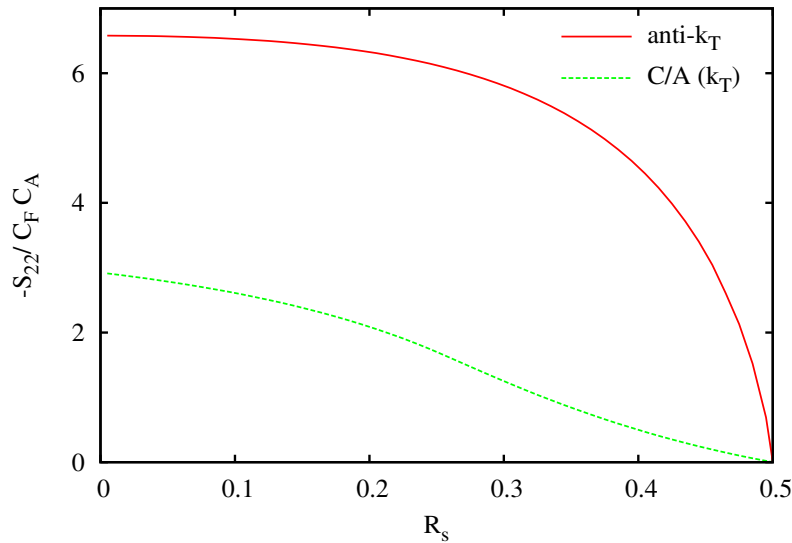, width= 0.75 \textwidth}
\caption{Leading non-global coefficient $\mc S_{22}$ in the \AKT and C/A algorithms.}
\label{fig.NG_coff_CA}
\end{figure}

Consider the gluonic configuration $(a)$ depicted in \fig{fig.NGLs_akt}.
Applying the C/A clustering algorithm on the latter yields two possibilities.
Namely the two gluons are either clustered or not. The former case completely
cancels against virtual corrections and thus does not contribute to NGLs. It is when the two gluons survive the clustering, the latter case, that a real--virtual mismatch takes place and NGLs are induced. The corresponding evolution parameter is equal to $t$ of the \AKT case, \eq{S2_akt_tot}. If we use the parametrisation \eqref{4-momenta_theta} then the clustering condition is simply one minus that in \eq{eq:EEJS1:KT-CA-ClusFunAlpha2}. The leading and next-to-leading NGLs coefficients can then be written, using the eikonal amplitude \eqref{W_S}, as
\begin{equation}\label{S2_CA_clus}
\mc S^{\ca}_{2i} = \mc S_{2i} + \mc S_{2i}^{\rm clus},\qquad i=1,2.
\end{equation}
The leading coefficient $\mc S_{22}$ is given in \eq{S2_akt_tot} and the clustering--induced correction $\mc S_{22}^{\rm clus}$ can be carried out along similar lines to $\mc S_{22}^{(a)}$ above, including the use of the symmetry of $W_S$ to reduce the phase space measure \eqref{PS_NNLL}. We have 
\begin{multline}\label{S2_clus_CA}
\mc S_{22}^{\rm clus} = 8\,\CF\CA
\int_{\sqrt{\Rs}}^{2\theta_{2}\cos\phi_{2}}\frac{\d\theta_{1}}{\sin\theta_{1}}
\int_{\frac{\sqrt{Rs}}{\cos\phi_{2}}}^{2\sqrt{\Rs}}\frac{\d\theta_{2}}{
\sin\theta_{2}} \int_{-\pi/3}^{\pi/3}\frac{\d\phi_{2}}{2\pi}
\left[\frac{1-\cos\theta_{1}\cos\theta_{2}}{1-\cos\theta_{12}} -1\right] \times
\\ \times \Theta\left(\frac{\Rs}{\te\cos\phi_{2}} -
\frac{Q}{2\omega_{2}}\right).
\end{multline}
We can perform the $\theta_{1}$--integral analytically and then resort to
numerical methods to evaluate the remaining $\theta_{2}$ and $\phi_{2}$
integrals. The result, in terms of the jet--radius $\Rs$, is depicted in
Fig.~\ref{fig.NG_coff_CA}.
$- \mc S^{\ca}_{22}$ saturates at around $0.44\times 2\pi^{2}/3\, \CF\CA\sim
2.92\, \CF\CA$, i.e., a reduction of more than $50\%$ in $\mc S_{22}$. This is due to the fact that for the two gluons to survive clustering they need to be
sufficiently far apart ($\theta_{12}^2 > 4\,\Rs$). The dominant
contribution to $\mc S_{22}$ comes, however, from the region of phase space where the gluons are sufficiently close. This corresponds to the collinear region of the matrix--element; $\theta_{1} \sim \theta_{2}$. Hence the further apart the two gluons get from each other, the less (collinear) singular the matrix becomes and thus the smaller the value of the NGLs coefficient.

In fact, we can numerically evaluate the full $\Rs$ dependence of $\mc S_{22}^\ca$ using the momentum parametrisation given in \app{Sec:F2}, just as we did with $C^P_{22}$. An alternative easier method, used in \cite{Hornig:2011tg}, is to write the polar C/A clustering condition $\theta_{2}^2 > \theta_{12}^2$, or equivalently $\cos\theta_{12} > \cos\theta_2$, in terms of $\eta$ and $\phi$. Recalling that $\eta = \ln\cot\theta/2$ and $\cos\theta_{12} = \cos\theta_1\cos\theta_2 + \sin\theta_1\sin\theta_2 \cos\phi$ we have
\begin{equation}
 \cos\theta_{12} > \cos\theta_{2} \Rightarrow \cos\phi > e^{-\eta_1} \sin\eta_2.
\label{CA_ClusCond-Theta2Eta}
\end{equation}
The coefficient $\mc S_{22}$ in the C/A algorithm is then given by
\begin{equation}
 \mc S_{22}^{\ca} = -\CF\CA \int_{-\frac{\deta}{2}}^{\frac{\deta}{2}} \d\eta_1 \int_{\frac{\deta}{2}}^{+\infty} \d\eta_2 \int_0^{2\pi}\frac{\d\phi}{2\pi}\, W_S\,\sbr{1 - \Theta\cbr{\cos\phi - e^{-\eta_1}\sin\eta_2 }},
\label{S22_CA_Full-R}
\end{equation}
where $W_S$ is given in \eq{W_S}. The result of integration is very close to that of \eq{S2_clus_CA} and hence we only show one curve in \fig{fig.NG_coff_CA}. The next-to-leading NGLs coefficient $\mc S_{21}^{(a), \ca}$ \eqref{S21a_akt} has an analogous form to \eqref{S22_CA_Full-R} with $W_S$ replaced by $W_S \sbr{\deta -\eta_2 -\ln(2\cosh\eta_1)}$. The integration result is plotted in \fig{fig.NGLs_DL_SL}.

\subsubsection{N$^3$LL non-global structure}

\begin{figure}[!t]
\centering
\epsfig{file=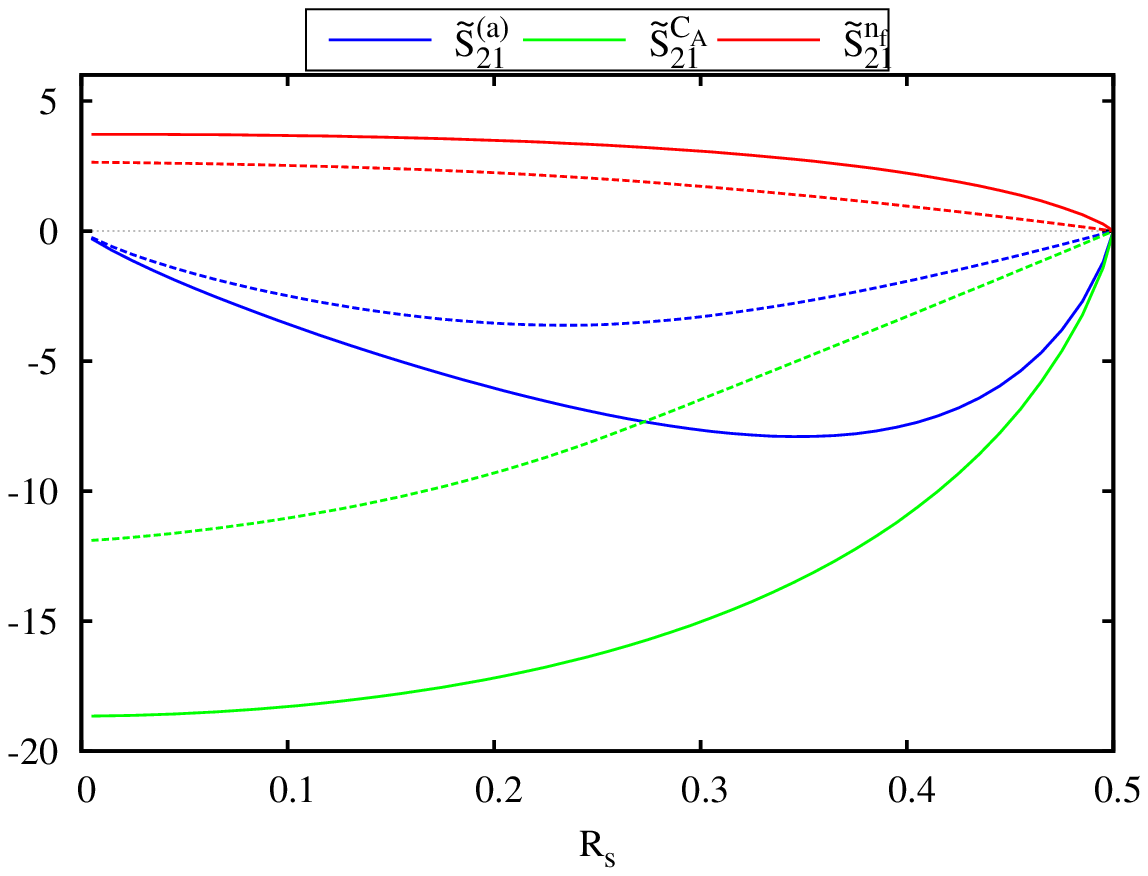, width= 0.75\textwidth}
\caption{N$^3$LL NGLs coefficients for both $\CF\CA$ and $\CF\TF\nf$ colour channels in \AKT (solid) and C/A (k$_{\rm T}$) (dashed). Note that $\wt{S}^{a}_{21} = \mc S^{(a)}_{21}/\CF\CA$, $\wt{S}^{\CA}_{21} = \mc S_{21}^{\CA}/\CF\CA$ and $\wt{S}^{\nf}_{21} = \mc S_{21}^{\nf}/\CF\TF\nf$.}
\label{fig.NGLs_DL_SL}
\end{figure}

On the same footing, the other N$^3$LL coefficients $\mc S_{21}$ can be evaluated numerically in the C/A algorithm. In \eq{S2_CA_clus}, the $\mc S_{21}$ coefficients in the \AKT algorithm are given in \eqs{S21_Hg}{S21_Hq} and the $\mc S_{21}^{\rm clus}$ coefficients are given by an integral analogous to \eqref{S21_gen_b} multiplied by the above clustering function. That is
\begin{equation}
 \mc S_{21}^{\rm clus} = \half\sbr{\int_{-\frac{\deta}{2}}^{\frac{\deta}{2}} \d\eta_1 \int_{\frac{\deta}{2}}^{+\infty} \d\eta_2 \int_0^{2\pi}\frac{\d\phi}{2\pi}  + 1 \lra 2}\, \mc A(\eta_1,\eta_2,\phi)\, \Theta\cbr{\cos\phi - e^{-\eta_1}\sin\eta_2 },
\label{S21_gen_b_CA}
\end{equation}
where the amplitude $\mc A$ is give in \eqss{Amp_Isq_term}{Amp_SJ_term}{Amp_w_term}. Carrying out the integration and adding up the various pieces one arrives at the results plotted in \fig{fig.NGLs_DL_SL}.

The fixed--order N$^3$LL logarithmic structure  of the $\te$ distribution should by now be clear for both jet algorithms. In order to assess the phenomenological impact of NGLs and clustering requirement on the final cross-section, it is necessary to perform an all--orders treatment, which we do below. Recall that for the resummed distribution, LL, NLL and so on refer to logarithms in the exponent.

\section{Resummation}
\label{sec.resummation_in_QCD} 

Resummation, which is essentially the organisation of large logs arising from
soft and/or collinear radiation to all--orders, is based on the factorisation
property of the pQCD matrix--element squared for multiple gluon radiation. This
is only true for independent primary emissions though. Including secondary
correlated emissions, the picture dramatically changes and the resummation can
only be performed in some limits, e.g., large-$\Nc$ limit \cite{Banfi:2005gj}. In the standard method \cite{Catani:1992ua, Collins:1984kg, Bonciani:2003nt} (see \ssec{ssec:Jets:Resummation}), resummation is carried out in Mellin (Laplace) space instead of momentum space. Only at the end does one transform the result back to the momentum space through (inverse Mellin transform),
\begin{equation}\label{inv_Mell_tauo}
\Sigma_{P}(\tauo, \Eo) = \int \frac{\d\nu}{2\imath\pi\nu}\; e^{\nu\tauo}\; \int
\frac{\d\mu}{2\imath\pi\mu}\; e^{\mu\Eo}\;\widetilde{\Sigma}_{P}(\nu^{-1},
\mu^{-1}),
\end{equation}
where $P$ stands for primary emission. With regard to non--global observables,
the important point to notice is that the resummation of NGLs is included as a
factor multiplying the single--gluon Sudakov form factor $\Sigma_{P}$ (as described in \ssec{ssec:Jets:NGLs});
\begin{equation}\label{resum_tot}
\Sigma\left(\tauo, \Eo\right) = \Sigma_{P}(\tauo,\Eo)\; \mc S\left( t\right),
\end{equation}
In this section, we first consider resummation of $\tauo$ distribution in events
where the final state jets are defined in the \AKT algorithm and, second, discuss the potential changes to the resummed result when the jets are defined in the C/A algorithm instead.

\subsection{The \AKT algorithm}
\label{subsec.resummation_QCD}

As stated in the introduction and proved in \sec{sec.fixed_order_1}, the
$\tauo$ observable is simply the sum of the invariant masses of the two
highest--energy (or highest--$p_T$ for hadron colliders) jets. Therefore the
$\tauo$ resummed Sudakov form factor is just double that computed
in \ssec{ssec:EEJS1:IndepResum} for $a=0$. That is, up to NLL level we have
\begin{equation}\label{resum_tot_akt}
\Sigma_{P}(\tauo,\Eo) = \frac{\exp\left[-2\left(\mc R_{\tauo}(\tauo) + \gamma_{E}
\mc R'_{\tauo}(\tauo) \right)\right]}{\Gamma\left(1+2\,\mc R'_{\tauo}(\tauo)\right)}\; \exp\left[ - \mc R_{\Eo}(\Eo)\right] .
\end{equation}
The full derivation of \eqref{resum_tot_akt} as well as the resultant
expressions of the various radiators are presented in the small jet--radius
limit in Chapter \ref{ch:EEJetShapes1}. To restore the full $\Rs$ dependence we make the replacement $R^{2}/\rho \to \Rs/(\tauo (1-\Rs))$ such that when expanded
\eq{resum_tot_akt} reproduces at $\Or(\as)$ the LO distribution \eqref{R1_full-b}.

To account for the NGLs at all--orders in $\as$, it is necessary to consider an
arbitrary ensemble of energy--ordered, soft wide--angle gluons that coherently
radiate a softest gluon into the vetoed region of phase
space \cite{Dasgupta:2001sh}\footnote{In our case the vetoed region is the jet region. Due to symmetry, we can choose one jet region and multiply the final answer by a factor of two.}. The analytical resummation of NGLs is then plagued with mathematical problems coming from the geometric and colour structure of the gluon ensemble. Two methods have been developed to address this issue: A numerical Monte Carlo evaluation \cite{Dasgupta:2001sh,Dasgupta:2002bw} and a non--linear evolution equation that resums single logs (SL) at all--orders \cite{Banfi:2002hw}. Both methods are only valid in the large-$\Nc$ limit. In the latter limit and for small values of the jet--radius $\Rs$, we argued in the previous chapter that the form of $\mc S(t)$ should be identical to that found in the hemisphere jet mass case \cite{Dasgupta:2001sh}. Since in the present chapter we are not confined to the small $\Rs$ limit, we need to modify and re--run the Monte Carlo algorithm of Ref. \cite{Dasgupta:2001sh}, for medium and large 
values of the jet--radius should we seek to resum the $\tauo$ NGLs distribution. We do this in \ssec{ssec.All-Orders}, where we consider the all--orders distribution in the \AKT and discuss modifications due to clustering.

However when comparing the analytical results of \sec{sec.fixed_order_1} and \sec{sec.fixed_order_2} with fixed--order NLO program \event\!, it suffices to simply exponentiate the fixed--order terms $\mc S_{22}$, \eq{S2_akt_tot}, and $\mc S_{21}$, Eqs.~\eqref{S21a_akt} $+$ \eqref{S21_Hg} $+$ \eqref{S21_Hq},
\begin{equation}\label{resum_NGLs_akt}
\mc S(t) = \exp\left(\mc S_{22}\; t^{2} + \mc S_{21}\,\asb\,t\right).
\end{equation}

The distribution \eqref{resum_tot} is of the generic form given in \eq{eq:Intro:exponentiation}. Explicitly, it reads
\begin{equation}\label{resum-form_QCD-b}
\Sigma(\tauo, \Eo) = \left(1+\sum_{k=1}^{\infty} C_{k}
\left(\frac{\as}{2\pi}\right)^{k} \right)
\exp\left[\sum_{n=1}^{\infty}\sum_{m=0}^{n+1} G_{nm}
\left(\frac{\as}{2\pi}\right)^{n} \widetilde{L}^{m} \right] +
D_{\mathrm{fin}}(\tauo),
\end{equation}
where $C_{k}$ is the $k^{th}$ loop--constant, $\widetilde{L} = \ln(1/\tauo)$ and $D_{\mathrm{fin}} \equiv D$ (where $D$ is given in \eqref{eq:Intro:exponentiation}), which vanishes in the limit $\tauo \rightarrow 0$. In order to determine the coefficients $G_{nm}$ at NLO and up to N$^3$LL, we need to expand the radiators, as well as the $\Gamma$ function, in \eq{resum_tot_akt} up to second order in the fixed coupling $\as$. The results are presented in \app{app.coeff_in_expansion}.
The only missing piece in the N$^3$LL coefficient, $G_{21}$ in \eq{G_nm-QCD}, are the coefficients of $\widetilde{L}$ which are independent of $\ln(\Rs/(1-\Rs))$ for all colour channels. These can be borrowed from thrust distribution \cite{Monni:2011gb, Kelley:2011ng, Hornig:2011iu}\footnote{All $\Rs$--dependent terms are captured as can be seen from comparison to the SCET result \eqref{app.tw_in_SCET}, which only contains the primary emission piece and is valid to N$^3$LL.}. It is worthwhile mentioning that the two--loop constant $C_{2}$ has also been computed for the latter variable as well as the thrust \cite{Monni:2011gb, Kelley:2011ng}. Moreover, to make contact with SCET calculations, we provide in \app{app.tw_in_SCET} the full formula of the Sudakov form factor for the $\tauo$ primary distribution including determination of $G_{nm}$ coefficients in SCET up to N$^3$LL. 

Next we comment on the form of resummation when final state jets are defined in
the C/A algorithm.

\subsection{The C/A algorithm}
\label{subsec.resum_CA}

With regard to primary emission  piece, resumming logs induced by clustering is
a cumbersome but doable task. It has been performed, for example,
in \cite{Delenda:2006nf} for interjet energy flow distribution where final state jets are defined in the inclusive \KT algorithm. The final result of the
resummed radiator was written as an expansion in the jet--radius and the first
three terms were determined. For secondary emissions, the resummation of NGLs has only been possible numerically and in the large-$\Nc$ limit. It has again been carried out for the above mentioned  energy flow distribution
in \cite{Appleby:2002ke}. We expect that analogous, to the interjet energy flow, analytical treatment and numerical evaluation can be achieved for the
resummation of CLs and NGLs, respectively, for the $\tauo$ variable. While we postpone the analytical resummation of CLs to the next chapter (Chapter \ref{ch:EEJetShapes3}), we provide in this section the all--orders numerical results for both logarithms.

Due to the fact that logarithmic contributions induced by clustering arise mainly from soft wide--angle gluons, we expect them, i.e., clustering--induced logs, to factorise from the primary Sudakov form factor at all--orders. Therefore, the resummed distribution, whereby clustering is imposed on the final state, may be written in the following factorised form
\begin{equation}\label{resum_tot_CA}
\Sigma(\tauo,\Eo) = \Sigma_{P}(\tauo, \Eo)\,\mc S^{\ca}\left(t\right)\, \mc C^{P}\left(t\right),
\end{equation}
where we recall that the evolution parameter of $\mc C^P$, \eq{t_params_p}, is identical to that of the non-global factor $\mc S$, \eq{t_param_a}. We investigate the numerical impact of the above resummed distribution in the next subsection, \ssec{ssec.All-Orders}. However, for the sake of comparison to \event\!, it is sufficient to simply consider the exponentiation of the fixed--order terms $\mc S^{\ca}_{22}, \mc S_{21}^{\ca}, C^P_{22}$ and $C_{21}^{P}$, just as we did with the \AKT algorithm case. In other words, we write $\mc S^{\ca}$ in the form \eqref{resum_NGLs_akt} with $\mc S_{22}, \mc S_{21}$ replaced by $\mc S_{22}^{\ca}, \mc S_{21}^{\ca}$ and, in analogy with the NGLs factor \eqref{resum_NGLs_akt}, we write for the $\te$ CLs factor
\begin{equation}\label{resum_C_P_CA}
\mc C^{P}(t) = \exp\left(C_{22}^{P}\,t^{2} + C^P_{21} \asb\, t\right).
\end{equation}
The resultant $G_{nm}$ coefficients in the exponent of \eq{resum_tot_CA} are presented in \app{app.coeff_in_expansion} (\eqs{G_nm-QCD}{G_nm-QCD_CA_alg}).

\subsection{All-orders numerical studies}
\label{ssec.All-Orders}

As stated above, non-global logs have hitherto only been resummed numerically using the program of \cite{Dasgupta:2001sh, Delenda:2006nf} in the large-$\Nc$ limit, both in \AKT and other jet algorithms. Although there have been analytical attempts to resum clustering logs at all--orders, which we shall discuss in detail in the next chapter, in this subsection we confine ourselves to only provide numerical estimates using the said program. To this end, we parametrise the non-global factor $\mc S(t)$ in any given jet algorithm by the formula \cite{Dasgupta:2001sh}
\begin{equation}
 \mc S(t) = \exp\sbr{-\CF\CA\,\mb I_{22}(\Rs) \cbr{\frac{1 + (at)^2}{1 + (bt)^c}} t^2},
\label{St_param_form}
\end{equation}
where $\mb I_{22}(\Rs) = \mc S_{22} /\CF\CA$.
\begin{table}[!t]
 \centering
\begin{tabular}{|c|c||c|c|c|}
\cline{2-5}
 \multicolumn{1}{c|}{}      &  \multicolumn{4}{|c|}{\AKT}
\\ \cline{2-5}
 \multicolumn{1}{c|}{$\Rs$} & $\mb I_{22}$& $a$  & $b$   & $c$ \\
\cline{1-5}
 $0.0025$ & $6.58$          & $0.72\,\CA$ & $0.71\,\CA$  & $1.78$ \\
\hline
 $0.04$   & $6.57$          & $0.20\,\CA$ & $0.36\,\CA$  & $1.85$ \\
\hline
 $0.12$   & $6.51$          & $0.61\,\CA$ & $0.83\,\CA$  & $1.85$ \\
\hline
 $0.3$    & $5.81$          & $0.83\,\CA$ & $1.27\,\CA$  & $1.85$ \\
\hline
\end{tabular}
\begin{tabular}{|c||c|c|c|}
\cline{1-4}
    \multicolumn{4}{|c|}{\KT}
\\ \cline{1-4}
 $\mb I_{22}$ & $a$ & $b$ & $c$ \\
\cline{1-4}
 $2.92$ & $1.80\,\CA$ & $3.04\,\CA$ & $1.00$ \\
\hline
 $2.81$ & $1.97\,\CA$ & $3.36\,\CA$ & $1.00$ \\
\hline
 $2.53$ & $1.69\,\CA$ & $2.74\,\CA$ & $1.00$ \\
\hline
 $1.25$ & $1.12\,\CA$ & $2.20\,\CA$ & $0.29$ \\
\hline
\end{tabular}
\caption{Fitting values of the parameters $a,b$ and $c$ for $\mc S(t)$ factor in both \AKT and \KT. The coefficient $\mb I_{22} = \mc S_{22}/\CF\CA$.}
\label{tab:St_fitting_values}
\end{table}
\begin{table}[!t]
 \centering
\begin{tabular}{|c|c||c|c|c|}
\cline{2-5}
\multicolumn{1}{c|}{}   &  \multicolumn{4}{|c|}{\KT}
\\ \cline{2-5}
 \multicolumn{1}{c|}{$\Rs$}    & $C^P_{22}/\CFsq$ & $a$    & $b$     & $c$ \\
\cline{1-5}
 $0.0025$ & $0.73$  & $0.00$      & $0.09\,\CA$  & $1.00$ \\
\hline
 $0.04$   & $0.74$  & $1.07\,\CA$ & $0.75\,\CA$ & $1.43$ \\
\hline
 $0.12$   & $0.76$  & $1.04\,\CA$ & $0.78\,\CA$ & $1.38$ \\
\hline
 $0.3$    & $0.89$  & $0.96\,\CA$ & $0.80\,\CA$ & $1.41$ \\
\hline
\end{tabular}
\caption{Fitting values of the parameters $a,b$ and $c$ for $\mc C^P$ factor.}
\label{tab:Ct_fitting_values}
\end{table}
In Table \ref{tab:St_fitting_values} we provide the corresponding values of $\mb I_{22}$ along with $a, b$ and $c$ for several jet radii in both \AKT and k$_{\rm T}$ (instead of C/A since it is currently the only available algorithm in the program besides the \AKT as stated at the outset) algorithms. Due to the similarity of CLs and NGLs; being only present for non-global observables and first appear at $\Or(\as^2)$, we opt to parametrise the all--orders factor $\mc C^P(t)$, \eq{resum_tot_CA}, by an analogous formula to \eq{St_param_form},
\begin{equation}
 \mc C^P(t) =  \exp\sbr{\CFsq\,C^P_{22} (\Rs) \cbr{\frac{1 + (at)^2}{1 + (bt)^c}} t^{2}}.
\label{Ct_param_form}
\end{equation}
The corresponding values of the fitting parameters are reported in Table \ref{tab:Ct_fitting_values} (notice that the values of $C_{22}^P/\CFsq$ have already been presented in Table \ref{tab:CLs_Coeffs_FullR}).

\begin{figure}[t]
\centering
 \epsfig{file=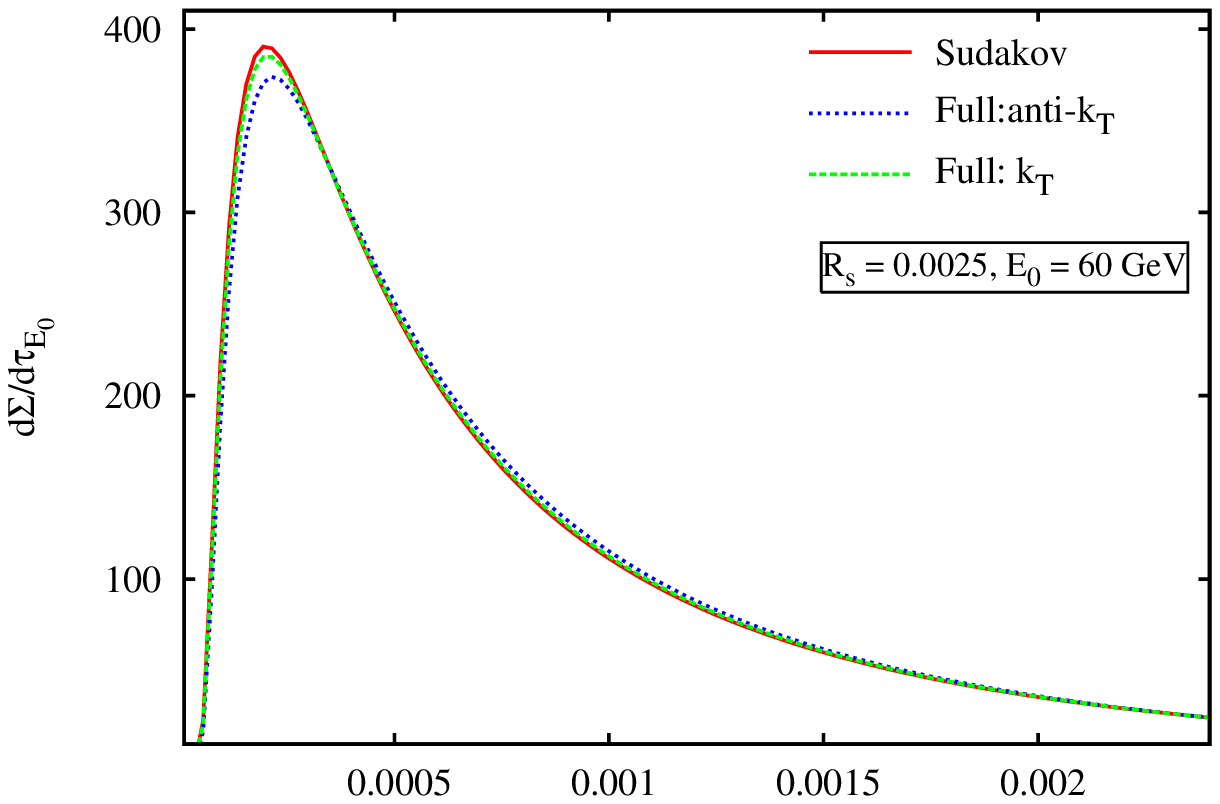, width= 0.49 \textwidth}
 \epsfig{file=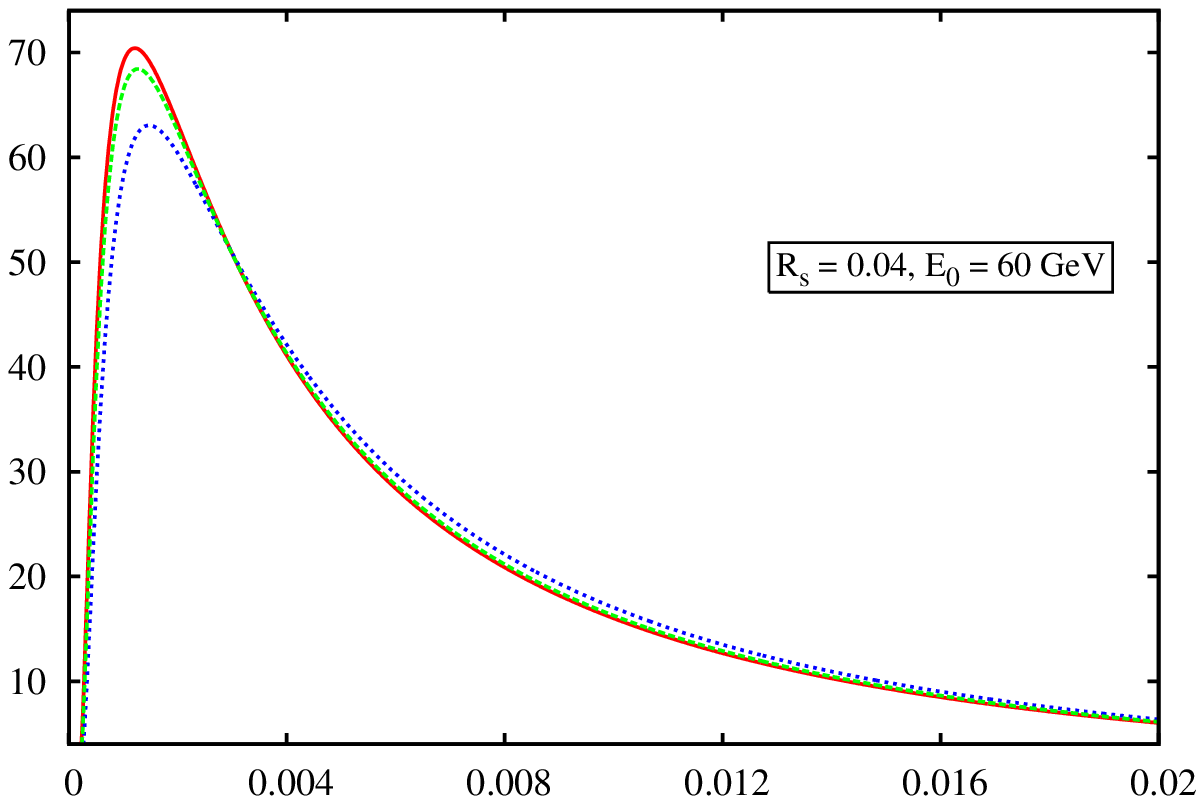, width= 0.49 \textwidth}
 \epsfig{file=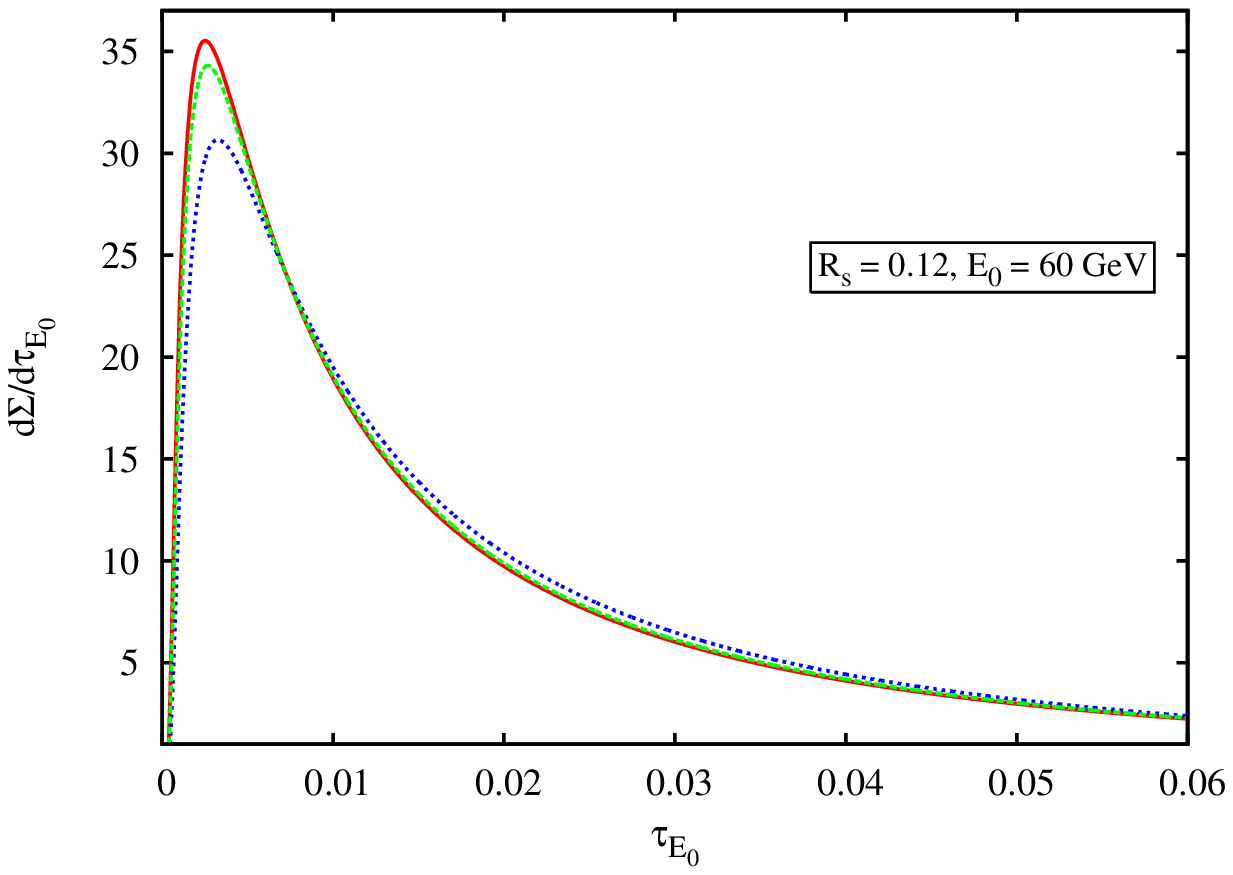, width= 0.49 \textwidth}
 \epsfig{file=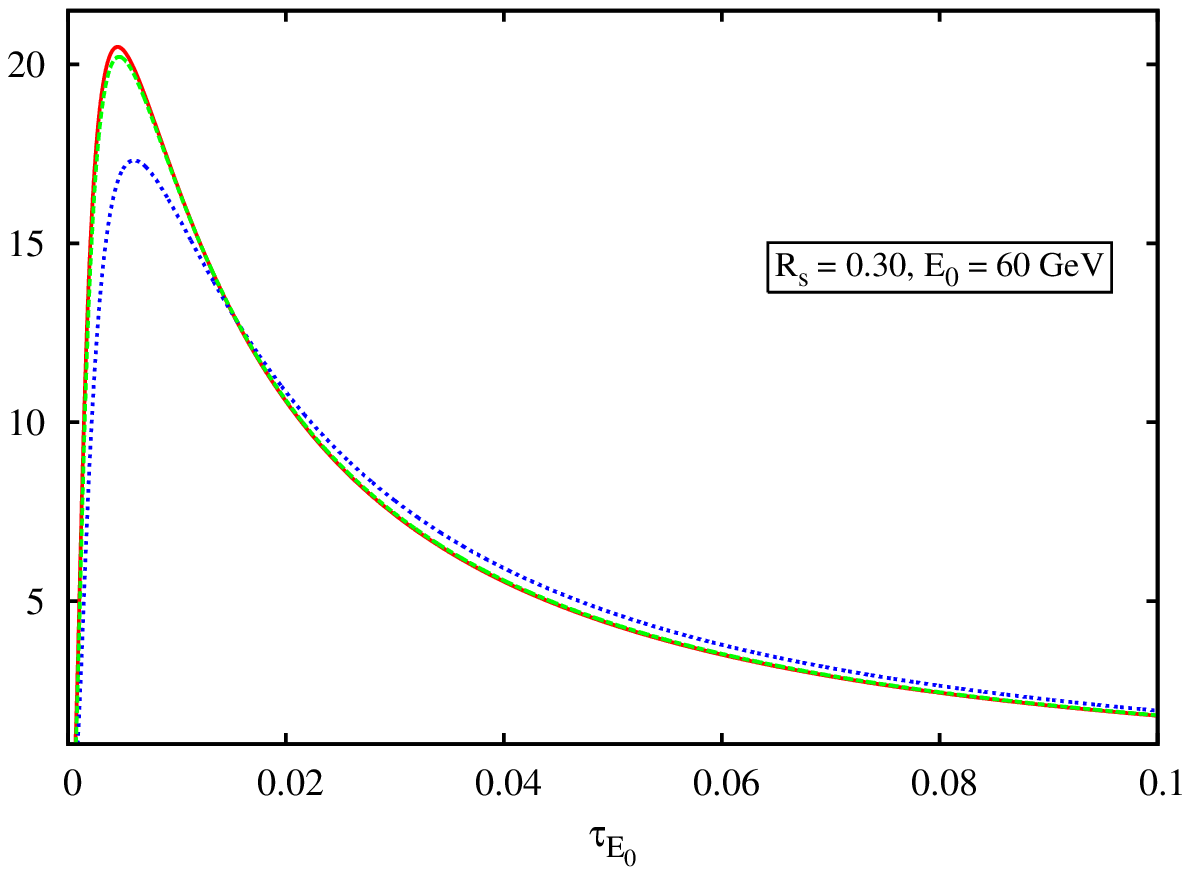, width= 0.49 \textwidth}	
\caption{Comparison of analytical resummed differential distribution
$\d\Sigma/\d\tauo$ where the Sudakov, full \AKT and full \KT distributions are given, respectively, in \eqss{resum_tot_akt}{resum_tot}{resum_tot_CA}. The plots are shown for a jet veto $\Eo = 60 \GeV$ and a hard scale $Q=500 \GeV$.}
\label{fig.resummed_CA_akt}
\end{figure}

In \fig{fig.resummed_CA_akt} we plot the resummed differential
distributions, $\d \Sigma(\tauo,\Eo)/\d\tauo = 1/\cSup{\s}{0}\d\sigma/\d\tauo$, for: Sudakov form factor \eqref{resum_tot_akt}, full form factor including NGLs factor $\mc S(t)$ for the \AKT algorithm \eqref{resum_tot} and full form factor including NGLs and CLs in the \KT algorithm \eqref{resum_tot_CA} at four different values of $\Rs$. We also plot in \fig{fig.St_Eo-R} the dependence of the non-global factor $\mc S(t)$ on the jet veto $E_0$ for the same values\footnote{These particular values of $\Rs$ will be used in our comparisons to the fixed--order Monte Carlo {\sc event}{\scriptsize 2}.} of $\Rs$ in both \AKT and \KT algorithms.
There are few points to note:
\begin{itemize}
 \item The effect of NGLs is a suppression of the (peak of) total cross-section relative to the primary result. This suppression, of the Sudakov form factor, increases as $\Rs$ increases. For example, for $\Eo=60\GeV$ the Sudakov peak is reduced due to NGLs by about $16\%$ at $\Rs=0.3\;(R=1.1)$, $10\%$ at $\Rs = 0.04\;(R=0.4)$ and only $4\%$ at $\Rs = 0.0025\;(R=0.1)$. From \fig{fig.NG_coff_CA}, one would expect that because $\mb I_{22}(\Rs)$ decreases with $\Rs$, $\mc S(t)$ becomes larger and hence the suppression would decrease with $\Rs$ (i.e., contrary to what we found above). The reason is due to the argument of the non-global logarithm $L_{\mr ng}$, and hence the logarithm itself, which grows larger with $\Rs$ and thus overcomes the reduction in $\mb I_{22}(\Rs)$. For instance, for the given values of $Q$ and $E_0$ we have $L_{\mr ng}(\Rs=0.04) = 2.8$ and $L_{\mr ng}(\Rs=0.3) = 4.15$.

 Note that for practical values of the jet radius ($\Rs \geq 0.04$), the uncertainty in the total cross-section due to neglecting NGLs is larger than the inherent uncertainty in all--orders QCD calculations due to large-$\Nc$ limit, and thus NGLs need to be fully accounted for when seeking precision calculations.
\begin{figure}[!t]
\centering
 \epsfig{file=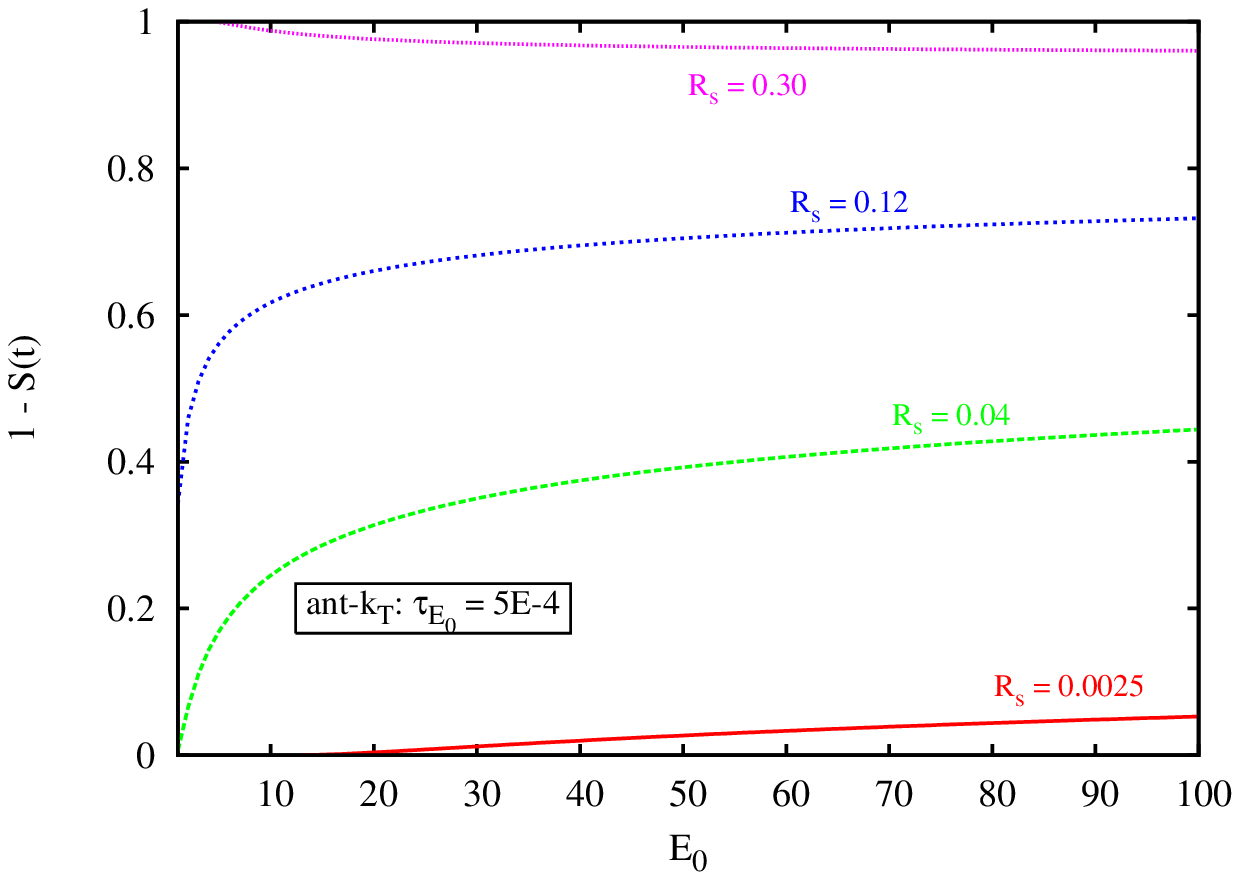, width= 0.49\textwidth}
 \epsfig{file=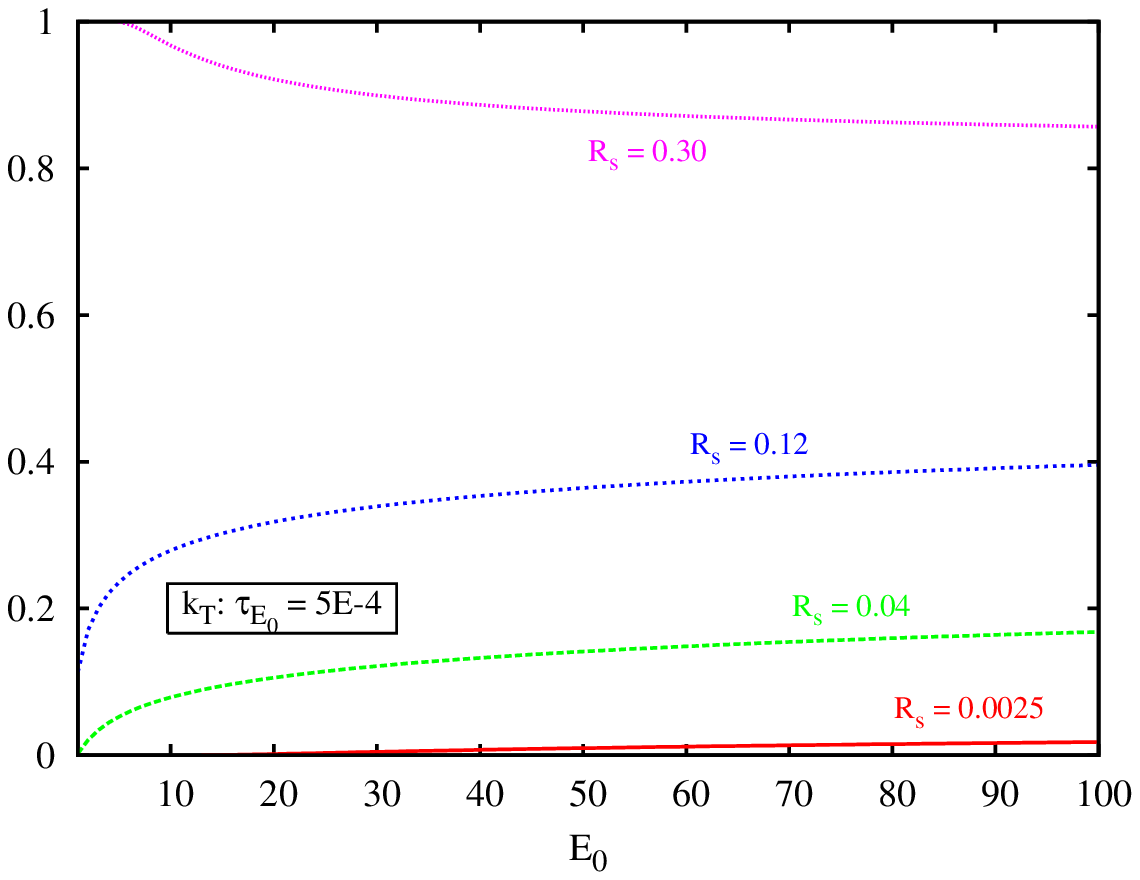,  width= 0.49\textwidth}	
\caption{Variation of $1 - S(t)$ as $E_0$ is varied for several jet radii. The smallest value of $E_0$ corresponds to the $E_0^{\min} = Q\te/(2\Rs)$, derived from \eqref{t_param_a} where $e^{\deta} \sim 1/\Rs$.}
\label{fig.St_Eo-R}
\end{figure}

 \item The effect of clustering is reducing the phenomenological significance
of NGLs. For instance, the Sudakov peak is reduced, for the same jet veto, by about $2.5\%$ for $\Rs = 0.04$ and less than $1\%$ for $\Rs = 0.3$. Compared to the unclustered (anti-k$_{\rm T}$) case, we see that the effect of NGLs has been reduced by more than $70\%$ for $\Rs= 0.04$ and more than $90\%$ for $\Rs = 0.3$. Therefore increasing the jet radius leads to a) larger NGLs impact in the unclustered (anti-k$_{\rm T}$) case and b) smaller impact in the clustering case. Note that in the case b) the whole NGLs effect on the $\te$ distribution is much less than $10\%$ for jet radii of $\Rs = 0.3\; (R=1.1)$ or above\footnote{Jets with large size are however more amenable to contaminations from underlying event and pile-up soft radiation.}.

 \item The impact of NGLs grows up with the jet veto up to jet radius $\Rs=0.3$ where the inverse (that is NGLs impact grows smaller) occurs, in both unclustered and clustered cases as shown in \fig{fig.St_Eo-R}. For instance at $E_0=100 \GeV$ and $\Rs=0.04$ the suppression in the Sudakov peak is about $13\%$ and $3\%$ for the unclustered and clustered case respectively. That is a further suppression of about $3\%$ and less than $1\%$ compared to $\Eo=60 \GeV$ for the same jet radius. 

 Notice that the behaviour mentioned above for $\Rs \geq 0.3$ is merely due to the choice of $\te = 5\times10^{-4}$. For larger values of $\te$ such a behaviour would disappear.
\end{itemize}

We conclude that clustering final states with jet algorithms other than the \AKT algorithm, and preferably the \KT or C/A algorithms, and picking up a moderate to low jet veto may yield reliable estimates (within the $10\%$ level) of the total cross-section or shape distribution even if non-global logarithms are totally ignored. This observation is valid for any jet radius $\Rs < 1/2$ (equivalent to $R$ up to and above unity).

In the next section, we compare our fixed--order analytical calculations to \event. In particular, we focus on establishing the presence of NGLs and CLs in the jet-thrust distribution at NLO.

\section{Comparison to {\sc event}{\normalsize 2}}
\label{sec.numerical_results}

The $\tauo$ numerical distribution has been computed using the fixed--order NLO
QCD program \event. The program implements the Catani--Seymour
subtraction formalism for NLO corrections to two-- and three--jet events
observables in $\EE$ annihilation. Final state partons have been clustered into
jets using the \fastjet library \cite{Cacciari:2011ma}. The latter provides an implementation of the longitudinally invariant k$_{\rm T}$, Cambridge--Aachen (CA) and \AKT jet finders along with many others. Cone algorithms such as SISCone \cite{Salam:2007xv} are also implemented as plugins for the package. It should be noted that the $\EE$ version of the aforementioned algorithms employs the following clustering condition for a pair of partons $(ij)$ 
\begin{equation}\label{clust_cond_FJ}
    1-\cos\theta_{ij} < 1-\cos(\widetilde{R}),
\end{equation}
where $\widetilde{R}$ is the jet--radius parameter used in \fastjet\footnote{In \fastjet's manual $\widetilde{R}$ is allowed to go up to $\pi$. Since we are interested in two-jet events the jet size cannot be wider than a hemisphere. Thus we restrict $\tilde{R}$ to be less than $\pi/2$.}. Compared to \eqs{clust_cond}{R_Rs_rel}, $\widetilde{R} = \cos^{-1}(1-2\Rs)$. The exact numerical distributions $(1/\cSup{\s}{0})(\d\sigma_{e}/\d L)$, with $L = -\widetilde{L} = \ln(\tauo)$, for the three colour channels, $\CFsq, \CF\CA$ and $\CF\TF\nf$, have been obtained with $10^{11}$ events in the bin range $-14 < L < 0$. We have used four values for the jet--radius: $\Rs = 0.30, \Rs=0.12, \Rs=0.04$ and $\Rs = 0.0025$, with an energy veto $\Eo = 0.01\,Q$. Standard deviations on individual bins range from $10^{-4}\% $ to $ 10^{-2}\%$.
\begin{figure}[!t]
 \centering
 \epsfig{file=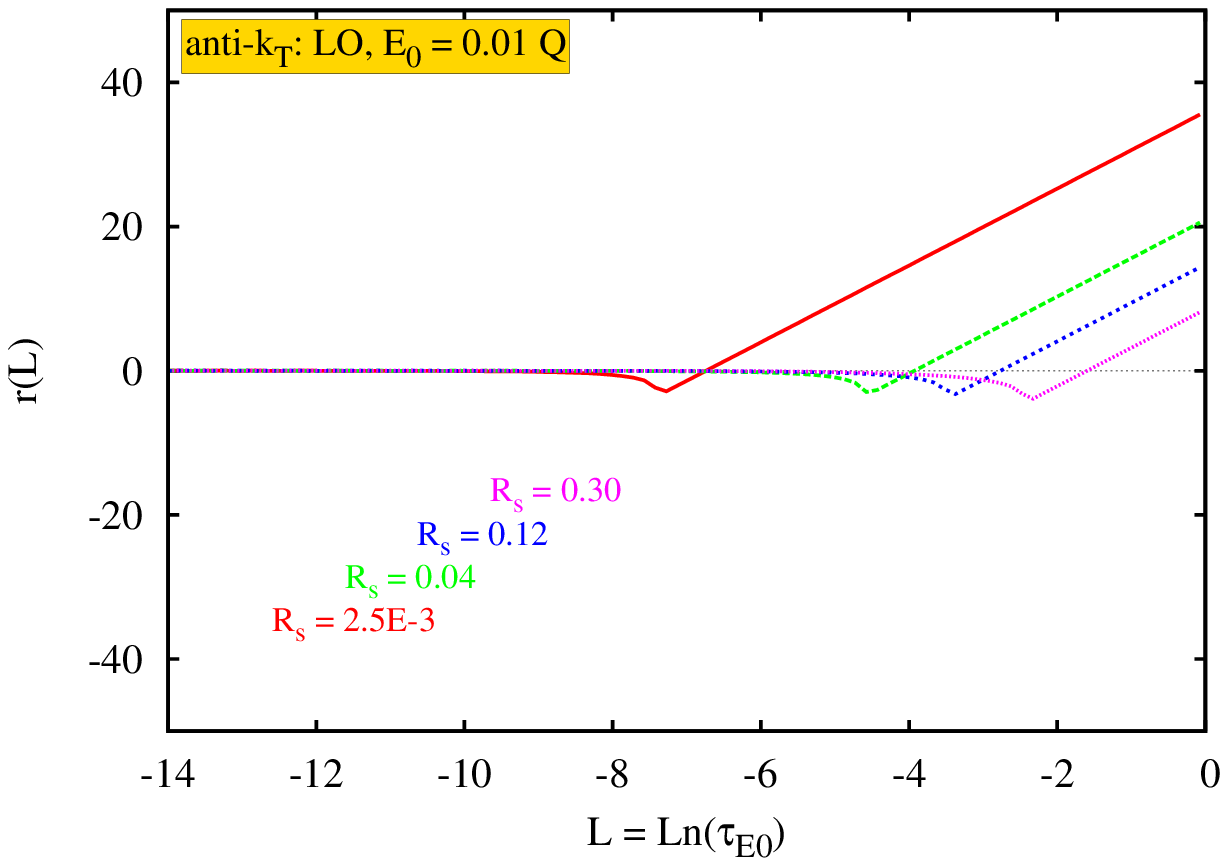, width= 0.49 \textwidth}
 \epsfig{file=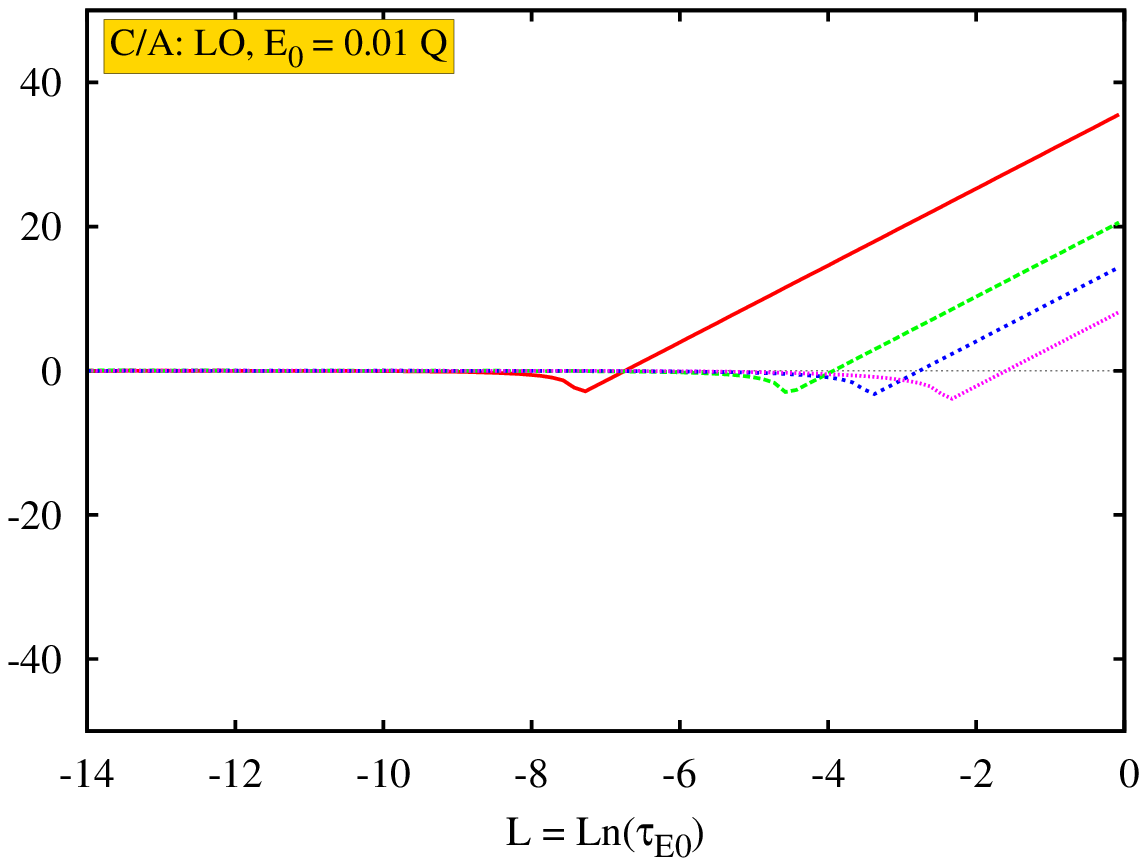, width= 0.49 \textwidth}
 \caption{The difference between \event and $\tauo$ LO distribution for various jet radii in both \AKT (left) and C/A (right) algorithms.}
 \label{fig.LO}
\end{figure}

\begin{figure}[!t]
 \centering
 \epsfig{file=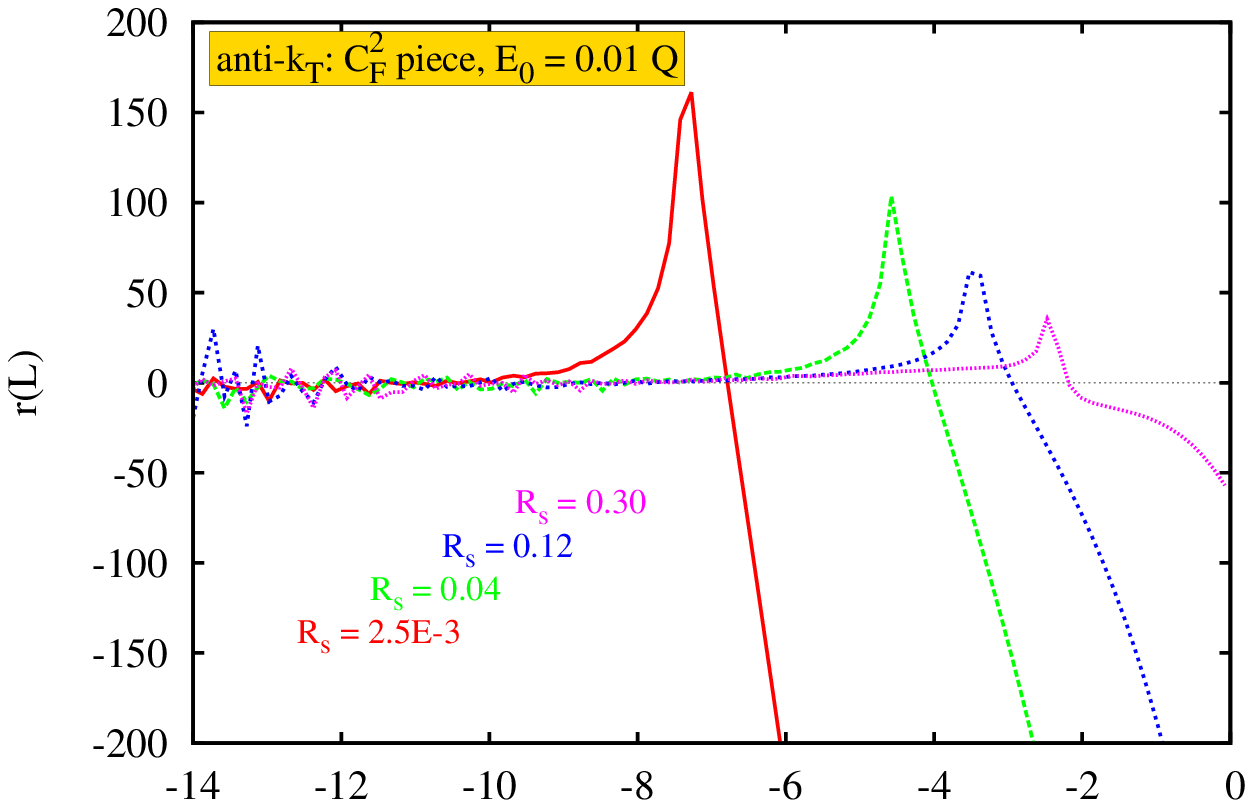, width= 0.49 \textwidth}
 \epsfig{file=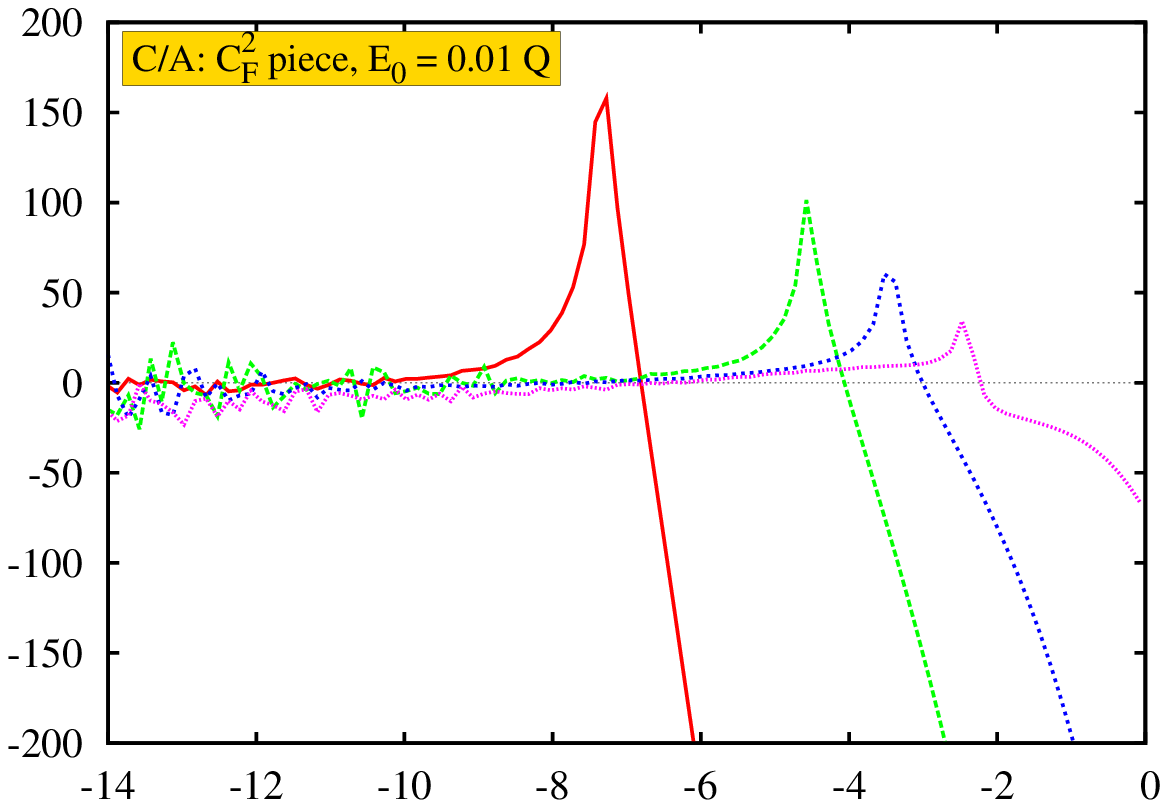,  width= 0.49 \textwidth}
 \epsfig{file=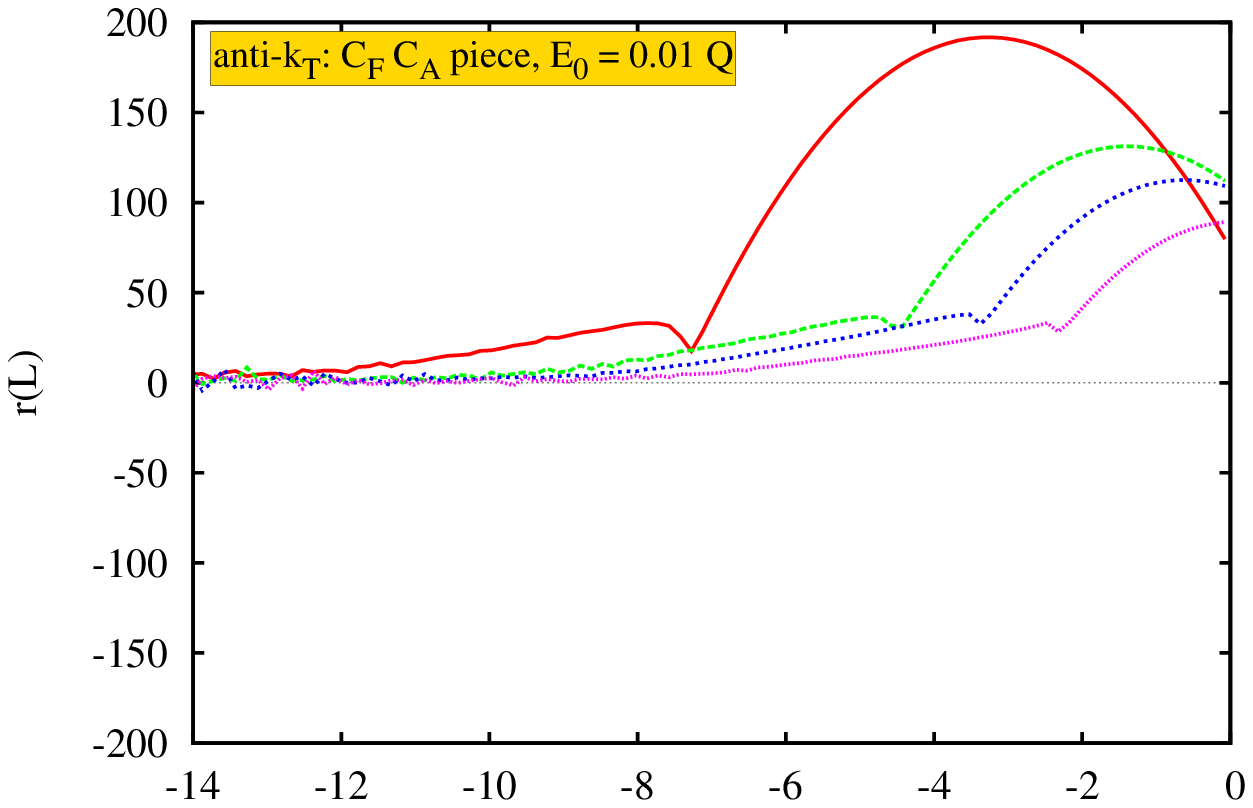, width= 0.49 \textwidth}
 \epsfig{file=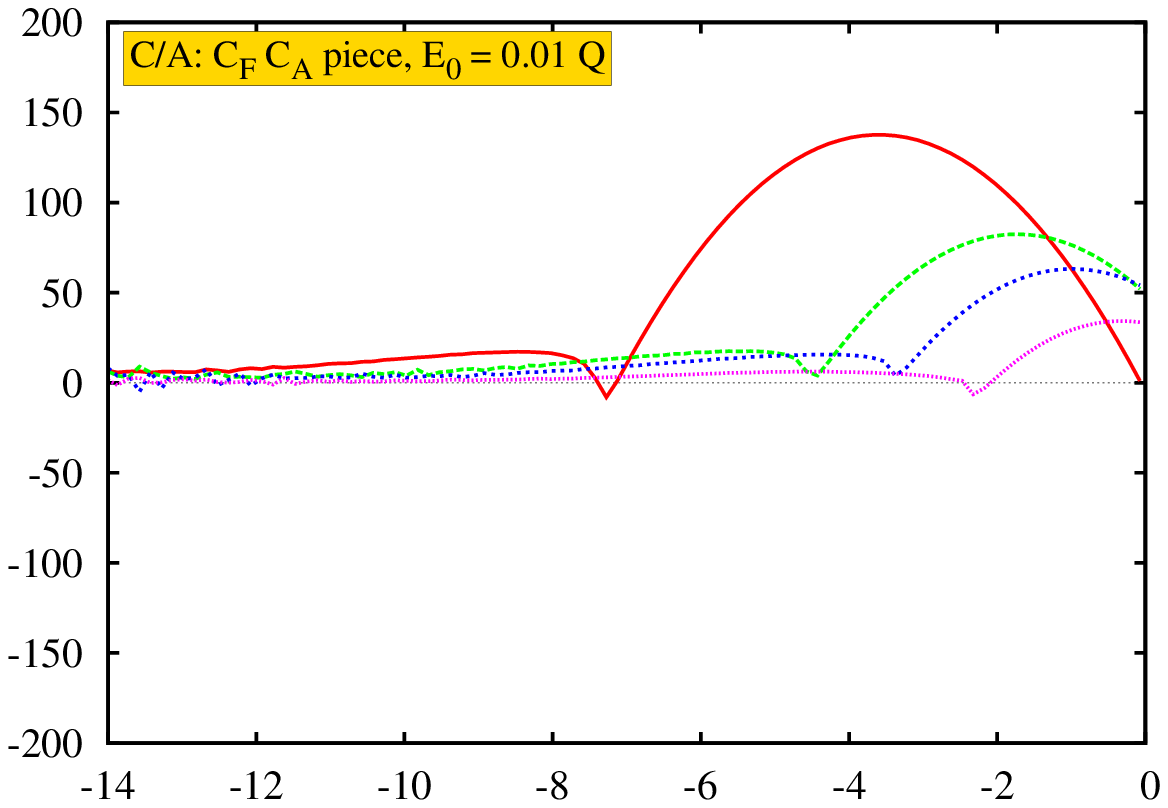,  width= 0.49 \textwidth}
 \epsfig{file=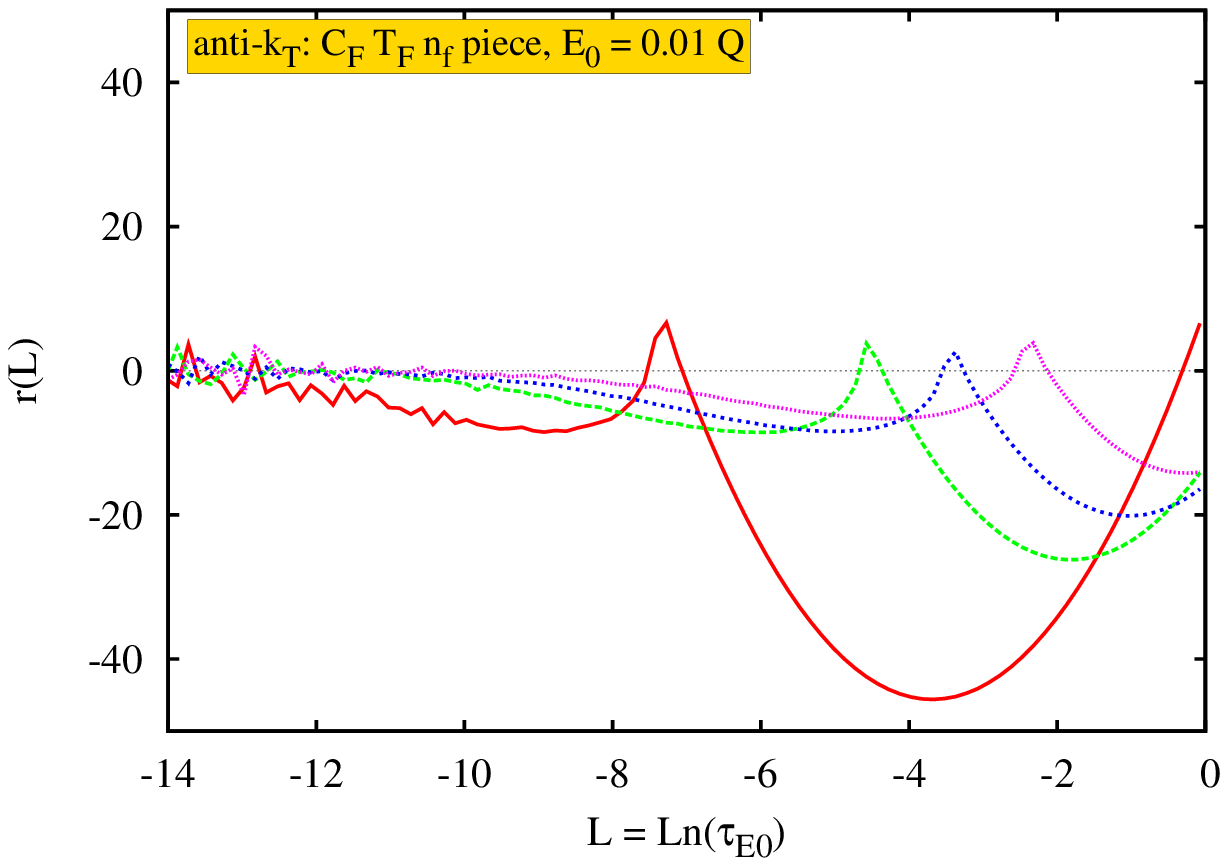, width= 0.49 \textwidth}
 \epsfig{file=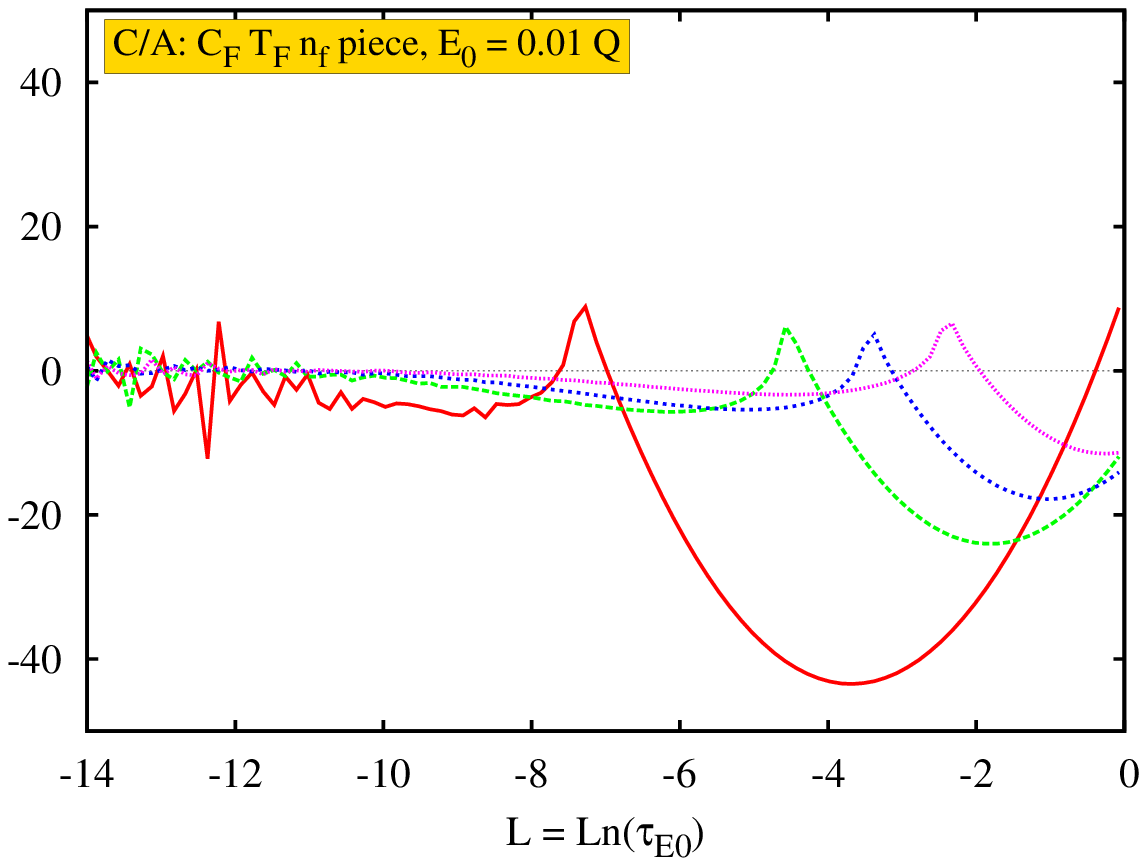,  width= 0.49 \textwidth}
 \caption{The difference between \event and $\tauo$ NLO distribution for various jet radii in both \AKT (left) and C/A (right) algorithms.}
 \label{fig.NLO}
\end{figure}

We plot the difference between the numerical and analytical distributions at
both LO and NLO,
\begin{equation}\label{remainder}
 r(L) = \frac{\d\sigma_{e}}{\cSup{\s}{0}\d \,L} - \frac{\d\sigma_{r,2}}{\cSup{\s}{0}\d L},
\end{equation}
where $1/\cSup{\s}{0}\d\sigma_{r,2}/\d L$ is given in \eq{resum_expanded-sig-diff}. Recall that at small values of the jet shape, $\tauo$, the finite remainder
function $D_{\mathrm{fin}}(\tauo)$ in \eq{resum-form_QCD-b} is vanishingly small and will thus be ignored. If our analytical calculations of the NGLs coefficients for both \AKT and C/A algorithms are correct then $r(L)$ should asymptotically vanish for all colour channels. We plot our results in Figs. \ref{fig.LO}, \ref{fig.NLO} and, for each radius $\Rs$ separately, in Figs.~\ref{fig.NLO_CF2},~\ref{fig.NLO_CFCA} and~\ref{fig.NLO_CFTF}. Overall, the cancellation in \eq{remainder} is, within statistical errors, reasonably good for all three colour channels and at most values of the jet radius considered.
Few comments are in order:
\begin{itemize}
 \item At LO, the distribution is independent of the jet definition. The kinks in \fig{fig.LO} correspond to the maximum values of $\te$ ($\te^{\max}(\Rs)$ given in \eq{eq:EEJS2:tau_E0_max}) that are kinematically allowed.
 Notice that the exact LO distribution vanishes for $\te > \te^{\max}$ and the straight lines for $\te > \te^{\max}$ are simply the analytical results.
 
 \item At NLO, the asymptotic region, i.e., the region where large logs are expected to dominate over non--logarithmic contributions, corresponds to $L$ less than about $-8$ and seems to decrease further as $\Rs$ becomes smaller. It also varies with the colour channels.

 \item Similar to the LO case, the exact NLO distribution vanishes for $\te > \te^{\max}$ and the curves in the latter region are merely the analytical results, $1/\cSup{\s}{0}\d\s_{r,2}/\d L$.

 \item For the $\CFsq$ channel (\fig{fig.NLO_CF2}) in the C/A algorithm, there seems to be no complete cancellation at $\Rs = 0.3$ (equivalent to $R = 1.1$). Hence contributions for jet radii ($R$) larger than unity are not controlled up to $\as^2\,L$\footnote{A more accurate conclusion may be drawn once the statistics are improved. We hope to achieve this in the near future.}.

 \item For the $\CF\CA$ channel (\fig{fig.NLO_CFCA}) in both algorithms, the asymptotic region starts at values of $L$ that are about or smaller than $-14$ for jet radii $\Rs < 0.04$. i.e., out of our reach. The shift of the asymptotic region towards smaller values of $L$ as $\Rs$ decreases is clear in the above mentioned figures (\fig{fig.NLO_CFCA}).

 \item The results for the $\CF\TF\nf$ channel (\fig{fig.NLO_CFTF}) in both algorithms are more pronounced (than for the other channels), again with the observation that the asymptotic region shifts towards smaller values of $L$ as $\Rs$ decreases.
\end{itemize}
In summary, we  have confirmed through explicit comparisons to exact numerical
distributions the existence of large NGLs and large CLs for the $\tauo$
distribution at N$^2$LL and beyond (in the expansion). We have also shown that clustering the final state partons with the C/A algorithm yielded a significant reduction in NGLs impact, both at N$^2$LL and beyond, albeit inducing large CLs, both at N$^2$LL and beyond, in the primary emission sector.

\section{Conclusion}
\label{sec.conclusion}

The jet mass with a jet veto, or simply the jet-thrust, is an example of a wider class of non--global observables. These have the characteristic of being sensitive to radiation into restricted regions of phase space, or sensitive to radiation into the whole phase space but differently in different regions. For such observables the universal Sudakov form factor fails to reproduce the full logarithmic structure even at NLL (in the exponent) accuracy. New contributions that are dependent on various variables such as the jet size and jet definition appear at this logarithmic level and beyond. In this chapter, we have elaborated on these very contributions for the aforementioned observable. 

Considering secondary emissions, we have computed the full analytical expressions of the first leading and next-to-leading terms, $\mc S_{22}$ and $\mc S_{21}$, in a series of missing large non-global logarithms. The coefficients depend, as anticipated, on the jet size and saturate at their maxima in the limit where the latter, i.e., jet size, vanishes. The saturation value of $\mc S_{22}$ was used in Chapter \ref{ch:EEJetShapes1} as an approximation to the full value in the small $\Rs$ limit. It turns out that the approximation is valid for quite a wide range of $\Rs$. The formulae for $\mc S_{22}$ and $\mc S_{21}$ have been checked against full exact numerical result obtained by the program \event. The difference between the analytical and numerical differential distributions was shown to asymptotically vanish signalling a complete cancellation of singular terms up to N$^3$LL (in the expansion) level. This has all been done for final states defined in the cone--like \AKT jet algorithm.

To illustrate the dependence of N$^2$LL (in the expansion) and beyond on the jet definition, we have investigated the effects of applying the C/A algorithm on $\EE$ final states.
The impact of soft partons clustering is two--fold. On one side, it reduces the
size of NGLs through shrinking the phase space region where the latter
dominantly come from. i.e., the region where the emitter and emitted soft partons are just in and just out of the jet. On the other side, it gives rise to new N$^2$LL, and beyond, logarithmic contributions, CLs, in the primary emission sector. We have computed the full $\Rs$ dependence of the corresponding coefficients, $C^P_{22}$ at N$^2$LL and $C^P_{21}$ at N$^3$LL, at second order, which were then verified against \event. With regard to jet definitions, the \KT and C/A algorithms behave identically at $\Or(\as^2)$ with respect to non-global and clustering logs.
 
Furthermore, we have provided a resummation of the jet-thrust up to NLL (in the exponent) accuracy with full jet radius dependence in both \AKT and \KT jet algorithms in $\EE$ annihilation events. Non-global as well as clustering logs logs in both algorithms  have been obtained with the aid of the numerical program of Ref. \cite{Dasgupta:2001sh} (with clustering added in \cite{Delenda:2006nf}), in the large-$\Nc$ limit. It has been observed that the size, and hence the impact of non-global logarithms on the Sudakov form factor for the jet-thrust is controlled by few factors, including: jet radius, jet veto and jet definition. Within our accuracy we have been able to conclude that using a jet definition other than the ant-k$_{\rm T}$, such as \KT or C/A, with reasonably low to medium jet vetos would diminish the effect of non-global logs for jet radii up to $1$, and above (such that the uncertainty from neglecting non-global logs is within the $10\%$ uncertainty level inherent from large-$\Nc$ approximation). 

Further, we note that the Monte Carlo program of \cite{Dasgupta:2001sh} only provides the leading non-global, and clustering, logs (which are next-to-leading logs in the jet-thrust). Since we have fully outlined the N$^3$LL structure of non-global, and clustering logs, at fixed--order, it would be interesting to extend the above program to include them at all--orders. i.e., resum non-global logs at NLL (in the exponent) accuracy (which is N$^2$LL accuracy for the jet-thrust). We investigate this more in future.

In the next chapter we present an attempt to analytically resum clustering logarithms that occur in the jet mass distribution at all--orders in both \KT and C/A algorithms.

%% file: ch6/chap6.tex

\chapter{On the resummation of clustering logarithms}
\label{ch:EEJetShapes3}

\section{Introduction}
\label{sec:EEJS3:Intro}

In the presence of a jet algorithm, other than \AKT~\cite{Cacciari:2008gp}, non-global observables suffer from large logs in the
Abelian part of the emission amplitude, as we have seen in the previous chapters. We referred to these as ``clustering logs'' (CLs). These logs 
were first computed analytically at fixed order and
numerically to all--orders in Ref.~\cite{Banfi:2005gj} and subsequently
\emph{partially} analytically resummed in Ref.~\cite{Delenda:2006nf}
for away-from--jets energy ($E_t$) flow. There, however, the resulting
exponent of the resummed distribution has been written as a
power-series in the radius parameter\footnote{It is actually a series in $t\,R$ where (at fixed--order) $t=\as/2\pi\,\ln(Q/Q_0)$, with $Q$ and $Q_0$ being the hard and veto scales, and is typically in the range $t\in[0,0.25]$.} of the jet algorithm $R$ starting from $R^3$. Thus for a typical jet radius ($R\sim 1$) it
was sufficient, for an accurate approximation, to compute the first
couple of terms ($\mathcal{O}(R^3) $ and $\mathcal{O}(R^5)$), as the
series rapidly converges. Excellent agreement was noticed when
compared to the output of the Monte Carlo program developed in
\cite{Dasgupta:2001sh}.

Clustering logs arise due to mis-cancellation between real emissions
and virtual corrections. This mis-cancellation results from re-clustering of
final-state configurations of soft gluons. Unlike the single-log
$E_t$ distribution, resummation of clustering logs in the jet mass
distribution cannot simply be written as a power-series in the jet
radius. Collinear singularities at the boundary of a small-$R$ jet
yields large logs in the radius parameter, which appear to all--orders in $\alpha_s$ (see Chapters \ref{ch:EEJetShapes1} and \ref{ch:EEJetShapes2}). Note that the jet veto distribution, studied in the latter chapters, disentangles from the jet mass distribution to all--orders \cite{Dokshitzer:2003uw} and has a non-global structure analogous to the $E_t$ distribution. That is, the coefficients of both non-global and clustering logs are identical for the jet veto and $E_t$ distributions. The arguments of the logs are different though, as explicitly shown in Chapter \ref{ch:EEJetShapes2}.

The clustering logs have recently been the subject of much study
both at fixed order and to all--orders. However there has been no
\emph{full} resummation to all--orders and it was recently suggested
that it is unlikely that such logs be fully resummed even to their
leading log level, which means NLL accuracy relative to leading
double logs (in jet mass distribution) \cite{Kelley:2012kj, Kelley:2012zs}. In the SCET framework \cite{Bauer:2000yr,Bauer:2001yt, Bauer:2000ew} the authors of Ref.~\cite{Kelley:2012kj} confirmed the findings of Ref.~\cite{Banfi:2010pa} (Chapter \ref{ch:EEJetShapes1} for $a=0$), regarding CLs coefficient at $\Or(\as^2)$, and computed the full $R$-dependence of the jet mass distribution at $\Or(\as^2)$. They also pointed out the unlikelihood of the resummation of the  clustering logs to all--orders.

In the preceding chapter (Chapter \ref{ch:EEJetShapes2}) the resummation of the jet mass distribution for $e^+e^-$ annihilation, in the \AKT and C/A (k$_{\mathrm{T}}$) algorithms, was performed to all--orders including the non-global and clustering components in large-$\Nc$ limit. Employing the \AKT clustering algorithm meant that the observable definition was linear in transverse momenta of soft emissions and therefore the resummation involved no clustering logs. In the same chapter we also carried out a fixed-order calculation for the jet mass distribution at
$\mathcal{O}(\alpha_s^2)$ employing the C/A (k$_{\mathrm{T}}$) algorithm
\cite{Catani:1993hr,Ellis:1993tq} and provided a numerical estimate of the all--orders factor with full $R$ dependence.

We present here an expression for the all--orders resummed result of
clustering logs to NLL accuracy in the \KT and C/A algorithms
\cite{Dokshitzer:1997in, Catani:1993hr} for the jet mass distribution. The logs that we control take the form $\mc{F}_n\,\as^n\,L^n$ ($n\geq 2$) in the
exponent of the resummed distribution, and we only compute
$\mc{F}_2$, $\mc{F}_3$ and $\mc{F}_4$ in this chapter. By comparing
our findings to the output of the Monte Carlo of Refs.~\cite{Dasgupta:2001sh, Delenda:2006nf} we show that missing higher-order terms are negligible. We also show that the impact of the primary-emission single clustering logs is of
maximum order 5\% for typical jet radii. Furthermore, we estimate the
non-global (clustering-induced) contribution to the jet mass
distribution in the large-$\Nc$ limit using the said Monte Carlo in 
the case of the \KT algorithm.

This chapter is organised as follows. In the next section we define
the jet mass and show how different algorithms affect its
distribution at NLL level. We then start with the impact of \KT
and C/A algorithms on the jet mass distribution at $\mc O
(\alpha_s^2)$, thus confirming the findings of Ref.~\cite{Banfi:2010pa} (Chapter \ref{ch:EEJetShapes1} for $a=0$). 
In \sec{sec:3-4loop} we compute higher-order clustering terms
and notice that they exhibit a pattern of exponentiation. By making
an anstaz of higher-order clustering-log terms we perform a
resummation of the clustering logs to all--orders in \sec{sec.all--orders} and compare our findings to the output of the said
Monte Carlo program in \sec{sec:MC}. We also perform a
numerical estimate of the non-global logs in the large-$\Nc$ limit
for the jet mass distribution in the \KT algorithm in \ssec{Sec:NG}. Finally we draw our conclusions and point to future work in \sec{sec.conc}.
\vfill

\section{The jet mass distribution and clustering algorithms}
\label{sec.jm distrib}

Consider for simplicity the $e^+e^-$ annihilation into two jets
produced back-to-back with high transverse momenta. We would like to
study the single inclusive jet mass distribution via measuring the
invariant mass of one of the final-state jets, $M_j^2$, while leaving
the other jet unmeasured. One can restrict inter-jet activity by
imposing a cut $Q_0$ on emissions in this region \cite{Ellis:2009wj,
Ellis:2010rwa}. For the purpose of this chapter we do not worry about
this issue because the effect of this cut has already been dealt with
in the previous chapters (as well as in \cite{Delenda:2006nf}) and can be included straightforwardly.

The normalised invariant jet-mass--squared fraction, $\rho$, is defined by:
\begin{equation}
\rho = \left(\sum_{i\in j} p_i\right)^2/ \left(\sum_i
E_{i}\right)^2, \label{eq:RhoDefinition}
\end{equation}
where $p_i$ is the four-momentum of the $i^{\rm th}$ particle and the sum in the numerator runs over all particles in the
measured jet, defined using an infrared and collinear (IRC)-safe
algorithm such as \KT or C/A algorithm. At born level the jet mass has the value zero and it departs from this value at higher orders.

A generic form of general sequential recombination algorithms
\cite{Cacciari:2008gp,Cacciari:2011ma,Dokshitzer:1997in,Catani:1993hr} was discussed in Chapter \ref{ch:Jets} (Algorithm~\ref{Alg:SRAlgInclusive}). For $\EE$ ahhihilation we employ the changes discussed in \sec{sec.fixed_order_1}. We may write $2(1-\cos\theta_{ij})\approx \theta_{ij}^2$
in the small-angles limit where jets are narrow and well-separated
to avoid correlations, and hence contamination, between various
jets.

In the \KT algorithm, softest partons are clustered first
according to the above-mentioned procedure (Algorithm~\ref{Alg:SRAlgInclusive}), while in the \AKT algorithm clustering starts with the hardest partons. In the C/A algorithm only angular separations between partons matter so
clustering starts with the geometrically closest partons. In effect,
clustering induces modifications to the mass of the measured jet due
to reshuffling of soft gluons. Only those gluons which end up in the
jet region would contribute to its mass. Different jet algorithms
would then give different values of the jet mass for the same event.

The global part of the integrated resummed jet mass distribution in
the \KT (C/A) algorithm is related to that in the \AKT
algorithm by:
\begin{equation}
\Sigma^{\kT (\ca)}\left(\frac{R^2}{\rho}\right) =
\Sigma^{\akt}\left(\frac{R^2}{\rho}\right)
\exp\left[{g_{2,A}^{\kT(\ca)}\left(\frac{R^2}{\rho}\right)}\right],
\label{eq:GeneralResummedFormkt-CA}
\end{equation}
where $\Sigma^{\akt}$ resums the leading double logs (DL) (due to
soft and collinear poles of the emission amplitude) as well as
next-to--leading single logs (SL) in the anti--$\rm k_T$ algorithm. The
reader is referred to Chapter \ref{ch:EEJetShapes1} for further details about
this piece. The function $g_{2,A}$ contains the new large clustering
single logs due to \KT(C/A) clustering.

In addition to the global part, each of the distributions
$\Sigma^{\akt, \kT, \ca}$ receives its own non-global NLL
contribution factor $\mc S(R^2/\rho)$. In the \AKT algorithm
the resummation of the non-global logs in the large-$\Nc$ limit for
the jet mass distribution was estimated in Chapters \ref{ch:EEJetShapes1} and \ref{ch:EEJetShapes2}. The result of Chapter \ref{ch:EEJetShapes1} was actually shown to coincide with that of the hemisphere jet mass case. The latter has been available for quite a while \cite{Dasgupta:2001sh, Dasgupta:2002bw}. For \KT clustering the effect of non-global logs has been dealt with numerically in Chapter \ref{ch:EEJetShapes2} and \cite{Appleby:2002ke} and analytically, for example, in the case of gaps-between--jets $E_t$ flow \cite{Delenda:2006nf}. We expect the gross features of the latter to hold
for the jet mass observable. Essentially, the impact of \KT
clustering is in such a way as to reduce the size of the \AKT
(non-clustering) non-global logs. Brief comments on such a reduction
will be given in \ssec{Sec:NG}. In this chapter we are, however,
mainly interested in the function $g_{2,A}(R^2/\rho)$.

\section{Two-gluon emission calculation}
\label{sec:2-gluon}

The effect of \KT and C/A clustering logs starts at
$\mathcal{O}(\alpha_s^2)$. At $\mathcal{O}(\alpha_s)$ there is
indeed a dependence on the jet radius but this dependence is also
present for the \AKT algorithm and has been dealt with in Ref.
\cite{Banfi:2010pa}. In this regard we begin with two-gluon emission case
and study the effect of clustering using both the C/A and \KT algorithms. This calculation has already been performed
in Ref.~\cite{Banfi:2010pa} for the \KT algorithm in the small-$R$ limit.
Here we perform a full-$R$ calculation of the logs coefficient.

\subsection{Calculation in the \KT algorithm}

Consider the independent emission of two soft energy-ordered gluons
$k_{t2}^2\ll k_{t1}^2\ll Q^2$, where $Q$ is the hard scale. This regime, i.e.,
strong energy ordering, is sufficient to extract the leading
clustering logs.

In the case of the \AKT algorithm discussed in Ref.
\cite{Banfi:2010pa} the only contribution from these gluons to the jet mass
differential distribution is when both of them are emitted within an
angle $R$ from the hard parton initiating the measured jet. This is
because the \AKT algorithm works in the opposite sense of the
\KT algorithm: clustering starts with the hardest particles. In
this sense the algorithm essentially works as a perfect cone around
the hard initiating parton with no dragging-in or dragging-out soft
effect.

In the \KT algorithm, on the other hand, the two gluons may be in
one of the following four configurations:
\begin{enumerate}
\item Both gluons $k_1$ and $k_2$ are initially, i.e., before applying
the algorithm, inside\footnote{We use inside, or simply ``in'', to
signify that the parton is within an angular separation of $R$ from
the hard triggered parton (jet axis), and outside, or simply
``out'', if it is more than $R$ away from it.} the triggered jet.
This configuration contributes to the jet shape (mass) regardless of
clustering. The corresponding contribution to the jet mass
distribution is identical to that of the \AKT case (accounted
for by $\Sigma^\akt$ -- see \eq{eq:C2generalFormula} below).
\item Both gluons are initially outside the triggered jet. This arrangement
does not contribute to the jet shape regardless of whether the two
gluons are clustered or not.
\item The harder gluon, $k_1$, is initially inside the triggered jet
and the softer gluon, $k_2$, is outside of it. The value of the jet
shape (mass) is not changed even if clustering takes place. This is
due to the strong-ordering condition stated above.
\item \label{Config} The harder gluon, $k_1$, is initially
outside the triggered jet and the softer gluon, $k_2$, is inside of
it. Applying the algorithm one finds that a real-virtual
mis-cancellation occurs only  if $k_2$ is pulled-out of the
triggered-jet vicinity by $k_1$, a situation which is only possible
if $k_2$ is ``closer'' (in terms of distances $d_{ij}$) to $k_1$
than to the axis of the jet.
\end{enumerate}
We translate the latter configuration mathematically into the step
function:
\begin{equation}
\Xi_{2}\left(k_1, k_2\right) = \Theta(\theta_1^2-R^2)
\Theta(R^2-\theta_2^2) \Theta(\theta_2^2-\theta_{12}^2),
\label{eq:ClustFun2}
\end{equation}
where $\theta_i$ is the angle between gluon $k_i$ and the jet axis
and $\theta_{12}$ is the relative angle of the two gluons. The above
condition is valid only in the small-$R$ limit and we extend this to
the full $R$-dependence in appendix \ref{Sec:F2}. Hence $\Xi_2 = 1$
when configuration \ref{Config} above is satisfied and $\Xi_2 = 0$
otherwise. While in the case where the gluon $k_1$ is virtual and
$k_2$ is real (meaning that $k_1$ cannot pull $k_2$ out) the
particle $k_2$ contributes to the jet mass. To the contrary, when
both $k_1$ and $k_2$ are real, then $k_1$ will not allow $k_2$ to
contribute to the jet mass as it pulls it out. Thus a real-virtual
mismatch occurs and a tower of large logarithms appears. To
calculate these large logarithms we insert the clustering condition
$\Xi_2$ above into the phase-space of the $\mc{O}(\alpha_s^2)$
integrated jet mass distribution. The latter, normalised to the Born
cross-section $\sigma_0$ is given, in the soft and collinear
approximation, by:
\renewcommand{\kt}{k_t}
\begin{eqnarray}
\nonumber \Sigma_2^{\kT}\left(R^2/\rho\right)
&=& \Sigma_2^\akt\left(R^2/\rho\right) + \Sigma_2^{\mathrm{clus}}\left(R^2/\rho\right),\\
\Sigma_2^{\mathrm{clus}}\left(R^2/\rho\right) &=&
\frac{1}{2!}\left(-\CF\frac{\alpha_s}{\pi}\right)^2 \int  dP_1 dP_2
\,\Xi_2\left(k_1,k_2\right), \label{eq:C2generalFormula}
\end{eqnarray}
with
\begin{equation}
dP_i = \frac{d\omega_i}{\omega_i}
\frac{d\cos\theta_i}{\sin^2\theta_i} \frac{d\phi_i}{\pi}
\Theta\left(\frac{4\omega_i}{Q} (1-\cos\theta_i)-\rho\right) \approx
\frac{dx_i}{x_i} \frac{d\phi_i}{2\pi} \frac{d\theta_i^2}{\theta_i^2}
\Theta\left[2 x_i (1-\cos\theta_i)-\rho\right], \label{eq:DiffPhaseSpace2}
\end{equation}
where $\theta_i$, $\phi_i$ and $x_i = 2\,\omega_i/Q$ are the polar
angles, with respect to the jet axis, and the energy fraction of the
$i^{\mathrm{th}}$ gluon. The factor $1/2!$ in Eq.
\eqref{eq:DiffPhaseSpace2} compensates for the fact that we are
considering both orderings: $x_2 \ll x_1 \ll 1$ and $x_1 \ll x_2 \ll
1$. Had we chosen to work with only one ordering, say $x_2 \ll x_1
\ll 1$ as stated at the beginning of this section, then the
said-factor would have not been included. In terms of the
coordinates $(\theta, \phi, x)$ the jet mass fraction defined in Eq.
\eqref{eq:RhoDefinition} reduces to (in the small-angles limit):
\begin{equation}
\rho = \frac{4\omega_i}{Q} (1-\cos\theta_i) \approx x_i\,\theta_i^2,
\label{eq:rhoDefined}
\end{equation}
where gluon $i$ is in the jet. In \eq{eq:DiffPhaseSpace2} we
used the step function to restrict the jet mass instead of
Dirac-$\delta$ function because we are considering the integrated
distribution instead of the differential one. Hence at two-gluon
level, the correction term due to \KT clustering (\eq{eq:C2generalFormula}) is given by:
\begin{multline}
\Sigma_2^{\mathrm{clus}} = \frac{1}{2!}
\left(-\frac{\CF\alpha_s}{\pi}\right)^2 \int^1 \frac{dx_1}{x_1}
\frac{dx_2}{x_2} \int_0
\frac{d\theta_1^2}{\theta_1^2}\frac{d\theta_2^2}{\theta_2^2}
\int_{-\pi}^{\pi} \frac{d\phi_1}{2\pi} \Theta(x_1\theta_1^2-\rho)
\times \\
\times \Theta(x_2\theta_2^2-\rho) \Theta(\theta_1^2-R^2)
\Theta(R^2-\theta_2^2) \Theta(\theta_2^2-\theta_{12}^2),
\label{eq:C2Distr}
\end{multline}
where we used our freedom to set $\phi_2$ to 0. We write the result
to single-log accuracy as:
\begin{equation}
\Sigma_2^{\mathrm{clus}} =
\frac{1}{2!}\left(-\frac{\CF\alpha_s}{\pi}\right)^2 \mathcal{F}_2\,
L^2, \label{eq:C2}
\end{equation}
where $L \equiv \ln\left(R^2/\rho\right)$. In the above equation we
ignored subleading logs and used the fact that in the small-angles
approximation: $\theta_{12}^2 = \theta_1^2
+\theta_2^2-2\theta_{1}\theta_2\cos\phi_1$. The two-gluon
coefficient $\mc F_2$ is given by:
\begin{equation}
\mathcal{F}_2 =  \frac{2}{\pi} \int _0^{\frac{\pi }{3}}  d\phi \ln^2
(2 \cos \phi) =\frac{\pi^2}{54} \approx 0.183. \label{eq:F2Coeff}
\end{equation}

As stated at the outset of this section, we can actually compute
$\mc F_2$ beyond the small-angles (thus small-$R$) limit. In
appendix \ref{Sec:F2} we present an analytic calculation of this
clustering coefficient as an expansion in the radius parameter. One
observes that the small-$R$ approximation of $\mc F_2$, given in \eq{eq:F2Coeff}, is actually valid for jet radii up to order
unity because of its slow variation with $R$ (first correction to
the small-$R$ result is of $\mc O (R^4)$). For instance, at $R =
0.7$ and $R = 1.0$ the coefficient is $\mc F_2 = 0.188$ and $0.208$
respectively, i.e., an increment of about $3\%$ and $15\%$. We
compare the analytical formula with the full numerical result in
\fig{fig:F2Full}.

\subsection{Calculation in the C/A algorithm}

At the two-gluon level, order $\as^2$ in the perturbative expansion
of the shape distribution, the C/A and \KT algorithms work
essentially in a similar manner. Although the C/A algorithm clusters
partons according only to the polar distances between the various
pairs (recall $\rm p=0$ in Algorithm~\ref{Alg:SRAlgInclusive}), at this
particular level energies do not seem
to play a role (once they are assumed strongly-ordered). The jet
shape is only altered if the softer gluon $k_2$ is (geometrically)
closer to $k_1$, or vice versa for the opposite ordering, than to
the jet axis, so as to escape clustering with the jet. This
similarity between the two algorithms does not hold to all--orders
though. We shall explicitly show, in the next section, that they
start differing at $\mc O(\as^3)$.

\section{Three and four-gluon emission} \label{sec:3-4loop}

\subsection{Three-gluon emission}

\subsubsection{\KT clustering case}

Consider the emission of three energy-ordered soft gluons $Q^2 \gg
k_{t1}^2\gg k_{t2}^2 \gg k_{t3}^2$. We proceed in the same way as for the
two-gluons case. First we write the step function $\Xi_3(k_1,
k_2,k_3)$, which describes the region of phase-space that gives rise
to clustering logs. To this end, applying the \KT clustering
algorithm yields the following expression for $\Xi_3$:
\begin{eqnarray}
\nonumber \Xi_3 (k_1,k_2,k_3)  &=&
\Theta(R^2-\theta_3^2)\Theta(\theta_2^2-R^2)\Theta(R^2-\theta_1^2)
\Theta(\theta_3^2-\theta_{23}^2)+ k_1 \leftrightarrow k_2+\\
\nonumber &&
+\Theta(R^2-\theta_3^2)\Theta(\theta_2^2-R^2)\Theta(\theta_1^2-R^2)
\Theta(\theta_3^2-\theta_{23}^2)\Theta(\theta_3^2-\theta_{13}^2)+\\
&&
\nn +\Theta(R^2-\theta_3^2)\Theta(R^2-\theta_2^2)\Theta(\theta_1^2-R^2)
\Theta(\theta_{13}^2-\theta_3^2)
\Theta(\theta_{23}^2-\theta_3^2)\times\\
&&
\times \Theta(\theta_{2}^2-\theta_{12}^2).
\label{eq:ClustFun3kt}
\end{eqnarray}
Hence the correction term due to \KT clustering at three-gluon
level, $\Sigma^{\mathrm{clus}}_3$, which is of an analogous form to
$\Sigma^{\mathrm{clus}}_2$ (\eq{eq:C2generalFormula}), is:
\begin{multline}
\Sigma^{\mathrm{clus}}_3 = \frac{1}{3!}
\left(-\frac{\CF\alpha_s}{\pi}\right)^3 \int^1 \frac{dx_1}{x_1}
\frac{dx_2}{x_2} \frac{dx_3}{x_3} \int_0
\frac{d\theta_1^2}{\theta_1^2} \frac{d\theta_2^2}{\theta_2^2}
\frac{d\theta_3^2}{\theta_3^2} \int_{-\pi}^{\pi}
\frac{d\phi_1}{2\pi} \frac{d\phi_2}{2\pi} \times \\ \times
\Theta(x_1\theta_1^2-\rho)\Theta(x_2\theta_2^2-\rho)\Theta(x_3\theta_3^2-\rho)
\times \Xi_3(\theta_1,\theta_2,\theta_3,\phi_1,\phi_2),
\label{eq:C3Distrkt}
\end{multline}
where, as in $\Sigma^{\mathrm{clus}}_2$, we used our freedom to set
$\phi_3=0$. Performing the integration, in the small-$R$ limit, the
final result may be cast, to single-log accuracy, in the form:
\begin{equation}
\Sigma^{\mathrm{clus}}_3 =
\frac{1}{3!}\left(-\frac{\CF\alpha_s}{\pi}\right)^3 \left[
\frac{3\times 2}{2} \frac{L^2}{2} \mathcal{F}_2 L^2 + \mathcal{F}_3
L^3\right], \label{eq:C3kt}
\end{equation}
where the three-gluon coefficient $\mathcal{F}_3 = -0.052$.
Extending the formalism developed in appendix \ref{Sec:F2} to the
case of three gluons, it is possible to write down an expansion of
$\mc F_3$ in terms of $R$, just as we did with $\mc F_2$. However,
since (a) the whole clustering logs correction to the \AKT
result is substantially small, as we shall see later in \sec{sec:MC}, and (b) $|\mc F_3|$ is much smaller than $\mc F_2$, we
do not perform such an analytical calculation here. We do perform a
numerical evaluation of the full-$R$ dependence of $\mc F_3$ for
various values of $R$, though. The final results are provided in
Table \ref{tab:ClusCoeffNumericalValues}.

The first leading term in \eq{eq:C3kt} is the product of the
one-gluon DL leading term in the \AKT algorithm, $\alpha_s
L^2$, and the two-gluon SL term of \eq{eq:C2}. The second NLL
term in \eq{eq:C3kt} is the new clustering log at $\mc
O(\alpha_s^3)$. We expect that at $n^{\mathrm{th}}$ order in
$\alpha_s$ new clustering logs of the form $\mc F_n L^n$ emerge.
Notice that $|\mc F_3| < \mc F_2$, indicating that the series $\mc
F_n L^n$ rapidly converges. Consequently, the two-gluon result
is expected to be the dominant contribution. This expectation will
be strengthened in the next section, where we compute the four-gluon
coefficient $\mc F_4$. Before doing so, we address the three-gluon
calculation in the C/A algorithm.

\subsubsection{C/A algorithm case}

We follow the same procedure, outlined above for the \KT
algorithm, to extract the large logarithmic corrections to the
\AKT result in the C/A algorithm. As we stated before the
algorithm deals with the angular separations of partons only. In
this regard we consider the various possibilities of the angular
configurations of gluons and apply the algorithm accordingly, to
find a mis-cancellation of real-virtual energy-ordered soft
emissions. The clustering condition step function in the C/A
algorithm can be expressed as:
\begin{equation}
\Xi_3^\ca = \Xi_3^{\kT} +  \widetilde{\Xi}_3, \label{eq:ClustFun3CA}
\end{equation}
where $\Xi_3^{\kT}$ is given in \eq{eq:ClustFun3kt} and the
extra function reads:
\begin{equation}
\widetilde{\Xi}_3 = \Theta(R^2-\theta_3^2) \Theta(\theta_1^2-R^2)
\Theta(\theta_{13}^2-\theta_3^2)\Theta(\theta_3^2-\theta_{23}^2)
\Theta(\theta_2^2-\theta_{12}^2)
\Theta(\theta_{23}^2-\theta_{12}^2). \label{eq:ClustFun3CAExtra}
\end{equation}
Inserting the step function  \eq{eq:ClustFun3CA} into the
equivalent of \eq{eq:C3Distrkt} for the C/A algorithm one
obtains:
\begin{equation}
\Sigma^{\mathrm{clus}, \ca}_3 = \Sigma^{\mathrm{clus}}_3 +
\frac{1}{3!} \left(-\frac{\CF\alpha_s}{\pi}\right)^3
\widetilde{\mathcal{F}}_3\,L^3, \label{eq:C3ca}
\end{equation}
where $\Sigma_3^\mathrm{clus}$ is given in \eq{eq:C3kt} and $ \widetilde{\mathcal{F}}_3 = 0.0236$. Hence the factor
$\mathcal{F}_3^{\mathrm{C/A}}$ which replaces the \KT clustering
term $\mathcal{F}_3$ is, in the small-$R$ limit,
$\mathcal{F}_3^{\mathrm{C/A}} = \mathcal{F}_3 +
\widetilde{\mathcal{F}}_3 = -0.028$. We provide the full-$R$
numerical estimates of $\mc F^{\ca}_3$ in table
\ref{tab:ClusCoeffNumericalValues}. This result illustrates that the
contribution to the shape distribution at this order ($\as^3$) in
the C/A clustering is approximately half that in the \KT
clustering  (although the values are, in both algorithms,
substantially small). Moreover, it confirms the conclusion
reached-at in the \KT algorithm case, for the C/A algorithm,
namely that the clustering logs series are largely dominated by the
two-gluon result. Next, we present the calculation of the four-gluon
coefficient $\mc F_4$.

\subsection{Four-gluon emission}

Consider  the emission of four energy-ordered soft primary gluons $Q^2
\gg k_{t1}^2 \gg k_{t2}^2 \gg k_{t3}^2 \gg k_{t4}^2$. The determination of the
clustering function $\Xi_4$ is more complex than previous lower
orders, particularly for the C/A algorithm. As a result of this we
only present, in this chapter, the findings for the \KT algorithm.
The four-gluon calculations for the jet mass variable are very much
analogous to those presented in Ref.~\cite{Delenda:2006nf} for the
energy flow distribution, to which the reader is referred for
further details. Here we confine ourselves to reporting on the final
answers. After performing the necessary phase-space integration,
which is again partially carried out using Monte Carlo integration
methods, and simplifying one arrives at the following expression for
the correction term, to the \AKT shape distribution, due to
\KT clustering:
\begin{equation}
\Sigma^{\mathrm{clus}}_4 = \frac{1}{4!}
\left(-\frac{\CF\alpha_s}{\pi}\right)^4\times\bigg \{
    6  \times \frac{L^4}{4}  \times  \mathcal{F}_2 L^2
+     4  \times    \frac{L^2}{2}\times \mathcal{F}_3 L^3+ 3 \times
(\mathcal{F}_2 L^2)^2 + \mathcal{F}_4 L^4 \bigg\}, \label{eq:C4kt}
\end{equation}
where, in the small-$R$ limit, $\mathcal{F}_4 = 0.0226$. The
full-$R$ numerical results are presented in table
\ref{tab:ClusCoeffNumericalValues}. As anticipated earlier we have
$\mc F_4 < |\mc F_3| < \mc F_2$, thus confirming the rapid
convergence behaviour of the clustering logs series.

We note that \eq{eq:C4kt} contains products of terms in the
expansion of the Sudakov \AKT form factor with the two- and
three-gluon results, Eqs. \eqref{eq:C2} and \eqref{eq:C3kt}, as well
as the new NLL clustering term at $\mc O(\as^4)$. The first leading
term in \eq{eq:C4kt} comes from the product $\as^2 L^4\times$
two-gluon result ($\as^2 \mc F_2 L^2$); the second term comes from
the product $\as L^2\times$ three-gluon result ($\as^3 \mc F_3
L^3$); and the third term is the square of the two-gluon result.
Therefore, the three- and four-gluon expressions, Eqs. \eqref{eq:C3kt} and \eqref{eq:C4kt}, seem to suggest a pattern of ``exponentiation''. 
Such behaviours give rise to the intriguing possibility of finding 
a reasonably good approximation to the full resummation of clustering 
logs to all--orders, the task to which we now turn.

\section{All-orders result}\label{sec.all--orders}

In analogy to the work of Ref.~\cite{Delenda:2006nf}, one can see
that the results obtained at 2, 3 and 4-gluon can readily be
generalised to $n$-gluon level, with new terms of the form
$\mathcal{F}_n^{(\ca)} L^n$ appearing at each order for the \KT
(C/A) algorithm. By similar arguments to those of Ref.
\cite{Delenda:2006nf} we can deduce the leading term in the
$n^{\mathrm{th}}$ order contribution due the \KT (C/A) clustering
to the shape distribution, $\Sigma_n^{\mathrm{clus}\,(\ca\!)}$. It
reads:
\begin{equation}
\Sigma_n^{\mathrm{clus}\;(\ca)} \propto\,
\frac{1}{(n-2)!}\,\left(-\frac{\CF\alpha_s }{\pi}
\frac{L^2}{2}\right)^{n-2} \frac{\mathcal{F}_2^{(\ca)}}{2}
\left(-\frac{\CF\alpha_s }{\pi} L\right)^{2}   ,\qquad n\geq 2
\label{eq:CnF2Leading}.
\end{equation}
Summing up the terms $\Sigma_n^{\mathrm{clus}\,(\ca\!)}$ to all--orders,
i.e., from $n=2$ to $n \rightarrow \infty$, yields the following
resummed expression:
\begin{equation}
\Sigma^{\mathrm{clus}\;(\ca)} \propto \, \exp\left\{
-\frac{\CF\alpha_s}{\pi} \frac{ L^2}{2} \right\}
\frac{\mathcal{F}_2^{(\ca)}}{2} \left(-\frac{\CF\alpha_s}{\pi}\,
L\right)^2. \label{eq:SigmaClusA}
\end{equation}
The first exponential in the above expression is the celebrated
Sudakov form factor, that one obtains when resumming the jet mass
distribution in the \AKT algorithm. Due to its Abelian nature,
the Sudakov is entirely determined by the first primary emission
result. There are also other pure $\mc F_2$ terms in
$\Sigma_n^{\mathrm{clus}\,(\ca\!)}$ for $ n \geq 4$ of the form:
\begin{equation}
\Sigma_n^{\mathrm{clus}\;(\ca)} \propto\,
\frac{1}{(n-4)!}\left(-\frac{\CF\alpha_s}{\pi}\frac{L^2}{2}
\right)^{n-4} \frac{\cbr{\mathcal{F}_2^{(\ca)}}^2}{8}
\left(-\frac{\CF\alpha_s}{\pi} L \right)^4 ,
\label{eq:CnF2SubLeading}
\end{equation}
which can be resummed to all--orders into:
\begin{equation}
\Sigma^{\mathrm{clus}\;(\ca)} \propto\, \exp\left\{ -\frac{\CF\alpha_s
}{\pi} \frac{L^2}{2} \right\} \frac{\cbr{\mathcal{F}_2^{(\ca)}}^2}{8}
\left(-\frac{\CF\alpha_s}{\pi} L \right)^4. \label{eq:SigmaClusB}
\end{equation}
From Eqs. \eqref{eq:SigmaClusA} and \eqref{eq:SigmaClusB}, one
anticipates the resummed result to all--orders in the clustering log,
$L$, to be of the form:
\begin{equation}
\Sigma^{\mathrm{clus}\;(\ca)} \propto  \exp\left\{ -\frac{\CF\alpha_s
}{\pi} \frac{L^2}{2} \right\} \left[ \exp\left\{\frac{1}{2}\,\mc
F_2^{(\ca\!)}\, \left(-\frac{\CF\as}{\pi}\, L\right)^2\right\}  -1 \right].
\label{eq:SigmaClusC}
\end{equation}

Furthermore we have the following expression in
$\Sigma_n^{\mathrm{clus}\,(\ca\!)}$:
\begin{equation}
\Sigma_n^{\mathrm{clus}\;(\ca)} \propto\,
\frac{1}{(n-3)!}\,\left(-\frac{\CF\alpha_s }{\pi}
\frac{L^2}{2}\right)^{n-3} \frac{1}{6} \mathcal{F}_3^{(\ca)}
\left(-\frac{\CF\alpha_s }{\pi} L\right)^{3}   ,\qquad n\geq 3
\label{eq:CnF3Leading}.
\end{equation}
which is resummed into:
\begin{equation}
\Sigma^{\mathrm{clus}\;(\ca)} \propto \, \exp\left\{
-\frac{\CF\alpha_s}{\pi} \frac{ L^2}{2} \right\}
\frac{\mathcal{F}_3^{(\ca)}}{6} \left(-\frac{\CF\alpha_s}{\pi}\,
L\right)^3. \label{eq:F3res}
\end{equation}

Similarly, one expects analogous expressions to Eq.
\eqref{eq:SigmaClusC} for the remaining $\mc F_3$, $\mc F_4$,
$\cdots$ terms, in addition to ``interference terms'' between these
coefficients, e.g., $\mc F_2 \mc F_3 $ which should first show up at
$\mc{O} (\alpha_s^5L^5)$. Recall that the shape distribution in the
\KT (and C/A) algorithm is, in the Abelian primary emission part,
the sum of the distribution in the \AKT algorithm
(clustering-free distribution) and a clustering-induced
distribution. Schematically:
\begin{equation}
\Sigma^{\kT(\ca\!)} = \Sigma^{\akt} + \Sigma^{\rm{clus}\,(\ca\!)},
\end{equation}
where $\Sigma^\akt$ is simply the Sudakov form factor mentioned
above. Thus gathering everything together and including the logs
which are present in the \AKT case the following
exponentiation is deduced:
\begin{equation}
\Sigma^{\kT (\ca)} =  \exp\left\{ -\frac{\CF\alpha_s}{\pi} \frac{
L^2}{2} \right\} \exp\left\{\sum_{n\geq 2} \frac{1}{n!}\,
\mathcal{F}_n^{(\ca)}\left(-\frac{\CF\alpha_s}{\pi} L \right)^n
\right\}, \label{eq:SigmaAbeliankt-CA}
\end{equation}
where $\mc F_n^{(\ca)}$ is the $n^{\mathrm{th}}$-gluon coefficient
in the $\kT$ (C/A) algorithm.

Confined to the \AKT jet algorithm, the authors in \cite{Banfi:2010pa}
computed the full resummed jet mass distribution up to NLL accuracy,
including the effect of the running coupling as well as hard
collinear emissions\footnote{Leaving non-global contributions aside
for now.}. Taking these and the fixed-order loop-constants into
account, \eq{eq:SigmaAbeliankt-CA} becomes:
\begin{equation}
\Sigma^{\kT (\ca)} = \left(1+\sum_n c_n \asb^n\right)\exp\left[ L g_1
(\alpha_s L) + g_2(\alpha_s L) \right] \exp
\left[g_{2,A}^{\kT(\ca)}(\alpha_s L)\right],
\label{eq:SigmaAbelianFullkt}
\end{equation}
where $\asb = \as/2\pi$ and the functions $g_1$ and   $g_2$ resum the
leading and next-to--leading logs occurring in the \AKT  case.
Their explicit formulae are given in \cite{Banfi:2010pa}. The new piece in
the resummation which is due to primary-emission clustering and
which contributes at NLL level is:
\begin{equation}
g_{2,A}^{\kT (\ca)}(t)=\sum_{n\geq 2} \frac{1}{n!}\,
\mathcal{F}_n^{(\ca)}\left(-2\,\CF\,t\right)^n,
\label{eq:g2AFinalForm}
\end{equation}
where we have introduced the evolution parameter $t$, which governs
the effect of the running coupling:
\begin{equation}
t = \frac{1}{2\pi} \int_{Q\sqrt{\rho}/R}^{Q} \frac{\d k_t}{k_t}\,
\as(k_t) = -\frac{1}{4\pi \beta_0} \ln(1-\as\beta_0 L),
\end{equation}
where the last equality is the one-loop expansion of $t$, and we
have $\beta_0 = (11\CA - 2n_f )/12\pi$.

In the present work, we have been able to compute, by means of brute
force, the first three coefficients, $\mc F_{i=2,3,4}$, for the
\KT algorithm and only the first two coefficients, $\mc
F_{i=2,3}^{\ca}$, for the C/A algorithm. They are, nonetheless,
sufficient to capture the behaviour of the all--orders result, for a
range of jet radii, as we shall show in the next section where we
compare our findings to the output of the numerical Monte Carlo of \cite{Dasgupta:2001sh, Delenda:2006nf}.

\section{Comparison to MC results} \label{sec:MC}

\begin{figure}
\centering
\epsfig{file=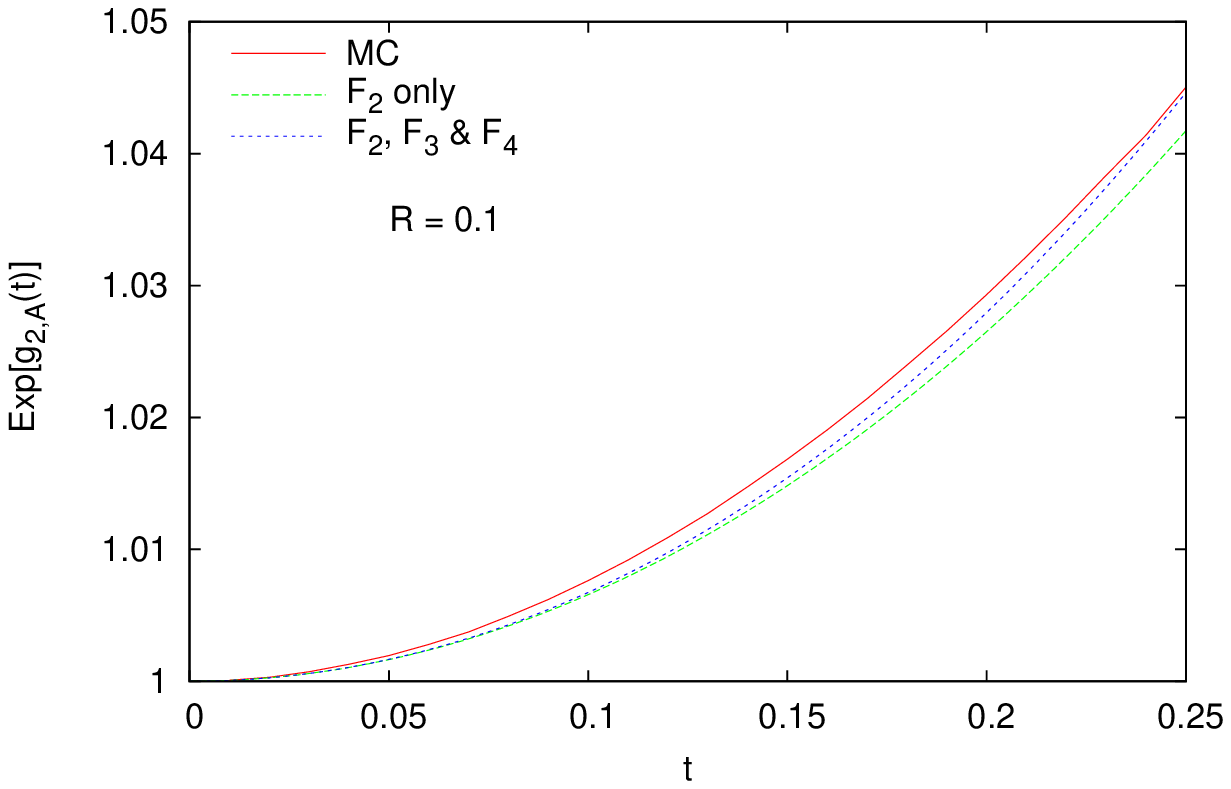,width=0.49\textwidth}
\epsfig{file=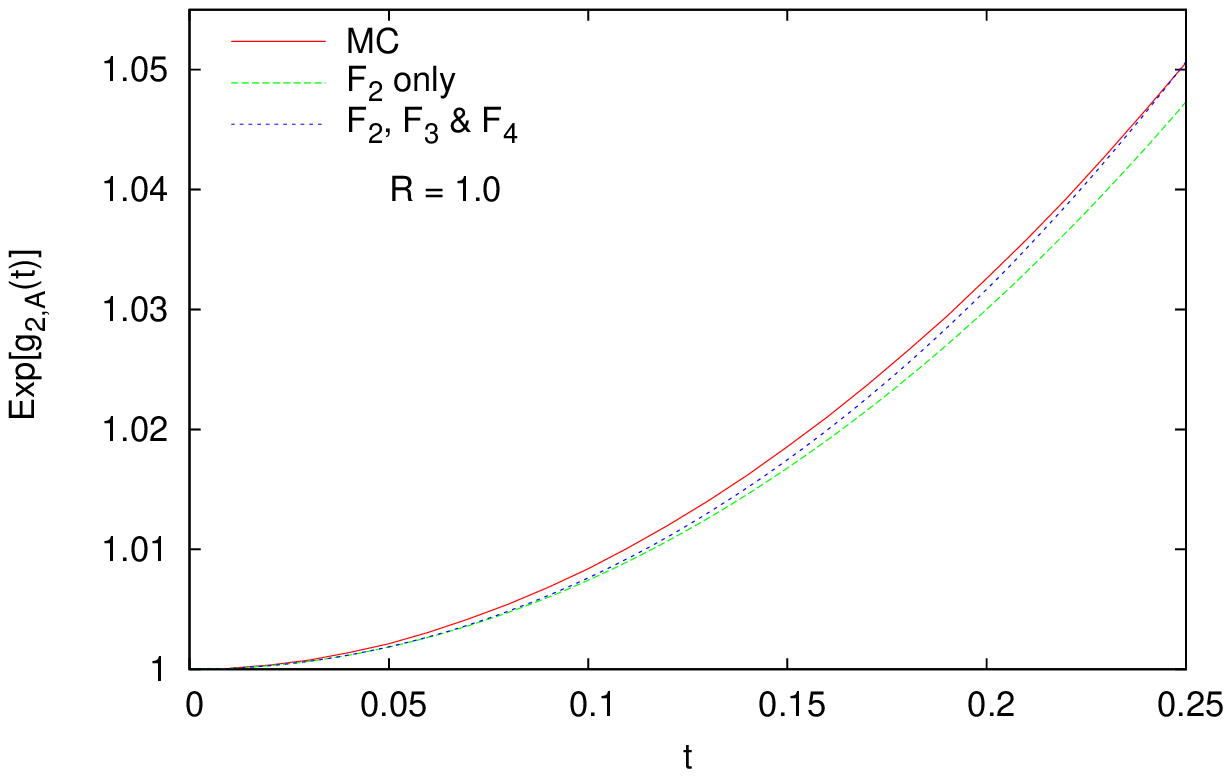,   width=0.49\textwidth}
\caption{Comparisons of the analytical result to the output of the
Monte Carlo program in the \KT algorithm for two values of the jet
radius.\label{fig:CLs_1jet_R1-R0.1}}
\end{figure}

The MC program we use was first developed in \cite{Dasgupta:2001sh} to resum
non-global logs in the large-$\Nc$ limit, and later modified to
include the \KT clustering in \cite{Appleby:2002ke,
Delenda:2006nf}. Since the MC program was originally designed to
resum soft wide-angle emissions to all--orders, it only resums single
logs. The leading logs in the jet mass distribution are, however,
double logs. As such one cannot produce the corresponding DL Sudakov
form factor with the MC. Hence it is not possible to directly
compare the output of the MC with \eq{eq:SigmaAbelianFullkt},
in order to verify the analytical calculations of $g_{2,A}$. One
can, however, extract the MC resummed clustering function,
$\exp\left[g_{2,A}^{\mr{MC}}\right]$, by subtracting off the result
with clustering ``switched off'' from that with clustering
``switched on''. The remainder is then directly compared to
$\exp\left[g_{2,A}\right]$, where $g_{2,A}$ is given in Eq.
\eqref{eq:g2AFinalForm}. Such comparisons are presented in Fig.
\ref{fig:CLs_1jet_R1-R0.1}.

The plots display the MC estimate of $\exp\left(g_{2,A}\right)$, the
analytical  result of the latter in the cases where (a) only the
first coefficient $\mc F_2$ is included in the sum
\eqref{eq:g2AFinalForm} and (b) the first three coefficients, $\mc
F_{i=2,3,4}$, are included. The dependence of the clustering
coefficients $\mc F_i$ on $R$, given in table
\ref{tab:ClusCoeffNumericalValues}, is taken into consideration.

One can clearly see that the function $\exp\left(g_{2,A}\right)$ is
largely dominated by the first coefficient $\mc F_2$, with minor
corrections from $\mc F_3$ and $\mc F_4$. For instance for $R=1.0$
the $\mc F_2$, $\mc F_3$ and $\mc F_4$ coefficients (put alone in
turn) induce a correction  to the \KT resummed distribution
of 1.6\%, 0.06\%, 0.002\% and 4.8\%, 0.3\% and 0.02\% for $t=0.15$
and $0.25$ respectively. This can be understood from Eq.
\eqref{eq:g2AFinalForm}: in addition to being smaller than $\mc
F_2$, the higher-gluon coefficients $\mc F_n$ ($n\geq 3$) are
suppressed by a factorial factor ($n!$), thus leading to a fast
convergence. Given the agreement between our analytical estimate and
the output of the Monte Carlo, and given that the function
$\exp(g_{2,A})$ contributes at most $\mc O(5\%)$, we conclude that
our results for the resummed clustering logs are phenomenologically
accurate for jet radii up to order unity, and that missing
higher-order coefficients $\mc F_n$ ($n\geq 5$) are unimportant.

\begin{figure}
\centering
\epsfig{file=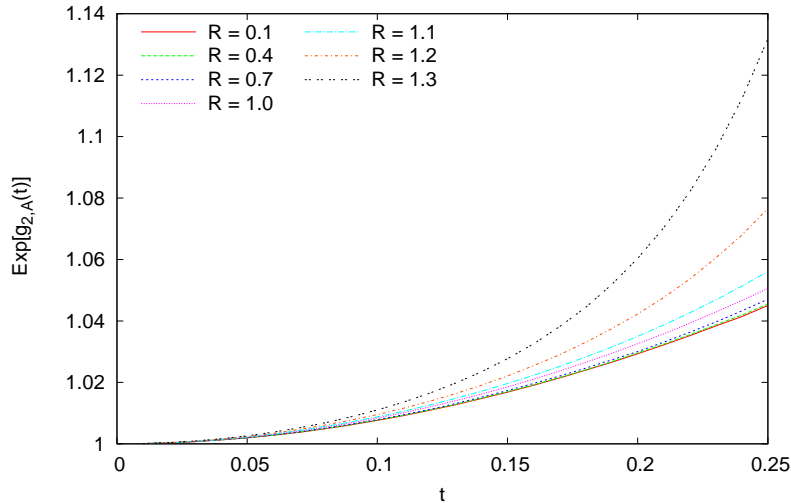,width=0.7\textwidth}
\caption{The output of the Monte Carlo program in the \KT
algorithm for various jet radii.
\label{fig:CLs_1jet_g2_AllR}}
\end{figure}

Lastly, we plot in \fig{fig:CLs_1jet_g2_AllR}, the MC results
of the function $\exp (g_{2,A})$ for various jet radii. The plots
unequivocally indicate that for a fixed $t$, say $0.15$, the
function $g_{2,A}$ varies very slowly with $R$ for $R\leq 1$ and
grows relatively rapidly as $R> 1$. Such a behaviour may be
explained by the analytical formula of $\mc F_2$, Eq.
\eqref{eq:F2RResult}, where the first correction to the small-$R$
result is proportional to $R^4$.

\subsection{Non-global logs}
\label{Sec:NG}

\begin{figure}
\centering
\epsfig{file=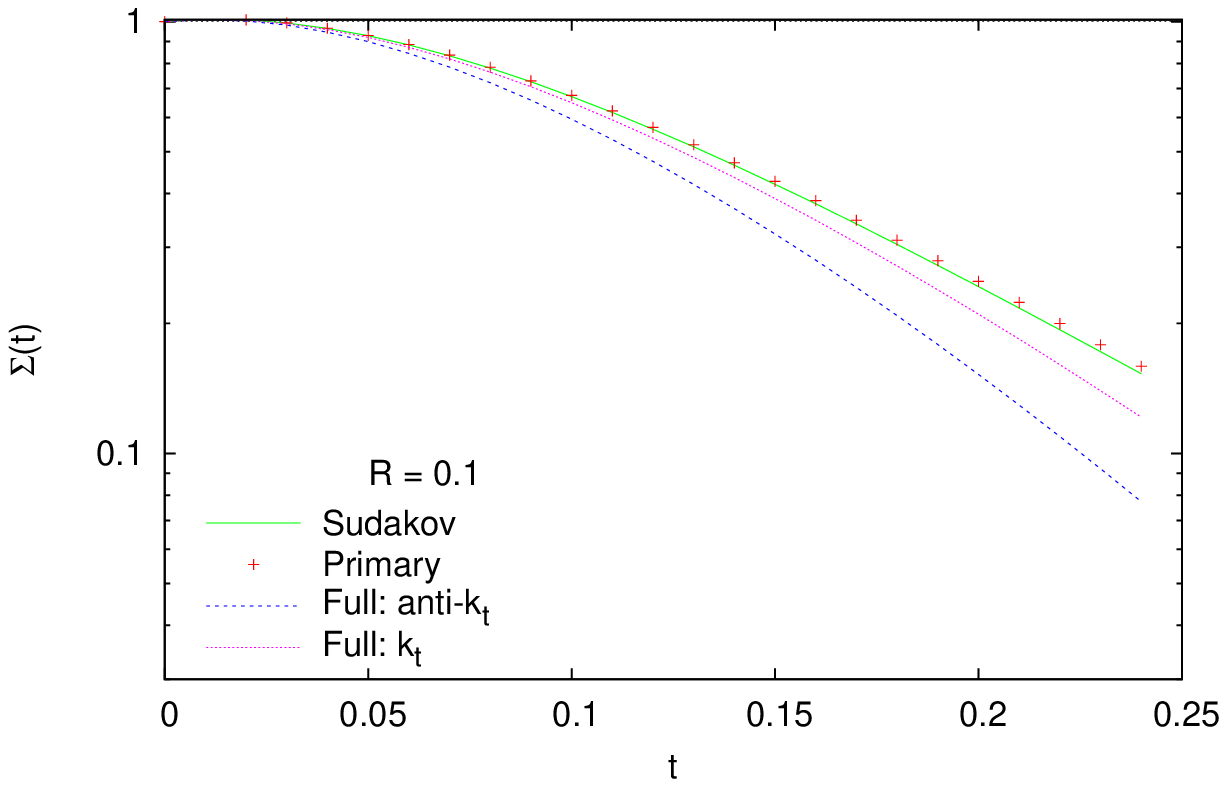,width=0.6\textwidth}
\epsfig{file=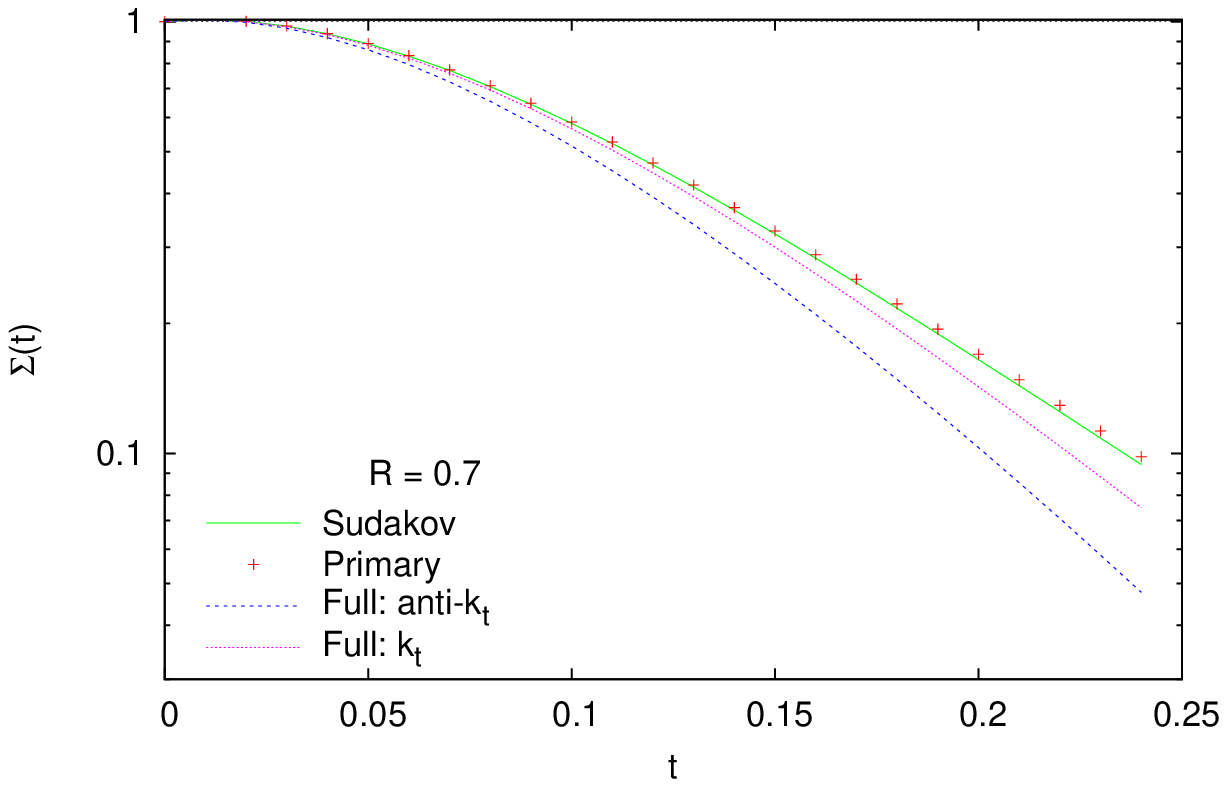,width=0.6\textwidth}
\epsfig{file=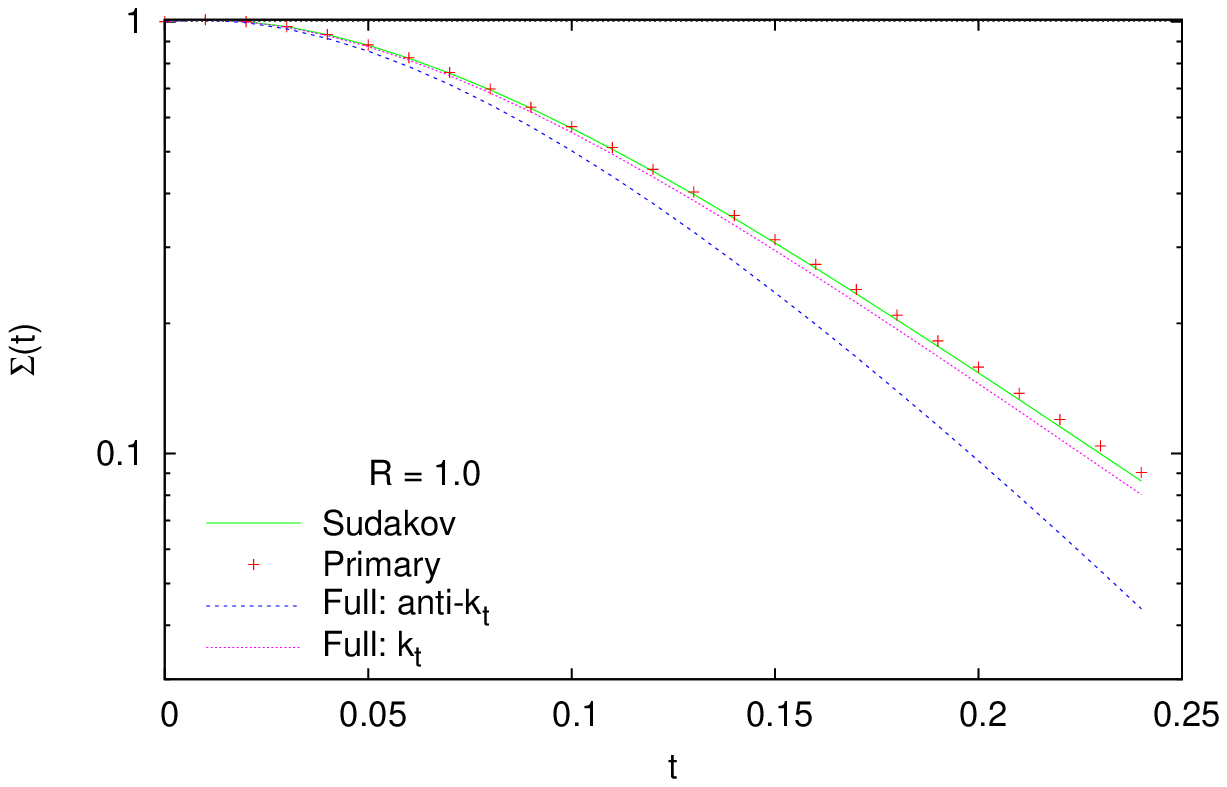,width=0.6\textwidth}
\caption{\label{fig:FullResumJetMassDistktB}Comparisons of the
Sudakov result, the correct primary result and the full result
including non-global logarithms with and without clustering, as
detailed in the main text.}
\end{figure}

As a final task, we plot in \fig{fig:FullResumJetMassDistktB}
the full resummed jet mass distribution in the \KT algorithm, Eq.
\eqref{eq:GeneralResummedFormkt-CA}, including the non-global factor
$\mc S(t)$ in the large-$\Nc$ limit, for various jet radii. In the
figure, ``Sudakov'' refers to the Sudakov \AKT form factor ($\Sigma^{\akt}$), ``primary'' refers to the primary form factor in the
\KT algorithm containing the clustering logarithms ($\Sigma^{\kT}$) in \eq{eq:SigmaAbelianFullkt}, ``Full: \AKT'' refers to
the full \AKT resummation, $\Sigma^\akt \, \mc S^\akt(t)$ and
``Full: \KT'' refers to $\Sigma^{\kT} \, \mc S^{\kT}(t)$. We
notice that the inclusion of non-global logs leads to a noticeably
large reduction of the full resummed result for the \AKT
algorithm case, while in the \KT algorithm the reduction is
moderate.

As is well-known by now (see Chapters \ref{ch:EEJetShapes1} and \ref{ch:EEJetShapes2} and Refs.~\cite{Delenda:2006nf, Appleby:2002ke, Hornig:2011tg}) clustering reduces the impact
of non-global logs through restricting the available phase-space for
their contribution. This is also the case with the jet mass
distribution where we note, for instance, that for $t = 0.15$ the
impact of non-global logs in the \KT algorithm (for $R=0.1,
0.7$ and $1.0$) is a 24\% reduction of the global part, while in the
\KT clustering case this is merely 7\% (for $R=0.1$ and $R=0.7$)
and 4\% for $R=1.0$. We further notice that for small values of the
jet radius, $\mc S(t)$ is independent of $R$. While this is true for
all values of $R$ in the \AKT algorithm\footnote{$\mc
S^\akt(t)$ for our jet mass is identical to that for the hemisphere
jet mass case considered in Ref.~\cite{Dasgupta:2001sh}.}, in the \KT algorithm $\mc S$ falls down as $R$ becomes larger. This is evident in the $R=1.0$ plot in \fig{fig:FullResumJetMassDistktB}.

\section{Conclusion}
\label{sec.conc}

In this chapter we have considered the possibility of exponentiation
of clustering logs in the \KT and C/A algorithms. We have found,
by explicit calculations of the first few orders (up to $\mc
O(\as^4)$, and including, for the first time in literature, the full
jet radius dependence), that the perturbative expansion of the
invariant jet mass distribution exhibits a pattern of an expansion
of an exponential. Consequently, we were able to write an all--orders
\emph{partially} resummed expression for primary emission clustering
logs. We further checked our formula against the output of the
numerical Monte Carlo of \cite{Dasgupta:2001sh} and found a good agreement, within the accuracy of our calculations. We have therefore concluded that
missing higher-order single-log terms in our resummed distribution
have a negligible impact on the total resummed distribution for
typical values of jet radii (up to order unity).

Furthermore, we have briefly discussed the impact of the inclusion
of non-global logs on the total resummed distribution. We confirmed
previous observations concerning the facts that (a) non-global logs
reduce the Sudakov peak of the distribution and that (b) such an
impact is diminished when clustering is imposed on final-state
particles.

We note that the calculations we performed here can readily be
generalised to a large class of non-global observables  defined
using the \KT or C/A algorithm, where the observable is sensitive
to soft emissions in a restricted region of phase space, e.g., angularities discussed in Chapter \ref{ch:EEJetShapes1}. Since the calculations of the coefficients $\mc F_n$ presented here are in fact independent of the jet shape and depend only on the angular configurations introduced by the clustering algorithm, then the effect of jet clustering can simply be included for any generic observable $v$ (being sensitive to soft and
collinear emissions inside the jet only) by introducing the
exponential function $\exp[g_{2,A}]$, with exactly the same
coefficients $\mc F_n$ we computed here. The only difference is the
argument of the logarithm essentially becoming $R^2/v$.

The end of this chapter closes the second part of this thesis, which has been devoted to perturbative calculations of jet shape distributions in $\EE$ annihilation processes. In the next chapter, which opens-up the third and final part, we extend our perturbative calculational machinery to more complex processes that occur at hadron colliders. We attempt at producing results that are competitive to Monte Carlo event generators and ultimately ready for potential comparisons to real experimental data.  

%% file: ch7/chap7.tex

\chapter{Resummation of jet mass at hadron colliders}
\label{ch:HHJetShapes1}

\section{Introduction}
\label{sec:Intro}

Studies of jet substructure have become an area of great interest and 
much activity in the context of LHC phenomenology \cite{Abdesselam:2010pt,Altheimer:2012mn}. The 
primary reason for this relatively recent explosion of interest has been the 
observation that massive particles at the LHC can emerge with large boosts 
either due to the fact that one has an unprecedentedly large centre-of--mass energy $\sqrt{s}$, which consequently enables access to a large range of particle transverse momenta, or due to the possibility of producing very heavy new BSM particles that could subsequently decay to lighter SM particles which thus emerge highly boosted. Such particles with energies or, equivalently, transverse momenta much greater than their masses $p_T \gg m$, would decay to products that are collimated and a significant fraction of the time would be recombined by  jet algorithms into single jets. One is then faced with the problem of  disentangling more interesting signal jets from those that are  purely QCD background jets. In this context, detailed studies of jet substructure had been proposed some time ago \cite{Seymour:1993mx} as having the potential to help separate signal jets (for instance, those arising from hadronic Higgs decays) from QCD background. More recent and detailed 
analyses \cite{Butterworth:2008iy,ATL-PHYS-PUB-2009-088,Plehn:2009rk} subsequently revealed the great potential offered by substructure studies in the specific case of boosted Higgs searches and paved the way for other substructure based studies in many different new physics contexts, which continue to emerge.

However, one important issue that is a significant hindrance to the success of studies that try to exploit detailed knowledge of jet substructure is the complexity of the LHC as a physics environment, particularly in the context of strong interactions. For example, while looking for signal mass-peaks in the vicinity of a certain value of mass, $m$, one is inevitably led to consider the distribution of background QCD jets around that value. In fact a precise knowledge of jet-mass distributions and the related role of cuts on jet masses are extremely crucial for a wide variety of substructure studies. At the LHC one faces formidable obstacles as far as an accurate description of such spectra is concerned. Firstly, from the viewpoint of QCD perturbation theory, the most reliable tool at our disposal, one is confronted with large logarithms due to the multi-scale nature of the distribution at hand, which are as singular as $\as^n \ln^{2n-1}(p_T/m_J)/m_J$, i.e., double logarithmic, plus subleading terms. For 
boosted studies involving values of $p_T \gg m_J$ such logarithms dominate the perturbative expansion and hence fixed-order tools like \nlojet \cite{Nagy:2003tz}, which are typically heavily relied upon for accurate QCD predictions, become at worst invalid or at best of severely limited use. To make matters worse, there are non-perturbative effects such as the underlying event and also the issue of pile-up, which leads to a contamination of jets resulting in an often very significant worsening of any perturbative description. 

In light of the above discussion, it is clear that the tools one can most 
readily use to estimate distributions of quantities such as jet masses are in 
fact Monte Carlo event generators. The parton showers encoded in these event generators derive from first principles of QCD and offer a partial resummation of the large logarithms we mentioned before. They can be combined with fixed-order NLO results \cite{Nason:2004rx,Frixione:2007vw,Frixione:2002ik} to yield descriptions that also describe accurately regions of phase space where the logarithms in question may not be entirely dominant. Moreover, they include hadronisation models as well as a model dependent treatment of other effects such as the underlying event where the parameters of the model are extensively tuned to various data sets, to render them more accurate descriptions. While such tools are very general and hence of immense value in addressing the complexity of the LHC environment, the level of precision they offer may be considered a still open question. 

On the perturbative side the accuracy of the logarithmic resummation represented by parton showers is not clear. While leading logarithms (double logarithms in the present case) are understood to be guaranteed, at the level of next-to--leading or single logarithms (NLL level) the showers are not expected to provide a complete description. For example, it is well-known that parton showers work in the leading colour (large-$\Nc$) approximation as far as large-angle soft radiation is concerned, while the jet mass distributions we discuss here receive single logarithmic contributions from such effects which start already at leading order in $\as$ and contain subleading colour terms. As we shall show, 
single-logarithmic terms on the whole have a large impact on the final result 
and thus it is a little disconcerting to note that they will not be fully accommodated in current shower descriptions. 

Moreover, the event generators we are considering have different models both for parton showers as well as for non-perturbative effects such as the underlying event (UE) and hadronisation. While it is always possible to tune the parameters in each event generator to experimental data, a comparison of the separate physics ingredients of each program often reveal large differences which, in certain cases, does not inspire much confidence in the accuracy of the final description, as far as QCD predictions are concerned. As one example of such differences, we can refer the reader to the studies of non-perturbative effects in jets at hadron colliders \cite{Dasgupta:2007wa}, where large differences were pointed out between the {\sc herwig} \cite{Corcella:2000bw} and {\sc pythia} \cite{Sjostrand:2006za} underlying event estimates at Tevatron energies. A perhaps even more directly relevant example is the recent comparison by the ATLAS collaboration of their data on jet masses and shapes 
\cite{ATLAS:2012am,  Aad:2012jf, Aad:2011kq} to predictions from {\sc herwig}{\fs++} \cite{Bahr:2008pv} and {\sc pythia} which for the jet mass case, for example, do not agree very well with one another, with {\sc pythia} describing the data better. An understanding of the origin of such differences is certainly an important issue in order to gain confidence in the use of Monte Carlo tools for future LHC phenomenology. 

Another avenue that can be explored in terms of theoretical predictions is the possibility for analytical resummation of large logarithms. While the 
techniques needed to achieve such resummation currently apply to a more limited number of observables (cf. event generators are far more general purpose tools) where possible, resummation alleviates some of the present difficulties inherent in an event generator based approach. The typical accuracy of analytical resummation is usually at least NLL with some results having been obtained even up to the $\mathrm{N^3LL}$ level \cite{Becher:2008cf}. It is also possible, and 
straightforward in principle, to match these resummed calculations to NLO estimates so as to have an accurate prediction over a wide range of observable values. Resummation has been a valuable tool in achieving high precision in several QCD studies ranging from pioneering studies of LEP event shapes \cite{Catani:1992ua} to current studies involving hadron collider event shapes \cite{Banfi:2010xy}. In this chapter we carry out such a resummation for the case of the jet mass in Z+jet and dijet processes at the LHC.

If one takes the study of jet-mass distributions as a particular case of studying shapes of jets produced in multi-jet events (see for instance Refs.~\cite{Ellis:2009wj, Ellis:2010rwa}) then it is clear that for substructure studies, resummation of the kind that we perform here for the jet mass, will be an important tool in achieving a precise description of the internal structure of jets. For the same reason, all the issues one encounters in the present study and the solutions we propose here to those problems, will also be of general relevance in the wider context of resummed calculations as substructure tools. There have in fact been several recent attempts to address the issue of resummation for jet masses and other substructure observables such as angularities \cite{Ellis:2009wj,Ellis:2010rwa, Li:2011hy,Li:2012bw}. In Refs.~\cite{Ellis:2009wj,Ellis:2010rwa} it was proposed to study angularities of one or more jets produced in multijet events but calculations were only carried out for jets produced in 
$e^{+}e^{-}$ annihilation. The calculations 
carried out in these references also omitted important contributions at single-logarithmic level, the so called non-global logarithms \cite{Dasgupta:2001sh,Dasgupta:2002bw}, as was explained in some detail in Chapter \ref{ch:EEJetShapes1}. Further calculations for jet-mass observable definitions in $e^{+}e^{-}$ annihilation were carried out in Chapter \ref{ch:EEJetShapes2} and Refs.~\cite{Kelley:2011tj,Kelley:2011aa}. 

Refs. \cite{Li:2011hy,Li:2012bw}, on the other hand, attempt to address the issue for the case of hadron collisions but employ calculations that are not complete to next-to--leading logarithmic accuracy, taking only into account the collinear branchings that generate the so called process independent jet function approximation to the resummed result. In this approach one does not treat the soft large-angle effects that arise from initial state radiation (ISR) or address the important issue of non-global logarithms or the dependence on the jet algorithm explained in Chapters \ref{ch:EEJetShapes1} and \ref{ch:EEJetShapes2}, and hence cannot be considered sufficient for a reasonable phenomenological description of data.  

In our current chapter we address both issues of ISR and non-global contributions and demonstrate their significance not just as formal NLL terms but also from the perspective of numerics and the accuracy of the final description for potential comparisons to data. We consider specifically the jet mass distribution of jets produced in two different hard processes: Z+jet production where it will be a background to the case of associated boosted Higgs production, with Higgs decay to $b \bar{b}$ and the case of jet production in dijet LHC events. We carry out a resummed calculation including the ISR contributions as a power series in jet radius $R$, while the non-global logarithms are calculated exactly at leading order (i.e., order $\as^2$)  and then resummed in the leading-$\Nc$ approximation as was the case for DIS single hemisphere event shapes studied phenomenologically in Ref.~\cite{Dasgupta:2002dc}. 

We demonstrate that developing calculations for $e^{+}e^{-}$ variables and carrying them over to the LHC with neglect of process dependent ISR and non-global logarithms can yield large differences with the full resummation which correctly includes these effects. Moreover, our calculations have an advantage also over Monte Carlo event generators in that we retain the full colour structure of the ISR terms resorting to the leading-$\Nc$ approximation only for the non-global terms starting from order $\as^3$. The accuracy that we achieve in our resummed exponent should then be comparable to that which yielded a good description of DIS event shape data \cite{Dasgupta:2002dc}. We also match our results to leading order QCD predictions so as to account for those terms which are not enhanced by large logarithms but may be important at larger values of jet mass, i.e., away from the peak of the distribution. 
The calculations of this chapter are valid for jets defined in the \AKT algorithm \cite{Cacciari:2008gp}. For jets defined in other algorithms such as the \KT \cite{Catani:1993hr,Ellis:1993tq}, and Cambridge--Aachen (C/A) algorithms \cite{Dokshitzer:1997in,Wobisch:1998wt} 
the role of gluon self-clustering effects greatly complicates the single-logarithmic resummation (see e.g., the discussions in Chapter \ref{ch:EEJetShapes2} and Refs. \cite{Delenda:2006nf, Banfi:2005gj, Kelley:2012kj, Kelley:2012zs}). For this reason we postpone the task of resumming the jet mass in such algorithms to future work.

The calculations in the present chapter also stop short of achieving the accuracy that was obtained for single-hemisphere DIS event shapes in one aspect. 
While we achieve the same accuracy as the DIS case for the resummed exponent, we do not yet obtain the NNLL accuracy in the expansion of the resummation as is achieved for most global event shapes in $e^{+}e^{-}$ and DIS in the leading-$\Nc$ limit \cite{Dasgupta:2003iq} as well as in hadron collisions \cite{Banfi:2010xy}. In other words, our current resummation would not guarantee obtaining the $\as^2 L^2$ terms in the expansion, that arise from a cross-talk between a constant coefficient function $ \as C_1$, which corrects the resummation off just the Born configuration, and the $\as L^2$ term of the Sudakov form factors one obtains for jet masses. A proper treatment of such constant coefficients requires further work. At that stage we will also be in a position to carry out an NLO matching which, at least from the perturbative viewpoint, will give us an answer that will represent the state-of-the-art for non-global observables. We do however estimate in this chapter the possible effect of correcting for 
the coefficient function on our present predictions for the case of Z+jet production. In order to proceed to full NLL accuracy one would also need to understand non-global logarithms beyond the leading-$\Nc$ approximation. This is however a much longer term goal. In the meantime we believe that the predictions we obtain here and certainly after forthcoming NLO matching will be sufficiently accurate so as to render them valuable for phenomenological studies. For the present moment we compare our resummed results to results from shower Monte Carlos and comment on the interesting features that emerge.

We organise this chapter as follows: in the following section we outline the 
general framework for our resummed results indicating the separation between 
the global piece and the non-global terms. Following this, we derive in more 
detail the results for the global terms for both the case of Z+jet and dijet production. In \sec{sec:nglogs} we detail the results of our calculation for the non-global component at fixed-order and at all orders in the large-$\Nc$ limit. In \sec{sec:Zjet} we plot our final results for the case of Z+jet production, having carried out a leading-order matching and commented on the impact of various contributions to the resummed exponent such as ISR and non-global logarithms. We also discuss the expected effect on our results of a proper treatment of the coefficient $C_1$, by treating its contribution in different approximations. Lastly, for the Z+jet case we compare our results to Monte Carlo estimates from a variety of event generators. In \sec{sec:dijets} we discuss final results for the case of dijet production with matching to leading-order calculations. We arrive at our conclusions in \sec{sec:conclusions}. Detailed calculations and explicit resummation formulae are collected in the Appendices.

\section{General framework} 
\label{sec:framework}

The purpose of this section is to outline the overall structure of the resummed results that we have computed for both jet production in association with a 
vector boson and dijet processes at hadron colliders. The notation we find most convenient to adopt is the one developed and used in Refs \cite{Banfi:2004yd, Banfi:2010xy} for the case of global event shape variables in hadron collisions. The extra ingredient specific to our calculation of jet masses is essentially that the observables we address are non-global, so that certain specifics shall of course differ from the case of global event shapes, which we shall highlight, where relevant.

For the case of jet production in association with a Z boson we shall thus 
examine a distribution of the form 
\begin{equation}
\frac{1}{\sigma} \frac{\d \sigma}{dm_J^2}
\end{equation}
where $m_J^2$ is the  jet-mass squared of the highest $p_T$ jet recoiling against the Z boson. For dijet production we shall instead adopt a different 
observable definition and study essentially the jet mass distribution averaged over the two-highest transverse momentum jets:
\begin{equation}
\frac{1}{\sigma} \left(\frac{\d \sigma}{dm_{J1}^2}+\frac{\d \sigma}{dm_{J2}^2} \right)
\end{equation}
where $m_{J1}^2$ and $m_{J2}^2$ are the squared masses of the highest and next-to-highest $p_T$ jets, respectively. 

At Born level, for the Z+jet case, we have a single parton recoiling against the Z boson. If we restrict ourselves to a single-jet final state, then the jet-mass distribution for small jet masses is generated by soft and collinear 
emission off the hard Born configuration whose colour content is provided by the two incoming partons and the final state parton that initiates the jet. 
In contrast, for the case of dijet production soft and collinear emissions 
around the Born configuration, involve an ensemble of four coloured particles, with two incoming partons and two outgoing partons corresponding to the jets. 

While for global observables, such as event shapes, resummation off the hard Born configuration is all that matters, in the present case it is obvious that small jet masses can be produced in events with any jet multiplicity, that represent higher order corrections to the basic Born configurations we address. To restrict oneself to addressing just the Born configuration one can, for example, impose a veto scale $p_{T0}$, as suggested in  Ref.~\cite{Ellis:2009wj}. However, depending on the value of this scale one may then need to also resum the consequent logarithms involving the veto scale (see Refs.~\cite{Ellis:2009wj, Ellis:2010rwa} for a discussion). Even if one chooses to adopt this procedure, the calculations we carry out and report in this chapter shall still form the basis of the resummed answer, but will need to be modified to account additionally for the imposition of a veto. In the present chapter, we do not impose a veto but note that any additional production of non-soft, non-collinear particles (
e.g., the Z+2 jet correction terms to the leading Z+jet process) will be associated with a suppression factor of $\as(p_T)$ relative to the Born term, for each additional jet, where $p_T$ is the typical transverse momentum of the additional jet. This means that to the accuracy of our present calculations (and indeed the accuracy of most current resummed calculations) we shall need to account for only the order $\as$ correction to the Born term supported by a form factor involving only the double logarithmic (soft {\emph{and}} collinear) component of the jet mass resummation. Thus, we never have to discuss, to our accuracy, the complex issue of soft wide-angle gluon resummation off an ensemble other than the Born configuration. The role of correction terms to the basic Born level resummation shall be discussed in more detail later in the chapter. 

Next, following the notation of Refs.~\cite{Banfi:2004yd, Banfi:2010xy} and denoting the Born kinematical configuration by ${\mathcal{B}}$ we write, for a fixed Born configuration, the cross-section for the squared jet mass to 
be below some value $v p_T^2 $, as 
\be \label{sigmadef}
\frac{\d\Sigma^{(\delta)}(v)}{\d \mathcal{B}}= \int  \d m_J^2 \frac{\d^2\sigma^{(\delta)}}{\d \mathcal{B} \d m_J^2 } \Theta(v p_T^2 -m_J^2) . 
\ee
The label $\delta$ corresponds to the relevant production channel at Born level, i.e., the flavour structure of the underlying $2 \to 2$ Born process. We have also introduced the dimensionless variable $v=m_J^2/p_T^2$, with $p_T$ the transverse momentum of the measured jet. One can integrate over the Born configuration with a set of kinematical cuts denoted by $\mathcal {{H}} (\mathcal{B})$ to obtain the integrated cross-section 
\begin{equation}
\Sigma^{(\delta)}(v) = \int d\mathcal{B} \frac{\d \Sigma^{(\delta)}(v)}{\d\mathcal{B}} {\mathcal{H}}(\mathcal{B}),
\end{equation} 
where, as should be clear from the notation, $\d\Sigma^{\left(\delta \right)}/\d \mathcal{B}$ is the fully differential Born cross-section (i.e., the leading order cross-section at fixed Born kinematics) for the subprocess labelled by $\delta$. 
We can then sum over Born channels $\delta$ to obtain $\Sigma(v)$, the integrated jet mass cross-section. However, the above result differs from the case of global observables considered in Refs.~\cite{Banfi:2004yd, Banfi:2010xy}, in that it is correct only as far as the resummation of logarithms off the Born configuration is concerned and not at the level of constant terms which can arise from the higher jet topologies we mentioned before, that are not related to the Born configuration. When addressing the issue of constant corrections we shall thus need to account in addition to the above, for jet production beyond the Born level. For the present we focus on the basic resummation and hence work only with the Born level production channels as detailed above.

Following Ref.~\cite{Banfi:2004yd} (for $v \ll 1$) we then write
\begin{equation}
\label{eq:fact}
\frac{\d\Sigma^{\left(\delta \right)}(v)}{\d \mathcal{B}} = \frac{\d\sigma_0^{\left(\delta \right)}}{\d \mathcal{B}} f_{\mathcal{B}}^{(\delta)}\left(1+\mathcal{O} \left(\as \right) \right).
\end{equation}
The resummation is included in the function $f_{\mathcal{B}} ^{(\delta)}$ and has the usual form \cite{Catani:1992ua}:
\begin{equation}
f_{\mathcal{B}}^{(\delta)} = \exp \left [Lg_1(\as L)+g_2(\as L)+\as g_3(\as L) +\cdots \right],
\end{equation} 
where $g_1$, $g_2$ and $g_3$ are leading, next-to--leading and next-to--next to leading logarithmic functions with further subleading terms indicated by the 
ellipses and $L=\ln 1/v$. 

For the observable we study here, namely non-global jet mass, the function $g_1$ is generated simply by the time-like soft and collinear branching of an outgoing parton and depends only on the colour Casimir operator of the parton initiating the jet, while being independent of the rest of the event. The function $g_2$ is much more complicated. It has a piece of  pure hard-collinear origin, which, like the leading logarithmic function $g_1$, only depends on the colour charge of the parton initiating the jet and factorises from the rest of the event. In the collinear approximation, combining the soft-collinear terms of $g_1$ and the hard-collinear terms included in $g_2$ we recover essentially the jet functions first computed for quark jets in $e^{+}e^{-}$ annihilation in \cite{Catani:1992ua}. However, for complete single logarithmic accuracy one has to consider also the role of soft wide-angle radiation. The function $g_2$ receives a pure soft large-angle contribution also due to emissions from hard partons 
other than the one initiating the jet. For the Z+jet case this piece 
would be generated by coherent soft wide-angle emission from a three hard parton ensemble, consisting of the incoming partons and the outgoing hard parton (jet). For the case of dijet production, we have instead to consider an ensemble of four hard partons and the consequent soft wide-angle radiation has a non-trivial colour matrix structure \cite{Kidonakis:1998nf}, as for global hadronic dijet event shapes. 

Other than the above effects, which are all present for global event shapes and which are all generated by a single soft emission, the function $g_2$ receives another kind of soft contribution, starting from the level of two soft gluons. 
Since we are looking into the interior of a jet rather than the whole of phase 
space, our observable is sensitive to soft gluons outside the jet region 
emitting a much softer gluon into the jet. While for a global observable such a much softer emission would cancel against virtual corrections, in the present case it makes an essential contribution  to the jet mass, triggering single logarithms in the jet mass distribution. These single logarithms (non-global logarithms) cannot be resummed by traditional methods which are based essentially on single gluon exponentiation. In fact a resummation of non-global terms, valid in the large-$\Nc$ limit, which corresponds to solving a non-linear evolution equation~\cite{Banfi:2002hw}, can be obtained by means of a dipole evolution code \cite{Dasgupta:2001sh}. We carry out such a resummation in this chapter but do not attempt to address the issue of the subleading-$\Nc$ corrections which are as yet an unsolved problem. Since non-global logarithms are next-to--leading and we are in fact able to obtain the full colour structure for them up to order $\as^2$, it is only single logarithmic terms starting at order $\as^3$ 
where one needs to use the large-$\Nc$ approximation. One may thus expect that for phenomenological purposes an adequate description of non-global effects will be provided by our treatment here, as was the case for DIS event shape variables studied in Ref.~\cite{Dasgupta:2002dc}.

In the next section we shall generate the entire result, except for the non-global terms, which we shall correct for in a subsequent section, \sec{sec:nglogs}. The results of the next section correspond to the answer that would be obtained if the observable were a global observable and include process independent soft and 
hard-collinear terms alluded to above, as well as a process dependent soft wide-angle piece, which also depends on the jet radius $R$. This soft wide-angle piece, which starts at order $\as$, is calculated with full colour structure whereas one would expect that in Monte Carlo event generators only the leading-$\Nc$ terms are retained, thus implying higher accuracy of the results we obtain here.

\section{The eikonal approximation and resummation} 
\label{sec:eikonal}

In the current section we shall consider the emission of a soft gluon by an ensemble of hard partons in the eikonal approximation (see \app{sec:app:QCD:EikonalApprox}). In this limit one can consider the radiation pattern to be a sum over dipole emission terms \cite{ellis2003qcd}. Our strategy is to calculate the individual dipole contributions to the jet mass distribution and then sum over dipoles to obtain results for both the 
Z+jet case as well as the dijet case. While the sum over dipoles shall 
generate both the soft-collinear and soft wide-angle terms we mentioned in the preceding section, we shall need to extend our answer 
to include also the relevant hard-collinear terms. Once this is done, the only remaining source of single logarithmic terms will be the non-global contribution to the single-logarithmic function $g_2$, which we shall address in detail in the following section.

We shall consider the most general situation that we need to address with all dipoles formed by two incoming partons and two outgoing hard partons. Clearly,
for the Z+jet case one would have only a single hard parton in the final state, with the other parton replaced by a massive vector boson and hence for this case we will exclude the dipole contributions involving the recoiling jet.

The squared matrix element for emission of a soft gluon $k$ by a system of hard dipoles is described, in the eikonal approximation, as a sum over contributions from all possible colour dipoles:
\be
 \label{eq:dipsum}
 \left| {\mathcal{M}}_{\delta}\right|^2 = \left| {\cal M}_{\mathcal{B},\delta} \right|^2 
 \sum_{(ij )\in \delta } C_{ij} \, W_{ij}(k)~,
\ee
where the sum runs over all distinct pairs $(ij)$ of hard partons present in the flavour configuration $\delta$, or equivalently, as stated before, over all dipoles. The quantity $\left| {\cal M}_{\mathcal{B,\delta}}\right|^2$ is the squared matrix element for the Born level hard scattering, which in our case has to be computed for each separate partonic subprocess $\delta$ contributing to the jet distribution and contains also the dependence on parton distribution functions. The contribution of each dipole $W_{ij}$ is weighted by the colour factor $C_{i j} $, which we shall specify later, while the kinematic factor $W_{ij} (k)$ is explicitly given by the classical antenna function
\be
 W_{ij} (k) = \frac{\as \left( \kappa_{t, i j} \right)}{2 \pi}
 \frac{p_i \cdot p_j}{(p_i \cdot k)(p_j \cdot k)}.
 \label{eikant}
\ee
In the above equation $\as$ is defined in the bremsstrahlung scheme \cite{Catani:1990rr}, and its argument is the invariant quantity $\kappa_{t, i j}^2 = 2 (p_i \cdot k)(p_j
\cdot k)/(p_i \cdot p_j)$, which is just the transverse momentum with respect
to  the dipole axis, in the dipole rest frame.
We note that in the eikonal approximation, as is well-known, the Born level production of hard partons in the relevant subprocess $\delta$, factorises from the production of soft gluons described by the antenna functions $W$. The squared matrix element $|M_{\mathcal{B,\delta}}|^2$ essentially produces the quantity 
$\d\Sigma_0^{\left(\delta \right)}/\d \mathcal{B}$ while the $W$ functions 
start to build up the exponential resummation factor $f_{\mathcal{B}}^{(\delta)}$ referred to in Eq.~\eqref{eq:fact}. In what follows below we shall focus on the various components of the resummation in more detail and in particular carry out the calculations for the individual dipole terms.

\subsection{Exponentiation: the Z+jet case}

We have mentioned above the antenna structure of soft gluon emission from a system of hard emitting dipoles. It is well understood by now that if one ignores configurations corresponding to non-global logarithms (in other words those that stem from emission regions where gluons are ordered in energy but not in angle), then to single-logarithmic accuracy there is an exponentiation of the 
one-gluon emission terms described in the preceding section, as well as the corresponding virtual corrections that have the same colour and dynamical 
structure but contribute with an opposite sign so as to cancel the divergences of real emission.  We note that for the case of Z+jet we are dealing with a hard parton ensemble with three partons, two incoming and one corresponding to the triggered jet. In this case the colour factors $C_{i j} = -2\left({\bf T}_i . {\bf T}_j \right)$ (with the ${\bf T}_i$ being SU(3) generators), that accompany the dipole contributions, can be straightforwardly expressed in terms of quark and gluon colour charges. Taking account of virtual corrections and the role of multiple emissions generating the jet mass one can write the result in a form that is familiar from the earliest studies of jet masses in $e^{+}e^{-}$ annihilation \cite{Catani:1992ua}
\begin{equation} \label{fglobalZ}
f_{\mathcal{B}, { \rm global}}^{(\delta )} = \frac{\exp[-\mc R_\delta-\gamma_E \mc R_\delta']}{\Gamma(1 + \mc R_\delta')},
\end{equation}
where the subscript ``global'' above denotes that we are considering the global term only i.e., ignoring all non-global corrections to $f_{\mathcal{B}}^{(\delta)}$.

The function $(-\mc R)$ represents the exponentiation of the single-gluon contribution after cancellation of real-virtual divergences, while $\mc R'$ is the logarithmic derivative of $\mc R$, $\partial_L {\mc R}$ to be evaluated to our accuracy simply by accounting for the leading logarithmic terms in $\mc R$. The terms involving $\mc R'$ arise due to the fact that direct exponentiation only occurs for the Mellin conjugate of the variable $v$. To single-logarithmic accuracy one can invert the Mellin transform analytically by \cite{Catani:1992ua} performing a Taylor expansion of the Mellin space result to first order and integrating over the Mellin variable, resulting in the form written above.

We have for the $\de$--channel ``radiator'', $\mc R_\delta$, the result:
\begin{equation}
\mc R_\delta = \sum_{(ij)\in \delta}\int C_{ij} \, {\d k_t} k_t \, \d\eta \, \frac{\d\phi}{2\pi} \, W_{ij}(k) \, \Theta \left(v(k)-v \right),
\end{equation} 
where we have introduced the integral over the momentum of the emitted gluon $k$ and the step function accounts for the fact that real-virtual cancellations  
occur below a value $v$ of the normalised squared jet mass, while uncancelled virtual corrections remain above $v$. The function $v(k)$ is just the dependence of jet-mass on the emission $k$. Letting the hard initiating parton have rapidity $y$ and transverse momentum $p_t$ and denoting by $k_t$, $\eta$ and $\phi$ the transverse momentum, rapidity and azimuth  of the soft gluon $k$, we have (when the hard parton and gluon are recombined to form a massive jet)
\begin{equation} \label{eikonalv}
v(k) = \frac{m_J^2}{|\underline{p}_t+\underline{k}_t|^2} = \frac{2 k_t}{p_t} \left [ \cosh \left ( 
\eta- y \right )- \cos \phi \right ]+\mathcal{O} \left ( \frac{k_t^2}{p_t^2} \right),
\end{equation}
where we neglect terms quadratic in the small quantity $k_t/p_t$.

It now remains to carry out the dipole calculations for the Z+jet case. We have a hard process with two coloured fermions and a gluon or a three-hard--particle antenna, irrespective of the Born channel $\delta$. Let us call $\delta_1$ the Born subprocess with an incoming quark (or anti-quark) and an incoming gluon, which results in a final state coloured quark or antiquark recoiling against the Z boson. Labelling the incoming partons as $1$ (fermion) and $2$ (gluon) and the measured jet as $3$ we have the following colour factors:
\begin{equation}
\label{eq:colfac}
C_{12} = \Nc, \; C_{23} = \Nc, \; C_{13} = -\frac{1}{\Nc}.
\end{equation} 
For the remaining Born channel with an incoming $q \bar{q}$ pair we obtain the same set of colour factors as above but with an interchange of $2$ and $3$, to correspond to the fact that it is always the quark-(anti)quark dipole which is colour suppressed.

The calculation of individual dipole terms is carried out in 
Appendix~\ref{app:global}. We use the results obtained there to construct the final answer. Let us focus on the Born channel $\delta_1$ corresponding to an incoming gluon and quark with a measured quark jet. In order to combine the various dipoles that contribute to the resummed exponent, we combine the pieces $\mc R_{ij}$ computed in the appendix weighting them appropriately by colour factors:
\begin{equation}
\mc R_{\delta_1} = C_{12} \, \mc R_{12}+C_{13} \, \left( \mc R_{13}^{\mathrm{soft}} +\mc R_{13}^{\mathrm{coll.}} 
\right)+C_{23} \, \left ( \mc R_{23}^{\mathrm{soft}} +\mc R_{23}^{\mathrm{coll}} \right).
\end{equation}

Using the appropriate dipole results generated by using the eikonal approximation, from Appendix~\ref{app:global} and using the colour factors mentioned in Eq.~\eqref{eq:colfac} we obtain 
\begin{equation}
\mc R_{\delta_1} = \frac{\Ncsq-1}{\Nc} \mc R^{\mathrm{coll.}} + \Nc \mc R^{\mathrm{soft}}_{12}+\frac{\Ncsq-1}{\Nc}\mc R^{\mathrm{soft}},
\end{equation}
where we used the fact that $\mc R_{13}^{\mathrm{coll.}}=\mc R_{23}^{\mathrm{coll.}}= \mc R^{\mathrm{coll.}}$ and $\mc R_{13}^{\mathrm{soft}}=\mc R_{23}^{\mathrm{soft}}= \mc R^{\mathrm{soft}}$. Writing the result in terms of the colour factors $\CF$ and $\CA$ and the explicit results for the various dipoles results in the following simple form:
\begin{multline} \label{radiatorCF}
\mc R_{\delta_1}(v)=2\CF \int \frac{\as \left( k_{t,J} \right)}{2\pi} \frac{\d k_{t,J}^2}{k_{t,J}^2} \ln \left (\frac{R p_te^{-3/2}}{k_{t,J}}   \right) \Theta \left (\frac{k_{t,J}^2}{p_t^2} -v \right) \Theta \left(R^2 -\frac{k_{t,J}^2}{p_t^2} \right)  
\\ 
+2 \CF \int \frac{\as \left( k_{t,J}\right)}{2\pi} \frac{\d k_{t,J}^2}{k_{t,J}^2} \ln \left (\frac{R k_{t,J}}{v p_{t}}   \right) \Theta \left (v-\frac{k_{t,J}^2}{p_t^2} \right ) \Theta \left(\frac{k_{t,J}^2}{p_t^2} -\frac{v^2}{R^2}\right) \\ 
+{R^2} \left(\CA+\frac{\CF}{2} \right) \int_v^1\frac{\d x}{x} \frac{\as(x p_t)}{2 \pi} + \frac{R^4}{144} \CF \int_v^1\frac{\d x}{x} \frac{\as(x p_t)}{2 \pi},
\end{multline}
where $k_{t,J}$ is the transverse momentum of the emitted gluon with respect to the jet.

The above result represents the decomposition of the resummed exponent into a collinear piece contained in the first two lines of the above equation and a soft wide-angle piece. We have included in the collinear piece a term $e^{-3/2}$ in the argument of the logarithm, which corrects the eikonal approximation for hard collinear splittings of the final state quark jet. This correction term emerges from replacing the IR singular (pole part) of the $q \to qg$ splitting function, treated by the eikonal approximation, by the full splitting function. As one would expect this collinear piece, which also contains the leading double logarithms, involves only the colour charge $\CF$ of the parton that initiates the measured massive jet, in this case a quark. The remaining part of the result above is a process dependent soft large angle piece that has a power series expansion in jet radius $R$, which we truncated at the $R^4$ term. We note that the $R^4$ term emerges with a numerically small coefficient and shall make 
a negligible impact on our final results which can thus be essentially obtained by considering the $\Or(R^2)$ corrections alone. We also note that the calculation of the soft large-angle component of the result should mean that our results are more accurate than those obtained from MC event generators, which would only treat such pieces in a leading $\Nc$ approximation.

The case of the other subprocess, which generates a gluon jet in the final state, is totally analogous. The result for the resummed exponent is
\begin{multline} \label{radiatorCA}
\mc R_{\delta_2}(v)=2\CA \int \frac{\as \left( k_{t,J} \right)}{2\pi} \frac{\d k_{t,J}^2}{k_{t,J}^2} \ln \left (\frac{R p_te^{-2 \pi \beta_0/\CA}}{k_{t,J}}   \right) \Theta \left (\frac{k_{t,J}^2}{p_t^2} -v \right) \Theta \left(R^2 -\frac{k_{t,J}^2}{p_t^2} \right)  \\ +2 \CA \int \frac{\as \left( k_{t,J}\right)}{2\pi} \frac{\d k_{t,J}^2}{k_{t,J}^2} \ln \left (\frac{R k_{t,J}}{v p_{t}}   \right) \Theta \left (v-\frac{k_{t,J}^2}{p_t^2} \right ) \Theta \left(\frac{k_{t,J}^2}{p_t^2} -\frac{v^2}{R^2}\right) \\ 
+{R^2} \left(2 \CF-\frac{\CA}{2} \right) \int_v^1\frac{\d x}{x} \frac{\as(x p_t)}{2 \pi}+\Or(R^4).
\end{multline}
In order to achieve NLL accuracy, the remaining integrals in Eq.~(\ref{radiatorCF}) and Eq.~(\ref{radiatorCA}) must be performed with the two-loop expression for the running coupling. We obtain:
\begin{eqnarray} \label{Zjet_rad}
\mc R_{\delta_1}&=& -\CF \left (L f_1+f_2 + f_{{\rm coll},q} \right) - R^2 f_{\rm l.a.} \left (\CA+\frac{\CF}{2} \right) + \Or(R^4), \nonumber \\
\mc R_{\delta_2}&=& -\CA \left (L f_1+f_2 + f_{{\rm coll},g} \right) -R^2 f_{\rm l.a.} \left ( 2 \CF -\frac{\CA}{2}\right) + \Or(R^4), \nonumber \\
\end{eqnarray}
with $L=\ln R^2/v$. Explicit expressions for the functions $f_i$ are collected in Appendix~\ref{app:resum}.

\subsection{Exponentiation: the dijet case}

We now turn our attention to the process
\begin{equation}
p(P_1) + p(P_2) \to J (p_3)+J(p_4) + X,
\end{equation}
where we want to measure the mass of the two leading jets. For convenience, we fix the kinematics of the two (back-to-back, in the eikonal limit) leading jets, i.e., their transverse momentum $p_T$ and their rapidity separation $|\Delta y|$.
The calculation of the dipoles in the eikonal limit proceeds in the same way as in the  Z+jet case that we have previously analysed. The main difference is 
the more complicated colour algebra, that leads to a matrix-structure of the resummed result. The formalism to perform the resummation in the presence of more than three hard partons was developed in~\cite{Kidonakis:1998nf}. For each partonic sub-process we need to fix a colour basis and find the corresponding representations of the colour factors ${\bf T}_i . {\bf T}_j$. For each partonic subprocess, the resummed exponent takes the form
\begin{equation} \label{dijets_rc}
f_{\mathcal{B}, { \rm global}}^{(\delta )} = \frac{1}{ {\rm tr} \, H_{\delta} } \sum_{J=3,4} {\rm tr} \left[ \frac{ H_{\delta} e^{-\left( \mathcal{G}_{\delta,J}+\gamma_E \mathcal{G}'_{\delta,J}\right)^{\dagger}}S_{\delta,J} e^{-\mathcal{G}_{\delta,J}-\gamma_E \mathcal{G}'_{\delta,J}}+(\Delta y \leftrightarrow - \Delta y)  }{\Gamma\left(1+2 \mathcal{G}'_{\delta,J} \right)}  \right].
\end{equation}
The matrices $H_\delta$ correspond to the different Born subprocesses and ${\rm tr } H_\de= \d \sigma_0^{(\delta)}/\d\mathcal{B}$. We note that the resummed expression in Eq.~(\ref{dijets_rc}) is written in terms of exponentials that describe the colour evolution of the amplitude~\footnote{In the literature this resummed expression is written in terms of an anomalous dimension $\Gamma$, where $\mathcal{G}= \Gamma \xi$ and $\xi$ is the appropriate evolution variable.}, rather than of the cross-section as in the Z+jet case. We obtain
\begin{eqnarray} \label{dijets_rad}
\mathcal{G}_{\delta,3}&=& -\frac{{\bf T}_3^2}{2}\left (L f_1+f_2 + f_{{\rm coll},3} \right) + {\bf T}_1 . {\bf T}_2 f_{\rm l.a.}(2 \pi i +R^2)
\nonumber \\ 
&&+R^2f_{\rm l.a.}  \left( \frac{1}{4} {\bf T}_3.{\bf T}_4 \tanh^2 \frac{\Delta y}{2}+   \frac{1}{4} ({\bf T}_1.{\bf T}_3+ {\bf T}_2.{\bf T}_3) \right.  \nonumber \\ 
&&+\frac{1}{2} {\bf T}_1.{\bf T}_4 \frac{e^{\Delta y}}{1+\cosh \Delta y}+ \left. \frac{1}{2} {\bf T}_2.{\bf T}_4 \frac{e^{-\Delta y}}{1+\cosh \Delta y} \right) \nonumber +\Or(R^4),  \nonumber \\
\mathcal{G}_{\delta,4}&=& -\frac{{\bf T}_4^2}{2}\left (L f_1+f_2 + f_{{\rm coll},4} \right) + {\bf T}_1 . {\bf T}_2 f_{\rm l.a.}(2 \pi i +R^2)
\nonumber \\ 
&&+R^2f_{\rm l.a.}  \left( \frac{1}{4} {\bf T}_3.{\bf T}_4 \tanh^2 \frac{\Delta y}{2}+   \frac{1}{4} ({\bf T}_1.{\bf T}_4+ {\bf T}_2.{\bf T}_4) \right.  \nonumber \\ 
&&+\frac{1}{2} {\bf T}_2.{\bf T}_3 \frac{e^{\Delta y}}{1+\cosh \Delta y}+ \left. \frac{1}{2} {\bf T}_1.{\bf T}_3 \frac{e^{-\Delta y}}{1+\cosh \Delta y} \right) \nonumber +\Or(R^4), \nonumber \\
 \end{eqnarray}
where the functions $f_i$ are reported in Appendix~\ref{app:resum} and, as before, $L=\ln R^2/v$, $\mathcal{G}'=\partial_L \mathcal{G}$. The collinear part of the result is diagonal in colour space, with a coefficient that is the Casimir of the jet. Large-angle radiation is instead characterised by a more complicated colour structure. We also note the presence of the imaginary phase due to Coulomb gluon exchange. We choose to work in the set of orthonormal bases specified in~\cite{Forshaw:2009fz}, to which we remind for the explicit expressions below. As a result, all the colour matrices are symmetric and the soft matrix appearing in Eq.~(\ref{dijets_rc}) is the identity $S_{\delta,J}=\mbb{1}$.

As an example, we report explicit results for the scattering of quarks with different flavours $q(i) q'(j) \to q(k) q'(l)$. We work in a normalised singlet-octet basis:
\begin{eqnarray} \label{qqqqbasis}
c_1 &= & \frac{1}{\Nc}\delta_{ik} \delta_{jl}\,, \nonumber \\
c_2 &=& \frac{1}{\sqrt{\Nc^2-1}} \left(\delta_{il} \delta_{jk}-\frac{1}{\Nc}\delta_{ik} \delta_{jl} \right).
\end{eqnarray}
In the $t$ channel ($\Delta y>0$), the hard scattering matrix is given by
\begin{equation} \label{qqpqqpH}  
H(t,u)= \frac{4}{\Nc^2}\left(\begin{array}{cc}
0 & 0\\
  0 &  \frac{u^2+s^2}{t^2}
 \end{array}
   \right)\,.
\end{equation}
We have that ${\bf T}_3^2={\bf T}_4^2=\CF$ and the other colour matrices are
\begin{eqnarray}
{\bf T}_1 . {\bf T}_2= {\bf T}_3. {\bf T}_4&=& \left(\begin{array}{cc}
0 & \frac{\sqrt{\Nc^2-}1}{2 \Nc}\\
 \frac{\sqrt{\Nc^2-1}}{2 \Nc} &-\frac{1}{\Nc}
 \end{array}   \right)\,, \nonumber \\
 {\bf T}_1 . {\bf T}_3= {\bf T}_2. {\bf T}_4&=& \left(\begin{array}{cc}
-\CF & 0\\
 0 &\frac{1}{2\Nc}
 \end{array}   \right)\,, \nonumber \\
 {\bf T}_2 . {\bf T}_3= {\bf T}_1. {\bf T}_4&=& \left(\begin{array}{cc}
0 & -\frac{\sqrt{\Nc^2-}1}{2 \Nc}\\
 -\frac{\sqrt{\Nc^2-1}}{2 \Nc} &\frac{1}{2\Nc}-\CF
 \end{array}   \right)\,.
 \end{eqnarray}

\subsection{The constant term $C_1$}
\label{sec:C1}

In order to achieve the NNLL accuracy in the perturbative expansion, that is common for most resummation studies of event shape variables, one must consider also the $\mathcal{O}(\as)$ corrections, which are not logarithmically enhanced in the small jet-mass limit, and their cross-talk with double logarithmic terms arising from the Sudakov form factors. The constant terms can be expressed as :
\begin{eqnarray} \label{C1def}
\as \mc C_{1}^{(\delta)}&=& \lim_{v \to 0} \left[ \Sigma_{\rm NLO}^{(\delta)}(v)-\Sigma^{(\delta)}_{{\rm NLL},\as}(v)\right] =
 \lim_{v \to 0} \left[ \int_0^{v}\frac{\d \sigma^{(\delta)}}{\d v} \d v-\Sigma^{(\delta)}_{{\rm NLL},\as}(v)\right] \nonumber 
\\ 
&=& \lim_{v \to 0} \left[ \sigma^{(\delta)}_{\rm NLO}-\int_{v}^{v_{\rm max}}\frac{\d \sigma^{(\delta)}}{\d v}\d v-\Sigma^{(\delta)}_{{\rm NLL},\as}(v)\right] \nonumber 
\\
&=& \sigma^{(\delta)}_{\rm NLO}+\lim_{v \to 0} \left[\int^{v}_{v_{\rm max}}\frac{\d \sigma^{(\delta)}}{\d v} \d v-\Sigma^{(\delta)}_{{\rm NLL},\as}(v)\right].
\end{eqnarray}
If $\delta$ is a partonic channel that is also present at Born level, then we can recover the usual definition of the constant term:
\be
{C}_{1}^{(\delta)}=\frac{\mc C_{1}^{(\delta)}}{\sigma_0^{(\delta)}}.
\ee

The general kinematic, colour and flavour structure of $\mc C_{1}^{(\delta)}$ can be rather complicated. However, as discussed for global event-shapes in Ref.~\cite{Banfi:2004yd, Banfi:2010xy}, one can simply multiply together this constant and the appropriate resummed exponent $f_{\mathcal{B}}^{(\delta)}$ previously discussed, essentially because the only relevant terms at NLL originate by the product of $C_{1}^{(\delta)}$  times the double logarithms (soft and collinear) in the exponent, which do not depend on the colour flow in the hard scattering nor on the parton distribution functions. This is also true in the case of the jet-mass we are considering in this chapter, with the further complication that we need to specify the flavour of the jet we are measuring, because quark and gluon jets will receive different suppression factors. In particular, new channels open up at relative order $\as$ which are not related to the Born channels and hence obtaining the contribution of the constant terms separately 
for these channels is an involved exercise. While we leave the complete determination of $C_1$ in different experimental set-ups, as well as an analysis of the dijet case, for future work, for the case of  the jet mass in Z+jet, where we measure the mass of the hardest jet, we shall provide later an estimate of the contribution of the constant terms $C_1$ to the resummed distribution.

\section{Non-global logarithms} 
\label{sec:nglogs}

\subsection{Fixed order}

As we emphasised before, the jet mass that we study here, is a non-global observable which means that the results presented in the previous section are not sufficient to obtain the next-to--leading logarithmic accuracy that is common for event shape variables in $e^{+}e^{-}$ annihilation. One is 
also required to correct the results for the effect of soft wide-angle emissions arising from gluon branching; in other words correlated gluon emission as opposed to independent emission of soft gluons by the hard particle ensemble. Calculations involving correlated gluon emission have been carried out at fixed-order and all-orders (in the leading-$\Nc$ limit), in the simpler cases of non-global event shapes in $e^{+}e^{-}$ annihilation (such as the light jet-mass) and DIS. Till date a full calculation even at fixed-order has not been carried out for hadron collisions, which we shall do below, in the context of the jet-mass distribution. We also note that in our previous work we carried out a calculation of non-global logarithms for jet masses of jets produced in $e^{+}e^{-}$ annihilation, both in the limit of small jet radius $R \ll 1$ in Chapter \ref{ch:EEJetShapes1} and including the full $R$ dependence in Chapter \ref{ch:EEJetShapes2}. In the current work we shall lift the approximation of small $R$, 
meaning that our calculations should be useful for jets of any radius . 

To address the correlated emission term at leading-order we need to consider 
the case where one has two energy-ordered soft gluons $k_1$ and $k_2$ such that, for instance,  $\omega_1 \gg \omega_2$, where $\omega_1$ and $\omega_2$ are the respective energies. In previous work involving $e^{+}e^{-}$ annihilation we have addressed this situation by using the fact that the emission of such a two-gluon system off a hard $q\bar{q}$ dipole is described by an independent 
emission term with colour factor $\CFsq$ and a correlated emission term with 
colour factor $\CF \CA$. However, in the present case of multiple hard partons, the emission of a two-gluon system is in principle more involved, since there are several emitting dipoles. In practice, the structure of two-parton emission for the cases we study in this chapter (i.e., up to $n=4$ hard legs) is known to be remarkably simple~\cite{Catani:1999ss,Banfi:2000si}.

As an illustrative example, we consider again the Z+jet case, with three hard legs and the Born channel $\delta_1$, with an outgoing quark jet. 
The squared matrix element once again contains an independent emission piece, which contributes to the exponentiation of the single gluon result, described by our function $\mc R$. This leaves the correlated parton emission term which has the structure
\begin{equation}
W^{\mathrm{corr.}}(k_1,k_2) = \frac{\Nc}{2} w_{12}^{\left(2\right)}+\frac{\Nc}{2} w_{23}^{\left(2\right)}-\frac{1}{2\Nc} w_{13}^{\left(2 \right)},
\end{equation}
where $w_{ij}^{\left(2 \right)}$ is the correlated two-gluon emission by a dipole $ij$, and which is the same as for the $q \bar{q}$ case studied in $e^{+}e^{-}$ annihilation. Hence the dipole emission and associated colour structure for a correlated two-parton system is precisely the same as for a single gluon emission \cite{Banfi:2000si}. 

Next we note that, as described for example in Ref.~\cite{Banfi:2000si}, a piece of the correlated two-parton emission contribution actually goes to build up 
the running coupling (which we have already considered in the exponentiated single gluon contribution $\mc R$). This piece comprises of gluon splittings into 
equally hard gluons or into a $q \bar q$ pair, which together produce the leading term of the QCD $\beta$ function \footnote{For a non-global observable, such hard splittings will also give rise to non-global logarithms below the single logarithmic level, as shown in Chapter \ref{ch:EEJetShapes2}, which are subleading from our point of view.}. This leaves us to consider only the soft part of the correlated emission $\mc S$~\cite{Dokshitzer:1997iz}, which describes the production of an energy-ordered two-gluon system. For a global 
observable, as is well-known, this term produces no effect at single-logarithmic accuracy whereas in the present case it provides us with the first term of the non-global contribution. For a general dipole $ij$ we can explicitly write
\begin{equation}
\label{eq:coll}
\mc S_{ij} = 2 \CA {\bf T}_i.{\bf T}_j \left (A_{ab}+A_{ba}-A_a A_b \right),
\end{equation}
where
\begin{equation}
\label{eq:ngdip}
A_{ab} = \frac{(ij)}{(i k_a)(k_a k_b)(k_b j)}\,, \quad A_{a} = \frac{(ij)}{(i k_a) (j k_a)}.
\end{equation} 
We note that this term is free of collinear singularities along the hard legs $a$ and $b$ due to cancellation between the various terms of Eq.~\eqref{eq:coll}. The remaining collinear singularity from the $1/\left( k_1.k_2 \right)$ , involving the soft gluons, shall turn out to be integrable for the non-global configurations considered below. Thus to obtain the leading non-global term for the jet mass under consideration, it suffices to carry out dipole calculations using the Eqs.~\eqref{eq:coll} and \eqref{eq:ngdip}, for each of the hard emitting dipoles, just as we did for the single gluon emission case. The results we 
obtain are mentioned and commented on below with the details on the calculation in Appendix~\ref{app:nonglobal}. We report below the coefficients $I_{ij}$ of the non-global single logarithms:
\be
\int [\d k_1] [\d k_2]\,  \mc S_{ij} \Theta_{k_1 \notin J } \Theta_{k_2 \in J } \equiv \CA {\bf T}_i.{\bf T}_j\, I_{ij}  \ln^2 \frac{1}{v}.
\ee

\begin{itemize}

\item {\bf{dipoles involving the measured jet}}

As was demonstrated in our previous work (Chapters \ref{ch:EEJetShapes1} and \ref{ch:EEJetShapes2}), for such dipoles one obtains\footnote{Notice that \eqs{eq:HHJS1:Ii3_NGLs}{eq:EEJS1:S2_Coeff} differ in the sign of the $R^4$ coefficient.}:
\begin{equation}
I_{13}=I_{23} \simeq I_{34} = \frac{\pi^2}{3} + \mc O(R^4).
\label{eq:HHJS1:Ii3_NGLs}
\end{equation}
The dipole $I_{34}$, which is relevant for the dijet calculation, depends, in principle, on the rapidity separation $\De y$ between the leading jets. However any $\De y$ dependence will be associated with powers of R that appear to make a negligible contribution.  

We note that the result $\pi^2/3$ is exact for $e^{+}e^{-}$ collisions, where one defines the jet in terms of a polar angle. This result is precisely the same as was obtained for hemisphere jet masses in $e^{+}e^{-}$ annihilation and implies that the result obtained does not depend on the position of the jet boundary, a feature that has also been observed in \cite{Hornig:2011tg}.

 \item {\bf{in-in dipole}}
 
Here we are considering a situation where the emitting dipole legs are away 
from the interior of the jet region into which the softest gluon $k_2$ is emitted. This situation is reminiscent of the much studied case of non-global logarithms in energy flow into gaps between jets \cite{Banfi:2002hw, Dasgupta:2002bw}. In Ref.~\cite{Dasgupta:2002bw} for instance, energy flow into an inter-jet region (i.e., a region between but away from the hard legs of an emitting dipole) was considered for specific choices of the geometry of this region, such as a rapidity slice and a patch in rapidity and azimuth. In the present case we are certainly studying the same problem but the region being considered here is a circle $(\eta-y)^2+\phi^2$, centred on the measured jet. The result we obtain for the non-global contribution due to the in-in dipole has, in the small-$R$ limit, the analytical form

\begin{equation}
I_{12} \approx 4 \left[1.17 R^2-R^2 \ln 2R +\mathcal{O} \left(R^4 \right)\right ].
\end{equation}
We note that the $R^2$ behaviour arises simply from integrating the emitted softest gluon $k_2$ over the interior of the jet region while the $\ln R$ behaviour is a reflection of the collinear singularity we mentioned above, between the two soft gluons $k_1$ and $k_2$. In practice the above small $R$ approximation is not a good approximation to the full result obtained numerically, for larger values of $R$ ($ R \gtrsim 1$), hence eventually we use the numerical answer rather than the form above.

\item {\bf{in-recoil dipole}}

Here again the situation is similar to the case of the interjet energy flow with the only difference from the previous dipole being the geometry of the hard dipole legs, which are now formed by an incoming parton and an outgoing parton at a finite rapidity with respect to the beam. The analytical result we find reads
\begin{equation}
I_{14} \approx\frac{\left(1+e^{\Delta y}\right)^2}{\left(1+\cosh \Delta y \right)^2} \left(1.17 R^2 -R^2 \ln 2 R\right)+\frac{(1+e^{\Delta y})}{1+\cosh \Delta y} \kappa (\Delta y) R^2.
\end{equation}
where $\kappa$ depends on rapidity difference $\Delta y =y-y_r$ and is evaluated numerically. In the limit $\Delta y \to \infty$, $\kappa(\Delta y) \to 0$ and one recovers the previous case.
The dipole $I_{24}$ is easily obtained as 
\begin{equation}
I_{24}= I_{14}(-\Delta y).
\end{equation}
Note that here too we use the numerical result instead of the formula given above.
\end{itemize}

\subsection{All-orders treatment}

In order to perform a complete (NLL) resummation of non-global logarithms, one would need to consider the emission of a soft gluon from an ensemble of any number of gluons. This problem can be treated only in the large-$\Nc$ limit~\cite{Dasgupta:2001sh,Banfi:2002hw}\footnote{An alternative approach that can be found in the literature consists of an expansion in the number of out-of-jet (in this case), gluons, keeping the full colour structure, see for instance~\cite{Forshaw:2009fz,Forshaw:2006fk,Forshaw:2008cq,DuranDelgado:2011tp}.}. For this study, we have adapted the dipole-evolution code used in~\cite{Dasgupta:2001sh} to perform the resummation in the case of jet-masses. 

The code developed in Ref.~\cite{Dasgupta:2001sh} handled the case of evolution off a hard primary dipole in the leading-$\Nc$ limit. The result for the non-global contribution $S(t)$ was obtained by dividing the resummed result by the contribution of primary emissions alone, off the hard emitting dipole. For our work here we have a situation with several possible emitting dipoles. In this situation one has to resort to the large-$\Nc$ limit in which one can treat the problem as independent evolution of only the {\emph{leading colour-connected dipoles in the hard process}}. Detailed formulae can be found in Ref.~\cite{Banfi:2002hw}.

In the Z+jet case, which is simpler, we have noted that there is no visible difference arising from considering just the leading colour-connected dipoles in 
the hard process relative to the case where one evolves \emph{all} hard dipoles using the evolution code. We choose the latter option here hence write the full resummed expression, with the exception of the contributions coming from the constant terms $C_1$, as
\be \label{masterZj}
\Sigma(v)= \sum_\delta \int d \mathcal{B} \frac{d \sigma^{(\delta)}_0}{d \mathcal{B}}  f^{(\delta)}_{\mathcal{B},{\rm global}} f^{(\delta)}_{\mathcal{B}, {\rm non-global}} \mathcal{H(B)}.
\ee
The resummation of non-global logarithms, including contributions from the colour suppressed hard dipoles, is encoded in the two terms
\begin{eqnarray} \label{NGZjet}
f^{(\delta_1)}_{\rm non-global}&=& \exp \left (-\CA  \CF \frac{\pi^2}{3}f_{13}(t) -\frac{\CAsq}{2} I_{12}(R )f_{12}(t)   \right), \\
f^{(\delta_2)}_{\rm non-global}&=& \exp \left (-\CAsq \frac{\pi^2}{3}f_{13}(t)-\CA \left( \CF -\frac{\CA}{2}\right) I_{12}( R ) f_{12}(t)  \right),
\end{eqnarray}
where we have used the fact that $I_{13}=I_{23}$; the other contribution $I_{12}$ does depend on $R$ and vanishes in the $R\to 0$ limit, where one recovers the picture of jets evolving independently. As we stated before, taking the large-$\Nc$ limit of the non-global contributions would amount to switching off the contribution from the colour suppressed dipoles or, equivalently, choosing $\CF =\CA/2$ in the above. This produces no significant difference to our results but the result written above has the advantage that it includes the the full contribution to $\Or(\as^2)$ non-global coefficient.
We have defined
\begin{equation}
f_{ij}(t)=  \frac{1+(a_{ij} t)^2}{1+(b_{ij}t)^{c_{ij}} } t^2\,, \quad t=\frac{1}{4 \pi \beta_0} \ln \left(1-2 \as(p_T) \beta_0 \ln\frac{R^2}{v}\right).
\end{equation}
The coefficients  $a_{ij}, b_{ij}, c_{ij}$ are obtained by fitting the functional form above to the numerical results from the large-$\Nc$ dipole-evolution code. Numerical results are reported in Appendix~\ref{app:resum}.

Following a similar method, we obtain the corresponding result for the dijet case:
\begin{multline} \label{masterdijets}
\Sigma(v) = \sum_{\delta,J=3}^{4} \int d \mathcal{B} \,{\rm tr} \left[ \frac{ H_{\delta} e^{-\left( \mathcal{G}_{\delta,J}+\gamma_E \mathcal{G}'_{\delta,J}+ \mc S_{\delta,J}\right)^{\dagger}} e^{-\mathcal{G}_{\delta,J}-\gamma_E \mathcal{G}'_{\delta,J}-\mc S_{\delta,J}}+(\Delta y \leftrightarrow - \Delta y)  }{\Gamma\left(1+ 2 \mathcal{G}'_{\delta,J} \right)} \right] \\ \times \mathcal{H(B)},
\end{multline}
with
\begin{eqnarray}
\mc S_{\delta,3}&=&  \Big [ \frac{\CA}{2} \Big (\frac{\pi^2}{3}{\bf T}_3^2 f_{13}(t)-  {\bf T}_1.{\bf T}_2 I_{12} ( R )f_{12}(t)-
 {\bf T}_1 . {\bf T}_4  I_{14}(R,\Delta y) f_{14}(t)   \nonumber \\ &&-{\bf T}_2.{\bf T}_4  I_{24}(R, \Delta y)f_{24}(t) \Big) \Big],\nonumber 
\\
\mc S_{\delta,4}&=& \Big [ \frac{\CA}{2} \Big (\frac{\pi^2}{3}{\bf T}_4^2 f_{13}(t)-  {\bf T}_1.{\bf T}_2 I_{12} ( R )f_{12}(t)-
 {\bf T}_2 . {\bf T}_3  I_{14}(R,\Delta y) f_{14}(t) \nonumber 
\\
 &&  -{\bf T}_1. {\bf T}_3  I_{24}(R, \Delta y)f_{24}(t) \Big) \Big].
\end{eqnarray}
As before the above expressions capture the full colour structure of the non-global contribution at $\Or(\as^2)$, but beyond that are valid only in the large-$\Nc$ limit.
\vfill

\section{Z+jet at the LHC} 
\label{sec:Zjet}

In this section we investigate the numerical impact of the different contributions which are relevant in order to achieve NLL accuracy. We decide to study the differential distribution $\d \sigma/\d \zeta$, where $\zeta =\sqrt{v} = m_{ J}/p_{ T}$, so as to study the jet mass distribution directly rather than the squared jet mass. We also find it useful to work with a dimensionless ratio to better separate soft physics contributions. In fact, a fairly large value of the jet mass can be generated by the emission of a very soft gluon, if the transverse momentum of the hard jet is large, while small values of $\zeta$ always correspond to the emission of soft and/or collinear gluons. If not stated otherwise, we normalise our curves to the Born cross-section. We use the matrix element generator {\sc comix}~\cite{Gleisberg:2008fv}, included in {\sc sherpa}~\cite{Gleisberg:2003xi,Gleisberg:2008ta}, to produce all the tree-level cross-sections and distributions. We consider proton-proton collisions at $\sqrt{s}=7\,\
mathrm{TeV}$ and select events requiring $p_{T}>200$~GeV;  the Z boson is produced on-shell and does not decay. Jets are defined using the \AKT algorithm~\cite{Cacciari:2008gp}.
In our calculation we use the set of parton distribution function {\sc cteq}6m~\cite{Pumplin:2002vw}, with renormalisation and factorisation scales fixed at $\mu_R=\mu_F=200$~GeV, to ease the comparison with different Monte Carlo parton showers, which we are going to perform in Section~\ref{sec:showers}.

\subsection{Different approximations to the resummed exponent} \label{sec:exponent}

\begin{figure} 
\begin{center}
\includegraphics[width=0.7\textwidth]{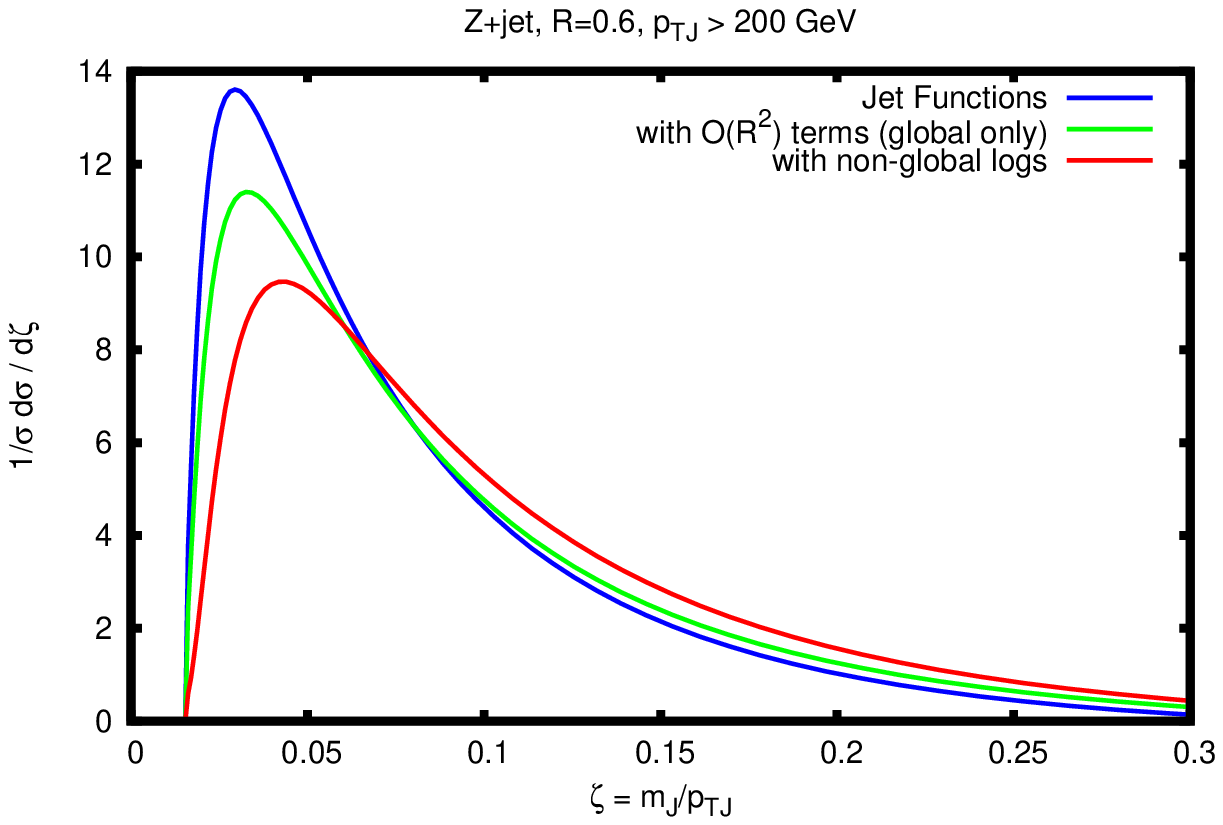}
\caption{Comparison between different approximations to the resummed exponent: jet functions (blue), with full resummation of the global contribution (green) and with non-global logarithms as well (red). The jet radius is $R=0.6$.} \label{fig:comparisonR06}
\end{center}
\end{figure}

\begin{figure} 
\begin{center}
\includegraphics[width=0.7\textwidth]{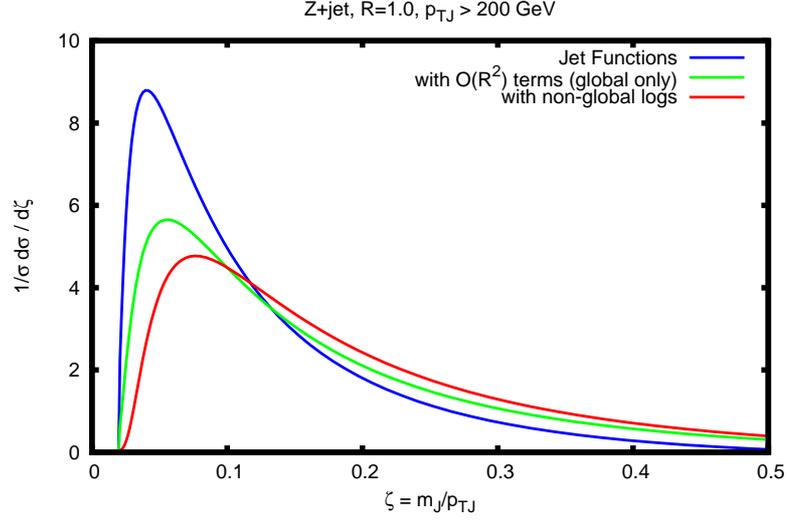}
\caption{Comparison between different approximations to the resummed exponent: jet functions (blue), with full resummation of the global contribution (green) and with non-global logarithms as well (red). The jet radius is $R=1.0$.} \label{fig:comparisonR10}
\end{center}
\end{figure}

We start by considering different approximations to the resummed exponent $f^{(\delta)}_{\mathcal{B}}$. 
We present our results for two different jet-radii: $R=0.6$, in Fig.~\ref{fig:comparisonR06},  and $R=1.0$, in Fig.~\ref{fig:comparisonR10}.
The blue curve corresponds to the most simple approximation to the NLL resummed exponent: the jet-function. This approximation correctly resums soft and collinear radiation as well as hard collinear, but does not capture all soft radiation at large angle. In particular this corresponds to neglecting terms that are suppressed by powers of the jet radius in the resummed exponent Eq.~(\ref{Zjet_rad}). These terms are included in the resummation of all global contributions (green line). We have checked that inclusion of $\Or(R^2)$ is enough because the $\Or(R^4)$ corrections are below the percent level even for $R=1.0$.  We stress that up to this point no approximation on the colour structure has been made, although we have checked that sub-leading colour corrections are small, once the collinear part has been properly treated. Finally, in the red curve we also take into account the resummation of non-global logarithms as described by Eq.~(\ref{masterZj}). The first, $\Or(\as^2)$, coefficient on the non-global 
contribution is computed exactly, while the subsequent resummation is performed using a numerical dipole-evolution code in the large-$\Nc$ limit. 
We note that the inclusion of $\Or(R^2)$ terms in the resummed exponent as well as non-global logarithms, noticeably corrects the simple jet-function picture, based on collinear evolution. The peak height is reduced by more than 30\% for $R=0.6$ and it is nearly halved for $R=1.0$. The effect of non-global logarithms is reduced in the latter case, but the $\Or(R^2)$ corrections to the jet-function approximation become bigger.

\subsection{Matching to fixed order} 
\label{sec:matching}

\begin{figure} 
\begin{center}
\includegraphics[width=0.49\textwidth]{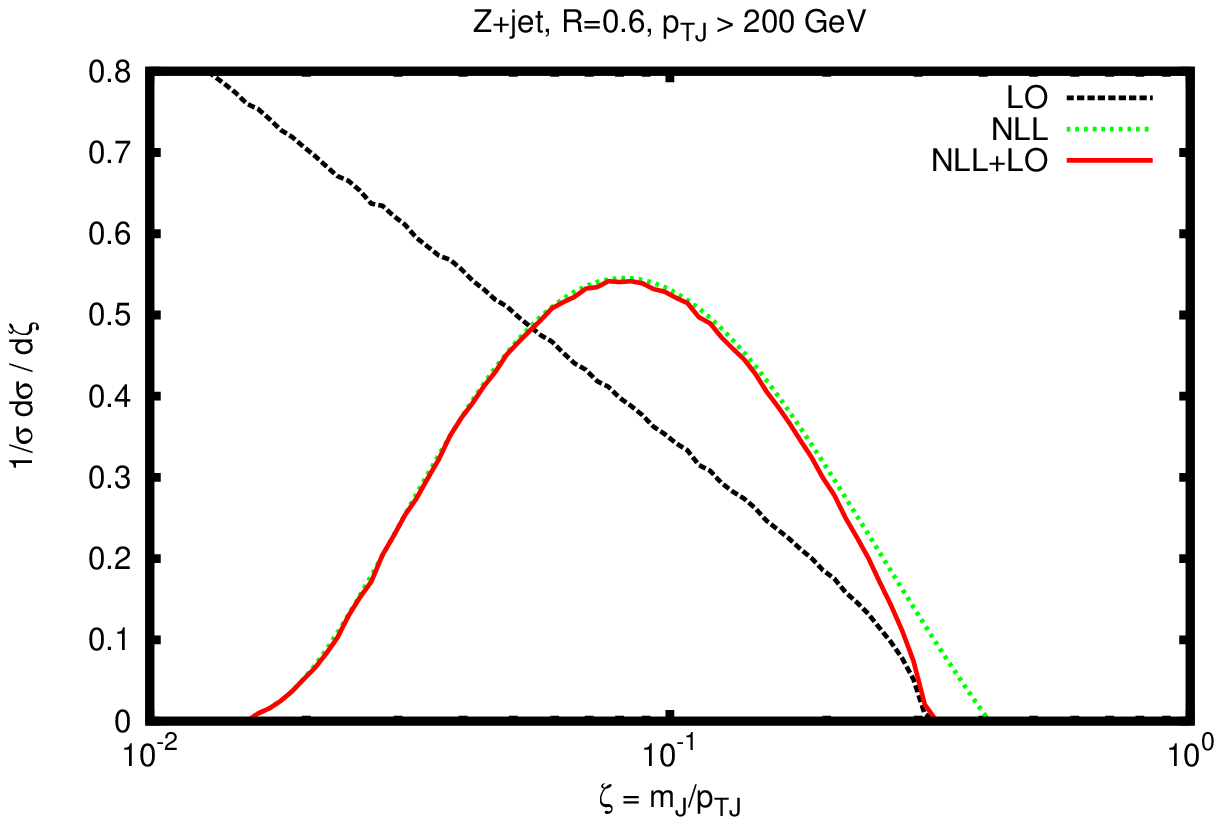}
\includegraphics[width=0.49\textwidth]{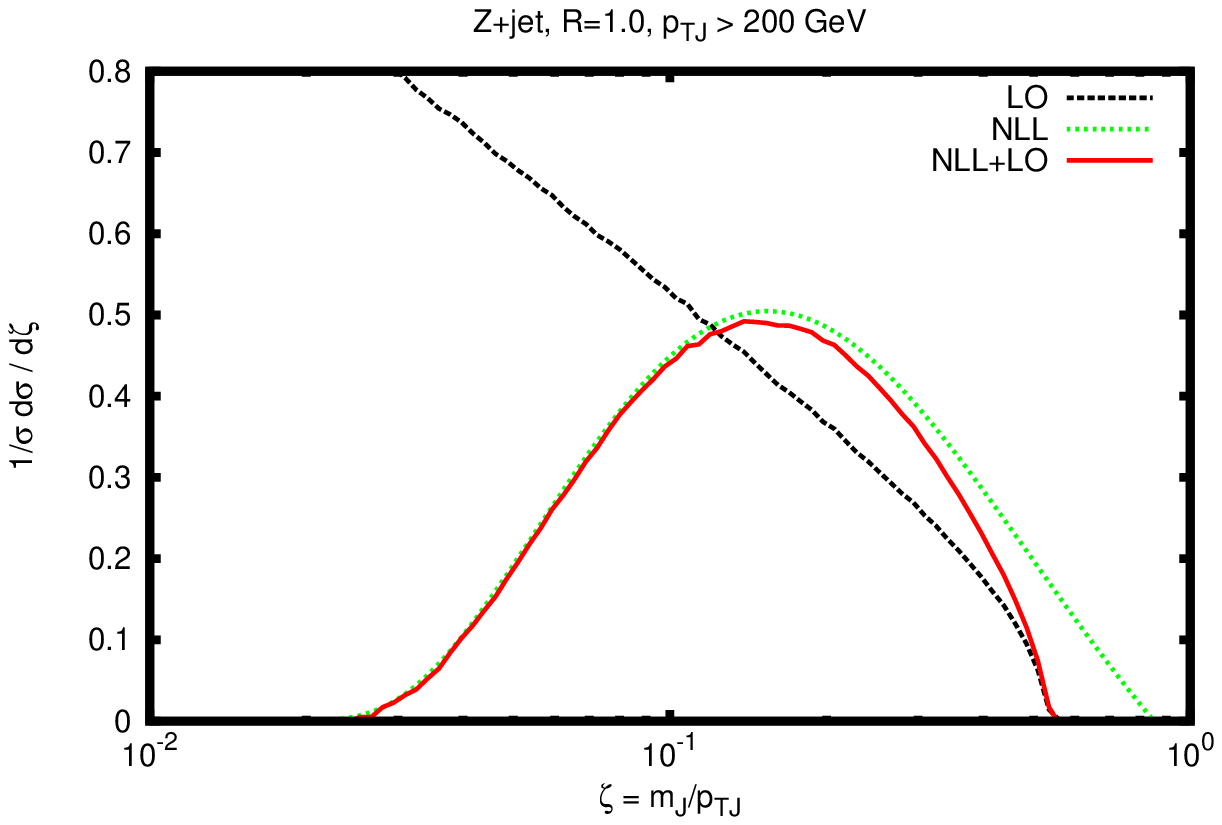} 
\caption{Matching of the NLL resummed distribution to the LO one for $R=0.6$ (on the left) and $R=1.0$ (on the right).}\label{fig:matching}
\end{center}
\end{figure}

We now turn our attention to obtaining a jet-mass distribution which is reliable for all values of $\zeta$. We achieve this by matching to a fixed-order (FO) calculation. 
Although a complete phenomenological analysis would require matching to a next-to leading order (NLO) QCD calculation, for the purpose of this chapter we limit ourselves to LO matching. 
We compute the differential jet-mass distribution at $\Or(\as)$ using {\sc comix}. The result is plotted in Fig.~\ref{fig:matching} (dashed black line): the differential distribution $\d\sigma/\d \ln\zeta$ diverges logarithmically in the small-mass limit.  The dotted green line instead represents the resummed result. The matched curve (shown in solid red) is obtained by straightforwardly adding the two contributions and removing the double-counted terms, i.e., the expansion of the resummation to $\Or(\as)$:
\be \label{matching}
\frac{1}{\sigma}\frac{{ \d \sigma}_{\rm NLL+LO}}{\d \ln \zeta}= \frac{1}{\sigma}\left[ \frac{\d \sigma_{\rm LO}}{\d \ln \zeta}+\frac{\d \sigma_{\rm NLL}}{\d \ln \zeta}-\frac{\d \sigma_{{\rm NLL},\as}}{\d \ln \zeta} \right].
\ee
We note that the matched result coincides with the resummation at small $\zeta$, because the logarithmically divergent contributions to the LO distribution are cancelled by the expansion of the resummation. Moreover, the matched distribution follows the LO order one in the opposite limit. In particular, we note that the LO result exhibits an end-point:
\begin{equation}
\zeta^2 =\frac{m_{ J}^2}{p_{T}^2} =  \frac{2 p_t k_t}{|\underline{p}_t+\underline{k}_t|^2} \left(\cosh y - \cos \phi \right)=\frac{2 p_t k_t}{p_t^2+k_t^2+2 p_t k_t \cos \phi} \left(\cosh y - \cos \phi \right),
\end{equation}
which leads to 
\begin{equation}
\zeta_{\rm max}=\sqrt{\max_{y^2+\phi^2\le R^2} \zeta^2}=\tan \frac{R}{2}=\frac{R}{2}+O(R^3).
\end{equation}
This feature is not present in the resummed distribution, because the jet does not recoil against the emission of the eikonal gluon, but it is restored thanks to matching.

\subsection{Numerical estimate of constant term $C_1$} 
\label{sec:C1num}

In this section we investigate the numerical impact of the constant $C_1$ defined in Eq.~(\ref{C1def}), in the case of Z+jet events, where we measure the mass of the highest $p_{\rm T}$ jet.
Different contributions build up the constant terms at $\mathcal{O}(\alpha_s)$: we have one-loop virtual corrections, terms that cancel the renormalisation and  factorisation scale dependence of the Born pieces and contributions with multiple partons in final state that may, or may not, end up in the same jet. 

One-loop virtual corrections have the same kinematical and flavour structure as the corresponding Born subprocess $\delta$ and, consequently, they need to be suppressed by the same resummed exponent. In order to cancel infrared divergences, one needs to consider real emissions in the soft and collinear limit as well. The final-state singularities are precisely the source of the logarithms we are resumming and these configurations can be mapped  onto one of the Born subprocesses $(\delta)$. Initial-state collinear singularities, which do not give rise to any logarithms of the jet mass, must be absorbed into the parton densities, leaving behind terms that depend on the factorisation scale. Finally, we also need to consider kinematic configurations where the two final-state partons are not recombined, resulting into Z+2 jet events, and suppress the hardest one with the appropriate exponent. Thus, we should perform the calculation of the NLO cross-section that appears in Eq.~(\ref{C1def}) keeping track of the 
kinematics of the final-state partons, in order to separate between $g$ and $q$ components. Although this can be clearly done by computing these corrections analytically, for the current analysis we use the program {\sc mcfm}~\cite{Campbell:2002tg}, which does not trivially allow us to do so. Nevertheless, we are able to compute the finite part of the virtual corrections, for the different initial-state channels. 
We suppress virtual corrections, as well as the integrated term in Eq.~(\ref{C1def})  with their appropriate form factor, $f^{(\delta)}$. We then multiply all the remaining real corrections by either a gluon or a quark form factor, producing a band. 

\begin{figure} 
\begin{center}
\includegraphics[width=0.49\textwidth]{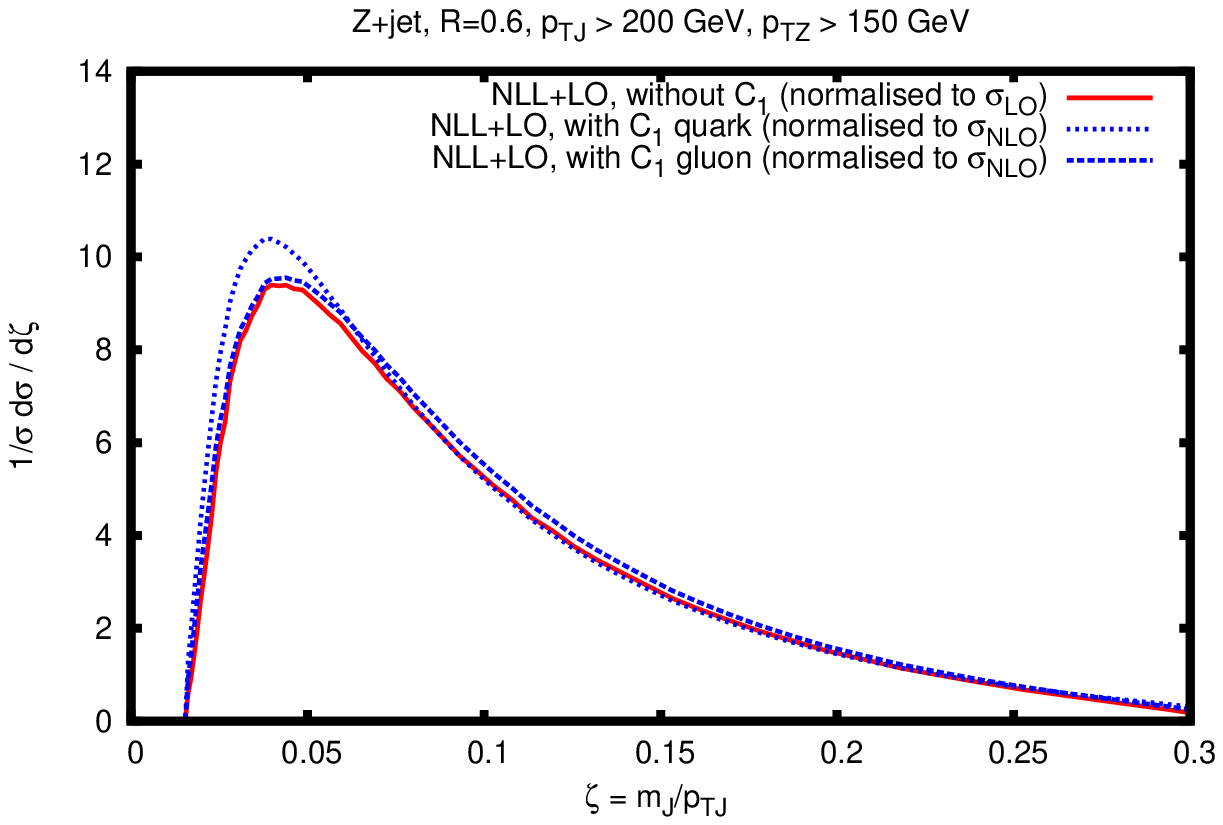}
\includegraphics[width=0.49\textwidth]{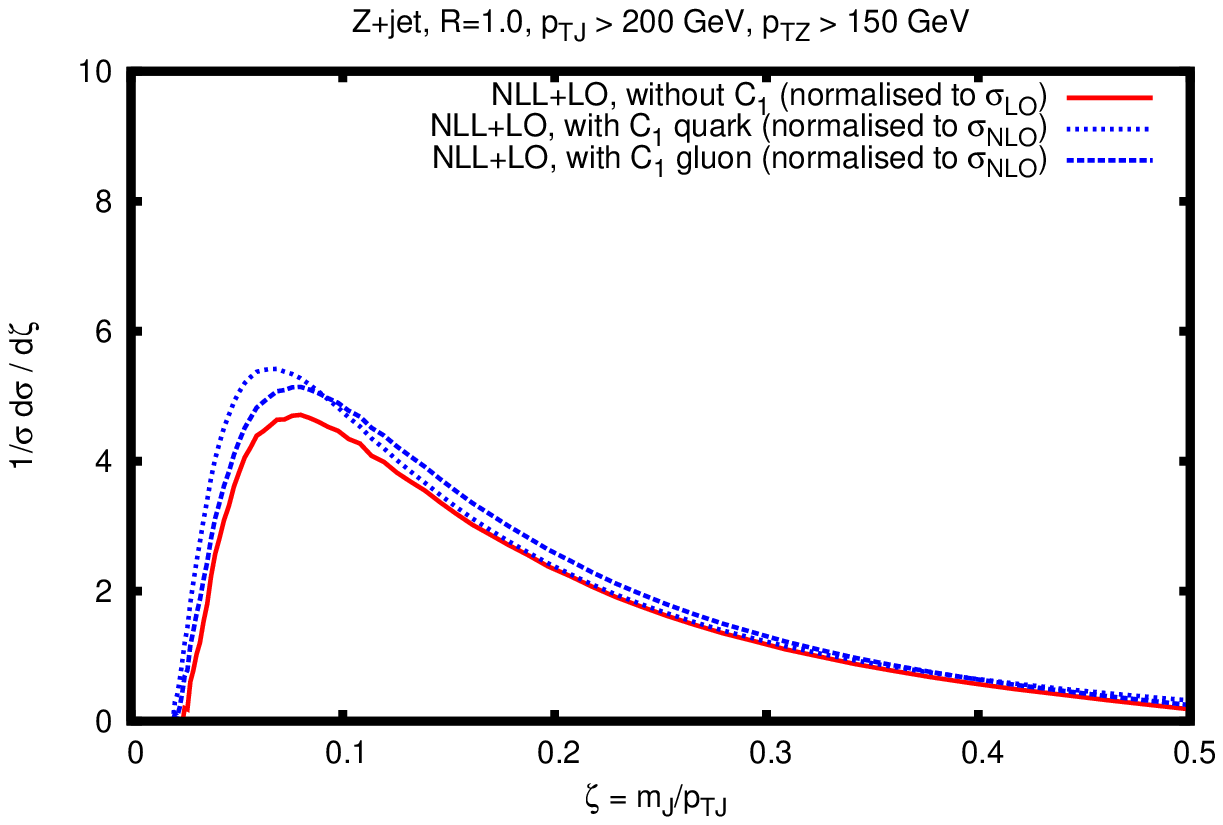}
\caption{The impact of the NLL constant $C_1$, for $R=0.6$ jets (on the left) and $R=1.0$ (on the right). The band is produced by suppressing the real radiation contributions with a quark or gluon jet form factor, as explained in the text.}\label{fig:C1}
\end{center}
\end{figure}

Our findings are plotted in Fig.~\ref{fig:C1}, for $R=0.6$, on the left, and $R=1.0$, on the right\footnote{We have suppressed $C_1$ with the full global resummed exponent, producing terms beyond our NLL accuracy, which we do not control. A more precise analysis would involve the complete determination of $C_1$, suppressed only with double-logarithmic terms, together with an uncertainty band, assessing the impact of higher logarithmic orders.}.
When $C_1$ is included, we normalise the distribution to the total NLO rate, rather than the usual Born cross-section. In order to avoid large NLO corrections~\cite{Rubin:2010xp}, we put a cut on the Z boson transverse momentum $p_{ TZ}>150$~GeV.  We have found that this leads to $K$-factor of $1.45$ and $1.57$, for $R=0.6$ and $R=1.0$, respectively.  With this set of cuts, the impact of $C_1$ is not too big, but definitely not negligible. The complete calculation of this constant is therefore necessary in order to be able to perform accurate phenomenology. 

\subsection{Comparison to Monte Carlo event generators}
\label{sec:showers}

In this section we compare our resummed and matched result NLL+LO to three standard Monte Carlo event generators: {\sc sherpa} \cite{Gleisberg:2008ta}, {\sc pythia}8 \cite{Sjostrand:2007gs} and {\sc herwig}{\fs++} \cite{Gieseke:2011na}. 
Monte Carlo parton showers are powerful tools to simulate complicated final states in particle collisions. When interfaced with hadronisation models, they are able to describe the transition between partons and hadrons. Moreover, they 
provide events which are fully differential in the particles' momenta. Further, 
they provide models of other non-perturbative effects at hadron colliders such as the underlying event. As stated before however, despite their usefulness and successes, it is quite difficult to assess the theoretical precision of these tools. For this reason, comparisons between parton showers and analytic calculations, which have a well-defined theoretical accuracy, form an important part of QCD phenomenology. 

For the case of jet-mass it has been noted in recent ATLAS studies~\cite{Aad:2012jf} for jet masses that {\sc pythia} with hadronisation and the underlying event gives a reasonably good description of the data. We would therefore expect our resummation to be in some accordance with {\sc pythia}, though we should stress that we do not include any non-perturbative effects. Hence we compare our results to the parton shower aspect of the various event generators on its own. While this is in principle correct, 
in practice one should be aware that there can be considerable interplay between the shower level and the non-perturbative effects in various event generators so that these programs may only return more meaningful results when all physical effects (perturbative and non-perturbative) are considered together. We should bear this caveat in mind while attempting to compare a resummed prediction with just the parton shower models in event generators. 

\begin{figure} 
\begin{center}
\includegraphics[width=0.7\textwidth]{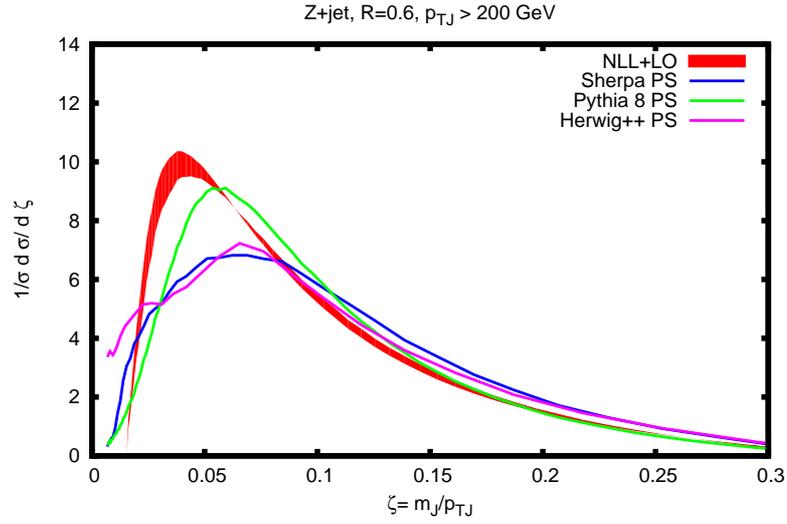}
\caption{Comparison of our resummed and matched result NLL+LO  (in red) to standard Monte Carlo event generators, at the parton level.} \label{fig:shower}
\end{center}
\end{figure}

The results of the comparison are shown in Fig.~\ref{fig:shower}.  Our NLL+LO result for $R=0.6$ is shown in red (the band represents the uncertainty due to the incomplete determination of $C_1$). The Monte Carlo results are obtained  with the same parton densities as in the resummed calculation and the same set of cuts. For {\sc sherpa} (blue line) and  {\sc pythia} (green line) we fix $\mu_R=\mu_F=200$~GeV, while for {\sc herwig}{\fs++} (magenta line) we use the default transverse mass of the Z boson. 
At the shower level, {\sc sherpa} and  {\sc herwig}{\fs++} appear to perform quite similarly. They produce fairly broad distributions, which 
are not very much suppressed as $\zeta \to 0$. 
The distribution obtained from {\sc pythia} instead produces a curve which is much closer to our resummed result. Although the position of the peaks differ by $ \delta \zeta =0.01$ ($\delta m_{\rm J}\sim2$~GeV), the height and general shapes appear in agreement. 

\begin{figure} 
\begin{center}
\includegraphics[width=0.7\textwidth]{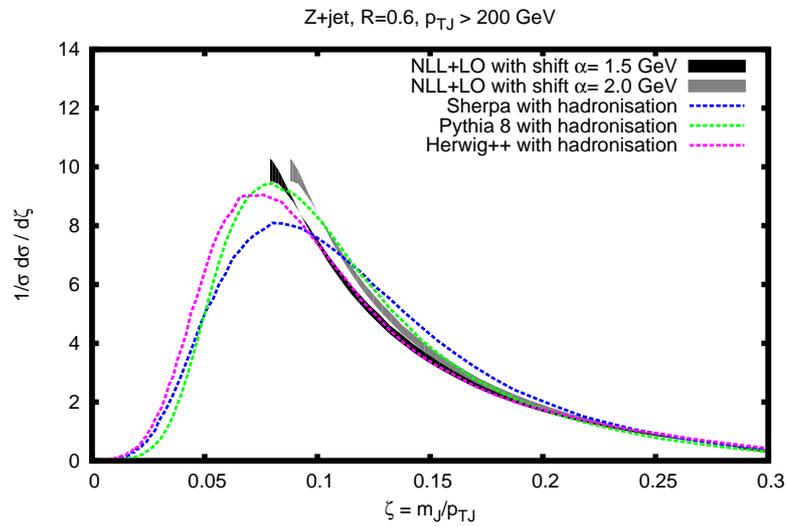}
\caption{Results for the $\zeta$ distributions obtained with standard Monte Carlo parton showers, with hadronisation corrections (dashed lines) compared to analytical resummation with non-perturbative shifts (shaded bands) as explained in the main text.} \label{fig:hadro}
\end{center}
\end{figure}

The agreement between the different Monte Carlo generators is restored when hadronisation corrections are included, as demonstrated in Fig.~\ref{fig:hadro}:  {\sc pythia} and {\sc herwig}{\fs++} produce very similar results, while the distribution obtained with {\sc sherpa} is broader, but not too different. Clearly, in order to compare to collider data, one must also include the contribution from the underlying event. 

It is also interesting to compare the resummed prediction we computed in this chapter with a shift to the right to account for hadronisation corrections to the event generator results after hadronisation. The shift approximation, initially suggested in \cite{Dokshitzer:1997ew}, should be valid to the right of the Sudakov peak but will certainly break down in the vicinity of the peak (see also the discussion in \cite{Dasgupta:2002dc} and references therein). The amount of the shift is related to the non-perturbative correction to mean values of jet masses derived in \cite{Dasgupta:2007wa}. From that reference, we note that the $v=m_J^2/p_T^2$ distribution should be shifted by an amount $\alpha R/p_T$ (the correction to the mean value) with a dimensionful coefficient $\alpha$ that can be taken to be of the order of a few times $\Lambda_{\rm QCD}$. The results of our calculations with a non-perturbative shift are shown in Fig.~\ref{fig:hadro}. From there one notes that a shift of $1.5\,\mathrm{GeV}\, R/p_T$ (
for the $v$ distribution which we carry over to the $\zeta$ distribution plotted in Fig.~\ref{fig:hadro}) where we take $p_T$ to be the value of the lower bound (200 GeV in this case) on transverse momentum, yields an excellent agreement with the {\sc herwig}{\fs++} result after hadronisation. On the other hand, a slightly larger shift of $2.0 \, \mathrm{GeV}\, R/p_T $ yields a good agreement with {\sc pythia}. We have truncated the shifted resummed result near the peak of the resummed distribution, as we would expect the shift not be meaningful beyond this region at the very best. Although we have made a crude estimate of hadronisation effects, and one may be able to better compute these corrections, we do note that within non-perturbative uncertainties our results are compatible with the most widely used event generator models. We therefore anticipate that after the further improvements we have in mind are accomplished, our results may be directly used for phenomenological purposes. 

\vfill

\section{Dijets at the LHC}
\label{sec:dijets}

In this section we provide numerical predictions for the jet mass distribution in dijet events. As before, we consider proton-proton collision at $\sqrt{s}=7$~TeV, with jets defined according to the \AKT algorithm~\cite{Cacciari:2008gp}.
The main complication with respect to the Z+jet case previously discussed is a more complicated colour algebra, which results into a matrix structure of large-angle soft gluon radiation. In order to simplify our resummed calculation, we work at fixed kinematics, i.e., we demand the jets' transverse momentum $p_T$ to be $200$~GeV and their rapidity separation to be $|\Delta y| = 2$ (at Born level we only have two jets).  We remind the reader that we consider
\be
\frac{1}{\sigma}\frac{{ \d \sigma}}{\d  \zeta}= \frac{1}{\sigma} \left( \frac{\d \sigma}{\d \zeta_1} +\frac{\d \sigma}{\d \zeta_2}\right)_{\zeta_1=\zeta_2=\zeta}.
\ee

\begin{figure} 
\begin{center}
\includegraphics[width=0.7\textwidth]{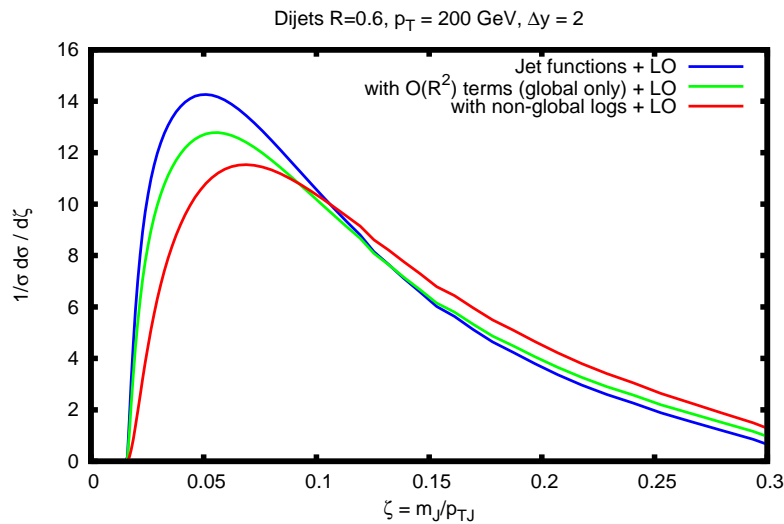}
\caption{The NLL+LO jet mass distribution for dijets, with different approximation to the resummed exponent.}\label{fig:dijets}
\end{center}
\end{figure}

We match the resummation to a LO calculation of the jet mass distribution obtained with {\sc nlojet}{\fs++}~\cite{Nagy:2003tz}, according to Eq~(\ref{matching}). In Fig.~\ref{fig:dijets} we plot our NLL+LO result with different approximation for the resummed exponent Eq.~(\ref{masterdijets}): jet-functions (blue), with the inclusion of $\Or(R^2)$ corrections (green) and non-global logarithms (red), as explained in detail for the Z+jet case. Although the corrections to the jet-function approximation are less pronounced than in the Z+jet case, they are still sizeable and must be taken into account. In our understanding, the perturbative part of the resummed result of Ref~\cite{Li:2011hy,Li:2012bw} is precisely the one captured by the jet-functions (plus an approximated treatment of the one-loop constant $C_1$).  

In order to obtain a more realistic prediction, one would need to integrate over the appropriate cuts in transverse momentum and rapidity and match to NLO, which, in principle, should not pose any difficulties. More delicate is instead the determination of the constant $C_1$, although this issue has been addressed in Ref.~\cite{Banfi:2004yd, Banfi:2010xy} for the case of global event-shapes.

\section{Conclusion}
\label{sec:conclusions}

In this chapter we have provided resummed expressions to NLL accuracy for jet-mass distributions in hadron-hadron collisions, for jets of arbitrary radius and defined in the \AKT algorithm. In particular, we have considered Z boson production in association with a jet and jet masses in dijet production. We have improved upon existing studies of jet masses and shapes (many of which use $e^{+}e^{-}$ annihilation as a model for jets produced in hadron collision processes) by incorporating initial state radiation effects as well as accounting for non-global logarithms in the large-$\Nc$ limit. We have matched our results to leading-order predictions to account for non-logarithmic terms to this order and have commented on the role of the coefficient function $C_1 \alpha_s$ (corrections of order $\alpha_s$ to the basic resummation off the Born configuration). Finally, we have compared our results to all the leading Monte Carlo event generators  {\sc pythia}, {\sc sherpa} and {\sc herwig}{\fs++}, with and without 
non-perturbative hadronisation corrections. 

We have found that: first, ISR and non-global logarithms play an important role even at relatively small values of jet radius such as $R=0.6$ and hence cannot be ignored while discussing inclusive jet shapes at hadron colliders. Our calculations for non-global logarithms both at fixed-order and at all orders represent to our knowledge the first attempt to calculate these terms beyond the simpler cases of $e^{+}e^{-}$ annihilation and DIS. Second, on comparing our results to event generators, widely used for phenomenology, we find that at the purely perturbative level the best agreement is with the {\sc pythia} shower, with an apparent small shift accounting for much of the difference between the analytical resummation and the parton shower estimate. The differences with  {\sc sherpa} and {\sc herwig}{\fs++} on the other hand are more marked especially towards smaller jet masses as one approaches the peak of the distributions. After hadronisation corrections are applied in the event generators and also 
applying a shift to the analytical resummation to account for hadronisation, we are able to obtain very good agreement with  {\sc pythia} and {\sc herwig}{\fs++} with slightly different shifts required in either case, for jet mass values to the right of the Sudakov peak of the distribution. For smaller jet masses we do not expect a simple shift of the analytical resummation to reproduce the correct result and we observe here discrepancy with all Monte Carlo generators which is to be expected. However, this region of very small jet masses of the order of a few GeV will not be of interest in LHC phenomenology in any case. 
We may thus expect that our results should be of direct phenomenological value even pending some of the improvements we intend to make in the near future. 

For the immediate future we aim to improve our results chiefly by taking proper account of the order $\alpha_s$ coefficient function $C_1$ and computing the various pieces of $C_1$ which originate from different regions and suppressing these by an appropriate form factor rather than the cruder treatment reported in the text, which was aimed at producing an uncertainty band associated to a lack of correct treatment of $C_1$. When this is done we will be in a position to carry out 
an NLO matching and estimate the uncertainty of our theoretical calculations accurately, which will be important in the context of phenomenology. 

We have not taken directly into account non-perturbative effects in this chapter, incorporating them as a simple shift of the perturbative spectra as for the case of global event shapes~\cite{Dasgupta:2003iq}. We can study non-perturbative corrections in more detail using the methods outlined in~\cite{Dasgupta:2007wa}, but for the moment we note that the Monte Carlo event generators we have studied, contain differences in their estimate of hadronisation which should be explored further along with studying the underlying event contributions. Since our predictions are valid for any value of jet radius $R$, one can hope that phenomenological studies selecting a small value of $R$ would help in better isolating the perturbative contributions, since the hadronisation and underlying event corrections for $m_J^2/p_T^2$ scale respectively as $R$ and $R^4$~\cite{Dasgupta:2007wa}. On the other hand moving to a larger $R$ after pinning down the perturbative content would help to more accurately constrain the non-
perturbative models in various approaches.

%% file: conc/conc.tex

\chapter{Conclusions}
\label{ch:conc}

Each chapter has its own conclusion section. In this chapter we tie together the most important points to be learned from the study presented in this thesis.

Due to the complex environment present at hadron colliders, such as the LHC, new physics discoveries are only feasible given a detailed understanding of the already established standard model, and QCD in particular as the strong interactions, governed by the latter, constitute the bulk of all possible interactions. At the LHC, the establishment of the latter understanding is best accessible via studies of jets, which are the final product of the aforementioned strong interactions and which populate the final state at the latter collider. High-$p_T$ jets, in particular, have some discernible features that make them ideal for perturbative analyses. Such analyses include exploration of the substructure of these jets, which is distinct from that of moderate- and low-$p_T$ jets and which would aid in discriminating new physics signals from QCD background.

Equipped with an infrared and collinear safe jet shape observable, angularities, we probed in Chapter \ref{ch:EEJetShapes1} the substructure of high-$p_T$ QCD jets in the simple process of dijet production in $\EE$ annihilation, where only one jet is measured while the other is left unmeasured. We identified large logarithms, non-global logs, that were missed out by previous studies and argued that in the limit of narrow well-separated jets their all--orders structure simplifies to that of the hemisphere jet mass. This is mainly due to the observation that non-global logs arise predominantly near the boundary of each jet and evolve independently of the other well-separated jets. Moreover, we commented on their phenomenological impact on the primary emission Sudakov form factor and put forward the ansatz that the above simplified picture of non-global structure holds even at hadron collider environments.

Given the fact that experimental measurements generally employ large values of the jet parameter $R$ in their jet finding, it proved necessary to extend the work of Chapter \ref{ch:EEJetShapes1} to include the effects of large jet radii. We have performed this very task in Chapter \ref{ch:EEJetShapes2}. Within the same context of dijet production in $\EE$ annihilation and taking the jet mass as an example we computed the full logarithmic structure of the latter shape distribution at $\Or(\as^2)$, i.e., up to $\as^2\,L$ (N$^3$LL accuracy in the expansion), while retaining the full $R$ dependence. The calculations were explicitly carried out for both \AKT and C/A jet algorithms and implicitly, through similarity to C/A, for the \KT algorithm, and represent one of the most accurate fixed-order calculations for non-global observables. Further, the NLL (in the exponent) resummation of Chapter \ref{ch:EEJetShapes1} has been extended to full jet radius, in the large-$\Nc$ limit, for both the \AKT and \KT algorithms.
 Very few non-global observables 
have been computed at fixed-order and/or resummed to all--orders to such an accuracy (In fact, the only observable the author is aware of is the energy flow in gaps-between-jets \cite{Delenda:2006nf, Dasgupta:2002bw, Banfi:2001aq}).

Based on our resummed results, which include both non-global and clustering logs, we concluded that it is recommended in jet phenomenological studies to use a variety of jet algorithms. Further, we found that ignoring non-global logs all together in events where the final state is clustered with a sequential recombination jet algorithm other than the \AKT induces an uncertainty that is well below that originated from the large-$\Nc$ approximation, for quite large jet radii. Our future directions regarding non-global observables include: extending the numerical program of Ref. \cite{Dasgupta:2001sh} to next-to-leading non-global logs and as a long-term project, going beyond the large-$\Nc$ approximation.

Following previous work on the energy flow in gaps-between-jets, we showed in Chapter \ref{ch:EEJetShapes3} that clustering logs occurring in jet mass calculations in \KT and C/A algorithms exponentiate. We explicitly computed the first few orders in the exponent with full $R$ dependence. Comparisons to the Monte Carlo program of \cite{Dasgupta:2001sh} revealed that the missing higher-order terms have negligible effect. Our findings settled the issue, raised by other research groups, of whether such logs, with highly non-trivial phase space, can exponentiate and hence resummed. The next obvious step is to extend the resummation to next-to-leading clustering logs.

Finally, we have exploited the wealth of techniques and experience gained in the work of the previous chapters to carry out, in the final chapter \ref{ch:HHJetShapes1}, a full NLL resummation of jet mass at hadron colliders for the \AKT jet algorithm. Explicitly studied were vector boson production in association with a single jet and dijet production. Parts of the resummed exponent, such as the soft wide-angle piece involves the full colour structure and thus exceeds the accuracy of standard Monte Carlo event generators (which work in the large-$\Nc$ approximation). Our calculation marks the first attempt in literature where non-global logs are computed at hadron colliders and with full $R$ dependence retained. Already with a simple shift in the exponent to account for hadronisation corrections and only leading-order matching, our results compared well to the output of the most widely used event generators, and even hinted noticeable differences between them.

Additionally, although in the last chapter \ref{ch:HHJetShapes1} we have treated a single variable, the jet mass, it is actually straightforward to extend our treatment to the entire range of angularities as for instance were explored for jets in $\EE$ annihilation in Chapter \ref{ch:EEJetShapes1} and Refs. \cite{Ellis:2009wj,Ellis:2010rwa}. The basic calculations we carried out here can be easily extended to include those variables as well as any variants of the jet mass itself such as studying the jet mass with an additional jet veto. Once we have accomplished an NLO matching we therefore intend to generalise our approach to accommodate a range of substructure variables in different hard processes, including final state clustering with other jet algorithms such as the k$_{\mathrm{T}}$, C/A and SISCone algorithms. Moreover, the {\sc boost}{\fs 11} review \cite{Altheimer:2012mn} recommended, in addition to the above two (vector boson + jet and dijets) processes, calculations for multijet events. We hope to 
perform this in the near future as 
well as writing a semi-analytical program for an automation of these calculations which will enable similar studies of other non-global jet shapes. We hope that our work will eventually lead to improved estimates of the accuracy and hence more confidence in a detailed understanding of jet substructure  than is the case presently, which could in turn be important for a variety of substructure applications at the LHC.
%

%% file: ch2/ch2app.tex

\renewcommand{\kt}[1]{k_{#1}}

\chapter{Colour algebra}
\label{app:ColorAlgebra}

\section{Fierz identities}
\label{ssec:QCD:FierzIdentities}

A matrix $\mb M$ in $\rm SU(\Nc)$ can be written as a linear combination of the basis matrices $\{v^a, a=1,\cdots,\Ncsq-1\}$. In the fundamental representation, $v^a = t^a$, this reads
\be
\mb M = m_0\, \mbb{1} + \sum_{a=1}^{\Ncsq-1} m_a\,t^a,
\label{eq:FI:LinearSeries} 
\ee
where $m_0$ and $m_a$ are real expansion coefficients. To determine the latter coefficients one can make use of the traceless property of $t^a$, \eq{eq:Color:Generatorsproperties}, together with \eq{eq:Color:GeneratorRepNormalisatn}. One can straightforwardly show that
\be
m_0 = \frac{1}{\Nc} \Tr\left(\mb M \right), \;\;\; m_a = 2\,\Tr\left(\mb M\,t^a \right).
\label{eq:FI:m0-ma}
\ee
If one takes, for example, the matrix $\mb M$ to be the commutator and anti--commutator\footnote{For two matrices $\mb A$ and $\mb B$, the anti--commutator is defined as $\{\mb A,\mb B\} = \mb A.\mb B + \mb B.\mb A$. Recall that the commutator is just the Lie product defined in \eq{eq:Color:LieProduct}.} of the basis matrices then one obtains, using \eqs{eq:FI:LinearSeries}{eq:FI:m0-ma},
\begin{subequations}
\begin{eqnarray}
 \imath f_{abc} &=& 2\,\Tr\left([t^a,t^b] t^c \right),
\label{eq:FI:M-examples-a}\\
 \{t^a,t^b\} &=& \frac{\delta^{ab}}{\Nc} \mbb{1} + d_{abc}\, t^c,\;\;\; d_{abc} = 2\,\Tr\left(\{t^a,t^b\} t^c \right).
\label{eq:FI:M-examples-b} 
\end{eqnarray}
\end{subequations}
It is worthwhile noting that \eq{eq:FI:M-examples-a} implies that the structure constants are totally antisymmetric in all three indices, $f_{abc} = -f_{acb}, f_{abc} = - f_{bac}$ etc. Moreover, \eq{eq:FI:M-examples-b} defines, in analogy to \eqref{eq:FI:M-examples-a}, the totally \emph{symmetric} constants $d_{abc}$, which form a symmetric adjoint representation of $\rm SU(\Nc)$\footnote{For $\rm SU(3)$, a system of three quarks (forming a baryon) can be in four representations. These are found using Young Tableaux techniques (see e.g., \cite{jones1998groups}): $3 \times 3 \times 3 = 1 \oplus 8 \oplus \br{8} \oplus 10$. $8$ and $\br{8}$ are respectively the symmetric and antisymmetric representations. Employing the same techniques one can also compute the dimension of a given representation.}. The corresponding casimir operator can be deduced from \eq{eq:FI:M-examples-b}. It reads \cite{dokshitzer1997}: $C_{As} = (\Ncsq-4)/\Nc$.

\begin{figure}[t]
 \centering
 \includegraphics[width=14cm]{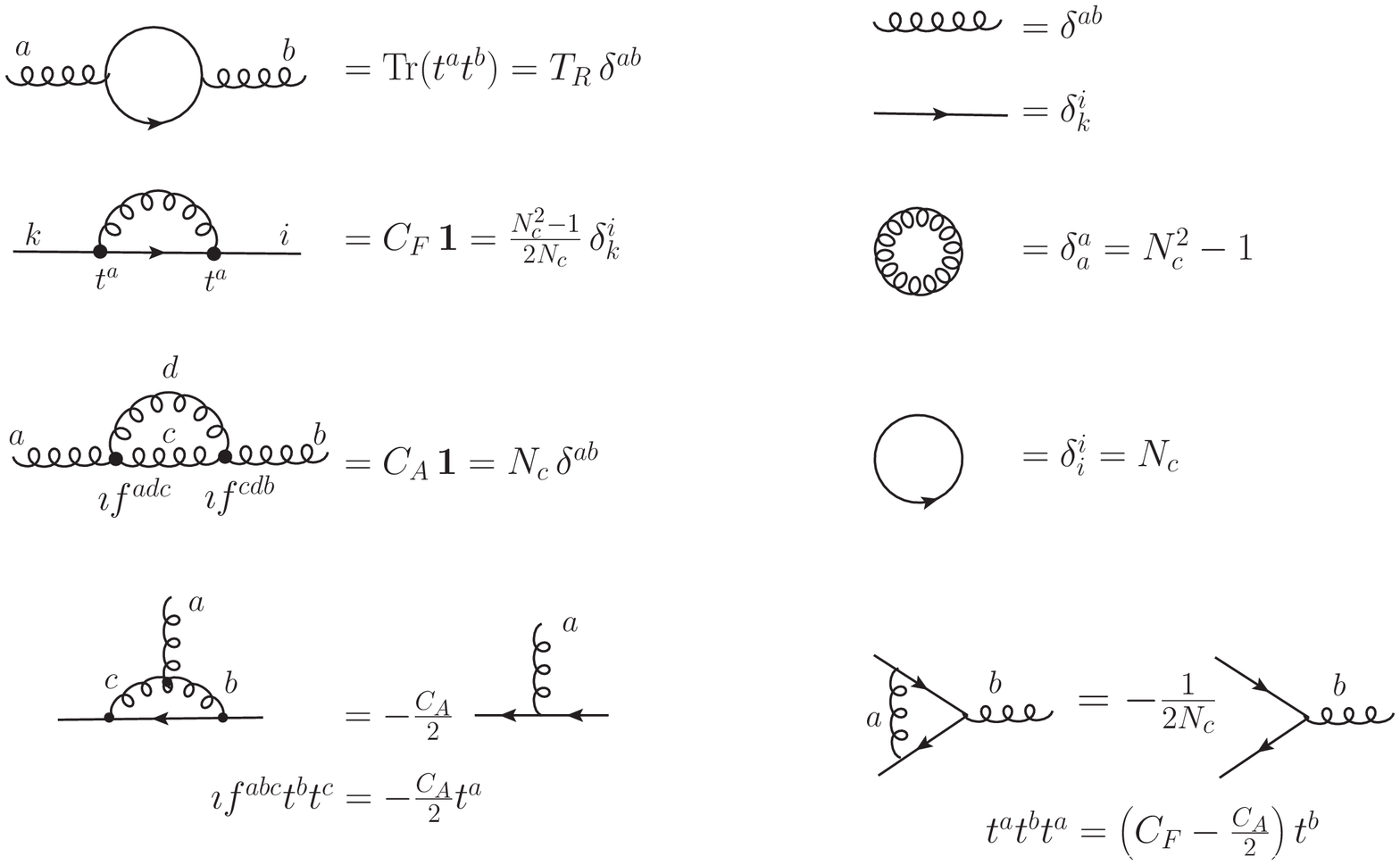}
 \caption{Graphical representation of some important colour relations.}
\label{fig:FI:ColourAlgebra}
\end{figure}

Adding up \eqs{eq:FI:M-examples-a}{eq:FI:M-examples-b} one obtains the following useful relation for the product of two generators, 
\be
t^a t^b = \frac{\imath}{2} f^{abc} t^c + \frac{1}{2} d^{abc} t^c + \frac{\de^{ab}}{2} \mbb 1.
\label{eq:FI:TaTbRelation}
\ee
Notice lastly that due to the fact that $t^e.\,t^e$ is a scalar (casimir) and $t^e$ are traceless, both $f_{abc}$ and $d_{abc}$ are traceless (easily seen from the corresponding expressions in \eqs{eq:FI:M-examples-a}{eq:FI:M-examples-b}). i.e., $f_{aac} = d_{aac} = 0$, etc. A number of relations that will be useful in later sections, and appendices, are depicted in \fig{fig:FI:ColourAlgebra}.

\begin{figure}[h]
 \centering
 \includegraphics[width=0.75 \textwidth]{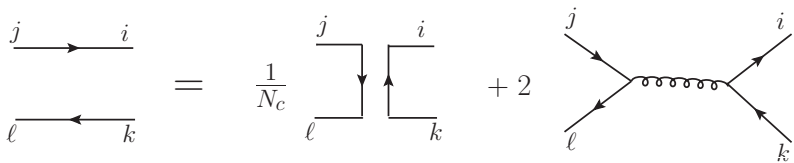}
 \caption{Graphical representation of the Fierz identity.}
\label{fig:FI:FierzID}
\end{figure}

Another interesting example \cite{dokshitzer1997} is when one chooses the matrix $\mb M$ to have only one non--zero component; $\mb M^i_k = \de^i_{(j)}\,\de^{(\ell)}_k$, where $j$ and $\ell$ are assumed to be fixed. Substituting into \eq{eq:FI:LinearSeries} yields
\be
\de^i_j\,\de^\ell_k = \frac{1}{\Nc}\, \de^i_k\,\de^\ell_j + 2\,\left(t^a\right)^i_k\,\left(t^a\right)^\ell_j. 
\label{eq:FI:FierzID}
\ee
The above relation is known as the Fierz identity and is shown pictorially in \fig{fig:FI:FierzID}.
If we rearrange \eq{eq:FI:FierzID} so that the colour matrices are written in terms of the delta functions then we can generalise the Fierz identity to the following useful formula \cite{Ross:HelicityMethod}
\be
L\,t^a\,Q\,t^a\,R = \half\!\Tr\left(Q\right) L\,R -\frac{1}{2 \Nc} L\,Q\,R,
\label{eq:FI:GeneralisedFierzID}
\ee
where $L, Q$ and $R$ are any colour matrices. In the special case where both $L$ and $R$ are unit matrices and $Q = t^b$ then
\be
t^a\,t^b\,t^a = -\frac{1}{2 \Nc}\,t^b,
\label{eq:FI:ID2-eg1}
\ee
or $Q = t^b\,t^c$ then
\be
t^a\,t^b\,t^c\,t^a = \frac{1}{4}\,\de^{bc} - \frac{1}{2 \Nc}\,t^b\,t^c.
\label{eq:FI:ID2-eg2}
\ee
Finally, the calculation of the cross section involves, as we shall see later in \sec{sec:QCD:Lagrangian}, products of traces of colour generators, with each colour summed over. It would thus prove useful to provide the following trace identity \cite{Ross:HelicityMethod}:
\be
\Tr\left(t^a\,Q \right) \Tr\left(t^a\,R \right) = \half\, \Tr\left(Q\,R \right) - \frac{1}{2 \Nc}\,\Tr\left(Q \right) \Tr\left(R \right).
\label{eq:FI:TraceOfGeneralisedFierzID}
\ee

\chapter{Perturbative calculations in $\EE$ annihilation}
\label{app:QCDReview}

In this appendix, we present detailed perturbative calculations of differential cross sections for one of the main two QCD processes discussed in this thesis, namely $\EE$ annihilation into hadrons. In addition to the classical Feynman diagrammatic approach of computing matrix elements, we consider the effective eikonal method. We start, in \sec{sec:app:QCD:EikonalApprox}, by laying out the general structure of the latter method, encountering in the way some of the main definitions and relations that will be essential to later calculations. We then compute the matrix element squared (and therefore the differential cross section) at Born level, in \ssec{ssec:app:EE:Born}. $\Or(\as)$ corrections stemming from both real emission and virtual loops are addressed in Subsecs. \ref{sssec:app:EE:RealGuonEmission} and \ref{sssec:app:EE:VirtualGluonCorrections}. Lastly we compute the more complicated $\Or\cbr{\as^2}$ real emission matrix element, taking into account hard emission contributions.

\section{Eikonal method}
\label{sec:app:QCD:EikonalApprox}

\begin{figure}[t]
 \centering
 \includegraphics[width=0.8\textwidth]{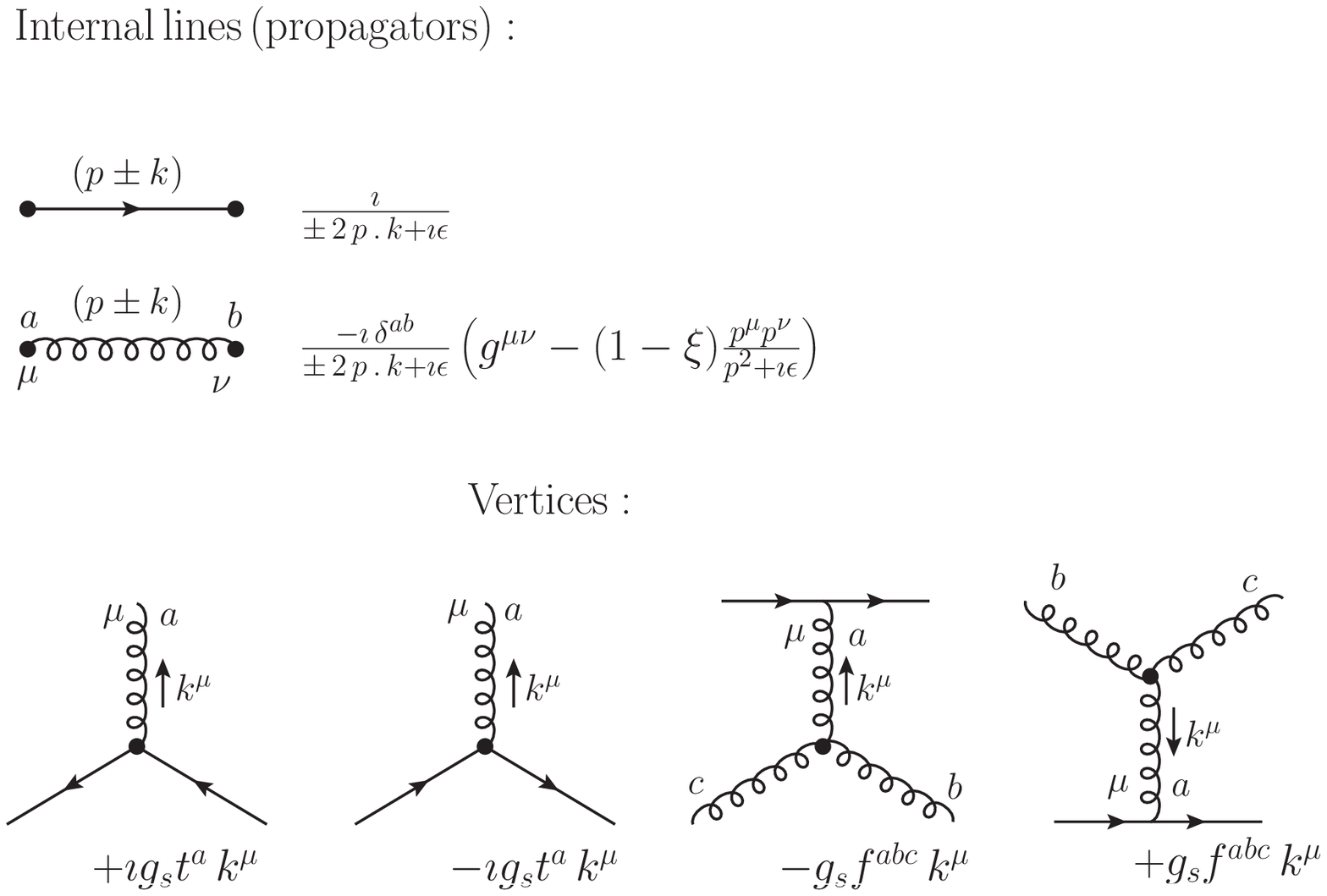}
 \caption{Eikonal Feynman rules for QCD: gluons are represented by curly lines and fermions by solid lines. For triple vertices, the colour is read clockwise.}
 \label{fig:app:Eik:EikonalFeynRules}
\end{figure}

The eikonal method \cite{PhysRev.186.1656}\footnote{The eikonal method was originally developed in the context of the non--relativistic theory of potential scattering \cite{1959lectures, PhysRev.103.443} (and references in \cite{PhysRev.186.1656}).} provides a reliable approximation to the full theory at high energies. The resultant all--orders eikonal matrix element may be written in a closed form. This method is based on the propagator approximation 
\be
 \frac{\imath}{\cbr{p \pm k}^2 +\imath\ep} \simeq \frac{\imath}{\pm 2\, p.k +\imath\ep},
\label{eq:app:Eik:FeynmanPropagator}
\ee
where $p$ is an external four--momentum, which is assumed to be on--shell ($p^2 = 0$)\footnote{The on--shell (or on mass shell) approximation will be used throughout unless stated otherwise. This approximation becomes almost exact when considering scattering at high energies, which are far larger than the masses of the light quarks (i.e., excluding the top quark).}, and $k$ may be a partial sum of internal four--momenta\footnote{In cases where $k$ is the four--momentum of a single parton, which is \emph{soft} relative to its emitter, this approximation may be referred to as the \emph{soft} approximation.}, thereby neglecting the $k^2$ term relative to $\inner{p}{k}$ ($k^2 \ll p.k$). The corresponding eikonal Feynman rules are shown in \fig{fig:app:Eik:EikonalFeynRules}. 

Consider the emission of a soft gluon $k=\cbr{\om, \vect{k}}$ off a system of $m$ (harder) partons, with momenta $p_i = \cbr{E_i, \vect{p}_i}, i=1,\cdots m$. One can write the corresponding emission amplitude as \cite{Catani:1996jh, Dokshitzer:2005ek, Catani:1996vz, dokshitzer1997}
\be
 \ket{m+1} = \gs \sum_{i=1}^{m} \frac{\inner{p_i}{\ep^{\ast}}}{\inner{p_i}{k}}\, \mb T^a_i \ket{m},
\label{eq:app:Eik:EikonalM+1State}
\ee
where $\ep$ is the polarisation vector of the emitted gluon and $\ket{m}$ represents the amplitude prior to emission and is a vector in the colour space. The set of ket vectors $\ket{\ell}$ is assumed to form an orthonormal basis, $\braket{\ell'}{\ell} = \de\cbr{\ell'-\ell}$, for the colour space. The operator $\mb T^a_i$ (generator of $\rm SU(3)$ group) maps the $m$--dimensional space onto the $(m+1)$--dimensional space due to the emission of one extra gluon off \emph{eikonal} leg $i$. The inverse map, i.e., $(m+1)$--dimensional space onto $m$--dimensional space, which corresponds to the absorption of a single gluon by an eikonal leg $i$ is represented by the hermitian conjugate operator $\cbr{\mb T^a_i}^\dagger$. We write the product of the two mapping operators as
\be
 \cbr{\mb T^a_i}^\dagger\, \mb T^a_j \equiv \inner{\mb T_i}{\mb T_j}.
\label{eq:app:Eik:EikMapsInnerProd}
\ee
In other words, $ \inner{\mb T_i}{\mb T_j}$ is symmetric under the interchange of $i$ and $j$. Notice that it is possible to reduce the product \eqref{eq:app:Eik:EikMapsInnerProd} into casimirs (scalars) when the dimension $m \leq 4$ (coloured partons). For $m > 4$ the product is matrix-valued. Colour conservation implies that $\ket{m}$ transforms as a colour singlet under $\rm SU(3)$ rotations. Schematically
\be
 \sum_{i=1}^m \mb T^a_i \ket{m} = 0.
\label{eq:app:Eik:ColourConservation}
\ee
The above relation will be of great help in simplifying the calculations of the colour factors associated with Feynman amplitudes.

If the $m$--parton amplitude $\ket{m}$ is represented by $\mc M_{i_1 i_2\cdots i_m}$, where the subscripts are the colour indices of the incoming or outgoing partons, then $\mb T^a$ is represented by (see \fig{fig:app:Eik:EikonalFeynRules}):
\begin{itemize}
 \item $+ t^a$ if the radiating parton is an outgoing quark (incoming antiquark).
 \item $- t^a$ if the radiating parton is an incoming quark (outgoing antiquark).
 \item $+(-)\,\imath\mb f^a$ if the radiating parton is an outgoing (incoming) gluon.
\end{itemize}
The operators $t^a$ and $-\imath \cbr{\mb f^a}_{bc} \equiv -\imath f^{abc}$ are, as before, the colour matrices in the fundamental and adjoint representations respectively. Computing the bra state $\bra{m+1}$ and multiplying out, one straightforwardly obtains for the braket, or equivalently the matrix element squared,
\be
 \abs{\mc M^2_{m+1}} = \braket{m+1}{m+1} = - \gs^2 \sum_{i<j =1}^{m} \frac{2\,\cinner{p_i}{p_j}}{ \cbr{\inner{p_i}{k}} \cbr{\inner{p_j}{k}} } \Tr\cbr{\bra{m} \inner{\mb T_i}{\mb T_j}\ket{m}},  
\label{eq:app:Eik:EikMESquared}
\ee
where the sum is over distinct $(ij)$ pairs and the factor $2$ is due to symmetry in $i \lra j$. Recall that we are working with on--shell partons. As such, all self--energy Feynman diagrams whereby $i=j$ vanish. Moreover, we have used the following sum over polarisations, in the Feynman gauge ($\xi = 1$):
\be
 \sum_{\ld = 0}^3 \cbr{-1}^{\de_{\ld 0}}\, \ep_\mu \cbr{\ld, \vect{k}}\, \ep_\nu^\ast\cbr{\ld, \vect{k}} = - g_{\mu\nu}.
\label{eq:app:Eik:SumGluonPolarisationVects}
\ee
We define the dipole \emph{antenna function}, $\om_{ij}(k)$, which corresponds to a soft gluon $k$ stretching between two eikonal (colour--connected) legs $i$ and $j$, as
\be
 \om_{ij}(k) = \frac{\om^2\,\cbr{\inner{p_i}{p_j}}}{ \cbr{\inner{p_i}{k}} \cbr{\inner{p_j}{k}}} = \frac{a_{ij}}{a_i\,a_j},
\label{eq:app:Eik:AntennaFun}
\ee
with $a_{ij} = 1-\cos\theta_{ij}$, $a_i = 1-\cos\theta_i$, $\theta_{ij}$ the angle between $(p_{i}, p_j)$ pair and $\theta_i$ the angle between $p_i$ and $k$.

In the eikonal approximation, the $(m+1)$--dimensional phase space factorises into an $m$--dimensional phase space factor for the radiating ensemble multiplying a one--dimensional phase space factor for the emitted gluon,
\be
 \d\Phi_{m+1} = \d\Phi_{m}\,\frac{\d^3 \vect{k}}{\cbr{2\pi}^3\,2 \om} 
\label{eq:app:Eik:EikM+1PhaseSpace}
\ee
Absorbing a factor of $4\pi^2$ into the definition of $\as/\pi$, we have
\be
 \frac{\d^3 \vect{k}}{4\pi\,\om}  = \frac{\om\d\om}{4\pi}\,\d\cos\theta\d\phi = \frac{\kt{t}\d\kt{t}}{4\pi}\d\eta\d\phi, 
\label{eq:app:Eik:GluonPhaseSpace}
\ee
where $(\theta, \phi)$ are the polar and azimuthal angles of the gluon with respect to (w.r.t.) a chosen axis. The (pseudo)rapidity $\eta$ and transverse momentum of the gluon $\kt{t}$ are related to its energy and angle, $\om$ and $\theta$, via
\be
 \eta = -\ln\tan\frac{\theta}{2},\;\;\; \kt{t} = \om\,\sin\theta.
\label{eq:app:Eik:Azimuth2Polar}
\ee

\section{pQCD: $\EE$ $\ra$ hadrons}
\label{sec:app:QCD:ee_example}

The purpose of this section is to provide a detailed derivation of the full differential cross section for the process $\EE$ annihilation into final state hadrons up to $\Or\cbr{\as^2}$ real radiative corrections. That is, we consider the the emission of up to two real gluons off the initial Born process $e^+e^- \ra q\, \br{q}$. We also consider the one--loop virtual correction in the eikonal limit. The latter is to demonstrate that for the accuracy sought in this thesis it is sufficient to simply take the virtual correction to be minus the real one. Detailed calculations of the total cross--section up to NNLO (two--loop corrections) can be found in, e.g., \cite{GehrmannDeRidder:2007ce, GehrmannDeRidder:2007hr}.

The perturbative expansion of the cross section for the said process, up to the $n^{th}$ order, may be cast in the form
\be
 \s_n^{\,e^+e^- \ra \rm hadrons} = \cSup{\s}{0} + \cSup{\s}{1} + \cSup{\s}{2} + \cdots + \cSup{\s}{n}. 
\label{eq:app:EE:TotalEE2HadronsXsec}
\ee
Below, we first present a general form for the $n$ gluon emission differential cross section, $\d\cSup{\s}{n}$, including the form of the $n$--body phase space. Using the resultant expressions, we then compute the Born term, $\cSup{\s}{0}$, the $\Or(\as)$ correction, $\cSup{\s}{1}$, including both real and virtual contributions and finally the $\Or\cbr{\as^2}$ real contribution, $\cSub{\s}{2}$. To simplify the calculations, we shall work at a centre--of--mass energy $s < m^2_Z$, so that contributions from the Z boson propagator can be neglected, and in the centre--of--mass (cm) frame\footnote{Recall that the whole problem is Lorentz invariant, and thus one can work at any chosen frame.}, such that the photon propagator's momentum is $q = \cbr{\sqs,\vect{0}}$.

\subsection{Generalities}
\label{ssec:app:Generalities}

The general form of the total cross section is given in \eq{eq:LQCD:2To2TotalCrossSection}. At order $\as^n$ in the perturbative expansion, \eq{eq:app:EE:TotalEE2HadronsXsec}, it reads
\be
\cSup{\s}{n} = \frac{1}{2 s}\int \abs{\br{\cSup{\mc M}{n}}}^2 \d\Pi_n.
\label{eq:app:EE:NthOrderIntegXsec}
\ee

\subsubsection{n$^{th}$ order matrix element squared}
\label{sssec:app:NthOrderMatrixElementSquared}

The invariant matrix element squared (averaged and summed over initial and final state quantum numbers) may be split into a leptonic part, $\EE \ra \g^\ast$, and a hadronic part, $\g^\ast \ra q\br{q} g\,\cdots$ :
\be
 \abs{\br{\cSup{\mc M}{n}}}^2 = \frac{1}{s^2}\, L^{\mu\nu}\,\cSup{H_{\mu\nu}}{n}.
\label{eq:app:EE:NthOrderMESquared}
\ee
The leptonic tensor, $L_{\mu\nu}$, is given by
\begin{eqnarray}
\nn L^{\mu\nu} &=& \frac{1}{4}\sum_{s_1,s_2}\,L^{\mu} \cbr{q_1,s_1;q_2.s_2}\, L^{\nu \dagger} \cbr{q_1,s_1;q_2,s_2},
\\ 
\nn &=&\frac{e^2}{4} \sum_{s_1,s_2} \sbr{\br{u}\cbr{s_1,q_1} \g^\mu v\cbr{s_2,q_2}} \sbr{\br{v}\cbr{s_2,q_2} \g^\nu u\cbr{s_1,q_1}},
\\
    &=& \frac{e^2}{4}\,\Tr\cbr{\slashed{q}{1}\, \g^\mu\, \slashed{q}{2}\, \g^\nu} = e^2 \cbr{ q_1^\mu q_2^\nu + q_1^\nu q_2^\mu - g^{\mu\nu} \inner{q_1}{q_2}}, 
 \label{eq:app:EE:LeptonicTensorBorn} 
\end{eqnarray}
where we have used the spinors completeness relations, for a four-momentum $p$,
\be
 \sum_{\ld = \pm 1} u_\ld (p) \br{u}_\ld (p) = \pslash,\;\;\;
 \sum_{\ld = \pm 1} v_\ld (p) \br{v}_\ld (p) = \pslash ,
\label{eq:app:Hel:SpinorCompletenessRelation}
\ee
along with the recursive relation for the trace of products of an even number, $N$, of Dirac matrices
\be
 \Tr\cbr{\g^{\mu_1} \g^{\mu_2} \cdots \g^{\mu_k}\cdots \g^{\mu_N}} = 4\,\sum_{k=2}^n \cbr{-1}^k g^{\mu_1\mu_k}\,\Tr\cbr{\g^{\mu_2} \cdots \g^{\mu_{k-1}} \g^{\mu_{k+1}} \cdots \g^{\mu_N}}.
\label{eq:app:EE:TraceOfDiracMatrices}
\ee
To simplify the calculations, particularly of $\cSup{H_{\mu\nu}}{n}$, it is instructive to employ the Ward identity \cite{peskin1995introduction}, which states that if a matrix element $\mc M(p) = \mc M^\mu(p)\,\ep_\mu(p)$ then upon the replacement $\ep_\mu \ra p_\mu$ the matrix element vanishes. This implies, in our case, that
\be
 q_\mu\,L^{\mu\nu}\cbr{\EE \ra \g^\ast} = 0,\;\;\; q^\nu\,H_{\mu\nu} \cbr{\g^\ast \ra q\br{q} g \cdots} = 0.
\label{eq:app:EE:WardIdentityAmpsAlphas1}
\ee
The hadronic tensor is thus proportional to the projector $\cbr{g^{\mu\nu} - q^\mu q^\nu/q^2}$\footnote{So is the leptonic tensor. However since one will be integrating over final state phase space then one only needs to extract the tensor structure of the final state term. i.e., the hadronic tensor.}. The matrix element squared \eqref{eq:app:EE:NthOrderMESquared} can then be written as
\begin{eqnarray}
\nn \abs{\br{\cSup{\mc M}{n}}}^2 &=& \frac{1}{s^2}\, L^{\mu\nu}\cbr{g_{\mu\nu} - \frac{q_\mu q_\nu}{q^2}} \cSup{\wt{H}}{n}, \\
\nn  &=& \frac{1}{s^2}\, g_{\mu\nu} L^{\mu\nu}\,\cSup{\wt{H}}{n},
\\
\nn  &=& \frac{1}{s^2}\, g_{\mu\nu} L^{\mu\nu}\, \frac{1}{3} g^{\ro\s} \cSup{H_{\ro\s}}{n},
\\
   &=& \frac{1}{3\,s^2}\, L\,\cSup{H}{n},
\label{eq:app:EE:MESAlphas1}
\end{eqnarray}
where we made use of \eq{eq:app:EE:WardIdentityAmpsAlphas1} to pass to the second line. We then multiplied both sides in the expansion of the hadronic tensor $\cSup{H_{\mu\nu}}{n}$ in terms of the aforementioned projector ($\cSup{H_{\mu\nu}}{n} = (g_{\mu\nu} - q_\mu q_\nu/q^2) \cSup{\wt{H}}{n}$) by the metric $g_{\mu\nu}$  and used the fact that $g_{\mu\nu} g^{\mu\nu} = 4$ (in four--dimensional spacetime) to pass to the third line. The last line in \eq{eq:app:EE:MESAlphas1} defines the scalars $L$ and $\cSup{H}{n}$:
\begin{eqnarray}
\nn L &=& g^{\nu\nu}\, L_{\mu\nu} = -e^2\,s, \label{eq:app:EE:LeptonicScalar}\\
 \cSup{H}{n} &=& g_{\mu\nu}\,\cSup{H_{\mu\nu}}{n}.
\label{eq:app:EE:NthOrderHadronicScalar}
\end{eqnarray}
Our task in the subsequent sections is simply to compute $\cSup{H}{n}$.

\subsubsection{n--body phase space}
\label{sssec:app:nBodyPhaseSpace}

The general form of the final state phase space factor for $n$ gluons emission is given in \eq{eq:LQCD:GeneralPhaseSpaceFactor}. There are two ways of simplifying the $n$--body form \eqref{eq:LQCD:GeneralPhaseSpaceFactor}. The first is to proceed by integrating out $\vect{p}_n$ then $\vect{p}_{n-1}$ and so on. The second method is to exclusively integrate out $\vect{p}_1$ and $\vect{p}_2$ leaving the rest intact. As we shall see below, this latter method results in an expression analogous to what would be found using the eikonal method of \sec{sec:app:QCD:EikonalApprox} (modulo a Jacobi factor). The first method is thoroughly presented in \cite{n-bodyPS}. We only explicitly derive the integrated two-- and three--body phase space factors for the on--shell case. The two--body phase space reads
\begin{eqnarray}
\nn \int \d\Pi_2 &=& \int \frac{\d^3 \vect{p}_1}{\cbr{2\pi}^3 2\,E_1} \frac{\d^3 \vect{p}_2}{\cbr{2\pi}^3 2\,E_2}\, \cbr{2\pi}^4 \cSup{\de}{4} \cbr{q - p_1 - p_2},
\\
\nn &=& \int \frac{\d^3 \vect{p}_1}{\cbr{2\pi}^3 2\,E_1}\,\cbr{2\pi}\, \frac{\de\cbr{\sqs - E_1 - E_2}}{2\,E_2}\Big|_{\vect{p}_1 = -\vect{p}_2} = \frac{1}{32\pi^2}\int\d^2\Om_1,\\
\label{eq:app:EE:TwoBodyPhaseSpaceFactorFinal}
\end{eqnarray}
Similarly the three--body phase space is given by
\begin{equation}
\int \d\Pi_3 = \int \frac{\d^3 \vect{p}_1}{\cbr{2\pi}^3 2\,E_1} \frac{\d^3 \vect{p}_2}{\cbr{2\pi}^3 2\,E_2} \frac{\d^3 \vect{p}_3}{\cbr{2\pi}^3 2\,E_3}\, \cbr{2\pi}^4 \cSup{\de}{4} \cbr{q - p_1 - p_2 - p_3}.
\label{eq:app:EE:ThreeBodyPhaseSpaceFactor_A}
\end{equation}
Integrating $\vect{p_3}$ out and using the fact that\footnote{This may be shown to hold by noting that $\cSup{\de}{4}(p^2) = \cSup{\de}{4}(p_0^2-\abs{\vect{p}}^2) = 1/2\abs{\vect{p}} \de(p_0 - \abs{\vect{p}})$ for $p_0 >0$.}
\begin{eqnarray}
\nn \frac{\de\cbr{\sqs - E_1 - E_2 - E_3}}{2 E_3} &=& \de\sbr{\cbr{q-p_1-p_2-p_3}^2} = \de\cbr{-s - 2\inner{p_1}{p_2} + 2\inner{q}{p_1} + 2\inner{q}{p_2}} 
\end{eqnarray}
\begin{eqnarray}
 &=& \frac{1}{s} \de\sbr{1 - x_1 -x_2 + x_1 x_2\,a_{12}/2},
 \label{eq:app:EE:ThreeBodyPhaseSpaceFactor_B}
\end{eqnarray}
where $a_{ij} = 1-\cos\theta_{ij}$ and the energy fraction $x_i$ is defined by
\be
 x_i = \frac{2\inner{q}{p_i}}{s} = \frac{2\,E_i}{\sqs}.
\label{eq:app:EE:EnergyFractionDef}
\ee
We have
\be
 \int\d\Pi_3 = \frac{s}{16(2\pi)^5} \int x_1\d x_1 x_2\d x_2 \d^2\Om_1\d^2\Om_2\, \de\sbr{1 - x_1 -x_2 + x_1 x_2\,a_{12}/2}
\label{eq:app:EE:ThreeBodyPhaseSpaceFactor_C}
\ee
Writing $\d^2\Om_2 = \d a_{12}\d\phi_{12}$, averaging over the latter azimuthal angle and simplifying yields
\be
 \int\d\Pi_3 = \frac{s}{16(2\pi)^5}\int \d x_1 \d x_2 \d^2\Om_1.
\label{eq:app:EE:ThreeBodyPhaseSpace}
\ee

The $n$--body phase space \eqref{eq:LQCD:GeneralPhaseSpaceFactor}, in the second method mentioned above, in $\EE$ annihilation is given by the general formula \cite{Dokshitzer1992675}
\be
\int\d\Pi_n = \frac{s^{n-2}}{2^{2n+2}\pi^3} \int \prod_{i=3}^n\, x_i\d x_i \frac{\d^2\Om_i}{2\pi}\, \cSup{J}{n},
\label{eq:app:EE:NthPhaseSpaceFactor2}
\ee
where $\cSup{J}{n}$ is the Jacobian resulting from integration over $p_1$ and $p_2$,
\be
\cSup{J}{n} = 2\int\d x_1 \d^3\vect{p}_2\, \frac{x_1}{x_2}\,\de\cbr{\sum_{i=1}^n x_i - 2} \de\cbr{\sum_{i=1}^n \vect{p}_i}.
\label{eq:app:EE:NthPhaseSpaceJacobian} 
\ee
Let us consider the cases $n=3$ and $n=4$. Integrating out $\vect{p}_2$ and using \eq{eq:app:EE:EnergyFractionDef}, together with the fact that $E_2 = \abs{\vect{p}_2}$, the $x_i$--$\de$ function corresponds, for $n=3$ and $n=4$ respectively, to
\begin{eqnarray}
\nn x_1 &=& \frac{2\cbr{1-x_3}}{2-x_3 a_{13}} \qquad (n=3),\\
x_1 &=& \frac{2\cbr{1-x_3-x_4} + x_3 x_4 a_{34}}{2 - x_3 a_{13} - x_4 a_{14}} \qquad (n=4),
\label{eq:app:EE:x2Tox1Andx3}
\end{eqnarray}
where $a_{ij} = 1-\cos\theta_{ij}$. Carrying out the remaining straightforward $\de$ integrals one obtains
\begin{eqnarray}
\nn \cSup{J}{3} &=& \frac{4\cbr{1-x_3}}{(2 - x_3 a_{13})^2 },
\\
\cSup{J}{4} &=& \frac{2\cbr{1-x_3-x_4} + 2 x_3 x_4 a_{34}}{(2 - x_3 a_{13} - x_4 a_{14})^2}.
\label{eq:app:EE:JacobianForn=3Andn=4}
\end{eqnarray}
%

\subsection{Born cross-section}
\label{ssec:app:EE:Born}

\begin{figure}[t]
 \centering
 \includegraphics[width=0.8\textwidth]{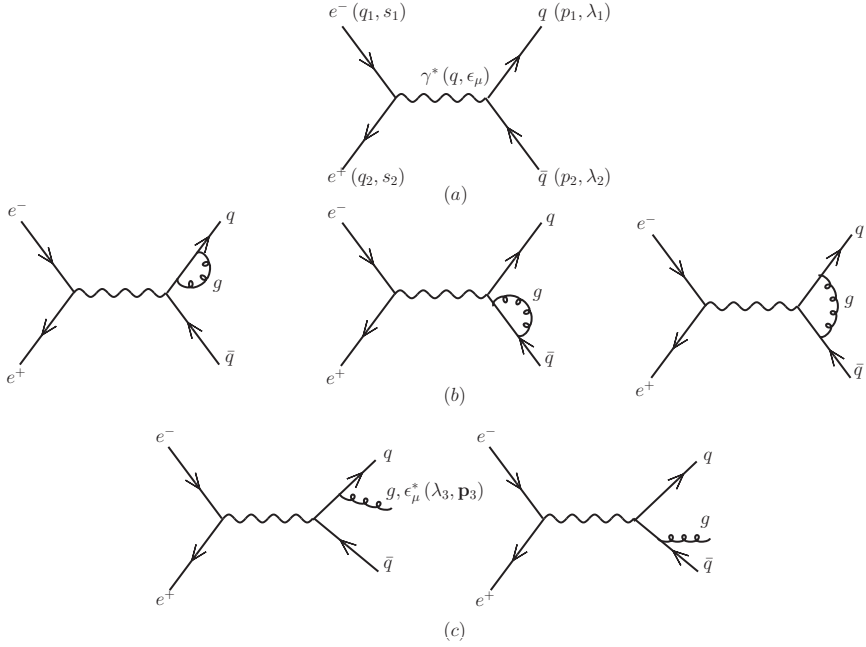}
 \caption{Feynman diagrams for $(a)\, \EE \ra q\,\br{q}$, Born amplitude, $(b)\, \EE \ra q\br{q}$, virtual corrections, and $(c)\, \EE \ra q\bar{q}g$, real emission.}
 \label{fig:app:EE:EEqqbar}
\end{figure}

Consider the Feynman diagram, depicted in \fig{fig:app:EE:EEqqbar}$(a)$, representing the Born amplitude.
The hadronic tensor $\cSup{H_{\mu\nu}}{0}$, given in \eq{eq:app:EE:NthOrderMESquared} for $n=0$, reads
\begin{eqnarray}
\nn \cSup{H_{\mu\nu}}{0} &=& \sum_{i,j} \sum_{\ld_1,\ld_2} \cSup{H_\mu}{0}\cbr{p_1,\ld_1, i;p_2,\ld_2, j}\, H_{\nu}^{(0) \dagger}\cbr{p_1,\ld_1, i;p_2,\ld_2, j},
\\
  &=& \sum_{f=1}^{\nf}\,e^2 e^2_{q_f} \de_{ij} \de_{ji} \sbr{\br{u}\cbr{\ld_1,p_1} \g^\mu v\cbr{\ld_2,p_2}} \sbr{\br{v}\cbr{\ld_2,p_2} \g^\nu u\cbr{\ld_1,p_1}}
\\ 
   &=& \sum_{f=1}^{\nf}\,4\,e^2 e^2_{q_f} \cbr{ p_{1\mu} p_{2\nu} + p_{1\nu} p_{2\mu} - g_{\mu\nu} \inner{p_1}{p_2}}\, \de_{ii},
\label{eq:app:EE:HadronicTensorBorn}
\end{eqnarray}
 where the quark--photon--antiquark vertex is associated with a factor $\imath e e_{q_f} \g^\mu \de_{ij}$, with $i, j$ being the colour indices of the quark and antiquark. We note that $\de_{ij} \de^{ji} = \de^i_i = \Nc = 3$ for $\rm SU(3)$. The corresponding hadronic scalar $\cSup{H}{0}$, defined in \eq{eq:app:EE:NthOrderHadronicScalar}, is
\be
\cSup{H}{0} = -\Nc\,\sum_{f=1}^{\nf}\,4\,e^2 e^2_{q_f}\, 2\,s.
\label{eq:app:EE:HadronicScalarBorn}
\ee
Substituting \eqs{eq:app:EE:LeptonicScalar}{eq:app:EE:HadronicScalarBorn} back into \eq{eq:app:EE:NthOrderMESquared}, which in turn is substituted together with the phase space factor \eqref{eq:app:EE:TwoBodyPhaseSpaceFactorFinal} into \eq{eq:app:EE:NthOrderIntegXsec} for $n=0$, yields the total $\EE \ra q\br{q}$ cross section:
\be
 \cSup{\s}{0} = \frac{4\pi\A^2}{3 s}\, \Nc\,\sum_{f=1}^{\nf} e_{q_f}^2,
\label{eq:app:EE:TotalIntegxsecBorn}
\ee
where $\A = e^2/4\pi$ is the QED coupling constant. In the above equation we have integrated out the solid angle $\int\d^2\Om_1 = 4\pi$.

\subsection{$\Or(\as)$ corrections}
\label{ssec:app:EE:AlphasCorrections}

We now consider the first order correction to the Born process. The radiative corrections at this order consist of two types of diagrams, shown in \fig{fig:app:EE:EEqqbar}$(b)$ and $(c)$: loop and tree level diagrams. The latter corresponds to real emissions of a gluon, leading to a three--body final state, and the former to virtual corrections to the self energies of the quark--antiquark pair as well as to the interaction vertex, keeping the final state identical to the Born case, i.e., a two--body final state. We first compute the contribution of the real emission diagrams.

\subsubsection{Real gluon emission}
\label{sssec:app:EE:RealGuonEmission}

\begin{figure}[t]
 \centering
 \includegraphics[width=0.5\textwidth]{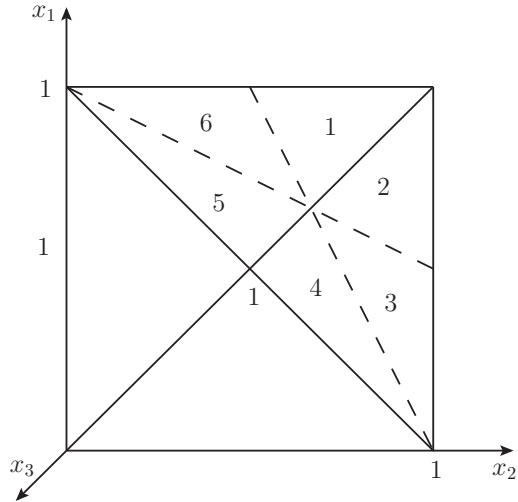}
 \caption{Dalitz plot for a three--body ``decay'' of a photon. All partons are assumed on--shell. The edges $x_i=1$ correspond to two partons being collinear and the corners $x_i = 0$ correspond to one parton being soft ($p_i \ra 0$). The different kinematical regions are: $(1)\, x_1>x_2>x_3, (2)\, x_2>x_1>x_3, (3)\,x_2>x_3>x_1, (4)\,x_3>x_2>x_1, (5)\,x_3>x_1>x_2$ and $(6)\, x_1>x_3>x_2$.}
 \label{fig:app:EE:DalitzPlot}
\end{figure}

The $\Or(\as)$ real integrated cross section is of the form \eqref{eq:app:EE:NthOrderIntegXsec} with $n=1$. The three--body phase space is given in \eq{eq:app:EE:ThreeBodyPhaseSpace}. The corresponding $\Or\cbr{\as}$ matrix element squared satisfies \eq{eq:app:EE:NthOrderMESquared} with the hadronic scalar $\cSup{H}{1}$ given by:
\begin{eqnarray}
\nn \cSup{H}{1} = g^{\mu\nu}\cSup{H_{\mu\nu}}{1} &=& g^{\mu\nu} \cbr{H_{1\mu}^{(1)\dagger} + H_{2\mu}^{(1)\dagger} } \cbr{H_{1\nu}^{(1)} + H_{2\nu}^{(1)} },\\
 &=& \cSup{H_{11}}{1} + H_{22}^{(1)} + \cSup{H_{12}}{1},
\label{eq:app:EE:HadronicTensorAlphas1}
\end{eqnarray}
where $H_{11}, H_{22}$ are the ``self--energy'' diagrams and $H_{12} = H_1^\dagger H_2 + H_2^{\dagger} H_1$ contains the ``interference'' diagrams. Applying Feynman rules (\fig{fig:LQCD:FeynmanRulesQCD}) on \fig{fig:app:EE:EEqqbar}$(c)$ we obtain
\begin{eqnarray}
\nn \cSup{H_{1\mu}}{1} &=& \imath e e_{q_f}\de_{i'j}\,\gs \mb t^a_{ii'}\sbr{\br{u}(p_1)\,\g^\ro\,\frac{\Slashed{p^+}{1}{3}}{ \cbr{p^+_{13}}^2}\;\g_\mu\,v(p_2)\, \ep^\ast_\ro},
\\
 H_{2\mu}^{(1)} &=& -\imath e e_{q_f}\de_{ij'}\,\gs \mb t^a_{jj'} \sbr{ \br{u}(p_1)\, \g_\mu\, \frac{\Slashed{p^+}{2}{3} }{ \cbr{p^+_{23}}^2}\,\g^\s\, v(p_2)\,\ep_\s^\ast },
\label{eq:app:EE:AmpAlpha1}
\end{eqnarray}
where $p^\pm_{ij} \equiv p_i \pm p_j$. In order to compute the hermitian conjugate we note that: $\cbr{\g_\mu}^\dagger = \g_\mu$ and $\cbr{\br{u} \g_\mu v}^\dagger = \br{v} \g_\mu u$.  The first term in the right hand side of \eqref{eq:app:EE:HadronicTensorAlphas1} reads
\begin{eqnarray}
\nn \cSup{H_{11}}{1} &=& -\sum_f e^2\,e_{q_f}^2\,\gs^2 \Tr\cbr{\mb t^a_{ij} \mb t^a_{ji}}\,\frac{8\,\inner{p_2}{p_3}}{\inner{p_1}{p_3}},\\
 &=& -\sum_f e^2\,e_{q_f}^2\,\Nc\,\gs^2\,\CF\,\frac{8 \cbr{1-x_1}}{\cbr{1-x_2}}.
\label{eq:app:EE:H11AmpAlphas1}
\end{eqnarray}
Note that the $(-)$ sign is due to summation over gluon's polarisations, \eq{eq:app:Eik:SumGluonPolarisationVects}. Similarly, the second and third terms are
\be
 \cSup{H_{22}}{1} = -\sum_f e^2\,e_{q_f}^2\,\Nc\,\gs^2\,\CF\,\frac{8 \cbr{1-x_2}}{\cbr{1-x_1}},
\label{eq:app:EE:H22AmpAlphas1}
\ee
and
\be
 \cSup{H_{12}}{1} = -\sum_f e^2\,e_{q_f}^2\,\Nc\,\gs^2\,\CF\,\frac{16 \cbr{1-x_3}}{\cbr{1-x_1} \cbr{1-x_2} }.
\label{eq:app:EE:H12AmpAlphas1}
\ee
Summing up the three terms, i.e., substituting Eqs. \eqref{eq:app:EE:H11AmpAlphas1}, \eqref{eq:app:EE:H22AmpAlphas1} and \eqref{eq:app:EE:H12AmpAlphas1} back into \eqref{eq:app:EE:HadronicTensorAlphas1}, yields
\be
 \cSup{H}{1} = -\sum_f e^2\,e_{q_f}^2\,\Nc\,\gs^2\,\CF\,\frac{8 \cbr{x_1^2 + x_2^2}}{\cbr{1-x_1} \cbr{1 - x_2} }.
\label{eq:app:EE:HadronicTensorAlpha1ContractedFinal}
\ee
Therefore, upon the substitution of \eqs{eq:app:EE:ThreeBodyPhaseSpace} {eq:app:EE:HadronicTensorAlpha1ContractedFinal} into \eq{eq:app:EE:NthOrderIntegXsec} (for $n=1$) one finally obtains for the $\Or\cbr{\as}$ differential cross section
\be
 \frac{\d^2\cSup{\s}{1}}{\d x_1\d x_2} = \cSup{\s}{0}\,\, \frac{\as}{2\pi}\,\CF\,\frac{x_1^2 + x_2^2}{\cbr{1-x_1} \cbr{1-x_2} }.
\label{eq:app:EE:IntegXsecAlphas1}
\ee
Examining the above equation one immediately observes that the differential cross section $\d\cSup{\s}{1}$ diverges whenever $x_1$ or $x_2$ approaches one. From \eq{eq:app:EE:EnergyFractionDef}, we have
\be
 \frac{s}{2}\cbr{1- x_k} = E_i\,E_j\cbr{1 - \cos\theta_{ij}}.
\label{eq:app:EE:EnergyFractionsRelations2}
\ee
The limit $x_k \ra 1$ corresponds to partons $p_i$ and $p_j$ being soft ($E_i\,\mathrm{or}\, E_j \ll 1$) and/or collinear ($\theta_{ij} \ll 1$). The former limit is termed soft (or infrared) divergence while the latter is referred to as the collinear (or mass) divergence. We show in \fig{fig:app:EE:DalitzPlot} a ``Dalitz'' plot of the various soft and collinear limits for the three parton final state.
These soft and collinear divergences are typical in QCD amplitude calculations (and QED as well). Since experimentally the answer, i.e., the value of the total cross section, is finite one ought to ``regularise'' the infinities that arise in the theoretical calculations. The final analytical answer should be independent of the ``regularisation procedure'', though. An example of a regularisation scheme would be the massive gluon scheme \cite{Basham:1977iq, Basham:1978bw}, whereby the gluon is given a fictitious mass $m_g$. Such a term would simultaneously prevent both infrared and mass singularities. Another much powerful scheme is the dimensional regularisation scheme \cite{'tHooft:1972fi}, whereby one works in dimension $d = 4-\vep$ and only at the end one takes the limit $\vep \ra 0$. The reader is referred to \cite{field1995applications} for more details.

It is possible to construct experimental observables that are insensitive to both infrared and collinear divergences. The distribution of such infrared and collinear (IRC) safe observables, which will be discussed in \chap{ch:Jets}, $\d\s/\d v$ produces, upon phase space integration, logarithmic enhanced terms of the form $\as^n \ln^m\cbr{v}$, where $v$ is a member of the class of IRC safe observables. Such terms break the convergence of the perturbative expansion, in terms of the coupling $\as$, and call for a ``resummation''.

As a final note, let us consider what would \eq{eq:app:EE:IntegXsecAlphas1} reduce to in the eikonal approximation. The latter approximation corresponds to neglecting the terms $\inner{p_1}{p_3}$ and $\inner{p_2}{p_3}$ relative to $\inner{p_1}{p_2}$ in the formulae of $\cSup{H_{11}}{1}, \cSup{H_{22}}{1}$ and $\cSup{H_{12}}{1}$. In other words, $\cSup{H}{1} \simeq \cSup{H_{12}}{1}$. Consequently the differential cross section \eqref{eq:app:EE:IntegXsecAlphas1} becomes
\begin{eqnarray}
\nn \s_{\mr r}^{(1),\mr{eik}} &=& \cSup{\s}{0}\,\astpi\,\CF\,\int \d x_1\d x_2\, \frac{2\cbr{1-x_3}}{\cbr{1-x_1} \cbr{1-x_2} },
\\
\nn &=& \frac{s}{2}\int \d x_1 \ x_2\, \d \cSup{\s}{0}\, \astpi\,\CF\,\frac{2\cbr{\inner{p_1}{p_2}}}{ \cbr{\inner{p_1}{p_3}} \cbr{\inner{p_2}{p_3}} },
\\
 &=& \cSup{\s}{0}\, \gs^2\,\CF\, \int \frac{\d^3\vect{p_3}}{\cbr{2\pi}^3\,2\,E_3}\,\frac{2\cbr{\inner{p_1}{p_2}}}{ \cbr{\inner{p_1}{p_3}} \cbr{\inner{p_2}{p_3}} }.
\label{eq:app:EE:IntegXsecAlphas1EikLimit}
\end{eqnarray}
The last equality can straightforwardly be verified using the eikonal techniques of \sec{sec:app:QCD:EikonalApprox}.

\subsubsection{$\Or(\as)$ virtual gluon corrections}
\label{sssec:app:EE:VirtualGluonCorrections}

\begin{figure}[h]
 \centering
 \includegraphics[width=8cm]{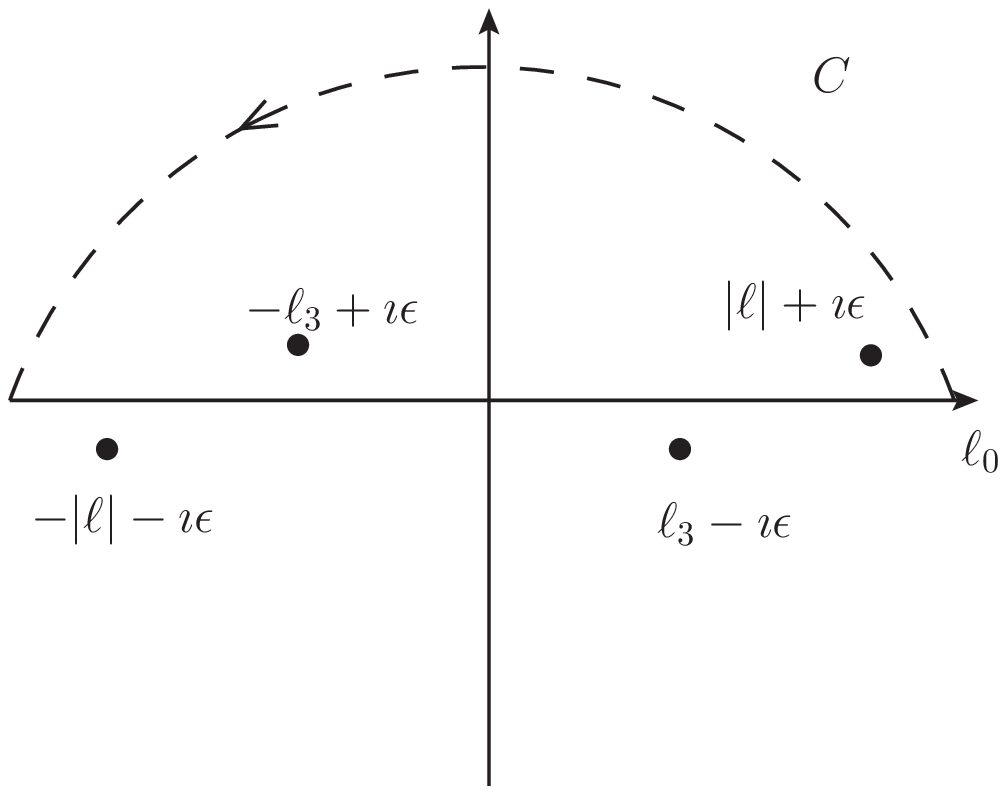}
 \caption{Poles of the integrand of \eq{eq:app:EE:VirtAmpAlphas1D} in the complex plane.}
 \label{fig:app:EE:VirtAmpPoles}
\end{figure}

The $\Or(\as)$ virtual corrections are depicted in \fig{fig:app:EE:EEqqbar}$(b)$ and correspond to a two--body final state. The associated differential cross section is again of the form
\be
 \d\cSup{\s_{\rm{vir}}}{1} = \frac{1}{2s}\, \abs{\br{\cSup{\mc M_{\rm{vir}}}{1}} }^2\, \d\Pi_2,
\label{eq:app:EE:VirDiffXsec}
\ee
where the two--body phase space factor has been computed in \eq{eq:app:EE:TwoBodyPhaseSpaceFactorFinal} and the matrix element squared is given by an expression analogous to \eqref{eq:app:EE:MESAlphas1}, with the real hadronic tensor replaced by the virtual one
\be
 \cSup{H_{\mr{vir}}}{1} = 2\,g^{\mu\nu}\,\Re\sbr{ H_\mu^{(0)\dagger}\cbr{ \cSup{H_{1\mr{vir},\nu}}{1} + \cSup{H_{2\mr{vir},\nu}}{1} + \cSup{H_{3\mr{vir},\nu}}{1}} }.
\label{eq:app:EE:HadronicTensorAlpha1ContractedVirt}
\ee
The pre-factor $2$ comes from the fact that the full $\Or(\as)$ Feynman amplitude is $\mc M_1 = \cSup{\mc M}{0} + \cSup{\mc M_{\mr r}}{1} + \cSup{\mc M_{\mr{vir}}}{1}$, which when squared is proportional to $2\,\Re \sbr{\mc M^{(0)\dagger} \cSup{\mc M_\mr{vir}}{1}} \equiv \abs{\br{\cSup{\mc M_\mr{vir}}{1}} }$. The amplitude $H_\mu^{(0)}$ is given in \eq{eq:app:EE:HadronicTensorBorn} and (see \fig{fig:app:EE:EEqqbar}$(b)$)
\begin{eqnarray}
\nn \cSup{H_{1\mr{vir},\mu}}{1} &=& -\imath e e_{q_f}\, \gs^2\,\CF\,\de_{ij}\, \br{u}(p_1) \sbr{ \int\frac{\d^4 \ell}{\cbr{2\pi}^4}\,\frac{-\imath}{\ell^2 + \imath\ep}\, \g_\ro \frac{ \slashed{p}{1} - \lslash}{\cbr{p_1 -\ell}^2 + \imath \ep}  \,\g^\ro} \frac{\slashed{p}{1}}{p_1^2 + \imath\ep}\,\g_\mu\, v(p_2),
\\
\nn\cSup{H_{2\mr{vir},\mu}}{1} &=& -\imath e e_{q_f}\,\gs^2\,\CF\,\de_{ij}\, \br{u}(p_1)\,\g_\mu\,\frac{\slashed{p}{2}}{p_2^2 + \imath\ep} \sbr{\int \frac{\d^4 \ell}{\cbr{2\pi}^4}\,\frac{-\imath}{\ell^2 + \imath\ep} \,\g_\ro\,\frac{ \slashed{p}{2} - \lslash}{ \cbr{p_2-\ell}^2 + \imath\ep}\,\g^\ro } v(p_2),
\\
\nn \cSup{H_{3\mr{vir},\mu}}{1} &=& \imath e e_{q_f}\,\gs^2\,\CF\,\de_{ij}\, \br{u}(p_1) \sbr{ \int\frac{\d^4 \ell}{\cbr{2\pi}^4}\,\frac{-\imath}{\ell^2 + \imath\ep} \,\g_\ro\, \frac{\slashed{p}{1} + \lslash}{\cbr{p_1 + \ell}^2 + \imath\ep}\,\g_\mu\,\frac{\slashed{p}{2} - \lslash}{\cbr{p_2 - \ell}^2 +\imath\ep}\,\g^\ro }\\
\nn && \times v(p_2).\\
\label{eq:app:EE:VirtAmpAlphas1}
\end{eqnarray}
Substituting \eq{eq:app:EE:HadronicTensorBorn} and the terms in \eq{eq:app:EE:VirtAmpAlphas1} into \eq{eq:app:EE:HadronicTensorAlpha1ContractedVirt} and simplifying the traces one finds, for the scalars $ \cSup{H_{i,\mr{vir}}}{1} = g^{\mu\nu}\,H_\mu^{(0)\dagger}\,\cSup{H_{i\mr{vir},\nu}}{1}$,
\begin{eqnarray}
\nn\cSup{H_{1,\mr{vir}}}{1} &=& -e^2 e^2_{q_f}\,\Nc\,\gs^2\,\CF\,\int\frac{\d^4 \ell}{\cbr{2\pi}^4}\,\frac{-\imath}{\ell^2 +\imath\ep}\, \frac{-16\,s}{p_1^2 + \imath\ep}\,\frac{\inner{p_1}{\ell}}{\cbr{p_1 - \ell}^2 + \imath\ep},
\\
\nn\cSup{H_{2,\mr{vir}}}{1} &=& \cSup{H_{1,\mr{vir}}}{1} \big|_{1 \ra 2},  
\\
\nn\cSup{H_{3,\mr{vir}}}{1} &=& e^2 e^2_{q_f}\,\Nc\,\gs^2\,\CF\,\int\frac{\d^4 \ell}{\cbr{2\pi}^4}\,\frac{-\imath}{\ell^2 +\imath\ep}\,\frac{-32\sbr{\inner{p_1}{\cbr{p_2-\ell}} } \sbr{\inner{p_2}{\cbr{p_1+ \ell}}} }{\sbr{(p_1+ \ell)^2 +\imath\ep} \sbr{(p_2- \ell)^2 + \imath\ep} }.\\
\label{eq:app:EE:VirtAmpAlphas1B}
\end{eqnarray}
\label{Thm:app:EE:BlockNordsieck}
The Block--Nordsieck theorem \cite{Bloch193754} states that the logarithmically enhanced (divergent) contributions due to real emission of collinear and soft gluons cancel against the corresponding virtual corrections. This can be explicitly checked for our example, $\EE \ra q \br{q} g$, using one of the aforementioned regularisation schemes along with Feynman parametrisation techniques. We refer the reader to \cite{field1995applications} for full details. Since the singular logarithmic contributions can be fully captured in the eikonal approximation, it is sufficient to verify the cancellation using the latter method. To this end, the self--energy terms in \eq{eq:app:EE:VirtAmpAlphas1B} vanish while the vertex term reduces to
\be
 \cSup{H_{3,\mr{vir}}}{1} = e^2 e^2_{q_f}\,\Nc\,\gs^2\,\CF\, \int\frac{\d^4 \ell}{\cbr{2\pi}^4}\,\frac{-\imath}{\ell^2 +\imath\ep}\, \frac{2\,s^2}{\cbr{\inner{p_1}{\ell} + \imath\ep} \cbr{\inner{p_2}{\ell} - \imath\ep} },
\label{eq:app:EE:VirtAmpAlphas1C}
\ee
where we have used $\inner{p_1}{p_2} = s/2$ (from momentum conservation $\cSup{\de}{4}\cbr{q - p_1 -p_2}$). We can simplify the integral in \eq{eq:app:EE:VirtAmpAlphas1C} using contour integration. We first specialise to the cm frame, $\vect{p_1} + \vect{p_2} = \vect{0}$, and write
\be
 \cSup{H_{3,\mr{vir}}}{1} \propto \int\frac{\d^3\vect{\ell}}{\cbr{2\pi}^3} \int\frac{\d \ell_0}{2\pi}\,\frac{-\imath}{\cbr{\ell_0 - \abs{\vect{\ell}} -\imath\ep} \cbr{\ell_0 + \abs{\vect{\ell}} +\imath\ep} } \frac{8\,s}{\cbr{\ell_0-\ell_3 +\imath\ep} \cbr{\ell_0 + \ell_3 - \imath\ep} }.
\label{eq:app:EE:VirtAmpAlphas1D}
\ee
The above integral has four simple poles in the $\ell_0$ complex plane (\fig{fig:app:EE:VirtAmpPoles}).
Closing the contour in the upper half, thus picking up only two poles, and applying the residue theorem\footnote{For more details on the residue theorem and complex integration techniques, see for instance \cite{ahlfors1966complex}.} one finds
\be
\cSup{H_{3,\mr{vir}}}{1} \propto 8\,s\,\int\frac{\d^3\vect{\ell}}{\cbr{2\pi}^3}\,\frac{1}{2 \abs{\vect{\ell_\perp}}^2}\cbr{\frac{1}{\abs{\vect{\ell}} } - \frac{1}{\ell_3 -\imath\ep} },
\label{eq:app:EE:VirtAmpAlphas1E}
\ee
where the invariant transverse momentum $\ell_\perp$ with respect to an $(ij)$ dipole cm frame is related to the antenna function, \eq{eq:app:Eik:AntennaFun}, through
\be
 \frac{2}{\abs{\vect{\ell_\perp}}^2} = \frac{\cbr{\inner{p_i}{p_j}}}{\cbr{\inner{p_i}{\ell}} \cbr{\inner{p_j}{\ell}} } = \frac{\om_{ij}(\ell)}{E^2_\ell},
\label{eq:app:EE:InvariantTranMoment}
\ee
with $E_\ell$ being the energy of the gluon $\ell$. Substituting \eq{eq:app:EE:VirtAmpAlphas1E} into \eq{eq:app:EE:HadronicTensorAlpha1ContractedVirt}, then plug into \eq{eq:app:EE:VirDiffXsec} results in the following expression for the (integrated) eikonal virtual cross section
\be
\s_{\mr{vir}}^{(1),\mr{eik}} = -\cSup{\s}{0}\,\gs^2\,\CF\,\Re\left\{\int\frac{\d^3\vect{\ell}}{\cbr{2\pi}^2 2\abs{\vect{\ell}}} \frac{2\cinner{p_1}{p_2}}{\cinner{p_1}{\ell} \cinner{p_2}{\ell}} \sbr{1 - \frac{\abs{\vect{\ell}}}{\ell_3 -\imath\ep}}\right\}.
\label{eq:app:EE:IntegXsecAlphas1VirEikLimit}
\ee
The term involving the $z$--component of the gluon momentum, $\ell_3$, in the square bracket of \eq{eq:app:EE:IntegXsecAlphas1VirEikLimit} can be simplified further. The important point is to explicitly keep the $\imath\ep$ prescription. That is, from \eq{eq:app:EE:VirtAmpAlphas1E} one has
\begin{eqnarray}
\nn \int\frac{\d^3\vect{\ell}}{\cbr{2\pi}^3 2\abs{\ell_\perp}^2}\, \frac{1}{\ell_3 - \imath\ep} &=& \int\frac{\d^2 \ell_\perp}{\cbr{2\pi}^3 2\abs{\ell_\perp}^2} \int_{-\infty}^{+\infty}\d\ell_3\, \frac{1}{\ell_3 - \imath\ep},
\\
\nn &=& \int\frac{\d^2 \ell_\perp}{\cbr{2\pi}^3 2\abs{\ell_\perp}^2}\,\lim_{a\ra +\infty} \int_{-a}^a \d\ell_3\,\frac{\ell_3 + \imath\ep}{\ell_3^2 + \ep^2},
\\
&=&  \int\frac{\d^2 \ell_\perp}{\cbr{2\pi}^3 2\abs{\ell_\perp}^2}\, \imath\pi.
\label{eq:app:EE:iPiTermA}
\end{eqnarray}
Thus the latter term is purely imaginary. It is worthwhile noting that the $\imath\pi$ factor is related to super--leading logarithms and the possibility of factorisation breaking \cite{Collins:2007jp, Collins:2007nk, Forshaw:2006fk, Forshaw:2008cq, Catani:2011qz}. The final form of the virtual cross section reads
\be
\s_{\mr{vir}}^{(1),\mr{eik}} = -\cSup{\s}{0}\,\gs^2\,\CF\int\frac{\d^3\vect{\ell}}{\cbr{2\pi}^2 2\abs{\vect{\ell}}} \frac{2\cinner{p_1}{p_2}}{\cinner{p_1}{\ell} \cinner{p_2}{\ell}} = -\s_{\mr r}^{(1),\mr{eik}}.
\label{eq:app:EE:IntegXsecAlphas1VirEikLimitReal}
\ee

\subsection{$\Or(\as^2)$ corrections}
\label{ssec:app:EE:AlphasSquareCorrections}

\begin{figure}[t]
  \centering
  \includegraphics[width=0.75\textwidth]{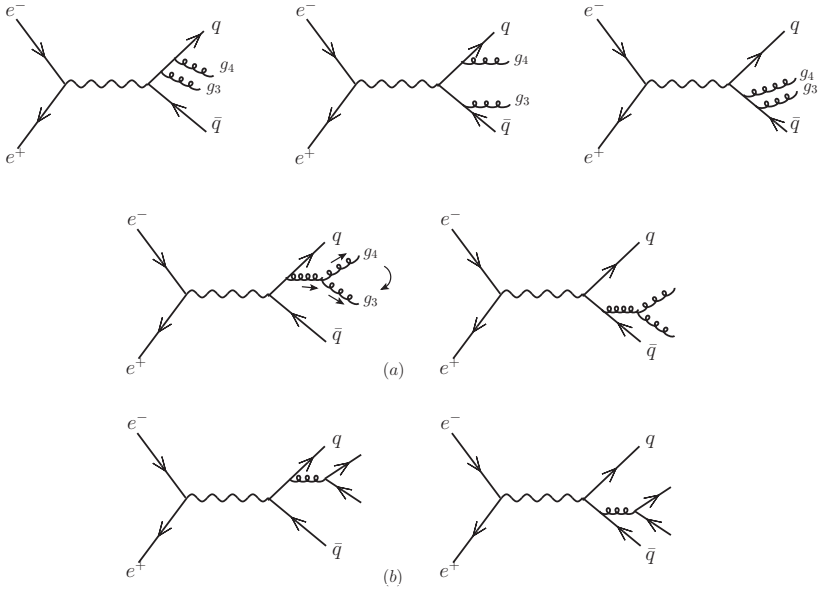}
  \caption{Feynman diagrams for $(a)\, \EE \ra q\br{q} g g$ and $(b)\, \EE \ra q \bar{q} q' \br{q}'$. There are three other diagrams corresponding to swapping the secondary gluons $g_3, g_4$ in the top three diagrams.}
  \label{fig:app:EE:EEqqbarAlphas2}
\end{figure}

There are two processes contributing to the amplitude at this order, as depicted in \fig{fig:app:EE:EEqqbarAlphas2}: $(a)\, \EE \ra q(p_1) \br{q}(p_2) g(p_3) g(p_4)$ and $(b)\, \EE \ra q(p_1)\bar{q}(p_2) q'(p_3) \br{q}'(p_4)$. The integrated cross section assumes the general form \eqref{eq:app:EE:NthOrderIntegXsec} ($n=2$) with the four--body phase space factor given in \eqs{eq:app:EE:NthPhaseSpaceFactor2}{eq:app:EE:JacobianForn=3Andn=4}. We begin by addressing the process $(a)$.
The hadronic scalar $\cSup{H}{2}$ is
\begin{eqnarray}
\nn \cSup{H}{2}_a &=& g^{\mu\nu} \sbr{\sum_{i=1}^8 \cSup{H_{a,i\mu}}{2}}^\dagger \sbr{\sum_{j=1}^8 \cSup{H_{a,j\nu}}{2}},
\\
 &=& \sum_{i,j=1}^8 \cSup{H_{a, ij}}{2},
\label{eq:app:EE:HadronicScalarAlphas2A}
\end{eqnarray}
where, starting from top left in \fig{fig:app:EE:EEqqbarAlphas2},
\begin{eqnarray}
\nn \cSup{H_{a,1\mu}}{2} &=& \imath ee_{q_f} \de_{i_1i_2}\,\gs^2\, \mb t^{a_4}\mb t^{a_3} \sbr{ \br{u}(p_1)\,\g_\A\,\frac{\Slashed{p^+}{1}{4} }{\cbr{p_{14}^+}^2}\,\g_\B\,  \frac{\SLashed{p^{++}}{1}{3}{4} }{\cbr{p^{++}_{134}}^2 }\, \g_\mu\,v(p_2) }\ep^{\ast\A}_4 \ep^{\ast\B}_3,
\\
\nn \cSup{H_{a,2\mu}}{2} &=& -\imath ee_{q_f} \de_{i_1i_2}\,\gs^2\, \mb t^{a_4}\mb t^{a_3} \sbr{\br{u}(p_1)\,\g_\A\, \frac{\Slashed{p^+}{1}{4}}{\cbr{p^+_{14}}^2 }\, \g_\mu\, \frac{\Slashed{p^+}{2}{3}}{\cbr{p^+_{23}}^2}\,\g_\B\,v(p_2) }\ep_4^{\ast\A}\ep_3^{\ast\B},
\\
\nn \cSup{H_{a,3\mu}}{2} &=& \imath ee_{q_f} \de_{i_1i_2}\,\gs^2\, \mb t^{a_4}\mb t^{a_3} \sbr{\br{u}(p_1)\,\g_\mu\,\frac{\SLashed{p^{++}}{2}{3}{4}}{\cbr{p^{++}_{234}}^2 }\, \g_\A\, \frac{\Slashed{p^+}{2}{3}}{\cbr{p^+_{23}}^2 }\,\g_\B\,v(p_2) }\ep^{\ast\A}_4 \ep^{\ast\B}_3,
\\
\nn \cSup{H_{a,4\mu}}{2} &=& -\imath ee_{q_f} \de_{i_1i_2}\,\gs^2\, \imath f^{a_4 a_3 a} \mb t^a \sbr{\br{u}(p_1)\,\g_\A\,\frac{\SLashed{p^{++}}{1}{3}{4}}{\cbr{p^{++}_{134}}^2 }\,\g_\mu\, v(p_2) } \ep^{\ast\s}_4 \ep^{\ast\ro}_3 \times
\\
\nn & & \times\;\; \frac{g^{\A\B}}{\cbr{p^+_{34}}^2} \sbr{g_{\B\s}\cbr{p^+_{34}+p_4}_\ro + g_{\s\ro}\cbr{p_3 - p_4}_\B - g_{\ro\B}\cbr{p^+_{34} + p_3}_\s },
\end{eqnarray}
\begin{eqnarray}
\nn \cSup{H_{a,5\mu}}{2} &=& \imath ee_{q_f} \de_{i_1i_2}\,\gs^2\, \imath f^{a_4 a_3 a} \mb t^a \sbr{\br{u}(p_1)\,\g_\mu\,\frac{\SLashed{p^{++}}{2}{3}{4}}{\cbr{p^{++}_{234}}^2 }\,\g_\A\, v(p_2) } \ep^{\ast\s}_4 \ep^{\ast\ro}_3 \times
\\
\nn & & \times\;\; \frac{g^{\A\B}}{\cbr{p^+_{34}}^2} \sbr{g_{\B\s}\cbr{p^+_{34}+p_4}_\ro + g_{\s\ro}\cbr{p_3 - p_4}_\B - g_{\ro\B}\cbr{p^+_{34} + p_3}_\s },
\\
\nn \cSup{H_{a,6\mu}}{2} &=& \cSup{H_{a,1\mu}}{2} \cbr{ 3 \lra 4 },
\\
\nn \cSup{H_{a,7\mu}}{2} &=& \cSup{H_{a,2\mu}}{2} \cbr{ 3 \lra 4 },
\\
    \cSup{H_{a,8\mu}}{2} &=& \cSup{H_{a,3\mu}}{2} \cbr{ 3 \lra 4 },
\label{eq:app:EE:EE2qqggFeynAmps}
\end{eqnarray}
where, in analogy to $p_{ij}^\pm$, we have defined $p_{ijk}^{\pm\pm} = p_i \pm p_j \pm p_k$ and used Feynman gauge ($\xi = 1$ in \fig{fig:LQCD:FeynmanRulesQCD}) for the gluon propagator. Let us further define $p^+_{34} \equiv p$. In the \emph{soft} limit\footnote{Unless stated otherwise, this limit is sufficient for the accuracy of most of the calculations performed in this thesis.}: $\inner{p_1}{p_2} \gg \inner{p_{1(2)}}{p_i}$ for $i=3,4$ (or equivalently $\pslash_1^2, \pslash_2^2 \gg \pslash_3^2, \pslash_4^2$), \eq{eq:app:EE:EE2qqggFeynAmps} can be greatly simplified. Employing the Dirac algebra relation, for massless spinors, 
\be
\br{u}(q)\,\g^\A\,\qslash = \br{u}(q)\, 2\,q^\A, 
\label{eq:app:EE:}
\ee
 and grouping together amplitudes with the same colour factor we write \eq{eq:app:EE:EE2qqggFeynAmps} as
\begin{eqnarray}
\nn \cSup{\mc H_{a,1\mu}}{2} &=& \cSup{H_\mu}{0}\, \gs^2 \mb t^{a_4} \mb t^{a_3} \sbr{ \frac{p_1^\A}{\inner{p_1}{p_4}} \frac{p_1^\B}{\inner{p_1}{p}} - \frac{p_1^\A}{\inner{p_1}{p_4}} \frac{p_2^\B}{\inner{p_2}{p_3}} + \frac{p_2^\A}{\inner{p_2}{p}} \frac{p_2^\B}{\inner{p_2}{p_3}} } \ep^\ast_{3,\B} \ep^\ast_{4,\A},
\\
\nn \cSup{\mc H_{a,2\mu}}{2} &=& \cSup{H_\mu}{0}\, \gs^2 \mb t^{a_3} \mb t^{a_4} \sbr{ \frac{p_1^\B}{\inner{p_1}{p_3}} \frac{p_1^\A}{\inner{p_1}{p}} - \frac{p_1^\B}{\inner{p_1}{p_3}} \frac{p_2^\A}{\inner{p_2}{p_4}} + \frac{p_2^\B}{\inner{p_2}{p}} \frac{p_2^\A}{\inner{p_2}{p_4}} } \ep^\ast_{3,\B} \ep^\ast_{4,\A},
\end{eqnarray}
\begin{eqnarray}
\nn \cSup{\mc H_{a,3\mu}}{2} &=& - \cSup{H_\mu}{0}\, \gs^2\,\imath f^{a_4 a_3 a} \mb t^a \cbr{\frac{p_1^\ld}{\inner{p_1}{p}} - \frac{p_2^\ld}{\inner{p_2}{p}} } \, \ep^{\ast\B}_{3} \ep^{\ast\A}_{4} \times \\
\nn & &  \times \frac{1}{p^2} \Big[g_{\ld\A}\cbr{p+p_4}_\B
+ g_{\A\B}\cbr{p_3 - p_4}_\ld  - g_{\B\ld}\cbr{p + p_3}_\A \Big],
\\
\nn &=& \cSup{H_\mu}{0}\, \gs^2\,\imath f^{a_3 a_4 a} \mb t^a\, \sbr{\frac{ \cinner{p_1}{p_3} \cinner{p_2}{p_4} - \cinner{p_1}{p_4}\cinner{p_2}{p_3} }{\cinner{p_3}{p_4} \cinner{p_1}{p} \cinner{p_2}{p}}\,g_{\A\B} + \cdots} \ep^{\ast\B}_{3} \ep^{\ast\A}_{4},\\
\label{eq:app:EE:EE2qqggFeynAmpsB}
\end{eqnarray}
where the $\cdots$ represent the following two terms
\begin{multline}
 \frac{1}{\inner{p_1}{p}} \sbr{p_{1\A}\cbr{p+p_4}_\B - p_{1\B}\cbr{p+p_3}_\A } - \frac{1}{\inner{p_2}{p}} \sbr{p_{2\A}\cbr{p+p_4}_\B - p_{2\B}\cbr{p+p_3}_\A }.
\label{eq:app:EE:EE2qqggFeynAmpsC}
\end{multline}
The above terms are linear in four--momenta, $p_{1\A(\B)}, p_{2\A(\B)}$, and will hence be ignored, as their contributions will be subleading.
 
We are now in a position to compute the \emph{soft} hadronic scalar $\cSup{H_a}{2}$ given in \eq{eq:app:EE:HadronicScalarAlphas2A}. Before proceeding we note the following colour products (see \sec{sec:QCD:Colour}):
\begin{eqnarray}
\nn \mb t^{a_3} \mb t^{a_4} \mb t^{a_4} \mb t^{a_3} &=& \CFsq,\\
\nn \mb t^{a_4} \mb t^{a_3} \mb t^{a_4} \mb t^{a_3} &=& \CFsq - \half\CF\CA,\\
\mb t^{a_3} \mb t^{a_4}\, \imath f^{a_3 a_4 a} \mb t^a &=& -\half\CF\CA.
\label{eq:app:EE:EE2qqggColorFactors}
\end{eqnarray}
We thus write $\cSup{H_a}{2}$ as a sum of \emph{abelian} and \emph{non--abelian} contributions:
\be
\frac{\cSup{H_a}{2}}{\gs^4} = \CFsq \sbr{ \cSup{\wt{\mc H}_{a,11}}{2} + 2\cSup{\wt{\mc H}_{a,12}}{2} + \cSup{\wt{\mc H}_{a,22}}{2} } +  \CF\CA \sbr{ \cSup{\wt{\mc H}_{a,33}}{2} + \cSup{\wt{\mc H}_{a,23}}{2} - \cSup{\wt{\mc H}_{a,13}}{2} - \cSup{\wt{\mc H}_{a,12}}{2} }.
\label{eq:app:EE:HadronicScalarAlphas2B}
\ee
where the colour--stripped hadronic scalars $\cSup{\wt{\mc H}_{a,ij}}{2}$ are:
\begin{eqnarray}
\nn \cSup{\wt{\mc H}_{a,11}}{2} &=& \cSup{H}{0}\,\gs^4\times \frac{2\,\cinner{p_1}{p_2}^2}{\cinner{p_1}{p} \cinner{p_2}{p} \cinner{p_1}{p_4} \cinner{p_2}{p_3}},
\\
\nn \cSup{\wt{\mc H}_{a,22}}{2} &=& \cSup{\wt{\mc H}_{a,11}}{2} \cbr{ 3 \lra 4},
\\
\nn \cSup{\wt{\mc H}_{a,12}}{2} &=& \cSup{H}{0}\,\gs^4 \Bigg[\cSup{\wt{\mc H}_{a,11}}{2}\,\frac{\cinner{p_2}{p_3}}{2\cinner{p_2}{p_4}} +  \cSup{\wt{\mc H}_{a,11}}{2}\,\frac{\cinner{p_1}{p_4}}{2\cinner{p_1}{p_3}} + \\
\nn && \hspace{4.3cm} + \frac{\cinner{p_1}{p_2}^2}{\cinner{p_1}{p_3}\cinner{p_2}{p_3} \cinner{p_1}{p_4}\cinner{p_2}{p_4} } \Bigg],
\label{eq:app:EE:HadronicScalarAlphas2C}\\
\end{eqnarray}
and
\begin{eqnarray}
\nn \cSup{\wt{\mc H}_{a,33}}{2} &=& \cSup{H}{0}\,\gs^4\times 4\,\sbr{\frac{  \cinner{p_1}{p_3}\cinner{p_2}{p_4} -  \cinner{p_1}{p_4} \cinner{p_2}{p_3} }{\cinner{p_3}{p_4} \cinner{p_1}{p} \cinner{p_2}{p}}}^2 +\cdots,
\end{eqnarray}
\begin{eqnarray}
\nn \cSup{\wt{\mc H}_{a,13}}{2} &=& -\cSup{H}{0}\,\gs^4 \frac{\cinner{p_1}{p_2}}{\cinner{p_1}{p_4} \cinner{p_2}{p_3}}\sbr{\frac{\cinner{p_1}{p_3}\cinner{p_2}{p_4}-  \cinner{p_1}{p_4} \cinner{p_2}{p_3} }{\cinner{p_3}{p_4} \cinner{p_1}{p} \cinner{p_2}{p}}},
\\
\nn \cSup{\wt{\mc H}_{a,23}}{2} &=& -\cSup{H}{0}\,\gs^4 \frac{\cinner{p_1}{p_2}}{\cinner{p_1}{p_3} \cinner{p_2}{p_4}}\sbr{\frac{ \cinner{p_1}{p_3}\cinner{p_2}{p_4} - \cinner{p_1}{p_4} \cinner{p_2}{p_3}}{\cinner{p_3}{p_4} \cinner{p_1}{p} \cinner{p_2}{p}}}.\\
\label{eq:app:EE:HadronicScalarAlphas2D}
\end{eqnarray}
Note that a $(-)$ sign has been absorbed into the Born hadronic tensor $\cSup{H}{0}$.  Substituting \eqs{eq:app:EE:HadronicScalarAlphas2C}{eq:app:EE:HadronicScalarAlphas2D} into \eq{eq:app:EE:HadronicScalarAlphas2B}, which is in turn substituted back into \eq{eq:app:EE:NthOrderMESquared}, for $n=2$ and process $a$, yields
\be
\abs{\br{\cSup{\mc M_a}{2}}}^2 = \abs{\br{\cSup{\mc M}{0}}}^2\,\gs^4 \sbr{4\,\CFsq\WP + \CF\CA \cbr{2\,S + \Hg}},
\label{eq:app:EE:MESAlphas2_qqbgg}
\ee
where the abelian (primary emission) amplitude is
\be
\WP = \frac{\cinner{p_1}{p_2}}{\cinner{p_1}{p_3} \cinner{p_2}{p_3}}\,\frac{\cinner{p_1}{p_2}}{\cinner{p_1}{p_4} \cinner{p_2}{p_4}} = \frac{\om_{12}(p_3) \,\om_{12}(p_4)}{E_3^2 E_4^2},
\label{eq:app:EE:AbelianAmpAlphas2}
\ee
and writing $\cinner{p_i}{p_j} = \cbr{ij}$ the non--abelian (secondary emission) amplitude is\footnote{The trick to simplify the non--abelian part of \eq{eq:app:EE:HadronicScalarAlphas2B} to the form presented in \eq{eq:app:EE:MESAlphas2_qqbgg} is to add and subtract a factor of $2$ in the square brackets of \eqs{eq:app:EE:HadronicScalarAlphas2C}{eq:app:EE:HadronicScalarAlphas2D}. } the sum of the soft, $S$, and gluon hard, $\Hg$, terms:
\begin{eqnarray}
\nn S &=& \frac{(12)}{(13)(34)(42)} + \frac{(12)}{(14)(43)(32)} - \frac{(12)}{(13)(32)}  \frac{(12)}{(14)(42)}, \\
 &=& \frac{1}{E_3^2 E_4^2}\, \om_{12}(p_3)\sbr{\om_{13}(p_4) + \om_{23}(p_4) - \om_{12}(p_4)},
\label{eq:app:EE:SoftNonAbelianAmpAlphas2}
\end{eqnarray}
and
\be
\Hg = 2\,\mc I^2 - S\,\mc J - 4 \frac{\om_{12}(p)}{(34) E_p^2},
\label{eq:app:EE:HardNonAbelianAmpAlphas2g}
\ee
where
\begin{eqnarray}
\nn \mc I &=& \frac{(13)(24) - (14)(23)}{(34)(1p)(2p)},\\
\mc J &=& \frac{(13)(24) + (14)(23)}{(1p)(2p)}.
\label{eq:app:EE:HardNonAbelianAmpAlphas2JandR}
\end{eqnarray}
%
%
%

The other process at $\Or(\as^2)$, as stated above, is represented by the Feynman diagrams in \fig{fig:app:EE:EEqqbarAlphas2}$(b)$. The corresponding hadronic scalar reads
\begin{eqnarray}
\cSup{H_b}{2} &=& g^{\mu\nu}\sbr{ \cSup{H_{b,1\mu}}{2} + \cSup{H_{b,2\mu}}{2} }^\dagger \sbr{ \cSup{H_{b,1\nu}}{2} + \cSup{H_{b,2\nu}}{2} },
\label{eq:app:EE:HadronicScalarAlphas2A_b}
\end{eqnarray}
where
\begin{eqnarray}
\nn \cSup{H_{b,1\mu}}{2} &=& \imath e e_{q_f} \de_{i'_1 i_2}\,\gs^2 \mb t^a_{i_1 i'_1} \mb t^a_{j_3 j_4} \sbr{\br{u}(p_1)\,\g_\A\,\frac{\SLashed{p^{++}}{1}{3}{4}}{\cbr{p^{++}_{134}}^2}\,\g_\mu\,v(p_2)} \frac{1}{\cbr{p^{+}_{34}}^2} \sbr{\br{u}(p_4)\,\g^\A\,v(p_3)},
\\
\nn \cSup{H_{b,2\mu}}{2} &=& -\imath e e_{q_f} \de_{i_1i'_2}\, \gs^2 \mb t^a_{i'_2 i_2} \mb t^a_{j_3 j_4} \sbr{\br{u}(p_1)\,\g_\mu\,\frac{\SLashed{p^{++}}{2}{3}{4}}{\cbr{p^{++}_{234}}^2}\,\g_\A\,v(p_2)} \frac{1}{\cbr{p^{+}_{34}}^2} \sbr{\br{u}(p_4)\,\g^\A\,v(p_3)}.\\
\label{eq:app:EE:HadronicScalarAlphas2B_b}
\end{eqnarray}
Similar to process $(a)$, \eq{eq:app:EE:HadronicScalarAlphas2B_b} simplifies significantly in the \emph{soft} limit. Adding up the two contributions (and using the above notation $p = p^+_{34}$) we have
\begin{eqnarray}
 \cSup{\mc H_{b,\mu}}{2} &=& \cSup{H_\mu}{0}\, \gs^2\, \mb t^a_{i_1 i_2} \mb t^a_{j_3 j_4} \cbr{ \frac{p_1^\A}{\cinner{p_1}{p}} - \frac{p_2^\A}{\cinner{p_2}{p}}} \frac{1}{p^2} \sbr{\br{u}(p_4)\,\g_\A\,v(p_3)}.
\label{eq:app:EE:HadronicScalarAlphas2C_b}
\end{eqnarray}
Squaring the above amplitude we find
\begin{eqnarray}
 \nn \cSup{H_b}{2} &\simeq& \cSup{H}{0}\,\gs^4\,\mb t^a_{i_1 i_2} \Tr\cbr{\mb t^a_{j_3 j_4} \mb t^b_{j_3 j_4}} \mb t^b_{i_1 i_2}\,\nf\,\cbr{\frac{p_1^\A}{(1p)} - \frac{p_2^\A}{(2p)}} \cbr{\frac{p_1^\B}{(1p)} - \frac{p_2^\B}{(2p)}} \times \\
\nn & & \frac{\Tr\sbr{\slashed{p}{4}\,\g_\A\,\slashed{p}{3}\,\g_\B}}{4\cinner{p_3}{p_4}^2},
\\
\nn &\simeq& \cSup{H}{0}\,\gs^4\,\CF\TR\nf\,\frac{2}{\sbr{(34)(1p)(2p)}^2}\sbr{(12)(34)(1p)(2p) - \cbr{(13)(24) - (14)(23)}^2}.\\
\label{eq:app:EE:HadronicScalarAlphas2D_b}
\end{eqnarray}
Therefore the matrix element squared for process $(b)$ can be written, following the steps outlined previously for process $(a)$, as
\be
 \abs{\br{\cSup{\mc M_b}{2}}}^2 = \abs{\br{\cSup{\mc M}{0}}}^2\, \gs^4\, \CF\TR\nf\, \Hq, 
\label{eq:app:EE:MESAlphas2_qqbqqb}
\ee
where $\TR = \TF = 1/2$ and  the quark hard term is
\be
\Hq = - 4\,\mc I^2 + 4\,\frac{\om_{12}(p)}{(34) E_p^2}.
\label{eq:app:EE:HardNonAbelianAmpAlphas2q}
\ee
Adding up the contributions of both processes $(a)$ and $(b)$ the final expression of the $\Or(\as^2)$ matrix element in the soft limit (which nonetheless includes the quark and gluon hard terms) is given by
\begin{equation}
\abs{\br{\cSup{\mc M}{2}}}^2 = \abs{\br{\cSup{\mc M}{0}}}^2\,\gs^4 \sbr{4\,\CFsq\,\WP + \CF\CA\cbr{2 S + \Hg} + \CF\TF\nf\, \Hq}.
\label{eq:app:EE:MESAlphas2}
\end{equation}
Notice that if one employs the eikonal method (described in \sec{sec:app:QCD:EikonalApprox}) then one would find an identical formula to \eq{eq:app:EE:MESAlphas2} without the hard terms $\Hg$ and $\Hq$.


%% file: ch3/ch3app.tex

\label{ch:app:Jets}

%% file: ch5/ch5app.tex

\chapter{Jet--thrust calculations}

\section{Derivation of the LO distribution}
\label{sec:app:EEJS2:LOShapeFrac}

In the present section we outline the necessary steps for the derivation of the full logarithmic part of the LO $\te$ integrated distribution, \eq{S1_tauo_dist}. For the emission of a single gluon, i.e, $\EE \to q\,\qb\,g$, we define the kinematic variables, $x_{i} = 2 \inner{p_i}{q}/Q^2 = 2 E_i/Q$ and $y_{ij} = 2\inner{p_i}{p_j}/Q^2 = 1 - x_{k}$ where $i,j,k = 1(q), 2(\qb), 3(g)$ and $q=p_1+p_2+p_3$. The leading--order ($q\qb g$) matrix--element squared is given in \eq{eq:app:EE:IntegXsecAlphas1}. It reads
\begin{equation}
\frac{\d^2 \cSup{\s}{1}}{\cSup{\s}{0}\d x_1 \d x_2} = \frac{\CF \as}{2\pi}
\frac{x_{1}^{2} + x_{2}^{2}}{(1-x_{1})(1-x_{2})},
\label{eq:app:EEJS2:DiffXsecAlpha1}
\end{equation} 
where $\cSup{\s}{0}$ is the Born cross--section. The integration region, which is originally $1\geq x_{1},x_{2}, x_3\geq 0$ (recall that $x_1+x_2+x_3=2$) and which leads to divergences, gets modified by introducing the jet shape variable. For three partons in the final state, $\te$ is zero unless two partons are clustered together. That is, the $\te$ shape distribution receives, at this order, contributions from two--jet configurations only. Three--jet final state configurations, which are also present at this order, do not contribute to the latter distribution. They contribute to the $E_0$ distribution instead, as we shall see later. 

It is a straightforward exercise to show that the thrust, or rather $\tau=1-T$, is, for small values of $\tau$, the sum of the masses of two \emph{hemisphere} jets. Precisely $\tau \simeq (p_{\rJ_1}^2 + p_{\rJ_2}^2)/Q^2 = 1-\max\{x_i\} = \te$ where $i=1,2,3$ \footnote{From conservation of momentum given above we have for a jet formed by a pair of partons $(i,j)$: $\te = (p_i+p_j)^2/Q^2 = 2\inner{p_i}{p_j}/ Q^2 = (1-x_k)$ where $x_k>x_i,x_j$. i.e, $\te = 1-\max\{x_i\}$.}. The two distributions are thus equal. However, here we are considering jets of size $\Rs < 1/2$ and the following (generic) jet algorithm condition for a pair $(i,j)$ to be recombined together to form a jet applies
\be
 \Rs > 2(1-\cos\theta_{ij}) = \frac{(1-x_k)}{x_i x_j} \simeq \frac{(1-x_k)}{x_j},\qquad k\neq i,j,
\label{eq:app:EEJS2:JAMergingCond}
\ee
where the last approximation follows from $x_k > x_i > x_j$ (we show in \sec{sec.Fixed-PT-antikt} that in the limit $\Rs \to 1/2 \LRa \theta_{ij} \to \pi/2 $ the latter approximation yields the exact $\tau$ distribution). 

The phase space of the two--jet/three--jet boundary for the $\te(p_1,p_2) < \te$ is given in terms of the energy fractions \footnote{Similar limits have been derived for the JADE variable $y$ in \cite{ellis2003qcd}, which were also used for $\tau$ in the same reference.} as
\be
 1 > x_1, x_2 > 1-\te, \qquad 1+\te > x_1+x_2, \qquad R^2 > \frac{4(1-x_2)}{2-x_1-x_3},
\label{eq:app:EEJS2:RhoPSIntegRegion} 
\ee
where we have chosen to recombine the gluon, $p_3$, with hard leg $p_1$. The other choice, i.e, recombining the gluon with hard leg $p_2$, produces identical result. Hence we multiply by a factor of two to account for the latter. Likewise, the phase space boundary for the three--jet configuration whereby the third jet, chosen to be the gluon jet, is vetoed to have energy less than a cut--off $E_0$ is given by
\be
 2 > x_1+x_2 > 2\cbr{1-\frac{E_0}{Q}},\qquad \frac{4(1-x_2)}{2-x_1-x_3} > R^2,\qquad \frac{4(1-x_1)}{2-x_2-x_3} > R^2.
\label{eq:app:EEJS2:E0PSIntegregion}
\ee
Due to the soft and collinear singularities of the matrix--element \eqref{eq:app:EEJS2:DiffXsecAlpha1}, $x_{1,2} \to 1$, we use the following completion relation in order to integrate the $\te$ and $E_0$ differential distributions
\begin{eqnarray}
\nn \cSup{\Sg}{0} + \sbr{\cSup{\Sg_{\rm in}}{1} + \cSup{\Sg_{\rm out}}{1}} 
 &=& 1 + \int_{0}^{\te} \d\te'  \frac{1}{\cSup{\s}{0}}\frac{\d\cSup{\s}{1}}{\d\te'} + \int_{0}^{E_0} \d E_3  \frac{1}{\cSup{\s}{0}}\frac{\d\cSup{\s}{1}}{\d E_3},
\\
\nn &=& \frac{\s^{\rm tot}}{\cSup{\s}{0}} - \int_{\te}^{\te^{\max}} \d\te'  \frac{1}{\cSup{\s}{0}}\frac{\d\cSup{\s}{1}}{\d\te'} - \int_{E_0}^{Q/2} \d E_3  \frac{1}{\cSup{\s}{0}}\frac{\d\cSup{\s}{1}}{\d E_3},\\
\label{eq:app:EEJS2:IntegralOfShapeFracAlpha1}
\end{eqnarray}
where ``in'' and ``out'' refer to the regions \eqref{eq:app:EEJS2:RhoPSIntegRegion} and \eqref{eq:app:EEJS2:E0PSIntegregion}, respectively, and the total cross--section $\s^{\rm tot}$, which includes both real (two--jet and three--jet contributions) and virtual corrections, is related to the Born cross--section by \cite{ellis2003qcd}
\be
 \frac{\s^{\rm tot}}{\cSup{\s}{0}} = 1 + \frac{3\CF}{2}\frac{\as}{2\pi} + \Or(\as^2).
\label{eq:app:EEJS2:ToT2BornXsec}
\ee
Thus inverting the conditions in \eqs{eq:app:EEJS2:RhoPSIntegRegion}{eq:app:EEJS2:E0PSIntegregion}, carrying out the integrals in \eq{eq:app:EEJS2:IntegralOfShapeFracAlpha1} separately and then adding up with the $\Or(\as)$ of \eq{eq:app:EEJS2:ToT2BornXsec} one obtains \eq{R1_full-b}.

\section{$G_{nm}$ coefficients}
\label{app.coeff_in_expansion}

The resultant coefficients from the expansion of the exponent in the resummed
integrated distribution, eq.~\eqref{resum-form_QCD-b}, for the \akt algorithm, are
\begin{eqnarray}\label{G_nm-QCD}
\nonumber G_{12} &=& -2\,\CF ,
\\
\nonumber G_{11} &=& \CF \left(3 - 4 L_{\Rs}\right),
\\
\nonumber G_{10} &=& \CF\left[-4 L_{\Rs} L_{\Eo} +
\frac{\bar{f}_{0}(\Rs)}{2}\right],
\\
\nonumber G_{23} &=&  \CF \left(\frac{4}{3} \TF \nf - \frac{11}{3}\CA \right),
\\
\nonumber G_{22} &=& -\frac{4 \pi^{2}}{3} \CFsq + \CF
\CA\left(\frac{\pi^{2}}{3} - \mb I_{22}(\Rs) -\frac{169}{36} - \frac{22}{3}
L_{\Rs}\right) + \CF \TF\nf \left(\frac{11}{9} + \frac{8}{3} L_{\Rs}\right),
\\
\nonumber G_{21} &=& - \CFsq\,\frac{8 \pi^{2}}{3} L_{\Rs} - \CF\CA\bigg[2 \mb I_{22}(\Rs) L_{\Eo} - \mb I^{a}_{21} (\Rs) + \mb I^{\CA}_{21}(\Rs) -
\\ 
\nn &-& \Big(\frac{2\pi^{2}}{3} - 2 \mb I_{22}(Rs) - \frac{134}{9} - \frac{11}{3} L_{\Rs}\Big) L_{\Rs}\bigg] + \\
&+& \CF\TF\nf \sbr{\mb I^{\nf}_{21}(\Rs) + \left(\frac{4}{3} L_{\Rs} + \frac{40}{9}\right) L_{\Rs}}.
\end{eqnarray}
where $L_{\Rs} = \ln(\Rs/(1-\Rs))$ and $L_{\Eo} = \ln(2\Eo/Q)$. The factor
$\bar{f}_{0}(\Rs)$ only captures the first term of $f_{0}$ given in
eq.~\eqref{f_omeg}. We simply replace $\bar{f}_{0} \mapsto f_{0}$ when comparing
to the numerical distribution. Moreover, we have
\begin{align}\label{I2i-S2i_relation}
\nn \mc S_{22} = -\CF\CA\;\mb I_{22}(\Rs),&\qquad \mc S_{21}^{a} = -\CF\CA\;\mb I_{21}^{a}(\Rs),\\ 
\mc S_{21}^{\CA} = -\CF\CA\;\mb I^{\CA}_{22}(\Rs),&\qquad \mc S_{21}^{\nf} = -\CF\CA\;\mb I^{\nf}_{22}(\Rs).
\end{align}
The one--loop constant is given by, eq.~\eqref{R1_full-b},
\begin{eqnarray}\label{C1_QCD}
C_{1} &=& \CF \left(-1+\frac{\pi^{2}}{3}\right),
\end{eqnarray}
For the \ca algorithm we have the following replacements in $G_{22}$ and $G_{21}$:
\begin{eqnarray}
\nn \mb I_{2i} &\to& \mb I_{2i}^{\ca},\\
\nn G_{22}:\qquad -\frac{4\pi^2}{3}\CFsq &\to& \CFsq\cbr{C^P_2 -\frac{4\pi^2}{3}},\\
G_{21}:\;- \CFsq\,\frac{8 \pi^{2}}{3} L_{\Rs} &\to& \CFsq\cbr{2\,C^P_2 -\frac{8 \pi^{2}}{3}} L_{\Rs} + C^P_{21}.
\label{G_nm-QCD_CA_alg}
\end{eqnarray}

Expanding the total resummed distribution in eq.~\eqref{resum-form_QCD-b} to
$\Or(\as^{2})$ and up to NLL we have
\begin{multline}\label{resum_tot_series}
\Sigma_{r,2}(\widetilde{L}) = 1 + \left(\frac{\as}{2\pi}\right)\left(H_{12}
\widetilde{L}^{2} + H_{11} \widetilde{L} + H_{10}\right) + 
\left(\frac{\as}{2\pi}\right)^{2} \Big(H_{24} \widetilde{L}^{4} + H_{23}
\widetilde{L}^{3} + H_{22} \widetilde{L}^{2} + \\+ H_{21} \widetilde{L} +
H_{20}\Big),
\end{multline}
where (recall that $\widetilde{L} = \ln(1/\tauo) \Rightarrow \tauo =
e^{-\widetilde{L}}$)
\begin{eqnarray}\label{H_nm_coeffs}
\nonumber D_{\mathrm{fin}} (e^{-\widetilde{L}}) &=&
\left(\frac{\as}{2\pi}\right)\,d_{1}(e^{-\widetilde{L}}) +
\left(\frac{\as}{2\pi}\right)^{2}\,d_{2}(e^{-\widetilde{L}}),\\ 
\nonumber H_{12} &=& G_{12},\\
\nonumber H_{11} &=& G_{11},\\
\nonumber H_{10} &=& G_{10} + C_{1} + d_{1}(\tauo),\\
\nonumber H_{24} &=& \frac{1}{2} G_{12}^{2}.\\
\nonumber H_{23} &=& G_{23} + G_{12} G_{11},\\
\nonumber H_{22} &=& G_{22} + (G_{10} + C_{1}) G_{12} + \frac{1}{2}
G_{11}^{2},\\
\nonumber H_{21} &=& G_{21} + (G_{10}+ C_{1}) G_{11},\\
H_{20} &=& G_{20} + \frac{1}{2} G_{10}^{2} + C_{1} G_{10} + C_{2} +
d_{2}(\tauo). 
\end{eqnarray}
Differentiating~\eqref{resum_tot_series} w.r.t. $\widetilde{L}$, the NLO
differential distribution reads
\begin{equation}\label{resum_expanded-sig-diff}
\frac{\d\Sigma_{r,2}}{\d \widetilde{L}} =
\frac{1}{\sigma_{0}}\frac{\d\sigma_{r,2}}{\d \widetilde{L}} =
\delta(\widetilde{L})\, D_{\delta} + \left(\frac{\as}{2\pi}\right)\,
D_{A}(\widetilde{L}) + \left(\frac{\as}{2\pi}\right)^{2}\, D_{B}(\widetilde{L}),
\end{equation}
where the singular (logarithmic) terms are given by
\begin{eqnarray}\label{singular_terms_gen}
\nonumber D_{\delta} &=& 1 + \left(\frac{\as}{2\pi}\right) \left[G_{10} +
C_{1}\right] + \left(\frac{\as}{2\pi}\right)^{2} \left[G_{20} + \frac{1}{2}
G_{10}^{2} + C_{1} G_{10} + C_{2}\right],
\\
\nonumber D_{A}(\widetilde{L}) &=& 2 H_{12} \widetilde{L} + H_{11} +
\frac{\d}{\d \widetilde{L}}\, d_{1}(e^{-\widetilde{L}}),
\\
 D_{B}(\widetilde{L}) &=& 4 H_{24} \widetilde{L}^{3} + 3 H_{23}
\widetilde{L}^{2} + 2 H_{22} \widetilde{L} + H_{21} + \frac{\d}{\d
\widetilde{L}}\, d_{2}(e^{-\widetilde{L}}).
\end{eqnarray}

\section{Jet-thrust distribution in SCET}\label{app.tw_in_SCET}

The resummation of the jet-thrust in SCET is presented in the current
section for comparison with pQCD. We shall only present the final form of the
resummed result taken from
Refs.~\cite{Kelley:2011tj,Kelley:2010qs,Becher:2008cf}. For a full derivation
and more in depth discussion one should consult the latter references. The only
task we have performed here is the expansion of the full resummed distribution
to $\Or(\as^{2})$.

\subsection{Resummation}
\label{subsec.resummation_SCET}

The general formula of the resummed distribution for the jet-thrust is
given by~\cite{Kelley:2011tj,Becher:2008cf}
\begin{multline}\label{resum-form_SCET-a}
\frac{\d\Sigma^{\rm SCET}(\tauo, R)}{\d\tauo} = \exp\Big[4S(\mu_{h},\mu_{j}) +
4S(\mu_{s},\mu_{j}) - 4A_{H}(\mu_{h},\mu_{s}) + 4A_{J}(\mu_{j},\mu_{s}) \Big]
\\
 \times\left(\frac{\Rs}{1- \Rs}\right)^{-2A_{\Gamma}(\mu_{\omega},\mu_{s})}
\left(\frac{Q^{2}}{\mu_{h}^{2}}\right)^{-2A_{\Gamma}(\mu_{h},\mu_{j})}
H(Q^{2},\mu_{h})
\\ 
\times S^{\outt}_{R}(\omega,\mu_{\omega}) 
\left[\tilde{j}\left(\ln\frac{\mu_{s}Q}{\mu_{j}^{2}} + \partial_{\eta},
\mu_{j}\right) \right]^{2} \tilde{s}^{\inn}_{\tauo}(\partial_{\eta}, \mu_{s})
\\
\times \frac{1}{\tauo} \left(\frac{\tauo Q}{\mu_{s}}\right)^{\eta} \frac{e^{-\gamma_{E}
\eta}}{\Gamma(\eta)}.
\end{multline}
See~\cite{Kelley:2011tj,Becher:2008cf} for full notation. In order to compute
the fixed--order expansion of~\eqref{resum-form_SCET-a} up to $\Or(\as^{2})$,
all scales should be set equal ($\mu_{h} = \mu_{j} = \mu_{s} = Q$). In this
limit, the evolution factors $S, A_{J}$ and $A_{H}$ vanish. The differentiation
w.r.t. $\eta$ is carried out using the explicit form of $\tilde{j}$ and
$\tilde{s}^{\inn}_{\tauo}$. The final result of the integrated distribution may
be cast in the generic form~\eqref{resum-form_QCD-b} with the constants and
coefficients of the logs given by
\begin{eqnarray}\label{eq.C1-C2_SCET}
C_{1} &=& \CF\left(-1+\frac{\pi^{2}}{3} \right),
\\
\nonumber C_{2} &=& \CFsq \left(1- \frac{3\pi^{2}}{8} +\frac{\pi^{4}}{72}
-6\zeta(3) \right) + \CF\CA \left(\frac{493}{324} + \frac{85\pi^{2}}{24}
-\frac{73\pi^{4}}{360} + \frac{283 \zeta(3)}{18} \right) +
\\
&+& \CF\TF\nf \left(\frac{7}{81} -\frac{7\pi^{2}}{6}
-\frac{22\zeta(3)}{9}\right) + C_{2}^{\inn} + C_{2}^{\outt},
\end{eqnarray}
and
\begin{eqnarray}\label{G_nm_SCET}
\nn G_{12} &=& -2\,\CF,
\\
\nn G_{11} &=& -\CF \left(3- 4 L_{\Rs}\right),
\\
\nn G_{10} &=& \CF\left(- 4 L_{\Rs} L_{\Eo} + \frac{f_{0}}{2}\right),
\\
\nn G_{23} &=& \CF \left(\frac{11}{3}\CA - \frac{4}{3} \TF \nf \right),
\\
\nn G_{22} &=& -\frac{4 \pi^{2}}{3} \CFsq + \CF
\CA\left(\frac{\pi^{2}}{3}-\frac{169}{36} - \frac{22}{3} L_{\Rs}\right) + \CF
\TF\nf \left(\frac{11}{9} + \frac{8}{3} L_{\Rs}\right),
\\
\nn G_{21} &=& \CFsq\left[-\frac{3}{4} - \pi^{2} + 4\zeta(3) +\frac{8
\pi^{2}}{3} L_{\Rs}\right] + \CF\TF\nf\left[5 - \left(\frac{4}{3} L_{\Rs}+ \frac{40}{9}\right)
L_{\Rs}\right]
\\
\nn &+& \CF\CA\bigg[-\frac{57}{4} + 6\zeta(3) - \left(\frac{2\pi^{2}}{3}
-\frac{134}{9} -\frac{11}{3} L_{\Rs}\right) L_{\Rs}\bigg].
\end{eqnarray}
\begin{eqnarray} 
\nn G_{20} &=& \CFsq  \Bigg[-\frac{f_{0}^{2}}{8} + \left(2\pi^{2}
-16\zeta(3)\right) L_{\Rs} - \Big(\frac{11\pi^{2}}{6} + \frac{f_{0}}{2}\Big)
L^{2}_{\Rs} - L_{\Rs}^{2} \Bigg]+
\\
\nonumber &+& \CF\CA \Bigg[\frac{11\pi^{2}}{9} L_{\Rs} - \frac{11}{6} L_{\Rs}
L^{2}_{\Eo} - L_{\Eo} \Bigg(\frac{11f_{0}}{12}
+\left[\frac{134}{9}-\frac{2\pi^{2}}{3}\right] L_{\Rs} +
\\
\nn &+& \frac{11}{6} L^{2}_{\Rs}\Bigg) \Bigg] +
 \CF\TF\nf\Bigg[-\frac{4\pi^{2}}{9} L_{\Rs} + \frac{2}{3} L_{\Rs}
L^{2}_{\Eo} + L_{\Eo}\Bigg(\frac{f_{0}(\Rs)}{3} +
\\
&+& \frac{40}{9} L_{\Rs} + \frac{2}{3} L^{2}_{\Rs}\Bigg) \Bigg],
\end{eqnarray}
where $f_0 \equiv f_o(\Rs)$. Considering primary emission, the only missing piece in the distribution is the two--loop constants in the soft function, namely $C_{2}^{\inn}$ and $C_{2}^{\outt}$.

\section{Comparisons to \event: single $\Rs$ plots}
\label{sec:app:ComparisonsToEvent2-Plots}

Below we provide plots for the difference between the analytical and numerical (exact) jet-thrust $\te$ distribution at NLO for each colour channel. The plots are shown for each value of $\Rs$ separately.
\begin{figure}[!t]
 \centering
 \epsfig{file=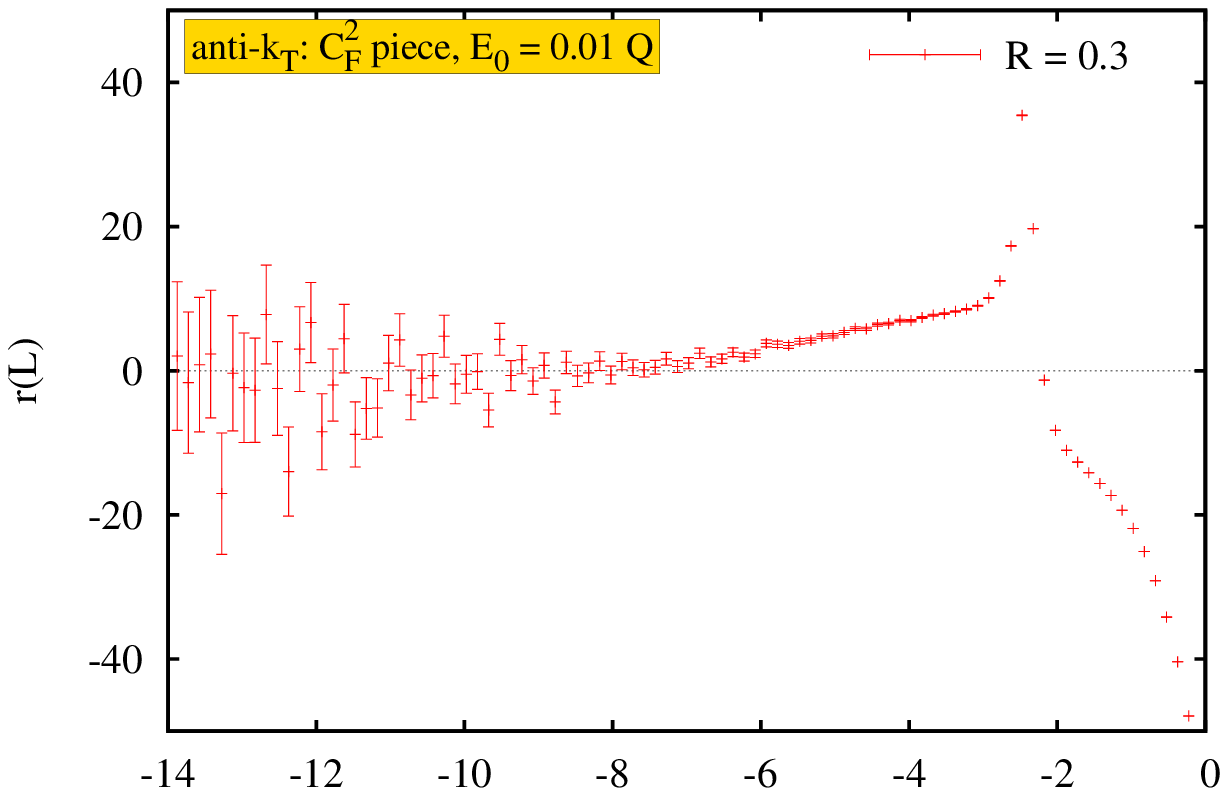,  width= 0.48 \textwidth}
 \epsfig{ file=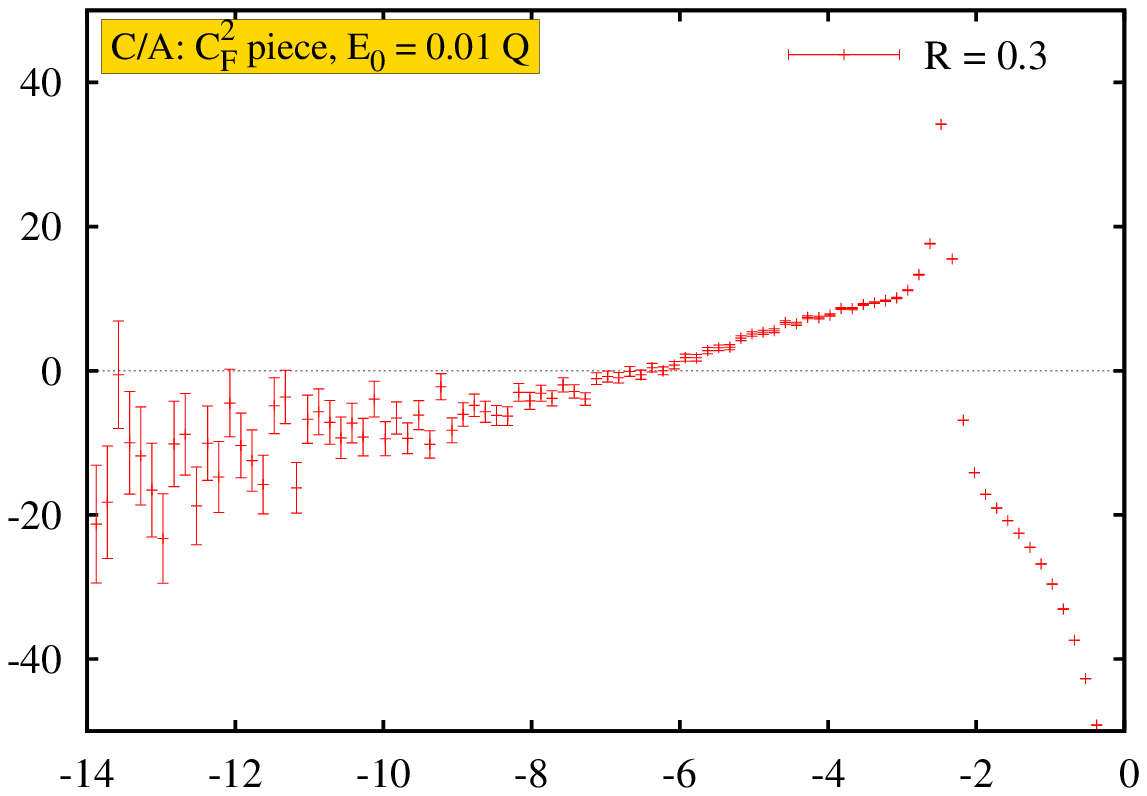,  width= 0.48 \textwidth}
 \epsfig{file=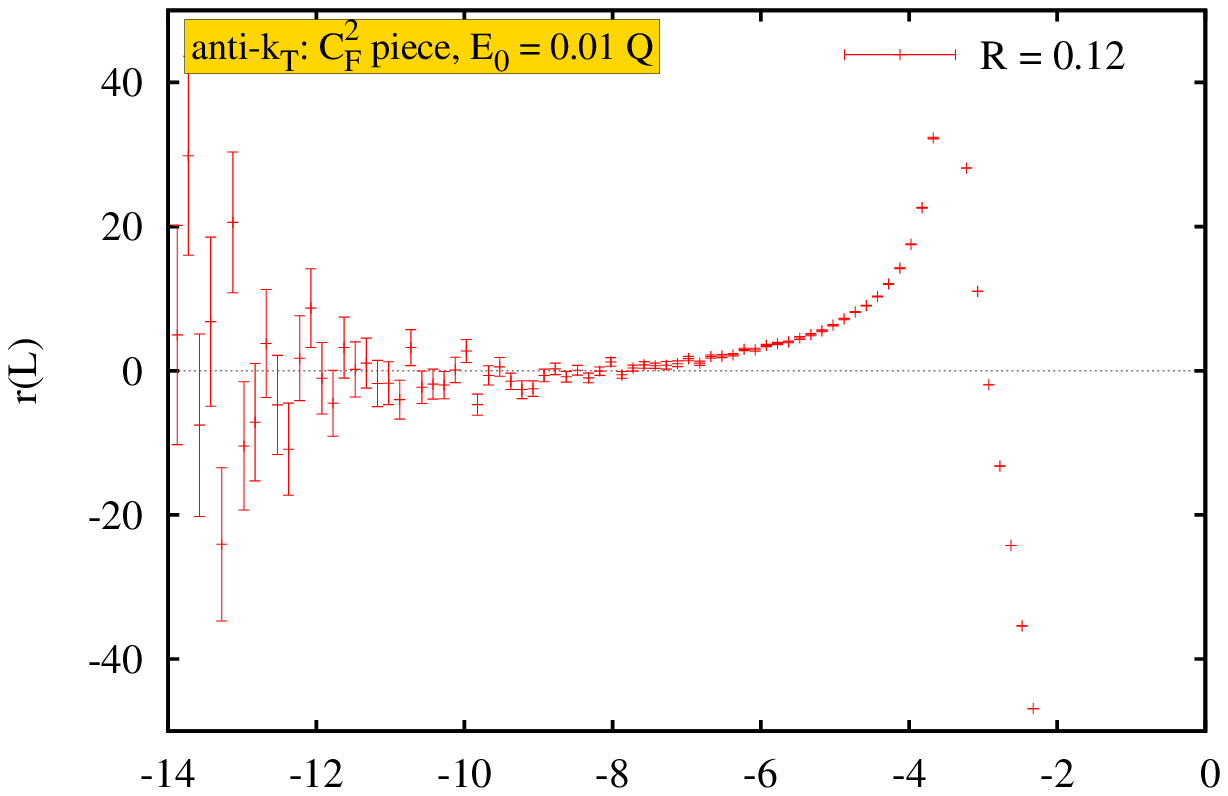, width= 0.48 \textwidth}
 \epsfig{ file=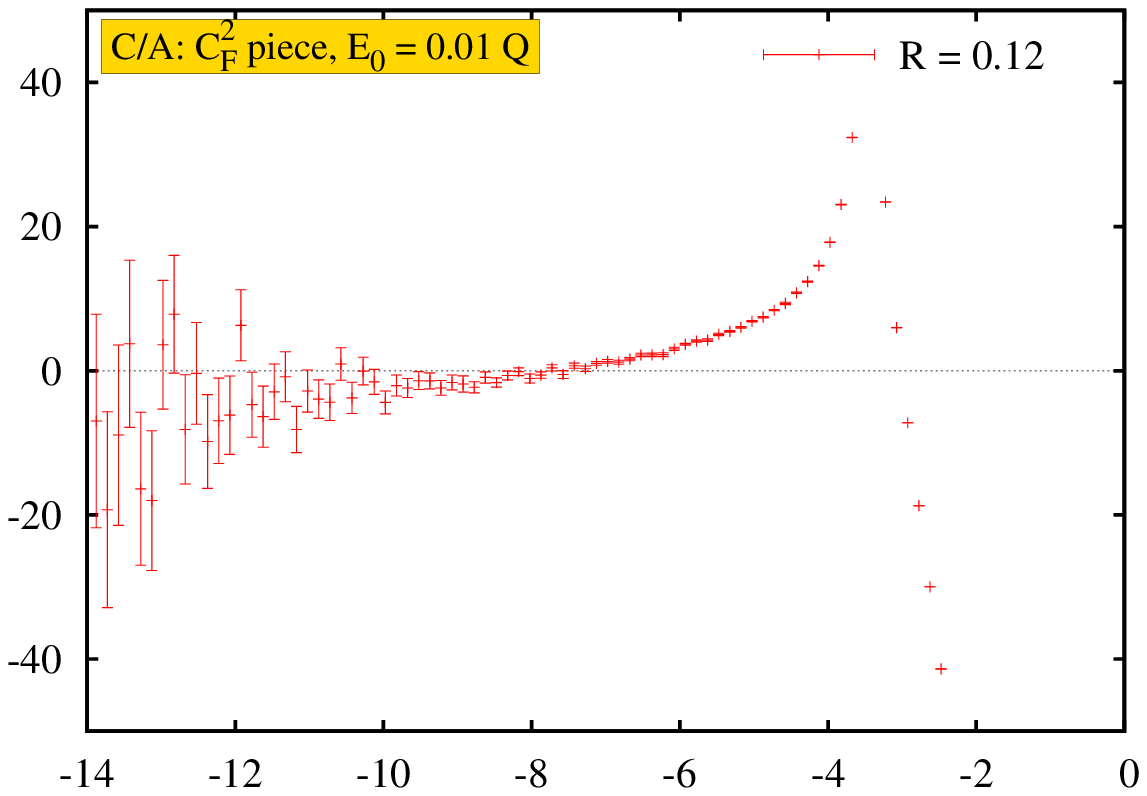, width= 0.48 \textwidth}
 \epsfig{file=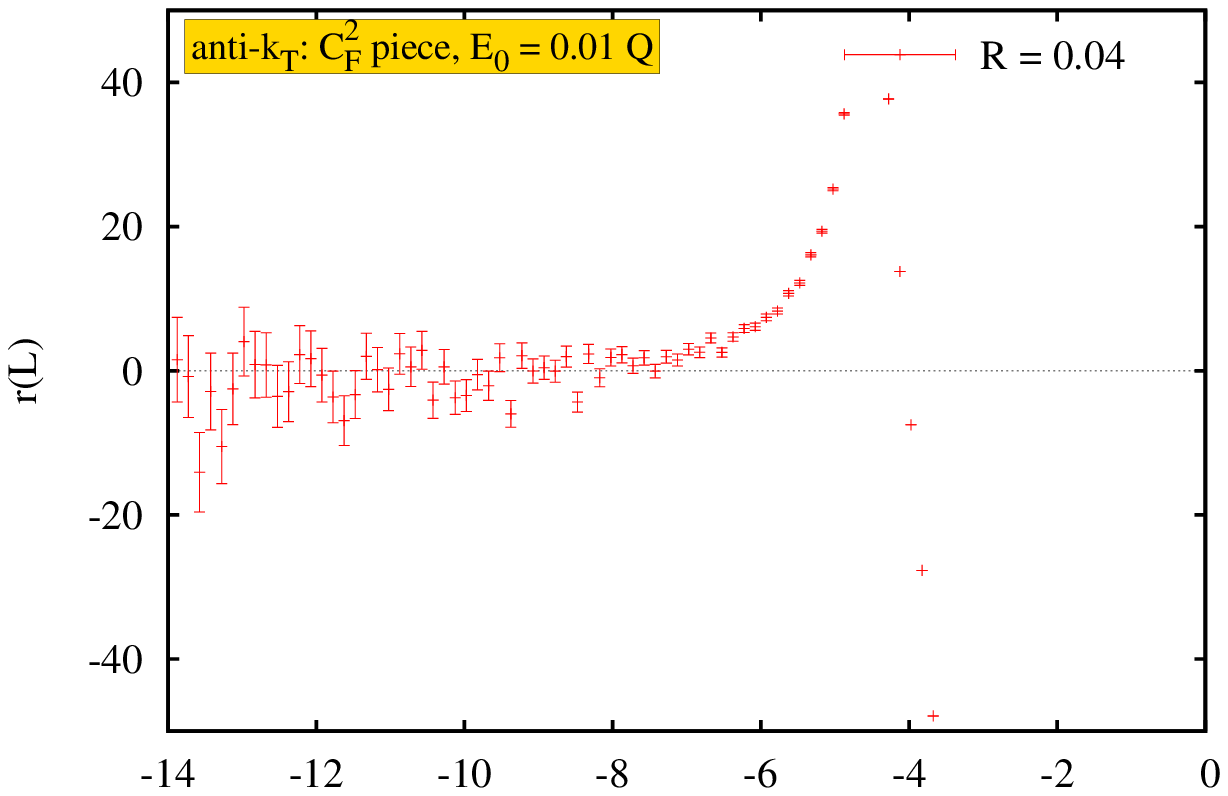, width= 0.48 \textwidth}
 \epsfig{ file=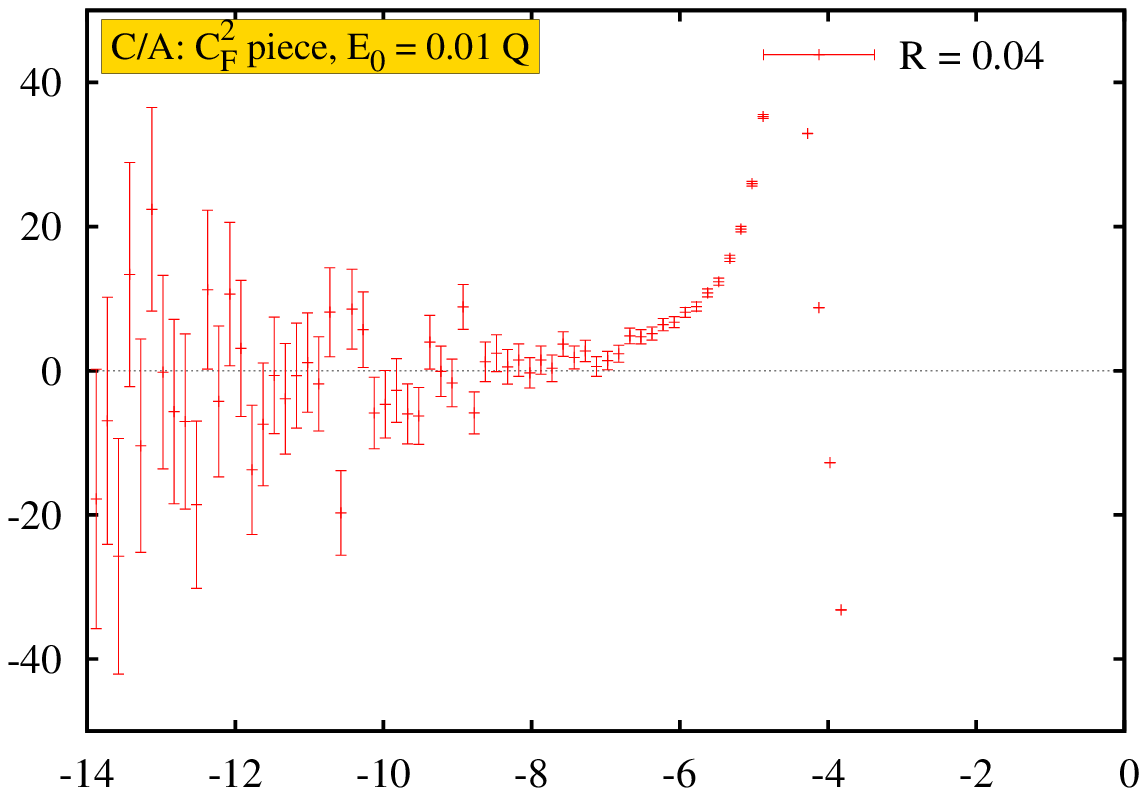, width= 0.48 \textwidth}
 \epsfig{file=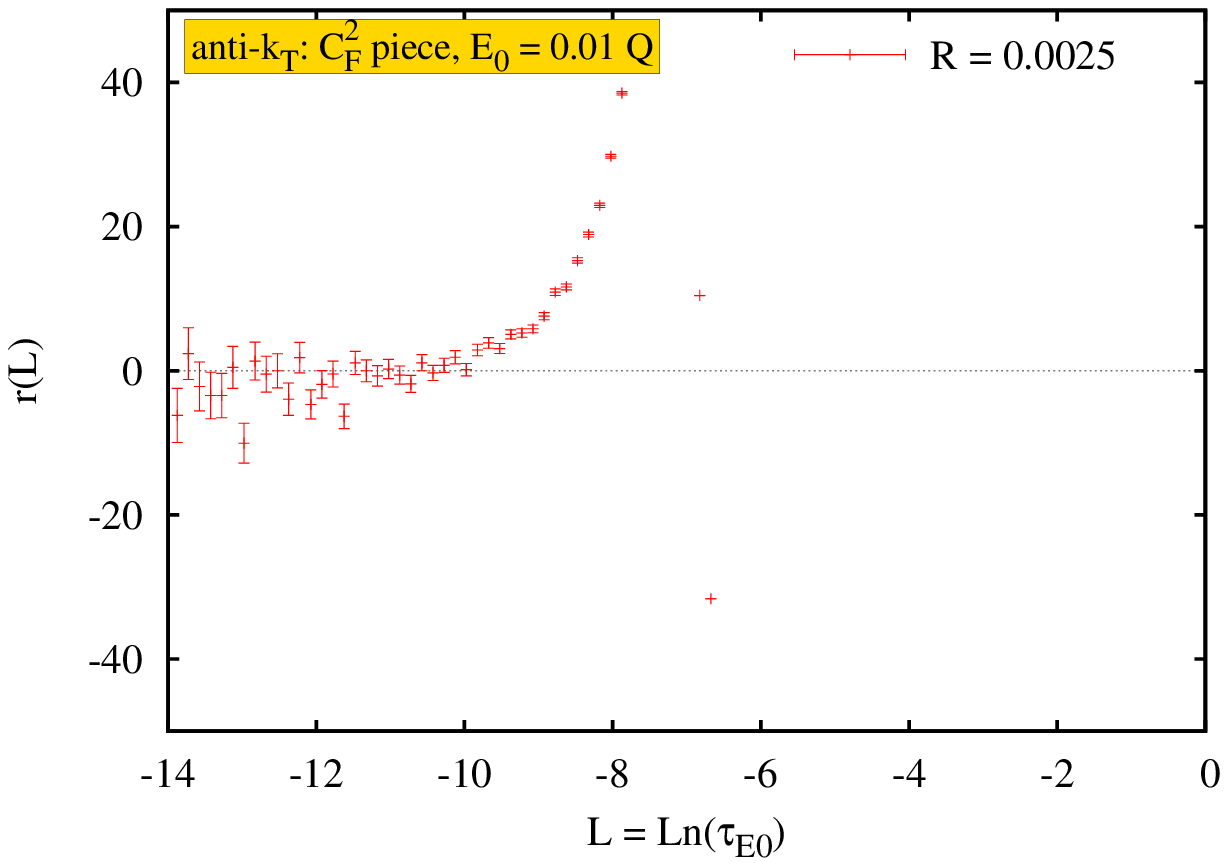, width= 0.48 \textwidth}
 \epsfig{ file=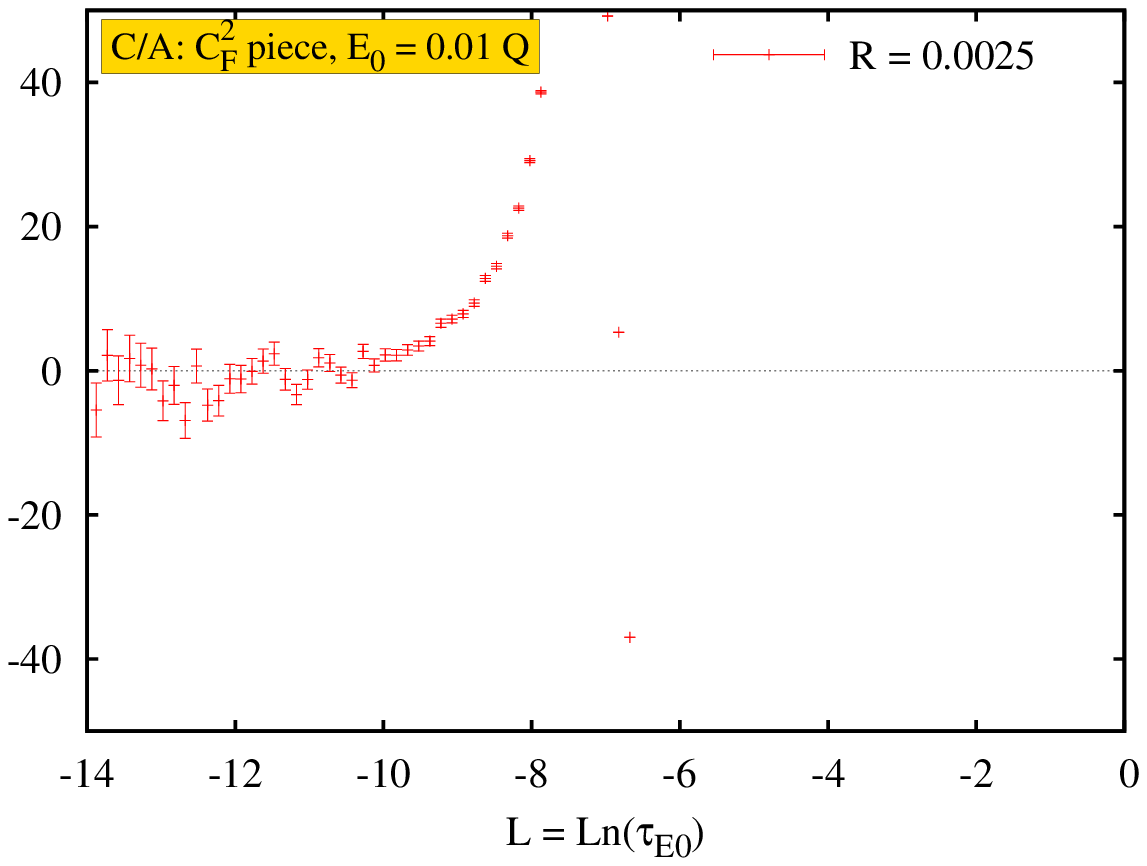, width= 0.48 \textwidth}
 \caption{The difference between \event and $\tauo$ NLO distribution for various jet radii in both \AKT (left) and C/A (right) algorithms. Plots are for $\CFsq$ channel.}
 \label{fig.NLO_CF2}
\end{figure}
\begin{figure}[!t]
 \centering
 \epsfig{file=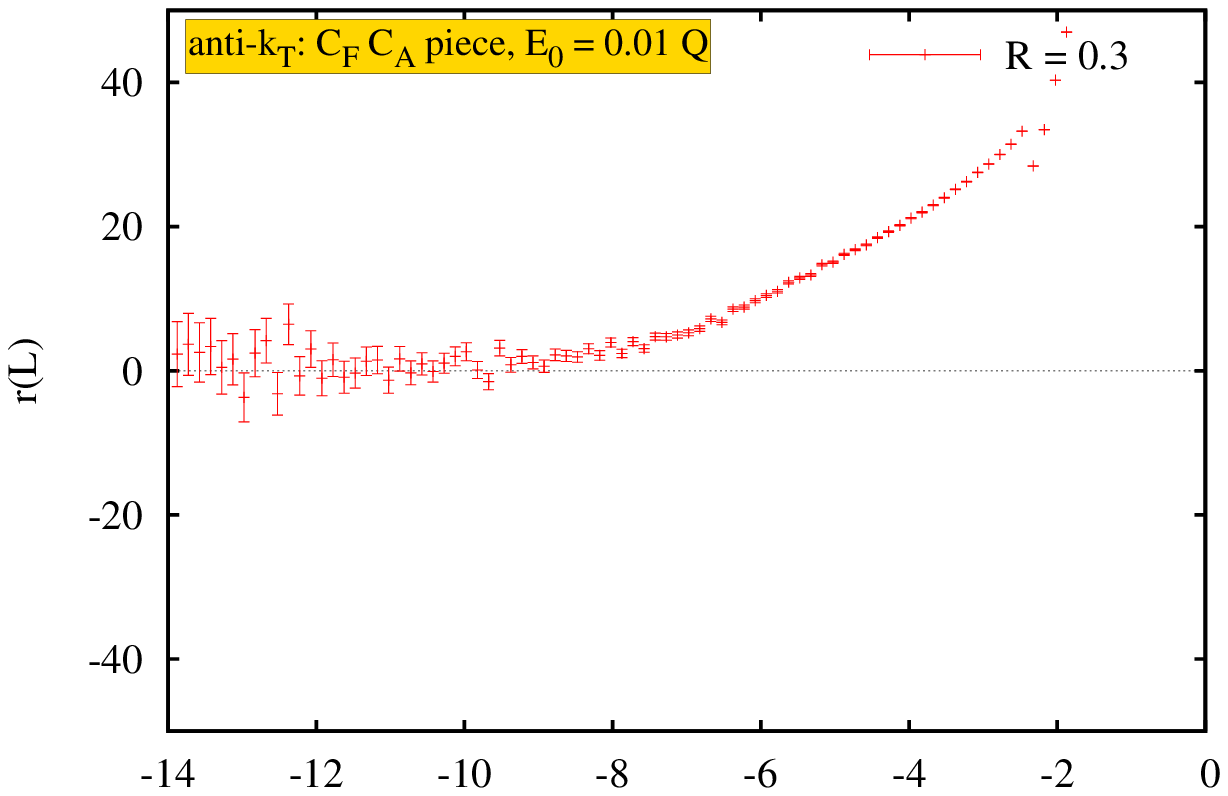,  width= 0.48 \textwidth}
 \epsfig{ file=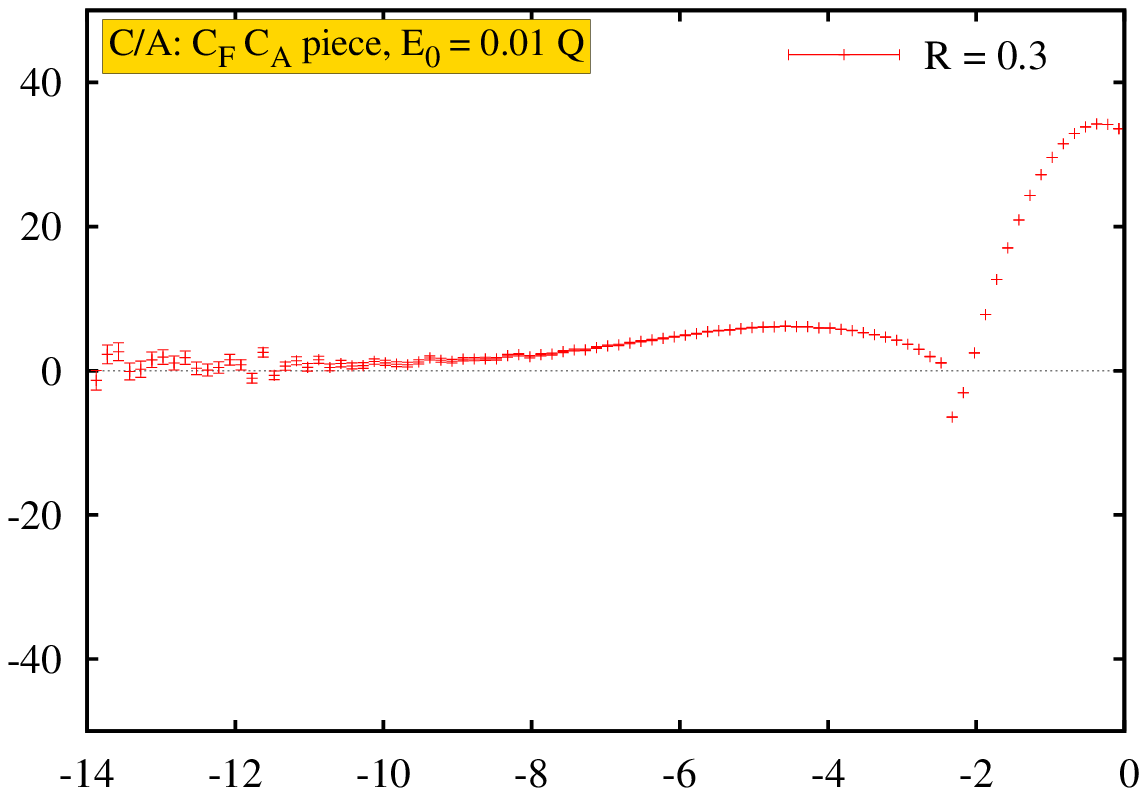,  width= 0.48 \textwidth}
 \epsfig{file=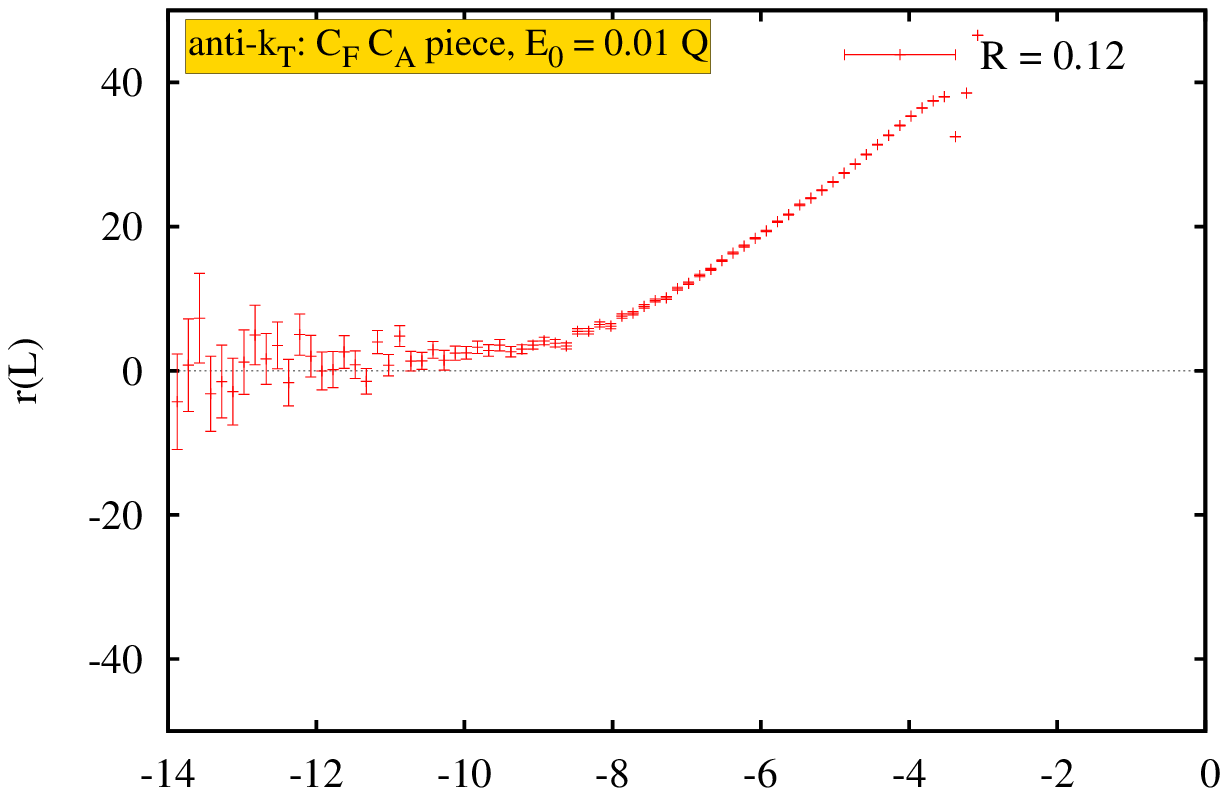, width= 0.48 \textwidth}
 \epsfig{ file=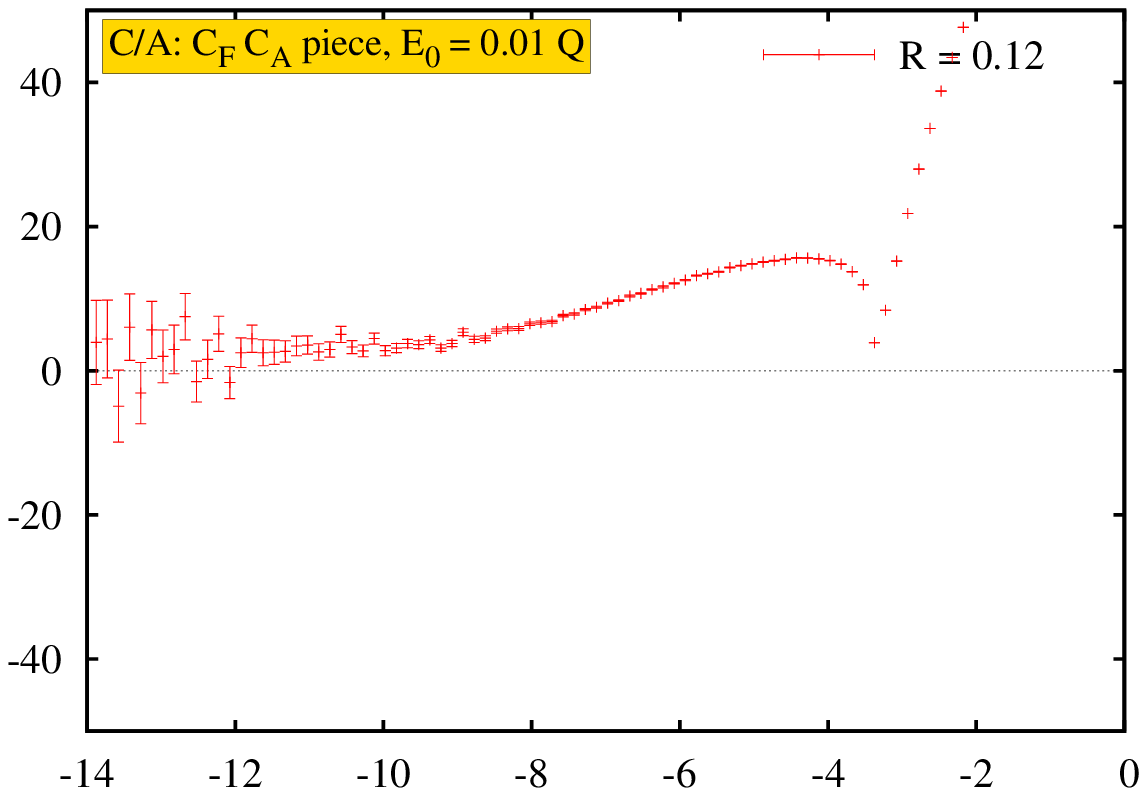, width= 0.48 \textwidth}
 \epsfig{file=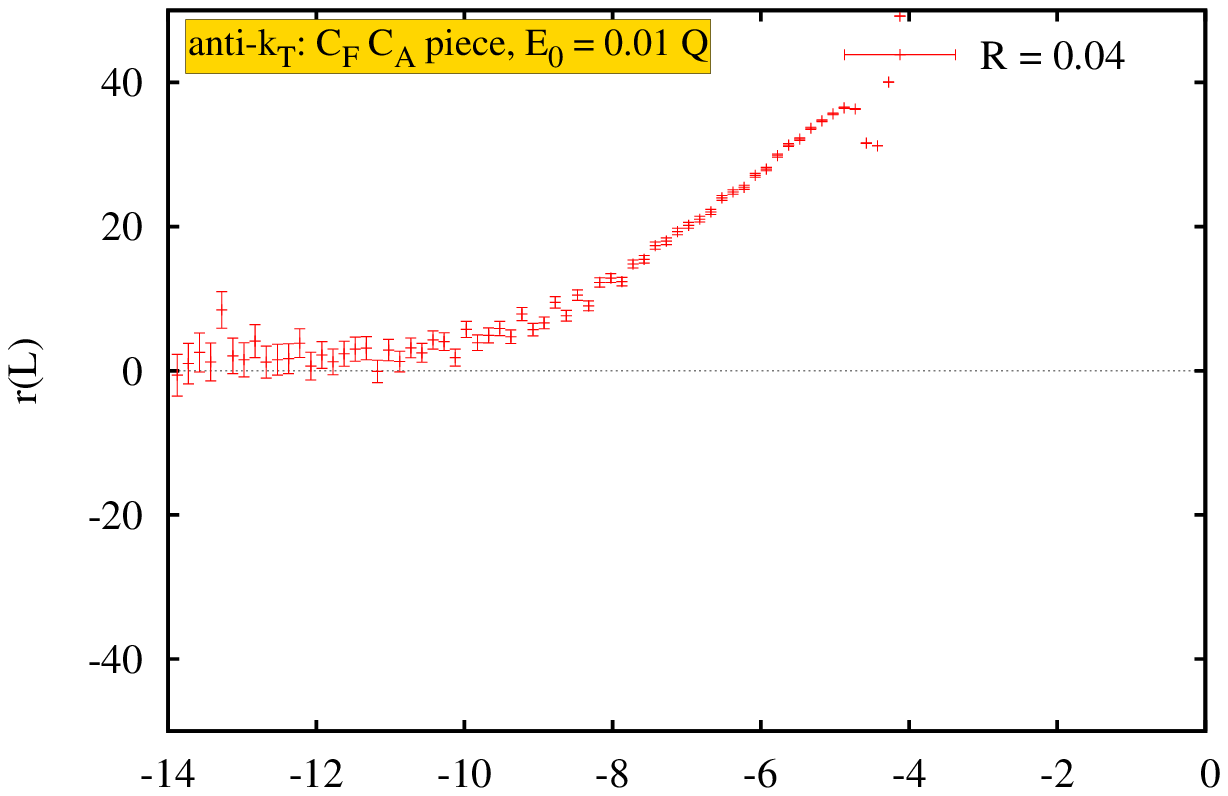, width= 0.48 \textwidth}
 \epsfig{ file=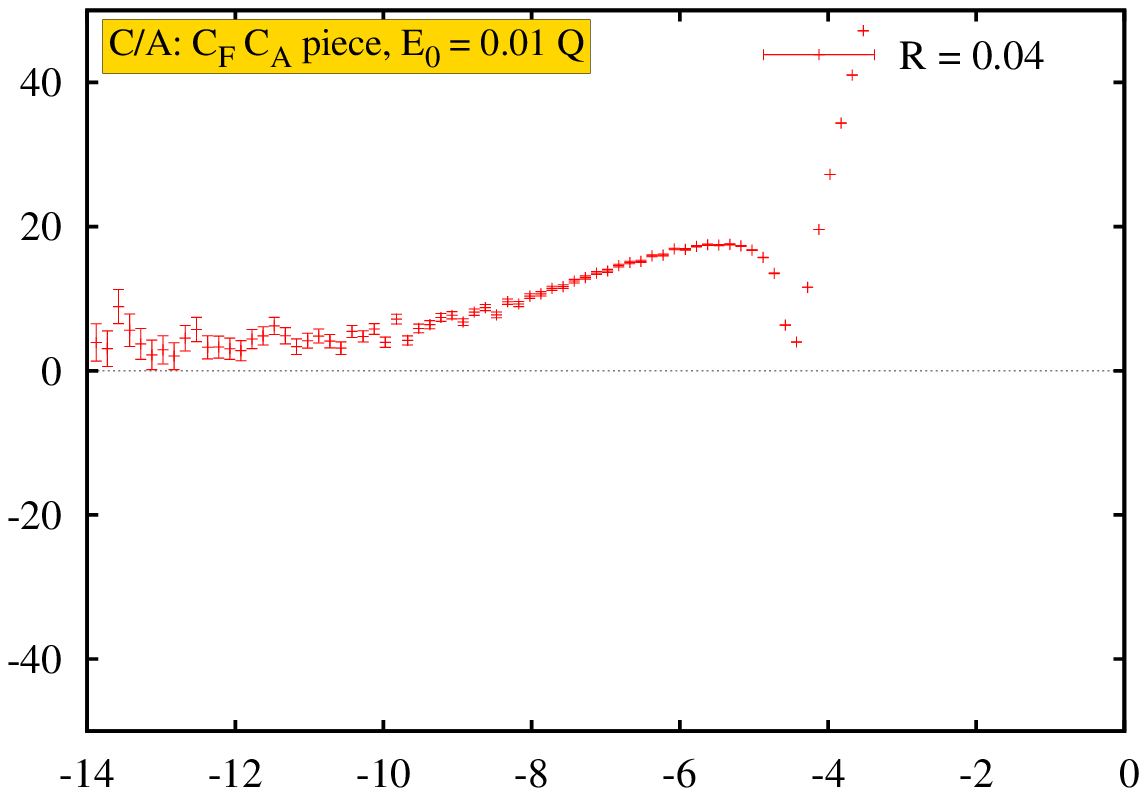, width= 0.48 \textwidth}
 \epsfig{file=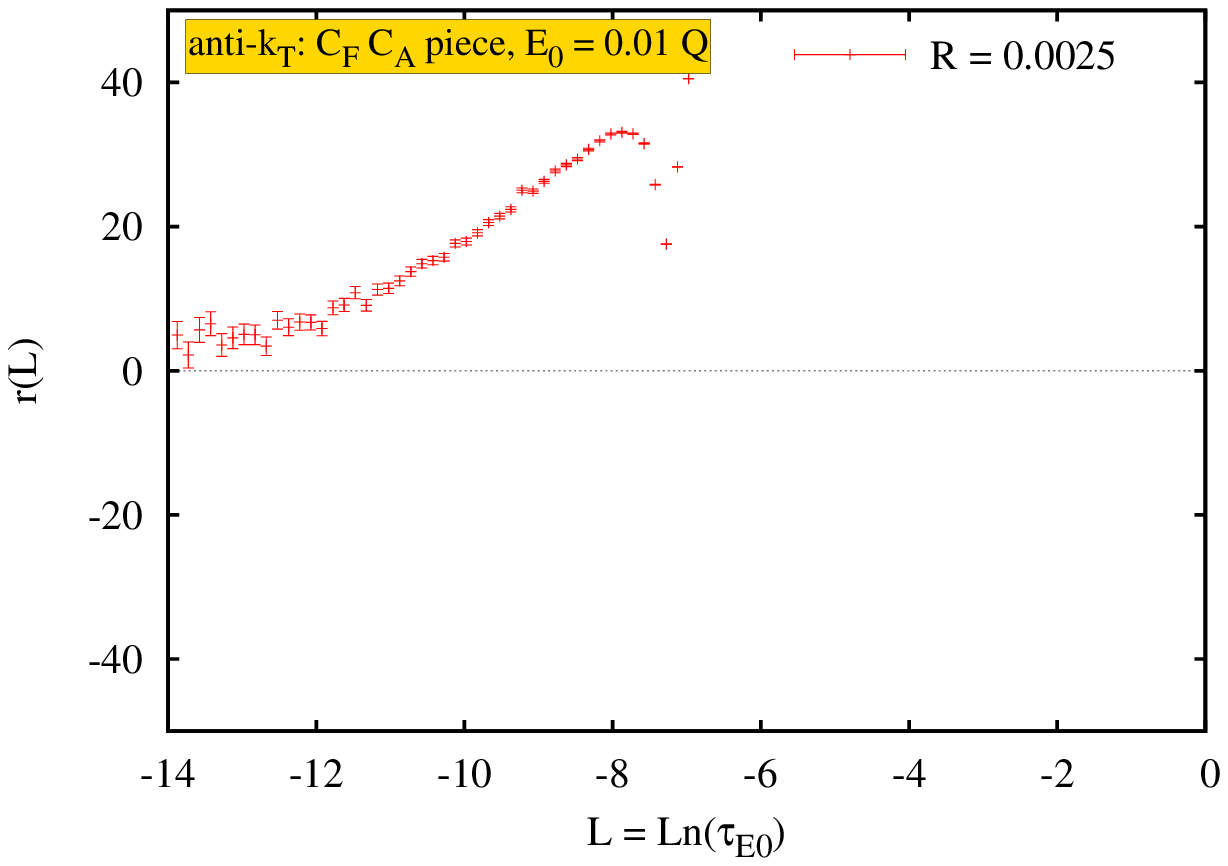, width= 0.48 \textwidth}
 \epsfig{ file=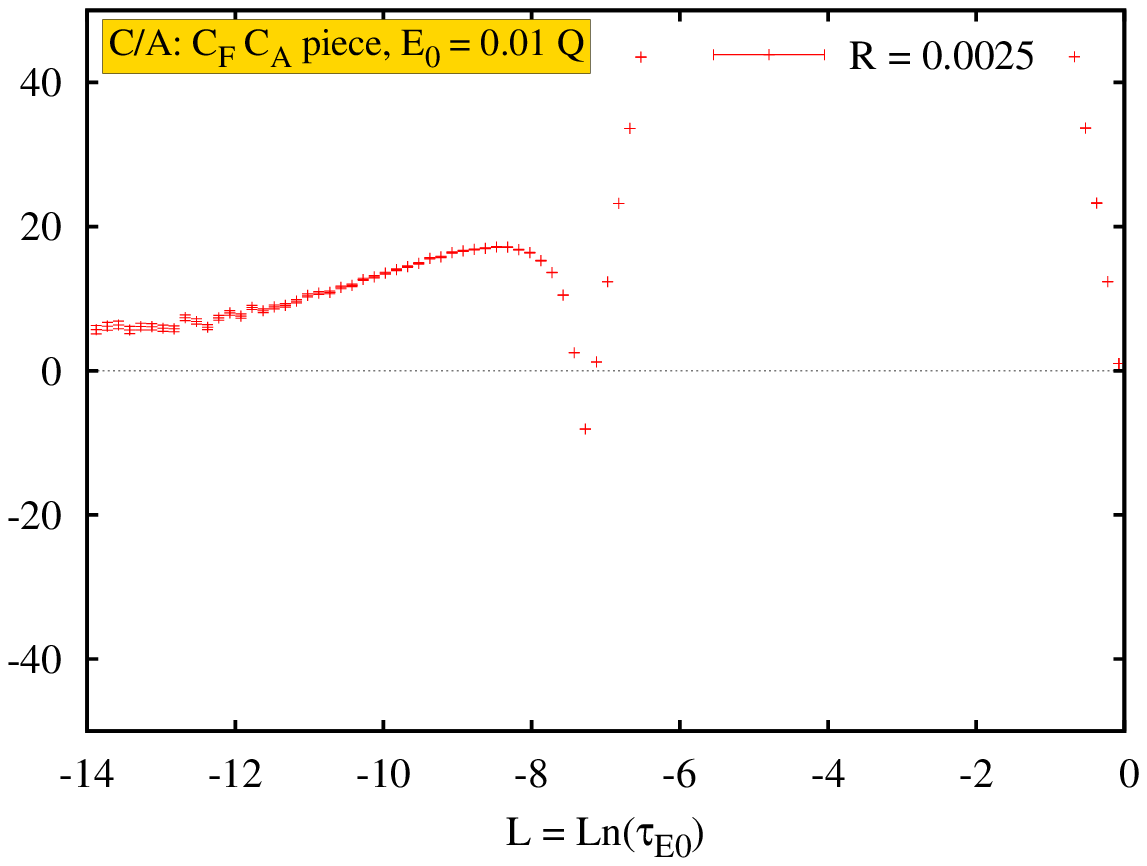, width= 0.48 \textwidth}
 \caption{The difference between \event and $\tauo$ NLO distribution for various jet radii in both \AKT (left) and C/A (right) algorithms. Plots are for $\CF\CA$ channel.}
 \label{fig.NLO_CFCA}
\end{figure}
\begin{figure}[!t]
 \centering
 \epsfig{file=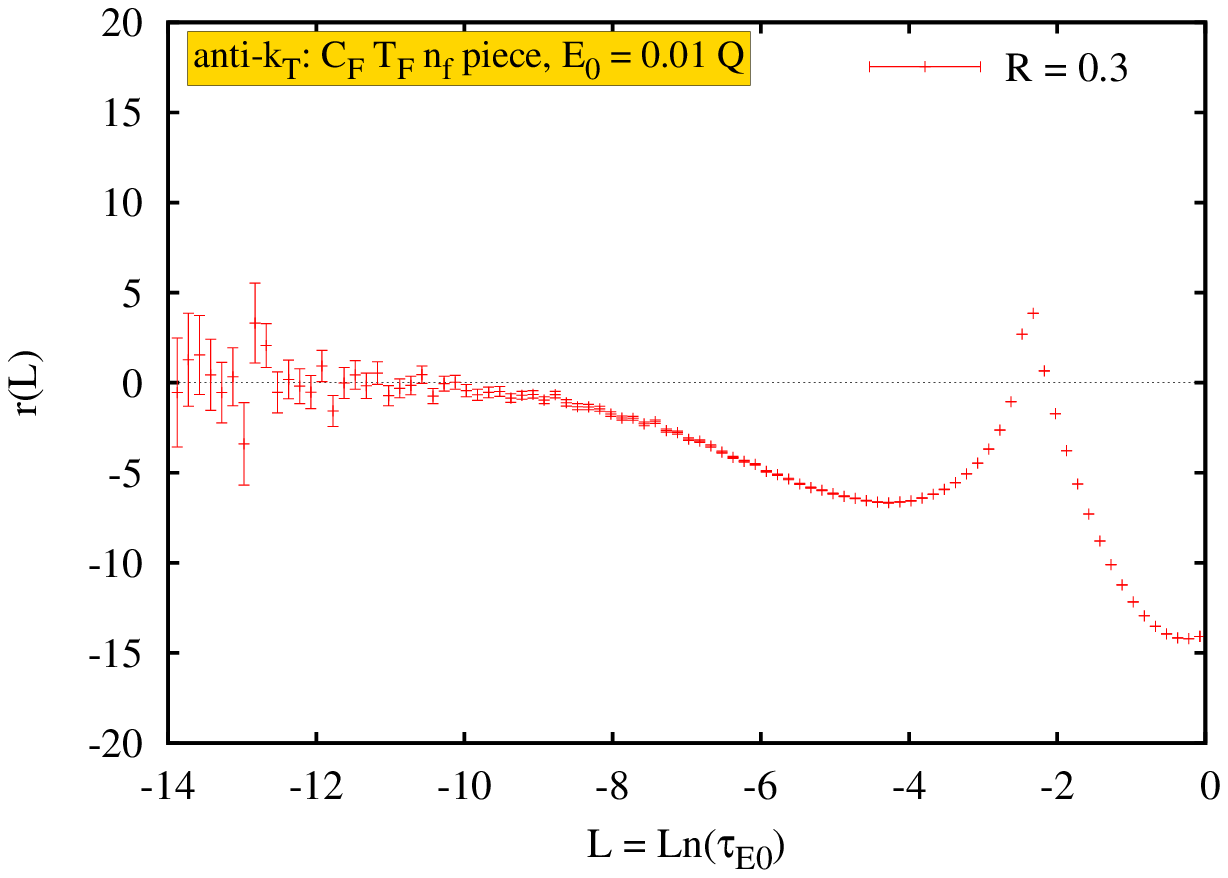,  width= 0.48 \textwidth}
 \epsfig{ file=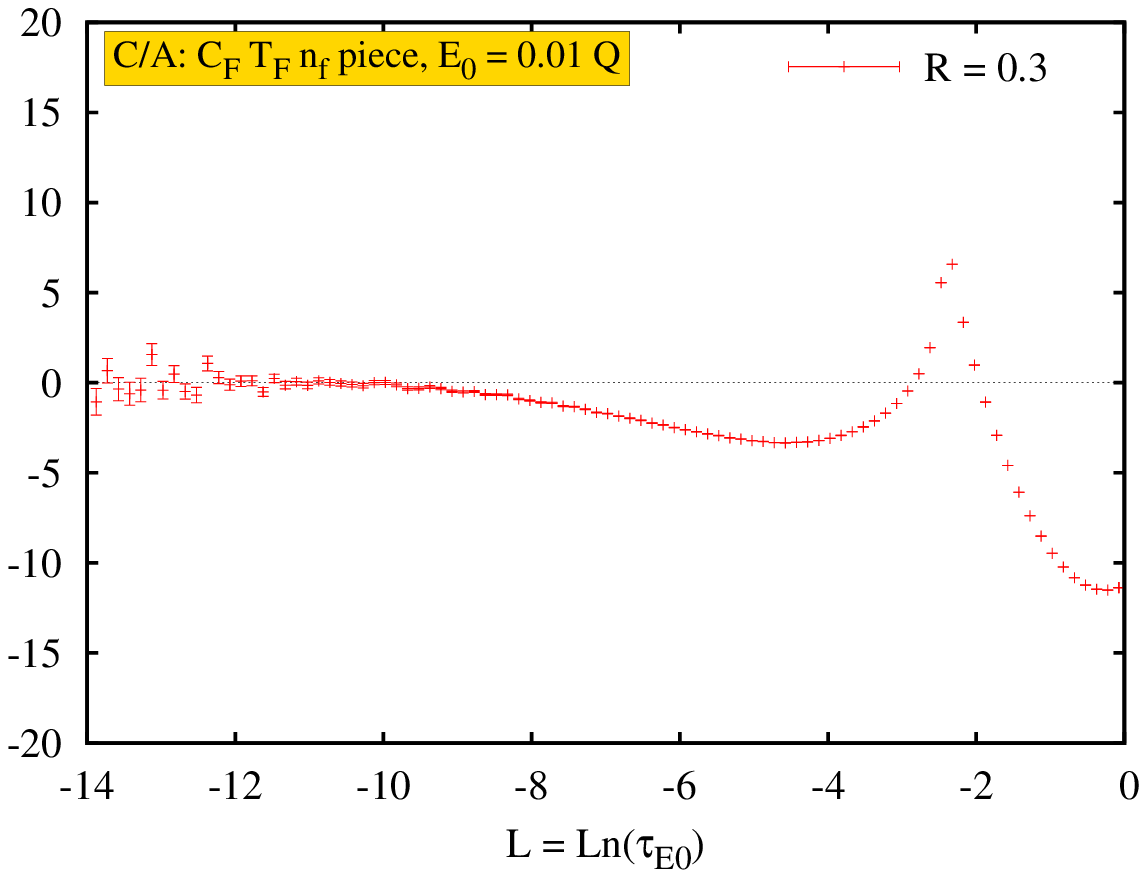,  width= 0.48 \textwidth}
 \epsfig{file=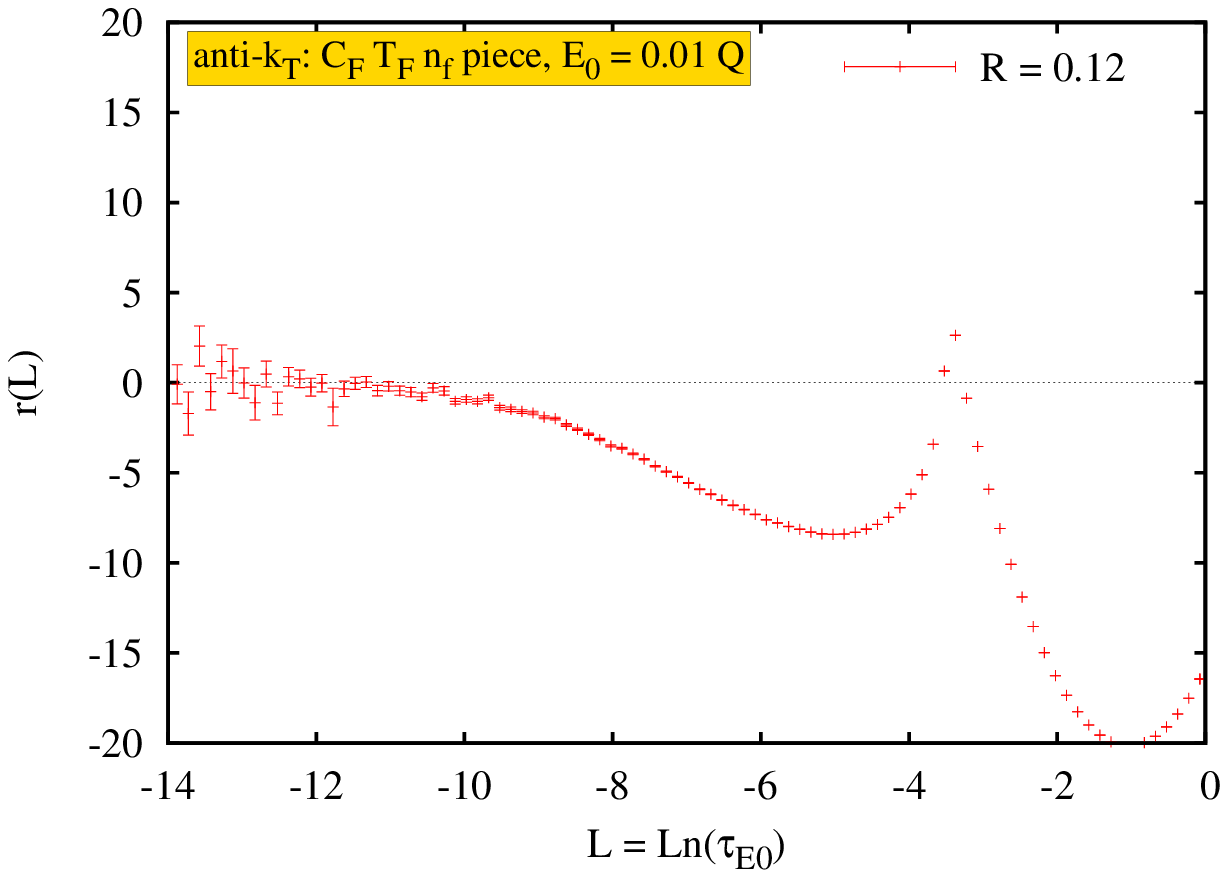, width= 0.48 \textwidth}
 \epsfig{ file=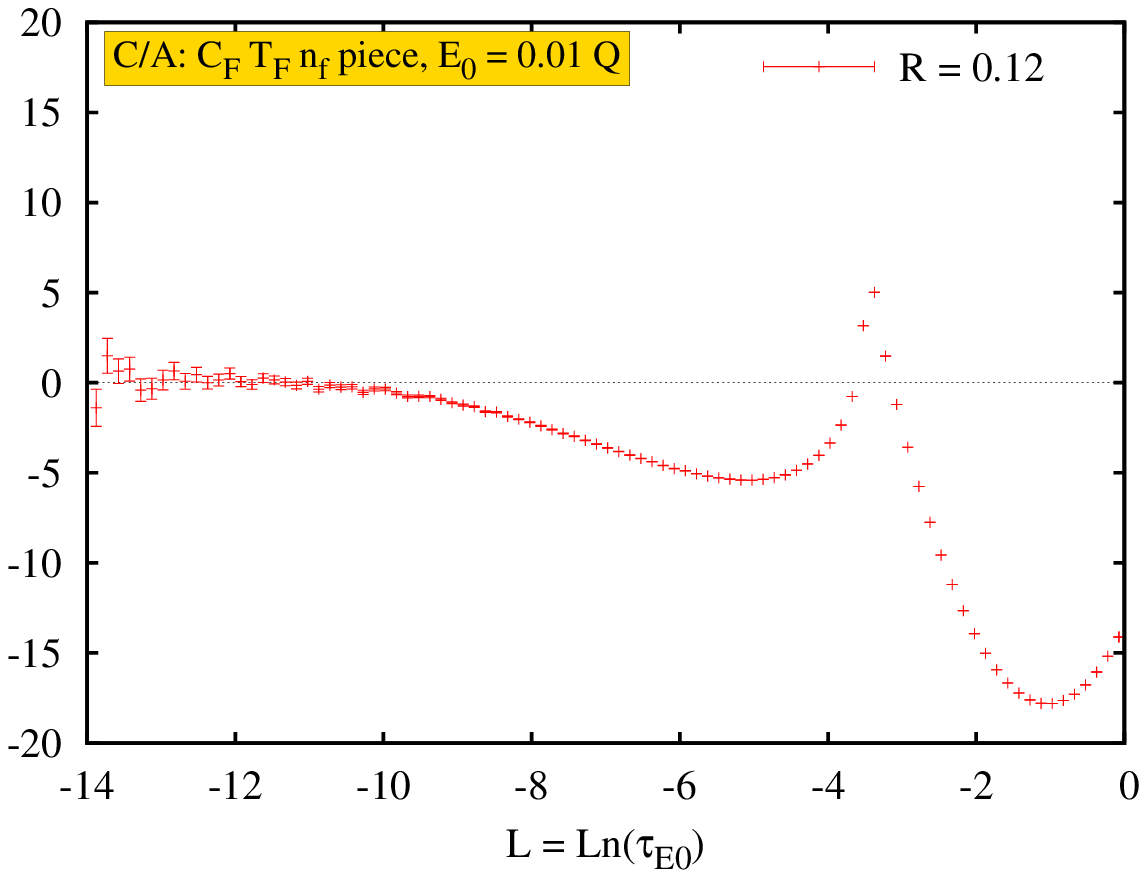, width= 0.48 \textwidth}
 \epsfig{file=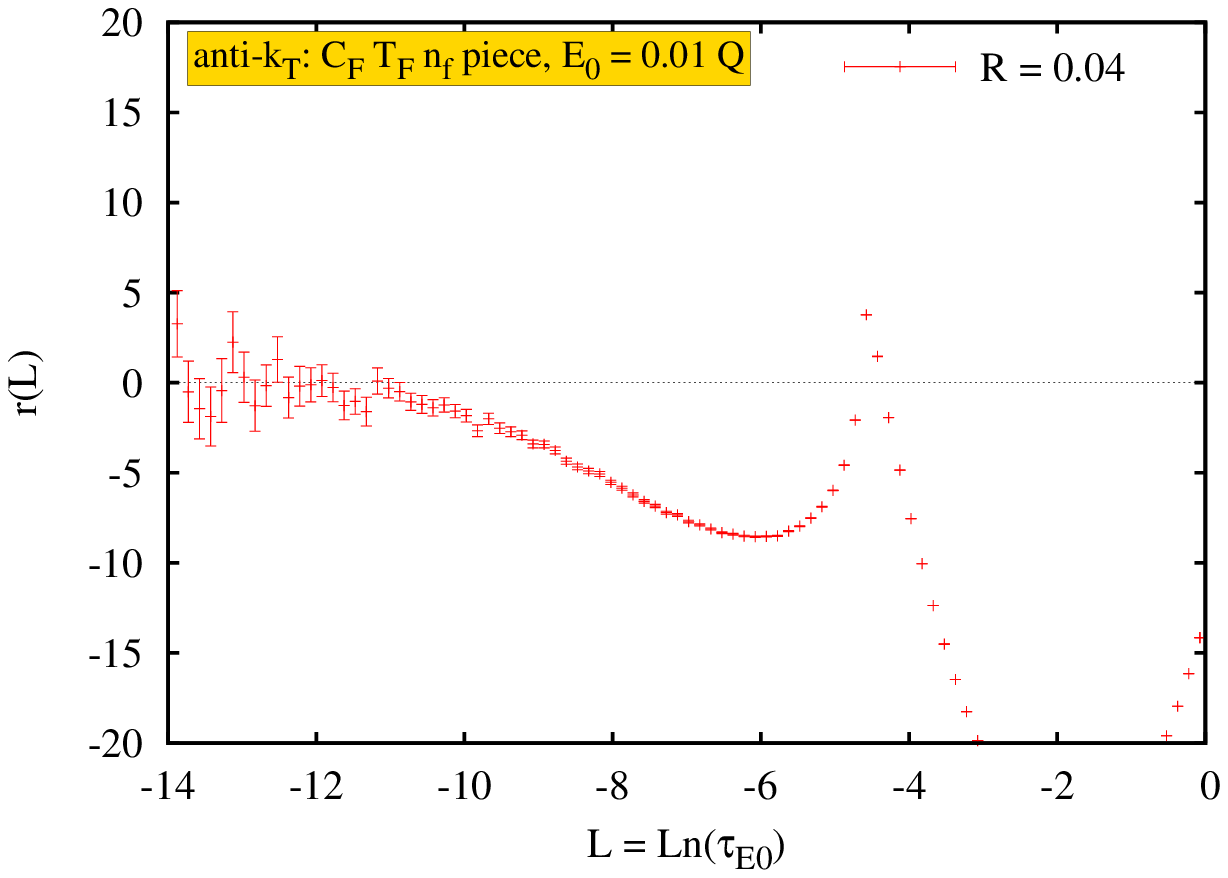, width= 0.48 \textwidth}
 \epsfig{ file=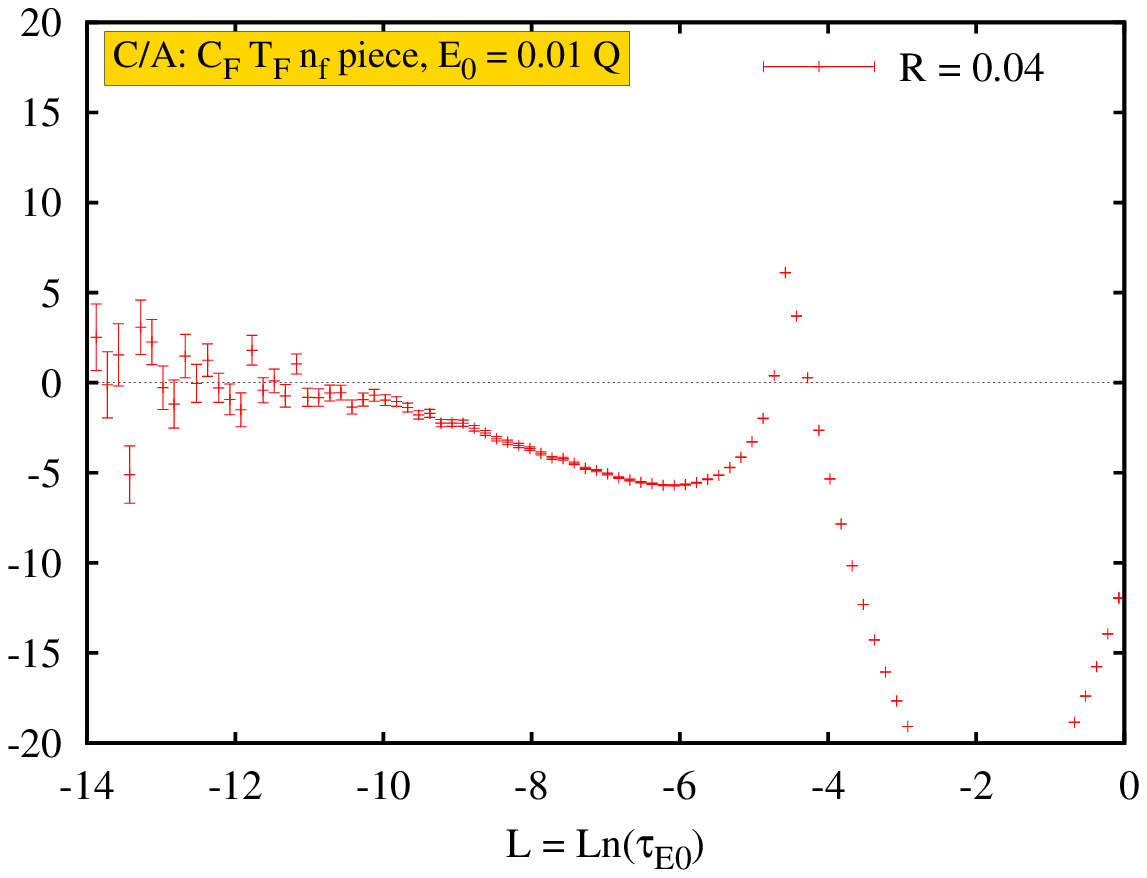, width= 0.48 \textwidth}
 \epsfig{file=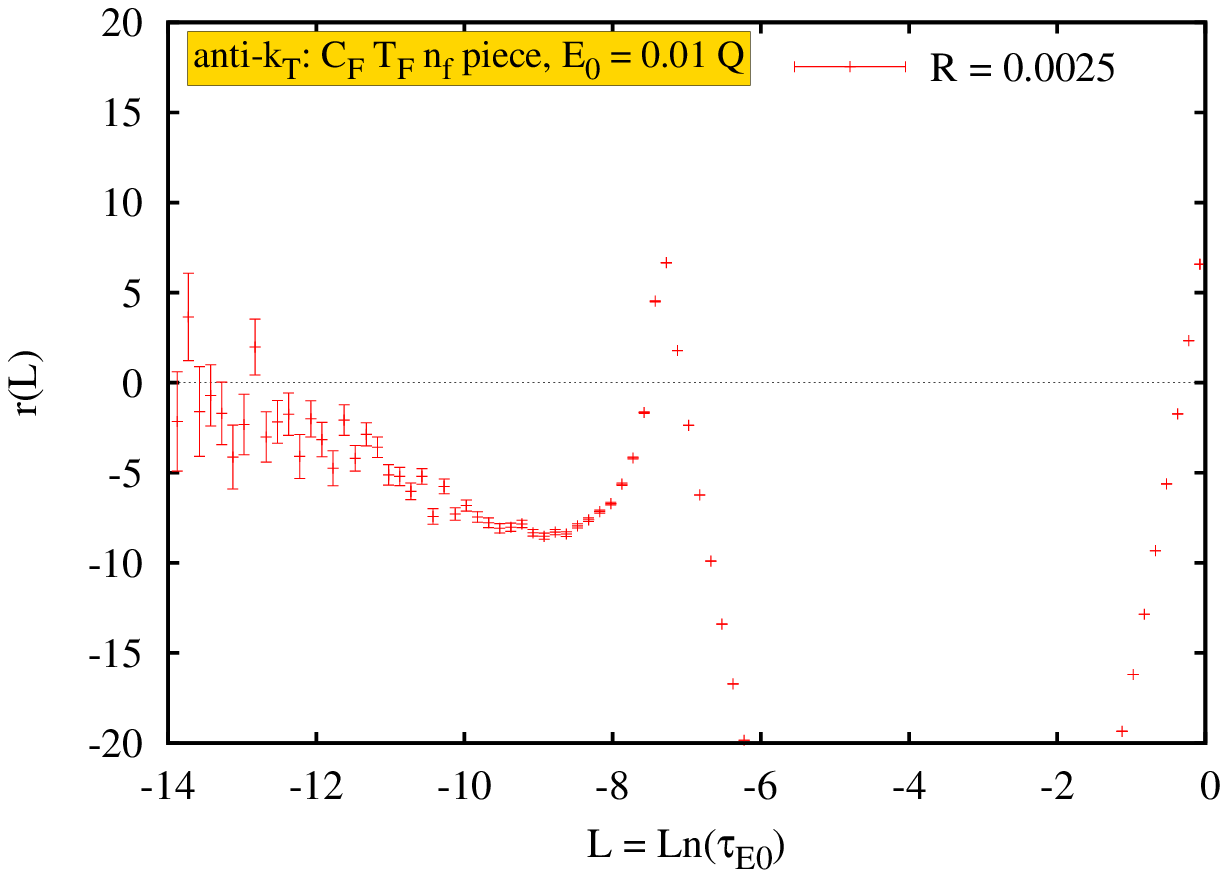, width= 0.48 \textwidth}
 \epsfig{ file=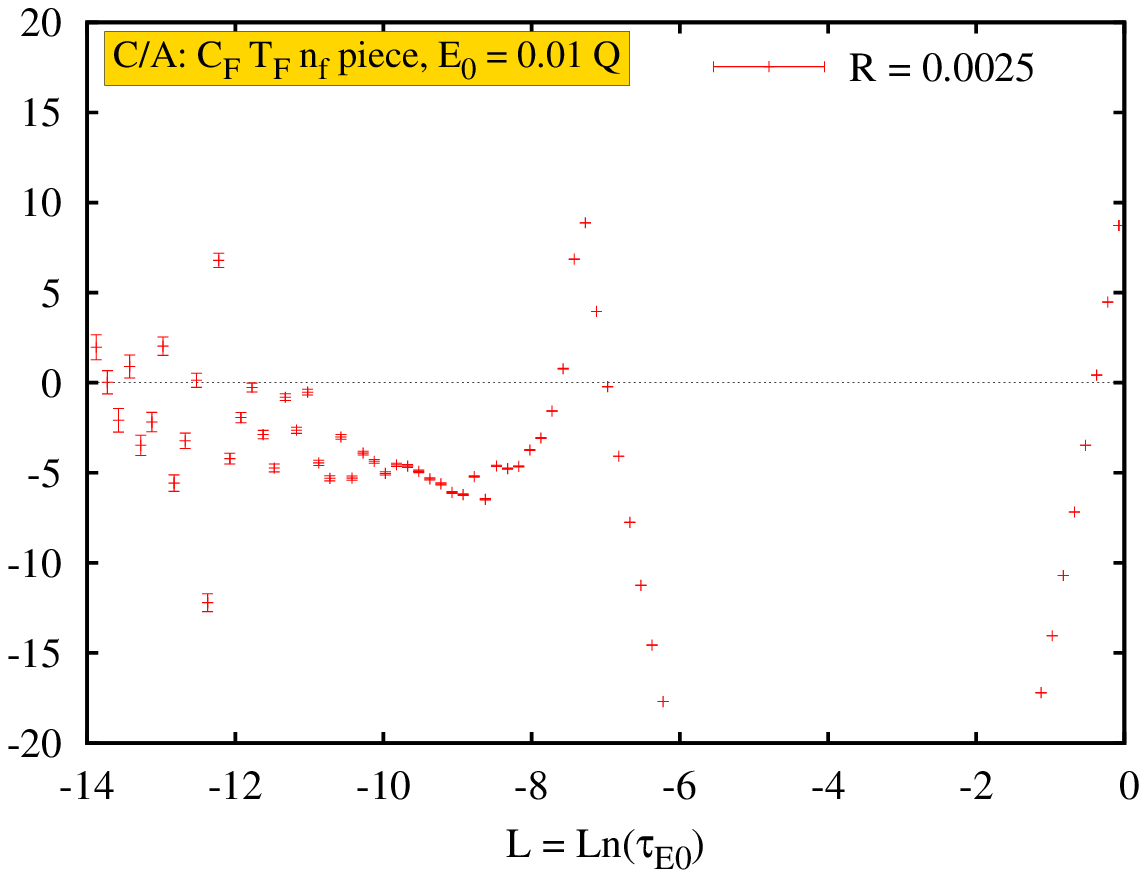, width= 0.48 \textwidth}
 \caption{The difference between \event and $\tauo$ NLO distribution for various jet radii in both \AKT (left) and C/A (right) algorithms. Plots are for $\CF\TF\nf$ channel.}
 \label{fig.NLO_CFTF}
\end{figure}
%

%% file: ch6/ch6app.tex

\chapter{Full $R$-dependence of clustering coefficients}

\label{Sec:F2}

Here we present a calculation of the dependence of the coefficient
of the clustering logs $\mc F_2$ away from the small-angles (thus
small-$R$) approximation. To do so, it is easier to work with
transverse momentum, (pseudo-)rapidity\footnote{Recall that $\eta =
-\ln \tan (\theta/2)$ and $k_t = \omega\sin\theta$.} and azimuthal
angle with respect to the \emph{beam} axis, $(k_t, \eta, \phi)$,
variables instead of energy and polar angles, as performed in sec.
\ref{sec:2-gluon}. We also specialise to the threshold limit in
which the trigged jet is created at $90^\circ$ to the beam (which is
along the $z$-axis). Our calculations can straightforwardly be
extended to the case where the triggered jet is at an arbitrary
rapidity. We parametrise the outgoing
four-momenta as:
\begin{eqnarray}
p_1 & = & \frac{Q}{2}(1,1,0,0),\nonumber\\
p_2 & = & \frac{Q}{2}(1,-1,0,0),\nonumber\\
k_i & = & k_{ti}(\cosh\eta_i,\cos\phi_i,\sin\phi_i,\sinh\eta_i),
\end{eqnarray}
with $i = 1, 2$ for the two gluons respectively. In terms of the new
variables, the clustering function (\eq{eq:ClustFun2}) reads:
\begin{equation}
\Xi_2 (k_1,k_2) = \Theta(d_{1j}-R^2)\Theta(R^2-d_{2j})
\Theta(d_{2j}-d_{12}),
\end{equation}
with
\begin{equation}
d_{1j} = \eta_1^2+\phi_1^2,\qquad d_{2j} = \eta_2^2+\phi_2^2,\qquad
d_{12} = (\eta_1-\eta_2)^2+(\phi_1-\phi_2)^2.
\end{equation}

In this coordinate system the jet mass becomes:
\begin{equation}
\rho = \frac{2 p_1.k_i}{Q^2/4} =
\frac{4k_{ti}}{Q}(\cosh\eta_i-\cos\phi_i) = 2x_i \left(1-\frac{ \cos
\phi_i}{\cosh \eta_i}\right),
\end{equation}
when particle $k_i$ is recombined with the triggered $p_1$ jet. In
the above we expressed the jet mass in terms of the energy fraction
$x_i = 2\omega_i/Q$, with $\omega_i = k_{ti}\cosh \eta_i$.

The probability of a single virtual soft gluon correction is given
by:
\begin{equation}
d\Gamma_i = - \frac{\d^3 \vec{k}_i}{2\omega_i(2\pi)^3} \gs^2 \CF
\frac{2(p_1.p_2)}{(p_1.k_i)(p_2.k_i)} = -\frac{\CF\alpha_s}{\pi}
\frac{\d x_i}{x_i}d\eta\frac{\d\phi_i}{\pi} \frac{1}{\cosh^2\eta_i-\cos^2\phi_i},
\end{equation}
where we note here that the collinear limit to the triggered jet
$p_1$ corresponds to $\eta_i\to 0$ \emph{and} $\phi_i \to 0$.

Thus we can write the correction term due to clustering (Eq.
\eqref{eq:C2generalFormula}) as:
\begin{equation}
\Sigma^{\rm{clus}}_2 = \frac{1}{2!}\int \prod_i^2 \d\Gamma_i
\Theta\left(x_i- \frac{\cosh\eta_i}{2(\cosh\eta_i-\cos\phi_i)}\rho\right)
\Xi_2(k_1,k_2).
\end{equation}
Performing the energy-fraction integration yields an expression
identical to \eq{eq:C2} with the full $R$-dependent
coefficient, $\mc F_2(R)$, given by:
\begin{equation}\label{eq:F2RInt}
\mc F_2(R) = \frac{1}{\pi^2}\int \d\eta_1\d\phi_1\d\eta_2\d\phi_2
\frac{1}{\cosh^2\eta_1-\cos^2\phi_1}
\frac{1}{\cosh^2\eta_2-\cos^2\phi_2} \Xi_2(k_1,k_2).
\end{equation}

\begin{figure}[!t]
\centering
\epsfig{file=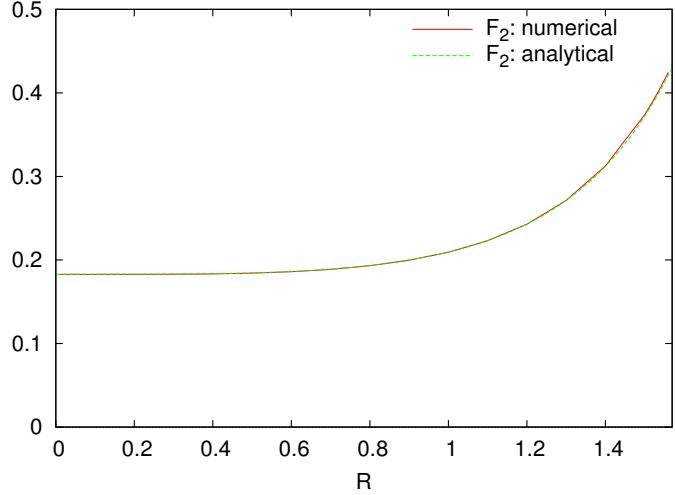,width=0.7\textwidth}
\caption{Full $R$-dependence of the two-gluon clustering
coefficient $\mc F_2$.\label{fig:F2Full}}
\end{figure}

This integration can be performed numerically to extract the value
of $\mc F_2$ for arbitrary $R$. However it proves useful, in the
simple case of $\mc F_2$ in order to obtain an analytic expression
at least as a power-series, to introduce the polar variables $r$ and
$\alpha$ defined by:
\begin{equation}
\eta = r \cos\alpha, \qquad \phi = r \sin\alpha.
\end{equation}
We rewrite \eq{eq:F2RInt} as:
\begin{multline}
\mc F_2 = \frac{1}{\pi^2}\int_R^\infty r_1\d r_1
\int_{-\pi}^{\pi}\d\alpha_1 \int_{0}^R r_2 \d r_2 \int_{-\pi}^{\pi}
\d\alpha_2 \Theta(2r_2\cos(\alpha_2-\alpha_1)-r_1)\times\\
\times \frac{1}{\cosh^2(r_1\cos\alpha_1)-\cos^2(r_1\sin\alpha_1)}
\,\,\frac{1}{\cosh^2(r_2\cos\alpha_2)-\cos^2(r_2\sin\alpha_2)}.
\label{eq:F2RInt2}
\end{multline}

We note that one can expand the second line of Eq. \eqref{eq:F2RInt2} in powers of $r_i$ so as to write the result as a
power-series in $R$. To first order, the second line expands as
$1/r_1^2 1/r_2^2$. We thus immediately identify this integral as the
one for the small-angles approximation, whose result is $\pi^2/54
\approx 0.183$. Performing higher-order integrals of the expansion
of the integrand yields the following result:
\begin{equation} \label{eq:F2RResult}
\mc F_2(R) = 0.183 + 0.0246 R^4 + 0.00183 R^8 + 0.000135 R^{12} +
\mc O(R^{16}).
\end{equation}
This expansion is actually valid for values of $R$ up to order
unity, a claim which is backed-up by fully performing the integral
\eqref{eq:F2RInt2} numerically via Monte Carlo methods and comparing
to the analytical estimate \eqref{eq:F2RResult}, Fig.
\ref{fig:F2Full}. We note that the clustering coefficient varies
very slowly with $R$.

\renewcommand{\kt}{k_t}
\begin{table}[!t]
\centering
\begin{tabular}{|c|c|c|c|c|c|c|}
\hline
 $R$         & $0$      & $0.1$    & $0.4$    & $0.7$    & $1.0$    & $1.2$ \\
\hline\hline
 $\mc F_2$   & $0.183$  & $0.183$  & $0.184$  & $0.188$  & $0.208$  & $0.242$ \\
\hline
$\mc F_3$    & $-0.052$ & $-0.053$ & $-0.053$ & $-0.055$ & $-0.061$ & $-0.072$ \\
\hline
$\mc F_3^\ca$& $-0.028$ & $-0.029$ & $-0.029$ & $-0.029$ & $-0.030$ & $-0.031$ \\
\hline
$\mc F_4$    & $0.022$  & $0.023$  & $0.023$  & $0.023$  & $0.024$  & $0.027$ \\
\hline
\end{tabular}
\label{tab:ClusCoeffNumericalValues} \caption{Estimates of the
clustering coefficients $\mc F_{n}$ for $n=2,3, 4$ at different
values of the jet radius $R$ in the $\kt$ and C/A algorithms. Recall
that $\mc F_2^\ca = \mc F_2^{\kt} (\equiv \mc F_2)$ and that $\mc
F_4^\ca$ has not been computed. Note that $\mc F_3^\ca$ and $\mc
F_4$ are more difficult to evaluate numerically than $\mc F_3$ (and
obviously $\mc F_2$). Thus their final answers are subjected to
larger errors.}
\end{table}

Finally we note that the calculation of $\mc F_3^{(\ca)}$ and $\mc
F_4$ may be performed in the same way. Schematically we have:
\begin{equation}\label{eq:FnRInt}
\mc F_n^{(\ca)} (R) = \frac{1}{\pi^n} \int \prod_i^n \d\eta_i\d\phi_i
\frac{1}{\cosh^2\eta_i-\cos^2\phi_i}
\Xi_n^{(\ca)}(k_1,k_2,\cdots,k_n),
\end{equation}
where the step function $\Xi_n^{(\ca)}$ is expressed in terms of the
distances $d_{i}=\eta_i^2+\phi_i^2$ and $d_{ij} = \delta \eta_{ij}^2
+ \delta \phi_{ij}^2$, respectively replacing $\theta_i$ and
$\theta_{ij}$ in, e.g., \eq{eq:ClustFun3kt}. We provide full
numerical estimates for all coefficients in table
\ref{tab:ClusCoeffNumericalValues}.

%% file: ch7/ch7app.tex

\chapter{Radiators}

\section{Dipole calculations for the global part}
\label{app:global}

In order to carry out the calculations for the various dipole terms required for the resummation detailed in the main text we shall make use of the following 
Born kinematics for the hard incoming and outgoing particles,
\begin{eqnarray}
p_1 &=& \frac{\sqrt{s}}{2} x_1 \left (1,0,0,1 \right), \nonumber \\
p_2 &=& \frac{\sqrt{s}}{2} x_2 \left (1,0,0,-1 \right), \nonumber \\
p_3 &=& p_t \left( \cosh y, 1,0,\sinh y \right), \nonumber \\
p_4 &=& p_t \left(\cosh y_r,-1,0,\sinh y_r \right),
\end{eqnarray}
where, as in the main, text $p_1$ and $p_2$ denote the four-momenta of the incoming hard partons, $p_3$ that of the jet whose mass is being considered and $p_4$ the four-momentum of a recoiling jet. In the case of a final state vector boson instead of a recoiling jet the contribution from $p_4$ to the dipole calculations is of course absent as the Z boson is colour neutral. Further we have indicated by $x_1$ and $x_2$ the fractional momenta of incoming partons relative to the colliding hadrons $P_1$ and $P_2$ respectively. Also $y$ denotes the rapidity of the measured jet, $y_r$ that of the recoiling jet and $p_t$ their transverse momenta. One can, by ignoring recoil, use the Born kinematics 
outlined above even in the presence of a soft emission, whose momentum can be expressed as 
\be
k=  k_t \left(\cosh \eta, \cos \phi, \sin \phi, \sinh \eta \right); 
\ee 
where transverse momentum $k_t$, rapidity $\eta$ and azimuth $\phi$ are all measured with respect to the beam direction. 

We shall now set up the various dipole contributions to the jet mass cross-section. We recall from the main text that eventually these dipole contributions will exponentiate with appropriate colour factors, so what follows below is in essence a calculation of various components of the resummed exponent.

\subsection{In-in dipole}

The in-in dipole is made from incoming partons $1$ and $2$ and provides the 
contribution 
\begin{eqnarray}
\mathcal{R}_{12} &=& \int k_t \d k_t \d\eta \frac{\d\phi}{2\pi} \frac{\alpha_s \left (\kappa_{t,12} \right) }{2 \pi} \frac{(p_1.p_2)}{(p_1.k)(p_2.k)}\Theta \left 
( \frac{2 k_t}{p_t} \left (\cosh (\eta-y) -\cos \phi \right) -v \right) \nonumber \\ &&
\Theta \left (R^2 -\left( (\eta-y)^2+\phi^2 \right) \right)
\end{eqnarray}
We note that $\kappa^2_{t,12} =  2 (p_1.k)(p_2.k)/(p_1.p_2 ) = k_t^2$ and carry out the integration over $\eta$ and $\phi$. In doing so we can neglect the dependence of the jet mass on $\eta,\phi$ and just retain its dependence on $k_t$ since the coefficient of $k_t$ which contains the $\eta,\phi$ dependence, produces terms that are below single logarithmic accuracy. Performing the integral over $\eta$ and $\phi$ over the interior of the jet region gives us to the relevant single-log accuracy that 
\begin{equation}
\mathcal{R}_{12} = {R^2} \int_{v p_t}^{p_t} \frac{\alpha_s (k_t)}{2 \pi} \frac{\d k_t}{k_t},
\end{equation}
where the lower limit of integration stems from the constraint on the jet-mass variable. 
We note that the dipole consisting of the two incoming partons gives rise to a pure single-logarithmic behaviour as is evident from the fact that the emitted gluon is inside the jet region, away from the hard legs constituting the dipole and hence there are no collinear enhancements. The soft wide-angle single logarithm we obtain is accompanied by an $R^2$ dependence on jet radius, reflecting the integration over the jet interior.

\subsection{In-recoil dipoles}

As before for the in-in case, for a dipole made up of an incoming parton and one recoiling against a measured jet, we have that the emitted gluon is in the interior of the jet away from the emitting legs,  and produces no collinear logarithm. We have that 
\begin{eqnarray}
\mathcal{R}_{14} &=& \int k_t \d k_t \d\eta \frac{\d\phi}{2\pi} \frac{\alpha_s \left (\kappa_{t,14} \right) }{2 \pi} \frac{(p_1.p_4)}{(p_1.k)(p_4.k)}\Theta \left 
( \frac{2 k_t}{p_t} \left (\cosh (\eta-y) -\cos \phi \right) -v \right)  \nonumber \\ &&
\Theta \left (R^2 -\left( (\eta-y)^2+\phi^2 \right) \right).
\end{eqnarray}
However , in this case we have that 
\begin{equation}
\kappa_{t,14}^2 = \frac{2(p_1.k)(p_4.k)}{(p_1.p_4)} =2 k_t^2 e^{y_r-\eta} \left( \cosh(\eta-y_r)+\cos \phi \right).
\end{equation}
To single logarithmic accuracy, we can write the result into the form
\begin{equation}
\int_{v}^{1} \frac{\d x}{x}\alpha_s (x p_t)  \int \frac{e^{\eta-y_r}}{\cosh(\eta-y_r)+\cos \phi} \Theta \left (R^2 -\left( (\eta-y)^2+\phi^2 \right) \right). 
\end{equation}
It is possible to evaluate the integral over $\eta,\phi$ as a power series in $R$, such that the overall result for this dipole reads
\begin{equation}
\int_v^1\frac{\d x}{x}\frac{\alpha_s (x p_t)}{2 \pi}  \left( \frac{R^2}{2} \frac{e^{\Delta  y}}{\left(1+\cosh \Delta y \right)}+\frac{R^4}{32} {\mathrm{sech}}^4 \frac{\Delta y}{2}+\mathcal{O} \left(R^6 \right) \right)
\end{equation}
where $\Delta y = y- y_r$
For the $\mathcal{R}_{24}$ dipole one obtains the same result as above, with the replacement of $\Delta y$ by $-\Delta y$.

\subsection{Jet-recoil dipole}

For this dipole we have also an involvement of the parton that initiates the 
measured jet as one of the hard emitting legs. The consequent collinear singularity, in addition to the pole due to soft emission, gives rise to 
double-logarithmic contributions. In this case one needs to evaluate the argument of the running coupling more carefully whereas for soft wide-angle pieces, dealt with previously,  all that is important at our accuracy is that the argument of the running coupling is proportional to the energy of the soft gluon being considered. 

In the present case we start with the integral 
\begin{eqnarray}
\label{eq:jetreco}
\mathcal{R}_{34} &=& \int k_t \d k_t \d\eta \frac{\d\phi}{2\pi} \frac{\alpha_s \left (\kappa_{t,34} \right) }{2 \pi} \frac{(p_3.p_4)}{(p_3.k)(p_4.k)} \Theta \left 
( \frac{2 k_t}{p_t} \left (\cosh (\eta-y) -\cos \phi \right) -v \right) \nonumber \\ &&
\Theta \left (R^2 -\left( (\eta-y)^2+\phi^2 \right) \right)
\end{eqnarray}
where
\begin{equation}
\kappa_{t,34}^2 = \frac{2(p_3.k)(p_4.k)}{(p_3.p_4)} = 2 k_t^2 
\frac{\left(\cosh \left(y_r-\eta \right) +\cos \phi \right) \left(\cosh \left(y-\eta \right)+\cos \phi) \right)}{1+\cosh \Delta y}.
\end{equation}
Next we introduce the variables $r$ and $\theta$ such that $y-\eta = r \cos \theta$ and $\phi = r \sin \theta$ such that the in jet condition $(y-\eta)^2+\phi^2<R^2$, simply produces $r < R$, while $\theta$ takes values between $0$ and $2\pi$. 
Thinking of \eq{eq:jetreco} as a power series in $r$ we can isolate the collinear-enhanced terms which produces double logarithms and separate it from 
terms that are finite as $r \to 0$, which upon integration produce terms that 
behave as powers of $R$ corresponding to soft wide-angle contributions.
The contribution $\mathcal{R}_{34}$ can thus be broken into terms that contain collinear enhancements (and hence leading double logarithms) and pure soft single-logarithmic pieces:
\begin{equation}
\mathcal{R}_{34} = \mathcal{R}_{34}^\mathrm{coll.}+\mathcal{R}_{34}^\mathrm{soft},
\end{equation}
where the pure soft terms can be expressed as a power series in $R$:
\begin{equation}
\mathcal{R}_{34}^\mathrm{soft} =  \int_v^1\frac{\alpha_s (x p_t)}{2 \pi} \frac{\d x}{x}
\left( \frac{R^2}{4} \tanh^2 \frac{\Delta y}{2}+\frac{R^4}{1152} \mathrm{sech}^4 \frac{\Delta y}{2} \left(-5+\cosh \Delta y \right)^2+\mathcal{O}(R^6) \right),
\end{equation}
where we have neglected terms that vanish as $v \to 0$.

The collinear contributions require us to be more precise in treating the 
argument of the running coupling. In particular, in the limit of small $r$, which is the collinear region, we obtain
\begin{equation}
\label{eq:fix}
 \mathcal{R}_{34}^{\mathrm{coll.}} = \int_v^{R^2}\frac{dr^2}{r^2} \frac{d\theta}{2 \pi} 
\int_{\frac{v}{R^2}}^1 \frac{\d x}{x} \frac{\alpha_s}{2 \pi} \left(x r p_t \right)
\end{equation}
where we have written the result in terms of a dimensionless energy fraction $x$ such that in the collinear limit $r \ll 1$, one has that $\kappa_{t,34}/(r p_t) = x$. The lower limits of the energy and angular integrals follow from the restriction on the jet mass to be greater than $v$, while the upper limit of the angular integral over $r^2$ corresponds to the jet boundary. 

It is straightforward to obtain the result corresponding to the fixed-coupling limit from Eq.~(\ref{eq:fix}), which gives: 
\begin{equation}
 \mathcal{R}_{34}^{\mathrm{coll.}}= \frac{\alpha_s}{2 \pi} \frac{1}{2} \ln^2 \frac{R^2}{v}.
\end{equation}

The running coupling result can be compared directly to that obtained in the $e^{+}e^{-}$ case in our previous work~\cite{Banfi:2010pa}. In order to do this we introduce the transverse momentum with respect to the emitting jet $k_{t,J} = x r p_t$, and changing variable we obtain

\begin{equation}
 \mathcal{R}_{34}^{\mathrm{coll.}}= \int \frac{\d k_{t,J}}{k_{t,J}} \frac{\alpha_s \left( k_{t,J} \right)}{2 \pi} 
 \frac{dr^2}{r^2} \Theta \left ( R^2-r^2\right) \Theta \left (r^2-v \right) \Theta \left (r^2 -\frac{k^2_{t,J}}{p_t^2} \right) \Theta \left (r^2 -\frac{v^2 p_t^2}{k_{t,J}^2} \right)
\end{equation}

Carrying out the integral over $r$, we are left to evaluate the integral 
over $k_{t,J}$ which reads 

\begin{multline}
 \mathcal{R}_{34}^{\mathrm{coll.}} = \int \frac{\d k_{t,J}^2}{k_{t,J}^2}  \frac{\alpha_s \left( k_{t,J} \right)}{2\pi}  \ln \left (\frac{R p_t}{k_{t,J}}   \right) \Theta \left (\frac{k_{t,J}^2}{p_t^2} -v \right) \Theta \left(R^2 -\frac{k_{t,J}^2}{p_t^2} \right)  \\ + \int  \frac{d k_{t,J}^2}{k_{t,J}^2} \frac{\alpha_s \left( k_{t,J}^2\right)}{2\pi} \ln \left (\frac{R k_{t,J}}{v p_{t}}   \right) \Theta \left (v-\frac{k_{t,J}^2}{p_t^2} \right ) \Theta \left(\frac{k_{t,J}^2}{p_t^2} -\frac{v^2}{R^2}\right)
\end{multline}

\subsection{In-jet dipoles}

Now we consider the dipole formed by the triggered jet and one of the incoming partons. Since here too the measured jet is involved as one of the legs of the hard dipoles, we can carry out precisely the same manipulations as for the $\mathcal{R}_{34}$ 
dipole immediately above. Doing so yields the same soft-collinear piece as before 
\begin{equation}
 \mathcal{R}_{13}^{\mathrm{coll.}}= \mathcal{R}_{23}^{\mathrm{coll.}}=\mathcal{R}_{34}^{\mathrm{coll.}},
\end{equation}
while the soft wide-angle piece reads
\begin{equation}
\mathcal{R}_{13}^{\mathrm{soft}} = \mathcal{R}_{23}^{\mathrm{soft}} = \int_{v}^{1} \frac{\alpha_s(x p_t)}{2 \pi} \frac{\d x}{x} \, \left( \frac{R^2}{4}+\frac{R^4}{288}+\mathcal{O}\left(R^6 \right) \right).
\end{equation}

\section{Dipole calculations for the non-global contribution} \label{app:nonglobal}

We now turn our attention to the evaluation of the different contributions arising from correlated-gluon emissions.

\subsection{Dipoles involving the measured jet}

We start by considering the the three dipoles involving the triggered jet $I_{13}$, $I_{23}$:
\begin{multline}
 I_{13} = 4 \int \d\eta_1 \frac{\d\phi_1}{2\pi} \int \d\eta_2 \frac{\d\phi_2}{2\pi} \\
\frac{1-e^{\eta_1}\cos\phi_1 - e^{\eta_2} \cos\phi_2 + e^{\eta_1 + \eta_2} \cos(\phi_1 - \phi_2)}{\cbr{\cosh(\eta_1 - \eta_2) - \cos(\phi_1-\phi_2)} \cbr{\cosh\eta_1 - \cos\phi_1} \cbr{\cosh\eta_2 - \cos\phi_2}} \\
\Theta\sbr{R^2 - (\eta_2^2 + \phi_2^2)} \Theta\sbr{\eta_1^2 + \phi_1^2 - R^2}.
\end{multline}
We evaluate the above integral in powers of the jet radius $R$. We find
\begin{equation}
 I_{13} = I_{23} = \frac{\pi^2}{3} + \A\,R^4 + \mc O(R^6).
\label{eq:app:HHJS1:I13_NGLs}
\end{equation}
As expected the $\mc O(R^0)$ term is the same result as in the hemisphere case. Moreover, we find that the $\mc O(R^2)$ is zero and the first non-vanishing correction to $\pi^2/3$ is $\mc O(R^4)$. A numerical evaluation of the integral leads to a small coefficient $\A = 0.013$.

The dipole $I_{34}$, which is relevant for the dijet calculation, depends, in principle, on the rapidity separation $\De y$ between the leading jets. However, any such dependence will be multiplied by, at least, $\mc O (R^4)$ and hence makes no significance contribution. Therefore $I_{34}$ assumes an analogous result to \eq{eq:app:HHJS1:I13_NGLs}.

\subsection{The remaining dipoles}
Let us consider the dipole from 2 incoming legs $I_{12}$.

We have to evaluate 
\begin{eqnarray}
I_{12}&=& 4\int \d\eta_1 \frac{\d\phi_1}{2\pi} \int \d \eta_2 \frac{\d\phi_2}{2\pi} 
\frac{\cos(\phi_1-\phi_2)}{\cosh(\eta_1-\eta_2)-\cos(\phi_1-\phi_2)} \Theta \left(R^2-(\eta_2^2+\phi_2^2)\right) \nonumber \\ &&\Theta \left(\eta_1^2+\phi_1^2-R^2 \right).
\end{eqnarray}
Let us change the variables to
\begin{eqnarray}
\eta_1-\eta_2 &=& \lambda \cos \beta \, , \, \phi_1-\phi_2 = \lambda \sin \beta \nonumber \\
\eta_2 &=& \rho \cos \alpha
\, , \, \phi_2 = \rho \sin \alpha.
\end{eqnarray}
Doing so and taking care of the fact that $-\pi < \phi_1 < \pi$ we get 
\begin{multline}
\int \frac{\cos \left(\lambda \sin \beta\right)}{\cosh \left( \lambda \cos \beta \right)-\cos \left(\lambda \sin \beta\right)}\Theta \left [\rho^2+\lambda^2+2 \rho \lambda\cos(\alpha-\beta) -R^2 \right] \Theta \left [R^2-\rho^2 \right]\\ 
\Theta\left [ \pi-(\rho \sin \alpha+\lambda \sin \beta) \right ]\Theta \left[ \pi+(\rho \sin \alpha+\lambda \sin \beta) \right]  \rho \d\rho \frac{\d\alpha}{2\pi} \lambda \d\lambda \frac{\d\beta}{2\pi} 
\end{multline}
The step function implies 
\begin{equation}
\cos (\alpha-\beta) > \frac{R^2-\rho^2-\lambda^2}{2\rho \lambda}
\end{equation}
This is automatically satisfied if $(R^2-\rho^2-\lambda^2)/(2 \rho \lambda) <-1$. It is never satisfied if $(R^2-\rho^2-\lambda^2)/(2 \rho \lambda) >1$. Hence, the step function in question is only active when $-1 <(R^2-\rho^2-\lambda^2)/(2 \rho \lambda) < 1$. For a non-zero result we must then have $\rho+\lambda>R$.

Let us consider first the region of integration where the first step function is active. We have two integrals
\begin{multline}
 \int_0^R\lambda \d\lambda\frac{\cos (\lambda \sin \beta)}{\cosh(\lambda \cos \beta)-\cos(\lambda \sin \beta)} \frac{\d\beta}{2\pi} \int_{R-\lambda}^{R}\rho \d\rho \frac{\d\alpha}{2\pi} \times \\ \times\Theta \left(\rho^2+\lambda^2+2\rho \lambda \cos(\alpha-\beta) -R^2 \right) +
\\
 + \int_R^{2 R}\lambda \d\lambda\frac{\cos (\lambda \sin \beta)}{\cosh(\lambda \cos \beta)-\cos(\lambda \sin \beta)} \frac{\d\beta}{2\pi} \int_{\lambda-R}^{R}\rho \d\rho \frac{\d\alpha}{2\pi} \times \\ \times \Theta \left(\rho^2+\lambda^2+2\rho \lambda \cos(\alpha-\beta) -R^2 \right),
\end{multline}
 where we have not written the other step functions that are automatically 
satisfied since $\lambda$ and $\rho$ are small here. The first integral yields 
$\approx 0.61 R^2$ while the second one is $\approx 0.37 R^2$.

 When the cosine condition is automatically satisfied we have the following range of values 
\begin{equation}
R< \lambda <2R , \, 0<\rho<\lambda-R 
\end{equation}
and 
\begin{equation}
 \lambda> 2R, \, 0< \rho< R.
\end{equation}
The first integral is
\begin{equation}
\int_R^{2R} \lambda \d\lambda\frac{\cos (\lambda \sin \beta)}{\cosh(\lambda \cos \beta)-\cos(\lambda \sin \beta)} \frac{\d\beta}{2\pi} \int_0^{\lambda-R}\rho \d\rho \frac{\d\alpha}{2\pi}
\end{equation}
Note that for small $R$ and the above range of values of $\rho$ all other theta functions are also satisfied as e.g it is always true that $\pi > \rho\sin \alpha+\lambda \sin \beta$. One can evaluate the integral above easily and the result is $\approx 0.19 R^2$.

The remaining integral is 
\begin{multline}
\int_{2R}^{\infty} \lambda \d\lambda\frac{\cos (\lambda \sin \beta)}{\cosh(\lambda \cos \beta)-\cos(\lambda \sin \beta)} \frac{\d\beta}{2\pi} \int_0^{R}\rho \d\rho \frac{d\alpha}{2\pi} 
\\ \Theta\left [ \pi-(\rho \sin \alpha+\lambda \sin \beta) \right ]\Theta \left[ \pi+(\rho \sin \alpha+\lambda \sin \beta) \right ]
\end{multline}
We note that we can split the $\lambda$ integral to go from $2 R$ to unity and then unity to infinity. The integral from $2R$ to unity produces the following result: 
$$-R^2 \ln 2R -0.12 R^2$$ which leaves to evaluate the integral 
\begin{multline}
\int_{1}^{\infty} \lambda \d\lambda\frac{\cos (\lambda \sin \beta)}{\cosh(\lambda \cos \beta)-\cos(\lambda \sin \beta)} \frac{\d\beta}{2\pi} \int_0^{R}\rho \d\rho \frac{\d\alpha}{2\pi} 
\\ \Theta\left [ \pi-(\rho \sin \alpha+\lambda \sin \beta) \right ]\Theta \left[ \pi+(\rho \sin \alpha+\lambda \sin \beta) \right ]
\end{multline}
This goes as $k R^2$ with $k$ a constant to be determined. To evaluate $k$ it proves simplest to first take the derivative with respect to$R^2$ which turns the theta function involving $\rho$ into a delta function. The $\rho$ integration is then trivial and the rest of the integral can be done numerically and yields $k=0.12$. We thus have that the overall result is $0.12 R^2$. Note that it is interesting that this cancels the $-0.12 R^2$ obtained previously. Hence the full result is 
\begin{multline}
I_{12}\approx 4\left[1.17R^2 -R^2 \ln 2R -0.12 R^2+0.12R^2 \right] + \Or(R^4) \\ \approx 4 \left[1.17 R^2 -R^2 \ln 2R\right]  + \Or(R^4).
\end{multline}
The above expansion well describes the full answer for $I_{12}$ up to values of $R\approx 0.6$. For larger values of the jet radius we use a full numerical evaluation of the coefficient $I_{12}$.

Next we consider the in-recoil dipole $I_{14}$. 
The result can be expressed in terms of the dipole $I_{12}$ and a rapidity dependent function $\kappa(\Delta y)$, which we evaluate numerically
\begin{equation}
\frac{\left(1+e^{\Delta y}\right)^2}{\left(1+\cosh \Delta y \right)^2} \left(1.17 R^2 -R^2 \ln 2 R\right)+\frac{(1+e^{\Delta y})}{1+\cosh \Delta y} \kappa (\Delta y) R^2.
\end{equation}
Finally, the dipole $I_{24}$ is easily obtained as 
\begin{equation}
I_{24}= I_{14}(-\Delta y).
\end{equation}

\section{Resummation formulae}
\ \label{app:resum}

In this section we collect the explicit expressions of the function $f_i$ which build up the resummed results:
\begin{eqnarray} \label{f_func}
f_1(\lambda) &=& - \frac{1}{2 \pi \beta_0 \lambda} \left [ \left(1-2 \lambda \right ) 
\ln \left(1-2\lambda \right)-2 \left ( 1-\lambda \right ) \ln \left
  (1-\lambda \right ) \right ],  \nonumber\\
f_2(\lambda) &=& - \frac{K}{4 \pi^2 \beta_0^2} \left [2 \ln \left 
(1-\lambda \right ) - \ln \left (1-2 \lambda \right )\right ] 
\nonumber \\  &&-\frac{ \beta_1}{2 \pi \beta_0^3} \left [ \ln \left (1-2\lambda \right )-2 \ln 
\left (1-\lambda \right ) + \frac{1}{2} \ln^2 \left (1- 2 \lambda \right ) 
- \ln^2 \left (1-\lambda \right ) \right ], \nonumber\\
f_{{\rm coll}, q}(\lambda)&=&- \frac{3}{4 \pi \beta_0} \ln \left ( 1-\lambda \right ), \nonumber \\
f_{{\rm coll}, g}(\lambda)&=&- \frac{1}{C_A} \ln \left ( 1-\lambda \right ), \nonumber \\
f_{{\rm l.a.}}(\lambda)&=&\frac{1}{4 \pi \beta_0} \ln (1-2 \lambda).
\end{eqnarray}
 $\lambda = \beta_0 \alpha_s L, \; L = \ln R^2/v$ and $\alpha_s =\alpha_s\left(p_T R \right)$ is the $\overline{\rm MS}$ strong coupling, which at aimed accuracy we need to consider at two-loops:
\begin{equation}
  \label{eq:twoloop-as}
  \alpha_s(k_t^2) = 
  \frac{\alpha_s(\mu^2)}{1-\rho}\left[1-\alpha_s(\mu^2)\frac{\beta_1}{\beta_0}
    \frac{\ln(1-\rho)}{1-\rho}\right] \,,\qquad
  \rho = \alpha_s(\mu^2) \beta_0 \ln\frac{\mu^2}{k_t^2}\,,
\end{equation}
where the coefficients of the QCD $\beta$-function are defined as
 \begin{equation}
\beta_0 = \frac{11 C_A - 2 n_f }{12 \pi}, \; \beta_1 = \frac{17 C_A^2 - 5 C_A n_f -3 C_F n_f}{24 \pi^2}\,,
\end{equation}
and the constant $K$ is given by
\begin{equation}
K = C_A \left (\frac{67}{18}- \frac{\pi^2}{6} \right ) - \frac{5}{9} n_f\,.
\end{equation}

Finally, we report the coefficients $a,b, c$ obtained  by fitting the functional form
\begin{equation}
f_{ij}(t)=  \frac{1+(a_{ij} t)^2}{1+(b_{ij}t)^{c_{ij}} } t^2
\end{equation}
to the numerical dipole evolution for non-global logarithms. The results for the other dipoles are reported in Tables~\ref{table1} and \ref{table2}. Note that the results for $I_{14}$ and $I_{24}$ depend on the rapidity separation between the jets and they have been computed for $|\Delta y|=2$.
\begin{table}[!h]
\begin{center}
\begin{tabular}{|c|c||c|c|c|}
\hline
 $R$   & $I_{12}$ & $a$     & $b$ & $c$ \\
\hline
 $0.1$ & $0.11$ & $1.90\,\CA$ & $0.31\,\CA$ & $1.66$ \\
\hline
 $0.2$ & $0.34$ & $1.35\,\CA$ & $0.10\,\CA$ & $1.66$ \\
\hline
 $0.3$ & $0.62$ & $1.07\,\CA$ & $0.00\,\CA$ & $1.66$ \\
\hline
 $0.4$ & $0.92$ & $0.90\,\CA$ & $0.00\,\CA$ & $1.66$ \\
\hline
 $0.5$ & $1.22$ & $1.45\,\CA$ & $1.02\,\CA$ & $1.66$ \\
\hline
 $0.6$ & $1.52$ & $0.85\,\CA$ & $0.00\,\CA$ & $1.33$ \\
\hline
 $0.7$ & $1.79$ & $0.88\,\CA$ & $0.19\,\CA$ & $1.33$ \\
\hline
 $0.8$ & $2.05$ & $1.09\,\CA$ & $0.53\,\CA$ & $1.33$ \\
\hline
 $0.9$ & $2.28$ & $1.20\,\CA$ & $0.73\,\CA$ & $1.33$ \\
\hline
 $1.0$ & $2.49$ & $1.12\,\CA$ & $0.66\,\CA$ & $1.33$ \\
\hline
 $1.1$ & $2.66$ & $1.13\,\CA$ & $0.70\,\CA$ & $1.33$ \\
\hline
 $1.2$ & $2.82$ & $1.26\,\CA$ & $0.89\,\CA$ & $1.33$ \\
\hline
\end{tabular}
\begin{tabular}{|c||c|c|c|}
\hline
 $I_{13}$& $a$         & $b$         & $c$ \\
\hline
 $3.289$   & $0.99\,\CA$ & $1.06\,\CA$ & $1.33$ \\
\hline
 $3.289$   & $0.99\,\CA$ & $1.06\,\CA$ & $1.33$ \\
\hline
 $3.289$   & $0.99\,\CA$ & $1.06\,\CA$ & $1.33$ \\
\hline
 $3.290$   & $0.94\,\CA$ & $0.99\,\CA$ & $1.33$ \\
\hline
 $3.290$   & $0.96\,\CA$ & $0.98\,\CA$ & $1.33$ \\
\hline
 $3.292$   & $0.86\,\CA$ & $0.87\,\CA$ & $1.33$ \\
\hline
 $3.293$   & $0.79\,\CA$ & $0.82\,\CA$ & $1.33$ \\
\hline
 $3.295$   & $0.79\,\CA$ & $0.81\,\CA$ & $1.33$ \\
\hline
 $3.299$   & $0.77\,\CA$ & $0.78\,\CA$ & $1.33$ \\
\hline
 $3.303$   & $0.86\,\CA$ & $0.85\,\CA$ & $1.33$ \\
\hline
 $3.310$   & $0.98\,\CA$ & $0.99\,\CA$ & $1.33$ \\
\hline
 $3.318$   & $0.91\,\CA$ & $0.92\,\CA$ & $1.33$ \\
\hline
\end{tabular}
\caption{Numerical results for the coefficients that parametrize the resummation of non-global logarithms.} \label{table1}
\end{center}
\end{table}
\begin{table}[!h]
\begin{center}
\begin{tabular}{|c|c||c|c|c|}
\hline
 $R$   & $I_{14}$ & $a$ & $b$ & $c$ \\
\hline
 $0.1$ & $0.09$ & $2.61\,\CA$ & $0.80\,\CA$ & $1.66$ \\
\hline
 $0.2$ & $0.28$ & $2.47\,\CA$ & $1.19\,\CA$ & $1.66$ \\
\hline
 $0.3$ & $0.51$ & $2.29\,\CA$ & $1.32\,\CA$ & $1.66$ \\
\hline
 $0.4$ & $0.77$ & $2.20\,\CA$ & $1.44\,\CA$ & $1.66$ \\
\hline
 $0.5$ & $1.04$ & $1.71\,\CA$ & $1.05\,\CA$ & $1.66$ \\
\hline
 $0.6$ & $1.31$ & $1.48\,\CA$ & $0.90\,\CA$ & $1.66$ \\
\hline
 $0.7$ & $1.57$ & $0.83\,\CA$ & $0.20\,\CA$ & $1.66$ \\
\hline
 $0.8$ & $1.81$ & $0.76\,\CA$ & $0.00\,\CA$ & $1.66$ \\
\hline
 $0.9$ & $2.04$ & $1.86\,\CA$ & $1.57\,\CA$ & $1.66$ \\
\hline
 $1.0$ & $2.25$ & $2.42\,\CA$ & $2.30\,\CA$ & $1.66$ \\
\hline
 $1.1$ & $2.44$ & $2.35\,\CA$ & $2.25\,\CA$ & $1.66$ \\
\hline
 $1.2$ & $2.61$ & $2.54\,\CA$ & $2.52\,\CA$ & $1.66$ \\
\hline
\end{tabular}
\begin{tabular}{|c||c|c|c|}
\hline
 $I_{24}$ & $a$ & $b$ & $c$ \\
\hline
 $0.003$& $1.12\,\CA$ & $0.91\,\CA$ & $1.66$ \\
\hline
 $0.01$ & $1.01\,\CA$ & $0.99\,\CA$ & $1.66$ \\
\hline
 $0.02$ & $3.18\,\CA$ & $0.49\,\CA$ & $1.66$ \\
\hline
 $0.03$ & $3.02\,\CA$ & $0.71\,\CA$ & $1.66$ \\
\hline
 $0.05$ & $2.85\,\CA$ & $0.72\,\CA$ & $1.66$ \\
\hline
 $0.07$ & $2.79\,\CA$ & $0.85\,\CA$ & $1.66$ \\
\hline
 $0.10$ & $2.66\,\CA$ & $0.88\,\CA$ & $1.66$ \\
\hline
 $0.13$ & $2.58\,\CA$ & $0.94\,\CA$ & $1.66$ \\
\hline
 $0.16$ & $2.58\,\CA$ & $1.06\,\CA$ & $1.66$ \\
\hline
 $0.20$ & $2.57\,\CA$ & $1.16\,\CA$ & $1.66$ \\
\hline
 $0.24$ & $2.53\,\CA$ & $1.22\,\CA$ & $1.66$ \\
\hline
 $0.30$ & $2.49\,\CA$ & $1.28\,\CA$ & $1.66$ \\
\hline
\end{tabular}
\caption{Numerical results for the coefficients that parametrize the resummation of non-global logarithms. Note that the above results are valid in the case $|\Delta y|=2$.} \label{table2}
\end{center}
\end{table}

%% file: Thesis.bbl
\begin{thebibliography}{100}
	
	\bibitem{Ellis:2007ib}
	S.~D. Ellis, J.~Huston, K.~Hatakeyama, P.~Loch, and M.~Tonnesmann, {\it {Jets
			in hadron-hadron collisions}},  {\em Prog.Part.Nucl.Phys.} {\bf 60} (2008)
	484--551, [\href{http://xxx.lanl.gov/abs/0712.2447}{{\tt arXiv:0712.2447}}].
	
	\bibitem{Dasgupta:2001sh}
	M.~Dasgupta and G.~P. Salam, {\it {Resummation of non-global QCD observables}},
	{\em Phys. Lett.} {\bf B512} (2001) 323--330,
	[\href{http://xxx.lanl.gov/abs/hep-ph/0104277}{{\tt hep-ph/0104277}}].
	
	\bibitem{Delenda:2006nf}
	Y.~Delenda, R.~Appleby, M.~Dasgupta, and A.~Banfi, {\it {On QCD resummation
			with $k_t$ clustering}},  {\em JHEP} {\bf 0612} (2006) 044,
	[\href{http://xxx.lanl.gov/abs/hep-ph/0610242}{{\tt hep-ph/0610242}}].
	
	\bibitem{Herten:2011qx}
	G.~Herten, {\it {The First Year of the Large Hadron Collider: A Brief Review}},
	{\em Mod.Phys.Lett.} {\bf A26} (2011) 843--855,
	[\href{http://xxx.lanl.gov/abs/1104.4205}{{\tt arXiv:1104.4205}}].
	
	\bibitem{Aad:2012gk}
	{\bf ATLAS Collaboration} Collaboration, G.~Aad et~al., {\it {Observation of a
			new particle in the search for the Standard Model Higgs boson with the ATLAS
			detector at the LHC}},  {\em Phys.Lett.B} (2012)
	[\href{http://xxx.lanl.gov/abs/1207.7214}{{\tt arXiv:1207.7214}}].
	
	\bibitem{Chatrchyan:2012gu}
	{\bf CMS Collaboration} Collaboration, S.~Chatrchyan et~al., {\it {Observation
			of a new boson at a mass of 125 GeV with the CMS experiment at the LHC}},
	{\em Phys.Lett.B} (2012) [\href{http://xxx.lanl.gov/abs/1207.7235}{{\tt
			arXiv:1207.7235}}].
	
	\bibitem{ellis2003qcd}
	R.~K. Ellis, W.~J. Stirling, and B.~R. Webber, {\em {QCD and collider
			physics}}.
	\newblock {Cambridge monographs on particle physics, nuclear physics, and
		cosmology}. {Cambridge University Press}, {2003}.
	
	\bibitem{Ellis:1988hv}
	S.~D. Ellis, Z.~Kunszt, and D.~E. Soper, {\it {The one jet inclusive
			cross-section at order $\alpha_s^3$: gluons only}},  {\em Phys.Rev.Lett.}
	{\bf 62} (1989) 726.
	
	\bibitem{Ellis:1989vm}
	S.~D. Ellis, Z.~Kunszt, and D.~E. Soper, {\it {The one jet inclusive
			cross-section at order $\alpha_s^3$. 1. gluons only}},  {\em Phys.Rev.} {\bf
		D40} (1989) 2188.
	
	\bibitem{Ellis:1990ek}
	S.~D. Ellis, Z.~Kunszt, and D.~E. Soper, {\it {The one jet inclusive
			cross-section at order $\alpha_s^{3}$ quarks and gluons}},  {\em
		Phys.Rev.Lett.} {\bf 64} (1990) 2121.
	
	\bibitem{Ellis:1992en}
	S.~D. Ellis, Z.~Kunszt, and D.~E. Soper, {\it {Two jet production in hadron
			collisions at order $\alpha_s^3$ in QCD}},  {\em Phys.Rev.Lett.} {\bf 69}
	(1992) 1496--1499.
	
	\bibitem{Dokshitzer:1995zt}
	Y.~L. Dokshitzer and B.~Webber, {\it {Calculation of power corrections to
			hadronic event shapes}},  {\em Phys.Lett.} {\bf B352} (1995) 451--455,
	[\href{http://xxx.lanl.gov/abs/hep-ph/9504219}{{\tt hep-ph/9504219}}].
	
	\bibitem{Dokshitzer:1995qm}
	Y.~L. Dokshitzer, G.~Marchesini, and B.~Webber, {\it {Dispersive approach to
			power behaved contributions in QCD hard processes}},  {\em Nucl.Phys.} {\bf
		B469} (1996) 93--142, [\href{http://xxx.lanl.gov/abs/hep-ph/9512336}{{\tt
			hep-ph/9512336}}].
	
	\bibitem{Butterworth:2008iy}
	J.~M. Butterworth, A.~R. Davison, M.~Rubin, and G.~P. Salam, {\it {Jet
			substructure as a new Higgs search channel at the LHC}},  {\em
		Phys.Rev.Lett.} {\bf 100} (2008) 242001,
	[\href{http://xxx.lanl.gov/abs/0802.2470}{{\tt arXiv:0802.2470}}].
	
	\bibitem{Agashe:2006hk}
	K.~Agashe, A.~Belyaev, T.~Krupovnickas, G.~Perez, and J.~Virzi, {\it {LHC
			Signals from Warped Extra Dimensions}},  {\em Phys.Rev.} {\bf D77} (2008)
	015003, [\href{http://xxx.lanl.gov/abs/hep-ph/0612015}{{\tt
			hep-ph/0612015}}].
	
	\bibitem{Lillie:2007yh}
	B.~Lillie, L.~Randall, and L.-T. Wang, {\it {The Bulk RS KK-gluon at the LHC}},
	{\em JHEP} {\bf 0709} (2007) 074,
	[\href{http://xxx.lanl.gov/abs/hep-ph/0701166}{{\tt hep-ph/0701166}}].
	
	\bibitem{Agashe:2007ki}
	K.~Agashe, H.~Davoudiasl, S.~Gopalakrishna, T.~Han, G.-Y. Huang, et~al., {\it
		{LHC Signals for Warped Electroweak Neutral Gauge Bosons}},  {\em Phys.Rev.}
	{\bf D76} (2007) 115015, [\href{http://xxx.lanl.gov/abs/0709.0007}{{\tt
			arXiv:0709.0007}}].
	
	\bibitem{Agashe:2007zd}
	K.~Agashe, H.~Davoudiasl, G.~Perez, and A.~Soni, {\it {Warped Gravitons at the
			LHC and Beyond}},  {\em Phys.Rev.} {\bf D76} (2007) 036006,
	[\href{http://xxx.lanl.gov/abs/hep-ph/0701186}{{\tt hep-ph/0701186}}].
	
	\bibitem{Brooijmans:2008se}
	G.~H. Brooijmans, A.~Delgado, B.~A. Dobrescu, C.~Grojean, M.~Narain, et~al.,
	{\it {New Physics at the LHC: A Les Houches Report. Physics at $\mathrm{TeV}$
			Colliders 2007 -- New Physics Working Group}},
	\href{http://xxx.lanl.gov/abs/0802.3715}{{\tt arXiv:0802.3715}}.
	
	\bibitem{Brooijmans:1077731}
	G.~Brooijmans, {\it High-$p_t$ hadronic top quark identification},  Tech. Rep.
	ATL-PHYS-CONF-2008-008. ATL-COM-PHYS-2008-001, CERN, Geneva, Jan, 2008.
	
	\bibitem{Butterworth:2002tt}
	J.~Butterworth, B.~Cox, and J.~R. Forshaw, {\it {$W W$ scattering at the CERN
			LHC}},  {\em Phys.Rev.} {\bf D65} (2002) 096014,
	[\href{http://xxx.lanl.gov/abs/hep-ph/0201098}{{\tt hep-ph/0201098}}].
	
	\bibitem{Flaugher:1990rv}
	B.~Flaugher and K.~Meier, {\it {A Compilation of jet finding algorithms}}, . To
	be publ. in Proc. of 1990 Summer Study on High Energy Physics, Research
	Directions for the Decade, Snowmass, CO, Jun 25 - Jul 13, 1990.
	
	\bibitem{PhysRevLett.39.1436}
	G.~Sterman and S.~Weinberg, {\it Jets from quantum chromodynamics},  {\em Phys.
		Rev. Lett.} {\bf 39} (Dec, 1977) 1436--1439.
	
	\bibitem{springerlink:10.1007/BF01410449}
	W.~Bartel and et~al, {\it {Experimental studies on multijet production in
			$e^{+} e^{-}$ annihilation at PETRA energies}},  {\em Zeitschrift für Physik
		C Particles and Fields} {\bf 33} (1986) 23--31. 10.1007/BF01410449.
	
	\bibitem{Bethke1988235}
	S.~Bethke, J.~Allison, and et~al, {\it {Experimental investigation of the
			energy dependence of the strong coupling strength}},  {\em Physics Letters B}
	{\bf 213} (1988), no.~2 235 -- 241.
	
	\bibitem{Abdesselam:2010pt}
	A.~Abdesselam et~al., {\it {Boosted objects: a probe of beyond the Standard
			Model physics}},  {\em Eur. Phys. J.} {\bf C71} (2011) 1661,
	[\href{http://xxx.lanl.gov/abs/1012.5412}{{\tt arXiv:1012.5412}}].
	
	\bibitem{Altheimer:2012mn}
	A.~Altheimer, S.~Arora, L.~Asquith, G.~Brooijmans, J.~Butterworth, et~al., {\it
		{Jet Substructure at the Tevatron and LHC: New results, new tools, new
			benchmarks}},  \href{http://xxx.lanl.gov/abs/1201.0008}{{\tt
			arXiv:1201.0008}}.
	
	\bibitem{Ellis:2009su}
	S.~D. Ellis, C.~K. Vermilion, and J.~R. Walsh, {\it {Techniques for improved
			heavy particle searches with jet substructure}},  {\em Phys.Rev.} {\bf D80}
	(2009) 051501, [\href{http://xxx.lanl.gov/abs/0903.5081}{{\tt
			arXiv:0903.5081}}].
	
	\bibitem{Ellis:2009me}
	S.~D. Ellis, C.~K. Vermilion, and J.~R. Walsh, {\it {Recombination Algorithms
			and Jet Substructure: Pruning as a Tool for Heavy Particle Searches}},  {\em
		Phys.Rev.} {\bf D81} (2010) 094023,
	[\href{http://xxx.lanl.gov/abs/0912.0033}{{\tt arXiv:0912.0033}}].
	
	\bibitem{Krohn:2009th}
	D.~Krohn, J.~Thaler, and L.-T. Wang, {\it {Jet Trimming}},  {\em JHEP} {\bf
		1002} (2010) 084, [\href{http://xxx.lanl.gov/abs/0912.1342}{{\tt
			arXiv:0912.1342}}].
	
	\bibitem{Banfi:2004yd}
	A.~Banfi, G.~P. Salam, and G.~Zanderighi, {\it {Principles of general
			final-state resummation and automated implementation}},  {\em JHEP} {\bf
		0503} (2005) 073, [\href{http://xxx.lanl.gov/abs/hep-ph/0407286}{{\tt
			hep-ph/0407286}}].
	
	\bibitem{Banfi:2010xy}
	A.~Banfi, G.~P. Salam, and G.~Zanderighi, {\it {Phenomenology of event shapes
			at hadron colliders}},  {\em JHEP} {\bf 1006} (2010) 038,
	[\href{http://xxx.lanl.gov/abs/1001.4082}{{\tt arXiv:1001.4082}}].
	
	\bibitem{ATLAS:2012am}
	{\bf ATLAS Collaboration} Collaboration, G.~Aad et~al., {\it {Jet mass and
			substructure of inclusive jets in $\sqrt{s} = 7\,\mathrm{TeV}$ pp collisions
			with the ATLAS experiment}},  {\em JHEP} {\bf 1205} (2012) 128,
	[\href{http://xxx.lanl.gov/abs/1203.4606}{{\tt arXiv:1203.4606}}].
	
	\bibitem{Aad:2012jf}
	{\bf ATLAS Collaboration} Collaboration, G.~Aad et~al., {\it {ATLAS
			measurements of the properties of jets for boosted particle searches}},
	\href{http://xxx.lanl.gov/abs/1206.5369}{{\tt arXiv:1206.5369}}.
	
	\bibitem{Hinzmann:2012zz}
	{\bf CMS Collaboration} Collaboration, A.~Hinzmann, {\it {Jet results from
			CMS}},  {\em Prog.Theor.Phys.Suppl.} {\bf 193} (2012) 249--253.
	
	\bibitem{ATL-PHYS-PUB-2009-088}
	{\bf ATLAS Collaboration} Collaboration, {\it Atlas sensitivity to the standard
		model higgs in the $\mathrm{HW}$ and $\mathrm{HZ}$ channels at high
		transverse momenta},  Tech. Rep. ATL-PHYS-PUB-2009-088, CERN, Geneva, Aug,
	2009.
	
	\bibitem{Plehn:2010st}
	T.~Plehn, M.~Spannowsky, M.~Takeuchi, and D.~Zerwas, {\it {Stop Reconstruction
			with Tagged Tops}},  {\em JHEP} {\bf 1010} (2010) 078,
	[\href{http://xxx.lanl.gov/abs/1006.2833}{{\tt arXiv:1006.2833}}].
	
	\bibitem{Berger:2002ig}
	C.~F. Berger, T.~Kucs, and G.~F. Sterman, {\it {Interjet energy flow / event
			shape correlations}},  {\em Int.J.Mod.Phys.} {\bf A18} (2003) 4159--4168,
	[\href{http://xxx.lanl.gov/abs/hep-ph/0212343}{{\tt hep-ph/0212343}}].
	
	\bibitem{Berger:2004xf}
	C.~F. Berger and L.~Magnea, {\it {Scaling of power corrections for angularities
			from dressed gluon exponentiation}},  {\em Phys.Rev.} {\bf D70} (2004)
	094010, [\href{http://xxx.lanl.gov/abs/hep-ph/0407024}{{\tt
			hep-ph/0407024}}].
	
	\bibitem{Catani:1992ua}
	S.~Catani, L.~Trentadue, G.~Turnock, and B.~R. Webber, {\it {Resummation of
			large logarithms in $e^+ e^-$ event shape distributions}},  {\em Nucl. Phys.}
	{\bf B407} (1993) 3--42.
	
	\bibitem{Bauer:2000yr}
	C.~W. Bauer, S.~Fleming, D.~Pirjol, and I.~W. Stewart, {\it {An effective field
			theory for collinear and soft gluons: Heavy to light decays}},  {\em Phys.
		Rev.} {\bf D63} (2001) 114020,
	[\href{http://xxx.lanl.gov/abs/hep-ph/0011336}{{\tt hep-ph/0011336}}].
	
	\bibitem{Bauer:2001yt}
	C.~W. Bauer, D.~Pirjol, and I.~W. Stewart, {\it {Soft-Collinear Factorization
			in Effective Field Theory}},  {\em Phys. Rev.} {\bf D65} (2002) 054022,
	[\href{http://xxx.lanl.gov/abs/hep-ph/0109045}{{\tt hep-ph/0109045}}].
	
	\bibitem{Beneke:2002ph}
	M.~Beneke, A.~Chapovsky, M.~Diehl, and T.~Feldmann, {\it {Soft collinear
			effective theory and heavy to light currents beyond leading power}},  {\em
		Nucl.Phys.} {\bf B643} (2002) 431--476,
	[\href{http://xxx.lanl.gov/abs/hep-ph/0206152}{{\tt hep-ph/0206152}}].
	
	\bibitem{Ellis:2009wj}
	S.~D. Ellis, A.~Hornig, C.~Lee, C.~K. Vermilion, and J.~R. Walsh, {\it
		{Consistent Factorization of Jet Observables in Exclusive Multijet
			Cross-Sections}},  {\em Phys. Lett.} {\bf B689} (2010) 82--89,
	[\href{http://xxx.lanl.gov/abs/0912.0262}{{\tt arXiv:0912.0262}}].
	
	\bibitem{Ellis:2010rwa}
	S.~D. Ellis, C.~K. Vermilion, J.~R. Walsh, A.~Hornig, and C.~Lee, {\it {Jet
			Shapes and Jet Algorithms in SCET}},  {\em JHEP} {\bf 11} (2010) 101,
	[\href{http://xxx.lanl.gov/abs/1001.0014}{{\tt arXiv:1001.0014}}].
	
	\bibitem{Becher:2008cf}
	T.~Becher and M.~D. Schwartz, {\it {A Precise determination of $\alpha_s$ from
			LEP thrust data using effective field theory}},  {\em JHEP} {\bf 07} (2008)
	034, [\href{http://xxx.lanl.gov/abs/0803.0342}{{\tt arXiv:0803.0342}}].
	
	\bibitem{Cheung:2009sg}
	W.~M.-Y. Cheung, M.~Luke, and S.~Zuberi, {\it {Phase Space and Jet Definitions
			in SCET}},  {\em Phys.Rev.} {\bf D80} (2009) 114021,
	[\href{http://xxx.lanl.gov/abs/0910.2479}{{\tt arXiv:0910.2479}}].
	
	\bibitem{Jouttenus:2009ns}
	T.~T. Jouttenus, {\it {Jet Function with a Jet Algorithm in SCET}},  {\em
		Phys.Rev.} {\bf D81} (2010) 094017,
	[\href{http://xxx.lanl.gov/abs/0912.5509}{{\tt arXiv:0912.5509}}].
	
	\bibitem{Chien:2010kc}
	Y.-T. Chien and M.~D. Schwartz, {\it {Resummation of heavy jet mass and
			comparison to LEP data}},  {\em JHEP} {\bf 08} (2010) 058,
	[\href{http://xxx.lanl.gov/abs/1005.1644}{{\tt arXiv:1005.1644}}].
	
	\bibitem{Kelley:2011tj}
	R.~Kelley, M.~D. Schwartz, and H.~X. Zhu, {\it {Resummation of jet mass with
			and without a jet veto}},  \href{http://xxx.lanl.gov/abs/1102.0561}{{\tt
			arXiv:1102.0561}}.
	
	\bibitem{Kelley:2011aa}
	R.~Kelley, M.~D. Schwartz, R.~M. Schabinger, and H.~X. Zhu, {\it {Jet Mass with
			a Jet Veto at Two Loops and the Universality of Non-Global Structure}},
	\href{http://xxx.lanl.gov/abs/1112.3343}{{\tt arXiv:1112.3343}}.
	
	\bibitem{Jouttenus:2011wh}
	T.~T. Jouttenus, I.~W. Stewart, F.~J. Tackmann, and W.~J. Waalewijn, {\it {The
			Soft Function for Exclusive N-Jet Production at Hadron Colliders}},  {\em
		Phys.Rev.} {\bf D83} (2011) 114030,
	[\href{http://xxx.lanl.gov/abs/1102.4344}{{\tt arXiv:1102.4344}}].
	
	\bibitem{Li:2011hy}
	H.-n. Li, Z.~Li, and C.-P. Yuan, {\it {QCD resummation for jet substructures}},
	{\em Phys.Rev.Lett.} {\bf 107} (2011) 152001,
	[\href{http://xxx.lanl.gov/abs/1107.4535}{{\tt arXiv:1107.4535}}].
	
	\bibitem{Li:2012bw}
	H.-n. Li, Z.~Li, and C.-P. Yuan, {\it {QCD resummation for light-particle
			jets}},  \href{http://xxx.lanl.gov/abs/1206.1344}{{\tt arXiv:1206.1344}}.
	
	\bibitem{Banfi:2001bz}
	A.~Banfi, G.~P. Salam, and G.~Zanderighi, {\it {Semi-numerical resummation of
			event shapes}},  {\em JHEP} {\bf 01} (2002) 018,
	[\href{http://xxx.lanl.gov/abs/hep-ph/0112156}{{\tt hep-ph/0112156}}].
	
	\bibitem{Catani:1996jh}
	S.~Catani and M.~H. Seymour, {\it {The Dipole Formalism for the Calculation of
			QCD Jet Cross Sections at Next-to-Leading Order}},  {\em Phys. Lett.} {\bf
		B378} (1996) 287--301, [\href{http://xxx.lanl.gov/abs/hep-ph/9602277}{{\tt
			hep-ph/9602277}}].
	
	\bibitem{Nason:2004rx}
	P.~Nason, {\it {A New method for combining NLO QCD with shower Monte Carlo
			algorithms}},  {\em JHEP} {\bf 0411} (2004) 040,
	[\href{http://xxx.lanl.gov/abs/hep-ph/0409146}{{\tt hep-ph/0409146}}].
	
	\bibitem{Nagy:2001fj}
	Z.~Nagy, {\it {Three jet cross-sections in hadron hadron collisions at
			next-to-leading order}},  {\em Phys.Rev.Lett.} {\bf 88} (2002) 122003,
	[\href{http://xxx.lanl.gov/abs/hep-ph/0110315}{{\tt hep-ph/0110315}}].
	
	\bibitem{Sterman:1995fz}
	G.~F. Sterman, {\it {Partons, factorization and resummation, TASI 95}},
	\href{http://xxx.lanl.gov/abs/hep-ph/9606312}{{\tt hep-ph/9606312}}.
	
	\bibitem{Gribov19831}
	L.~Gribov, E.~Levin, and M.~Ryskin, {\it Semihard processes in qcd},  {\em
		Physics Reports} {\bf 100} (1983), no.~1–2 1 -- 150.
	
	\bibitem{Bassetto1983201}
	A.~Bassetto, M.~Ciafaloni, and G.~Marchesini, {\it Jet structure and infrared
		sensitive quantities in perturbative qcd},  {\em Physics Reports} {\bf 100}
	(1983), no.~4 201 -- 272.
	
	\bibitem{Collins:1985ue}
	J.~C. Collins, D.~E. Soper, and G.~F. Sterman, {\it {Factorization for Short
			Distance Hadron - Hadron Scattering}},  {\em Nucl.Phys.} {\bf B261} (1985)
	104.
	
	\bibitem{Collins:1989gx}
	J.~C. Collins, D.~E. Soper, and G.~F. Sterman, {\it {Factorization of Hard
			Processes in QCD}},  {\em Adv.Ser.Direct.High Energy Phys.} {\bf 5} (1988)
	1--91, [\href{http://xxx.lanl.gov/abs/hep-ph/0409313}{{\tt hep-ph/0409313}}].
	To be publ. in 'Perturbative QCD' (A.H. Mueller, ed.) (World Scientific
	Publ., 1989).
	
	\bibitem{Collins:1988ig}
	J.~C. Collins, D.~E. Soper, and G.~F. Sterman, {\it {Soft Gluons and
			Factorization}},  {\em Nucl.Phys.} {\bf B308} (1988) 833.
	
	\bibitem{Banfi:2004nk}
	A.~Banfi, G.~P. Salam, and G.~Zanderighi, {\it {Resummed event shapes at hadron
			- hadron colliders}},  {\em JHEP} {\bf 0408} (2004) 062,
	[\href{http://xxx.lanl.gov/abs/hep-ph/0407287}{{\tt hep-ph/0407287}}].
	Erratum added online, nov/29/2004.
	
	\bibitem{Dasgupta:2003iq}
	M.~Dasgupta and G.~P. Salam, {\it {Event shapes in $e^+ e^-$ annihilation and
			deep inelastic scattering}},  {\em J. Phys.} {\bf G30} (2004) R143,
	[\href{http://xxx.lanl.gov/abs/hep-ph/0312283}{{\tt hep-ph/0312283}}].
	
	\bibitem{Dokshitzer1979234}
	Y.~Dokshitzer and D.~Dyakonov, {\it Angular distribution of energy in jets},
	{\em Physics Letters B} {\bf 84} (1979), no.~2 234 -- 236.
	
	\bibitem{Parisi1979427}
	G.~Parisi and R.~Petronzio, {\it Small transverse momentum distributions in
		hard processes},  {\em Nuclear Physics B} {\bf 154} (1979), no.~3 427 -- 440.
	
	\bibitem{Brown1990657}
	N.~Brown and W.~Stirling, {\it Jet cross sections at leading double logarithm
		in $e^+ e^-$ annihilation},  {\em Physics Letters B} {\bf 252} (1990), no.~4
	657 -- 662.
	
	\bibitem{Dasgupta:2002dc}
	{Dasgupta, Mrinal and Salam, Gavin P.}, {\it {Resummed event shape variables in
			DIS}},  {\em JHEP} {\bf 0208} (2002) 032,
	[\href{http://xxx.lanl.gov/abs/hep-ph/0208073}{{\tt hep-ph/0208073}}].
	
	\bibitem{Banfi:2002hw}
	A.~Banfi, G.~Marchesini, and G.~Smye, {\it {Away-from-jet energy flow}},  {\em
		JHEP} {\bf 08} (2002) 006,
	[\href{http://xxx.lanl.gov/abs/hep-ph/0206076}{{\tt hep-ph/0206076}}].
	
	\bibitem{Banfi:2003jj}
	A.~Banfi and M.~Dasgupta, {\it {Dijet rates with symmetric E$_t$ cuts}},  {\em
		JHEP} {\bf 0401} (2004) 027,
	[\href{http://xxx.lanl.gov/abs/hep-ph/0312108}{{\tt hep-ph/0312108}}].
	
	\bibitem{Cacciari:2008gp}
	M.~Cacciari, G.~P. Salam, and G.~Soyez, {\it {The anti-$k_t$ jet clustering
			algorithm}},  {\em JHEP} {\bf 04} (2008) 063,
	[\href{http://xxx.lanl.gov/abs/0802.1189}{{\tt arXiv:0802.1189}}].
	
	\bibitem{Banfi:2005gj}
	A.~Banfi and M.~Dasgupta, {\it {Problems in resumming interjet energy flows
			with $k_t$ clustering}},  {\em Phys. Lett.} {\bf B628} (2005) 49--56,
	[\href{http://xxx.lanl.gov/abs/hep-ph/0508159}{{\tt hep-ph/0508159}}].
	
	\bibitem{Appleby:2002ke}
	R.~B. Appleby and M.~H. Seymour, {\it {Non-global logarithms in inter-jet
			energy flow with $k_t$ clustering requirement}},  {\em JHEP} {\bf 12} (2002)
	063, [\href{http://xxx.lanl.gov/abs/hep-ph/0211426}{{\tt hep-ph/0211426}}].
	
	\bibitem{Buckley:2011ms}
	A.~Buckley, J.~Butterworth, S.~Gieseke, D.~Grellscheid, S.~Hoche, et~al., {\it
		{General-purpose event generators for LHC physics}},  {\em Phys.Rept.} {\bf
		504} (2011) 145--233, [\href{http://xxx.lanl.gov/abs/1101.2599}{{\tt
			arXiv:1101.2599}}].
	
	\bibitem{Gardi:2001di}
	E.~Gardi, {\it {Dressed gluon exponentiation}},  {\em Nucl.Phys.} {\bf B622}
	(2002) 365--392, [\href{http://xxx.lanl.gov/abs/hep-ph/0108222}{{\tt
			hep-ph/0108222}}].
	
	\bibitem{Ball:1995ni}
	P.~Ball, M.~Beneke, and V.~M. Braun, {\it {Resummation of $(\beta_0\,
			\alpha_s)^n$ corrections in QCD: Techniques and applications to the tau
			hadronic width and the heavy quark pole mass}},  {\em Nucl.Phys.} {\bf B452}
	(1995) 563--625, [\href{http://xxx.lanl.gov/abs/hep-ph/9502300}{{\tt
			hep-ph/9502300}}].
	
	\bibitem{Beneke:1998ui}
	M.~Beneke, {\it {Renormalons}},  {\em Phys. Rept.} {\bf 317} (1999) 1--142,
	[\href{http://xxx.lanl.gov/abs/hep-ph/9807443}{{\tt hep-ph/9807443}}].
	
	\bibitem{Dasgupta:2007wa}
	M.~Dasgupta, L.~Magnea, and G.~P. Salam, {\it {Non-perturbative QCD effects in
			jets at hadron colliders}},  {\em JHEP} {\bf 02} (2008) 055,
	[\href{http://xxx.lanl.gov/abs/0712.3014}{{\tt arXiv:0712.3014}}].
	
	\bibitem{Banfi:2001aq}
	A.~Banfi, G.~Marchesini, G.~Smye, and G.~Zanderighi, {\it {Out of plane QCD
			radiation in hadronic Z0 production}},  {\em JHEP} {\bf 0108} (2001) 047,
	[\href{http://xxx.lanl.gov/abs/hep-ph/0106278}{{\tt hep-ph/0106278}}].
	
	\bibitem{Dasgupta:2007hr}
	M.~Dasgupta and Y.~Delenda, {\it {Aspects of power corrections in hadron-hadron
			collisions}},  {\em JHEP} {\bf 0711} (2007) 013,
	[\href{http://xxx.lanl.gov/abs/0709.3309}{{\tt arXiv:0709.3309}}].
	
	\bibitem{Dokshitzer:1997ew}
	Y.~L. Dokshitzer and B.~Webber, {\it {Power corrections to event shape
			distributions}},  {\em Phys.Lett.} {\bf B404} (1997) 321--327,
	[\href{http://xxx.lanl.gov/abs/hep-ph/9704298}{{\tt hep-ph/9704298}}].
	
	\bibitem{Banfi:2010pa}
	A.~Banfi, M.~Dasgupta, K.~Khelifa-Kerfa, and S.~Marzani, {\it {Non-global
			logarithms and jet algorithms in high-$p_T$ jet shapes}},  {\em JHEP} {\bf
		08} (2010) 064, [\href{http://xxx.lanl.gov/abs/1004.3483}{{\tt
			arXiv:1004.3483}}].
	
	\bibitem{KhelifaKerfa:2011zu}
	K.~Khelifa-Kerfa, {\it {Non-global logs and clustering impact on jet mass with
			a jet veto distribution}},  {\em JHEP} {\bf 1202} (2012) 072,
	[\href{http://xxx.lanl.gov/abs/1111.2016}{{\tt arXiv:1111.2016}}].
	
	\bibitem{Delenda:2012mm}
	Y.~Delenda and K.~Khelifa-Kerfa, {\it {On the resummation of clustering
			logarithms for non-global observables}},  {\em JHEP} {\bf 1209} (2012) 109,
	[\href{http://xxx.lanl.gov/abs/1207.4528}{{\tt arXiv:1207.4528}}].
	
	\bibitem{Dasgupta:2012hg}
	M.~Dasgupta, K.~Khelifa-Kerfa, S.~Marzani, and M.~Spannowsky, {\it {On jet mass
			distributions in Z+jet and dijet processes at the LHC}},  {\em JHEP} {\bf
		1210} (2012) 126, [\href{http://xxx.lanl.gov/abs/1207.1640}{{\tt
			arXiv:1207.1640}}].
	
	\bibitem{Fermi19491739}
	Y.-C. Fermi, E., {\it Are mesons elementary particles?},  {\em Physical Review}
	{\bf 76} (1949), no.~12 1739--1743. cited By (since 1996) 58.
	
	\bibitem{PhysRev.186.1656}
	M.~L\'evy and J.~Sucher, {\it Eikonal approximation in quantum field theory},
	{\em Phys. Rev.} {\bf 186} (Oct, 1969) 1656--1670.
	
	\bibitem{RevModPhys.67.157}
	G.~Sterman~\emph{et al}, {\it {Handbook of perturbative QCD}},  {\em Rev. Mod.
		Phys.} {\bf 67} (Jan, 1995) 157--248.
	
	\bibitem{sterman1993introduction}
	G.~Sterman, {\em {An Introduction to Quantum Field Theory}}.
	\newblock Cambridge University Press, 1993.
	
	\bibitem{zeidler2008quantum}
	E.~Zeidler, {\em {Quantum Field Theory II: Quantum Electrodynamics: A Bridge
			Between Mathematicians and Physicists}}.
	\newblock Quantum Field Theory: A Bridge Between Mathematicians and Physicists.
	Springer, 2008.
	
	\bibitem{peskin1995introduction}
	M.~Peskin and D.~Schroeder, {\em {An Introduction To Quantum Field Theory}}.
	\newblock Advanced Book Program. Addison-Wesley Publishing Company, 1995.
	
	\bibitem{weinberg2000quantum}
	S.~Weinberg, {\em {The Quantum Theory of Fields: Supersymmetry}}.
	\newblock No.~v. 3 in The Quantum Theory of Fields. Cambridge University Press,
	2000.
	
	\bibitem{zinn2002quantum}
	J.~Zinn-Justin, {\em {Quantum Field Theory and Critical Phenomena}}.
	\newblock International Series of Monographs on Physics. Clarendon Press, 2002.
	
	\bibitem{Sterman:2004pd}
	G.~F. Sterman, {\it {QCD and jets}},
	\href{http://xxx.lanl.gov/abs/hep-ph/0412013}{{\tt hep-ph/0412013}}.
	
	\bibitem{Sterman:2008kj}
	G.~F. Sterman, {\it {Some Basic Concepts of Perturbative QCD}},  {\em Acta
		Phys.Polon.} {\bf B39} (2008) 2151--2172,
	[\href{http://xxx.lanl.gov/abs/0807.5118}{{\tt arXiv:0807.5118}}].
	
	\bibitem{Soper:2000kt}
	D.~E. Soper, {\it {Basics of QCD perturbation theory}},
	\href{http://xxx.lanl.gov/abs/hep-ph/0011256}{{\tt hep-ph/0011256}}.
	
	\bibitem{georgi1999lie}
	H.~Georgi, {\em Lie Algebras in Particle Physics}.
	\newblock Frontiers in Physics. Perseus Books, Advanced Book Program, 1999.
	
	\bibitem{jones1998groups}
	H.~Jones, {\em Groups, Representations And Physics}.
	\newblock Insitute of Physics Pub., 1998.
	
	\bibitem{cornwell1997group}
	J.~Cornwell, {\em Group Theory in Physics: An Introduction}.
	\newblock Techniques of Physics. Academic Press, 1997.
	
	\bibitem{Hooft1974461}
	G.~Hooft, {\it A planar diagram theory for strong interactions},  {\em Nuclear
		Physics B} {\bf 72} (1974), no.~3 461 -- 473.
	
	\bibitem{Lonnblad:1992tz}
	L.~Lonnblad, {\it {ARIADNE version 4: A Program for simulation of QCD cascades
			implementing the color dipole model}},  {\em Comput.Phys.Commun.} {\bf 71}
	(1992) 15--31.
	
	\bibitem{Sterman:2005vn}
	G.~F. Sterman, {\it {Quantum chromodynamics}},
	\href{http://xxx.lanl.gov/abs/hep-ph/0512344}{{\tt hep-ph/0512344}}.
	
	\bibitem{bjorken1964relativistic}
	J.~Bjorken and S.~Drell, {\em Relativistic quantum mechanics}.
	\newblock International series in pure and applied physics. McGraw-Hill, 1964.
	
	\bibitem{Faddeev196729}
	L.~Faddeev and V.~Popov, {\it {Feynman diagrams for the Yang-Mills field}},
	{\em Physics Letters B} {\bf 25} (1967), no.~1 29 -- 30.
	
	\bibitem{Abers:1973qs}
	E.~Abers and B.~Lee, {\it {Gauge Theories}},  {\em Phys.Rept.} {\bf 9} (1973)
	1--141.
	
	\bibitem{collins1986renormalization}
	J.~Collins and J.~Collins, {\em {Renormalization: An Introduction to
			Renormalization, the Renormalization Group and the Operator-Product
			Expansion}}.
	\newblock Cambridge Monographs on Mathematical Physics. Cambridge University
	Press, 1986.
	
	\bibitem{Kataev:2012rf}
	A.~Kataev and S.~Larin, {\it {Analytical five-loop expressions for the
			renormalization group QED $\beta$-function in different renormalization
			schemes}},  \href{http://xxx.lanl.gov/abs/1205.2810}{{\tt arXiv:1205.2810}}.
	
	\bibitem{Bethke:2009jm}
	S.~Bethke, {\it {The 2009 World Average of $\alpha_s$}},  {\em Eur.Phys.J.}
	{\bf C64} (2009) 689--703, [\href{http://xxx.lanl.gov/abs/0908.1135}{{\tt
			arXiv:0908.1135}}].
	
	\bibitem{PhysRevD.8.3633}
	D.~J. Gross and F.~Wilczek, {\it {Asymptotically Free Gauge Theories. I}},
	{\em Phys. Rev. D} {\bf 8} (Nov, 1973) 3633--3652.
	
	\bibitem{Politzer:1974fr}
	H.~D. Politzer, {\it {Asymptotic Freedom: An Approach to Strong Interactions}},
	{\em Phys.Rept.} {\bf 14} (1974) 129--180.
	
	\bibitem{Bodwin:1984hc}
	G.~T. Bodwin, {\it {Factorization of the Drell-Yan Cross-Section in
			Perturbation Theory}},  {\em Phys.Rev.} {\bf D31} (1985) 2616.
	
	\bibitem{Collins:1987pm}
	J.~C. Collins and D.~E. Soper, {\it {The Theorems of Perturbative QCD}},  {\em
		Ann.Rev.Nucl.Part.Sci.} {\bf 37} (1987) 383--409.
	
	\bibitem{Gribov:1972ri}
	V.~Gribov and L.~Lipatov, {\it {Deep inelastic e p scattering in perturbation
			theory}},  {\em Sov.J.Nucl.Phys.} {\bf 15} (1972) 438--450.
	
	\bibitem{Lipatov:1974qm}
	L.~Lipatov, {\it {The parton model and perturbation theory}},  {\em
		Sov.J.Nucl.Phys.} {\bf 20} (1975) 94--102.
	
	\bibitem{Altarelli:1977zs}
	G.~Altarelli and G.~Parisi, {\it {Asymptotic Freedom in Parton Language}},
	{\em Nucl.Phys.} {\bf B126} (1977) 298.
	
	\bibitem{Dokshitzer:1977sg}
	Y.~L. Dokshitzer, {\it {Calculation of the Structure Functions for Deep
			Inelastic Scattering and $e^+ e^-$ Annihilation by Perturbation Theory in
			Quantum Chromodynamics.}},  {\em Sov.Phys.JETP} {\bf 46} (1977) 641--653.
	
	\bibitem{Moch:2004pa}
	S.~Moch, J.~Vermaseren, and A.~Vogt, {\it {The Three loop splitting functions
			in QCD: The Nonsinglet case}},  {\em Nucl.Phys.} {\bf B688} (2004) 101--134,
	[\href{http://xxx.lanl.gov/abs/hep-ph/0403192}{{\tt hep-ph/0403192}}].
	
	\bibitem{Vogt:2004mw}
	A.~Vogt, S.~Moch, and J.~Vermaseren, {\it {The Three-loop splitting functions
			in QCD: The Singlet case}},  {\em Nucl.Phys.} {\bf B691} (2004) 129--181,
	[\href{http://xxx.lanl.gov/abs/hep-ph/0404111}{{\tt hep-ph/0404111}}].
	
	\bibitem{Salam:2008qg}
	G.~P. Salam and J.~Rojo, {\it {A Higher Order Perturbative Parton Evolution
			Toolkit (HOPPET)}},  {\em Comput.Phys.Commun.} {\bf 180} (2009) 120--156,
	[\href{http://xxx.lanl.gov/abs/0804.3755}{{\tt arXiv:0804.3755}}].
	
	\bibitem{Salam:2010zt}
	G.~P. Salam, {\it {Elements of QCD for hadron colliders}},
	\href{http://xxx.lanl.gov/abs/1011.5131}{{\tt arXiv:1011.5131}}.
	
	\bibitem{Soper:1996sn}
	D.~E. Soper, {\it {Parton distribution functions}},  {\em
		Nucl.Phys.Proc.Suppl.} {\bf 53} (1997) 69--80,
	[\href{http://xxx.lanl.gov/abs/hep-lat/9609018}{{\tt hep-lat/9609018}}].
	
	\bibitem{Martin:2009iq}
	A.~Martin, W.~Stirling, R.~Thorne, and G.~Watt, {\it {Parton distributions for
			the LHC}},  {\em Eur.Phys.J.} {\bf C63} (2009) 189--285,
	[\href{http://xxx.lanl.gov/abs/0901.0002}{{\tt arXiv:0901.0002}}].
	
	\bibitem{Thorne:2009ky}
	R.~Thorne, A.~Martin, W.~Stirling, and G.~Watt, {\it {Status of MRST/MSTW PDF
			sets}},  \href{http://xxx.lanl.gov/abs/0907.2387}{{\tt arXiv:0907.2387}}.
	
	\bibitem{Catani:2011st}
	S.~Catani, D.~de~Florian, and G.~Rodrigo, {\it {Space-like (versus time-like)
			collinear limits in QCD: Is factorization violated?}},  {\em JHEP} {\bf 1207}
	(2012) 026, [\href{http://xxx.lanl.gov/abs/1112.4405}{{\tt
			arXiv:1112.4405}}].
	
	\bibitem{Collins:2007jp}
	J.~Collins, {\it {2-soft-gluon exchange and factorization breaking}},
	\href{http://xxx.lanl.gov/abs/0708.4410}{{\tt arXiv:0708.4410}}.
	
	\bibitem{Collins:2007nk}
	J.~Collins and J.-W. Qiu, {\it {$k_{T}$ factorization is violated in production
			of high-transverse-momentum particles in hadron-hadron collisions}},  {\em
		Phys.Rev.} {\bf D75} (2007) 114014,
	[\href{http://xxx.lanl.gov/abs/0705.2141}{{\tt arXiv:0705.2141}}].
	
	\bibitem{Forshaw:2012bi}
	J.~R. Forshaw, M.~H. Seymour, and A.~Siodmok, {\it {On the breaking of
			collinear factorization in QCD}},
	\href{http://xxx.lanl.gov/abs/1206.6363}{{\tt arXiv:1206.6363}}.
	
	\bibitem{Dokshitzer:2005ek}
	Y.~Dokshitzer and G.~Marchesini, {\it {Hadron collisions and the fifth
			form-factor}},  {\em Phys.Lett.} {\bf B631} (2005) 118--125,
	[\href{http://xxx.lanl.gov/abs/hep-ph/0508130}{{\tt hep-ph/0508130}}].
	
	\bibitem{Seymour:1995gq}
	M.~H. Seymour, {\it {Jets in QCD}},  {\em AIP Conf.Proc.} {\bf 357} (1996)
	568--587, [\href{http://xxx.lanl.gov/abs/hep-ph/9506421}{{\tt
			hep-ph/9506421}}].
	
	\bibitem{Campbell:2006wx}
	J.~M. Campbell, J.~Huston, and W.~Stirling, {\it {Hard Interactions of Quarks
			and Gluons: A Primer for LHC Physics}},  {\em Rept.Prog.Phys.} {\bf 70}
	(2007) 89, [\href{http://xxx.lanl.gov/abs/hep-ph/0611148}{{\tt
			hep-ph/0611148}}].
	
	\bibitem{Salam:2009jx}
	G.~P. Salam, {\it {Towards Jetography}},  {\em Eur.Phys.J.} {\bf C67} (2010)
	637--686, [\href{http://xxx.lanl.gov/abs/0906.1833}{{\tt arXiv:0906.1833}}].
	
	\bibitem{Zhang:2012rz}
	B.-W. Zhang, Y.~He, R.~Neufeld, I.~Vitev, and E.~Wang, {\it {Probing nuclear
			matter with jets}},  \href{http://xxx.lanl.gov/abs/1207.6558}{{\tt
			arXiv:1207.6558}}.
	
	\bibitem{ATLAS-TDR-015}
	{\em ATLAS detector and physics performance: Technical Design Report, 2}.
	\newblock Technical Design Report ATLAS. CERN, Geneva, 1999.
	\newblock Electronic version not available.
	
	\bibitem{Froidevaux:2006rg}
	D.~Froidevaux and P.~Sphicas, {\it {General-purpose detectors for the Large
			Hadron Collider}},  {\em Ann.Rev.Nucl.Part.Sci.} {\bf 56} (2006) 375--440.
	
	\bibitem{Aad:2009wy}
	{\bf ATLAS Collaboration} Collaboration, G.~Aad et~al., {\it {Expected
			Performance of the ATLAS Experiment - Detector, Trigger and Physics}},
	\href{http://xxx.lanl.gov/abs/0901.0512}{{\tt arXiv:0901.0512}}.
	
	\bibitem{Bayatian:2006zz}
	{\bf CMS Collaboration} Collaboration, G.~Bayatian et~al., {\it {CMS physics:
			Technical design report}}, .
	
	\bibitem{Buttar:2008jx}
	C.~Buttar, J.~D'Hondt, M.~Kramer, G.~Salam, M.~Wobisch, et~al., {\it {Standard
			Model Handles and Candles Working Group: Tools and Jets Summary Report}},
	\href{http://xxx.lanl.gov/abs/0803.0678}{{\tt arXiv:0803.0678}}.
	
	\bibitem{Huth:1990mi}
	J.~E. Huth, N.~Wainer, K.~Meier, N.~Hadley, F.~Aversa, et~al., {\it {Toward a
			standardization of jet definitions}}, .
	
	\bibitem{Sterman:1977wj}
	G.~F. Sterman and S.~Weinberg, {\it {Jets from Quantum Chromodynamics}},  {\em
		Phys.Rev.Lett.} {\bf 39} (1977) 1436.
	
	\bibitem{Salam:2007xv}
	G.~P. Salam and G.~Soyez, {\it {A practical Seedless Infrared-Safe Cone jet
			algorithm}},  {\em JHEP} {\bf 05} (2007) 086,
	[\href{http://xxx.lanl.gov/abs/0704.0292}{{\tt arXiv:0704.0292}}].
	
	\bibitem{Cacciari:2011ma}
	M.~Cacciari, G.~P. Salam, and G.~Soyez, {\it {FastJet user manual}},  {\em
		Eur.Phys.J.} {\bf C72} (2012) 1896,
	[\href{http://xxx.lanl.gov/abs/1111.6097}{{\tt arXiv:1111.6097}}].
	
	\bibitem{Bartel:1986ua}
	{\bf JADE Collaboration} Collaboration, W.~Bartel et~al., {\it {Experimental
			Studies on Multi-Jet Production in $e^+ e^-$ Annihilation at PETRA
			Energies}},  {\em Z.Phys.} {\bf C33} (1986) 23.
	
	\bibitem{Dorfan:1980gc}
	J.~Dorfan, {\it {A cluster algorithm for the study of jets in high-energy
			physics}},  {\em Z.Phys.} {\bf C7} (1981) 349.
	
	\bibitem{Sjostrand:1982am}
	T.~Sjostrand, {\it {The Lund Monte Carlo for $e^+ e^-$ Jet Physics}},  {\em
		Comput.Phys.Commun.} {\bf 28} (1983) 229.
	
	\bibitem{Dokshitzer:1997in}
	Y.~L. Dokshitzer, G.~Leder, S.~Moretti, and B.~Webber, {\it {Better jet
			clustering algorithms}},  {\em JHEP} {\bf 9708} (1997) 001,
	[\href{http://xxx.lanl.gov/abs/hep-ph/9707323}{{\tt hep-ph/9707323}}].
	
	\bibitem{Wobisch:1998wt}
	M.~Wobisch and T.~Wengler, {\it {Hadronization corrections to jet
			cross-sections in deep inelastic scattering}},
	\href{http://xxx.lanl.gov/abs/hep-ph/9907280}{{\tt hep-ph/9907280}}.
	
	\bibitem{Catani:1993hr}
	S.~Catani, Y.~L. Dokshitzer, M.~H. Seymour, and B.~R. Webber, {\it
		{Longitudinally invariant k$_t$ clustering algorithms for hadron hadron
			collisions}},  {\em Nucl. Phys.} {\bf B406} (1993) 187--224.
	
	\bibitem{Banfi:2006hf}
	A.~Banfi, G.~P. Salam, and G.~Zanderighi, {\it {Infrared safe definition of jet
			flavor}},  {\em Eur.Phys.J.} {\bf C47} (2006) 113--124,
	[\href{http://xxx.lanl.gov/abs/hep-ph/0601139}{{\tt hep-ph/0601139}}].
	
	\bibitem{Krohn:2009zg}
	D.~Krohn, J.~Thaler, and L.-T. Wang, {\it {Jets with Variable R}},  {\em JHEP}
	{\bf 0906} (2009) 059, [\href{http://xxx.lanl.gov/abs/0903.0392}{{\tt
			arXiv:0903.0392}}].
	
	\bibitem{Lonnblad:1992qd}
	L.~Lonnblad, {\it {ARCLUS: A New jet clustering algorithm inspired by the color
			dipole model}},  {\em Z.Phys.} {\bf C58} (1993) 471--478.
	
	\bibitem{Rubin:2010fc}
	M.~Rubin, {\it {Non-Global Logarithms in Filtered Jet Algorithms}},  {\em JHEP}
	{\bf 1005} (2010) 005, [\href{http://xxx.lanl.gov/abs/1002.4557}{{\tt
			arXiv:1002.4557}}].
	
	\bibitem{Kaplan:2008ie}
	D.~E. Kaplan, K.~Rehermann, M.~D. Schwartz, and B.~Tweedie, {\it {Top Tagging:
			A Method for Identifying Boosted Hadronically Decaying Top Quarks}},  {\em
		Phys.Rev.Lett.} {\bf 101} (2008) 142001,
	[\href{http://xxx.lanl.gov/abs/0806.0848}{{\tt arXiv:0806.0848}}].
	
	\bibitem{Giurgiu:2009wv}
	{\bf CMS collaboration} Collaboration, G.~Giurgiu, {\it {Reconstruction of High
			Transverse Momentum Top Quarks at CMS}},
	\href{http://xxx.lanl.gov/abs/0909.4894}{{\tt arXiv:0909.4894}}.
	
	\bibitem{Kribs:2009yh}
	G.~D. Kribs, A.~Martin, T.~S. Roy, and M.~Spannowsky, {\it {Discovering the
			Higgs Boson in New Physics Events using Jet Substructure}},  {\em Phys.Rev.}
	{\bf D81} (2010) 111501, [\href{http://xxx.lanl.gov/abs/0912.4731}{{\tt
			arXiv:0912.4731}}].
	
	\bibitem{Kribs:2010hp}
	G.~D. Kribs, A.~Martin, T.~S. Roy, and M.~Spannowsky, {\it {Discovering Higgs
			Bosons of the MSSM using Jet Substructure}},  {\em Phys.Rev.} {\bf D82}
	(2010) 095012, [\href{http://xxx.lanl.gov/abs/1006.1656}{{\tt
			arXiv:1006.1656}}].
	
	\bibitem{Krohn:2011zp}
	D.~Krohn, L.~Randall, and L.-T. Wang, {\it {On the Feasibility and Utility of
			ISR Tagging}},  \href{http://xxx.lanl.gov/abs/1101.0810}{{\tt
			arXiv:1101.0810}}.
	
	\bibitem{Soper:2011cr}
	D.~E. Soper and M.~Spannowsky, {\it {Finding physics signals with shower
			deconstruction}},  {\em Phys.Rev.} {\bf D84} (2011) 074002,
	[\href{http://xxx.lanl.gov/abs/1102.3480}{{\tt arXiv:1102.3480}}].
	
	\bibitem{Thaler:2008ju}
	J.~Thaler and L.-T. Wang, {\it {Strategies to Identify Boosted Tops}},  {\em
		JHEP} {\bf 0807} (2008) 092, [\href{http://xxx.lanl.gov/abs/0806.0023}{{\tt
			arXiv:0806.0023}}].
	
	\bibitem{Delsart:2012jm}
	P.-A. Delsart, K.~L. Geerlings, J.~Huston, B.~T. Martin, and C.~K. Vermilion,
	{\it {SpartyJet 4.0 User's Manual}},
	\href{http://xxx.lanl.gov/abs/1201.3617}{{\tt arXiv:1201.3617}}.
	
	\bibitem{Seymour:1997kj}
	M.~Seymour, {\it {Jet shapes in hadron collisions: Higher orders, resummation
			and hadronization}},  {\em Nucl.Phys.} {\bf B513} (1998) 269--300,
	[\href{http://xxx.lanl.gov/abs/hep-ph/9707338}{{\tt hep-ph/9707338}}].
	
	\bibitem{Thaler:2010tr}
	J.~Thaler and K.~Van~Tilburg, {\it {Identifying Boosted Objects with
			N-subjettiness}},  {\em JHEP} {\bf 1103} (2011) 015,
	[\href{http://xxx.lanl.gov/abs/1011.2268}{{\tt arXiv:1011.2268}}].
	
	\bibitem{Chekanov:2010vc}
	S.~Chekanov and J.~Proudfoot, {\it {Searches for $\mathrm{TeV}$-scale particles
			at the LHC using jet shapes}},  {\em Phys.Rev.} {\bf D81} (2010) 114038,
	[\href{http://xxx.lanl.gov/abs/1002.3982}{{\tt arXiv:1002.3982}}].
	
	\bibitem{Gallicchio:2010sw}
	J.~Gallicchio and M.~D. Schwartz, {\it {Seeing in Color: Jet Superstructure}},
	{\em Phys.Rev.Lett.} {\bf 105} (2010) 022001,
	[\href{http://xxx.lanl.gov/abs/1001.5027}{{\tt arXiv:1001.5027}}].
	
	\bibitem{Hook:2011cq}
	A.~Hook, M.~Jankowiak, and J.~G. Wacker, {\it {Jet Dipolarity: Top Tagging with
			Color Flow}},  {\em JHEP} {\bf 1204} (2012) 007,
	[\href{http://xxx.lanl.gov/abs/1102.1012}{{\tt arXiv:1102.1012}}]. 8 pages, 6
	figures (updated to JHEP version).
	
	\bibitem{Almeida:2008yp}
	L.~G. Almeida, S.~J. Lee, G.~Perez, G.~F. Sterman, I.~Sung, et~al., {\it
		{Substructure of high-$p_T$ Jets at the LHC}},  {\em Phys.Rev.} {\bf D79}
	(2009) 074017, [\href{http://xxx.lanl.gov/abs/0807.0234}{{\tt
			arXiv:0807.0234}}].
	
	\bibitem{Almeida:2008tp}
	L.~G. Almeida, S.~J. Lee, G.~Perez, I.~Sung, and J.~Virzi, {\it {Top Jets at
			the LHC}},  {\em Phys.Rev.} {\bf D79} (2009) 074012,
	[\href{http://xxx.lanl.gov/abs/0810.0934}{{\tt arXiv:0810.0934}}].
	
	\bibitem{Chekanov:2010gv}
	S.~Chekanov, C.~Levy, J.~Proudfoot, and R.~Yoshida, {\it {New approach for
			jet-shape identification of $\mathrm{TeV}$-scale particles at the LHC}},
	{\em Phys.Rev.} {\bf D82} (2010) 094029,
	[\href{http://xxx.lanl.gov/abs/1009.2749}{{\tt arXiv:1009.2749}}].
	
	\bibitem{Almeida:2010pa}
	L.~G. Almeida, S.~J. Lee, G.~Perez, G.~Sterman, and I.~Sung, {\it {Template
			Overlap Method for Massive Jets}},  {\em Phys.Rev.} {\bf D82} (2010) 054034,
	[\href{http://xxx.lanl.gov/abs/1006.2035}{{\tt arXiv:1006.2035}}].
	
	\bibitem{Jankowiak:2011qa}
	M.~Jankowiak and A.~J. Larkoski, {\it {Jet Substructure Without Trees}},  {\em
		JHEP} {\bf 1106} (2011) 057, [\href{http://xxx.lanl.gov/abs/1104.1646}{{\tt
			arXiv:1104.1646}}].
	
	\bibitem{Gallicchio:2011xq}
	J.~Gallicchio and M.~D. Schwartz, {\it {Quark and Gluon Tagging at the LHC}},
	{\em Phys.Rev.Lett.} {\bf 107} (2011) 172001,
	[\href{http://xxx.lanl.gov/abs/1106.3076}{{\tt arXiv:1106.3076}}].
	
	\bibitem{Rakow:1981qn}
	P.~E. Rakow and B.~Webber, {\it {Transverse momentum moments of hadron
			distributions in qcd jets}},  {\em Nucl.Phys.} {\bf B191} (1981) 63.
	
	\bibitem{Dokshitzer:1998kz}
	Y.~L. Dokshitzer, A.~Lucenti, G.~Marchesini, and G.~Salam, {\it {On the QCD
			analysis of jet broadening}},  {\em JHEP} {\bf 9801} (1998) 011,
	[\href{http://xxx.lanl.gov/abs/hep-ph/9801324}{{\tt hep-ph/9801324}}].
	
	\bibitem{Berger:2003iw}
	C.~F. Berger, T.~Kucs, and G.~F. Sterman, {\it {Event shape / energy flow
			correlations}},  {\em Phys.Rev.} {\bf D68} (2003) 014012,
	[\href{http://xxx.lanl.gov/abs/hep-ph/0303051}{{\tt hep-ph/0303051}}].
	
	\bibitem{Dasgupta:2002bw}
	M.~Dasgupta and G.~P. Salam, {\it {Accounting for coherence in interjet E$_t$
			flow: A case study}},  {\em JHEP} {\bf 03} (2002) 017,
	[\href{http://xxx.lanl.gov/abs/hep-ph/0203009}{{\tt hep-ph/0203009}}].
	
	\bibitem{Dokshitzer:2003uw}
	Y.~Dokshitzer and G.~Marchesini, {\it {On large angle multiple gluon
			radiation}},  {\em JHEP} {\bf 0303} (2003) 040,
	[\href{http://xxx.lanl.gov/abs/hep-ph/0303101}{{\tt hep-ph/0303101}}].
	
	\bibitem{Kelley:2012kj}
	R.~Kelley, J.~R. Walsh, and S.~Zuberi, {\it {Abelian Non-Global Logarithms from
			Soft Gluon Clustering}},  \href{http://xxx.lanl.gov/abs/1202.2361}{{\tt
			arXiv:1202.2361}}.
	
	\bibitem{Catani:1996vz}
	S.~Catani and M.~Seymour, {\it {A General algorithm for calculating jet
			cross-sections in NLO QCD}},  {\em Nucl.Phys.} {\bf B485} (1997) 291--419,
	[\href{http://xxx.lanl.gov/abs/hep-ph/9605323}{{\tt hep-ph/9605323}}].
	
	\bibitem{GehrmannDeRidder:2007jk}
	A.~Gehrmann-De~Ridder, T.~Gehrmann, E.~Glover, and G.~Heinrich, {\it {Infrared
			structure of $e^+ e^-$ $\rightarrow$ 3 jets at NNLO}},  {\em JHEP} {\bf 0711}
	(2007) 058, [\href{http://xxx.lanl.gov/abs/0710.0346}{{\tt
			arXiv:0710.0346}}]. 62 pages, LaTeX, 11 Figures, additional information on
	the precise nature of the infrared cancellations can be found in the paper's
	source.
	
	\bibitem{GehrmannDeRidder:2008ug}
	A.~Gehrmann-De~Ridder, T.~Gehrmann, E.~Glover, and G.~Heinrich, {\it {Jet rates
			in electron-positron annihilation at $\mathcal{O} (\alpha_s^3)$ in QCD}},
	{\em Phys.Rev.Lett.} {\bf 100} (2008) 172001,
	[\href{http://xxx.lanl.gov/abs/0802.0813}{{\tt arXiv:0802.0813}}].
	
	\bibitem{Campbell:2002tg}
	J.~M. Campbell and R.~K. Ellis, {\it {Next-to-leading order corrections to $W +
			2$ jet and $Z + 2$ jet production at hadron colliders}},  {\em Phys.Rev.}
	{\bf D65} (2002) 113007, [\href{http://xxx.lanl.gov/abs/hep-ph/0202176}{{\tt
			hep-ph/0202176}}].
	
	\bibitem{Dokshitzer:2008ia}
	Y.~Dokshitzer and G.~Marchesini, {\it {Monte Carlo and large angle gluon
			radiation}},  {\em JHEP} {\bf 0903} (2009) 117,
	[\href{http://xxx.lanl.gov/abs/0809.1749}{{\tt arXiv:0809.1749}}].
	
	\bibitem{Amati:1980ch}
	D.~Amati, A.~Bassetto, M.~Ciafaloni, G.~Marchesini, and G.~Veneziano, {\it {A
			Treatment of Hard Processes Sensitive to the Infrared Structure of QCD}},
	{\em Nucl.Phys.} {\bf B173} (1980) 429.
	
	\bibitem{Lindelof}
	E.~Lindelöf and R.~H. Mellin {\em Adv. Math.} {\bf 61} (1993) i--vi.
	
	\bibitem{Catani:1990rr}
	S.~Catani, B.~Webber, and G.~Marchesini, {\it {QCD coherent branching and
			semiinclusive processes at large x}},  {\em Nucl.Phys.} {\bf B349} (1991)
	635--654.
	
	\bibitem{Korchemsky:1987wg}
	G.~Korchemsky and A.~Radyushkin, {\it {Renormalization of the Wilson Loops
			Beyond the Leading Order}},  {\em Nucl.Phys.} {\bf B283} (1987) 342--364.
	
	\bibitem{Korchemskaya:1992je}
	I.~Korchemskaya and G.~Korchemsky, {\it {On lightlike Wilson loops}},  {\em
		Phys.Lett.} {\bf B287} (1992) 169--175.
	
	\bibitem{Korchemsky:1993uz}
	G.~Korchemsky and G.~Marchesini, {\it {Resummation of large infrared
			corrections using Wilson loops}},  {\em Phys.Lett.} {\bf B313} (1993)
	433--440.
	
	\bibitem{Kodaira:1981nh}
	J.~Kodaira and L.~Trentadue, {\it {Summing Soft Emission in QCD}},  {\em
		Phys.Lett.} {\bf B112} (1982) 66.
	
	\bibitem{Davies:1984sp}
	C.~Davies, B.~Webber, and W.~J. Stirling, {\it {Drell-Yan Cross-Sections at
			Small Transverse Momentum}},  {\em Nucl.Phys.} {\bf B256} (1985) 413.
	
	\bibitem{Catani:1988vd}
	S.~Catani, E.~D'Emilio, and L.~Trentadue, {\it {The gluon form-factor to higher
			orders: gluon gluon annihilation at small q-transverse}},  {\em Phys.Lett.}
	{\bf B211} (1988) 335--342.
	
	\bibitem{Catani:1989ne}
	S.~Catani and L.~Trentadue, {\it {Resummation of the QCD Perturbative Series
			for Hard Processes}},  {\em Nucl.Phys.} {\bf B327} (1989) 323.
	
	\bibitem{Monni:2011gb}
	P.~F. Monni, T.~Gehrmann, and G.~Luisoni, {\it {Two-Loop Soft Corrections and
			Resummation of the Thrust Distribution in the Dijet Region}},  {\em JHEP}
	{\bf 08} (2011) 010, [\href{http://xxx.lanl.gov/abs/1105.4560}{{\tt
			arXiv:1105.4560}}].
	
	\bibitem{Webber:1994cp}
	B.~Webber, {\it {Estimation of power corrections to hadronic event shapes}},
	{\em Phys.Lett.} {\bf B339} (1994) 148--150,
	[\href{http://xxx.lanl.gov/abs/hep-ph/9408222}{{\tt hep-ph/9408222}}].
	
	\bibitem{Beneke:1995pq}
	M.~Beneke and V.~M. Braun, {\it {Power corrections and renormalons in Drell-Yan
			production}},  {\em Nucl.Phys.} {\bf B454} (1995) 253--290,
	[\href{http://xxx.lanl.gov/abs/hep-ph/9506452}{{\tt hep-ph/9506452}}].
	
	\bibitem{Akhoury:1995sp}
	R.~Akhoury and V.~I. Zakharov, {\it {On the universality of the leading, 1/Q
			power corrections in QCD}},  {\em Phys.Lett.} {\bf B357} (1995) 646--652,
	[\href{http://xxx.lanl.gov/abs/hep-ph/9504248}{{\tt hep-ph/9504248}}].
	
	\bibitem{Korchemsky:1994is}
	G.~P. Korchemsky and G.~F. Sterman, {\it {Nonperturbative corrections in
			resummed cross-sections}},  {\em Nucl.Phys.} {\bf B437} (1995) 415--432,
	[\href{http://xxx.lanl.gov/abs/hep-ph/9411211}{{\tt hep-ph/9411211}}].
	
	\bibitem{Korchemsky:1997sy}
	G.~P. Korchemsky, G.~Oderda, and G.~F. Sterman, {\it {Power corrections and
			non-local operators}},  \href{http://xxx.lanl.gov/abs/hep-ph/9708346}{{\tt
			hep-ph/9708346}}.
	
	\bibitem{Korchemsky:1999kt}
	G.~P. Korchemsky and G.~F. Sterman, {\it {Power corrections to event shapes and
			factorization}},  {\em Nucl.Phys.} {\bf B555} (1999) 335--351,
	[\href{http://xxx.lanl.gov/abs/hep-ph/9902341}{{\tt hep-ph/9902341}}].
	
	\bibitem{Gardi:2001ny}
	E.~Gardi and J.~Rathsman, {\it {Renormalon resummation and exponentiation of
			soft and collinear gluon radiation in the thrust distribution}},  {\em
		Nucl.Phys.} {\bf B609} (2001) 123--182,
	[\href{http://xxx.lanl.gov/abs/hep-ph/0103217}{{\tt hep-ph/0103217}}].
	
	\bibitem{Salam:2001bd}
	G.~Salam and D.~Wicke, {\it {Hadron masses and power corrections to event
			shapes}},  {\em JHEP} {\bf 0105} (2001) 061,
	[\href{http://xxx.lanl.gov/abs/hep-ph/0102343}{{\tt hep-ph/0102343}}].
	
	\bibitem{Marchesini:1987cf}
	G.~Marchesini and B.~Webber, {\it {Monte Carlo Simulation of General Hard
			Processes with Coherent QCD Radiation}},  {\em Nucl.Phys.} {\bf B310} (1988)
	461.
	
	\bibitem{Bahr:2008pv}
	M.~Bahr, S.~Gieseke, M.~Gigg, D.~Grellscheid, K.~Hamilton, et~al., {\it
		{Herwig++ Physics and Manual}},  {\em Eur.Phys.J.} {\bf C58} (2008) 639--707,
	[\href{http://xxx.lanl.gov/abs/0803.0883}{{\tt arXiv:0803.0883}}]. 143 pages,
	program and additional information available from
	http://projects.hepforge.org/herwig.
	
	\bibitem{Sjostrand:2006za}
	T.~Sjostrand, S.~Mrenna, and P.~Z. Skands, {\it {PYTHIA 6.4 Physics and
			Manual}},  {\em JHEP} {\bf 0605} (2006) 026,
	[\href{http://xxx.lanl.gov/abs/hep-ph/0603175}{{\tt hep-ph/0603175}}].
	
	\bibitem{Sjostrand:2007gs}
	T.~Sjostrand, S.~Mrenna, and P.~Z. Skands, {\it {A Brief Introduction to PYTHIA
			8.1}},  {\em Comput.Phys.Commun.} {\bf 178} (2008) 852--867,
	[\href{http://xxx.lanl.gov/abs/0710.3820}{{\tt arXiv:0710.3820}}].
	
	\bibitem{Gleisberg:2003xi}
	T.~Gleisberg, S.~Hoeche, F.~Krauss, A.~Schalicke, S.~Schumann, et~al., {\it
		{SHERPA 1.$\alpha$: A Proof of concept version}},  {\em JHEP} {\bf 0402}
	(2004) 056, [\href{http://xxx.lanl.gov/abs/hep-ph/0311263}{{\tt
			hep-ph/0311263}}].
	
	\bibitem{Gleisberg:2008ta}
	T.~Gleisberg, S.~Hoeche, F.~Krauss, M.~Schonherr, S.~Schumann, et~al., {\it
		{Event generation with SHERPA 1.1}},  {\em JHEP} {\bf 0902} (2009) 007,
	[\href{http://xxx.lanl.gov/abs/0811.4622}{{\tt arXiv:0811.4622}}].
	
	\bibitem{Seymour:1993mx}
	M.~H. Seymour, {\it {Searches for new particles using cone and cluster jet
			algorithms: A Comparative study}},  {\em Z.Phys.} {\bf C62} (1994) 127--138.
	
	\bibitem{Ellis:1993tq}
	S.~D. Ellis and D.~E. Soper, {\it {Successive combination jet algorithm for
			hadron collisions}},  {\em Phys. Rev.} {\bf D48} (1993) 3160--3166,
	[\href{http://xxx.lanl.gov/abs/hep-ph/9305266}{{\tt hep-ph/9305266}}].
	
	\bibitem{Skiba:2007fw}
	W.~Skiba and D.~Tucker-Smith, {\it {Using jet mass to discover vector quarks at
			the LHC}},  {\em Phys.Rev.} {\bf D75} (2007) 115010,
	[\href{http://xxx.lanl.gov/abs/hep-ph/0701247}{{\tt hep-ph/0701247}}].
	
	\bibitem{Holdom:2007nw}
	B.~Holdom, {\it {t-prime at the LHC: The Physics of discovery}},  {\em JHEP}
	{\bf 0703} (2007) 063, [\href{http://xxx.lanl.gov/abs/hep-ph/0702037}{{\tt
			hep-ph/0702037}}].
	
	\bibitem{Krohn:2009wm}
	D.~Krohn, J.~Shelton, and L.-T. Wang, {\it {Measuring the Polarization of
			Boosted Hadronic Tops}},  {\em JHEP} {\bf 1007} (2010) 041,
	[\href{http://xxx.lanl.gov/abs/0909.3855}{{\tt arXiv:0909.3855}}].
	
	\bibitem{Plehn:2009rk}
	T.~Plehn, G.~P. Salam, and M.~Spannowsky, {\it {Fat Jets for a Light Higgs}},
	{\em Phys.Rev.Lett.} {\bf 104} (2010) 111801,
	[\href{http://xxx.lanl.gov/abs/0910.5472}{{\tt arXiv:0910.5472}}].
	
	\bibitem{Butterworth:2009qa}
	J.~M. Butterworth, J.~R. Ellis, A.~R. Raklev, and G.~P. Salam, {\it
		{Discovering baryon-number violating neutralino decays at the LHC}},  {\em
		Phys.Rev.Lett.} {\bf 103} (2009) 241803,
	[\href{http://xxx.lanl.gov/abs/0906.0728}{{\tt arXiv:0906.0728}}].
	
	\bibitem{Baur:2008uv}
	U.~Baur and L.~Orr, {\it {Searching for $t \bar{t}$ Resonances at the Large
			Hadron Collider}},  {\em Phys.Rev.} {\bf D77} (2008) 114001,
	[\href{http://xxx.lanl.gov/abs/0803.1160}{{\tt arXiv:0803.1160}}].
	
	\bibitem{FileviezPerez:2008ib}
	P.~Fileviez~Perez, R.~Gavin, T.~McElmurry, and F.~Petriello, {\it {Grand
			Unification and Light Color-Octet Scalars at the LHC}},  {\em Phys.Rev.} {\bf
		D78} (2008) 115017, [\href{http://xxx.lanl.gov/abs/0809.2106}{{\tt
			arXiv:0809.2106}}].
	
	\bibitem{Bai:2008sk}
	Y.~Bai and Z.~Han, {\it {Top-antitop and Top-top Resonances in the Dilepton
			Channel at the CERN LHC}},  {\em JHEP} {\bf 0904} (2009) 056,
	[\href{http://xxx.lanl.gov/abs/0809.4487}{{\tt arXiv:0809.4487}}].
	
	\bibitem{Brooijmans:2009xa}
	G.~Brooijmans, {\it {After the Standard Model: New Resonances at the LHC}},
	{\em Mod.Phys.Lett.} {\bf A24} (2009) 1--15,
	[\href{http://xxx.lanl.gov/abs/0901.3911}{{\tt arXiv:0901.3911}}].
	
	\bibitem{QCD_coherence}
	B.~I. Ermolaev and V.~S. Fadin, {\it {Log - Log Asymptotic Form of Exclusive
			Cross-Sections in Quantum Chromodynamics}},  {\em JETP Lett.} {\bf 33} (1981)
	269--272. [Pisma Zh.Eksp.Teor.Fiz.33:285-288,1981].
	
	\bibitem{Dokshitzer1992675}
	Y.~Dokshitzer, G.~Marchesini, and G.~Oriani, {\it Measuring colour flows in
		hard processes: beyond leading order},  {\em Nuclear Physics B} {\bf 387}
	(1992), no.~3 675 -- 714.
	
	\bibitem{deFlorian:2007fv}
	D.~de~Florian and W.~Vogelsang, {\it {Resummed cross-section for jet production
			at hadron colliders}},  {\em Phys.Rev.} {\bf D76} (2007) 074031,
	[\href{http://xxx.lanl.gov/abs/0704.1677}{{\tt arXiv:0704.1677}}].
	
	\bibitem{Kelley:2010qs}
	R.~Kelley and M.~D. Schwartz, {\it {Threshold Hadronic Event Shapes with
			Effective Field Theory}},  {\em Phys. Rev.} {\bf D83} (2011) 033001,
	[\href{http://xxx.lanl.gov/abs/1008.4355}{{\tt arXiv:1008.4355}}].
	
	\bibitem{Oderda:1998en}
	G.~Oderda and G.~F. Sterman, {\it {Energy and color flow in dijet rapidity
			gaps}},  {\em Phys. Rev. Lett.} {\bf 81} (1998) 3591--3594,
	[\href{http://xxx.lanl.gov/abs/hep-ph/9806530}{{\tt hep-ph/9806530}}].
	
	\bibitem{olver2010nist}
	F.~Olver, D.~Lozier, R.~Boisvert, and C.~Clark, {\em {NIST Handbook of
			Mathematical Functions}}.
	\newblock Cambridge University Press, 2010.
	
	\bibitem{Hornig:2011tg}
	A.~Hornig, C.~Lee, J.~R. Walsh, and S.~Zuberi, {\it {Double Non-Global
			Logarithms In-N-Out of Jets}},  \href{http://xxx.lanl.gov/abs/1110.0004}{{\tt
			arXiv:1110.0004}}.
	
	\bibitem{Kelley:2011ng}
	R.~Kelley, M.~D. Schwartz, R.~M. Schabinger, and H.~X. Zhu, {\it {The two-loop
			hemisphere soft function}},  {\em Phys. Rev.} {\bf D84} (2011) 045022,
	[\href{http://xxx.lanl.gov/abs/1105.3676}{{\tt arXiv:1105.3676}}].
	
	\bibitem{Collins:1984kg}
	J.~C. Collins, D.~E. Soper, and G.~F. Sterman, {\it {Transverse Momentum
			Distribution in Drell-Yan Pair and W and Z Boson Production}},  {\em Nucl.
		Phys.} {\bf B250} (1985) 199.
	
	\bibitem{Bonciani:2003nt}
	R.~Bonciani, S.~Catani, M.~L. Mangano, and P.~Nason, {\it {Sudakov resummation
			of multiparton QCD cross sections}},  {\em Phys. Lett.} {\bf B575} (2003)
	268--278, [\href{http://xxx.lanl.gov/abs/hep-ph/0307035}{{\tt
			hep-ph/0307035}}].
	
	\bibitem{Hornig:2011iu}
	A.~Hornig, C.~Lee, I.~W. Stewart, J.~R. Walsh, and S.~Zuberi, {\it {Non-global
			Structure of the $\mathcal{O} (\alpha_s^2)$ Dijet Soft Function}},  {\em
		JHEP} {\bf 08} (2011) 054, [\href{http://xxx.lanl.gov/abs/1105.4628}{{\tt
			arXiv:1105.4628}}].
	
	\bibitem{Kelley:2012zs}
	R.~Kelley, J.~R. Walsh, and S.~Zuberi, {\it {Disentangling Clustering Effects
			in Jet Algorithms}},  \href{http://xxx.lanl.gov/abs/1203.2923}{{\tt
			arXiv:1203.2923}}.
	
	\bibitem{Bauer:2000ew}
	C.~W. Bauer, S.~Fleming, and M.~E. Luke, {\it {Summing Sudakov logarithms in B
			$\rightarrow$ X$_s$ + $\gamma$ in effective field theory}},  {\em Phys.Rev.}
	{\bf D63} (2000) 014006, [\href{http://xxx.lanl.gov/abs/hep-ph/0005275}{{\tt
			hep-ph/0005275}}].
	
	\bibitem{Nagy:2003tz}
	Z.~Nagy, {\it {Next-to-leading order calculation of three jet observables in
			hadron hadron collision}},  {\em Phys. Rev.} {\bf D68} (2003) 094002,
	[\href{http://xxx.lanl.gov/abs/hep-ph/0307268}{{\tt hep-ph/0307268}}].
	
	\bibitem{Frixione:2007vw}
	S.~Frixione, P.~Nason, and C.~Oleari, {\it {Matching NLO QCD computations with
			Parton Shower simulations: the POWHEG method}},  {\em JHEP} {\bf 0711} (2007)
	070, [\href{http://xxx.lanl.gov/abs/0709.2092}{{\tt arXiv:0709.2092}}].
	
	\bibitem{Frixione:2002ik}
	S.~Frixione and B.~R. Webber, {\it {Matching NLO QCD computations and parton
			shower simulations}},  {\em JHEP} {\bf 0206} (2002) 029,
	[\href{http://xxx.lanl.gov/abs/hep-ph/0204244}{{\tt hep-ph/0204244}}].
	
	\bibitem{Corcella:2000bw}
	G.~Corcella, I.~Knowles, G.~Marchesini, S.~Moretti, K.~Odagiri, et~al., {\it
		{HERWIG 6: An Event generator for hadron emission reactions with interfering
			gluons (including supersymmetric processes)}},  {\em JHEP} {\bf 0101} (2001)
	010, [\href{http://xxx.lanl.gov/abs/hep-ph/0011363}{{\tt hep-ph/0011363}}].
	
	\bibitem{Aad:2011kq}
	{\bf ATLAS Collaboration} Collaboration, G.~Aad et~al., {\it {Study of Jet
			Shapes in Inclusive Jet Production in pp Collisions at $\sqrt{s} =
			7\,\mathrm{TeV}$ using the ATLAS Detector}},  {\em Phys.Rev.} {\bf D83}
	(2011) 052003, [\href{http://xxx.lanl.gov/abs/1101.0070}{{\tt
			arXiv:1101.0070}}].
	
	\bibitem{Kidonakis:1998nf}
	N.~Kidonakis, G.~Oderda, and G.~F. Sterman, {\it {Evolution of color exchange
			in {QCD} hard scattering}},  {\em Nucl. Phys.} {\bf B531} (1998) 365--402,
	[\href{http://xxx.lanl.gov/abs/hep-ph/9803241}{{\tt hep-ph/9803241}}].
	
	\bibitem{Forshaw:2009fz}
	J.~Forshaw, J.~Keates, and S.~Marzani, {\it {Jet vetoing at the LHC}},  {\em
		JHEP} {\bf 0907} (2009) 023, [\href{http://xxx.lanl.gov/abs/0905.1350}{{\tt
			arXiv:0905.1350}}].
	
	\bibitem{Catani:1999ss}
	S.~Catani and M.~Grazzini, {\it {Infrared factorization of tree level QCD
			amplitudes at the next-to-next-to-leading order and beyond}},  {\em
		Nucl.Phys.} {\bf B570} (2000) 287--325,
	[\href{http://xxx.lanl.gov/abs/hep-ph/9908523}{{\tt hep-ph/9908523}}].
	
	\bibitem{Banfi:2000si}
	A.~Banfi, G.~Marchesini, Y.~L. Dokshitzer, and G.~Zanderighi, {\it {QCD
			analysis of near-to-planar three jet events}},  {\em JHEP} {\bf 0007} (2000)
	002, [\href{http://xxx.lanl.gov/abs/hep-ph/0004027}{{\tt hep-ph/0004027}}].
	
	\bibitem{Dokshitzer:1997iz}
	Y.~L. Dokshitzer, A.~Lucenti, G.~Marchesini, and G.~P. Salam, {\it
		{Universality of 1/Q corrections to jet-shape observables rescued}},  {\em
		Nucl. Phys.} {\bf B511} (1998) 396--418,
	[\href{http://xxx.lanl.gov/abs/hep-ph/9707532}{{\tt hep-ph/9707532}}].
	
	\bibitem{Forshaw:2006fk}
	J.~R. Forshaw, A.~Kyrieleis, and M.~H. Seymour, {\it {Super-leading logarithms
			in non-global observables in QCD}},  {\em JHEP} {\bf 08} (2006) 059,
	[\href{http://xxx.lanl.gov/abs/hep-ph/0604094}{{\tt hep-ph/0604094}}].
	
	\bibitem{Forshaw:2008cq}
	J.~Forshaw, A.~Kyrieleis, and M.~Seymour, {\it {Super-leading logarithms in
			non-global observables in QCD: Colour basis independent calculation}},  {\em
		JHEP} {\bf 0809} (2008) 128, [\href{http://xxx.lanl.gov/abs/0808.1269}{{\tt
			arXiv:0808.1269}}].
	
	\bibitem{DuranDelgado:2011tp}
	R.~M. Duran~Delgado, J.~R. Forshaw, S.~Marzani, and M.~H. Seymour, {\it {The
			dijet cross section with a jet veto}},  {\em JHEP} {\bf 1108} (2011) 157,
	[\href{http://xxx.lanl.gov/abs/1107.2084}{{\tt arXiv:1107.2084}}].
	
	\bibitem{Gleisberg:2008fv}
	T.~Gleisberg and S.~Hoeche, {\it {Comix, a new matrix element generator}},
	{\em JHEP} {\bf 0812} (2008) 039,
	[\href{http://xxx.lanl.gov/abs/0808.3674}{{\tt arXiv:0808.3674}}].
	
	\bibitem{Pumplin:2002vw}
	J.~Pumplin, D.~Stump, J.~Huston, H.~Lai, P.~M. Nadolsky, et~al., {\it {New
			generation of parton distributions with uncertainties from global QCD
			analysis}},  {\em JHEP} {\bf 0207} (2002) 012,
	[\href{http://xxx.lanl.gov/abs/hep-ph/0201195}{{\tt hep-ph/0201195}}].
	
	\bibitem{Rubin:2010xp}
	M.~Rubin, G.~P. Salam, and S.~Sapeta, {\it {Giant QCD K-factors beyond NLO}},
	{\em JHEP} {\bf 1009} (2010) 084,
	[\href{http://xxx.lanl.gov/abs/1006.2144}{{\tt arXiv:1006.2144}}].
	
	\bibitem{Gieseke:2011na}
	S.~Gieseke, D.~Grellscheid, K.~Hamilton, A.~Papaefstathiou, S.~Platzer, et~al.,
	{\it {Herwig++ 2.5 Release Note}},
	\href{http://xxx.lanl.gov/abs/1102.1672}{{\tt arXiv:1102.1672}}.
	
	\bibitem{dokshitzer1997}
	Y.~L. Dokshitzer, {\em Lecture notes on QCD for beginners}.
	\newblock Unpublished, 1997.
	
	\bibitem{Ross:HelicityMethod}
	D.~Ross, {\it {Lecture notes on Modern Methods in Perturbative QCD}},
	http://www.hep.phys.soton.ac.uk/hepwww/staff/D.Ross/.
	
	\bibitem{1959lectures}
	W.~E. Brittin and B.~W. Downs, {\em {Lectures in theoretical physics}}.
	\newblock {Summer Institute for Theoretical Physics (University of Colorado
		(Boulder Campus))}. {Interscience}, {1960}.
	
	\bibitem{PhysRev.103.443}
	L.~I. Schiff, {\it Approximation method for high-energy potential scattering},
	{\em Phys. Rev.} {\bf 103} (Jul, 1956) 443--453.
	
	\bibitem{GehrmannDeRidder:2007ce}
	A.~Gehrmann-De~Ridder, T.~Gehrmann, E.~Glover, and G.~Heinrich, {\it {First
			results on $e^+ e^- \rightarrow$ 3 jets at NNLO}},  {\em eConf} {\bf
		C0705302} (2007) TOP10, [\href{http://xxx.lanl.gov/abs/0709.1608}{{\tt
			arXiv:0709.1608}}].
	
	\bibitem{GehrmannDeRidder:2007hr}
	A.~Gehrmann-De~Ridder, T.~Gehrmann, E.~W.~N. Glover, and G.~Heinrich, {\it
		{NNLO corrections to event shapes in $e^+ e^-$ annihilation}},  {\em JHEP}
	{\bf 12} (2007) 094, [\href{http://xxx.lanl.gov/abs/0711.4711}{{\tt
			arXiv:0711.4711}}].
	
	\bibitem{n-bodyPS}
	H.~Murayama, {\it Lecture notes on phase space},
	\url{http://hitoshi.berkeley.edu/233B}.
	
	\bibitem{Basham:1977iq}
	C.~L. Basham, L.~S. Brown, S.~Ellis, and S.~Love, {\it {Electron - Positron
			Annihilation Energy Pattern in Quantum Chromodynamics: Asymptotically Free
			Perturbation Theory}},  {\em Phys.Rev.} {\bf D17} (1978) 2298.
	
	\bibitem{Basham:1978bw}
	C.~L. Basham, L.~S. Brown, S.~D. Ellis, and S.~T. Love, {\it {Energy
			Correlations in electron - Positron Annihilation: Testing QCD}},  {\em
		Phys.Rev.Lett.} {\bf 41} (1978) 1585.
	
	\bibitem{'tHooft:1972fi}
	G.~'t~Hooft and M.~Veltman, {\it {Regularization and Renormalization of Gauge
			Fields}},  {\em Nucl.Phys.} {\bf B44} (1972) 189--213.
	
	\bibitem{field1995applications}
	R.~Field and D.~Pines, {\em Applications Of Perturbative Qcd}.
	\newblock Frontiers in physics. Addison-Wesley, 1995.
	
	\bibitem{Bloch193754}
	N.~A. Bloch, F., {\it Note on the radiation field of the electron},  {\em
		Physical Review} {\bf 52} (1937), no.~2 54--59. cited By (since 1996) 224.
	
	\bibitem{ahlfors1966complex}
	L.~Ahlfors, {\em Complex analysis: an introduction to the theory of analytic
		functions of one complex variable}.
	\newblock International series in pure and applied mathematics. McGraw-Hill,
	1966.
	
	\bibitem{Catani:2011qz}
	S.~Catani, L.~Cieri, D.~de~Florian, G.~Ferrera, and M.~Grazzini, {\it {Diphoton
			production at hadron colliders: a fully-differential QCD calculation at
			NNLO}},  {\em Phys.Rev.Lett.} {\bf 108} (2012) 072001,
	[\href{http://xxx.lanl.gov/abs/1110.2375}{{\tt arXiv:1110.2375}}].
	
\end{thebibliography}
